\documentclass[11pt]{article}

\usepackage{epsfig}

\newcommand{\half}{ {\textstyle\frac{1}{2}} }

\newcommand{\lrD}{\raisebox{0.09em}{$
\stackrel{\raisebox{-0.03em}{$\scriptstyle\leftrightarrow$}}{D}$}{}}
\newcommand{\lD}{\raisebox{0.09em}{$
\stackrel{\raisebox{-0.03em}{$\scriptstyle\leftarrow$}}{D}$}{}}
\newcommand{\rD}{\raisebox{0.09em}{$
\stackrel{\raisebox{-0.03em}{$\scriptstyle\rightarrow$}}{D}$}{}}

\newcommand{\lrpartial}{\raisebox{0.1em}{$
\stackrel{\raisebox{-0.03em}{$\scriptstyle\leftrightarrow$}}{\partial}$}{}}
\newcommand{\lpartial}{\raisebox{0.1em}{$
\stackrel{\raisebox{-0.03em}{$\scriptstyle\leftarrow$}}{\partial}$}{}}
\newcommand{\rpartial}{\raisebox{0.1em}{$
\stackrel{\raisebox{-0.03em}{$\scriptstyle\rightarrow$}}{\partial}$}{}}

\newcommand{\lsim}{\raisebox{-4pt}{$
\,\stackrel{\textstyle <}{\sim}\,$}}
\newcommand{\gsim}{\raisebox{-4pt}{$
\,\stackrel{\textstyle >}{\sim}\,$}}

\newcommand{\re}{\mathrm{Re}\,}
\newcommand{\im}{\mathrm{Im}\,}

\renewcommand{\slash}[1]{#1 \hspace{-0.45em} / }
\newcommand{\Slash}[1]{#1 \hspace{-0.66em} / }

\newcommand{\pv}[1]{\setlength{\unitlength}{1pt}
		    \begin{picture}(0,0)
                      \put(#1,0){\line(2,1){10}}
                    \end{picture}}

\newcommand{\tvec}[1]{{\mathbf{#1}}}

\newcommand{\xB}{x_{\! B}}

\newcommand{\smallcite}[1]{\mbox{\scriptsize\protect\cite{#1}}}

\begin{document}

\date{}

\title{
\hfill \parbox{9em}{\small
	DESY-THESIS-2003-018 \\ hep-ph/0307382 \\ \\} \\[\baselineskip]
Generalized Parton Distributions}

\author{M.~Diehl \\[\baselineskip]
\textit{Deutsches Elektronen-Synchroton DESY, 22603 Hamburg, Germany}}

\maketitle

\begin{abstract}
We give an overview of the theory for generalized parton
distributions.  Topics covered are their general properties and
physical interpretation, the possibility to explore the
three-dimensional structure of hadrons at parton level, their
potential to unravel the spin structure of the nucleon, their role in
small-$x$ physics, and efforts to model their dynamics.  We review our
understanding of the reactions where generalized parton distributions
occur, to leading power accuracy and beyond, and present strategies
for phenomenological analysis.  We emphasize the close connection
between generalized parton distributions and generalized distribution
amplitudes, whose properties and physics we also present.  We finally
discuss the use of these quantities for describing soft contributions
to exclusive processes at large energy and momentum transfer.
\end{abstract}

\newpage

\tableofcontents

\newpage


\section{Introduction}

Among the main open questions in the theory of strong interactions is
to understand how the nucleon and other hadrons are built from quarks
and gluons, the fundamental degrees of freedom in QCD.  An essential
tool to investigate hadron structure is the study of deep inelastic
scattering processes, where individual quarks and gluons are resolved.
The parton densities one can extract from such processes encode the
distribution of longitudinal momentum and polarization carried by
quarks, antiquarks and gluons within a fast moving hadron.  They have
provided much to shape our physical picture of hadron structure.  Yet
important pieces of information are missed out in these quantities, in
particular how partons are distributed in the plane transverse to the
direction in which the hadron is moving, or how important their
orbital angular momentum is in making up the total spin of a nucleon.
In recent years it has become clear that appropriate exclusive
scattering processes may provide such information, encoded in
generalized parton distributions (GPDs).

GPDs have been discovered and rediscovered in different contexts, and
the processes where they appear have been considered early on.  Let us
try to give the main lines of development here, with apologies to
those whose contributions we may have overlooked.
\begin{itemize}
\item One of the principal reactions involving GPDs is 
Compton scattering with nonzero momentum transfer to the proton and at
least one photon off-shell (see Section~\ref{sec:factor} for
specification of the relevant kinematics).  The nonforward Compton
amplitude has been considered in various contexts, as well as the
physical processes where it may be measured such as $ep \to ep
\gamma$, $\gamma p\to \mu^+\mu^-\, p$, or $ep\to ep\, \mu^+\mu^-$
\cite{Altarelli:1972sw,Gatto:1972sy,DeRujula:1973aa,Wieczorek:1974zu}.
A rather detailed treatment of nonforward Compton scattering in the
operator product expansion has been given by Watanabe
\cite{Watanabe:1981ce,Watanabe:1982ue}, where quantities resembling
double distributions (see Section~\ref{sec:double-d}) appear, but this
work has not been pursued.
\item A systematic investigation of GPDs and their appearance in
virtual Compton scattering has been performed by the Leipzig group in
a series of papers
\cite{Geyer:1985vw,Braunschweig:1986nr,Dittes:1988xz} culminating in
\cite{Muller:1994fv}.  Using the nonlocal
operator product expansion, a unified framework was developed to
describe both the DGLAP evolution
\cite{Gribov:1972ri,Lipatov:1975qm,Altarelli:1977zs,Dokshitzer:1977sg}
of parton densities and the ERBL evolution
\cite{Efremov:1980qk,Lepage:1979zb} of meson distribution amplitudes.
GPDs appeared as functions whose scale evolution ``interpolates''
between these two limiting cases.
\item A different line of research concerns diffractive processes
induced by highly virtual photons.  Bartels and Loewe
\cite{Bartels:1982jh} considered $\gamma^* p\to \gamma p$ and
$\gamma^* p\to Z p$ in terms of nonforward gluon ladder diagrams,
without recourse to the concept of parton distributions.  Ryskin
\cite{Ryskin:1993ui} pointed out that $\gamma^* p\to J/\Psi\, p$
may provide a sensitive access to the gluon distribution, which
appears \emph{squared} in the cross section.  Electroproduction of
light vector mesons was later investigated by Brodsky et
al.~\cite{Brodsky:1994kf}.  Somewhat ironically, these studies did not
emphasize that it is the \emph{generalized} gluon distribution one
probes in these reactions: in the leading $\log \frac{1}{x}$
approximation there is indeed no difference between the usual gluon
distribution and the generalized one at vanishing invariant momentum
transfer $t$ to the proton (see Section~\ref{sub:beyond-collinear}).
\item The interest of a wide community in GPDs was raised in 1996, when
the nonforward nature of the parton distributions entering virtual
Compton scattering \cite{Ji:1997ek,Radyushkin:1996nd,Ji:1997nm} and
meson production \cite{Radyushkin:1996ru} was emphasized by Ji and by
Radyushkin, and when Collins et al.~\cite{Collins:1997fb} provided a
proof of factorization of meson electroproduction in diffractive and
nondiffractive kinematics.  Ji's work \cite{Ji:1997ek} pointed out the
potential of such studies to unravel the spin structure of the
nucleon: GPDs fulfill a sum rule which may provide access to the
\emph{total} angular momentum carried by partons, comprising the spin
and the orbital part (see Section~\ref{sec:spin}).  It also showed how
GPDs provide a connection between ordinary parton densities and
elastic form factors, and hence between the principal quantities which
so far have provided information on nucleon structure.  This
connection had previously been pointed out by Jain and Ralston
\cite{Jain:1993jf}.
\item The potential of GPDs to study hadron structure in three
dimensions (instead of the one-dimensional projection inherent in the
ordinary parton densities) has been fully recognized only recently,
starting with the work by Burkardt \cite{Burkardt:2000za} on the
impact parameter representation (see Section~\ref{sec:impact}).
\end{itemize}
Experimental measurement of the exclusive processes involving GPDs is
a challenge, requiring high luminosity to compensate for small cross
sections and detectors capable of ensuring the exclusivity of the
final state.  Data on electroproduction of vector mesons have been
available for some time, whereas virtual Compton scattering in the
appropriate kinematics has only been measured recently (see
Sections~\ref{sec:small-x-mesons}, \ref{sec:small-x-dvcs},
\ref{sub:dvcs-data}, \ref{sec:meson-pheno}).

A profound property of quantum field theory is crossing symmetry.
This symmetry  leads to
generalized distribution amplitudes (GDAs) as the analogs of GPDs in
the crossed channel.  GDAs describe the transition from a
quark-antiquark or gluon pair to a hadronic system such as $p\bar{p}$
or $\pi^+\pi^-$ and thus involve the physics of hadronization in a
very specific situation.  Early studies of multi-hadron distribution
amplitudes were performed by Baier and Grozin
\cite{Grozin:1983aa,Baier:1985aa}.  The close analogy between the
short-distance regimes of Compton scattering and of $\gamma^*
\gamma^*$ annihilation into two hadrons has been noted by Watanabe
\cite{Watanabe:1982ue} and by M\"uller et al.~\cite{Muller:1994fv}.
The dedicated study of GDAs, with their connections both to GPDs and
to the usual distribution amplitudes of single mesons, was initiated
by \cite{Diehl:1998dk}.  Although much physics looks quite different
in the GPD and GDA channels, there are important similarities and
links between them.  To profit from these links we will discuss the
two types of quantities together rather than treat GDAs in a separate
section.  Subsections concerning the physics of GDAs are
\ref{sec:gda}, \ref{sub:radon}, \ref{sub:gda-impact},
\ref{sec:gda-dynamics}, \ref{sub:meson-pair-prod},
\ref{sub:two-photon}, \ref{sec:gaga}, \ref{sub:meson-pair-pheno}, and
\ref{sec:large-s}.

In the following section we will set the stage with a non-technical
overview.  This will introduce several concepts and questions this
review is concerned with and anticipate different points to be
elaborated on in later sections.  In Section~\ref{sec:proper} we
present the general properties of GPDs and GDAs, and discuss different
aspects of the physics information contained in these quantities.  The
status of efforts to understand their dynamics and to model them in
practice is reviewed in Section~\ref{sec:dynamics}.

The theory of leading-twist observables where GPDs or GDAs occur is in
a quite advanced state and will be described in
Section~\ref{sec:leading-power}.  Much less complete is our grasp of
dynamics beyond the leading-twist approximation, which we attempt to
sketch in Sections~\ref{sec:beyond} and
\ref{sec:beyond-twist-two}.  The appearance of GPDs in small-$x$
physics involves many aspects specific to the high-energy regime,
which are discussed in Section~\ref{sec:small-x}.  The highly
developed techniques to extract information about GPDs from
experimental observables will be presented in Section~\ref{sec:pheno},
as well as the results of phenomenological analyses of available data.

The developments discussed so far concern scattering at small
invariant momentum transfer $t$ to the target, or the production of
hadronic systems with small invariant mass $s$.  GPDs and GDAs also
appear in the description of large-angle scattering processes at high
$s$ and $t$, where they parameterize soft dynamics akin to the Feynman
mechanism in form factors at large momentum transfer.  For large-angle
Compton scattering this was first explored in
\cite{Radyushkin:1998rt} and \cite{Diehl:1998kh}, whereas the
crossed-channel processes have only been considered recently
\cite{Diehl:2001fv,Diehl:2002yh,Freund:2002cq}.  The driving physics
questions in this regime are different from those at small $s$ or $t$,
as is the status of the theory, and both will be described in
Section~\ref{sec:large-t}.  We conclude in Section~\ref{sec:con}.

A number of dedicated reviews about GPDs have already appeared in the
literature.  The work of Ji~\cite{Ji:1998pc} presents the state of the
art in an early phase of development of the field.  Detailed
information about double distributions can be found in a contribution
by Radyushkin \cite{Radyushkin:2000uy}.  The review by Goeke, Polyakov
and Vanderhaeghen \cite{Goeke:2001tz} pays particular attention to the
dynamics of GPDs in the context of chiral symmetry, the large-$N_c$
limit, and their implementation in the chiral quark-soliton model.  It
also develops detailed models for GPDs and shows their quantitative
effects in the cross sections for exclusive processes.

It is a good sign for the liveliness of the field (although a source
of occasional despair for the reviewer) that new results have steadily
appeared while this article was being written.  With some exceptions
we have limited ourselves to work that appeared before the end of
2002.  More recent work will often be pointed out, but not be
discussed in detail.


\section{In a nutshell}
\label{sec:nutshell}

\begin{figure}[b]
\begin{center}
     \epsfxsize=0.88\textwidth
     \epsffile{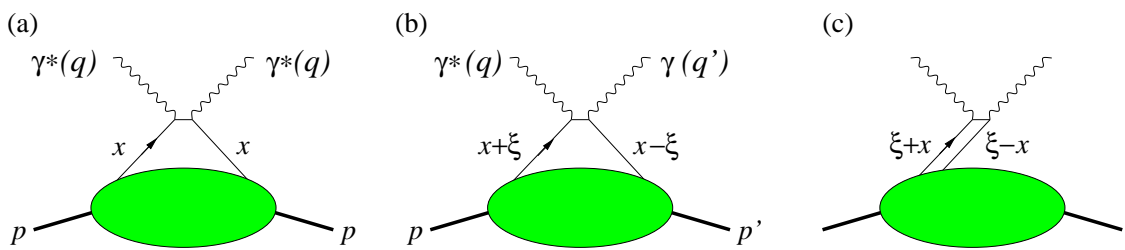}
\caption{\label{fig:handbag}(a) Handbag diagram for the forward
Compton amplitude $\gamma^*p \to \gamma^*p$, whose imaginary part
gives the DIS cross section. (b) Handbag diagram for DVCS in the
region $\xi<x<1$.  (c) The same in the region $-\xi<x<\xi$.  Momentum
fractions $x$ and $\xi$ refer to the average hadron momentum
$\frac{1}{2}(p+p')$.  A second diagram is obtained in each case by
interchanging the photon vertices.}
\end{center}
\end{figure}

A convenient starting point to introduce GPDs is the description of
inclusive deep inelastic scattering (DIS), $e p \to e X$.  In the
Bjorken limit, i.e., when the photon virtuality $Q^2 = -q^2$ and the
squared hadronic c.m.\ energy $(p+q)^2$ both become large with the
ratio $x_B = Q^2/ (2 p\cdot q)$ fixed, the dynamics factorizes into a
hard partonic subprocess, calculable in perturbation theory, and a
parton distribution, which represents the probability density for
finding a parton of specified momentum fraction $x$ in the target.
Using the optical theorem to relate the inclusive $\gamma^* p$ cross
section to the imaginary part of the forward Compton amplitude
$\gamma^* p \to \gamma^* p$, the relevant Feynman diagrams at leading
order in the strong coupling have the handbag form shown in
Fig.~\ref{fig:handbag}a.  Note that the parton densities appear
linearly in the cross section.  They can be thought of as the squared
amplitudes for the target fluctuating into the parton with momentum
fraction $x$ and any remnant system---this gives them their meaning as
probabilities in the classical sense, at least to leading logarithmic
accuracy in $Q^2$.

The simple factorization of dynamics into short- and long-distance
parts is not only valid for the forward Compton amplitude, but also
for the more general case where there is a finite momentum transfer to
the target, provided at least one of the photon virtualities is large.
A particular case is where the final photon is on shell, so that it
can appear in a physical state.  To be more precise, one has to take
the limit of large initial photon virtuality $Q^2$, with the Bjorken
variable (defined as before) and the invariant momentum transfer
$t=(p'-p)^2$ remaining fixed.  One then speaks of deeply virtual
Compton scattering (DVCS) and has again handbag diagrams as shown in
Fig.~\ref{fig:handbag}b, which can be accessed in the exclusive
process $e p \to e \gamma p$.  The long-distance part, represented by
the lower blob, is now called a generalized parton distribution (GPD).

\begin{figure}
\begin{center}
     \epsfxsize=0.6\textwidth
     \epsffile{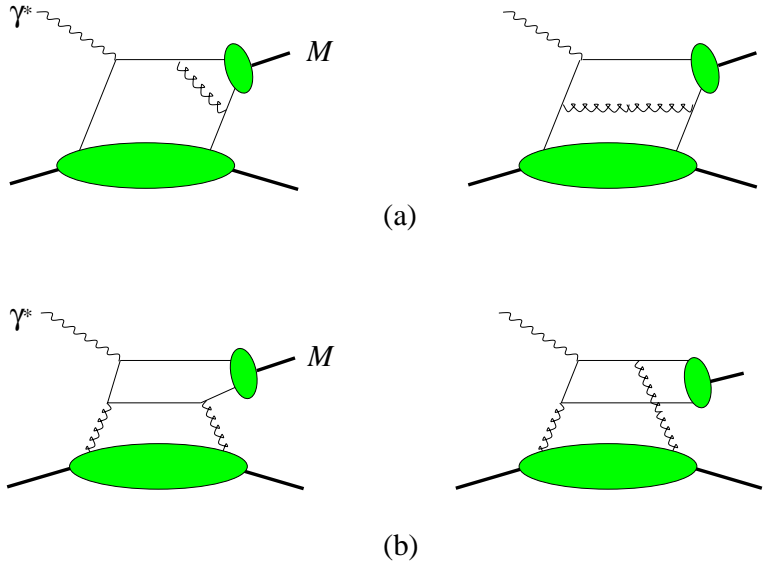}
\caption{\label{fig:mesons}  Diagrams for hard meson production
$\gamma^* p \to M p$ with (a) quark and (b) gluon GPDs.}
\end{center}
\end{figure}

An important class of other processes where GPDs occur is the
production of a light meson instead of the $\gamma$.  If the meson
quantum numbers permit, the GPDs for gluons enter at the same order in
$\alpha_s$ as those for quarks, see Fig.~\ref{fig:mesons}.  A~second
nonperturbative quantity in these processes is the meson distribution
amplitude, which describes the coupling of the meson to the $q\bar{q}$
(or gluon) pair produced in the hard scattering.

The transformation of a virtual photon into a real photon or a meson
requires a finite transfer of longitudinal momentum, where
``longitudinal'' refers to the direction of the initial proton
momentum in a frame where both $p$ and $p'$ move fast (an appropriate
frame is for instance the c.m.\ of the $\gamma^* p$ collision).  One
easily sees that the fraction of momentum lost by the proton is
determined by~$\xB$.  If momentum fractions are parameterized in the
symmetric way shown in Fig.~\ref{fig:handbag}b, one has
\begin{equation}
  \label{xi-xB}
\xi \approx \frac{\xB}{2-\xB} 
\end{equation}
in the Bjorken limit.  Proton and parton momenta now are no longer the
same on the right- and left-hand sides of the diagrams.  Therefore a
GPD no longer represents a squared amplitude (and thus a probability),
but rather the \emph{interference} between amplitudes describing
different quantum fluctuations of a nucleon.  This becomes explicit
when representing GPDs in terms of light-cone wave functions
(Section~\ref{sec:overlap}).

Apart from the longitudinal momentum, various degrees of freedom can
differ between the incoming and outgoing hadron state, each revealing
a particular aspect of hadron structure.
\begin{itemize}
\item The momentum transfer can have a transverse component (which has
to be small to fulfill the condition that $t$ should not be large).
This leads to information about the transverse structure of the
target, in addition to probing the longitudinal momentum of partons.
An intuitive physical picture is obtained in the impact impact
parameter representation (Section~\ref{sec:impact}), where GPDs
describe the spatial distribution of quarks and gluons in the plane
transverse to the momentum of a fast moving hadron.  Together with the
information on longitudinal parton momentum one thus obtains a fully
three-dimensional description of hadron structure.
\item Not only the momentum but also the polarization of the target
can be changed by the scattering, which leads to a rich spin structure
of GPDs.  We will see in Section~\ref{sec:spin} how this provides ways
to study aspects of the nucleon spin difficult to come by otherwise,
in particular the orbital angular momentum carried by partons.
\item Because of the finite momentum transfer, the handbag diagrams
admit a second kinematical regime, shown in Fig.~\ref{fig:handbag}c.
Instead of a parton being emitted and reabsorbed by the target, we
have emission of a quark-antiquark (or gluon) pair.  In this regime,
which has no counterpart in the usual parton densities, GPDs probe
$q\bar{q}$ and gluon pairs in the hadron wave function and are thus
sensitive to physics associated with the dynamics of sea quarks and of
meson degrees of freedom (see Section~\ref{sec:dynamics}).
\item Finally, the proton can be scattered inelastically into a
different baryon, or a multi-particle state.  The factorization just
discussed still holds, as long as the invariant mass of that state is
small compared with the photon virtuality $Q^2$.  In this way,
information can be obtained about the partonic structure of hadrons
not available as targets, like nucleon resonances, the $\Delta$, or
hyperons (see Section~\ref{sec:transition}).
\end{itemize}

Factorized amplitudes are written as a convolution of a GPD with a
hard scattering kernel depending on the momentum fractions $x$ and
$\xi$, with $x$ being integrated over as a loop variable.  The typical
integral occurring at lowest order in $\alpha_s$ is
\begin{equation}
  \label{Born-convolution}
\int_{-1}^1 dx\, f(x,\xi,t) \frac{1}{x - \xi + i\epsilon} 
= \pv{2}\int_{-1}^1  dx\,
  f(x,\xi,t) \frac{1}{x - \xi} - i \pi f(\xi,\xi,t) ,
\end{equation}
in both DVCS and light meson production, where $f$ stands for a
generic GPD and $\pv{0.8}\int\,$ for Cauchy's principal value
integral.  Note that whereas GPDs are real valued quantities due to
time reversal invariance, the hard scattering kernel is complex and
leads to both a real and an imaginary part of the process amplitude.

If one takes the forward limit of zero $\xi$ and $t$, the GPDs
describing equal polarization of the initial and final state hadron
reduce to the usual parton densities.  This smooth limit is to be
contrasted with the very different ways GPDs and usual parton
densities occur in physical observables---the former in exclusive, and
the latter in inclusive processes.  The forward limit also exists for
GPDs with different polarization of the incoming and outgoing hadron,
but the corresponding quantities cannot be accessed in inclusive
processes where parton densities always occur via the optical theorem.

Taking moments of GPDs in $x$ one obtains form factors of local
currents and thus a connection to the quantities which apart from
parton densities have historically been one of the main tools to
investigate hadron structure.  These relations generalize the sum
rules for ordinary parton densities, which provide form factors at
$t=0$, i.e., charges.  This connection provides constraints on GPDs
for those moments where the corresponding form factors are known from
direct measurement.  On the other hand, knowledge of GPDs can provide
access to form factors of currents that do not couple to an elementary
probe easily accessible to experiment.  Important examples are the
energy-momentum tensor (otherwise requiring graviton scattering on
hadrons) and currents involving gluon fields.

\begin{figure}
\begin{center}
     \epsfxsize=0.6\textwidth
     \epsffile{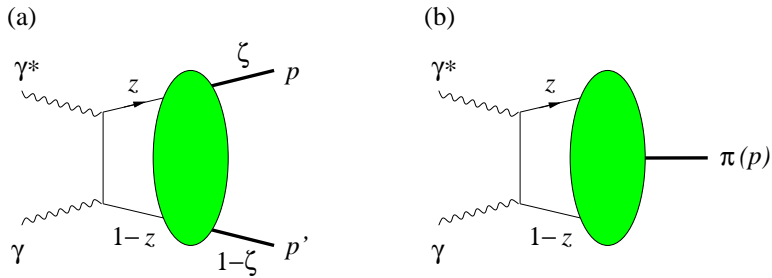}
\caption{\label{fig:gda} Handbag diagrams for the annihilation of
$\gamma^*\gamma$ into a $p\bar{p}$-pair (a) and into a single pion
(b).  Momentum fractions $z$ and $\zeta$ refer to the sum of momenta
in the final state.}
\end{center}
\end{figure}

Crossing symmetry relates Compton scattering $\gamma^* p\to \gamma^*
p$ with two-photon annihilation $\gamma^* \gamma^* \to p \bar{p}$.  In
kinematics where at least one of the photon virtualities is large, in
particular compared to the invariant mass of the hadron pair, one
finds a factorized structure analogous to the one in DVCS, with
crossed handbag diagrams as shown in Fig.~\ref{fig:gda}a.  In a
short-distance process the two photons annihilate into a
quark-antiquark (or gluon) pair, and the hadronization of this pair
into the final state is described by generalized distribution
amplitudes.  These quantities are the crossed-channel analogs of GPDs,
and although the physics they describe looks quite different, we will
see important connections between GDAs and GPDs in
Sections~\ref{sec:proper} and \ref{sec:dynamics}.  Of course one may
consider final states other than $p\bar{p}$, like pion pairs or
systems with more than two hadrons.  {}From this point of view GDAs
naturally generalize the familiar distribution amplitudes of
\emph{single} hadrons, say of a pion.  Indeed the factorization in
Fig.~\ref{fig:gda}a is completely analogous to the one for
$\gamma^*\gamma\to \pi$ in Fig.~\ref{fig:gda}b, which is among the
simplest processes where the pion distribution amplitude can be
accessed \cite{Lepage:1980fj}.

\section{Properties of GPDs and GDAs}
\label{sec:proper}

In this section we present the general properties of generalized
parton distributions and generalized distribution amplitudes, and
provide tools for their physical interpretation.  Throughout we will
limit ourselves to the distributions of twist two; a discussion of
twist three (and of twist in general) will be given in
Section~\ref{sec:twist-three}.


\subsection{Notation and conventions}
\label{sec:conventions}

The description of parton densities and related quantities is
naturally given in light-cone coordinates, 
\begin{equation}
v^\pm = \frac{1}{\sqrt{2}} (v^0 \pm v^3) ,
\qquad\qquad
\tvec{v} = (v^1, v^2)
\end{equation}
for any four-vector $v$.  To avoid proliferation of subscripts like
$\perp$ we will reserve boldface for two-dimensional transverse
vectors throughout (but write ``$k_T$'' to refer to ``transverse
momentum'' in the text).  It is often useful to work with two
light-like four-vectors $n_+ = (1,0,0,1)/ \sqrt{2}$ and $n_- =
(1,0,0,-1)/\sqrt{2}$, in terms of which the light-cone coordinates are
given by
\begin{equation}
  v^\mu  = v^+ n_+^\mu + v^- n_-^\mu + v_T^\mu
\end{equation}
with $v^+ = v n_-$, $v^- = v n_+$ and $v_T = (0,\tvec{v},0)$.  The
invariant product of two four-vectors is given as $v w = v^+ w^- + v^-
w^+ - \tvec{v} \tvec{w}$, and the invariant integration element reads
$d^4z = dz^+ dz^- d^2 \tvec{z}$.  The derivative operators along the
light-cone directions are $\partial^+ = \partial / (\partial z^-)$ and
$\partial^- = \partial / (\partial z^+)$.

The physical picture of the parton model holds in frames where hadrons
and partons move fast.  The light-cone momentum $p^+$ becomes
proportional to the momentum (or energy) of a particle in the infinite
momentum frame where $p^3 \to +\infty$, but can be used to calculate
in any convenient reference frame (including the one where a hadron is
at rest).  For simplicity we often refer to plus-momentum fractions as
``momentum fractions''.

\begin{figure}
\begin{center}
     \leavevmode
     \epsfxsize=0.65\textwidth 
     \epsffile{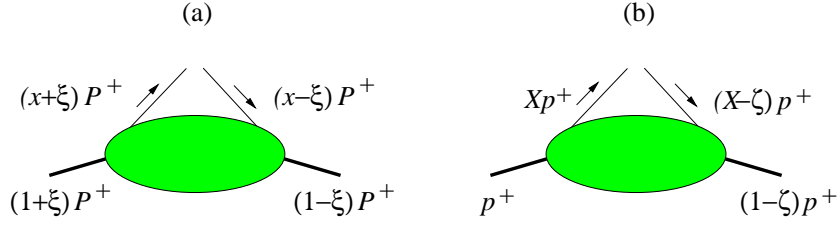}
\end{center}
\caption{\label{fig:Ji-vs-Rad} Hadron and parton plus-momenta in the
parameterization of (a): Ji \protect\cite{Ji:1998pc} and (b):
Radyushkin \protect\cite{Radyushkin:1997ki}.}
\end{figure}

For GPDs and the processes where they appear we use the notation
\begin{equation}
  \label{basic-vectors}
P = \frac{p+p'}{2} , \qquad \Delta = p'-p , \qquad t= \Delta^2 ,
\end{equation}
with $p$ for the incoming and $p'$ for the outgoing hadron momentum,
and $m$ for the target mass.  For $\gamma^* p$ collisions we use the
standard variables
\begin{equation}
Q^2 = - q^2 , \qquad W^2 = (p+q)^2 , \qquad \xB = Q^2 /(2 p\cdot q) ,
\end{equation}
where $q$ is the momentum of the incident $\gamma^*$.  At least two
parameterizations of momentum fractions for a GPD are common in the
literature, shown in Fig.~\ref{fig:Ji-vs-Rad}.  Their relation is
\begin{equation}
X = \frac{x+\xi}{1+\xi} , \qquad \zeta = \frac{2\xi}{1+\xi} .
\end{equation}
The variables $X$ and $\zeta$ are closer to the ones used in forward
kinematics, with the simple relation $\zeta \approx x_B$ for DVCS and
light meson production in the Bjorken limit.  Ji's
variables~\cite{Ji:1998pc} make symmetry properties between the
initial and final hadron more transparent and will be used here.  We
note that the definition of $\xi$ in Ji's original papers
\cite{Ji:1997ek,Ji:1997nm} differs from that in
\cite{Ji:1998pc}  and subsequent work, with $\xi_{\smallcite{Ji:1998pc}} =
\xi_{\smallcite{Ji:1997ek}} /2$.  The variable
$\xi_{\smallcite{Ji:1997ek}}$ is occasionally found in the literature,
but most work (including this review) uses
$\xi_{\smallcite{Ji:1998pc}}$ given by
\begin{equation}
\xi= \frac{p^+-p'^+}{p^+ +p'^+} .
\end{equation}
Since plus-momenta of physical states are non-negative, the physical
region of $\xi$ is the interval $[-1,1]$.  In all known processes
where GPDs may be measured one has $\xi\ge 0$, which we will tacitly
assume unless we specifically discuss symmetry properties under
$\xi\to -\xi$ or analytical continuation in $\xi$.

In the following we specify further conventions used throughout this
review.  For the momentum fractions of distribution amplitudes (see
Fig.~\ref{fig:gda}) we often use the ``bar'' notation
\begin{equation}
	\bar{z} = 1-z .
\end{equation}
The electric charges of quarks with flavor $q$ are $e_q$ in units of
the positron charge $e$.  For vectors in Hilbert space we work with
the usual relativistic normalization.  Our convention is
\begin{equation}
\epsilon_{0123} = 1
\end{equation}
for the totally antisymmetric tensor, and $\gamma_5 = i \gamma^0
\gamma^1 \gamma^2 \gamma^3$.  We also use the transverse tensors
\begin{equation}
g_T^{\alpha\beta} = g^{\alpha\beta} - n_{+}^\alpha n_{-}^\beta -
n_{-}^\alpha n_{+}^\beta ,
\qquad
\epsilon_T^{\alpha\beta} = \epsilon^{\alpha\beta\gamma\delta}\, 
                             n_{-\gamma} n_{+\delta}  ,
\end{equation}
whose only nonzero components are $g_T^{11} = g_T^{22} = -1$ and
$\epsilon_T^{12} = - \epsilon_T^{21} = 1$.

Quark fields will be denoted by $q(x)$, the gluon field by $A^\mu(x)$,
the gluon field strength by $G^{\mu\nu}(x)$, and the dual field
strength by
\begin{equation}
\tilde{G}^{\alpha\beta}(x) = \frac{1}{2}
\epsilon^{\alpha\beta\gamma\delta} G_{\gamma\delta}(x) .
\end{equation}
Finally, we use the abbreviations
\begin{equation}
\lrpartial^\mu = \half (\rpartial^\mu - \lpartial^\mu) , \qquad
\lrD^\mu = \half (\rD^\mu - \lD^\mu) ,
\end{equation}
where $D^\mu = \partial^\mu - ig A^\mu$ is the covariant derivative.


\subsection{Definition of GPDs}
\label{sec:definitions}

In full analogy to the usual parton densities, GPDs can be defined
through matrix elements of quark and gluon operators at a light-like
separation.  In this section we will restrict ourselves to the case
where the partons do not transfer helicity; parton helicity flip will
be discussed in Section~\ref{sub:helicity-flip}.

Following the conventions of \cite{Ji:1998pc} we define generalized
quarks distributions
\begin{eqnarray}
  \label{quark-gpd}
F^q &=&
\frac{1}{2} \int \frac{d z^-}{2\pi}\, e^{ix P^+ z^-}
  \langle p'|\, \bar{q}(-\half z)\, \gamma^+ q(\half z) 
  \,|p \rangle \Big|_{z^+=0,\, \tvec{z}=0}
\nonumber \\
&=& \frac{1}{2P^+} \left[
  H^q(x,\xi,t)\, \bar{u}(p') \gamma^+ u(p) +
  E^q(x,\xi,t)\, \bar{u}(p') 
                 \frac{i \sigma^{+\alpha} \Delta_\alpha}{2m} u(p)
  \, \right] ,
\nonumber \\
\tilde{F}^q &=&
\frac{1}{2} \int \frac{d z^-}{2\pi}\, e^{ix P^+ z^-}
  \langle p'|\, 
     \bar{q}(-\half z)\, \gamma^+ \gamma_5\, q(\half z)
  \,|p \rangle \Big|_{z^+=0,\, \tvec{z}=0}
\nonumber \\
&=& \frac{1}{2P^+} \left[
  \tilde{H}^q(x,\xi,t)\, \bar{u}(p') \gamma^+ \gamma_5 u(p) +
  \tilde{E}^q(x,\xi,t)\, \bar{u}(p') \frac{\gamma_5 \Delta^+}{2m} u(p)
  \, \right] ,
\end{eqnarray}
where for legibility we have not displayed the polarization dependence
of the hadron states and spinors.  Because of Lorentz invariance the
GPDs $H^q$, $E^q$, $\tilde{H}^q$, $\tilde{E}^q$ can only depend on the
kinematical variables $x$, $\xi$, and $t$.  To see this we rewrite the
definitions in a frame-independent form,
\begin{eqnarray}
  \label{quark-gpd-lorentz}
F^q &=&
\frac{1}{2} \int \frac{d \lambda}{2\pi}\, e^{ix (P z)}
  \langle p'|\, \bar{q}(-\half z)\, \slash{n}_- \, q(\half z)
  \,|p \rangle \Big|_{z = \lambda n_-}
\nonumber \\
&=& \frac{1}{2(P n_-)} \left[
  H^q(x,\xi,t)\, \bar{u}(p') \slash{n}_- u(p) +
  E^q(x,\xi,t)\, 
     \bar{u}(p') \frac{i \sigma^{\alpha\beta} 
	n_{-\alpha} \Delta_\beta}{2m} u(p) \, \right]
\end{eqnarray}
with an analogous expression for $\tilde{H}^q$ and $\tilde{E}^q$,
where $n_-$ can be any light-like vector.  The GPDs are allowed to
depend on $x$ and on Lorentz invariant products of the vectors $p$,
$p'$ and $n_-$, which one may choose as $\Delta n_-$, $P n_-$ and $t$.
Under a boost along the $z$ axis the light-cone vectors $n_-$ and
$n_+$ transform as
\begin{equation}
n_- \to \alpha\, n_- , \qquad n_+ \to \alpha^{-1}\, n_+ ,
\end{equation}
and we readily see from (\ref{quark-gpd-lorentz}) that the GPDs are
independent under such a boost.  Therefore they depend on $\Delta n_-$
and $P n_-$ only via the ratio $\xi = - (\Delta n_-) / (2 P n_-)$.  In
other words they depend only on plus-momentum \emph{fractions}, but
not on individual plus-momenta, which are rescaled under boosts.

The above definitions hold in the light-cone gauge $A^+=0$ for the
gluon field.  In other gauges a Wilson line $W[-\half z^-, \half z^-]$
along a light-like path appears between the two fields at positions
$-\half z$ and $\half z$, where
\begin{equation}
W[a,b] =  P\exp\left(ig \hspace{-0.5ex}
        \int_{b}^{a} d x^- A^+(x^- n_-) \right)
\end{equation}
and $P$ denotes ordering along the path from $a$ to $b$.  Most
formulae and statements of this review are readily generalized to this
case, with exceptions we will indicate.

\begin{figure}
\begin{center}
     \leavevmode
     \epsfxsize=0.82\textwidth
     \epsffile{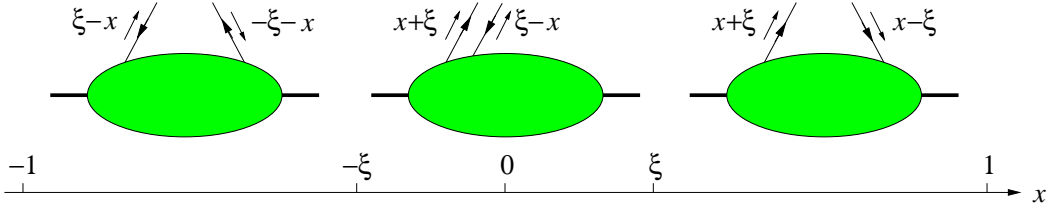}
\end{center}
\caption{\label{fig:regions} The parton interpretation of GPDs in
the three $x$-intervals $[-1,-\xi]$, $[-\xi,\xi]$, and $[\xi,1]$.}
\end{figure}

The distributions we have defined have support in the interval $x \in
[-1,1]$, which falls into the three regions shown in
Fig.~\ref{fig:regions}:
\begin{enumerate}
\item for $x \in [\xi,1]$ both momentum fractions $x+\xi$ and $x-\xi$
are positive; the distribution describes emission and reabsorption of
a quark.
\item for $x \in [-\xi,\xi]$ one has $x+\xi\ge 0$ but $x-\xi \le 0$.
The second momentum fraction is now interpreted as belonging to a an
antiquark with momentum fraction $\xi-x$ emitted from the initial
proton.
\item for $x \in [-1,-\xi]$ both $x+\xi$ and $x-\xi$ are negative;
one has emission and reabsorption of antiquarks with respective
momentum fractions $\xi-x$ and $-\xi-x$.
\end{enumerate}
The first and third case are commonly referred to as DGLAP regions and
the second as ERBL region, following the pattern of evolution in the
factorization scale (Section~\ref{sec:evolution}).  Why the support of
GPDs is restricted to $|x| \le 1$ will be discussed in
Section~\ref{sec:light-cone}.

The above interpretation can be made explicit in the framework of
light-cone quantization.  As we will see in Section
\ref{sec:light-cone} one can then decompose the field operators
$\bar{q}$ and $q$ in the definitions (\ref{quark-gpd}) in terms of
annihilation and creation operators $b, b^\dag$ for quarks and $d,
d^\dag$ for antiquarks
\cite{Ji:1998pc,Golec-Biernat:1998ja,Diehl:2000xz}.   With the
constraint that parton states must have positive plus-momentum the
above three cases then respectively select the combinations $b^\dag
b$, $d b$, and $d d^\dag$.

For gluons we define
\begin{eqnarray}
  \label{gluon-gpd}
F^g &=&
\frac{1}{P^+} \int \frac{d z^-}{2\pi}\, e^{ix P^+ z^-}
  \langle p'|\, 
     G^{+\mu}(-\half z)\, G_{\mu}{}^{+}(\half z)\, 
%
%
  \,|p \rangle \Big|_{z^+=0,\, \tvec{z}=0}
\nonumber \\
&=& \frac{1}{2P^+} \left[
  H^g(x,\xi,t)\, \bar{u}(p') \gamma^+ u(p) +
  E^g(x,\xi,t)\, \bar{u}(p') 
                 \frac{i \sigma^{+\alpha} \Delta_\alpha}{2m} u(p)
  \, \right] ,
\nonumber \\
\tilde{F}^g &=&
- \frac{i}{P^+} \int \frac{d z^-}{2\pi}\, e^{ix P^+ z^-}
  \langle p'|\, 
     G^{+\mu}(-\half z)\, \tilde{G}_{\mu}{}^{+}(\half z)\, 
  \,|p \rangle \Big|_{z^+=0,\, \tvec{z}=0}
\nonumber \\
&=& \frac{1}{2P^+} \left[
  \tilde{H}^g(x,\xi,t)\, \bar{u}(p') \gamma^+ \gamma_5 u(p) +
  \tilde{E}^g(x,\xi,t)\, \bar{u}(p') \frac{\gamma_5 \Delta^+}{2m} u(p)
  \, \right] .
\end{eqnarray}
These distributions differ by a factor of $2 x$ from those of Ji
\cite{Ji:1998pc} and by a factor of $2$ from those of Goeke
et al.~\cite{Goeke:2001tz,Polyakov:2003pr}:
\begin{equation}
2 x H_g(x) \Big|_{\smallcite{Ji:1998pc}} 
= 2 H^g(x) \Big|_{\smallcite{Goeke:2001tz}}
  = H^g(x) \Big|_{\mbox{\scriptsize here}} ,
\end{equation}
with analogous relations for $E^g$, $\tilde{H}^g$, and $\tilde{E}^g$.
Taking out a factor of $x$ from our gluon GPDs leads to a more direct
relation with the usual gluons densities in the forward limit (see
Section~\ref{sub:forward-limit}).  As remarked in
\cite{Radyushkin:1997ki} this introduces however an additional
singularity of the GPDs at $x=0$ (the point where two gluons with
equal plus-momenta are emitted), since at this point the matrix
elements in (\ref{gluon-gpd}) are in general finite but nonzero.  In
physical processes it is in fact the distributions defined in
(\ref{gluon-gpd}) that appear in the amplitude, without any factor of
$1/x$.

The number of GPDs for spin-zero hadrons, say pions or spin-zero
nuclei, is smaller.  The pion has often been considered in theoretical
investigations to avoid the complications of spin (but may also be
accessible experimentally, see Section~\ref{sub:spin-0-1}).  One
defines
\begin{eqnarray}
  \label{pion-gpd}
H_\pi^q(x,\xi,t) &=&
\frac{1}{2} \int \frac{d z^-}{2\pi}\, e^{ix P^+ z^-}
  \langle \pi^+(p')|\, \bar{q}(-\half z)\, \gamma^+ q(\half z)\, 
  \,|\pi^+(p) \rangle \Big|_{z^+=0,\, \mathbf{z}=0} \: ,
\nonumber \\
H_\pi^g(x,\xi,t) &=&
\frac{1}{P^+} \int \frac{d z^-}{2\pi}\, e^{ix P^+ z^-}
  \langle \pi^+(p')|\, G^{+\mu}(-\half z)\, G_{\mu}{}^{+}(\half z)\, 
%
%
  \,|\pi^+(p) \rangle \Big|_{z^+=0,\, \mathbf{z}=0} \: .
\end{eqnarray}
The analogs of $\tilde{H}^q$ and $\tilde{H}^g$ are zero in this case
because of parity invariance.  The case of targets with spin 1 will be
discussed in Section~\ref{sec:deuteron}.  

GPDs have also be introduced for scalar partons, in order to
facilitate the study of certain theoretical issues.  The relevant
operator is $\phi^\dag(- \half z) \phi(\half z)$ instead of $\bar{q}(-
\half z) \gamma^+ q(\half z)$, where $\phi$ is the scalar field
representing the partons, for the definitions see
\cite{Radyushkin:1997ki,Pobylitsa:2002vi}.  Alternatively the operator
$\phi^\dag(- \half z)\, i\smash{\lrpartial^+} \phi(\half z)$ has been
used, which in forward matrix elements gives the number density of the
partons \cite{Tiburzi:2002kr}.

Other parameterizations of GPDs given in the literature are
$\widetilde{\mathcal{F}}^q_\zeta(X)$ by Radyushkin
\cite{Radyushkin:1997ki}, and $f_{q/p}(x_1,x_2)$, $f_{g/p}(x_1,x_2)$
by Collins et al.~\cite{Collins:1997fb}.  They are related to the
distributions given here by normalization factors that can be found in
\cite{Radyushkin:1997ki,Ji:1998pc}, in addition to using different
parameterizations of the momentum fractions.  Golec-Biernat and Martin
\cite{Golec-Biernat:1998ja} have introduced separate distributions
$\hat\mathcal{F}_q(X,\zeta)$ and $\hat\mathcal{F}_{\bar{q}}(X,\zeta)$,
which respectively correspond to Ji's distributions in the regions
$x\in [-\xi,1]$ and $x\in [-1,\xi]$.  A different separation into
``quark'' and ``antiquark'' GPDs $\mathcal{F}^q_\zeta(X)$ and
$\mathcal{F}^{\bar{q}}_\zeta(X)$ has been introduced by Radyushkin on
the basis of double distributions and will be briefly explained in
Section~\ref{sec:double-d}.  The relation between the different
parameterizations is discussed in detail
in~\cite{Golec-Biernat:1998ja}.  We finally remark that in the course
of history generalized parton distributions have also been termed
``off-diagonal'', ``off-forward'', ``nondiagonal'', ``nonforward'', or
``skewed'' parton distributions.


\subsection{Basic properties}
\label{sec:basic}

\subsubsection{Forward limit}
\label{sub:forward-limit}

For $p=p'$ and equal helicities of the initial and final state
hadrons, the matrix elements in (\ref{quark-gpd}) and
(\ref{gluon-gpd}) reduce to the ones defining the ordinary spin
independent or spin dependent densities $q(x)$, $\Delta q(x)$ for
quarks and $g(x)$, $\Delta g(x)$ for gluons.  One thus obtains
reduction formulas
\begin{eqnarray}
  \label{forward-quark}
H^q(x,0,0) &=& \phantom{-} q(x) , \phantom{-} \qquad
\tilde{H}^q(x,0,0) \;=\; \Delta q(x) \phantom{-}
\qquad \mbox{for}~x>0, 
\nonumber \\
H^q(x,0,0) &=& -\bar{q}(-x) , \qquad
\tilde{H}^q(x,0,0) \;=\; \Delta \bar{q}(-x)
\qquad \mbox{for}~x<0
\end{eqnarray}
in the quark sector.  For gluons one has
\begin{eqnarray}
  \label{forward-gluon}
H^g(x,0,0) &=& x g(x) , \qquad
\tilde{H}^g(x,0,0) \;=\; x \Delta g(x)
\qquad \qquad \mbox{for}~x>0
\end{eqnarray}
and corresponding relations for $x<0$ (see Section \ref{sub:symm}).
The forward limits of the pion GPDs $H_\pi^q$ and $H_\pi^g$ are as in
(\ref{forward-quark}) and (\ref{forward-gluon}).

No corresponding relations exist for the quark or gluon distributions
$E$ and $\tilde{E}$ in the nucleon: in their defining equations they
are multiplied with factors proportional to $\Delta$ and therefore
decouple in the forward limit.  The information residing in these
functions at zero $\xi$ and $t$ can thus not be accessed in processes
where parton distributions appear in the cross section via the optical
theorem; it is only visible in exclusive processes with a finite
momentum transfer to the target.  We shall see in
Section~\ref{sec:spin} that the forward limit of $E^q$ and $E^g$
carries information about the orbital angular momentum of partons.

In analogy to the forward limit, the distributions $H$ and $E$ are
sometimes referred to as ``unpolarized'' and $\tilde{H}$ and
$\tilde{E}$ as ``polarized''.  Note that this refers to the spin of
the partons and not of the target.  More precisely, $H$ and $E$
correspond to the sum over parton helicities, and $\tilde{H}$ and
$\tilde{E}$ to the difference.

\subsubsection{Symmetry properties}
\label{sub:symm}

Because gluons are their own antiparticles the distributions
$H^g(x,\xi,t)$ and $E^g(x,\xi,t)$ are even functions of $x$, and
$\tilde{H}^g(x,\xi,t)$ and $\tilde{E}^g(x,\xi,t)$ are odd functions of
$x$.  Quark distributions are in general neither even nor odd in $x$,
and it is often useful to consider the combinations
\begin{eqnarray}
H^{q(+)}(x,\xi,t) &=& H^q(x,\xi,t) - H^q(-x,\xi,t) ,
\nonumber \\
\tilde{H}^{q(+)}(x,\xi,t) &=& 
                \tilde{H}^q(x,\xi,t) + \tilde{H}^q(-x,\xi,t)
  \label{C-even}
\end{eqnarray}
and their analogs for $E^q$ and $\tilde{E}^q$, which correspond to the
exchange of charge conjugation $C= +1$ in the $t$-channel.  They are
sometimes referred to as ``singlet'' combinations (even when not
summed over quark flavors).  In the forward limit one simply has
$H^{q(+)}(x,0,0) = q(x) + \bar{q}(x)$ and $\tilde{H}^{q(+)}(x,0,0) =
\Delta q(x) + \Delta\bar{q}(x)$ for $x>0$.  The ``nonsinglet'' or
``valence'' combinations
\begin{eqnarray}
H^{q(-)}(x,\xi,t) &=& H^q(x,\xi,t) + H^q(-x,\xi,t) , 
\nonumber \\
\tilde{H}^{q(-)}(x,\xi,t) &=& 
                \tilde{H}^q(x,\xi,t) - \tilde{H}^q(-x,\xi,t) .
  \label{C-odd}
\end{eqnarray}
and their counterparts for $E^q$ and $\tilde{E}^q$ correspond to $C=
-1$ exchange.  In the forward limit one has $H^{q(-)}(x,0,0) = q(x) -
\bar{q}(x)$ and $\tilde{H}^{q(-)}(x,0,0) = \Delta q(x) -
\Delta\bar{q}(x)$ for $x>0$.  The combinations (\ref{C-even}) and
(\ref{C-odd}) are typically accessible in different processes: the
transition from a $\gamma^*$ to a photon or a vector meson selects for
instance the $C$-even combinations, whereas the transition from a
$\gamma^*$ to a pseudoscalar meson selects the $C$-odd ones.
Furthermore, the combinations (\ref{C-even}) and (\ref{C-odd}) do not
mix under evolution.  For targets with definite charge conjugation
parity like the $\pi^0$ the $C$-odd combinations (\ref{C-odd}) are
zero.

Further symmetry properties follow from time reversal invariance.
Inserting $V V^{-1}$ in the matrix elements defining the GPDs, where
$V$ is the antiunitary operator implementing time reversal in Hilbert
space, one obtains
\begin{equation}
H(x,-\xi,t) = H(x,\xi,t)
  \label{time-rev}
\end{equation}
and analogous relations for $E$, $\tilde{H}$, $\tilde{E}$, both for
quark and for gluon distributions.  That time reversal changes the
sign of $\xi$ is not surprising: interchanging initial and final
states of the matrix elements in particular interchanges the momenta
$p$ and $p'$ and thus reverses the sign of $\xi = (p-p')^+ /
(p+p')^+$.  Taking the complex conjugate of the defining matrix
elements gives on the other hand
\begin{equation}
\Big[ H(x,-\xi,t) \Big]^* = H(x,\xi,t)
  \label{complex-conjug}
\end{equation}
and corresponding relations for $E$, $\tilde{H}$, $\tilde{E}$.
Together with (\ref{time-rev}) this implies that the GPDs defined here
are real valued functions.  For the pion GPDs $H^q_{\pi}$ and
$H^g_{\pi}$ one finds the same relations (\ref{time-rev}) and
(\ref{complex-conjug}).  Note that in deriving the constraints from
time reversal we have used that the hadron states in the distribution
are stable.  For unstable particles such as the $\Delta$ resonance the
situation is different (see Section~\ref{sec:transition}).  The
relation (\ref{time-rev}) is a good example for the usefulness of
parameterizing the parton momentum fractions by the the symmetric
variables $x$ and $\xi$.  In terms of Radyushkin's variables $X$ and
$\zeta$ (see Fig.~\ref{fig:Ji-vs-Rad}) the corresponding symmetry is
less explicit.

For pions isospin invariance relates the GPDs for $\pi^+$, $\pi^-$ and
$\pi^0$.  In particular, one has for the isosinglet combinations
\begin{equation}
H_{\pi^+}^{u + d} = H_{\pi^-}^{u + d} = H_{\pi^0}^{u + d}
\end{equation}
and for the isotriplet combinations
\begin{equation}
H_{\pi^+}^{u - d} = - H_{\pi^-}^{u - d} ,
\qquad
H_{\pi^0}^{u - d} = 0 ,
\end{equation}
where we have introduced the notation $H^{u \pm d} = H^u \pm H^d$.
The gluon distributions are of course isosinglet and hence equal for
the three pion states.  For simplicity we will henceforth reserve the
notation $H_\pi$ for the $\pi^+$.  {}From charge conjugation
invariance one has in addition
\begin{equation}
H_\pi^{u+d}(x,\xi,t) = - H_\pi^{u+d}(-x,\xi,t) ,
\qquad
H_\pi^{u-d}(x,\xi,t) = H_\pi^{u-d}(-x,\xi,t) ,
\end{equation}
so that the isosinglet sector corresponds to $C=+1$ and the isotriplet
sector to $C=-1$.

\subsubsection{Sum rules and polynomiality}
\label{sub:polynom}

Moments in the momentum fraction $x$ play an important role in the
theory of GPDs, as they do for the ordinary parton distributions.
Integrating the matrix elements (\ref{quark-gpd}) or (\ref{gluon-gpd})
over $x$ gives matrix elements of local quark-antiquark or gluon
operators, so that $x$-integrals of GPDs are related with the form
factors of these local currents.  In particular one has
\begin{eqnarray}
\int_{-1}^1 dx\, H^q(x,\xi,t) = F^q_1(t) , \qquad 
\int_{-1}^1 dx\, E^q(x,\xi,t) = F^q_2(t) , 
\nonumber \\
\int_{-1}^1 dx\, \tilde{H}^q(x,\xi,t) = g^q_A(t) , \qquad 
\int_{-1}^1 dx\, \tilde{E}^q(x,\xi,t) = g^q_P(t) ,
  \label{basic-sum-rules}
\end{eqnarray}
where the Dirac and Pauli form factors
\begin{equation}
  \label{vector-ff}
\langle p'|\, \bar{q}(0) \gamma^\mu q(0) \,|p \rangle
= \bar{u}(p') 
  \left[\, F_1^{q}(t)\, {\gamma}^{\mu} +
           F_2^{q}(t)\, \frac{i {\sigma}^{\mu\alpha}
                              \Delta_\alpha}{2m} \, \right] u(p) ,
\end{equation}
and the axial and pseudoscalar ones,
\begin{equation}
  \label{axial-ff}
\langle p'|\, \bar{q}(0) \gamma^\mu \gamma_5 q(0) \,|p \rangle
= \bar{u}(p') 
  \left[\, g_A^{q}(t)\, {\gamma}^{\mu} \gamma_5 +
           g_P^{q}(t) \frac{\gamma_5 \Delta^\mu}{2m} \right] u(p) ,
\end{equation}
are defined for each separate quark flavor.  Remarkably, the integrals
in (\ref{basic-sum-rules}) are independent of $\xi$.  This is a
consequence of Lorentz invariance: integrating the matrix elements
(\ref{quark-gpd}) over $x$ removes all reference to the particular
light-cone direction with respect to which $\xi$ is defined, so that
the result must be $\xi$-independent.

Higher Mellin moments in $x$ lead to derivative operators between the
two fields according to the relation
\begin{eqnarray}
 \label{higher-moments}
\lefteqn{
(P^+)^{n+1} \int dx\, x^n  \int \frac{d z^-}{2\pi}\, e^{ix P^+ z^-}
  \Big[ \bar{q}(-\half z)\, \gamma^+ q(\half z) \Big]_{z^+=0,\,
  \tvec{z}=0} 
}
\nonumber \\
 &=& \left. \Big(i \frac{d}{dz^-}\Big)^n
        \Big[ \bar{q}(-\half z)\, \gamma^+ q(\half z) \Big]
     \right|_{z=0}
  =  \bar{q}(0)\, \gamma^+ (i \lrpartial^+)^n\, q(0)
\end{eqnarray}
and its analogs for the other operators defining GPDs.  In a general
gauge, where a Wilson line appears between the operators at positions
$-\half z$ and $\half z$, one obtains the covariant derivative $D$
instead of $\partial$ on the right-hand side.  Higher $x$-moments of
GPDs thus are form factors of the local twist-two operators
\begin{eqnarray}
\mathcal{O}_q^{\mu \mu_1 \ldots \mu_n} &=& 
  \mathbf{S}\, \bar{q} \gamma^\mu\,
       i \lrD^{\mu_1} \ldots i \lrD^{\mu_n}\, q \, ,
\nonumber \\
\widetilde{\mathcal{O}}_{q}^{\mu \mu_1 \ldots \mu_n} &=& 
  \mathbf{S}\, \bar{q} \gamma^\mu\gamma_5\,
       i \lrD^{\mu_1} \ldots i \lrD^{\mu_n}\, q \, ,
\nonumber \\
\mathcal{O}_g^{\mu \mu_1 \ldots \mu_n \nu} &=& 
  \mathbf{S}\, G^{\mu\alpha}\,
       i \lrD^{\mu_1} \ldots i \lrD^{\mu_n}\, 
        G_{\alpha}{}^{\nu} \, ,
\nonumber \\
\widetilde{\mathcal{O}}_{g}^{\mu \mu_1 \ldots \mu_n \nu} &=& 
  \mathbf{S}\, (-i) G^{\mu\alpha}\,
       i \lrD^{\mu_1} \ldots i \lrD^{\mu_n}\, 
        \tilde{G}_{\alpha}{}^{\nu}
  \label{twist-two}
\end{eqnarray}
familiar from deep inelastic scattering.  Here $\mathbf{S}$ denotes
symmetrization in all uncontracted Lorentz indices and subtraction of
trace terms.  A form factor decomposition of $\mathcal{O}_q$ can be
written as \cite{Ji:1998pc}
\begin{eqnarray}
  \label{Ji-decomposition}
\lefteqn{
\langle p'|\, \mathcal{O}_q^{\mu \mu_1 \ldots \mu_n}(0) \,|p \rangle 
= \mathbf{S}\, \bar{u}(p') \gamma^\mu u(p) 
  \sum_{i=0 \atop \scriptstyle{\rm even}}^n A^q_{n+1,i}(t)\, 
  \Delta^{\mu_1} \ldots \Delta^{\mu_i} P^{\mu_{i+1}} \ldots P^{\mu_n}
}
\nonumber \\
&+& \mathbf{S}\, \bar{u}(p') \frac{i {\sigma}^{\mu\alpha}
                              \Delta_\alpha}{2m} u(p)
  \sum_{i=0 \atop \scriptstyle{\rm even}}^n B^q_{n+1,i}(t)\, 
  \Delta^{\mu_1} \ldots \Delta^{\mu_i} P^{\mu_{i+1}} \ldots P^{\mu_n}
\nonumber \\
&+& \mathbf{S}\,\frac{\Delta^{\mu}}{m} \bar{u}(p') u(p) \, 
  \mbox{mod}(n,2)\, C_{n+1}^q(t)\, 
  \Delta^{\mu_1} \ldots \Delta^{\mu_n} ,
\end{eqnarray}
where $\mbox{mod}(n,2)$ is 1 for odd $n$ and $0$ for even $n$.  The
Mellin moments of the ordinary quark distribution give forward matrix
elements and hence reduce to the $A^q_{n,0}(0)$.  Due to the Gordon
identities $\bar{u}(p') u(p)$ can be traded for the proton vector and
tensor currents when multiplied with $P^\alpha$, so that it only
appears as an independent structure in the form of the last line of
(\ref{Ji-decomposition}).  The vector $\Delta^\alpha$ only appears in
even powers because of time reversal invariance.  According to
(\ref{higher-moments}) the $x$-moments of GPDs involve the $+$ tensor
components of the twist-two operators, and using $\Delta^+ = -2\xi
P^+$ one finds
\begin{eqnarray}
  \label{general-moments}
\int_{-1}^1 dx\, x^n H^q(x,\xi,t) &=&
  \sum_{i=0 \atop \scriptstyle{\rm even}}^n (2\xi)^i A^q_{n+1,i}(t)
  + \mbox{mod}(n,2)\, (2\xi)^{n+1} C_{n+1}^q(t) ,
\nonumber \\
\int_{-1}^1 dx\, x^n E^q(x,\xi,t) &=&
  \sum_{i=0 \atop \scriptstyle{\rm even}}^n (2\xi)^i B^q_{n+1,i}(t)
  - \mbox{mod}(n,2)\, (2\xi)^{n+1} C_{n+1}^q(t) .
\end{eqnarray}
This is the remarkable \emph{polynomiality} property of GPDs: the
$x$-integrals of $x^n H^q$ and of $x^n E^q$ are polynomials in $\xi$
of order $n+1$.  Like the $\xi$-independence of the lowest moments
(\ref{basic-sum-rules}) this property is a consequence of the Lorentz
invariance encoded in the form factor decomposition
(\ref{Ji-decomposition}).  A particularity is that the terms with the
highest power $\xi^{n+1}$ are equal and opposite for the moments of
$H^q$ and $E^q$, so that the $x^n$ moment of the combination $H^q+E^q$
is only a polynomial in $\xi$ of degree $n$.  The contributions to
$H^q$ and $E^q$ going with $\xi^{n+1} C_{n+1}^q(t)$ only occur for odd
$n$ and thus in the $C= +1$ sector.  They lead to the so-called $D$
term in GPDs and will be discussed further in
Section~\ref{sub:d-term}.  They have no analog for the spin dependent
quark distributions, where the relevant form factor decomposition is
\begin{eqnarray}
  \label{tilde-decomposition}
\lefteqn{
\langle p'|\, \widetilde{\mathcal{O}}_q^{\mu \mu_1 \ldots \mu_n}(0) \,
        |p \rangle 
= \mathbf{S}\, \bar{u}(p') \gamma^\mu\gamma_5 u(p) 
  \sum_{i=0 \atop \scriptstyle{\rm even}}^n \tilde{A}^q_{n+1,i}(t)\, 
  \Delta^{\mu_1} \ldots \Delta^{\mu_i} P^{\mu_{i+1}} \ldots P^{\mu_n}
}
\nonumber \\
&+& \mathbf{S}\, \frac{\Delta^\mu }{2m}\, \bar{u}(p') \gamma_5 u(p)
  \sum_{i=0 \atop \scriptstyle{\rm even}}^n \tilde{B}^q_{n+1,i}(t)\, 
  \Delta^{\mu_1} \ldots 
        \Delta^{\mu_i} P^{\mu_{i+1}} \ldots P^{\mu_n} .
\end{eqnarray}
Here time reversal invariance constrains $\bar{u}(p') \gamma_5 u(p)$
to come with odd powers of $\Delta^\alpha$.  One then has sum rules
\begin{eqnarray}
  \label{general-moments-tilde}
\int_{-1}^1 dx\, x^n \tilde{H}^q(x,\xi,t) &=&
  \sum_{i=0 \atop \scriptstyle{\rm even}}^n (2\xi)^i 
        \tilde{A}^q_{n+1,i}(t) ,
\nonumber \\
\int_{-1}^1 dx\, x^n \tilde{E}^q(x,\xi,t) &=&
  \sum_{i=0 \atop \scriptstyle{\rm even}}^n (2\xi)^i 
        \tilde{B}^q_{n+1,i}(t) ,
\end{eqnarray}
where the highest power in $\xi$ is $n$ instead of $n+1$.  Notice that
the moments considered here correspond to integration over the full
$x$ range from $-1$ to $1$.  Integrals $\int dx\, x^{2n+1} H^q$ and
$\int dx\, x^{2n} \tilde{H}^q$ therefore involve the $C$-even
combinations $H^{q(+)}$ and $\tilde{H}^{q(+)}$, whereas integrals
$\int dx\, x^{2n} H^q$ and $\int dx\, x^{2n+1} \tilde{H}^q$ involve
the $C$-odd combinations $H^{q(-)}$ and $\tilde{H}^{q(-)}$, in
accordance with the $C$ parity of the corresponding local operators
(\ref{twist-two}).  The same holds of course for $E^q$ and
$\tilde{E}^q$.

For gluon operators the form factor decompositions are analogous, with
\begin{eqnarray}
  \label{glue-decomposition}
\lefteqn{
\langle p'|\, \mathcal{O}_g^{\mu \mu_1 \ldots \mu_{n-1} \nu}(0)
	\,|p \rangle 
}
\nonumber \\
&=& \mathbf{S}\, \bar{u}(p') \gamma^\mu u(p) 
  \sum_{i=0 \atop \scriptstyle{\rm even}}^n A^g_{n+1,i}(t)\, 
  \Delta^{\mu_1} \ldots \Delta^{\mu_i} P^{\mu_{i+1}} \ldots
  P^{\mu_{n-1}} P^\nu
\nonumber \\
&+& \mathbf{S}\, \bar{u}(p') \frac{i {\sigma}^{\mu\alpha}
                              \Delta_\alpha}{2m} u(p)
  \sum_{i=0 \atop \scriptstyle{\rm even}}^n B^g_{n+1,i}(t)\, 
  \Delta^{\mu_1} \ldots \Delta^{\mu_i} P^{\mu_{i+1}} \ldots
                              P^{\mu_{n-1}} P^\nu
\nonumber \\
&+& \mathbf{S}\,\frac{\Delta^{\mu}}{m} \bar{u}(p') u(p) \, 
  \mbox{mod}(n,2)\, C_{n+1}^g(t)\, 
  \Delta^{\mu_1} \ldots \Delta^{\mu_{n-1}} \Delta^\nu
\end{eqnarray}
and
\begin{eqnarray}
  \label{glue-tilde-decomposition}
\lefteqn{
\langle p'|\, 
    \widetilde{\mathcal{O}}_g^{\mu \mu_1  \ldots \mu_{n-1} \nu}(0)
        \,|p \rangle 
}
\nonumber \\
&=& \mathbf{S}\, \bar{u}(p') \gamma^\mu\gamma_5 u(p) 
  \sum_{i=0 \atop \scriptstyle{\rm even}}^n \tilde{A}^g_{n+1,i}(t)\, 
  \Delta^{\mu_1} \ldots \Delta^{\mu_i} P^{\mu_{i+1}} \ldots
   P^{\mu_{n-1} } P^\nu
\nonumber \\
&+& \mathbf{S}\, \frac{\Delta^\mu }{2m}\, \bar{u}(p') \gamma_5 u(p)
  \sum_{i=0 \atop \scriptstyle{\rm even}}^n \tilde{B}^g_{n+1,i}(t)\, 
  \Delta^{\mu_1} \ldots 
        \Delta^{\mu_i} P^{\mu_{i+1}} \ldots P^{\mu_{n-1}} P^\nu .
\end{eqnarray}
The corresponding sum rules read
\begin{eqnarray}
\int_{0}^1 dx\, x^{n-1} H^g(x,\xi,t) &=&
  \sum_{i=0 \atop \scriptstyle{\rm even}}^n (2\xi)^i A^g_{n+1,i}(t)
  + \mbox{mod}(n,2)\, (2\xi)^{n+1} C_{n+1}^g(t) ,
\nonumber \\
\int_{0}^1 dx\, x^{n-1} E^g(x,\xi,t) &=&
  \sum_{i=0 \atop \scriptstyle{\rm even}}^n (2\xi)^i B^g_{n+1,i}(t)
  - \mbox{mod}(n,2)\, (2\xi)^{n+1} C_{n+1}^g(t) 
\end{eqnarray}
for odd $n$ and 
\begin{eqnarray}
\int_{0}^1 dx\, x^{n-1} \tilde{H}^g(x,\xi,t) &=&
  \sum_{i=0 \atop \scriptstyle{\rm even}}^n (2\xi)^i 
        \tilde{A}^g_{n+1,i}(t) ,
\nonumber \\
\int_{0}^1 dx\, x^{n-1} \tilde{E}^g(x,\xi,t) &=&
  \sum_{i=0 \atop \scriptstyle{\rm even}}^n (2\xi)^i 
        \tilde{B}^g_{n+1,i}(t) 
\end{eqnarray}
for even $n$, where we have used the symmetry properties of the gluon
distributions in $x$ to restrict the integration to $x\ge 0$.  Note
the correspondence between the $(n-1)$st moment of a gluon GPD and the
$n$th moment of a GPD for quarks.  In fact the corresponding quark and
gluon operators, and hence their form factors, mix under evolution
(see Section~\ref{sec:evolution}).


\subsection{Parton interpretation and the light-cone}
\label{sec:light-cone}

Let us discuss in more detail how GPDs can be interpreted as
quantities giving us information about partons in hadrons.  A useful
interpretation of parton distributions (ordinary or generalized) is as
amplitudes for the scattering of a parton on a proton; this is the
basis of the covariant parton model of Landshoff et
al.~\cite{Landshoff:1971ff}.  To be more precise, parton distributions
are Green functions integrated over the $k^-$ and $\tvec{k}$ of the
partons (and thus over their off-shellness).  This is seen by writing
\begin{eqnarray}
  \label{off-shell}
F^q &=&
\frac{1}{2} \int \frac{d z^-}{2\pi}\, e^{ix P^+ z^-} 
\langle p'|\, \bar{q}(-\half z)\, \gamma^+ q(\half z)
   \,|p \rangle \Big|_{z^+=0,\,\tvec{z}=0}
\nonumber \\
&=& \frac{1}{2} \int \frac{dk^-\, d^2 \tvec{k}}{(2\pi)^4}
\left[ \, \int d^4z\, e^{i (k z)}\,
       \langle p'|\, \bar{q}(-\half z)\, \gamma^+ q(\half z)
   \,|p \rangle 
     \right]_{k^+ = x P^+} 
\nonumber \\
&=& \frac{1}{2} \int \frac{dk^-\, d^2 \tvec{k}}{(2\pi)^4}
\left[ \, \int d^4z\, e^{i (k z)}\,
       \langle p'|\, T \bar{q}(-\half z)\, \gamma^+ q(\half z)
   \,|p \rangle 
     \right]_{k^+ = x P^+}
\end{eqnarray}
and the analogous relations for $\tilde{F}^q$ and the gluon
distributions.  The expression in square brackets in the last line is
a Green function $G(p,p',k)$ with on-shell hadrons and off-shell
partons.  In going from the second to the third line of
(\ref{off-shell}) we have used that the time ordering of operators may
be omitted under the integral over $k^-$.  The key ingredient to show
this is to locate the singularities of $G(p,p',k)$ in the complex
$k^-$ plane according to the general analyticity properties of Green
functions.  By complex contour deformation one can then replace the
$k^-$ integral of $G(p,p',k)$ by the integral of its discontinuity in
appropriate external invariants. The resulting cut Green functions are
associated with operators that are not time ordered but only normal
ordered, as in the second line of (\ref{off-shell}).  For ordinary
parton densities this argument was already given in the original work
of Landshoff et al.~\cite{Landshoff:1971ff}.  The situation for
nonforward kinematics was discussed in
\cite{Frankfurt:1998ha,Radyushkin:1997ki} and studied in detail in
\cite{Diehl:1998sm}.  A different line of argument was pursued by
Jaffe \cite{Jaffe:1983hp} and also applied to higher-twist parton
distributions.

The cut Green functions require on-shell intermediate states, and this
requirement gives the support of GPDs as $|x|\le 1$.  In physical
terms, no parton in a hadron can have a larger plus-momentum than the
hadron itself since this would force other partons to have negative
plus-momentum, which an on-shell particle cannot have.  We remark that
the above argument assumes that Green functions with colored external
lines have the same analyticity properties as Green functions in
perturbation theory, although the ``on-shell states'' across a cut are
colored and hence not part of the physical spectrum.

To obtain this simple interpretation of parton distributions in terms
of four-point functions one must take the light-cone gauge $A^+= 0$,
otherwise there is a Wilson line with additional $A^+$ gluons between
the two parton fields.  We will therefore adopt this gauge in the rest
of this section.  There has recently been work pointing out subtle
issues of this gauge in connection with parton densities, and we will
comment on this in Section~\ref{sub:light-cone-problems}.

An alternative to the covariant representation (\ref{off-shell})
involves partons on their mass shell, which is in particular helpful
if we want to assign helicities to them.  This can be achieved in a
noncovariant framework of light-cone quantization, whose essentials
for our context we shall now briefly recall.  More detail can be found
in the original work by Kogut and Soper \cite{Kogut:1970xa} or in the
reviews \cite{Brodsky:1989pv,Brodsky:1998de}.  A discussion with focus
on spin has been given by Jaffe \cite{Jaffe:1996zw}.  In light-cone
quantization, the canonical (anti)commutation relations defining the
quantum theory are imposed not at a given time, but at a given
light-cone time $z^+$, which we choose to be $z^+ =0$.  If one
introduces projection operators $P_\pm = \half \gamma^\mp \gamma^\pm$
for Dirac fields and projects the Dirac equation on its $P_-$
component, one obtains
\begin{equation}
  \label{dirac-constraint}
i \partial^+ q_-(z) = - \half \gamma^+ (i \Slash{D}_T -m)\, q_+(z) .
\end{equation}
with $q_- = P_-q$ and $q_+ = P_+q$.  This equation does not involve
derivatives with respect to light-cone time $z^+$ (remember that
$\partial^+ = \partial / (\partial z^-)$).  The ``bad'' field
components $q_-$ are therefore dynamically dependent on the ``good''
components $q_+$, and can be eliminated by solving the differential
equation (\ref{dirac-constraint}).  Canonical anticommutation
relations for free fields are then imposed on the dynamically
independent field $q_+$ at $z^+ =0$, which can be Fourier decomposed
into
\begin{eqnarray}
  \label{good-quark}
q_+(z) \Big|_{z^+=0} &=&
  \int \frac{d^2 \tvec{k}\, dk^+}{16\pi^3 k^+}\, \theta(k^+) \,
\sum_{\mu} \Big[ 
        b(k^+, \tvec{k},\mu)\, u_+(k^+, \mu)\, e^{-i (kz)}
\\
&& \hspace{8.5em}
    {}+ d^\dag(k^+, \tvec{k},\mu)\, v_+(k^+, \mu)\, e^{i (kz)} 
	\Big]_{z^+=0} \; ,
\nonumber 
\end{eqnarray}
where $\mu = \pm \half$ denotes the light-cone helicity of the parton.
(For better legibility we omit color indices throughout.)  Light-cone
helicity will be discussed in Section~\ref{sub:light-cone-hel}; for
the moment it is sufficient to know that it is identical to usual
helicity for massless particles, and approximately the same if
particles move fast in the positive $z$-direction.  Notice that the
good components of the corresponding spinors $u_+ = P_+ u$ and $v_+ =
P_+ v$ are independent on $\tvec{k}$ (this can be easily seen from the
matrix representation of the projector $P_+$ and the explicit spinors
in Appendix~\ref{app:spinors}).  An important consequence is that we
can associate a definite light-cone helicity to the partons described
by GPDs, which are integrated over $\tvec{k}$ rather than referring to
a particular transverse parton momentum.

For gluons the steps are analogous. One component of the Yang-Mills
equations (in $A^+=0$ gauge) does not involve derivatives $\partial/
(\partial z^+)$ and can be used to eliminate the ``bad'' component
$A^-$ as dynamically dependent.  Free-field commutation relations are
then imposed on the ``good'' transverse components of ${A}^\mu$ at
$z^+=0$, whose Fourier decomposition reads
\begin{eqnarray}
  \label{good-gluon}
A^i(z) \Big|_{z^+=0} &=&
  \int \frac{d^2 \tvec{k}\, dk^+}{16\pi^3 k^+}\, \theta(k^+) \, 
\sum_{\mu} \Big[ 
        a(k^+, \tvec{k},\mu)\, \epsilon^i(\mu)\, e^{-i (kz)}
\nonumber \\
&& \hspace{8.5em}
    {}+ a^\dag(k^+, \tvec{k},\mu)\, \epsilon^{i *}(\mu)\, 
                e^{i (kz)} \Big]_{z^+=0} \; ,
\end{eqnarray}
where $i=1,2$ is a transverse index, and the polarization vectors for
states with definite light-cone helicity $\mu= \pm 1$ are given by
\begin{equation}
\mbox{\boldmath{$\epsilon$}}(+) = - \frac{1}{\sqrt{2}} 
              \left( \begin{array}{c} 1 \\ i \end{array} \right) ,
\hspace{3em}
\mbox{\boldmath{$\epsilon$}}(-) = \frac{1}{\sqrt{2}}
              \left( \begin{array}{c} 1 \\ -i \end{array} \right) .
\end{equation}
The operators $b$, $d$ and $a$ respectively annihilate quarks,
antiquarks and gluons, and (anti)commute with the creation operators
$b^\dag$, $d^\dag$ and $a^\dag$, as do the operators describing free
particles.  Notice that the good components of quark and gluon fields
behave as free \emph{only} at light-cone time $z^+=0$, which is just
where these operators enter the definition of parton distributions.
This reflects the physics of the parton model, where partons are
treated as free just during the short time where they are ``seen'' by
a probe like a highly virtual photon in a hard process.  Notice also
that the parton states obtained by acting with $b^\dag$, $d^\dag$ and
$a^\dag$ on the vacuum are defined with reference to the time of this
hard interaction, and not with reference to the far past or far
future.  In this sense we do not use that partons correspond to
observable particles, which of course they are not because of
confinement.  Let us add that light-cone quantization involves subtle
issues well beyond the scope of this review.  Among these is the
problem of modes with zero plus-momentum (``zero modes'') and their
relevance in the physical vacuum; for a discussion we refer to
\cite{Yamawaki:1998cy}.

We now have the ingredients necessary to take a closer look at the
operators defining GPDs.  As an example we take the operator
\begin{eqnarray}
  \label{quark-helicities}
\lefteqn{
\int \frac{d z^-}{2\pi}\, e^{ix P^+ z^-}
  \frac{1}{4} \bar{q}(-\half z)\, \gamma^+ (1+\gamma_5) 
        q(\half z)\, \Big|_{z^+=0,\, \tvec{z}=0}
}
\nonumber \\[0.4em]
&=& 
\int\frac{d^2 \tvec{k}\, dk^+}{16\pi^3 k^+}\, \theta(k^+)
\int\frac{d^2 \tvec{k}'\, dk'^+}{16\pi^3 k'^+}\, \theta(k'^+) \;
\frac{\sqrt{k^+ k'^+}}{P^+}
\nonumber \\
&& \Bigg\{ \hspace{1.3ex}
     \delta\Big(x - \frac{k^+ + k'^+}{2P^+}\Big)\;
     b^\dag(k'^+,\tvec{k}', +)\, b^{\phantom{\dag}}(k^+,\tvec{k}, +)\, 
\nonumber \\
&& {}+ \delta\Big(x + \frac{k^+ + k'^+}{2P^+}\Big)\;
     d^{\phantom{\dag}}(k'^+,\tvec{k}', -)\, d^\dag(k^+,\tvec{k}, -)\, 
\nonumber \\[0.4em]
&& {}- \delta\Big(x - \frac{k^+ - k'^+}{2P^+}\Big)\;
     d^{\phantom{\dag}}(k'^+,\tvec{k}', -)\, 
        b^{\phantom{\dag}}(k^+,\tvec{k}, +)\, 
\nonumber \\[0.4em]
&& {}- \delta\Big(x + \frac{k^+ - k'^+}{2P^+}\Big)\;
       b^\dag(k'^+,\tvec{k}', +)\, d^\dag(k^+,\tvec{k}, -) 
\Bigg\} .
\end{eqnarray}
For better legibility we will henceforth give fermion helicities as
$\pm$ instead of $\pm \half$ when they just appear as labels.  In
matrix elements of the operator (\ref{quark-helicities}) between
hadron states, momentum conservation fixes $(k-k')^+$ and
$\tvec{k}-\tvec{k}'$ in the terms with $b^\dag b$ and $d d^\dag$,
whereas in the terms with $d b$ and $b^\dag d^\dag$ it fixes
$(k+k')^+$ and $\tvec{k}+\tvec{k}'$.  The parton plus-momenta are
hence completely fixed, whereas one still has to integrate over their
overall transverse momentum $\tvec{k}$.  Compared with the
representation (\ref{off-shell}) of the covariant parton model there
is however no longer an integration over $k^-$, because we have
expressed the operators defining parton distributions in terms of
creation and annihilation operators for \emph{on-shell} partons.
Which term of the decomposition (\ref{quark-helicities}) contributes
in a matrix element is determined by the fraction $x$, the constraint
of positive parton plus-momenta $k^+$ and $k'^+$, and the constraint
from momentum conservation.  We thus find the different partonic
regimes discussed in Section~\ref{sec:definitions}.  For $\xi>0$ the
combination $b^\dag d^\dag$ does not appear; it replaces $d\, b$ in
the case $\xi<0$.

Looking at the helicity structure we observe the correspondence of
right-helicity quarks and left-helicity antiquarks.  In particular,
this ensures that the net \emph{transfer} of helicity is always zero,
either by emitting a parton and reabsorbing it with the same helicity,
or by emitting (or absorbing) a pair of partons with opposite
helicities.  The formula for the operator $\frac{1}{4}\bar{q} \gamma^+
(1-\gamma_5) q$ is analogous, with all fermion helicities reversed.
We also observe that the operator in (\ref{quark-helicities}) is
invariant under rotations around the $z$ axis and thus carries no
angular momentum component $J^3$.

The normalization of the operator in (\ref{quark-helicities}) is such
that taking the forward matrix element with a proton state we obtain
the density of positive-helicity quarks $\half[q(x) + \Delta q(x)]$
for $x>0$.  The operators in (\ref{quark-helicities}) should be read
as normal ordered, and bringing the combination $d\, d^\dag$ in the
second term into the form $d^\dag d$ appropriate for a counting
operator one obtains a minus sign.  The density of positive-helicity
quarks thus goes with the operator $\frac{1}{4} \bar{q} \gamma^+
(1+\gamma_5) q$, and the density for positive-helicity antiquarks with
$- \frac{1}{4} \bar{q} \gamma^+ (1-\gamma_5) q$.

For gluon operators we have in analogy to (\ref{quark-helicities})
\begin{eqnarray}
  \label{gluon-helicities}
\lefteqn{
\frac{1}{P^+} \int \frac{d z^-}{2\pi}\, e^{ix P^+ z^-}
  \frac{1}{2}\, \Big[ G^{+\mu}(-\half z)\, G_{\mu}{}^{+}(\half z)
           -i G^{+\mu}(-\half z)\, \tilde{G}_{\mu}{}^{+}(\half z)
  \Big]_{z^+=0,\, \tvec{z}=0}
} 
\nonumber \\
&=&
\int\frac{d^2 \tvec{k}\, dk^+}{16\pi^3 k^+}\, \theta(k^+)
\int\frac{d^2 \tvec{k}'\, dk'^+}{16\pi^3 k'^+}\, \theta(k'^+) \;
\frac{k^+ k'^+}{(P^+)^2} 
    \hspace{8em}
\nonumber \\
&& \Bigg\{ \hspace{1.3ex}
     \delta\Big(x - \frac{k^+ + k'^+}{2P^+}\Big)\;
     a^\dag(k'^+,\tvec{k}', +)\, a^{\phantom{\dag}}(k^+,\tvec{k}, +)\, 
\nonumber \\
&& {}+ \delta\Big(x + \frac{k^+ + k'^+}{2P^+}\Big)\;
     a^{\phantom{\dag}}(k'^+,\tvec{k}, -)\, a^\dag(k^+,\tvec{k}, -)\, 
\nonumber \\[0.4em]
&& {}+ \delta\Big(x - \frac{k^+ - k'^+}{2P^+}\Big)\;
     a^{\phantom{\dag}}(k'^+,\tvec{k}', -)\, 
        a^{\phantom{\dag}}(k^+,\tvec{k}, +)\, 
\nonumber \\[0.4em]
&& {}+ \delta\Big(x + \frac{k^+ - k'^+}{2P^+}\Big)\;
       a^\dag(k'^+,\tvec{k}', +)\, a^\dag(k^+,\tvec{k}, -) 
\Bigg\} .
\end{eqnarray}
Notice that because $G^{\mu\nu}$ and $\tilde{G}^{\mu\nu}$ are
antisymmetric tensors, only transverse indices $\mu = 1,2$ contribute
in the operators on the left-hand side.  Since in light-cone gauge one
has $G^{+\mu} = \partial^+ A^\mu$ we see that only good field
components appear in the operators.  This is characteristic of both
quark and gluon distributions of twist two.  Comparing
(\ref{quark-helicities}) and (\ref{gluon-helicities}) we notice an
extra factor of $\sqrt{k^+ k'^+} /P^+$ for gluons.  The forward matrix
element of (\ref{gluon-helicities}) is therefore given by $\half [x
g(x) + x \Delta g(x)]$ for $x>0$, with an extra factor of $x$ compared
to the quark case.


\subsection{Helicity structure}
\label{sec:helicity}

Let us now take a closer look at the spin structure of GPDs for a spin
$\half$ target.  The structure for spin 0 targets is much simpler and
will be briefly discussed at the end of Section~\ref{sub:counting}.
Spin 1 targets and transitions between hadron states with different
spin will be discussed in Sections \ref{sec:transition} and
\ref{sec:deuteron}.

\subsubsection{Lorentz symmetry on the light-cone}
\label{sub:light-cone-hel}

Parton distributions probe hadron structure with respect to a
light-like direction, which is physically singled out by the hard
probe in a process where the distributions appear, and technically
reflected in the light-like separation of the operators in their
definitions (\ref{quark-gpd}) and (\ref{gluon-gpd}).  An essential
feature of GPDs is that the hadron states are not both ``aligned''
along this axis, so that there is a nontrivial choice to be made in
defining their helicities.  As a guide to our choice we will first
discuss an important symmetry of light-cone physics.

We have already seen the special role played by plus-momentum in the
definition of GPDs.  The Lorentz transformations which leave
plus-momenta invariant are rotations around the $z$-axis and the
so-called \emph{transverse boosts}, which transform a four-vector $k$
according to
\begin{equation}
k^+ \to k^+ , \qquad  \tvec{k} \to \tvec{k} - k^+ \tvec{v} ,
  \label{transverse-boost}
\end{equation}
where $\tvec{v}$ is a two-dimensional vector parameterizing the
transformation.  $k^-$ transforms in a more complicated way, ensuring
that $k^2$ remains invariant.  Because of Lorentz invariance the GPDs
can only depend on the transverse components of $p$ and of $p'$ via
the transverse components of the vector
\begin{equation}
  \label{D-def}
D = \frac{p'}{1-\xi} - \frac{p}{1+\xi} .
\end{equation}
Under a transverse boost $\tvec{D}$ is invariant because by
construction $D^+ = 0$.  In fact, the invariant $t$ can be written as
\begin{equation}
  \label{t-by-D}
t = t_0 - (1-\xi^2) \tvec{D}^2  ,
\end{equation}
where 
\begin{equation}
t_0 = - \frac{4\xi^2 m^2}{1-\xi^2}
\end{equation}
is the maximum value of $t$ at given $\xi$.  Note that this implies an
upper bound $\xi \le \sqrt{-t\phantom{^1}} /\sqrt{4m^2 -
t\phantom{^1}}$ on $\xi$ at given $t$.  Only in particular frames is
$\tvec{D}$ proportional to $\tvec{\Delta}$.  Examples are the
symmetric case where $\tvec{p}' = - \tvec{p} = \half\tvec{\Delta}$ so
that $\tvec{D} = (1-\xi^2)^{-1} \tvec{\Delta}$, or the case $\tvec{p}
= 0$ where $\tvec{D} = (1-\xi)^{-1}\tvec{\Delta}$.

Given the special role played by transverse boosts it is natural to
discuss proton helicity in terms of light-cone helicity states
\cite{Soper:1972xc,Brodsky:1998de}.  A light-cone helicity state for a
particle with momentum components $(k^+, \tvec{k})$ is obtained from
the one with momentum components $(k^+, \tvec{0})$ via the transverse
boost specified by $\tvec{v} = - \tvec{k} /k^+$.  Such states have
simple properties under the Lorentz transformations important in
light-cone physics.  Consider a transformation which changes the
momentum $k$ into $k'$.  A state with light-cone helicity $\lambda$
transforms as $|k, \lambda\rangle \to |k', \lambda\rangle$ under a
longitudinal boost along $z$ and under a transverse boost
(\ref{transverse-boost}), and as $|k, \lambda\rangle \to e^{-i\lambda
\varphi} |k', \lambda\rangle$ under a rotation by $\varphi$ around the
$z$-axis.  In contrast, the usual helicity state for a particle with
$(k^+, \tvec{k})$ is obtained from a state with zero transverse
momentum by a spatial rotation, and has rather complicated
transformation properties under longitudinal and transverse boosts.
The distinction between the two types of states is however unimportant
for particles having a large positive momentum $p^3$, with their
difference being of order $m /p^+$ (and strictly zero if $\tvec{p}=0$
and $p^3>0$).  The situation is hence the same as for light-cone
plus-momentum versus usual three-momentum and reflects that the
light-cone quantities correspond to the limit of the infinite-momentum
frame.  We will henceforth use ``helicity'' in the sense of
``light-cone helicity'' unless stated otherwise.  The explicit spinors
used in our subsequent calculations are given in
Appendix~\ref{app:spinors}, as well as their relation with usual
helicity spinors.

{}From our discussion it follows that if we consider the matrix
elements (\ref{quark-gpd}) and (\ref{gluon-gpd}) for states with
definite light-cone helicity, they are invariant under both transverse
and longitudinal boosts, and can therefore only depend on the
light-cone momentum fractions $x$ and $\xi$ and on $\tvec{D}$.  For
the matrix elements $F_{\lambda'\lambda}$ and
$\tilde{F}_{\lambda'\lambda}$ defined in (\ref{quark-gpd}) and
(\ref{gluon-gpd}) we explicitly find
\begin{eqnarray}
  \label{hel}
F_{++}         &=& \phantom{-} F_{--} \hspace{1.2em} =\,
      \sqrt{1-\xi^2} \Bigg[ H - \frac{\xi^2}{1-\xi^2}\, E \Bigg] ,
\nonumber \\
\tilde{F}_{++} &=& - \tilde{F}_{--} \hspace{1.2em} =\,
       \sqrt{1-\xi^2} \Bigg[ \tilde{H} - 
          \frac{\xi^2}{1-\xi^2}\, \tilde{E} \Bigg],
\nonumber \\
F_{-+}         &=& - \Big[F_{+-}\Big]^* =\;
       e^{i\varphi}\, \frac{\sqrt{t_0 - t}}{2m}\; E \,,
\nonumber \\[0.4em]
\tilde{F}_{-+} &=& \phantom{-} \Big[\tilde{F}_{+-}\Big]^* =\;
       e^{i\varphi}\, \frac{\sqrt{t_0 - t}}{2m}\; \xi \tilde{E} 
\end{eqnarray}
for both quarks and gluons, where the first index $\lambda'$ refers to
the outgoing proton with momentum $p'$ and the second index $\lambda$
to the incoming proton with momentum $p$.  These relations are of
fundamental importance for phenomenology since they specify which GPDs
are selected by particular proton helicity transitions.  Given the
transformation properties of helicity states under rotations about the
$z$ axis, the matrix elements (\ref{hel}) depend on $\tvec{D}$ only
via its length $|\tvec{D}| = \sqrt{t_0 - t \rule{0pt}{1.8ex}}
/\sqrt{1-\xi^2}$ and a phase factor of the form $e^{i n\varphi}$,
where the azimuthal angle $\varphi$ of $\tvec{D}$ is given by
$e^{i\varphi} |\tvec{D}| = D^1+i D^2$.

\subsubsection{Parton helicity flip} 
\label{sub:helicity-flip}

So far we have only discussed quark and gluon operators with total
helicity transfer zero, which in the DGLAP regions do not change the
helicity of the emitted and reabsorbed parton.  There are however
twist-two distributions which flip the parton helicity in the DGLAP
regions.  Both in the DGLAP and ERBL regions they transfer a net
helicity of one unit in the case of quarks and of two units in the
case of gluons.  These distributions are off-diagonal in the parton
helicity basis.  They become diagonal if one changes basis from
eigenstates of helicity to eigenstates of ``transversity''
\cite{Jaffe:1996zw}, which is related to (but not the same as)
transverse spin, and are hence also called transversity distributions.
They are constructed from the tensor operators
\begin{equation}
  \label{tensor-operators}
\bar{q} \sigma^{+i} q , \qquad
\half \Big[ G^{+i} G^{j+} + G^{+j} G^{i+} \Big] 
        - \half g_T^{ij}\, G^{+\alpha} G_\alpha{}^{+}  , \qquad
i,j = 1,2
\end{equation}
with the fields separated by a light-like distance, supplemented by
Wilson lines in gauges other than $A^+=0$.  The Fourier transforms of
these non-local operators have representations analogous to
(\ref{quark-helicities}) and (\ref{gluon-helicities}).  Notice that
the operator defining quark transversity is chiral-odd, i.e., it flips
quark chirality, in contrast to the chiral-even operators $\bar{q}
\gamma^{+} q$ and $\bar{q} \gamma^{+}\gamma_5\, q$. 

In a similar manner as for parton helicity conserving GPDs one can
define the helicity-flip distributions by a form factor decomposition.
Both for quarks and for gluons one has four distributions, given by
\cite{Diehl:2001pm} 
\begin{eqnarray}
  \label{flip-gpd}
\lefteqn{ 
\frac{1}{2} \int \frac{d z^-}{2\pi}\, e^{ix P^+ z^-}
  \langle p'|\, 
     \bar{q}(-\half z)\, i \sigma^{+i}\, q(\half z)\, 
  \,|p \rangle \Big|_{z^+=0,\, \tvec{z}=0} 
} 
\nonumber \\
&=& \frac{1}{2P^+} \bar{u}(p') \left[
  H_T^q\, i \sigma^{+i} +
  \tilde{H}_T^q\, \frac{P^+ \Delta^i - \Delta^+ P^i}{m^2} \right.
\nonumber \\
&& \left. \hspace{5em} {}+
  E_T^q\, \frac{\gamma^+ \Delta^i - \Delta^+ \gamma^i}{2m} +
  \tilde{E}_T^q\, \frac{\gamma^+ P^i - P^+ \gamma^i}{m}
  \right] u(p) , 
\nonumber \\
\lefteqn{ 
\frac{1}{P^+} \int \frac{d z^-}{2\pi}\, e^{ix P^+ z^-}
  \langle p'|\, {\mathbf S}
     G^{+i}(-\half z)\, G^{j+}(\half z)
  \,|p \rangle \Big|_{z^+=0,\,\tvec{z}=0} 
}
\nonumber \\
&=& {\mathbf S}\,
\frac{1}{2 P^+}\, \frac{P^+ \Delta^j - \Delta^+ P^j}{2 m P^+}
\nonumber \\
&& {}\times \bar{u}(p') \left[
  H_T^g\, i \sigma^{+i} +
  \tilde{H}_T^g\, \frac{P^+ \Delta^i - \Delta^+ P^i}{m^2} \right.
\nonumber \\
&& \left. \hspace{2.8em} {}+
  E_T^g\, \frac{\gamma^+ \Delta^i - \Delta^+ \gamma^i}{2m} +
  \tilde{E}_T^g\, \frac{\gamma^+ P^i - P^+ \gamma^i}{m}\, 
   \right] u(p) ,
\hspace{2em} 
\end{eqnarray}
where $\mathbf{S}$ stands for symmetrization of uncontracted Lorentz
indices and removal of the trace (here in the two transverse
dimensions).  The notation has been chosen to mirror the one in the
non-flip sector, but distributions with and without tilde in
(\ref{flip-gpd}) do \emph{not} correspond to different quantum
numbers.  The definition for quark distributions can alternatively be
expressed in terms of $\sigma^{+\beta}
\gamma_5 = -\half i\, \epsilon^{+\beta\gamma\delta}
\sigma_{\gamma\delta}$ and is given in~\cite{Diehl:2001pm}.  Our
functions $H_T^g$ and $E_T^g$ respectively coincide with $H_G^T$ and
$E_G^T$ defined in \cite{Belitsky:2000jk}, and the relation of our
convention with the distributions originally introduced by Hoodbhoy
and Ji \cite{Hoodbhoy:1998vm} is
\begin{eqnarray}
H_T^q &=& H_{Tq} \Big|_{\smallcite{Hoodbhoy:1998vm}} , 
\qquad\qquad
E_T^q = -E_{Tq} \Big|_{\smallcite{Hoodbhoy:1998vm}} ,
\nonumber \\
H_T^g &=& -2x H_{Tg} \Big|_{\smallcite{Hoodbhoy:1998vm}} , 
\qquad
E_T^g = -2x E_{Tg} \Big|_{\smallcite{Hoodbhoy:1998vm}} .
\end{eqnarray}
The existence of the distributions $\tilde{H}_T$ and $\tilde{E}_T$ has
been overlooked in the pioneering paper \cite{Hoodbhoy:1998vm} due to
a nontrivial error, see the footnote in Section~\ref{sub:counting}.

For both quarks and gluons, time reversal invariance and hermiticity
of the operators give constraints on $H_T$, $\tilde{H}_T$ and $E_T$,
which read as (\ref{time-rev}) and (\ref{complex-conjug}).  For
$\tilde{E}_T$ one has
\begin{equation}
  \label{E-tilde-T}
\tilde{E}_T(x,\xi,t) = - \tilde{E}_T(x,-\xi,t) , \qquad
\Big[\tilde{E}_T(x,\xi,t)\Big]^* = - \tilde{E}_T(x,-\xi,t) 
\end{equation}
with extra minus signs, again for both quarks and gluons.  The gluon
distributions in (\ref{flip-gpd}) are even functions of $x$ because
gluons are their own antiparticles.  Taking Mellin moments in $x$ of
the transversity distributions gives form factors of local operators
constructed in analogy to (\ref{twist-two}).  Because of polynomiality
the lowest moment of $\tilde{E}_T^q$ must be independent of $\xi$, so
that the time reversal constraint in (\ref{E-tilde-T}) leads to
$\int_{-1}^1 dx\, \tilde{E}^q_T = 0$.  In other words, time reversal
invariance reduces the number of form factors of the matrix element
$\langle p'| \bar{q}(0) \sigma^{\mu\nu} q(0) | p\rangle$ from four to
three \cite{Diehl:2001pm}.

An important property of helicity-flip distributions is that they
evolve independently in the renormalization scale $\mu$, i.e., quarks
do not mix with gluons and vice versa.  In particular, gluon
transversity probes glue in the target which is ``intrinsic'' in the
sense that it cannot be generated from quarks by the perturbative
DGLAP parton splitting processes.

{}From (\ref{E-tilde-T}) and the factors of $\Delta$ in
(\ref{flip-gpd}) we immediately see that only the quark GPDs $H^q_T$
can be measured in the forward limit.  There they become equal to the
quark transversity distributions, usually denoted by $\delta q(x)$ or
$h_1(x)$.  These distributions have been introduced long ago
\cite{Ralston:1979ys}, but so far not much is known experimentally
about them.  This will hopefully change with the progress of the spin
programs at HERMES, COMPASS, and RHIC
\cite{Jaffe:1997yz,Barone:2001sp}.  Concerning gluons it is clear that
the helicity-flip distributions must all decouple in the forward limit
for a spin $\half$ target, since the change of gluon helicity by two
units cannot be compensated by the target.  For targets with spin 1 or
higher, such as a deuteron or a photon, gluon transversity is however
visible in forward matrix elements \cite{Jaffe:1989xy,Artru:1990zv}.

As we will see in Section~\ref{sub:selection}, gluon transversity GPDs
appear to leading-twist accuracy in DVCS.  Prospects of their
measurement will be discussed in
Sections~\ref{sub:compton-corrections} and
\ref{sec:transversity-access}.   Experimental access to 
quark transversity GPDs seems much less trivial and will also be
discussed in Section~\ref{sec:transversity-access}.

\subsubsection{Helicity amplitudes and GPDs}
\label{sec:helicity-transitions}

\begin{figure}
\begin{center}
     \leavevmode
     \epsfxsize=0.75\hsize
     \epsffile{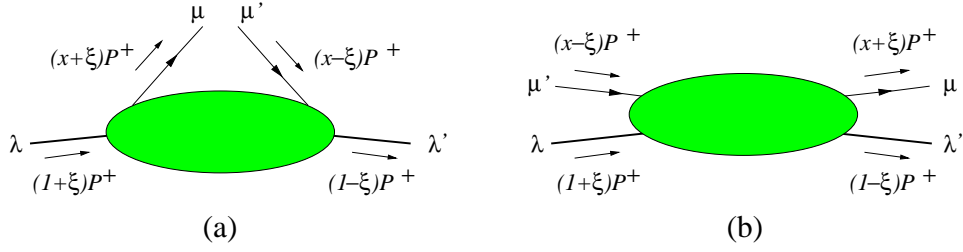}
\end{center}
\caption{\label{fig:order} Representation of a generalized parton
distribution in the region $x \in[\xi,1]$. The flow of plus-momentum
is indicated by arrows, and the labels $\lambda$, $\lambda'$, $\mu$,
$\mu'$ denote helicities. (a) shows the ordering of lines as ``proton
in, quark out, quark back in, proton out'' that is common for parton
distributions. (b) displays the order ``proton in, quark in, quark
out, proton out'' appropriate for a scattering amplitude.}
\end{figure}

We can now give a systematic account of the helicity structure of GPDs
in terms of the matrix elements
\begin{eqnarray}
  \label{on-shell}
A^q_{\lambda'\mu', \lambda\mu} &=&
\int \frac{d z^-}{2\pi}\, e^{ix P^+ z^-}
  \langle p',\lambda'|\, \mathcal{O}^q_{\mu' \mu}(z)
  \,|p,\lambda \rangle \Big|_{z^+=0,\, \mathbf{z}=0} \: ,
\nonumber \\
A^g_{\lambda'\mu', \lambda\mu} &=&
\frac{1}{P^+} \int \frac{d z^-}{2\pi}\, e^{ix P^+ z^-}
  \langle p',\lambda'|\, \mathcal{O}^g_{\mu' \mu}(z)
  \,|p,\lambda \rangle \Big|_{z^+=0,\, \mathbf{z}=0} \: .
\end{eqnarray}
In Table~\ref{tab:helicities} we list the relevant operators
$\mathcal{O}_{\mu'\mu}$ for parton transitions in the region $x
\in[\xi,1]$, where the combinations $b^\dag(\mu') b(\mu)$ or
$a^\dag(\mu') a(\mu)$ are active.  The helicity transitions in the
other regions of $x$ are easily deduced from the correspondence
between outgoing left-helicity partons and incoming right-helicity
partons, as we saw in (\ref{quark-helicities}) and
(\ref{gluon-helicities}).  Note that the ordering of helicity labels
in (\ref{on-shell}) corresponds to Fig.~\ref{fig:order}a and differs
from the conventions for the familiar $s$-channel helicity amplitudes
in elastic scattering (where helicities are defined in the collision
c.m.).  The matrix elements (\ref{on-shell}) share important features
with helicity amplitudes, in particular they obey constraints
\begin{equation}
  \label{parity}
A_{-\lambda'-\mu', -\lambda-\mu} = (-1)^{\lambda'-\mu'-\lambda+\mu}\,
                 \Big[ A_{\lambda'\mu', \lambda\mu} \Big]^*
\end{equation}
from parity invariance.  The sign factor on the right-hand side
corresponds to our choice of spinors given in
Appendix~\ref{app:spinors}.

\begin{table}[b]
\caption{\label{tab:helicities} Operators $\mathcal{O}^{q}_{\mu'\mu}$
and $\mathcal{O}^{g}_{\mu'\mu}$ for leading-twist GPDs, and the parton
helicity transitions they describe in the region $x \in[\xi,1]$.  It
is implied that the first field is taken at $-\half z$ and the second
at $\half z$. The label $\mu$ refers to the emitted parton and $\mu'$
to the reabsorbed one.}
\vspace{1em}
\begin{center}
\leavevmode
{\renewcommand{\arraystretch}{1.3}
\begin{tabular}{cccc} 
\hline
 \multicolumn{2}{c}{operator} & \multicolumn{2}{c}{helicities} \\
 $\mathcal{O}^{q}_{\mu'\mu}$ & $\mathcal{O}^{g}_{\mu'\mu}$ &
 $\mu'$ & $\mu$ 
\\ \hline
 $\frac{1}{4}\bar{q} \gamma^+ (1+\gamma_5) q$ & 
   $\half [ G^{+\alpha}\, G_{\alpha}{}^{+}
         -i G^{+\alpha}\, \tilde{G}_{\alpha}{}^{+} ]$ & $+$ & $+$ \\
 $\frac{1}{4}\bar{q} \gamma^+ (1-\gamma_5) q$ &
   $\half [ G^{+\alpha}\, G_{\alpha}{}^{+}
         +i G^{+\alpha}\, \tilde{G}_{\alpha}{}^{+} ]$ & $-$ & $-$ \\
 $-\frac{1}{4}i \, \bar{q} (\sigma^{+1} - i\sigma^{+2}) q$ &
   $\half [ G^{+1} G^{1+} - G^{+2} G^{2+}       
         -i G^{+1} G^{2+} -i G^{+2} G^{1+} ]$ & $-$ & $+$ \\
 $\phantom{-}\frac{1}{4}i \, \bar{q} (\sigma^{+1} + i\sigma^{+2}) q$ &
   $\half [ G^{+1} G^{1+} - G^{+2}\, G^{2+}     
         +i G^{+1} G^{2+} +i G^{+2} G^{1+} ]$ & $+$ & $-$ \\
\hline
\end{tabular}
}
\end{center}
\end{table}

Explicitly we obtain parton helicity conserving matrix elements
\begin{eqnarray}
  \label{no-flip-amplitudes}
A_{++,++} &=& \sqrt{1-\xi^2} \left( \frac{H+\tilde{H}}{2} - 
            \frac{\xi^2}{1-\xi^2}\, \frac{E+\tilde{E}}{2} \right) , 
\nonumber \\
A_{-+,-+} &=& \sqrt{1-\xi^2} \left( \frac{H-\tilde{H}}{2} - 
            \frac{\xi^2}{1-\xi^2}\, \frac{E-\tilde{E}}{2} \right) , 
\nonumber \\
A_{++,-+} &=& - e^{-i\varphi} \,
              \frac{\sqrt{t_0-t}}{2m}\, \frac{E-\xi\tilde{E}}{2} , 
\nonumber \\
A_{-+,++} &=& e^{i\varphi} \,
              \frac{\sqrt{t_0-t}}{2m}\, \frac{E+\xi\tilde{E}}{2} ,
\end{eqnarray}
for both quarks and gluons, where $\varphi$ is the azimuthal angle of
the vector $D$ as before.  In the parton helicity flip sector we have
\cite{Diehl:2001pm}
\begin{eqnarray}
  \label{flip-amplitudes}
A^q_{++,+-} &=&  e^{i\varphi} \, 
                \frac{\sqrt{t_0-t}}{2m} \left( \tilde{H}_T^q
              + (1-\xi)\, \frac{E_T^q + \tilde{E}_T^q}{2} \right) , 
\nonumber \\
A^q_{-+,--} &=&  e^{i\varphi} \, 
                \frac{\sqrt{t_0-t}}{2m} \left( \tilde{H}_T^q
              + (1+\xi)\, \frac{E_T^q - \tilde{E}_T^q}{2} \right) , 
\nonumber \\
A^q_{++,--} &=& \sqrt{1-\xi^2} \left(H_T^q +
              \frac{t_0-t}{4 m^2}\, \tilde{H}_T^q -
              \frac{\xi^2}{1-\xi^2}\, E_T^q +
              \frac{\xi}{1-\xi^2}\, \tilde{E}_T^q \right) , 
\nonumber \\
A^q_{-+,+-} &=& - e^{2i\varphi}\,
                 \sqrt{1-\xi^2}\; \frac{t_0-t}{4 m^2}\, \tilde{H}_T^q
\end{eqnarray}
for quarks, and analogous expressions with an additional global factor
$e^{i\varphi} \, \sqrt{1-\xi^2}\, \sqrt{t_0-t \rule{0pt}{1.8ex}}
/(2m)$ on the right-hand side for gluons.

An essential feature of GPDs is that they can mediate transitions
where overall helicity is not conserved.  This is because angular
momentum along the $z$ axis is not given by the helicities alone, but
also receives an orbital contribution as soon as there is a transfer
of transverse momentum between the hadrons.  That the helicity
mismatch is compensated by orbital angular momentum is in fact
signaled by the prefactors $\sqrt{t_0-t}{}^{\,
|\lambda'-\mu'-\lambda+\mu|}$ in the above matrix elements, where each
factor of $\sqrt{t_0-t} \propto |\tvec{D}|$ corresponds to one unit of
angular momentum along $z$.  This observation makes it clear why GPDs
are sensitive to orbital angular momentum, which we will see again in
Sections~\ref{sec:spin} and
\ref{sub:overlap-formulae}.

\subsubsection{Counting GPDs}
\label{sub:counting}

The connection between GPDs and helicity amplitudes offers a
convenient method to determine the number of independent GPDs for
targets with arbitrary spin.  The method is well-known for the usual
parton distributions \cite{Jaffe:1996zw}, but must be modified in the
nonforward case \cite{Diehl:2001pm}.  Taking the proton GPDs as an
example, we start with $2^4 = 16$ helicity amplitudes for quark-proton
scattering, and an equal number for gluon-proton scattering.  The
parity constraints (\ref{parity}) relate amplitudes pairwise, which
for each parton species brings us to 8 independent amplitudes
parameterized by eight GPDs.  One easily sees that 4 of them flip
parton helicity and 4 do not.  It is important to realize that,
whereas time reversal invariance provides two further constraints for
ordinary $s$-channel helicity amplitudes, it does \emph{not} further
reduce the number of GPDs.\footnote{This was incorrectly assumed in
\protect\cite{Hoodbhoy:1998vm}, whose authors looked for 6 independent
GPDs and thus missed $\tilde{H}_T$ and $\tilde{E}_T$.}
The reason for this difference is that $s$-channel helicity amplitudes
kinematically depend on Lorentz invariants $s$ and $t$, which do not
change under time reversal.  In contrast, the helicity matrix elements
(\ref{on-shell}) refer to a particular light-cone direction, given by
the vector $n_-$ in the definition (\ref{quark-gpd-lorentz}), and
depend on light-cone fractions $x$ and $\xi$.  Time reversal changes
the sign of $\xi$ as seen in (\ref{time-rev}) and hence constrains
GPDs to be even or odd in $\xi$.  Together with the hermiticity
constraints (\ref{complex-conjug}) this fixes the phase of GPDs, but
does not set any of them to zero.  This happens only in the case
$\xi=0$ (which still allows for nonzero $t$).  {}From
(\ref{no-flip-amplitudes}) and (\ref{flip-amplitudes}) together with
(\ref{E-tilde-T}) and (\ref{parity}) we see that for vanishing $\xi$
one has indeed the time reversal constraints $A^q_{++,-+} = -
[A^q_{-+,++}]^*$ and $A^q_{++,+-} = - [A^q_{+-,++}]^*$ familiar for
$s$-channel helicity amplitudes.

The general ``counting rules'' for GPDs are then $(i)$ to determine
the total number of helicity transitions for a given parton species
and given hadron states, and then $(ii)$ to eliminate those
transitions which are dependent due to the parity constraints
(\ref{parity}).  This counting can be done separately for parton
helicity conserving or changing transitions.  Time reversal invariance
results in certain properties of the GPDs but does not further reduce
their number.  The same holds for charge conjugation invariance for
hadrons of definite charge parity like neutral pions: the $x$
dependence of the GPDs then corresponds to $C= +1$ exchange in the
$t$-channel as discussed in Section~\ref{sub:symm}.

Applying these rules to spin-zero targets we find that for each parton
species there is one parton helicity conserving GPD, given in
(\ref{pion-gpd}), and one parton helicity flip GPD, defined from the
tensor operators (\ref{tensor-operators}).  For a spin-zero target
both quark and gluon helicity flip GPDs decouple of course in the
forward limit.

The counting for the \emph{form factors} of the local twist-two matrix
elements associated with GPDs is yet different.  A general method has
been constructed by Ji and Lebed and applied to the nucleon
\cite{Ji:2000id}.  Instead of parity and time reversal invariance for
matrix elements $\langle p'| \mathcal{O} | p\rangle$ the authors used
parity and charge conjugation invariance for the matrix elements
$\langle \bar{p} p| \mathcal{O} |0 \rangle$ in the crossed channel.
Because of overall $CPT$ invariance, the two types of constraints are
equivalent and lead to the same structure for the allowed form factor
decompositions.  We note that the explicit results for the number of
form factors given in \cite{Ji:2000id} hold for the parity-even
operators $\mathcal{O}_q^{\mu \mu_1 \ldots \mu_n}$ and
$\mathcal{O}_g^{\mu \mu_1 \ldots \mu_n \nu}$ in (\ref{twist-two}) and
have to be appropriately modified for the other twist-two operators.


\subsection{The energy-momentum tensor and the spin of the nucleon}
\label{sec:spin}

A fundamental question about nucleon structure is how its total spin
is made up from quarks and gluons.  The polarized quark densities
measurable in DIS specify the distribution of quark spins aligned or
antialigned with the spin of the nucleon.  Evaluating their lowest
moments $\int dx\, [\Delta q(x) + \Delta\bar{q}(x)]$ and summing over
flavors one finds that the fraction of the total proton spin carried
by quarks and antiquarks is moderate, not more that 30\% to 50\%
depending on the renormalization scheme defining the quark
distributions, see e.g.~\cite{Windmolders:1999ra,Filippone:2001ux}.
The corresponding contribution $\int dx\, \Delta g(x)$ from the gluon
helicity is presently not well constrained by measurements.  An
important contribution of the proton spin may however be due to the
orbital angular momentum of partons, as proposed long ago by Sehgal
\cite{Sehgal:1974rz}.  That the parton splitting processes $q \to
q g$ and $g \to q\bar{q}$ responsible for the DGLAP evolution of
parton densities generate orbital angular momentum was observed by
Ratcliffe \cite{Ratcliffe:1987dp}.

The possibility to access orbital angular momentum through GPDs has
been realized by Ji~\cite{Ji:1997ek}.  Starting point of the argument
is the Belinfante improved energy-momentum tensor
\begin{equation}
T^{\mu\nu} = \sum_{i=q,g} T_i^{\mu\nu}
\end{equation}
with individual quark and gluon contributions
\begin{equation}
  \label{en-mom-ten}
T_q^{\mu\nu} = \bar{q} \gamma^{(\mu} i \lrD^{\nu)}\, q ,
\qquad\qquad
T_g^{\mu\nu} = G^{\mu\alpha} G_{\alpha}{}^{\nu}
  + {\textstyle\frac{1}{4}} g^{\mu\nu} G^{\alpha\beta} G_{\alpha\beta}
  \, ,
\end{equation}
where $t^{(\mu\nu)} = \half ( t^{\mu\nu} + t^{\nu\mu} )$ denotes
symmetrization for any tensor.  The angular momentum density is given
by
\begin{equation}
  \label{ang-mom-dens}
M^{\alpha\mu\nu} = T^{\alpha\nu} x^\mu - T^{\alpha\mu} x^\nu ,
\end{equation}
and the operator for angular momentum along the $z$ direction by
\begin{equation}
  \label{ang-mom-three}
\int d^3 x\, M^{012}(x) ,
\end{equation}
with analogous expressions for the individual quark and gluon
contributions.  Let us consider the expectation values $\langle
J^3_{q} \rangle$ and $\langle J^3_{q} \rangle$ of the operator
(\ref{ang-mom-three}) for a proton state at rest with spin along the
positive $z$ axis.  Ji has shown \cite{Ji:1998pf} that one obtains
identical expectation values if the proton moves along the positive or
negative $z$ direction, and also if instead of (\ref{ang-mom-three})
one takes the light-cone helicity operator
\begin{equation}
  \label{ang-mom-lc}
J^3 = \int d x^- d^2 \tvec{x}\, M^{+12}(x) ,
\end{equation}
which naturally arises in the context of light-cone quantization
\cite{Kogut:1970xa} and hence of the parton interpretation 
presented in Section \ref{sec:light-cone}.  The quark and gluon parts
of the energy-momentum tensor can be decomposed on form factors as
\begin{eqnarray}
	\label{en-mom-decomp}
\langle p' | T_{q,g}^{\mu\nu} | p \rangle &=&
  A_{q,g}(t)\, \bar{u} P^{(\mu} \gamma^{\nu)} u
  + B_{q,g}(t)\, \bar{u} 
    \frac{P^{(\mu}\, i\sigma^{\nu) \alpha} \Delta_\alpha}{2m} u
\nonumber \\
 && {}+ C_{q,g}(t)\, \frac{\Delta^\mu \Delta^\nu 
		- g^{\mu\nu} \Delta^2}{m} \bar{u} u
      + \bar{C}_{q,g}(t)\, m g^{\mu\nu}\, \bar{u} u ,
\end{eqnarray}
where for legibility we have omitted the momentum labels in the
spinors $\bar{u}$ and $u$ for the outgoing and incoming proton.  $A$,
$B$ and $C$ coincide with the form factors $A_{2,0}$, $B_{2,0}$ and
$C_2$ introduced in (\ref{Ji-decomposition}) for quarks and in
(\ref{glue-decomposition}) for gluons. The angular momentum carried by
each parton species is found to be
\begin{equation}
  \label{Ji-ang-mom}
\langle J^3_{q} \rangle = \half [ A_{q}(0) + B_{q}(0) ] ,
\qquad
\langle J^3_{g} \rangle = \half [ A_{g}(0) + B_{g}(0) ] ,
\end{equation}
and can be represented in terms of the moments of GPDs according to
our discussion in Section~\ref{sub:polynom} since
\begin{eqnarray}
  \label{Ji-sum}
A_{q}(t) + B_{q}(t) &=& 
	\int_{-1}^1 dx\, x \Big[ H_q(x,\xi,t) + E_q(x,\xi,t) \Big] ,
\nonumber \\
A_{g}(t) + B_{g}(t) &=& 
	\int_{0}^1 dx\, \Big[ H_g(x,\xi,t) + E_g(x,\xi,t) \Big] .
\end{eqnarray}
The combination of (\ref{Ji-ang-mom}) and (\ref{Ji-sum}) is often
referred to as Ji's sum rules \cite{Ji:1997ek}.  Notice their
similarity to the momentum sum rules
\begin{eqnarray}
  \label{momentum-sum}
\langle P^+_q \rangle &=& A_q(0) 	
	= \int_0^1 dx\, x \Big[ q(x) + \bar{q}(x) \Big] ,
\nonumber \\
\langle P^+_g \rangle &=& A_g(0) 	
	= \int_0^1 dx\, x g(x) 
\end{eqnarray}
where $P^+_{q,g} = \int d x^- d^2\, \tvec{x}\, T^{++}_{q,g}$ is the
plus-momentum operator.  Evaluating Ji's sum rules from data is highly
demanding: it requires knowledge of GPDs as functions of $x$ at fixed
$\xi$ and $t$, including the separation of the spin combinations $H$
and $E$, and of individual parton species.  These sum rules are
however the only known way to experimentally access the total angular
momentum carried by quarks or by gluons in the nucleon.

Several further remarks on Ji's sum rules can be made.
\begin{itemize}
\item  The individual contributions $\langle J^3_q \rangle$, $\langle
J^3_g \rangle$ depend on the scale $\mu$ at which the operators are
renormalized, i.e., on the scale at which quarks and gluons are
resolved.  Their evolution is exactly as the one for the momentum
fractions $\langle P^+_q \rangle$, $\langle P^+_g \rangle$, since the
$A$ and $B$ form factors belong to the same operator.  In particular
one asymptotically has
\begin{equation}
  \label{equipartition}
\langle J^3_q \rangle : \langle J^3_g \rangle  =
\langle P^+_q \rangle : \langle P^+_g \rangle  = 3 : 16 
\qquad \qquad \mbox{as $\mu\to \infty$}
\end{equation}
for any single quark flavor $q$.  At asymptotic scale the partition of
angular momentum between quarks and gluons hence becomes the same as
the partition of plus-momentum.
\item The sum rules for the total spin and momentum of the proton,
\begin{equation}
\sum_q \langle P^+_q \rangle + \langle P^+_g \rangle = 1 ,
\qquad
\sum_q \langle J^3_q \rangle + \langle J^3_g \rangle = \half ,
\end{equation}
imply together with (\ref{Ji-ang-mom}) and (\ref{momentum-sum}) that
\begin{equation}
  \label{magnetic-zero}
\sum_q B_q(0) + B_g(0) 
  = \sum_q \int_{-1}^1 dx\, x E^{q}(x,0,0) + \int_0^1 dx\, E^g(x,0,0)
  = 0 
\end{equation}
at any renormalization scale.  Brodsky et al.~\cite{Brodsky:2000ii}
have derived the same result using the wave function representation of
form factors discussed in Section~\ref{sec:overlap}.  Explicitly
calculating the $B$ form factors for an electron to one-loop accuracy
in QED (with electrons and photons taking the roles of quarks and
gluons) they confirmed the sum rule (\ref{magnetic-zero}).
Furthermore, they found that $\sum_q B_q(t) + B_g(t)$ at finite $t$
was nonzero, as well as the individual contributions from quarks and
gluons at $t=0$.  Nonzero values of $\sum_q B_q(0)$ and $B_g(0)$ are
hence not in conflict with general symmetries, contrary to a claim
made in \cite{Teryaev:1999su}.  Only at asymptotically large $\mu$
must $B_q(0)$ and $B_g(0)$ evolve to zero according
to~(\ref{equipartition}).

We note on the other hand that recent calculations in lattice QCD
\cite{Gockeler:2003jf,Hagler:2003jd} obtain small values for the
flavor summed combination $B_u(t) + B_d(t)$, due to large
cancellations between $B_u(t)$ and $B_d(t)$.  Such a behavior is also
predicted in the large-$N_c$ limit of QCD (see
Section~\ref{sub:large-Nc}).
\item An explicit decomposition $\langle J_q \rangle = \langle S_q
\rangle + \langle L_q \rangle$ into spin and orbital angular momentum
has been given by Ji~\cite{Ji:1997ek,Ji:1998pc}, with 
\begin{equation}
S^3_q = \half \int d^3 x\, {q}^\dag \sigma^{12}\, q , 
\qquad
L^3_q = - \int d^3 x\, {q}^\dag \Big( x^1 iD^2 - x^2 iD^1\Big) q .
\end{equation}
For the spin part one finds $\langle S^3_q \rangle = \half \int dx\,
[\Delta q(x) + \Delta \bar{q}(x)]$ as expected.  For gluons there is
no gauge invariant separation into spin and orbital angular momentum
operators, and Ji and collaborators \cite{Ji:1997ek,Hoodbhoy:1998bt}
have proposed to \emph{define} the orbital angular momentum part as
$\langle L^3_g \rangle = \langle J^3_g \rangle - \int dx\, \Delta
g(x)$, assuming the usual interpretation of $\int dx\, \Delta g(x)$ as
the total helicity carried by gluons in the proton.

We note that there is a controversy in the literature concerning the
question whether a decomposition of angular momentum operators needs
to be gauge invariant in the context of a parton interpretation, and
that there is a decomposition of $J^3$ into separate spin and orbital
contributions for both quarks and gluons, going back to Jaffe and
Manohar \cite{Jaffe:1990jz}.  For the arguments of the different sides
we refer to \cite{Hoodbhoy:1998yb,Hoodbhoy:1998bt} and
\cite{Jaffe:2000kr}.
\item  Using QCD sum rules, Balitsky and Ji estimated $\langle J^3_{g}
\rangle \approx 0.25$ at $\mu=1$~GeV for the gluon angular momentum in
the proton \cite{Balitsky:1997rs}.  Barone et al.\ obtained $\langle
J^3_{g} \rangle \approx 0.24$ at $\mu=0.5$~GeV in a quark model
\cite{Barone:1998dx}.  The quark angular momentum $\langle J^3_{q}
\rangle$ in the proton has been studied in lattice QCD by several
groups
\cite{Mathur:1999uf,Gadiyak:2001fe,Gockeler:2003jf,Hagler:2003jd}.
Comparing with lattice results for the contribution $\langle S^3_{q}
\rangle$ from quark helicity, one gets information about the
orbital part $\langle L^3_{q} \rangle$.  Whereas an early study by
Mathur et al.~\cite{Mathur:1999uf} suggests a substantial contribution
from quark orbital angular momentum, recent results by G\"ockeler et
al.~\cite{Gockeler:2003jf} and H\"agler et al.~\cite{Hagler:2003jd}
find its contribution to be very small when summing over flavors.  To
settle this issue will require better control over systematic errors
in the lattice calculations.
\item The axial anomaly of QCD leads to notorious problems in
separating quark and gluon helicity beyond leading order in
$\alpha_s$.  Bass \cite{Bass:2001dg} has pointed out that these
problems are yet more subtle in nonforward matrix elements, and that
nonforward analogs of the AB and JET schemes for spin dependent parton
densities cannot be defined in a gauge invariant way.
On the other hand, the operators and matrix elements for the total
angular momentum components $J^3_q$ and $J^3_g$ are \emph{not}
affected by the problems related with the axial anomaly, since they
involve the parity-even operators $T_q^{\alpha\beta}$ and
$T_g^{\alpha\beta}$.  This implies that anomaly related ambiguities in
renormalizing $\langle S^3_q \rangle$ must be compensated by
corresponding ambiguities in renormalizing $\langle L^3_q \rangle$, so
that they cancel in the sum $\langle J^3_q \rangle$.
\item As mentioned above the matrix elements $\langle J_{q,g}^3
\rangle$ can be understood as the expectation values of the light-cone
helicity (\ref{ang-mom-lc}) carried by quarks and gluons in a fast
moving proton.  Polyakov and Shuvaev
\cite{Polyakov:2002wz,Polyakov:2002yz} have elaborated on interpreting
the same quantities as the usual orbital angular momentum
(\ref{ang-mom-three}) in a proton at rest, using the same framework
that leads to the interpretation of the Sachs form factors $G_E$ and
$G_M$ at $t=0$ as the electric charge and magnetic moment of the
nucleon \cite{Sachs:1962aa}.  In this framework one can in addition
relate the form factors $C_{q,g}$ at $t=0$ with the traceless part of
the three-dimensional stress tensor $T^{ij}_{q,g}$ and thus to shear
forces experienced by quarks and gluons in a hadron
\cite{Polyakov:2002wz,Polyakov:2002yz}.  Representations of the form
factors $A_{q,g}$, $B_{q,g}$, $C_{q,g}$ at nonzero $t$ have also been
given in \cite{Polyakov:2002yz}.  It is not obvious whether a
light-cone equivalent can be found to illuminate the physical meaning
of the form factors $C_{q,g}(t)$.
\end{itemize}


\subsection{Generalized distribution amplitudes}
\label{sec:gda}

Let us now investigate generalized distribution amplitudes, which are
the crossed-channel analogs of GPDs \cite{Muller:1994fv,Diehl:1998dk}.
GDAs parameterize the matrix elements of the same light-cone operators
as GPDs, but not between two different hadron states, but between the
vacuum and a system of hadrons.  Much of the literature has focused on
pion pairs, which have a simpler spin structure that for instance the
$p\bar{p}$ system.  The quark and gluon GDAs for the $\pi^+\pi^-$
system are defined as \cite{Diehl:2000uv}
\begin{eqnarray}
  \label{pion-gda}
\Phi^q(z,\zeta,s) &=&
\int \frac{d x^-}{2\pi}\, e^{i (2z-1)\, (p+p')^+ x^- /2}
\nonumber \\
 && {}\times 
  {}_{\mathrm{out}} \langle 
	\pi^+(p) \pi^-(p')|\, \bar{q}(-\half x)\, \gamma^+ q(\half x)\, 
  \,|0 \rangle \Big|_{x^+=0,\, \mathbf{x}=0}  \: ,
\nonumber \\
\Phi^g(z,\zeta,s) &=&
\frac{1}{(p+p')^+} \int \frac{d x^-}{2\pi}\, 
        e^{i (2z-1)\, (p+p')^+ x^- /2}
\nonumber \\
 && {}\times 
  {}_{\mathrm{out}} \langle \pi^+(p) \pi^-(p')|\, 
     G^{+\mu}(-\half x)\, G_{\mu}{}^{+}(\half x)\, 
%
%
  \,|0 \rangle \Big|_{x^+=0,\, \mathbf{x}=0} \: ,
\end{eqnarray}
they are also called ``two-pion distribution amplitudes'' in the
literature.  Due to Lorentz invariance these functions depend on the
plus-momentum fraction $z$ of the quark with respect to $p+p'$, on the
fraction $\zeta = p^+ / (p+p')^+$ describing how the pions share their
plus-momentum (see Fig.~\ref{fig:gda}a), and on the invariant mass
$s=(p+p')^2$.  Our convention for $\Phi^g$ coincides with $\Phi^G$
defined in \cite{Kivel:1999sd}; the relation of our $\Phi^q$ with the
GDAs $\Phi^{I=0,1}$ with definite isospin of Polyakov
\cite{Polyakov:1998ze} is given in Section~III.A of
\cite{Diehl:2000uv}.

The close connection between GDAs and GPDs is obvious from their
respective definitions (\ref{pion-gda}) and (\ref{pion-gpd}).
Crossing the final-state $\pi^-(p')$ to an initial-state $\pi^+(p)$ we
must respectively identify $p$ and $p'$ in the GDA with $p'$ and $-p$
in the GPD, and thus identify the corresponding pairs of light-cone
variables in GDAs and GPDs as
\begin{equation}
  \label{gpd-gda-cross}
1-2\zeta \leftrightarrow \frac{1}{\xi} ,
\qquad
1-2z \leftrightarrow \frac{x}{\xi} .
\end{equation}
Note that in order to have the simple forward limit
(\ref{forward-quark}) one needs a prefactor $\half$ in the definition
of $H^q$, which conventionally does not appear in $\Phi^q$.  Notice
also different factors of $\half$ between the definitions in the gluon
sector.

Due to parity invariance there are no corresponding matrix elements of
the operators $\bar{q}\,\gamma^+ \gamma_5 q$ and $G^{+\mu}\,
\tilde{G}_{\mu}{}^{+}$.  There are however GDAs $\Phi^q_T$ and
$\Phi^g_T$ defined from the tensor operators $\bar{q} \sigma^{+i} q$
and $\mathbf{S} G^{+i} G^{j+}$ in (\ref{tensor-operators}), whose GPD
analogs flip parton helicity.  The GDAs $\Phi^q$ and $\Phi^g$ describe
a hadronic system with total light-cone helicity $J^3 =0$ (the
light-cone operators defining them carry zero $J^3$ as discussed in
Section~\ref{sec:light-cone}).  In contrast, $\Phi^q_T$ describes
hadrons in a $J^3 = \pm 1$ state, and $\Phi^g_T$ describes a $J^3 =
\pm 2$ state.  No process has so far been identified where one could
access the chiral-odd quantity $\Phi^q_T$.  The tensor gluon GDA
$\Phi^g_T$ does however appear in several processes, see
Sections~\ref{sub:meson-pair-prod} and
\ref{sec:transversity-access}.  For the two-pion system it was
introduced by Kivel et al.~\cite{Kivel:1999sd}, to which we refer for
its definition and general properties.

The definitions (\ref{pion-gda}) are reminiscent of those for the
leading-twist distribution amplitudes (DAs) of a single meson $M$ with
natural parity $P = (-1)^J$,
\begin{eqnarray}
  \label{vector-da}
\Phi^q(z) &=&
\int \frac{d x^-}{2\pi}\, e^{i (2z-1)\, p^+ x^- /2}
  \langle M(p) |\, \bar{q}(-\half x)\, \gamma^+ q(\half x)\, 
  \,|0 \rangle \Big|_{x^+=0,\, \mathbf{x}=0}  \: ,
\nonumber \\
\Phi^g(z) &=&
\frac{1}{p^+} \int \frac{d x^-}{2\pi}\, 
        e^{i (2z-1)\, p^+ x^- /2}
  \langle M(p)|\, G^{+\mu}(-\half x)\, G_{\mu}{}^{+}(\half x)\, 
  \,|0 \rangle \Big|_{x^+=0,\, \mathbf{x}=0} \: ,
\end{eqnarray}
where $M$ is in the polarization state with light-cone helicity zero.
Of course, the gluon DA only exists for $C = +1$ mesons, e.g. for an
$f_0$ or $f_2$ but not for a $\rho$.  All that changes between a DA
and a GDA is the replacement of a single meson by the state of several
hadrons, and the introduction of variables describing the internal
degrees of freedom of this state.  We already pointed out in
Section~\ref{sec:nutshell} the close analogy between $\gamma^*
\gamma^* \to \pi\pi$ and the simpler annihilation process $\gamma^*
\gamma^* \to \pi$.  In this case the relevant distribution amplitude
is defined by
\begin{eqnarray}
  \label{pseudo-da}
\Phi^q(z) &=&
\int \frac{d x^-}{2\pi}\, e^{i (2z-1)\, p^+ x^- /2}
  \langle M(p) |\, \bar{q}(-\half x)\, \gamma^+ \gamma_5\, q(\half x)\, 
  \,|0 \rangle \Big|_{x^+=0,\, \mathbf{x}=0}
\end{eqnarray}
as appropriate for states with unnatural parity $P = (-1)^{J+1}$.
Note that meson DAs are often defined with an additional factor $i$ in
(\ref{pseudo-da}), in order to match the standard definition of
pseudoscalar decay constants while having a real-valued function
$\Phi^q$.  The same convention is also sometimes taken for mesons with
other quantum numbers.  This factor $i$ corresponds to redefining the
phase of the meson state and is hence unobservable.

GDAs can also be defined for more complicated systems.  The case of
three pions has been studied in \cite{Pire:2000ky} and the case of two
$\rho$ mesons in \cite{Anikin:2003fr}.  The spin structure of GDAs for
baryon-antibaryon pairs can be parameterized as \cite{Diehl:2002yh}
\begin{eqnarray} 
\lefteqn{\int \frac{dx^-}{2\pi}\, e^{i (2z-1)\, (p+p')^+ x^- /2}\, 
{}_{\mathrm{out}} \langle\, p(p) \bar{p}(p') \,\big|\, 
   \bar{q}(-\half x)\, \gamma^+ q(\half x)\, 
  \,|0 \rangle \Big|_{x^+=0,\, \mathbf{x}=0}
} 
\nonumber\\
&=& \frac{1}{(p+p')^+} 
  \left[ \Phi^{q}_{V}\, \bar{u}(p) \gamma^+ v(p')
       + \Phi^{q}_{S}\, \frac{(p+p')^+}{2m}\, 
                                   \bar{u}(p) v(p') \right] , 
\nonumber\\
\lefteqn{\int \frac{dx^-}{2\pi}\, e^{i (2z-1)\, (p+p')^+ x^- /2}\, 
{}_{\mathrm{out}} \langle\, p(p) \bar{p}(p') \,\big|\, 
   \bar{q}(-\half x)\, \gamma^+\gamma_5\, q(\half x)\, 
  \,|0 \rangle \Big|_{x^+=0,\, \mathbf{x}=0}
}  
\nonumber\\
&=& \frac{1}{(p+p')^+} 
  \left[ \Phi^{q}_{A}\, \bar{u}(p) \gamma^+\gamma_5\, v(p')
       + \Phi^{q}_{P}\, \frac{(p+p')^+}{2m}\,\bar{u}(p) 
                          \gamma_5\, v(p') \right],
\label{GDA-proton}
\end{eqnarray}
for quarks, where for simplicity we have omitted helicity labels for
the baryons and the arguments $z,\zeta,s$ of the GDAs.  In analogy one
defines gluon GDAs, with $\bar{q} \gamma^+ q$ replaced by $G^{+\mu}\,
G_{\mu}{}^{+} /(p+p')^+$ and $\bar{q} \gamma^+\gamma_5\, q$ by $-i
G^{+\mu}\, \tilde{G}_{\mu}{}^{+} /(p+p')^+$.  We should remark on a
subtle point in the relation between proton GPDs and the $p\bar{p}$
distribution amplitudes just introduced.  Comparing the Lorentz
structures in (\ref{GDA-proton}) and (\ref{quark-gpd}), and taking
into account the change of momentum variables under crossing, we
recognize $\Phi_A$ and $\Phi_P$ as the respective counterparts of
$\tilde{H}$ and $\tilde{E}$.  In the vector channel one may use the
Gordon decomposition to trade the scalar current for the tensor one.
By crossing the defining relation for $H$ and $E$ one would obtain the
scalar current $\bar{u}(p) v(p')$ multiplied with $(p'-p)^+ =
(1-2\zeta) (p+p')^+$ instead of $(p+p')^+$.  A distribution amplitude
$\widehat\Phi_S$ defined with such a prefactor would however have an
artificial singularity at $\zeta=\frac{1}{2}$, since there is no
symmetry by which $(p'-p)^+
\,\widehat\Phi_S(z,\zeta,s) = (p+p')^+ \,\Phi_S(z,\zeta,s)$ has to
vanish at $p^+ = p'^+$.  This is in contrast to the definition of
$\tilde{E}$, where due to time reversal invariance the product
$(p'-p)^+ \tilde{E}$ \emph{is} zero for $p^+ = p'^+$.

\subsubsection{General properties}

Parity invariance has the same consequences for GPDs and for GDAs,
restricting the number of independent functions that parameterize a
matrix element and, as in the case of pions, constraining the matrix
elements of certain operators to vanish altogether.  If they describe
a particle-antiparticle system, GDAs are subject to further
constraints from charge conjugation invariance.  In the case of the
two-pion system they read
\begin{eqnarray}
  \label{C-conjugation}
\Phi^q(1-z,1-\zeta,s) &=& - \Phi^q(z,\zeta,s) , 
\nonumber \\
\Phi^g(1-z,1-\zeta,s) &=& \Phi^g(z,\zeta,s) .
\end{eqnarray}
In a similar fashion as for quark GPDs, it is useful to separate out
the quark GDA combinations
\begin{equation}
\Phi^{q (\pm)}(z,\zeta,s) = \frac{1}{2} \Big[ 
        \Phi^q(z,\zeta,s) \pm \Phi^q(z,1-\zeta,s) \Big] ,
\end{equation}
which respectively describe a $\pi\pi$ system in its $C= \pm 1$ state.
These combinations do not mix under evolution, and they are selected
in certain hard processes (although in some cases both combinations
contribute, see Section~\ref{sec:meson-lt}).  Since gluons are their
own antiparticles we further have $\Phi^g(z,1-\zeta,s) =
\Phi^g(z,\zeta,s)$, so that the two gluons only couple to the $C= +1$
state of the $\pi\pi$ system.  In the case of pions further
constraints are provided by isospin invariance
\cite{Polyakov:1998ze,Diehl:2000uv}, which relate the GDAs for
$\pi^+\pi^-$ and $\pi^0\pi^0$, as well as those for $u$ and $d$
quarks.  These relations are the exact analogs of the relations
between the GPDs of charged and neutral pions in
Section~\ref{sub:symm}.  The charge conjugation properties of
$p\bar{p}$ distribution amplitudes are given in \cite{Diehl:2002yh}.

In contrast to the case of GPDs, time reversal invariance does not
provide constraints on GDAs.\footnote{Notice however the similarity
between the time reversal constraints (\protect\ref{time-rev}) for
GPDs and the charge conjugation constraints
(\protect\ref{C-conjugation}) for GDAs, taking into account the
relation (\protect\ref{gpd-gda-cross}) between momentum fractions.
See also our comment at the end of
Section~\protect\ref{sub:counting}.}
In the matrix elements (\ref{pion-gda}) the two-pion state appears as
${}_{\mathrm{out}}\langle \pi^+\pi^+|$, in accordance to the physical
process where these matrix elements appear.  Time reversal transforms
this into $| \pi^+\pi^- \rangle_{\mathrm{in}}$, whereas complex
conjugation of the matrix element gives $| \pi^+\pi^-
\rangle_{\mathrm{out}}$.  ``In'' and ``out'' states are only
identical for states with a single stable particle, whereas a two-pion
state interacts and changes between the far past and the far future.
GDAs are hence not constrained to be real-valued functions, rather they
have dynamical phases reflecting the interactions within the
multiparticle system.  What time reversal tells us is that the GDA for
the transition $q\bar{q}\to \pi\pi$ is the same as for the transition
$\pi\pi \to q\bar{q}$.  Analogous statements hold for
proton-antiproton GDAs (with some fine print concerning minus signs,
see \cite{Diehl:2002yh}).  It is not obvious to find processes where
GDAs with ``in'' states appear, but they do have applications as we
will see in Section~\ref{sec:large-s}.

According to the relation (\ref{higher-moments}), polynomial moments
in $(2z-1)^n$ of the GDAs in (\ref{pion-gda}) give matrix elements of
the local twist-two operators (\ref{twist-two}) between the vacuum and
the two-pion state.  These moments are polynomials in $(2\zeta -1)$
whose coefficients are form factors corresponding to the ones
describing the $x^n$ moments of the GPDs (\ref{pion-gpd}). The form
factors of local operators have known analyticity properties in their
Mandelstam variables, given by $s$ for GDAs and by $t$ for GPDs.  This
makes explicit how GDAs and GPDs are related by crossing: they are
connected by analytic continuation in $s$ or $t$ of their polynomial
moments in $2z-1$ or $x$ \cite{Polyakov:1998ze,Polyakov:1999gs}.  In
particular, the lowest moments $\int_0^1 dz\, \Phi^q(z,\zeta,s)$ and
$\int_{-1}^1 dx\, H^q(x,\xi,t)$ are respectively proportional to the
pion form factor $F_\pi$ in the timelike and in the spacelike region.

\subsubsection{Evolution and expansion on polynomials}
\label{sub:gda-evolution}

As any matrix element of light-cone separated operators, GDAs depend
on a factorization scale $\mu^2$.  Their evolution in this scale is
the same as the one for singe-particle DAs with the corresponding
quantum numbers, and therefore given by the ERBL equations
\cite{Efremov:1980qk,Lepage:1979zb}.  We will 
discuss these in Section~\ref{sec:evolution} and give here only some
results of specific interest for GDAs.  We restrict ourselves to
evolution at leading logarithmic accuracy, i.e., to the leading-order
approximation in $\alpha_s$ of evolution kernels and anomalous
dimensions.  The variables $s$ and $\zeta$ concern the hadronic system
but not the partons and are passive in the evolution equations, which
involve only $\mu^2$ and $z$.  The evolution can be explicitly solved
by an expansion in Gegenbauer polynomials $C^k_n(2z-1)$.  For GDAs a
double expansion is useful \cite{Polyakov:1998ze,Kivel:1999sd}:
\begin{eqnarray}
  \label{Polyakov-expansion}
\Phi^{q(-)}(z,\zeta,s;\mu^2) &=&
  6z(1-z) \sum_{n=0 \atop \mathrm{even}}^\infty 
        \sum_{l=1 \atop \mathrm{odd}}^{n+1} 
        B^q_{nl}(s,\mu^2) C_n^{3/2}(2z-1) P_l(2\zeta-1)
\nonumber \\
\Phi^{q(+)}(z,\zeta,s;\mu^2) &=&
  6z(1-z) \sum_{n=1 \atop \mathrm{odd}}^\infty 
        \sum_{l=0 \atop \mathrm{even}}^{n+1} 
        B^q_{nl}(s,\mu^2) C_n^{3/2}(2z-1) P_l(2\zeta-1)
\nonumber \\
\Phi^g(z,\zeta,s;\mu^2) &=&
  9z^2 (1-z)^2 \sum_{n=1 \atop \mathrm{odd}}^\infty 
        \sum_{l=0 \atop \mathrm{even}}^{n+1} 
        B^g_{nl}(s,\mu^2) C_{n-1}^{5/2}(2z-1) P_l(2\zeta-1) .
\end{eqnarray}
Our normalization of the gluon coefficients $B^g_{nl}$ has been chosen
to obtain a simple form of the energy-momentum relation in
(\ref{pion-fraction}) and differs from the one in \cite{Kivel:1999sd}
by a factor of $\frac{3}{10}$.  The coefficients $B^{q,g}_{nl}$ used
here should not be confused with the form factors $B^{q,g}_{n,i}$
introduced in (\ref{Ji-decomposition}) and (\ref{glue-decomposition}).

Before discussing the evolution properties of the expansion
coefficients $B^q_{nl}$ and $B^g_{nl}$ let us elaborate on the
expansion itself.  The restrictions on $n$ and $l$ to be odd or even
implement the charge conjugation constraints discussed above, and the
restriction $l \le n+1$ reflects the polynomiality property of GDAs,
which is derived in analogy to the one for GPDs.  The expansion in
Legendre polynomials $P_l(2\zeta-1)$ is motivated by the following: in
the rest frame of the two-pion system one has $2\zeta-1 = \beta
\cos\theta$, where $\beta = (1-4 m_\pi^2 /s)^{1/2}$ is the relativistic
velocity of each pion and $\theta$ the polar angle of the $\pi^+$ with
respect to the $z$-axis defining the light-cone direction in the GDAs.
In the limit $\beta\to 1$ the coefficients $B^q_{nl}$ and $B^g_{nl}$
are therefore associated with the $l$th partial wave of the two-pion
system.  An exact partial wave decomposition is obtained by
rearranging the finite sum $\sum_l B_{nl} P_l(2\zeta-1)$ for given $n$
into $\sum_l \tilde{B}_{nl} P_l(\cos\theta)$,
see~\cite{Polyakov:1998ze,Diehl:2000uv}.  One then finds in particular
the usual phase space suppression $\tilde{B}_{nl} \propto
\beta^l$ for each partial wave.

The expansion (\ref{Polyakov-expansion}) also allows one to obtain
information about the phases of GDAs
\cite{Polyakov:1998ze,Diehl:2000uv}.  In the $s$ region where $\pi\pi$
scattering is strictly elastic, ``in'' and ``out'' two-pion states in
a definite partial wave $l$ are simply related by the phase shifts as
$|\pi\pi(l)\rangle_{\mathrm{in}} = e^{2i \delta_l}
|\pi\pi(l)\rangle_{\mathrm{out}}$.  The relations from time reversal
discussed above then fix the phase of the expansion coefficients to be
$\tilde{B}_{nl} = \eta_{nl}\, e^{i \delta_l} |\tilde{B}_{nl}|$, up to
a sign factor $\eta_{nl} =\pm 1$.  In analogy to a different context
this is called Watson's theorem \cite{Watson:1954uc}.  Since
low-energy $\pi\pi$ scattering is rather well studied experimentally
and known to be approximately elastic up to $s\sim 1$~GeV$^2$,
information about the phases of the two-pion DAs is available in this
region.  The dynamical phases of GDAs can have characteristic
signatures in experimental observables (see Sections~\ref{sec:gaga}
and \ref{sub:meson-pair-pheno}).

To leading logarithmic accuracy, the coefficients $B^q_{nl}$ of
$\Phi^{q(-)}$ renormalize multiplicatively, so that one has
\begin{equation}
  \label{mult-renorm}
B^q_{nl}(s,\mu^2) = B^q_{nl}(s,\mu_0^2) 
  \left(\frac{\alpha_s(\mu^2)}{\alpha_s(\mu_0^2)}
        \right)^{2\gamma_n /\beta_0} 
\qquad\qquad \mbox{for even $n$}
\end{equation}
where $\beta_0 = 11 - \frac{2}{3} n_f$ for $n_f$ quark flavors, and
where the $\gamma_n$ are the same anomalous dimensions which govern
the evolution of the DA for a single $\rho$ or $\pi$
\cite{Efremov:1980qk,Lepage:1979zb},
\begin{equation}
  \label{quark-gamma}
\gamma_n = \frac{4}{3} \left( \frac{1}{2} - \frac{1}{(n+1) (n+2)}
           + 2 \sum_{k=2}^{n+1} \frac{1}{k} \right) .
\end{equation}
Numerically one has $\gamma_n \approx \frac{8}{3} \log(n+1)$ within at
most 6\% for all $n$.  For the difference of $\Phi^{q(+)}$ between any
two quark flavors $q_1$ and $q_2$ one has accordingly
\begin{equation}
\Big[ B^{q_1}_{nl}(s,\mu^2)   - B^{q_2}_{nl}(s,\mu^2) \Big]= 
\Big[ B^{q_1}_{nl}(s,\mu_0^2) - B^{q_2}_{nl}(s,\mu_0^2) \Big]
  \left(\frac{\alpha_s(\mu^2)}{\alpha_s(\mu_0^2)}
        \right)^{2\gamma_n /\beta_0} 
\end{equation}
for odd $n$.  The flavor singlet combination $\sum_q \Phi^{q(+)}$ and
$\Phi^g$ mix under evolution, and for odd $n$
\begin{eqnarray}
  \label{mult-renorm-singlet}
\frac{1}{n_f} \sum_{q}^{n_f} B^q_{nl}(s,\mu^2) &=& 
   B_{nl}^{+} \left(\frac{\alpha_s(\mu^2)}{\alpha_s(\mu_0^2)}
        \right)^{2\gamma^+_n /\beta_0} +
   B_{nl}^{-} \left(\frac{\alpha_s(\mu^2)}{\alpha_s(\mu_0^2)}
        \right)^{2\gamma^-_n /\beta_0} ,
\nonumber \\
B^g_{nl}(s,\mu^2) &=& 
   a^+_n B_{nl}^{+} \left(\frac{\alpha_s(\mu^2)}{\alpha_s(\mu_0^2)}
        \right)^{2\gamma^+_n /\beta_0} +
   a^-_n B_{nl}^{-} \left(\frac{\alpha_s(\mu^2)}{\alpha_s(\mu_0^2)}
        \right)^{2\gamma^-_n /\beta_0} .
\end{eqnarray}
The anomalous dimensions $\gamma^\pm$ and mixing coefficients
$a^\pm_n$ were first calculated in \cite{Chase:1980hj,Baier:1982aa},
results corresponding to the the definitions of GDAs used here are
given in \cite{Diehl:2000uv}.  For $n=1$ one explicitly has
\begin{eqnarray}
\gamma^-_1 = 0 , & \qquad & \gamma^+_1 = \frac{16}{9} + \frac{1}{3} n_f
\nonumber \\
a^-_1 = \frac{16}{3} , & & a^+_1 = - n_f .  
\end{eqnarray}
Note that the superscripts $\pm$ in this notation do not refer to the
$C$ parity, but label the two eigenvalues of the anomalous dimension
matrix for the mixing between the quark flavor singlet and gluons.

All of the above anomalous dimensions above are positive except for
$\gamma_0^{\phantom{-}} = \gamma^-_1 =0$, so the asymptotic forms for
the GDAs are
\begin{eqnarray}
  \label{asy-gda}
\Phi^{q(-)}(z,\zeta,s;\mu^2) &\stackrel{\mu\to\infty}{\to}&
  6z(1-z) (2\zeta-1) B^q_{01}(s),
\phantom{\Big[ \Big]}
\nonumber \\
\Phi^{q(+)}(z,\zeta,s;\mu^2) &\stackrel{\mu\to\infty}{\to}&
  18z(1-z)(2z-1) \Big[  B^-_{10}(s) + B^-_{12}(s) P_2(2\zeta-1) \Big] ,
\nonumber \\
\Phi^g(z,\zeta,s;\mu^2) &\stackrel{\mu\to\infty}{\to}&
  48z^2 (1-z)^2 \Big[  B^-_{10}(s) + B^-_{12}(s) P_2(2\zeta-1) \Big] ,
\end{eqnarray}
where $P_2(2\zeta-1) = 1 - 6\zeta(1-\zeta)$.  The coefficients
$B^q_{01}$ and $B^-_{1l}$ are form factors of conserved currents and
hence $\mu$-independent.  $B^q_{01}$ belongs to the vector current
$\bar{q}\gamma^\mu q$ and is just given by the timelike
electromagnetic pion form factor as $B^u_{01} = - B^d_{01} =
F_\pi(s)$.  The coefficients $B^-_{1l}$ belong to the total
energy-momentum tensor $T^{\mu\nu}$ (summed over all quark flavors and
gluons).  In contrast, $B^q_{1l}$ and $B^g_{1l}$ are associated with
the individual contributions of quarks and gluons to $T^{\mu\nu}$ and
do evolve with $\mu$.  Further information can be obtained by analytic
continuation to the point $s=0$, which corresponds to the forward
matrix element $\langle\pi^+ |T^{\mu\nu}|
\pi^+\rangle$.  The normalization of the energy-momentum tensor
provides the identification 
\begin{equation}
  \label{pion-fraction}
B^q_{12}(0,\mu^2) = \frac{9}{10}\, A_q(0,\mu^2) , \qquad
B^g_{12}(0,\mu^2) = \frac{9}{10}\, A_g(0,\mu^2) ,
\end{equation}
where $A_q(0) = \int_0^1 dx\, x [ q(x) + \bar{q}(x) ]$ is the fraction
of plus-momentum carried by quarks and antiquarks of flavor $q$ in the
pion and $A_g(0) = \int_0^1 dx\, x g(x)$ its analog for gluons.
Asymptotically one has the well-known limit $A^q(0,\mu^2)
\stackrel{\mu\to\infty}{\to} 3 /(3n_f + 16)$.

The proton GDAs $\Phi^q_V$ and $\Phi^q_S$ and their analogs for gluons
are defined with the same operators as the corresponding two-pion GDAs
and hence follow the same evolution equations.  Their expansion in
$p\bar{p}$ partial waves is similar to (\ref{Polyakov-expansion}) but
differs in detail because of spin (see Section~\ref{sub:moments}).
The evolution of the parity-odd GDAs $\Phi^q_A$ and $\Phi^q_P$ in the
nonsinglet sector (i.e.\ for the $C$-odd combinations or for the
difference of two quark flavors) is the same as the nonsinglet
evolution of $\Phi^q_V$ and $\Phi^q_S$, since the anomalous dimensions
for vector and axial vector quark operators are identical at leading
order in $\alpha_s$.  For gluons and their mixing with quarks the
anomalous dimensions differ in the parity-even and odd sectors.  In
particular, the only vanishing anomalous dimension in the parity-odd
case is $\gamma_0$, belonging now to the axial current
$\bar{q}\gamma^+\gamma_5 q$.  This leads to an asymptotic form going
like $z(1-z)$ for the $C= +1$ combinations of quark GDAs.  The $C= -1$
combinations of quark GDAs and the GDAs for gluons tend to zero when
evolved to $\mu\to \infty$ (the same holds for the tensor GDAs for
quarks or gluons briefly discussed above).  The anomalous dimensions
and mixing coefficients in the parity-odd sector can e.g.\ be found in
\cite{Kroll:2002nt}, which includes a comparison with earlier results
in the literature.


\subsection{Evolution of GPDs}
\label{sec:evolution}

GPDs depend on a factorization scale $\mu^2$, which technically arises
as the renormalization scale of the bilocal operators in their
definitions and physically represents the scale at which the partons
are resolved.  Early results on evolution kernels for nonforward
kernels were given by Gribov, Levin and Ryskin (Section 6.2.2.1 of
\cite{Gribov:1983tu}), but their formulae contain mistakes
\cite{Martin:1998wy}.  Detailed investigations of the leading-order
(LO) evolution equations and kernels for GPDs have been carried out by
the Leipzig group
\cite{Geyer:1985vw,Braunschweig:1986nr,Dittes:1988xz} and led to the
classical paper \cite{Muller:1994fv} on GPDs and nonforward Compton
scattering.  Later work was done by Ji \cite{Ji:1997nm}, by Radyushkin
and Balitsky
\cite{Radyushkin:1997ki,Balitsky:1997mj,Radyushkin:1998es}, and by
Bl\"umlein et al.~\cite{Blumlein:1997pi,Blumlein:1999sc}.  Anomalous
dimensions and kernels at next-to-leading order (NLO) accuracy have
been obtained in a series of papers by Belitsky et al.\
\cite{Belitsky:1998vj,Belitsky:1998gc,Belitsky:1999gu,Belitsky:1999fu,Belitsky:1999hf}.
The evolution of gluon transversity GPDs was studied in
\cite{Hoodbhoy:1998vm,Belitsky:2000jk}, and the kernels for gluon and
quark transversity at NLO accuracy have been given in
\cite{Belitsky:2000yn}.  We note that the leading-order nonforward
kernels relevant for GPDs were also studied by Bukhostov et
al.~\cite{Bukhvostov:1985rn} as building blocks for the evolution of
higher-twist parton distributions.

\subsubsection{Evolution of light-cone operators}
\label{sub:light-cone-evolution}

To understand the intimate relationship between the evolution for
GPDs, the one of usual parton distributions, and the one for meson DAs
or GDAs, it is useful to consider the renormalization group evolution
of the defining operators themselves.  This can be done for local
operators (see Section~\ref{sub:solve-evolution}) but also directly
for the nonlocal operators
\begin{eqnarray}
  \label{string-operators}
\mathcal{O}^q(\kappa_1,\kappa_2) &=& 
  z_\alpha\, \bar{q}(\kappa_1 z)\, 
	\gamma^\alpha q(\kappa_2 z) \Big|_{z^2 =0} \: ,
\nonumber \\
\mathcal{O}^g(\kappa_1,\kappa_2) &=& z_\alpha z_\beta\,
   G^{\alpha\mu}(\kappa_1 z)\, 
	G_{\mu}{}^{\beta}(\kappa_2 z) \Big|_{z^2 =0} \: ,
\nonumber \\
\tilde\mathcal{O}^q(\kappa_1,\kappa_2) &=& 
  z_\alpha\, \bar{q}(\kappa_1 z)\, \gamma^\alpha \gamma_5\, 
	q(\kappa_2 z) \Big|_{z^2 =0} \: ,
\nonumber \\
\tilde\mathcal{O}^g(\kappa_1,\kappa_2) &=& z_\alpha z_\beta\, (-i)
   G^{\alpha\mu}(\kappa_1 z)\, 
	\tilde{G}_{\mu}{}^{\beta}(\kappa_2 z) \Big|_{z^2 =0} \: .
\end{eqnarray}
The $C$-odd quark sector evolves without mixing with gluons and one
has the nonsinglet evolution equation \cite{Balitsky:1989bk}
\begin{eqnarray}
	\label{op-evol-ns}
\mu^2 \frac{d}{d \mu^2} \Big[ \mathcal{O}^q(\kappa_1,\kappa_2) 
   + \mathcal{O}^q(\kappa_2,\kappa_1) \Big] 
&=&   \int_0^1 d\alpha_1 \int_0^1 d\alpha_2 \, 
	K_{\mathrm NS}(\alpha_1,\alpha_2)\, 
	\Big[ \mathcal{O}^q(\kappa_1',\kappa_2') +
	  \mathcal{O}^q(\kappa_2',\kappa_1') \Big]
\end{eqnarray}
with
\begin{equation}
\kappa_1' = \kappa_1 (1-\alpha_1) + \kappa_2\, \alpha_1 ,
\qquad
\kappa_2' = \kappa_1 \alpha_2 + \kappa_2\, (1-\alpha_2) ,
\end{equation}
where for simplicity we omit $\mu$ as an explicit argument of the
light-cone operators or of their matrix elements.  An analogous
equation holds for the $C$-odd operators
$\tilde\mathcal{O}^q(\kappa_1,\kappa_2) -
\tilde\mathcal{O}^q(\kappa_2,\kappa_1)$, and also for flavor  
differences $\mathcal{O}^{q_1} - \mathcal{O}^{q_2}$ and
$\tilde\mathcal{O}^{q_1} - \tilde\mathcal{O}^{q_2}$ of quark operators
in the $C$-even sector.  To leading order in $\alpha_s$ the evolution
kernel is the same for all cases and reads
\cite{Balitsky:1989bk,Blumlein:1997pi}
\begin{eqnarray}
K_{\mathrm NS}(\alpha_1,\alpha_2) &=&
\frac{\alpha_s}{2\pi} C_F\, \theta(\alpha_1+\alpha_2 \le 1) \,
  \Bigg\{ 1 + \frac{3}{2}\, \delta(\alpha_1) \delta(\alpha_2) 
  - \delta(\alpha_1) - \delta(\alpha_2)
\nonumber \\
&& \hspace{3em} {}+ 
  \delta(\alpha_1) \Bigg[ \frac{1}{\alpha_2} \Bigg]_+
+ \delta(\alpha_2) \Bigg[ \frac{1}{\alpha_1} \Bigg]_+
\Bigg\} ,
\end{eqnarray}
where $C_F = \frac{4}{3}$ and the plus distribution is defined as
\begin{equation}
\int_0^1 d\alpha\, f(\alpha) \Bigg[ \frac{1}{\alpha} \Bigg]_+ 
= \int_0^1 d\alpha\, \frac{f(\alpha) - f(0)}{\alpha} .
\end{equation}
In the singlet sector quarks mix with gluons, and one has a matrix
equation
\begin{equation}
\mu^2 \frac{d}{d \mu^2} 
	\mbox{\boldmath{$\mathcal{O}$}}(\kappa_1,\kappa_2) 
= \int_0^1 d\alpha_1 \int_0^1 d\alpha_2 \, 
	K(\alpha_1,\alpha_2,\kappa_2-\kappa_1)\, 
  \mbox{\boldmath{$\mathcal{O}$}}(\kappa_1',\kappa_2') 
\end{equation}
with
\begin{eqnarray}
\mbox{\boldmath{$\mathcal{O}$}}(\kappa_1,\kappa_2) &=&
\left( \begin{array}{c}
	(2 n_f)^{-1}\, {\displaystyle \sum_{q}^{n_f}} 	
	   \Big[ \mathcal{O}^q(\kappa_1,\kappa_2) 
               + \mathcal{O}^q(\kappa_2,\kappa_1) \Big] \\
	\mathcal{O}^g(\kappa_1,\kappa_2) 
\end{array} \right) ,
\end{eqnarray}
and
{\renewcommand{\arraystretch}{1.3}
\begin{equation}
  \label{evol-matrix-s}
K =
\left( \begin{array}{cc}
	K^{qq}(\alpha_1,\alpha_2) & 
  	   -i (\kappa_2-\kappa_1) K^{qg}(\alpha_1,\alpha_2)  \\
        i (\kappa_2-\kappa_1)^{-1} K^{gq}(\alpha_1,\alpha_2) &
	   K^{gg}(\alpha_1,\alpha_2)
\end{array} \right) .
\end{equation}
}
The dependence of the off-diagonal elements on the light-cone distance
$\kappa_2-\kappa_1$ can be traced back to the different powers of $z$
appearing in the quark and gluon operators (\ref{string-operators}).
An analogous equation describes the singlet evolution for the
parity-odd operators $\tilde\mathcal{O}^q$ and $\tilde\mathcal{O}^g$.
At LO (but not beyond) one has $\tilde{K}^{qq} = K^{qq} =
K_{\mathrm{NS}}$, whereas the kernels ${K}^{gg}$, ${K}^{gq}$,
${K}^{qg}$ differ from those in the parity-odd sector already at LO.
Explicit forms of the LO kernels for the parity-even operators can be
found in \cite{Balitsky:1989bk,Radyushkin:1997ki}, and for the
parity-odd ones in \cite{Balitsky:1997mj,Blumlein:1999sc}.  Evolution
equations analogous to (\ref{op-evol-ns}) but with different kernels
hold for the operators describing quark or gluon transversity (see
Section~\ref{sub:helicity-flip}), which again evolve by themselves
without mixing.

\begin{figure}
\begin{center}
     \leavevmode
     \epsfxsize=0.2\textwidth
     \epsffile{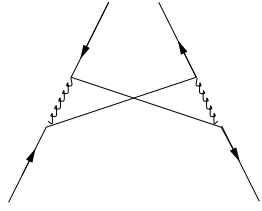}
\end{center}
\caption{\label{fig:q-qbar-mixing} A diagram contributing
to the region $\alpha_1 + \alpha_2 >1$ of the evolution kernel
$K_{\mathrm{NS}}$ at $O(\alpha_s^2)$.}
\end{figure}

The support properties of the evolution kernels can be determined
using the $\alpha$-parameter representation of Feynman diagrams and
are $0\le \alpha_1 \le 1$ and $0\le \alpha_2 \le 1$
\cite{Muller:1994fv}.  The region $\alpha_1 + \alpha_2
>1$ (which corresponds to opposite orientation along the light-cone of
the operators on the two sides of the evolution equation) only appears
in $K_{\mathrm{NS}}$ at two-loop order and comes from diagrams
``mixing'' quarks and antiquarks as in Fig.~\ref{fig:q-qbar-mixing}
\cite{Belitsky:1998uk}.

\subsubsection{Evolution of GPDs}
\label{sub:gpd-evolution}

The evolution equations of the light-cone operators readily generate
the various evolution equations for their matrix elements and thus for
GPDs.  {}From the definitions (\ref{quark-gpd}) one obtains in the
$C$-odd sector
\begin{equation}
  \label{gpd-evol-ns}
\mu^2 \frac{d}{d \mu^2} H^{q(-)}(x,\xi,t) =
  \int_{-1}^1 dx'\, \frac{1}{|\xi|} 
     V_{\mathrm{NS}}\Big( \frac{x}{\xi}, \frac{x'}{\xi} \Big) 
	H^{q(-)}(x',\xi,t) .
\end{equation}
Analogous equations hold for $E^{q(-)}$, $\tilde{H}^{q(-)}$,
$\tilde{E}^{q(-)}$ and for the difference of quark distributions for
two flavors in the $C$-even sector, with identical kernels at
$O(\alpha_s)$.  Notice that both $x$ and $\xi$ are active variables in
the evolution equation since they specify the parton momenta, whereas
the momentum fraction $\zeta$ for GDAs is passive in the evolution.
The evolution kernel is obtained as
\begin{eqnarray}
  \label{kernel-relation}
\lefteqn{
\frac{1}{|\xi|} V_{\mathrm{NS}}
	\Big( \frac{x}{\xi}, \frac{x'}{\xi} \Big) 
}
\nonumber \\
&=& \int_0^1 d\alpha_1 \int_0^1 d\alpha_2\, 
	\delta\Big(x - x' (1-\alpha_1-\alpha_2)
	- \xi (\alpha_1-\alpha_2) \Big) 
		K_{\mathrm{NS}}(\alpha_1,\alpha_2)
\end{eqnarray}
and to LO accuracy reads
\begin{eqnarray}
  \label{LO-quark-kernel}
\lefteqn{
V_{\mathrm{NS}}(x,x')
}
\nonumber \\
 &=& \frac{\alpha_s}{4\pi} C_F\, \Bigg[ \rho(x,x')
	\Bigg\{ \frac{1+x}{1+x'} \Bigg( 1 + \frac{2}{x'-x} \Bigg)
			\Bigg\}
 + \{ x\to -x, x'\to -x' \} \Bigg]_{+}
\end{eqnarray}
with its support controlled by
\begin{equation}
  \rho(x,x') = \theta(x'\ge x \ge -1) - 
			\theta(x'\le x \le -1) ,
\end{equation}
where we use a shorthand notation $\theta(x \le y \le z) =
\theta(y-x)\, \theta(z-y)$.  The plus-distribution is here defined
with respect to the first argument, i.e.,
\begin{equation}
  \label{plus-def}
\Big[ f(x,x') \Big]_+ = f(x,x') 	
  - \delta(x-x') \int dx''\, f(x'', x') ,
\end{equation}
where the integral on the right-hand side is over the full support
region in $x''$, which is finite thanks to the function $\rho$ in the
kernels.  One finds that this prescription also makes the integral
over the second argument finite, which is needed in the evolution
equation (\ref{gpd-evol-ns}).  {}From (\ref{gpd-evol-ns}),
(\ref{LO-quark-kernel}) and (\ref{plus-def}) one directly sees that
$\int dx\, H^q(x)$ is independent of $\mu^2$, as befits the form
factors of conserved currents.

\begin{figure}
\begin{center}
     \leavevmode
     \epsfxsize=0.32\textwidth
     \epsffile{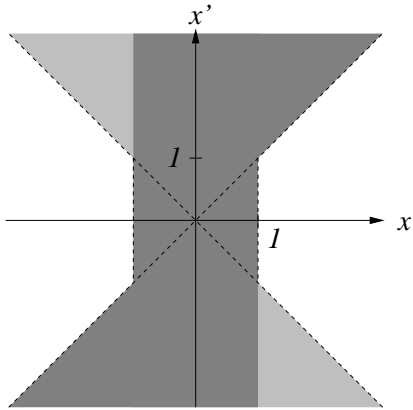}
\end{center}
\caption{\label{fig:evol-support} Support region of the quark
evolution kernel $V_{\mathrm{NS}}(x,x')$.  The dark shaded region
corresponds to the LO kernel, and the light shaded region starts
contributing at NLO.}
\end{figure}

The support region of the kernels for GPDs follows from the support of
the kernels for light-cone operators and is as shown in
Fig.~\ref{fig:evol-support} \cite{Muller:1994fv}.  The leading-order
result (\ref{LO-quark-kernel}) corresponds to the dark-shaded region.
The light-shaded region starts contributing at NLO and corresponds to
$\alpha_1+\alpha_2>1$ in the position-space kernels, i.e., to diagrams
like in Fig.~\ref{fig:q-qbar-mixing} which in momentum space describe
radiation of an antiquark from a quark \cite{Belitsky:1998uk}.

The evolution equation in the singlet sector is obtained from
(\ref{gpd-evol-ns}) by replacing
\begin{equation}
H^{q(-)} \to
\left( \begin{array}{c}
	(2 n_f)^{-1}\, {\displaystyle \sum_{q}^{n_f}} H^{q(+)} \\
	H^g
\end{array} \right)
\end{equation}
and \cite{Belitsky:1999hf}
\begin{equation}
V_{\mathrm{NS}} \Big(\frac{x}{\xi}, \frac{x'}{\xi}\Big) \to
\left( \begin{array}{cc} 
	\displaystyle
        V^{qq}\Big(\frac{x}{\xi}, \frac{x'}{\xi}\Big) &
          \displaystyle
	  \frac{1}{\xi}\, V^{qg}\Big(\frac{x}{\xi}, \frac{x'}{\xi}\Big) 
         	\\[1em]
	\displaystyle
	\xi\, V^{gq}\Big(\frac{x}{\xi}, \frac{x'}{\xi}\Big) &
          \displaystyle
          V^{gg}\Big(\frac{x}{\xi}, \frac{x'}{\xi}\Big) 
\end{array} \right) .
\end{equation}
The kernels $V^{qq}$ and $V^{gg}$ are obtained as in
(\ref{kernel-relation}), whereas due to the factors
$(\kappa_2-\kappa_1)$ and $(\kappa_2-\kappa_1)^{-1}$ in the matrix
kernel (\ref{evol-matrix-s}) the corresponding relations for $V^{qg}$
and $V^{gq}$ respectively involve the derivative and the integral of
the $\delta$ function in (\ref{kernel-relation}).  Compilations of
these kernels at LO, as well as the ones for the singlet evolution of
$\tilde{H}^{q(+)}$ and $\tilde{H}^g$ can be found in
\cite{Blumlein:1999sc}.  The corresponding results for Radyushkin's
distributions $\mathcal{F}_\zeta(X)$ are given in
\cite{Radyushkin:1998es} and those for the distributions
$\hat\mathcal{F}_q(X,\zeta)$ of Golec-Biernat and Martin in
\cite{Golec-Biernat:1998ja}.  The NLO kernels are listed in
\cite{Belitsky:1999hf} (with some misprints in the skewed DGLAP
kernels pointed out in \cite{Freund:2001hd}).  Different forms of the
off-diagonal kernels are given in the literature, but note that only
the even part in $x'$ of $V^{qg}$ and the odd part in $x'$ of $V^{gq}$
survive the convolution with the corresponding GPDs in the evolution
equations.  An analogous statement (with opposite symmetry properties
in $x'$) holds in the parity-odd sector.

The evolution kernels $V(x,x')$ for GPDs include the ERBL and DGLAP
kernels as limiting cases.  The limit $\xi\to \pm 1$ of a GPD
corresponds to an initial or final state with zero plus-momentum,
which for the evolution equations is as good as the vacuum state
appearing in DAs or GDAs.  The ERBL equations have hence the same form
as (\ref{gpd-evol-ns}) and its singlet counterparts, with $\xi$ set to
$\pm 1$.  The momentum fraction $x$ we use for GPDs is proportional to
the difference of quark and antiquark momenta in the ERBL region and
corresponds to $2z-1$ in terms of the quark momentum fraction $z$ we
use for distribution amplitudes.  Taking this correspondence into
account, the ERBL evolution kernels are just given by $V(x,x')$
restricted to the interval $[-1,1]$ in both arguments.  Indeed one
recognizes in (\ref{LO-quark-kernel}) the familiar ERBL kernel with
$\rho(x,x')$ instead of $\theta(x'-x)$.  M\"uller et al.\
\cite{Muller:1994fv} have shown that one can  obtain the
evolution kernels for GPDs as a unique extension of the usual ERBL
kernels to the full $(x,x')$-plane.  The argument consists in showing
that the Fourier transform $\int dx\, \exp(i \lambda x)\, V(x,x')$ is
analytic in both $\lambda$ and $x'$ and can hence be obtained by
analytic continuation from the region $|x'| \le 1$, where $V(x,x')$
reduces to the usual ERBL kernels according to
Fig.~\ref{fig:evol-support}.  The kernel $V(x,x')$ in the full
$(x,x')$-plane is then obtained by inverting the Fourier transform.
At $O(\alpha_s)$ this procedure reduces to replacing $\theta(x'-x)$ in
the ERBL kernels by $\rho(x,x')$.

The DGLAP evolution kernels are obtained by taking the forward limit
$\xi\to 0$ of (\ref{gpd-evol-ns}) and its singlet counterparts.
Comparing with the conventional form of the DGLAP equations in the
nonsinglet sector,
\begin{equation}
\mu^2 \frac{d}{d \mu^2} q(z) =
\int_{z}^1 \frac{dz'}{z'}\, 
	P_{\mathrm{NS}}\Big( \frac{z}{z'} \Big)\, q(z') ,
\end{equation}
where $z>0$, one finds
\begin{eqnarray}
P_{\mathrm{NS}}(z)  &=& \lim_{\xi \to 0} \frac{1}{|\xi|} 
     V_{\mathrm{NS}}\Big( \frac{z}{\xi}, \frac{1}{\xi} \Big) 
 \;=\; \frac{\alpha_s}{2\pi} C_F\, 
	\Bigg[ \theta(0\le z \le 1)\,
		\frac{1+z^2}{1-z} \Bigg]_+ + O(\alpha_s^2) .
\end{eqnarray}
Analogous relations hold for the other kernels.  Notice that the DGLAP
evolution equations are relevant not only for the usual parton
distributions but also for GPDs at $\xi=0$ but nonzero $t$.

For nonzero skewness $\xi$ the evolution in the DGLAP region $\xi \le
|x|$ is very similar to the usual DGLAP evolution.  In particular we
see from Fig.~\ref{fig:evol-support} that the change of a GPD
$f^i(x,\xi,t)$ with $\mu$ is influenced only by GPDs $f^j(x',\xi,t)$
at values $|x'| \ge |x|$, i.e.\ by partons with the same or larger
plus-momentum.  Qualitatively, evolution to higher scales $\mu$
``shifts'' partons from larger to smaller momentum fractions $|x|$, as
suggested by the physical picture of partons losing momentum by
radiating other partons before they are probed in the hard process.

In the ERBL region, the change of $f^i(x,\xi,t)$ with $\mu$ depends on
GPDs $f^j(x',\xi,t)$ in the full range of $x'$.  Rewriting
(\ref{gpd-evol-ns}) as
\begin{eqnarray}
\lefteqn{
\mu^2 \frac{d}{d \mu^2} H^{q(-)}(y \xi,\xi,t) =
  \int_{-1}^1 dy'\, V_{\mathrm{NS}}(y,y') \,
	H^{q(-)}(y' \xi,\xi,t) 
}
\nonumber \\
 && + \int dy' \Big[ \theta(y' <-1) + \theta(y'>1) \Big]\, 
	V_{\mathrm{NS}}(y,y') \, H^{q(-)}(y' \xi,\xi,t)
\end{eqnarray}
we see that the evolution equation in the ERBL region can be regarded
as an inhomogeneous equation, with the homogeneous part identical to
the usual ERBL equations for distribution amplitudes and an
inhomogeneous term due to the GPDs in the DGLAP region (Belitsky et
al.\ have proposed to use this decomposition in the actual solution of
the evolution equations \cite{Belitsky:2000yn}).  In the ERBL region,
evolution tends to ``equalize'' the plus-momenta of the two emitted
partons towards a symmetric or antisymmetric shape in $x$ (depending
on the channel) and eventually to their asymptotic shapes discussed
below.  Dedicated numerical studies of leading-order evolution have
been performed by various groups in
\cite{Frankfurt:1998ha,Freund:1998uf},
\cite{Martin:1998wy,Golec-Biernat:1998vf,Golec-Biernat:1999ib}, 
\cite{Belitsky:1998pc}, \cite{Musatov:1999xp}.  Belitsky et
al.~\cite{Belitsky:1998pq,Belitsky:1998uk} numerically studied the
two-loop corrections to evolution at moderate $\xi$ and found small to
moderate effects, depending on how small the starting scale $\mu_0$
was taken.  Similar results were reported by Freund and McDermott
\cite{Freund:2001bf}.  As is to be expected, two-loop effects become
more important when starting evolution at low scales, where the
running coupling is large.

\subsubsection{Solving the evolution equations}
\label{sub:solve-evolution}

As in the case of forward parton densities, the evolution of GPDs may
also be treated in terms of the ultraviolet renormalization of the
\emph{local} operators that correspond to their moments in $x$
(Section~\ref{sub:polynom}).  What complicates the nonforward case is
that the set of leading-twist operators in (\ref{twist-two}) mixes
under renormalization with operators having additional overall
derivatives.  Projecting all Lorentz indices on the plus-direction for
simplicity, one has
\begin{equation}
  \label{mellin-evol}
\mu^2 \frac{d}{d \mu^2}\, \Big[ 
	\bar{q} \gamma^+ (\lrD^{+})^n\, q \Big]
= \sum_{m=0 \atop \scriptstyle{\rm even}}^n   \Gamma_{nm}\,
	\Big[ (\partial^+)^{n-m}\,
        \bar{q} \gamma^+ (\lrD^{+})^m q  \Big] 
\end{equation}
for even $n$.  In matrix elements between states with momenta $p'$ and
$p$ the overall derivatives $\partial^+$ translate into factors of $i
\Delta^+$.  It is therefore only for $\Delta^+=0$ that one has the
simple multiplicative renormalization of Mellin moments familiar from
the usual parton densities.  To leading logarithmic accuracy the
system (\ref{mellin-evol}) is diagonalized by the so-called conformal
operators \cite{Efremov:1980rn,Makeenko:1981bh,Ohrndorf:1982qv}
\begin{equation}
   \label{conf-op}
\mathcal{O}^q_n = (\partial^+)^{n}\, \bar{q} \gamma^+  
  C_n^{3/2}\Bigg( 
      \frac{\rD^{+} \! - \lD^{+}}{\rpartial^{+} \! + \lpartial^{+}} 
           \Bigg) q 
\end{equation}
with even $n$ (where the inverse derivatives in the polynomial cancel
against derivatives $\partial^+$ in front of the operator).  To
leading logarithmic accuracy the conformal moments
\begin{eqnarray}
  \label{conf-mom}
\mathcal{C}^q_n(\xi,t;\mu^2) &=& \xi^{n}
\int_{-1}^1 dx\, C^{3/2}_n\Big(\frac{x}{\xi}\Big) H^q(x,\xi,t;\mu^2) 
\end{eqnarray}
of GPDs thus renormalize multiplicatively for even $n$, as written in
(\ref{mult-renorm}) for the Gegenbauer coefficients of GDAs.  For
gluon distributions the corresponding moments are
\begin{eqnarray}
  \label{conf-mom-glue}
\mathcal{C}^g_n(\xi,t;\mu^2) &=& \xi^{n-1}
\int_{-1}^1 dx\, 
	C^{5/2}_{n-1}\Big(\frac{x}{\xi}\Big) H^g(x,\xi,t;\mu^2) ,
\end{eqnarray}
and the mixing between $\mathcal{C}^g_n$ and the flavor singlet
combination $\sum_q \mathcal{C}^q_n$ for odd $n$ is solved as in
(\ref{mult-renorm-singlet}) for GDAs (with different mixing
coefficients due to the different relative normalization of
$B^q_{nl}$, $B^g_{nl}$ and $\mathcal{C}^q_n$, $\mathcal{C}^g_n$).  The
conformal moments of the parton helicity dependent GPDs $\tilde{H}^q$,
$\tilde{H}^g$ and of quark or gluon transversity distributions evolve
in a similar manner: the relevant Gegenbauer polynomials are always
$C_n^{3/2}$ for quarks and $C_{n-1}^{5/2}$ for gluons, and the
differences in the evolution arise from the differences in the
anomalous dimensions.  The moments of $E$ and $\tilde{E}$ evolve of
course as their counterparts for $H$ and $\tilde{H}$.  An explicit
form of conformal moments sometimes used in the literature is based on
the relation between Gegenbauer and Jacobi polynomials:
\begin{eqnarray}
\xi^{n} C^{3/2}_n \Big(\frac{x}{\xi}\Big)
&=& \frac{n+1}{2^{n+1}} \, \sum_{k=0}^n
\left( \begin{array}{c} n \\ k \end{array} \right)
\left( \begin{array}{c} n+2 \\ k+1 \end{array} \right)
(x+\xi)^k\, (x-\xi)^{n-k} ,
\nonumber \\
\xi^{n} C^{5/2}_n \Big(\frac{x}{\xi}\Big)
&=& \frac{(n+1)(n+2)}{3 \cdot 2^{n+2}} \, \sum_{k=0}^n
\left( \begin{array}{c} n \\ k \end{array} \right)
\left( \begin{array}{c} n+4 \\ k+2 \end{array} \right)
(x+\xi)^k\, (x-\xi)^{n-k} .
\end{eqnarray}
Notice that for $\xi=0$ the conformal moments reduce to the usual
Mellin moments up to a normalization factor.

At NLO the conformal operators (\ref{conf-op}) mix under
renormalization, and one has 
\begin{equation}
  \label{nlo-mix}
\mu^2 \frac{d}{d \mu^2}\, \mathcal{O}^q_n =
   \sum_{m=0 \atop \scriptstyle{\rm even}}^n   \hat{\Gamma}_{nm}\,
	\Big[ (\partial^+)^{n-m}\, \mathcal{O}^q_m \Big]
\end{equation}
for even $n$.  At NLO the change with $\mu$ of the $n$th conformal
moment $\mathcal{C}^q_n$ thus depends also on the lower moments.
Corresponding results hold for the quark and gluon operators in the
$C$-even sector.  The off-diagonal entries of the anomalous dimension
matrix $\hat{\Gamma}_{nm}$ start at order $\alpha_s^2$ and are
specific to nonforward kinematics, whereas the diagonal entries
coincide with the anomalous dimensions known from DGLAP evolution.
The solution of the coupled system of equations (\ref{nlo-mix}) and
its analogs in the favor singlet sector is discussed in
\cite{Belitsky:1998gc,Belitsky:1998pq,Belitsky:1998uk}.

When evolving to asymptotically large $\mu^2$ the running coupling
decreases, and the asymptotic forms of the GPDs are controlled by the
leading-order anomalous dimensions as already discussed in
Sect.~\ref{sub:gda-evolution}.  Only the Gegenbauer moments (or their
quark-gluon combinations) corresponding to zero anomalous dimensions
survive and one finds \cite{Goeke:2001tz}
\begin{eqnarray}
  \label{asy-gpd-s}
H^{q(+)}(x,\xi,t) &\stackrel{\mu\to\infty}{\to}& \theta(|x| < \xi)\,
   \frac{3}{3 n_f +16}\; \frac{15 x}{4 \xi} 
	\left( 1 - \frac{x^2}{\xi^2} \right) 
		\left( \frac{A(t)}{\xi^2} + 4 C(t) \right) ,
\nonumber \\
E^{q(+)}(x,\xi,t) &\stackrel{\mu\to\infty}{\to}& \theta(|x| < \xi)\,
   \frac{3}{3 n_f +16}\; \frac{15 x}{4 \xi} 
	\left( 1 - \frac{x^2}{\xi^2} \right) 
		\left( \frac{B(t)}{\xi^2} - 4 C(t) \right) ,
\nonumber \\
H^{g}(x,\xi,t) &\stackrel{\mu\to\infty}{\to}& \theta(|x| < \xi)\,
   \frac{16}{3 n_f +16}\; \frac{15\xi}{8} 
	\left( 1 - \frac{x^2}{\xi^2} \right)^2 
		\left( \frac{A(t)}{\xi^2} + 4 C(t) \right) ,
\nonumber \\
E^{g}(x,\xi,t) &\stackrel{\mu\to\infty}{\to}& \theta(|x| < \xi)\,
   \frac{16}{3 n_f +16}\; \frac{15\xi}{8} 
	\left( 1 - \frac{x^2}{\xi^2} \right)^2
		\left( \frac{B(t)}{\xi^2} - 4 C(t) \right) 
\end{eqnarray}
in the $C$-even sector, where $A = \sum_q A_q + A_g$ and its analogs
$B$ and $C$ are the form factors of the total energy-momentum tensor
discussed in Section~\ref{sec:spin}.  Note that at $t=0$ one has $A=1$
and $B=0$ due to the momentum and angular momentum sum rules.  One
further has
\begin{eqnarray}
  \label{asy-gpd-ns}
{H}^{q(-)}(x,\xi,t)
	 &\stackrel{\mu\to\infty}{\to}& \theta(|x| < \xi)\,
  \frac{3}{4 \xi} \left( 1 - \frac{x^2}{\xi^2} \right) F^q_1(t) ,
\nonumber \\
{E}^{q(-)}(x,\xi,t)
	 &\stackrel{\mu\to\infty}{\to}& \theta(|x| < \xi)\,
  \frac{3}{4 \xi} \left( 1 - \frac{x^2}{\xi^2} \right) F^q_2(t) ,
\nonumber \\
\tilde{H}^{q(+)}(x,\xi,t)
	 &\stackrel{\mu\to\infty}{\to}& \theta(|x| < \xi)\,
  \frac{3}{4 \xi} \left( 1 - \frac{x^2}{\xi^2} \right) g^q_A(t)  ,
\nonumber \\
\tilde{E}^{q(+)}(x,\xi,t)
	 &\stackrel{\mu\to\infty}{\to}& \theta(|x| < \xi)\,
  \frac{3}{4 \xi} \left( 1 - \frac{x^2}{\xi^2} \right) g^q_P(t) ,
\end{eqnarray}
whereas the remaining combinations $\tilde{H}^{q(-)}$,
$\tilde{E}^{q(-)}$ and $\tilde{H}^g$, $\tilde{E}^g$ asymptotically
evolve to zero, as well as the quark and gluon transversity
distributions.  In the limit $\xi\to 0$ the functions in
(\ref{asy-gpd-s}) and (\ref{asy-gpd-ns}) become proportional to either
$\delta(x)$ or $\delta'(x)$ and one recovers the singular asymptotic
forms of the forward densities.

For distribution amplitudes, whose kinematics corresponds to $\xi=\pm
1$, the conformal moments (\ref{conf-mom}) and (\ref{conf-mom-glue})
are readily inverted and thus provide a solution of the evolution
equations in momentum space.  This is however not the case for $|\xi|
<1$, where the Gegenbauer polynomials $C_n^{k}(x /\xi)$ do not form an
orthogonal set of functions on the interval $x \in [-1,1]$.  The
reconstruction of GPDs from the moments (\ref{conf-mom}) and
(\ref{conf-mom-glue}) is then nontrivial, as is the inversion of
Mellin moments in the forward limit.  A strategy proposed in
\cite{Shuvaev:1999fm} leads to the Shuvaev transform  discussed in the
next section.  Another method is to expand the GPDs in a set of
orthogonal polynomials on the interval $x \in [-1,1]$.  Choosing
Gegenbauer polynomials $C_n^{3/2}(x)$ one finds a structure
\begin{equation}
 H^q(x,\xi,t) = (1-x^2) \sum_{n=0}^\infty C_n^{3/2}(x)
     \sum_{m=0 \atop \scriptstyle{\rm even}}^n a_{nm}(\xi)\, 
	\mathcal{C}^q_{n-m}(\xi)
\end{equation}
where the $a_{nm}(\xi)$ are known even polynomials in $\xi$ of order
$m$ \cite{Belitsky:1998pc,Kivel:1999wa}.  The method is not restricted
to expanding on $C_n^{3/2}(x)$, and the relevant coefficients for
Jacobi polynomials have been given in
\cite{Belitsky:1998pq,Belitsky:1998uk}.  Solutions based on polynomial
expansion have also been presented in
\cite{Balitsky:1989bk,Kivel:1999wa} for the position space operators
(\ref{string-operators}) or their hadronic matrix elements (termed
``coordinate space distributions'').

Expansion on different polynomials has been used for the numerical
solution of the evolution equations at leading order by several groups
\cite{Belitsky:1998pc,Mankiewicz:1998uy,Golec-Biernat:1998vf,Kivel:1999wa}.
Belitsky et al.~\cite{Belitsky:1998pq,Belitsky:1998uk,Belitsky:1999sg}
have treated NLO evolution in this way.  Expanding on Legendre
polynomials, the authors report that a rather large number of terms
(several tens to hundreds) was necessary to achieve a reliable
accuracy because of the oscillatory behavior of the polynomials, and
found the method intractable or unreliable for $x$ close to $\xi$ or
for small $\xi$, when the GPDs can have rapid variations in $x$.  It
is not known whether there is an adequate choice of polynomials for
which a smaller number of terms would be sufficient.  Direct numerical
integration of the evolution equations has been used as an alternative
method; some technical information is given in
\cite{Frankfurt:1998ha,Freund:2001hd}.  For the case where evolution
is performed over a limited range in $\mu$, Musatov and Radyushkin
\cite{Musatov:1999xp} have proposed an iterative solution of the
evolution equations.

\subsubsection{Shuvaev transform}
\label{sub:Shuvaev}

Shuvaev \cite{Shuvaev:1999fm} has proposed a method to reduce the
leading logarithmic evolution of GPDs to the usual DGLAP evolution,
making use of the fact that the Gegenbauer moments of GPDs evolve in
the same way as the Mellin moments of forward distributions.  The
method introduces ``effective forward distributions'' $f^q_{\xi}(x)$,
$f^g_{\xi}(x)$, whose Mellin moments are equal to the Gegenbauer
moments of a given GPD $H^{q,g}(x,\xi)$, up to a normalization factor
which ensures that in the forward limit one has $f^q_0(x) =
H^q(x,0,0)$ and $x f_0^g(x) = H^g(x,0,0)$:
\begin{equation}
\int_{-1}^1 dx\, x^n f^{q,g}_{\xi}(x) = 
        c^{q,g}_n\, \mathcal{C}^{q,g}_n(\xi) 
\end{equation}
with 
\begin{equation}
c^q_n = \frac{2^n\, [n!]^2}{(2n+1)!} , \qquad
c^g_n = \frac{3 \cdot 2^n\, [n!]^2}{n\, (2n+1)!} .
\end{equation}
For simplicity we suppress the dependence of both $H$ and $f$ on $t$
and on $\mu^2$ here.  The leading order evolution of a nonforward
distribution can then be performed by transforming $H$ to $f$, solving
the usual DGLAP evolution for $f$, and transforming back to $H$ at the
new factorization scale.

A detailed investigation of the Shuvaev transform has been given by
Noritzsch \cite{Noritzsch:2000pr}.  The Shuvaev transform is an
integral transform
\begin{equation}
H^q(x,\xi) = \int_{-y_\xi}^{y_\xi} dy\, \mathcal{K}_q(x,\xi,y)
  f^q_\xi(y) , 
\label{shuv}
\end{equation}
with $y_\xi = \half (1 + \sqrt{1-\xi^2})$.  In the DGLAP regions $|x|
\ge \xi$ the kernel further restricts the integration to the two
intervals $y \ge \half (x + \sqrt{x^2-\xi^2})$ and $y \le \half (x -
\sqrt{x^2-\xi^2})$, whereas no such restriction occurs for $x$ in the
ERBL region \cite{Noritzsch:2000pr}.  The inverse transform is given
by
\begin{eqnarray}
f^q_\xi(y) &=& \int_{|x|\ge \xi} dx\, 
        \mathcal{K}_q^{-1}(y,\xi,x) H^q(x,\xi) 
\nonumber \\
&+& \sum_{n=0}^\infty \frac{(-1)^n}{n!} \, 
	\delta^{(n)}(y) \,  c^q_n \int_{|x|\le \xi} dx\,
        \xi^n C^{3/2}_n\Big(\frac{x}{\xi}\Big) H^q(x,\xi) .
  \label{shuv-inverse}
\end{eqnarray}
Analogous expressions hold for gluons.  Different equivalent
expressions for the integral kernels $\mathcal{K}_{q,g}$ and
$\mathcal{K}_{q,g}^{-1}$ can be found in
\cite{Shuvaev:1999fm,Noritzsch:2000pr,Shuvaev:1999ce}.  For $|x| \gg
\xi$ one finds $\mathcal{K}_q(x,\xi,y) = \delta(x-y) + x^{-1} O(\xi^2
/x^2)$ and $\mathcal{K}_g(x,\xi,y) = x \delta(x-y) + O(\xi^2 /x^2)$ so
that in this limit GPDs and effective forward distributions almost
coincide~\cite{Noritzsch:2000pr}.

In practice the Shuvaev transform has not been used to solve the
evolution of GPDs, possibly because of the infinite sum over $\delta$
functions and their derivatives in the inverse transform
(\ref{shuv-inverse}), overlooked in \cite{Shuvaev:1999fm} and pointed
out in \cite{Noritzsch:2000pr}.  The transform has however been useful
for approximating GPDs at small $\xi$, by replacing $f_\xi(x)$ with
the forward densities times a $t$ dependent factor in (\ref{shuv}).
This will be discussed in Section~\ref{sec:small-x-gpd}.


\subsection{Double distributions}
\label{sec:double-d}

Double distributions are an alternative way to parameterize the
hadronic matrix elements which define GPDs and which appear in several
hard processes.  Their study provides insight into the properties of
these matrix elements, and it provides a strategy to make ans{\"a}tze
for GPDs, which we will discuss in Section~\ref{sec:ansatz}.  Double
distributions have been introduced by M\"uller et
al.~\cite{Muller:1994fv}, who called them ``spectral functions''.
They were rediscovered and studied by Radyushkin, who has given a
detailed review on the subject~\cite{Radyushkin:2000uy}.

Starting point for introducing double distributions are the matrix
elements of the light-cone operators we already encountered when
discussing evolution in Section~\ref{sec:evolution}, namely
\begin{equation}
  \label{lc-matrix-element}
  \langle p'|\, \bar{q}(-\half z)\, \slash{z} q(\half z)\, 
  \,|p \rangle \Big|_{z^2 =0} 
\end{equation}
and its analogs in the parity-odd sector and for gluons.  Because of
Lorentz invariance these matrix elements only depend on scalar
products $pz$, $p'z$, $p p'$ or alternatively on $Pz$, $\Delta z$, $t$
once appropriate spinor structures have been taken off.  Double
distributions are defined through a Fourier transform in the two
independent variables $P z$ and $\Delta z$,
\begin{eqnarray}
  \label{dd-def}
\langle p'|\, \bar{q}(-\half z)\, \slash{z} q(\half z)\, 
  \,|p \rangle \Big|_{z^2 =0} 
&=& \bar{u}(p') \slash{z} u(p) \int d\beta\, d\alpha\, 
        e^{-i \beta (Pz) + i \alpha (\Delta z) /2}\, f^q(\beta, \alpha,t)
\nonumber \\
&+& \bar{u}(p') \frac{i\sigma^{\mu\alpha} z_\mu \Delta_\alpha}{2m} u(p)
 \int d\beta\, d\alpha\, 
        e^{-i \beta (Pz) + i \alpha (\Delta z) /2}\, k^q(\beta, \alpha,t)
\nonumber \\
&-& \bar{u}(p') \frac{\Delta z}{2m} u(p)  \int d\alpha\, 
        e^{i \alpha (\Delta z) /2}\, D^q(\alpha,t) .
\end{eqnarray}
The analphabetic order of the variables $\beta$ and $\alpha$ follows
the convention of~\cite{Goeke:2001tz}.\footnote{Radyushkin
\protect\cite{Radyushkin:1998bz} had earlier introduced the variables
$(x,\alpha)$ instead of $(\beta,\alpha)$, but we reserve $x$ for the
first argument of GPDs.}
The so-called $D$-term in the last line had for a long time been
overlooked and was only introduced by Polyakov and Weiss
\cite{Polyakov:1999gs}.  For the moment we will leave it aside and
discuss it in Section~\ref{sub:d-term}.

\begin{figure}
\begin{center}
     \leavevmode
     \epsfxsize=0.68\textwidth
     \epsffile{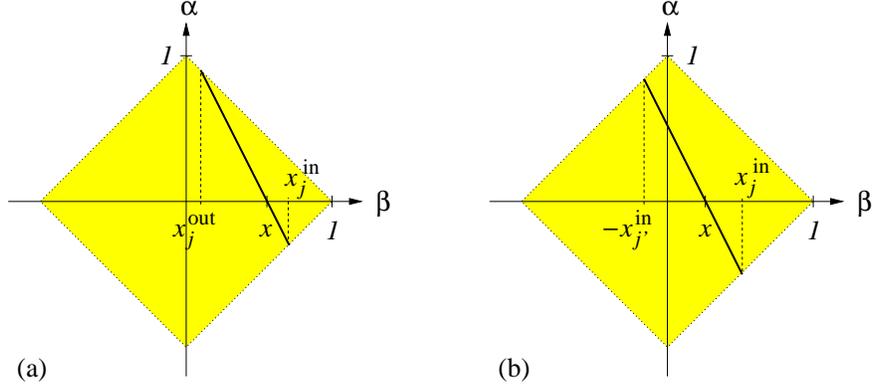}
\end{center}
\caption{\label{fig:dd-sup} Support region for double distributions,
and lines of integration to obtain GPDs in the regions (a) $x\in
[\xi,1]$ and (b) $x\in [-\xi,\xi]$ according to
(\protect\ref{dd-red}).  The momentum fractions $x_j^{\mathrm{in}}$,
$x_j^{\mathrm{out}}$, $x_{j'}^{\mathrm{in}}$ are given in
(\protect\ref{fractions-DGLAP}) and (\protect\ref{fractions-ERBL})
with $x_j = x$.}
\end{figure}

The support region of the double distributions $f$ and $k$ is given by
the rhombus $|\beta| + |\alpha| \le 1$ shown in Fig~\ref{fig:dd-sup},
and $D$ has support for $|\alpha| \le 1$ (for simplicity we do not
write out the corresponding limits in the integrals over $\beta$ and
$\alpha$).  These spectral properties can be shown using the
$\alpha$-representation of Feynman diagrams
\cite{Radyushkin:1997ki,Radyushkin:1983wh}.  Note that double
distributions need not be smooth functions; one explicitly encounters
$\delta$ distributions and their derivatives in meson exchange
contributions (see Section~\ref{sub:connect-forward}).

To obtain the relation between double distributions and GPDs one can
take a given lightlike $z$ and choose a frame where $z^+=0$ and
$\tvec{z}=0$.  Dividing the matrix element (\ref{lc-matrix-element})
by $z^-$ and taking the Fourier transform with respect to $z^-$ one
finds the relations
\begin{eqnarray}
  \label{dd-red}
H^q(x,\xi,t) &=& 
  \int d\beta\, d\alpha \;
        \delta(x - \beta - \xi \alpha)\, f^q(\beta, \alpha,t) 
+ \mbox{sgn}(\xi) D^q\Big( \frac{x}{\xi}, t \Big) ,
\nonumber \\
E^q(x,\xi,t) &=& 
  \int d\beta\, d\alpha \;
        \delta(x - \beta - \xi \alpha)\, k^q(\beta, \alpha,t) 
- \mbox{sgn}(\xi) D^q\Big( \frac{x}{\xi}, t \Big) .
\end{eqnarray}
Notice that because of its support property, the $D$-term only
contributes in the region $|x| < |\xi|$.  The lines of integration
over the double distributions in the $\beta$-$\alpha$ plane are shown
in Fig.~\ref{fig:dd-sup}.  They have slope $-1/\xi$ and cut the $\beta$
axis at $x$.  Furthermore, they intersect the edges of the support
rhombus at points where $|\beta|$ is the momentum fraction of one or
the other parton with respect to its parent hadron in the GPD (given
in (\ref{fractions-DGLAP}) and (\ref{fractions-ERBL})).  The forward
quark density is simply obtained by integration over the vertical line
$\beta = x$,
\begin{equation}
q(x) = \int_{x-1}^{1-x}  d\alpha\, 
        f^q(x, \alpha,0)
\end{equation}
for $x>0$, with an analogous relation for $x<0$.  

\begin{figure}
\begin{center}
     \leavevmode
     \epsfxsize=0.42\textwidth
     \epsffile{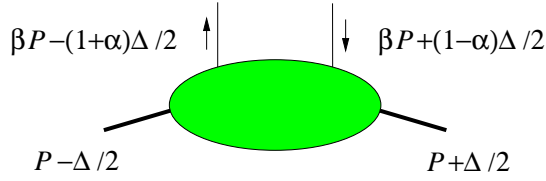}
\end{center}
\caption{\label{fig:dd-kin} Momenta associated with the partons and
hadrons in a double distribution, as explained in the text.}
\end{figure}

{}From the arguments of the Fourier transform in (\ref{dd-def}) one
can associate the two variables $\beta$ and $\alpha$ with a flow of
momentum as shown in Fig.~\ref{fig:dd-kin}.  For $\Delta=0$ one
clearly obtains the kinematics of a forward parton density, whereas
the opposite case $P=0$ is reminiscent of a distribution amplitude
(this will become yet more explicit in Sect.~\ref{sub:radon}).  This
suggests that the dependence of the double distribution on $\beta$
will resemble the pattern of a parton density, and the dependence on
$\alpha$ that of a DA.  It is however important to realize that the
picture in Fig.~\ref{fig:dd-kin} is only a guide.  $pz$ and $p'z$ are
treated as \emph{independent} variables when defining the double
distribution through a two-dimensional Fourier transform.  As a
consequence, a double distribution refers to fixed $t=(p-p')^2$ but
\emph{not} to fixed individual hadron momenta $p$ and $p'$ (and hence
does not depend on the plus-momentum fraction $\xi$).  This is in
contrast to GPDs, whose definition (\ref{quark-gpd}) is readily turned
into a one-dimensional Fourier transform
\begin{eqnarray}
  \label{gpd-redef}
\langle p'|\, \bar{q}(-\half z)\, \slash{z} q(\half z)\, 
  \,|p \rangle \Big|_{z^2 =0} 
&=& \bar{u}(p') \slash{z} u(p) \int dx\,
        e^{-i x (Pz)}\, H^q(x, \xi,t)
\nonumber \\
&+& \bar{u}(p') \frac{i\sigma^{\mu\alpha} z_\mu \Delta_\alpha}{2m} u(p)
    \int dx\,e^{-i x (Pz)}\, E^q(x, \xi,t) .
\end{eqnarray}
Here $pz$ and $p'z$ are understood as dependent variables subject to
the constraint $\xi = - (\Delta z) /(2 Pz)$, so that $p$ and $p'$ can
both be taken fixed.  The above feature of double distributions leads
to their simple properties regarding Lorentz invariance (see
Section~\ref{sub:dd-evolution}), but at the same time makes their
physical interpretation less immediate than the one of GPDs.  In
particular, no interpretation comparable to the ones we will develop
in Sections~\ref{sec:impact} and \ref{sec:overlap} has been found so
far.

For gluons double distributions can be defined by
\begin{eqnarray}
  \label{dd-glue}
\lefteqn{
\langle p'|\, z_\alpha z_\beta
   G^{\alpha\mu}(-\half z)\, G_{\mu}{}^{\beta}(\half z)\, 
  \,|p \rangle \Big|_{z^2 =0}
}
\nonumber \\
&=& \bar{u}(p') \slash{z} u(p)\, \frac{Pz}{2}\,  \int d\beta\, d\alpha\, 
        e^{-i \beta (Pz) + i \alpha (\Delta z) /2}\, 
	\beta f^g(\beta, \alpha,t)
\nonumber \\
&+& \bar{u}(p') \frac{i\sigma^{\mu\alpha} z_\mu \Delta_\alpha}{2m} u(p)\,
    \frac{Pz}{2} \,\int d\beta\, d\alpha\, 
        e^{-i \beta (Pz) + i \alpha (\Delta z) /2}\, 
	\beta k^g(\beta, \alpha,t)
\nonumber \\
&+& \bar{u}(p') \frac{\Delta z}{2m} u(p)\, \frac{\Delta z}{4}\, 
	\int d\alpha\, e^{i \alpha (\Delta z) /2}\, D^g(\alpha,t) .
\end{eqnarray}
This definition of $f^g$ follows the convention of
\cite{Musatov:1999xp,Radyushkin:2000uy}; definitions with other
prefactors have also been used in the literature.  With the above
choice one obtains the forward gluon density as $g(x) = \int d\alpha\,
f^g(x, \alpha,0)$.  The reduction formulae to the gluon GPDs $H^g$ and
$E^g$ read
\begin{eqnarray}
  \label{dd-red-glue}
H^g(x,\xi,t) &=& 
  \int d\beta\, d\alpha \;
        \delta(x - \beta - \xi \alpha)\, \beta f^g(\beta, \alpha,t) 
+ |\xi| D^g\Big( \frac{x}{\xi}, t \Big) , 
\nonumber \\
E^g(x,\xi,t) &=& 
  \int d\beta\, d\alpha \;
        \delta(x - \beta - \xi \alpha)\, \beta k^g(\beta, \alpha,t) 
- |\xi| D^g\Big( \frac{x}{\xi}, t \Big) .
\end{eqnarray}
Double distributions corresponding to the quark or gluon helicity
dependent GPDs $\tilde{H}$ and $\tilde{E}$ are defined as
\begin{eqnarray}
  \label{dd-pol-def}
\lefteqn{
\langle p'|\, \bar{q}(-\half z)\, \slash{z} \gamma_5\, q(\half z)\, 
  \,|p \rangle \Big|_{z^2 =0} 
}
\nonumber \\
&=& \bar{u}(p') \slash{z} \gamma_5\, u(p) \int d\beta\, d\alpha\, 
        e^{-i \beta (Pz) + i \alpha (\Delta z) /2}\, 
	\tilde{f}^q(\beta, \alpha,t)
\nonumber \\
&+& \bar{u}(p') \frac{\gamma_5 (\Delta z)}{2m} u(p)
 \int d\beta\, d\alpha\, 
        e^{-i \beta (Pz) + i \alpha (\Delta z) /2}\, 
	\tilde{k}^q(\beta, \alpha,t) ,
\nonumber \\[0.4em]
\lefteqn{
\langle p'|\, z_\alpha z_\beta \, (-i)
   G^{\alpha\mu}(-\half z)\, \tilde{G}_{\mu}{}^{\beta}(\half z)\, 
  \,|p \rangle \Big|_{z^2 =0}
}
\nonumber \\
&=& \bar{u}(p') \slash{z} \gamma_5\, u(p)\, \frac{Pz}{2}\,  
	\int d\beta\, d\alpha\, 
        e^{-i \beta (Pz) + i \alpha (\Delta z) /2}\, 
	\beta \tilde{f}^g(\beta, \alpha,t)
\nonumber \\
&+& \bar{u}(p') \frac{\gamma_5 (\Delta z)}{2m} u(p)\,
    \frac{Pz}{2} \,\int d\beta\, d\alpha\, 
        e^{-i \beta (Pz) + i \alpha (\Delta z) /2}\, 
	\beta \tilde{k}^g(\beta, \alpha,t) .
\end{eqnarray}
Except for the absence of $D$-terms in this sector (see
Section~\ref{sub:d-term}) the corresponding reduction relations read
as in (\ref{dd-red}) and (\ref{dd-red-glue}).

There are different variants of the double distributions defined
above.  Radyushkin has discussed double distributions
$\tilde{F}^q(x,y) = 2 f^q(\beta,\alpha)$ using different variables
\begin{equation}
x=\beta , \qquad y=\half (1-\beta+\alpha) ,
\end{equation}
which match more naturally the variables $X$ and $\zeta$ for GPDs
mentioned in Section~\ref{sec:conventions}.  He has also introduced
separate ``quark'' and ``antiquark'' double distributions $F^q(x,y)$
and $F^{\bar{q}}(x,y)$, which respectively describe the regions $x>0$
and $x<0$ of $\tilde{F}^q(x,y)$.  Separate ``quark'' and ``antiquark''
GPDs $\mathcal{F}^q_\zeta(X)$ and $\mathcal{F}^{\bar{q}}_\zeta(X)$ are
then defined by integrals in the $x$-$y$ plane.  In the DGLAP regions
they respectively correspond to Ji's $H^q(x,\xi)$ in the quark and
antiquark regions $x>\xi$ and $x<-\xi$, whereas in the ERBL region
$H^q(x,\xi)$ is a linear combination of $\mathcal{F}^q_\zeta(X)$ and
$\mathcal{F}^{\bar{q}}_\zeta(X)$.  This type of distributions has been
discussed in detail in earlier work but not much been used in the
recent literature.  Details on their relation with Ji's GPDs can be
found in \cite{Radyushkin:1998bz,Golec-Biernat:1998ja}.

\subsubsection{General properties and evolution}
\label{sub:dd-evolution}

In a similar way as for GPDs one finds that time reversal invariance
and hermiticity constrains double distributions to be real valued and
even functions of $\alpha$.  This is sometimes referred to as ``Munich
symmetry'' with reference to \cite{Mankiewicz:1998uy}, where it was
discovered as a much less obvious symmetry in terms of the variables
$x$ and $y$ mentioned above.  As we did for GPDs in
Section~\ref{sub:symm}, one can form combinations with definite
$C$-parity,
\begin{eqnarray}
  \label{dd-definite-charge}
f^{q(\pm)}(\beta,\alpha,t) &=& 
	f^q(\beta,\alpha,t) \mp f^q(-\beta,\alpha,t) ,
\nonumber \\[0.3ex]
\tilde{f}^{q(\pm)}(\beta,\alpha,t) &=& 
	\tilde{f}^q(\beta,\alpha,t) \pm \tilde{f}^q(-\beta,\alpha,t)
\end{eqnarray}
and their analogs for $k^q$ and $\tilde{k}^q$.  The gluon double
distributions $f^g$ and $k^g$ are odd in $\beta$, and $\tilde{f}^g$
and $\tilde{k}^g$ are even in $\beta$.

Probably the most important aspect of double distributions for
practical purposes is that they generate GPDs which automatically
satisfy the polynomiality conditions discussed in
Section~\ref{sub:polynom}.  This is readily checked by taking Mellin
moments in $x$ of the reduction formulae (\ref{dd-red}) and their
analogs for the other distributions.  In other words, Lorentz
invariance restricts the functional form of the $x$ and $\xi$
dependence for GPDs, but not the dependence on double distributions on
$\beta$ and $\alpha$.  To see how this difference comes about one can
choose $z^+=0$, $\tvec{z}=0$ in the definitions (\ref{dd-def}) and
(\ref{gpd-redef}), divide by $z^-$, take the $n$-th derivative $(d/d
z^-)^n$, and then set $z^-$ to zero.  On the left-hand sides one then
obtains the local twist-two operators (\ref{twist-two}).  The
right-hand side of the double distribution definition (\ref{dd-def})
automatically has the required polynomial dependence on $\Delta^+$,
but not the right-hand side of the GPD definition (\ref{gpd-redef}),
where the dependence on $\Delta^+$ is implicit in the $\xi$ dependence
of $H$ and $E$.

Double distributions satisfy evolution equations of the form
\begin{equation}
  \label{evol-dd}
\mu^2 \frac{d}{d \mu^2}\, f(\beta,\alpha,t;\mu^2) 
= \int d\beta' \, d\alpha' \, 
        R_{\mathrm{NS}}(\beta,\alpha;\beta' \alpha') \, 
		f(\beta',\alpha',t;\mu^2)
\end{equation}
for the quark non-singlet combination, and corresponding matrix
equations for the mixed evolution of the quark singlet and gluon
distributions.  The kernels $R$ readily follow from the kernels $K$
for the evolution of the light-cone operators discussed in
Section~\ref{sub:light-cone-evolution}.  Up to global factors
$R_{\mathrm{NS}}$, $R^{qq}$ and $R^{gg}$ are obtained by a simple
change of arguments, whereas the quark-gluon mixing kernels require
differentiation or integration with respect to $\beta$ because of the
factors $(\kappa_2-\kappa_1)$ and $(\kappa_2-\kappa_1)^{-1}$ in the
matrix kernels (\ref{evol-matrix-s}).  Explicit results for the LO
kernels can be found in \cite{Blumlein:1997pi,Blumlein:1999sc}, and
for the double distribution variables $x,y$ of Radyushkin in
\cite{Radyushkin:1998es}.  After suitable integration along a line in
the $\beta$-$\alpha$ plane the kernels $R$ reduce to the evolution
kernels for GPDs, including the limiting cases of the DGLAP or ERBL
kernels, as shown in detail in
\cite{Radyushkin:1997ki,Radyushkin:1998es}.

A strategy for solving the evolution equations is to connect the
conformal moments (\ref{conf-mom}) of GPDs with the moments of double
distributions (see \cite{Radyushkin:2000uy} for details)
\begin{eqnarray}
  \label{dd-conformal}
\mathcal{C}^{q,g}_n(\xi,t;\mu) &=&
\sum_{m=0 \atop n-m~\mathrm{even}}^n a^{q,g}_{nm} \, \xi^{n-m}
        \int d\beta\, d\alpha \;
        \beta^{m}\, C_{n-m}^{3/2+m}(\alpha)\, f^{q,g}(\beta,\alpha,t;\mu)
\nonumber \\
 && {}+ \mbox{($D$-term contribution)}
\end{eqnarray}
with known numerical coefficients $a^{q,g}_{nm}$.  The $D$-term
contribution goes with $\xi^{n+1}$ and only occurs for odd $n$.  We
see that the quantities that renormalize multiplicatively (after
orthogonalization in the case of quark-gluon mixing) are combined
Mellin moments in $\beta$ and Gegenbauer moments in $\alpha$ of double
distributions.  Notice that quarks and gluons come with the same
Gegenbauer polynomials here.  Using their orthogonality one finds from
(\ref{dd-conformal}) that the Mellin moments have the form
\begin{eqnarray}
  \label{dd-mellin}
\int_{|\alpha|-1}^{1-|\alpha|} d\beta \, 
        \beta^{m}\, f^{q,g}(\beta,\alpha,t;\mu) 
= (1-\alpha^2)^{m+1} \sum_{k=0 \atop \mathrm{even}}^\infty 
        A^{q,g}_{mk}(t;\mu) \; C_k^{3/2+m}(\alpha) ,
\end{eqnarray}
where $A_{mk}$ is related to the conformal moments by
\begin{equation}
  \label{dd-relate}
\mathcal{C}^{q,g}_n(\xi,t;\mu) = \sum_{m=0 \atop n-m~\mathrm{even}}^n 
        b^{q,g}_{nm}\, \xi^{n-m}\, A^{q,g}_{m, n-m}(t;\mu)  
+ \mbox{($D$-term contribution)}
\end{equation}
with known numerical coefficients $b^{q,g}_{nm}$.  In the above sums
$n$ and $m$ are even for the quark nonsinglet and odd for the quark
singlet and the gluon distributions.  The quark nonsinglet
coefficients $A^q_{mk}$ are renormalized multiplicatively with
anomalous dimension $\gamma_{m+k}$, in analogy to the moments of GDAs
in (\ref{mult-renorm}).  Appropriate linear combinations of quark
singlet and gluon coefficients evolve with $\gamma_{m+k}^+$ or
$\gamma_{m+k}^-$ instead of $\gamma_{m+k}$.  One can now easily deduce
the behavior of the Mellin moments (\ref{dd-mellin}) under evolution
to $\mu\to \infty$.  In all cases the term with smallest anomalous
dimension is the $k=0$ term in the sum, so that the moments
(\ref{dd-mellin}) asymptotically go like $(1-\alpha^2)^{m+1}$.
Furthermore, all Mellin moments vanish asymptotically except those
associated with the anomalous dimensions $\gamma_0^{\phantom{-}} =
\gamma^-_1 =0$.  This leads to asymptotic forms of double
distributions going like
\begin{eqnarray}
  \label{dd-asy}
f^{q(-)}(\beta,\alpha) &\;\stackrel{\mu\to\infty}{\sim}\;&
(1-\alpha^2)\; \delta(\beta) , 
\nonumber \\
 f^g(\beta,\alpha) , \; f^{q(+)}(\beta,\alpha)
 &\;\stackrel{\mu\to\infty}{\sim}\;&
        (1-\alpha^2)^2\; \delta'(\beta) .
\end{eqnarray}
The resulting GPDs are those given in
Section~\ref{sub:solve-evolution}.  The evolution of the polarized
double distributions $\tilde{f}^{q,g}$ can be solved in analogy to
(\ref{dd-conformal}) to (\ref{dd-relate}).  Apart from the absence of
$D$-terms the difference to the unpolarized case is due to the
anomalous dimensions.  The asymptotic form for the $C$-even
combinations is thus given by
\begin{eqnarray}
  \label{dd-pol-asy}
\tilde{f}^{q(+)}(\beta,\alpha) &\;\stackrel{\mu\to\infty}{\sim}\;&
(1-\alpha^2)\; \delta(\beta) , 
\end{eqnarray}
whereas $\tilde{f}^g$ and $\tilde{f}^{q(-)}$ asymptotically evolve to
zero.  Analogous forms are of course obtained for $k^{q,g}$ and
$\tilde{k}^{q,g}$.

\subsubsection{The $D$-term}
\label{sub:d-term}

We now discuss in more detail the functions $D^q$ and $D^g$ in
(\ref{dd-def}) and (\ref{dd-glue}).  As we have seen in
Section~\ref{sub:polynom} the Mellin moments $\int dx\, x^n H^{q}$ and
$\int dx\, x^{n-1} H^{g}$ with odd $n$ are polynomials in $\xi$ of
degree $n+1$.  As one readily sees from the reduction formulae
(\ref{dd-red}) and (\ref{dd-red-glue}), the double distributions
$f^{q,g}$ and $k^{q,g}$ give maximal powers of $\xi^n$ for these
moments, so that the general decompositions of the light-cone matrix
elements in (\ref{dd-def}) and (\ref{dd-glue}) require extra terms.
It has been shown in \cite{Polyakov:1999gs} that these extra terms can
be chosen to depend only on $\alpha$ but not on $\beta$, and that the
resulting decomposition in spectral functions $f$, $k$, and $D$ is
unique.

Important properties of the $D$-terms are
\begin{itemize}
\item The $D$-term contributions to GPDs have support only in the ERBL
region $x\in [-\xi,\xi]$.  This makes them an example of information
unaccessible in the forward parton distributions.
\item There is a $D$-term for each separate quark flavor and for the
gluon.  It contributes to either $H^g$ and $E^g$ or to $H^{q(+)}$ and
$E^{q(+)}$, but not to the $C$-odd combinations of quark GPDs.  There
is no analog of the $D$ term for the helicity dependent distributions
$\tilde{H}^{q,g}$ or $\tilde{E}^{q,g}$: as we discussed in
Section~\ref{sub:polynom} their appropriate Mellin moments do not have
maximal power $\xi^{n+1}$ but only $\xi^{n}$, which is generated by
double distributions alone.
\item The same function $D^{q,g}$ contributes with opposite
sign to $H^{q,g}$ and to $E^{q,g}$ as seen in the reduction formulas
(\ref{dd-red}) and (\ref{dd-red-glue}).  In a sense this is a
particularity of decomposing the GPD matrix elements (\ref{quark-gpd})
and (\ref{gluon-gpd}) on the proton vector and tensor currents.  If
instead one were to take the vector and scalar current of the proton,
the corresponding set of GPDs would be $(H+E)$ and $E$, and the
$D$-term would only contribute to $E$.
\item $D$-terms also appear for targets with spin zero.  For a
pion one has
\begin{eqnarray}
  \label{dd-pion}
\lefteqn{
\langle \pi^+(p')|\, \bar{q}(-\half z)\, \slash{z} q(\half z)\, 
  \,|\pi^+(p) \rangle \Big|_{z^2 =0} 
}
\nonumber \\
&=& 2 (P z) \int d\beta\, d\alpha\, 
        e^{-i \beta (Pz) + i \alpha (\Delta z) /2}\, f^q(\beta, \alpha,t)
\nonumber \\
&-& (\Delta z)  \int d\alpha\, 
        e^{i \alpha (\Delta z) /2}\, D^q(\alpha,t)
\end{eqnarray}
and its analog for gluons.  The reduction formulae to the GPDs
$H^q_\pi$ and $H^g_\pi$ read as in (\ref{dd-red}) and
(\ref{dd-red-glue}).  Due to isospin invariance the $C= +1$ quark
combinations have isospin $I=0$ and the $C= -1$ combinations have
isospin $I=1$.  Hence there is a $D$-term contribution to the
isosinglet combination $H^{u+d}_\pi$ but none to the isotriplet
combination $H^{u-d}_\pi$.  A corresponding statement does \emph{not}
hold for the nucleon GPDs, \emph{except} if one considers the large
$N_c$ limit (see Section~\ref{sub:large-Nc}).
\item {}From time reversal one obtains that $D^q(\alpha,t)$ is odd and
$D^g(\alpha,t)$ even in $\alpha$, in accordance with the properties of
the corresponding GPDs under the replacement $x \to -x$.
\item $D^q$ and $D^g$ evolve according to the ERBL equations, just
like distribution amplitudes with the same quantum numbers.  Their
Gegenbauer coefficients $d^q_n$ and $d^g_n$, defined by\footnote{The
moments $d_n^g$ defined here coincide with the $d_n^G$ of
\protect\cite{Goeke:2001tz}.  The different factors in their  eq.~(24)
and our eq.~(\protect\ref{D-Gegenbauer}) reflect the different
definitions of $H^g$.}
\begin{eqnarray}
  \label{D-Gegenbauer} D^q(x,t;\mu) &=& (1-x^2)\, \sum_{n=1 \atop
\mathrm{odd}}^\infty d_n^q(t;\mu) \, C^{3/2}_n(x) ,
\nonumber \\
D^g(x,t;\mu) &=& \frac{3}{2}\, (1-x^2)^2\, 
        \sum_{n=1 \atop \mathrm{odd}}^\infty
        d_n^g(t;\mu) \, C^{5/2}_{n-1}(x) ,
\end{eqnarray}
thus evolve multiplicatively with anomalous dimension $\gamma_n$ for
the difference of any two quark flavors, and with $\gamma_n^+$ and
$\gamma_n^-$ after appropriate diagonalization for the quark flavor
singlet and the gluon.  Asymptotically one has~\cite{Goeke:2001tz}
\begin{equation}
d_1^q(t;\mu) \stackrel{\mu\to\infty}{=} \frac{3}{3 n_f + 16}\, d(t) ,
\qquad
d_1^g(t;\mu) \stackrel{\mu\to\infty}{=} \frac{16}{3 n_f + 16}\, d(t) ,
\end{equation}
with all higher moments going to zero.  Note that $d(t) = 5 C(t)$ is
given by one of the form factors of the total energy-momentum tensor
(see Section~\ref{sec:spin}), with individual contributions $d_1^q = 5
C_q$ and $d^g_1 = 5 C_g$ from quarks and gluons.
\item  As we will see in Section~\ref{sec:factor}, the contribution of a
$D$-term to hard process amplitudes is energy independent at fixed
photon virtuality.
\end{itemize}

As already mentioned, the $D$ term is uniquely defined for a given
hadronic matrix element (\ref{lc-matrix-element}), and one may ask
about its direct relation to a given GPD.  The $D$-term contribution
is fixed by the coefficients of the appropriate highest power in $\xi$
of the Mellin moments.  This can be used to obtain the Mellin moments
of $D$ as
\begin{eqnarray}
  \label{d-from-mellin}
\int_{-1}^1 dx\, x^{n-1} D^q(x) &=& \frac{1}{n!}\,
        \Big( \frac{\partial}{\partial\xi} \Big)^{n}
        \int_{-1}^1 dx\, x^{n-1} H^q(x,\xi,t) ,
\nonumber \\
\int_{-1}^1 dx\, x^{n-2} D^g(x) &=& \frac{1}{n!}\,
        \Big( \frac{\partial}{\partial\xi} \Big)^{n}
        \int_{-1}^1 dx\, x^{n-2} H^g(x,\xi,t) ,
\end{eqnarray}
where the right-hand sides can be evaluated at any $\xi$ due to the
polynomiality condition.  In analogy, the Gegenbauer moments of the
$D$ term are given by
\begin{eqnarray}
  \label{d-from-gegen}
\int_{-1}^1 dx\, C^{3/2}_{n-1}(x)\, D^{q}(x) = \frac{1}{n!}\,
        \Big( \frac{\partial}{\partial\xi} \Big)^{n}
        \mathcal{C}^q_{n-1}(\xi) ,
\nonumber \\
\int_{-1}^1 dx\, C^{5/2}_{n-2}(x)\, D^{g}(x) = \frac{1}{n!}\,
        \Big( \frac{\partial}{\partial\xi} \Big)^{n}
        \mathcal{C}^g_{n-1}(\xi)
\end{eqnarray}
in terms of the conformal moments (\ref{conf-mom}) of the GPDs.
Another consequence of polynomiality, reported in~\cite{Goeke:2001tz},
\begin{equation}
  \label{maxim}
\int_{-1}^1 dx\, \frac{H^q(x, \xi+x z) - H^q(x, \xi)}{x} =
\sum_{n=1}^\infty z^n \int_{-1}^1 dx\, x^{n-1} D^q(x)
\end{equation}
and its analog for gluons is readily checked against
(\ref{d-from-mellin}) by taking the derivative $\partial^{n}
/(\partial z)^n$ and setting $z$ to zero.  Resumming the geometric
series on the right-hand side we get
\begin{equation}
  \label{maxim-plus} \int_{-1}^1 dx\, \frac{D^q(x)}{1- xz} =
\int_{-1}^1 dx\, \frac{H^q(x, \xi+x z) - H^q(x, \xi)}{xz} ,
\end{equation}
where one may take any $\xi$ and $|z| < 1 - |\xi|$, thus staying
within the physical region for the second argument of the GPD.

We note that there are alternative ways to generate GPDs with the
correct polynomiality properties.  For definiteness let us consider
the quark distributions in a pion.  As observed in the pioneering
paper of Polyakov and Weiss \cite{Polyakov:1999gs} a more general
parameterization than (\ref{dd-pion}) is
\begin{eqnarray}
  \label{general-pion-dd}
\langle \pi^+(p')|\, \bar{q}(-\half z)\, \slash{z} q(\half z)\, 
  \,|\pi^+(p) \rangle \Big|_{z^2 =0} 
&=& 2 (P z) \int d\beta\, d\alpha\, 
        e^{-i \beta (Pz) + i \alpha (\Delta z) /2}\, f^q(\beta,
        \alpha,t) 
\nonumber \\
&-& (\Delta z)  \int  d\beta\, d\alpha\, 
        e^{-i \beta (Pz) + i \alpha (\Delta z) /2}\, g^q(\beta,
        \alpha,t) ,
\end{eqnarray}
where $2 Pz$ and $-\Delta z$ are treated in a symmetric way.  This
leads to a reduction formula
\begin{equation}
  \label{general-red}
H_\pi^q(x,\xi,t) = 
  \int d\beta\, d\alpha \, \delta(x - \beta - \xi \alpha)\, 
  \Big[ f^q(\beta, \alpha,t) + \xi g^q(\beta, \alpha,t) \Big] .
\end{equation}  
In analogy one may generalize the $D$-term in the decompositions
(\ref{dd-def}) for a spin $\half$ target to a double distribution
depending on both $\beta$ and $\alpha$ \cite{Polyakov:1999gs}.  The
decomposition (\ref{general-pion-dd}) into functions $f^q$ and $g^q$
is not unique.  Teryaev \cite{Teryaev:2001qm} has pointed out its
invariance under the combined transformations
\begin{eqnarray}
  \label{oleg-gauge-trf}
f^q(\beta,\alpha,t)  &\to&  f^q(\beta,\alpha,t)
	+ \frac{\partial}{\partial \alpha} \varphi(\beta,\alpha,t) ,
\nonumber \\
g^q(\beta,\alpha,t)  &\to&  g^q(\beta,\alpha,t)
	- \frac{\partial}{\partial \beta} \varphi(\beta,\alpha,t) ,
\end{eqnarray}
for a sufficiently regular function $\varphi$ vanishing at the
boundary $|\beta| + |\alpha| =1$ of the support region.  He has also
noted the analogy with a gauge transformation in two-dimensional
magnetostatics, with $(f,g)$ being the analog of the potential $(A^x,
A^y)$ and $(\alpha, -\beta)$ playing the role of the spatial
coordinates $(x,y)$.  The freedom (\ref{oleg-gauge-trf}), which
readily follows from (\ref{general-pion-dd}) via integration by parts,
may be used to transform $g^q$ to a function independent of $\beta$.
This function is then just the $D$-term of (\ref{dd-pion}) in the
$C$-even sector, and zero in the $C$-odd sector.  The invariance
property (\ref{oleg-gauge-trf}) of the general decomposition
(\ref{general-pion-dd}) has important consequences for constructing
models (see Section~\ref{sub:t-depend}).

Suggestions have also been made to parameterize the matrix element in
(\ref{general-pion-dd}) by a single function.  In
\cite{Belitsky:2000vk} it was proposed to define double distributions
that reduce to $x^{-1}\, H_\pi^q(x, \xi,t)$ instead of $H_\pi^q$ as in
(\ref{dd-red}).  This clearly raises the highest power in $\xi$ of the
Mellin moments by 1.  A different possibility
\cite{Teryaev:2001qm} is to take double distributions to generate
$\partial /(\partial x)\, H_\pi^q(x, \xi,t)$.  Neither suggestion has
much been pursued in the literature, because the $\xi=0$ limits of
$x^{-1} H_\pi^q$ and $\partial /(\partial x) H_\pi^q$ have stronger
singularities than $H_\pi^q$ at $x=0$, which turns out to be
unpleasant for modeling purposes, see Section~\ref{sec:ansatz}.  More
recently Pobylitsa \cite{Pobylitsa:2002vi} has considered double
distributions which reduce to $(1-x)^{-1}\, H_\pi^q(x, \xi,t)$ instead
of $H_\pi^q$.

\subsubsection{The crossed channel and the inversion formula}
\label{sub:radon}

It is straightforward to obtain a GPD from the corresponding double
distribution (and the associated $D$-term if appropriate), and it is
natural to ask how to compute a double distribution from a given GPD.
This inversion problem has been solved in
\cite{Belitsky:2000vk,Teryaev:2001qm}, with an earlier partial result
in \cite{Radyushkin:1998bz}.  Unfortunately this solution is difficult
to use in practice; it is however instructive to see why that is the
case.

Let us first try to invert the double Fourier transform (\ref{dd-def})
defining our double distributions, after appropriately projecting out
a chosen proton spinor structure.  Given the finite support of the
double distributions (assumed to be sufficiently well-behaved), the
Fourier integrals in (\ref{dd-def}) are defined for all $Pz$ and
$\Delta z$.  The matrix element on the left-hand side is however
restricted to the region $|\Delta z| < 2 |P z|$.  To see this, choose
a frame with $z^+=0$ with $\tvec{z}=0$, where this condition reads
$|p'^+ - p^+| < p'^+ + p^+$, i.e.\ $|\xi| < 1$ and follows from the
positivity of the plus-momentum for physical states.  To invert the
double Fourier transform in (\ref{dd-def}) one thus needs an
appropriate continuation of the light-cone matrix elements to the
unphysical region $|\Delta z| \ge 2 |P z|$.

\begin{figure}
\begin{center}
     \leavevmode
     \epsfxsize=0.68\textwidth
     \epsffile{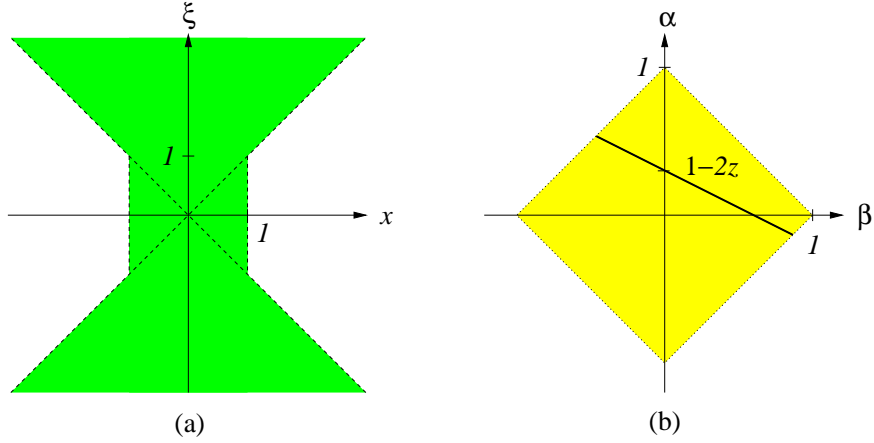}
\end{center}
\caption{\label{fig:dd-extension} (a) Support region of GPDs
following from the reduction formula (\protect\ref{dd-red}).  (b)~Line
of integration to obtain a GDA from a double distribution according to
the reduction formula (\protect\ref{cross-red}).}
\end{figure}

Similarly, the right-hand side of the reduction formula (\ref{dd-red})
gives a nonzero result outside the physical region $|x| < 1$ and
$|\xi| < 1$.  Given the support properties of double distributions and
the $D$-term this additional region is given by $|x| < |\xi|$ for
$|\xi| > 1$.  The full region of support of the functions $H$, $E$
defined by the representation (\ref{dd-red}) is the hour-glass shaped
region shown in Fig.~\ref{fig:dd-extension}a, and for $|\xi| > 1$ the
corresponding lines in the $\beta$-$\alpha$ plane have a slope
$-1/\xi$ between $-1$ and $1$ as shown in
Fig.~\ref{fig:dd-extension}b.

Clearly $|\xi| > 1$ requires one of the plus-momentum fractions $p^+$
or $p'^+$ to be negative.  This should remind us of the crossing
relation between GPDs and GDAs discussed in Section~\ref{sec:gda}.  In
fact, one can represent the crossed matrix element relevant for GDAs
by double distributions and a $D$-term in full analogy to
(\ref{dd-def}).  For the simpler case of pions one has
\cite{Teryaev:2001qm,Kivel:2002ia}
\begin{eqnarray}
  \label{dd-pion-time}
\lefteqn{
\langle \pi^+(p) \pi^-(p')|\, \bar{q}(-\half x)\, \slash{x} q(\half x)\, 
  \,|0 \rangle \Big|_{x^2 =0} 
}
\nonumber \\
&=& (p-p')x \,  \int d\beta\, d\alpha\, 
        e^{-i \beta (p-p')x /2 
		+ i \alpha (p+p')x /2}\, f^q(\beta, \alpha,s)
\nonumber \\
&-& (p+p')x \,  \int d\alpha\, 
        e^{i \alpha (p+p')x /2}\, D^q(\alpha,s)
\end{eqnarray}
as crossed version of (\ref{dd-pion}), and a reduction formula
\begin{eqnarray}
  \label{cross-red}
-\frac{1}{2} \Phi^q(z,\zeta,s) &=& (1-2\zeta) \int d\beta\, d\alpha\, 
  \delta\Big( (1-2z) - \beta (1-2\zeta) - \alpha \Big)
     f^q(\beta,\alpha,s) 
\nonumber \\
 & & {}+ D^q(1-2z, s) .
\end{eqnarray}
Analogous expressions hold for gluons.  The spectral functions
$f^q(\beta,\alpha,s)$ and $D^q(\alpha,s)$ are the analytic
continuation of the double distributions to positive values of the
invariant $t$ (which in the GDA channel we call $s$).  For positive
$t$, integrals of double distributions over the lines with slope
between $-1$ and $1$ in Fig.~\ref{fig:dd-extension}b thus generate
GDAs in their physical region, with the slope equal to $2\zeta-1$ and
the intercept with the $\alpha$-axis at $1-2z$.  The case of zero
slope, i.e.\ integration over $\beta$ at fixed $\alpha$ corresponds to
a GDA at the particular point $\zeta=\half$.  This confirms the
interpretation of $\alpha$ as a DA-like variable, which underlies most
intuition about what the functional dependence of double distribution
may be.

Let us assume that one has obtained $H$ or $E$ in the unphysical
region $|\xi| \ge 1$, for instance by analytically continuing the
corresponding $p\bar{p}$ distribution amplitudes from their physical
region to $s<0$.  It is then indeed possible to invert the reduction
formula (\ref{dd-red}): the particular integral over the double
distribution there is known as a Radon transform (for references
see~\cite{Belitsky:2000vk,Teryaev:2001qm}), whose inversion can be
obtained via its relation to Fourier transformation.  One finds
\cite{Teryaev:2001qm}
\begin{eqnarray}
f^q(\beta,\alpha) &=&
- \frac{1}{2\pi^2} \, \int_{-\infty}^{\infty} d\xi
        \pv{3}\int_{-\infty}^{\infty} dz\, 
        \frac{ H^q_{DD}(z + \beta+\alpha \xi, \xi) - 
               H^q_{DD}(\beta+\alpha \xi, \xi) }{z^2}
\end{eqnarray}
where $H^q_{DD}(x,\xi) = H^q(x,\xi) - \mbox{sgn}(\xi) D^q(x /\xi)$ has
the $D$-term subtracted.  Analogous relations exist for the gluon
distributions.  For the case where $H^q(x,\xi)$ has a non-integrable
singularity at $x=0$ we refer to \cite{Belitsky:2000vk}.

A similar task is to calculate the $D$-term for a given GPD.  Also
here there is no known closed expression which would not either
involve a moment inversion or knowledge of $H$ or $E$ outside their
physical regions.  As remarked in \cite{Belitsky:2000vk} one has
$D^q(z) = \half \lim_{\xi\to \infty} H^{q(+)}(z \xi, \xi)$, which can
be seen from the reduction formula (\ref{dd-red}).  For gluons one has
correspondingly $D^g(z) = \lim_{\xi\to \infty} \xi^{-1}\, H^g(z \xi,
\xi)$.  We remark that the relation (\ref{maxim-plus}) for $D$ has the
form of a Hilbert transform and admits an explicit inversion, which
requires one to know the right-hand side of (\ref{maxim-plus}) for all
values of $z$.  This needs again a continuation of the GPDs to
skewness $|\xi| >1$.  It is amusing to note that in the crossed
channel, the $D$ term is easily reconstructed from the GDA at
$\zeta=\half$, as seen in (\ref{cross-red}).


\subsection{Impact parameter space}
\label{sec:impact}

Our discussion and interpretation of GPDs so far was in momentum
space.  An intuitive physical picture can also be obtained in position
space, which we will now present.  More precisely, we will use a
``mixed representation'', keeping momentum in the light-cone
plus-direction, but Fourier transforming from transverse momentum to
transverse position, called ``impact parameter'' in this context.
This representation is useful in a variety of contexts, such as
high-energy scattering (see Section~\ref{sec:small-x}) or the
resummation of Sudakov logarithms in hard processes
\cite{Collins:1981uk}.  Its use in the context of GPDs has been
pioneered by Burkardt with emphasis on the case $\xi=0$, where a
density interpretation is possible, see the original
work~\cite{Burkardt:2000za} and the detailed
review~\cite{Burkardt:2002hr}.  This framework has been extended to
nonzero $\xi$ in \cite{Diehl:2002he}.  {}From a more general
perspective, Ralston and Pire \cite{Ralston:2001xs} have highlighted
the analogy with imaging techniques: the Fourier transform occurs in
geometrical optics, with light rays corresponding to momentum space
and an image plane to transverse position, or in $X$-ray diffraction
of crystals, where the diffraction pattern has to be Fourier
transformed to recover spatial information.

The first step in constructing impact parameter GPDs is to form hadron
states with definite plus-momentum and definite position $\tvec{b}$ in
the transverse plane (and hence undetermined transverse momentum):
\begin{equation}
|p^+, \tvec{b}, \lambda\rangle 
  = \int \frac{d^2 \tvec{p}}{16\pi^3}\, e^{-i \tvec{p}\tvec{b}}\,
    |p^+, \tvec{p}, \lambda\rangle .
 \label{proton-sharp}
\end{equation}
Choosing states of definite light-cone helicity on the right-hand side
we find that under a rotation by $\varphi$ about the $z$ axis this
state transforms as $|p^+, \tvec{b}, \lambda\rangle \to e^{-i\lambda
\varphi} |p^+, \tvec{b}', \lambda\rangle$, where $\tvec{b}'$ is
related to $\tvec{b}$ by the same rotation.  Under a spatial
translation by $\tvec{a}$ the state transforms like $|p^+, \tvec{b},
\lambda\rangle \to |p^+, \tvec{b} + \tvec{a}, \lambda\rangle$, as one
would expect.

Of course, the proton is an extended object, and we should make it
more explicit what ``localized in the transverse plane'' means.  The
key to this is the analogy of transverse boosts
(\ref{transverse-boost}) with Galilean transformations in
two-dimensional nonrelativistic mechanics.  In this analogy,
$\tvec{v}$ in (\ref{transverse-boost}) corresponds to the velocity
characterizing the transformation, and $k^+$ to the mass of the
particle.  The conserved quantity following from Galilean invariance
is the center-of-mass coordinate, $\tvec{r} = \sum m_i
\tvec{r}_i /\sum m_i$ of a many-body system, and by analogy the
conserved quantity following from invariance under transverse boosts
is the ``center of plus-momentum'' \cite{Soper:1977jc} or ``transverse
center of momentum'' \cite{Burkardt:2000wq} $\tvec{b} = \sum p^+_i
\tvec{b}_i^{\phantom{+}} /\sum p^+_i$ of the partons in the proton
state (\ref{proton-sharp}).  We will explicitly obtain this result in
the wave function representation, see Section~\ref{sec:overlap}.  The
field-theoretical operator for the center of plus-momentum is
constructed from the generators $B^i$ of transverse boosts as
\cite{Soper:1972xc}
\begin{equation}
R^i = - (P^+)^{-1} B^i , \qquad i=1,2.
\end{equation}
It has commutation relations $[R^i , P^j] = i \delta^{ij}$ and $[J^3 ,
R^i] = i\epsilon^{3ij} R^j$ as befits a position operator in the
transverse plane.  {}From the explicit form $B^i = \int dx^- d^2
\tvec{x} \, M^{++i}(x)$, where $M^{\alpha\mu\nu}$ is the
angular-momentum density (\ref{ang-mom-dens}), we get
\begin{equation}
R^i = (P^+)^{-1} \int dx^- d^2 \tvec{x}\; x^i\, 
        T^{++}(x) \Big|_{x^+=0}
\end{equation}
in terms of the energy-momentum tensor $T^{++}$.  Having taken
$T^{++}$ at $x^+ =0$ we can rewrite it in terms of the parton creation
and annihilation operators in light-cone quantization, and confirm our
interpretation of $\tvec{R}$: it is determined by the transverse
positions of the partons, weighted by their plus-momentum fractions
with respect to the total plus-momentum $P^+$ of the hadron
\cite{Burkardt:2002hr}.

Taking matrix elements of the precisely localized states $|p^+,
\tvec{b}, \lambda\rangle$ can lead to infinities due to their
normalization.  This can be avoided by forming wave packets,
integrating over $\tvec{p}$ in (\ref{proton-sharp}) with a weight that
falls off sufficiently fast at large $|\tvec{p}|$.  There is a further
condition if we wish to identify the state $|p^+, \tvec{b},
\lambda\rangle$ of definite $p^+$ as a hadron moving along the $z$
axis with a definite three-momentum, since $p^3 = (p^+ - p^-)
/\sqrt{2}$ and $p^- = (\tvec{p}^2 + m^2) /(2p^+)$.  To prevent $p^3$
from varying too much at given $p^+$ (and from becoming negative for
too large $|\tvec{p}|$) we must restrict the relevant range of
integration in (\ref{proton-sharp}) to $|\tvec{p}| \ll p^+$.  By the
uncertainty principle we can then achieve a transverse localization
$\Delta b_T \sim 1 /\Delta{p}_T \gg 1 /p^+$.  This is however not a
strong restriction in a frame where the proton moves sufficiently fast
in the $z$-direction, and it is anyway in such a frame that the parton
picture is most adequate.  With this in mind we will use the sharply
localized states $|p^+, \tvec{b}, \lambda\rangle$ in the following;
explicit calculations with wave packets can be found in
\cite{Burkardt:2000za,Diehl:2002he}.

The impact parameter representation of GPDs is obtained by Fourier
transforming the light-cone matrix elements $F_{\lambda'\lambda}$ and
$\tilde{F}_{\lambda'\lambda}$ with respect to the vector $\tvec{D}$
defined in (\ref{D-def}).  The result reads \cite{Diehl:2002he}
\begin{eqnarray}
  \label{matrix-final}
\lefteqn{
\int \frac{d^2 \tvec{D}}{(2\pi)^2}\; e^{-i\, \tvec{D} \tvec{b}}\, 
    \Big[ H - \frac{\xi^2}{1-\xi^2}\, E \,\Big]
}
\nonumber \\
 &=& \mathcal{N}^{-1}\,  \frac{1+\xi^2}{(1-\xi^2)^{5/2}} \;
     \Bigg\langle p'^+, -\frac{\xi \tvec{b}}{1-\xi}, +\half \,\Bigg|\,
          \mathcal{O}(\tvec{b}) \,\Bigg|\,
     p^+, \frac{\xi \tvec{b}}{1+\xi} , +\half \Bigg\rangle ,
\nonumber \\[0.3em]
\lefteqn{
\frac{i}{2m} \left( 
  \frac{\partial}{\partial b^1} + i \frac{\partial}{\partial b^2} 
\right)
\int \frac{d^2 \tvec{D}}{(2\pi)^2}\; e^{-i\, \tvec{D} \tvec{b}}\, E
}
\nonumber \\
 &=& \mathcal{N}^{-1}\,  \frac{1+\xi^2}{(1-\xi^2)^{5/2}} \;
    \Bigg\langle p'^+, -\frac{\xi \tvec{b}}{1-\xi}, -\half \,\Bigg|\,
         \mathcal{O}(\tvec{b}) \,\Bigg|\,
    p^+, \frac{\xi \tvec{b}}{1+\xi} , +\half \Bigg\rangle
\end{eqnarray}
with the quark-antiquark operator
\begin{equation}
\mathcal{O}(\tvec{b}) = \int \frac{d z^-}{4\pi}\, e^{ix P^+ z^-}
  \bar{q}(0,-\half z^-, \tvec{b})\, \gamma^+ 
	q(0, \half z^-, \tvec{b}) ,
\end{equation}
where position arguments of fields are given in the form $q(z) =
q(z^+,z^-,\tvec{z})$.  Analogous expressions exist for the Fourier
transforms of helicity dependent and gluon GPDs.  Note that the
Fourier transform is to be taken at fixed $x$ and $\xi$, so that the
$t$ dependence of the GPDs has to be rewritten using (\ref{t-by-D}).
Under this transform the prefactor $e^{i \varphi} |\tvec{D}| = D^1 + i
D^2$ in the helicity flip matrix element $F_{-+}$ has turned into
derivatives with respect to the components of $\tvec{b}$.  The factor
$\mathcal{N} = (64\pi^4)^{-1} \int d^2 \tvec{p}$ in
(\ref{matrix-final}) comes from the normalization of the states $|p^+,
\tvec{b}, \lambda\rangle$ and is singular like
$\delta^{(2)}(\tvec{0})$ as hinted at above.  It becomes finite for
proton states whose transverse position is smeared out.

\begin{figure}
\begin{center}
  \leavevmode
  \epsfxsize=0.53\textwidth
  \epsfbox{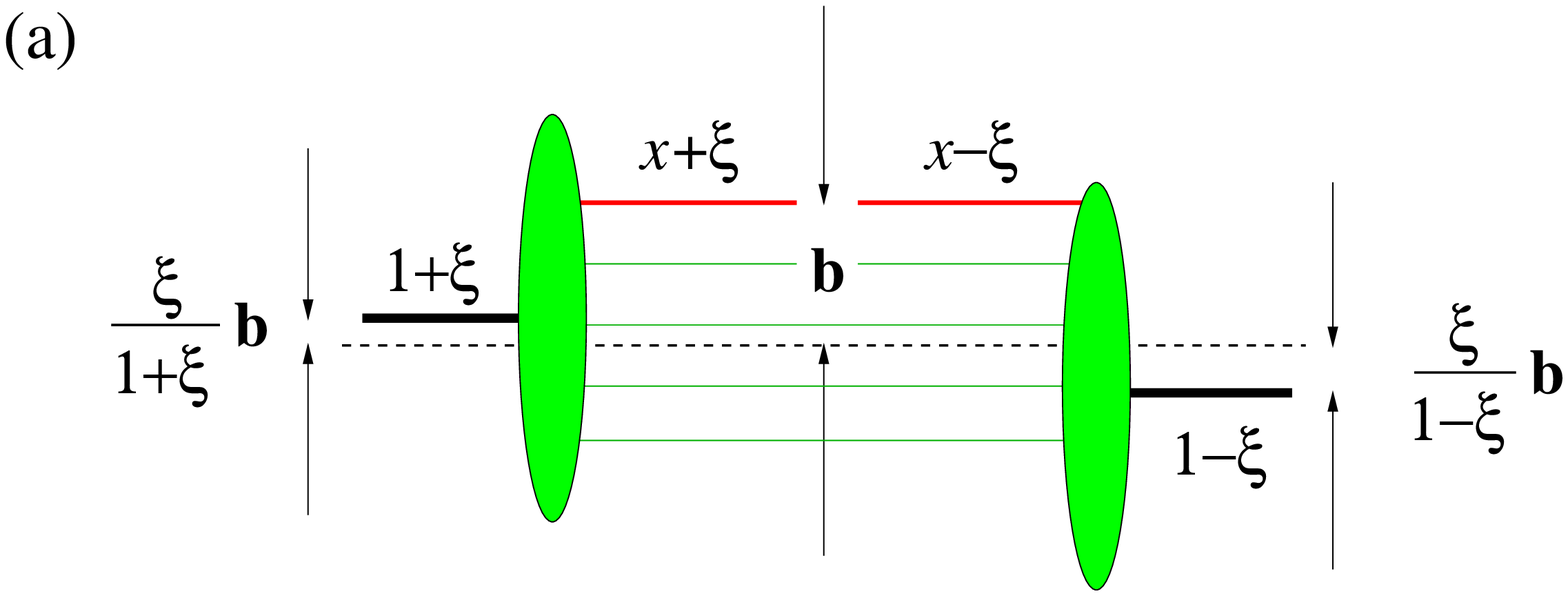}
\end{center}

\vspace{1em}

\begin{center}
  \epsfxsize=0.53\textwidth
  \epsfbox{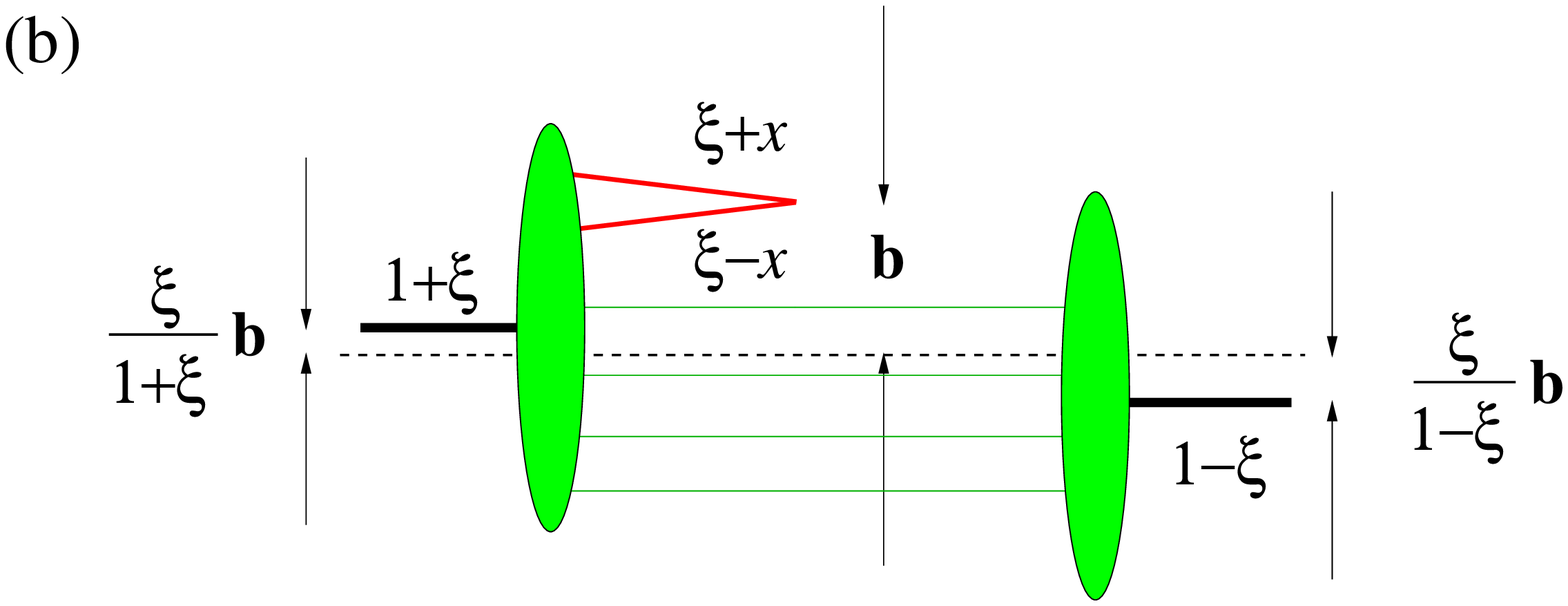}
\end{center}
\caption{\label{fig:impact} Representation of a GPD in impact
parameter space.  Plus-momentum fractions refer to the average proton
momentum $\frac{1}{2}(p+p')$ and are indicated above or below lines.
The region $x \in[\xi,1]$ is shown in (a), and the region $x
\in[-\xi,\xi]$ in (b).}
\end{figure}

Let us now discuss several aspects of the representation
(\ref{matrix-final}), visualized in Fig.~\ref{fig:impact}.
\begin{itemize}
\item The impact parameter representation offers an intuitive picture
of the information encoded in GPDs. According to the range of $x$ the
quark-antiquark operator either describes the emission and
reabsorption of a quark (or antiquark) at transverse position
$\tvec{b}$, or the emission of a quark-antiquark pair at position
$\tvec{b}$.  The transverse locations of the initial and final state
proton are shifted relative to each other by an amount of order $\xi
\tvec{b}$.  In the DGLAP region the struck quark suffers a loss of
plus-momentum proportional to $\xi$, so that the weighting of
transverse parton positions in the proton is different in the initial
and the final state.  Similarly, in the ERBL region the proton
``loses'' a quark-antiquark pair, which leads to a shift of its center
of plus-momentum.

\item Since this shift is proportional to $\xi$, it makes sense to say
that for small $\xi$ the impact parameter GPDs are ``almost'' diagonal
in transverse space.  This is in contrast to the longitudinal degrees
of freedom, because the plus-momenta are still very asymmetric between
the initial and final state if $x$ is of order $\xi$, even if $\xi$ is
small.

We also notice that the transverse shift depends on $\xi$ but not on
$x$.  In this sense, the transverse information is not ``washed out''
if one integrates the GPDs over $x$ with a weight depending on $x$ and
$\xi$.  This is of practical importance, because it is in such a
convolution that GPDs appear in the amplitudes of hard processes, and
to reconstruct functions of $x$ from the convolution is a delicate
problem.  The information that \emph{is} affected by integration over
$x$ is the separation of the DGLAP and ERBL regimes, which correspond
to different physical configurations in terms of quarks and
antiquarks.  Note however that for processes like DVCS and meson
electroproduction the imaginary part of the amplitude comes only from
the point $x=\xi$ to lowest order in $\alpha_s$, and only from the
DGLAP region beyond tree approximation.  Similarly, the separation of
the two partonic regimes of Fig.~\ref{fig:impact} is lost when one
goes from the mixed representation with its definite plus-momenta to
the position representation in full three-dimensional space.  This was
proposed in~\cite{Ralston:2001xs} but has not been very much explored
yet.  The position space operators appearing in the matrix elements
defining GPDs contain both quark and antiquark modes, and the
distinction of the two is naturally made in the space of frequencies
or (on the light-cone) of plus-momenta.

\item In practice one will only ever have experimental information on
GPDs in a limited range of $t$ and thus of $\tvec{D}$, so that the
integrals in (\ref{matrix-final}) can only be evaluated if one either
cuts off the integration range or extrapolates the integrand.  One may
also weight the integrand in (\ref{matrix-final}) with a smooth
function that suppresses values $|\tvec{D}|$ above some $\Delta D_T$.
As shown in \cite{Diehl:2002he} this corresponds to smoothly smearing
out the relative transverse positions of hadrons and partons in
Fig.~\ref{fig:impact} by an amount of order $\Delta b_T
\sim 1 /\Delta D_T$.  Whatever the procedure, with information on GPDs
up to a momentum transfer $|t|_{\mathrm{max}}$ one can only localize
partons up to order $(|t|_{\mathrm{max}} - |t_0|)^{-1/2}$.

It is important to distinguish this resolution limit from the
resolution described by the renormalization scale $\mu$ of the GPDs,
set by the hardness of the probe in the reaction where GPDs are
measured.  Since the evolution equations involve GPDs at fixed $\xi$
and $t$, their Fourier transform with respect to $\tvec{D}$ is
straightforward, and impact parameter GPDs evolve exactly like their
$t$-dependent counterparts.  As for ordinary parton distributions one
may think of $\mu$ as a cutoff on transverse momenta in the theory
(with a sufficient amount of caution concerning the effects of a
cutoff regularization on Lorentz and gauge symmetries).  The partons
described by the bilocal operators then have an effective size or
extension of order $1/\mu$, and may reveal themselves as consisting of
several partons at finer spatial resolution \cite{Kogut:1974ub}.  The
scale $\mu$ thus specifies \emph{what} the partons are that are probed
in the two regimes of Fig.~\ref{fig:impact}, whereas the maximum $|t|$
of the measurement limits the ability to determine \emph{where} these
partons are located in the proton.
\end{itemize}

In the preceding discussion we have again assumed the $A^+ =0$ gauge,
which is natural for the parton interpretation.  We remark that
(\ref{matrix-final}) holds in any other gauge, with a Wilson line
along a light-like path inserted between the two quark fields.
Although this will modify the partonic interpretation in a way that
has not much been explored so far, it should not alter the basic
picture we have obtained on the transverse structure, since the
additional $A^+$ gluons in the Wilson line are all at the same impact
parameter $\tvec{b}$ as the struck partons in Fig.~\ref{fig:impact}.

A brief comment is in order concerning the choice of the momentum
variable $\tvec{D}$ and the Fourier conjugate impact parameter
$\tvec{b}$.  In \cite{Pobylitsa:2002iu,Belitsky:2002ep} the impact
parameter was defined with respect to $\tilde{\tvec{D}} = (1-\xi^2)
\tvec{D}$, which equals $\tvec{\Delta}$ in frames where
$\tvec{p}^{\,\prime} = - \tvec{p} = \half\tvec{\Delta}$.  With this
choice of variables one has e.g.
\begin{eqnarray}
  \label{matrix-altern}
\int \frac{d^2 \tilde{\tvec{D}}}{(2\pi)^2}\; 
        e^{-i\, \tilde{\tvec{D}} \tilde{\tvec{b}}}\, 
    \Big[ H - \frac{\xi^2}{1-\xi^2}\, E \Big]
 \;=\; \mathcal{N}^{-1}\,  \frac{1+\xi^2}{(1-\xi^2)^{1/2}}\;
	\Bigg\langle p'^+, -\xi \tilde{\tvec{b}}, +\half \,\Bigg|\,
           \mathcal{O}(\tilde{\tvec{b}})
        \,\Bigg|\, p^+, \xi \tilde{\tvec{b}} , +\half \Bigg\rangle ,
\end{eqnarray}
which looks even simpler than (\ref{matrix-final}).  We will however
find that the variable $\tvec{b}$ leads to more natural expressions in
the context of the wave function representation, see
Section~\ref{sec:overlap}.  In fact, the different transverse
variables in (\ref{matrix-final}) reflect the difference of
longitudinal momenta.  The variables $\tvec{b}$ and
$\widetilde\tvec{b}$ describe of course the same physics, just as the
different sets of plus-momentum fractions common in the literature.

\subsubsection{The case $\xi=0$}
\label{sub:impact-forward}

We have seen that for nonzero skewness $\xi$ GPDs are nondiagonal both
in longitudinal momenta and in impact parameter.  For $\xi=0$ one
recovers a symmetric configuration in both variables, and if one
chooses the same polarization for the initial and final state proton,
one obtains the interpretation as a density of quarks or gluons with
momentum fraction $x$ and transverse distance $\tvec{b}$ from the
proton center \cite{Burkardt:2000za}.  More precisely, for $x>0$ the
Fourier transform of $H^q$ gives the density of quarks with any
helicity, whereas the transform of $\tilde{H}^q$ is the difference of
densities for positive-helicity and negative-helicity quarks, with
$x<0$ corresponding to antiquarks as usual.  This joint information
about longitudinal momentum and transverse location can be reduced in
two ways.  Taking the integral over $\tvec{b}$ one recovers the case
$t=0$ in momentum space and thus the usual parton distributions with
their familiar density interpretation.

Integrating over $x$ one obtains the Fourier transforms of elastic
form factors, $F_1^q$ from $H^q$ and $\smash{g_A^q}$ from
$\tilde{H}^q$, as \emph{two-dimensional} densities (or density
differences) \cite{Soper:1977jc}.  As emphasized in
\cite{Burkardt:2000za} this is different from the more familiar
interpretation of Fourier transformed form factors as
\emph{three-dimensional} densities in the non-relativistic limit,
which goes back to Sachs \cite{Sachs:1962aa} and underlies the work of
Polyakov and Shuvaev \cite{Polyakov:2002wz,Polyakov:2002yz} briefly
discussed at the end of Section~\ref{sec:spin}.  The three-dimensional
density interpretation refers to a hadron at rest and receives
relativistic corrections, see the discussion in
\cite{Burkardt:2000za}.  These corrections reflect that one cannot
localize the hadron in all three dimensions more accurately than
within its Compton wavelength.  An extension of this framework using
the concept of phase space distributions has recently been proposed by
Belitsky \cite{Belitsky:2003tm} and Ji~\cite{Ji:2003ak}.  In contrast,
the interpretation obtained here refers to a hadron moving fast.  It
gives a density in the plane transverse to the direction of motion and
is fully relativistic.  Even the Fourier transforms from momentum to
position space are different in the two cases, being two-dimensional
in one and three-dimensional in the other.

One may wonder how form factors like $F_1(t)$ at small to moderate $t$
can carry any information about partons, which can only be resolved by
probes with large virtuality.  The solution of this apparent paradox
is that $F_1$ is the form factor of the conserved current $\bar{q}(x)
\gamma^\mu q(x)$ and thus does not depend on a renormalization scale
$\mu$: it is the same whether directly measured in elastic scattering
at moderate $t$ or obtained via a sum rule from GPDs measured in hard
processes.  In more physical terms, $F_1$ measures the transverse
distribution of charge, which is insensitive to whether one resolves
individual partons or not.

Whereas the integrals of quark GPDs over $x$ have lost any reference
to the longitudinal parton momentum, some information about it is
retained when taking higher moments in $x$.  Fourier transforming the
integral $\int dx\, x H^q$ to impact parameter space gives the
transverse distribution of partons weighted with their momentum
fraction.  Such higher moments do depend on a resolution scale $\mu$,
which is physically plausible since for instance the DGLAP-type
splitting of a quark into a quark and a gluon transfers momentum from
quarks to gluons.  Only the sum $\sum_q \int_{-1}^1 dx\, x H^q +
\int_0^1 dx\,H^g$ over all partons, corresponding to the conserved
total energy-momentum tensor, gives a transverse density of
plus-momentum which does not evolve with the resolution scale $\mu$.

A density interpretation of the distributions $E^q$ and $E^g$ at
$\xi=0$ is more subtle, since the corresponding matrix elements are
still non-diagonal in the proton helicity, see~(\ref{matrix-final}).
One can however diagonalize by the usual trick of changing basis from
helicity states $|+\rangle_z$ and $|-\rangle_z$ to transversity states
$|\pm \rangle_x = \frac{1}{\sqrt{2}} (|+\rangle_z \pm |-\rangle_z)$.
For a particle at rest the new basis states are polarized along the
positive or negative $x$-axis, but for our fast-moving and
transversely localized protons (\ref{proton-sharp}) they are not
eigenstates of the angular momentum operator $J^1$ along the $x$-axis,
and their physical meaning is not quite clear.  This reflects the
notorious difficulty of defining transverse spin for relativistic
particles and the complicated nature of transverse spin operators in
the light-cone framework, see e.g.~\cite{Brodsky:1998de}.  Proceeding
nevertheless along this line, Burkardt has obtained several physically
rather intuitive results~\cite{Burkardt:2002hr}.  The Fourier
transforms of $E^{q}$ and $E^{g}$ describe a relative shift in the
transverse density of partons along the $y$ direction between the
polarization states $|+\rangle_x$ and $|-\rangle_x$, or between the
states $|+\rangle_x$ and $|+\rangle_z$.  The moments $F_2^q= \int dx\,
E^q$ and $\int dx\, x E^q$ at $\xi=0$ and $t=0$ are related with the
corresponding averages $\langle b^y \rangle$ and $\langle x b^y
\rangle$ for quarks.  Conservation of the transverse center of
momentum implies that the sum of $\langle x b^y \rangle$ over all
parton species is zero in a hadron with zero center of momentum, which
provides another derivation of the sum rule (\ref{magnetic-zero}) for
the distributions $E^q$ and $E^g$ in the forward limit.  Burkardt also
notes that the shift in the transverse distribution of partons in a
polarized state $|+\rangle_x$ is consistent with the classical picture
of a polarized proton as a rotating sphere.

The distributions $\tilde{E}^q$ and $\tilde{E}^g$ have no
representation analogous to the ones we have just discussed, since at
$\xi=0$ they decouple from the matrix elements
$\tilde{F}_{\lambda'\lambda}$ in (\ref{hel}).  As discussed in
Section~\ref{sub:counting} this is a constraint from time reversal
invariance.  In the present context it implies that the densities in
$x$ and $\tvec{b}$ one can form with the light-cone operators
$\bar{q}\gamma^+\gamma_5 q$ and $G^{+\mu}\tilde{G}_{\mu}{}^{+}$ are
described by the Fourier transforms of $\tilde{H}^{q}$ and
$\tilde{H}^{g}$ alone, for any superposition of the helicity states
$|+\rangle_z$ and $|-\rangle_z$.

To conclude this subsection let us point out that there is another
class of hadronic matrix elements that carries information on both the
transverse and longitudinal structure of partons, namely $k_T$
dependent or $k_T$ unintegrated parton distributions (see
e.g.~\cite{Mulders:1996dh,Mulders:1999mc} and also
Section~\ref{sub:beyond-collinear}).  In momentum space they specify
the transverse momentum of the struck parton.  This corresponds to
different transverse positions of the emitted and the reabsorbed
parton in the DGLAP regime, and to different transverse positions of
the two emitted partons in the ERBL region.  The centers of
plus-momentum of the incoming and outgoing hadron are shifted
accordingly.  The Fourier transforms of unintegrated parton
distributions with respect to the parton ${k}_T$ thus describe
correlations of the transverse location of partons within a hadron,
and thus never represent densities in impact parameter space.  A more
detailed discussion is given in \cite{Diehl:2002he}.  The connection
between parton distributions in impact parameter space and those
depending on parton $k_T$ has recently been explored by Burkardt
\cite{Burkardt:2002ks,Burkardt:2003uw} with focus on the spin and
angular momentum structure of the nucleon.

\subsubsection{GDAs in impact parameter space}
\label{sub:gda-impact}

\begin{figure}
\begin{center}
  \leavevmode
  \epsfxsize=0.32\textwidth
  \epsfbox{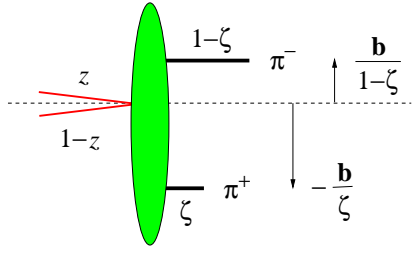}
\end{center}
\caption{\label{fig:impact-gda} Impact parameter representation of the
two-pion distribution amplitude.  Plus-momentum fractions $z$ and
$\zeta$ refer to the sum of the pion plus-momenta}
\end{figure}

In close analogy to the case of GPDs one can derive an impact
parameter representation for GDAs, as shown by Pire and Szymanowski
\cite{Pire:2002ut}.  One starts by  introducing two-pion states in
the impact parameter representation,
\begin{eqnarray}
| \pi^+(p^+, \tvec{b})\, \pi^-(p'^+, \tvec{b}') \rangle_{\mathrm{out}}
&& = 
  \int \frac{d^2 \tvec{p}}{16\pi^3}\, \frac{d^2 \tvec{p}'}{16\pi^3}\, 
	e^{-i \tvec{p}\tvec{b} - i \tvec{p}' \tvec{b}'}\,
  |\pi^+(p^+, \tvec{p}) \pi^-(p'^+, \tvec{p}') \rangle_{\mathrm{out}} .
 \label{pions-sharp}
\end{eqnarray}
Strictly speaking, one has to form appropriate wave packets to ensure
that in the far future one has two pions with a large separation along
the $z$ axis (even though their separation in the transverse plane may
be small, even compared with the pion radius) .  Only then may one
neglect their interactions, treat them as individual free particles,
and interpret (\ref{pions-sharp}) as two separate pions with fixed
plus-momenta and transverse positions.  One proceeds by decomposing
the invariant mass of the pair into terms coming from longitudinal and
transverse momentum components,
\begin{equation}
s = \frac{m^2}{\zeta (1-\zeta)} + \zeta (1-\zeta) \tvec{D}^2 ,
\end{equation}
where the vector $D$ is defined in the GDA channel by
\begin{equation}
{D} = \frac{{p}}{\zeta} - \frac{{p}'}{1-\zeta} .
\end{equation}
One then finds for the GDA in impact parameter space
\begin{eqnarray}
\lefteqn{
\int \frac{d^2 \tvec{D}}{(2\pi)^2}\; e^{-i\, \tvec{D} \tvec{b}}\, 
    \Phi^q(z,\zeta,s)
}
\nonumber \\
 &=& \mathcal{N}^{-1}\,  
	\frac{\zeta^2 + (1-\zeta)^2}{\zeta^2 (1-\zeta)^2} \;
    {\phantom{\Bigg|}}_{\mathrm{out}\!\!}\Bigg\langle 
		\pi^+\Big( p^+, -\frac{\tvec{b}}{\zeta} \Big)\,
                \pi^-\Big( p'^+, \frac{\tvec{b}}{1-\zeta} \Big)
    \,\Bigg|\, \mathcal{O}(\tvec{0}) \,\Bigg|\,0 \Bigg\rangle
\end{eqnarray}
with the same quark-antiquark operator $\mathcal{O}$ and normalization
factor $\mathcal{N}$ as above.  The physical interpretation is shown
in Fig.~\ref{fig:impact-gda}: a quark-antiquark pair at transverse
position zero hadronizes into two pions with transverse separation
$\zeta^{-1} (1-\zeta)^{-1} \tvec{b}$.  The overall center of
plus-momentum is conserved in this process.  As discussed in
\cite{Pire:2002ut} one can for instance see how hadronization into
a pion pair with $s$ close to a squared resonance mass $m_R^2$
translates into typical transverse distances of order $1/m_R$.
Hadronization into pairs of very large invariant mass $s$ corresponds
to small transverse distances between the initial $q\bar{q}$ pair and
the center of momentum of each final-state pion, which provides a
physical picture for the perturbative mechanism discussed in
Section~\ref{sub:very-large-s}.

\subsection{The wave function representation}
\label{sec:overlap}

GPDs can be represented in terms of the wave functions for the target.
This offers a further way to make explicit which kind of information
on hadron structure is contained in these quantities.  Certain
properties of GPDs, in particular the positivity constraints discussed
in Section~\ref{sec:positivity} are naturally derived from this
representation, which furthermore has been used as a strategy to model
GPDs, see Section \ref{sub:soft-overlap-gpd}.  It can equally be given
in momentum space and in the impact parameter representation.

\subsubsection{Fock space decomposition}
\label{fock-space}

The wave function representation is most naturally obtained in the
framework of light-cone quantization and of $A^+ =0$ gauge, and has
been derived and discussed in \cite{Diehl:2000xz,Diehl:2002he} (whose
notation we will follow here) and in \cite{Brodsky:2000xy}.  Starting
point is the Fock state expansion in QCD quantized on the light-cone
\cite{Brodsky:1989pv}, where a hadron state is represented in terms of
the partonic Fock states created from the vacuum by the operators
$b^\dag$, $d^\dag$, $a^\dag$, which appear in the decompositions
(\ref{good-quark}), (\ref{good-gluon}) of the good components of quark
and gluon fields.  This expansion reads
\begin{eqnarray}
  \label{Fock-mom}
\lefteqn{
|p^+, \tvec{p}, \lambda\rangle =  \sum_{N, \beta}\, 
  \frac{1}{\sqrt{f_{N\beta}}} \;
        \int [d x]_N \, [d^2 \tvec{k}]_{N} \;
  \Psi^\lambda_{N\beta}(x_i, \tvec{k}_{i}) \,
  }
\nonumber \\
&\times&   
  \prod_{i=1}^{N_q} \,
   \frac{1}{\sqrt{x_i}}\, b^\dag(x_i\, p^+, \tvec{k}_{i} + x_i \tvec{p})
  \prod_{i=N_q+1}^{N_q+N_{\bar{q}}}
   \frac{1}{\sqrt{x_i}}\, d^{\,\dag}(x_i\, p^+, \tvec{k}_{i} + x_i \tvec{p})
\nonumber \\
&& \hspace{9em} {}\times
  \prod_{i=N_q+N_{\bar{q}}+1}^{N}
   \frac{1}{\sqrt{x_i}}\, a^\dag(x_i\, p^+, \tvec{k}_{i} + x_i \tvec{p})
  \, |\, 0 \rangle ,
\end{eqnarray}
where $N$ specifies the number of partons in a Fock state, and $\beta$
collectively labels its parton composition and the discrete quantum
numbers (flavor, color, helicity) for each parton.  $f_{N\beta}$ is a
normalization constant providing a factor $n!$ for each subset of $n$
partons with identical discrete quantum numbers (not explicitly
labeled in the operators for brevity).  For the $N$-parton integration
elements we have used the shorthand notation
\begin{eqnarray}
[d^2 \tvec{k}]_{N} &=& (16\pi^3)^{-N+1} \prod_{i=1}^N 
   d^2\tvec{k}_i\; \delta^{(2)} \Big( \sum_{i=1}^N \tvec{k}_i \Big) ,
\nonumber \\{}
[d x]_N &=& 
   \prod_{i=1}^N d x_i\; \delta\Big( 1 - \sum_{i=1}^N x_i \Big)  .
\end{eqnarray}

The wave functions $\Psi$ depend on the momentum variables of each
parton only via its light-cone momentum fraction $x^{\phantom{+}}_i =
k_i^+ /p^+$ and via its transverse momentum $\tvec{k}_{i}$ relative to
the transverse momentum of the hadron.  This is an important
consequence of Lorentz invariance on the light-cone: the state $|p^+,
\tvec{p}, \lambda\rangle$ can be obtained from a proton state at rest
by a combination of a longitudinal boost (to the state $|p^+,
\tvec{0}, \lambda\rangle$) and a transverse boost.  In the literature
different conventions for light-cone wave functions are used as far as
kinematical and overall factors are concerned.  Our wave functions
here are normalized such that
\begin{eqnarray}
P^\lambda_{N\beta}
&=& \int [d x]_N \, [d^2 \tvec{k}]_{N} \;
 \Big| \Psi^\lambda_{N\beta}(x_i, \tvec{k}_i) \Big|^2
\end{eqnarray}
is the probability to find the
corresponding Fock state in the proton, so that in total
$\sum_{N,\beta} P^\lambda_{N\beta} = 1$. 

Light-cone wave functions can be represented as matrix elements
between the hadron state and the vacuum, with operators formed from
the relevant quark or gluon field components at vanishing light-cone
time $z^+$.  As a simple example consider the $q\bar{q}$ Fock state in
a pion, with opposite parton helicities.  Its wave function can be
projected out as
\begin{eqnarray}
  \label{wf-from-operator}
\lefteqn{
\Psi_{+-}(x,\tvec{k} -x \tvec{p})
}
\\
&=&  {}- \frac{1}{2\sqrt{3}}\, 
  \int dz^- d^2 \tvec{z}\; e^{i (2x-1) p^+ z^- /2}\,
  e^{- i (2\tvec{k} - \tvec{p}) \tvec{z} /2}\,
  \langle \pi(p) \,|\, \bar{q}(-\half z)\, \gamma^+ \gamma_5\,
        q(\half z)\, |\, 0\rangle \Big|_{z^+=0} \: ,
\nonumber
\end{eqnarray}
where for simplicity we have only given the momentum arguments of the
quark.\footnote{Following common usage we have summed over quark
colors on the right-hand side of (\protect\ref{wf-from-operator}),
corresponding to the wave function for a $q\bar{q}$ state coupled to
zero color.}
The subscripts $+$ and $-$ respectively refer to the quark and
antiquark helicities, and we have used the parity constraint
$\Psi_{+-} = -\Psi_{-+}$ for a pion.  Configurations with equal parton
helicities are obtained from the tensor operator $\bar{q}
\sigma^{+i} q$ we have encountered in the discussion of helicity flip
GPDs, see~\cite{Burkardt:2002uc}.  The operators projecting out
three-quark wave functions in a proton have been given in
\cite{Ji:2002xn}.

Integrating a light-cone wave function over the transverse momenta of
all partons (after weighting with appropriate factors of
$\tvec{k}_{i}$ if the partons carry nonzero orbital angular momentum)
one obtains matrix elements of operators with light-like separation
between the fields, and thus distribution amplitudes.  For our simple
example we have
\begin{equation}
  \label{pion-wf-da}
\int \frac{d^2 \tvec{k}}{16\pi^3}\, \Psi_{+-}(z,\tvec{k})
  = {}- \frac{1}{4\sqrt{3}}\, \Phi^q_\pi(z)
\end{equation}
as easily seen from the definition (\ref{pseudo-da}).  Leading-twist
DAs correspond to the lowest Fock states $q\bar{q}$ of mesons and
$qqq$ for baryons, involve only the good components of the quark
fields, and describe vanishing orbital angular momenta of the partons.
In particular, the twist-two meson DAs are generated from the same
operators as the twist-two GPDs.  Beyond leading twist the situation
is more involved; results on meson and baryon DAs can respectively be
found in \cite{Ball:1998fj,Ball:1998je} and
\cite{Braun:2000kw} and the references therein.

\subsubsection{The overlap formulae}
\label{sub:overlap-formulae}

The wave function representation of GPDs is obtained by
(\textit{i})~inserting the Fock state expansions (\ref{Fock-mom}) into
the matrix elements defining GPDs, (\textit{ii})~representing the
quark-antiquark or two-gluon operators in the matrix elements by
parton creation and annihilation operators as in
(\ref{quark-helicities}), (\ref{gluon-helicities}), and
(\textit{iii})~using the (anti)commutation relations of these
operators.  The resulting formulae may appear somewhat involved
because of the parton kinematics, but their physics content and
interpretation is rather simple.

In the region $x \in [\xi,1]$ the result for the combinations of $H^q$
and $E^q$ in (\ref{hel}) is
\begin{eqnarray}
  \label{DGLAP-mom}
\lefteqn{
F^q_{\lambda'\lambda}(x,\xi,t) \;=\; \sum_{N,\beta}
  \sqrt{1-\xi}^{\, 2-N} \sqrt{1+\xi}^{\, 2-N}\, 
	\sum_{j=q} [d x]_N \, [d^2 \tvec{k}]_{N} \;
}
\\ \nonumber  
 &\times&  \delta(x-x_j)\;
  \Psi_{N\beta}^{*\, \lambda'}(x_i^{\rm out}, \tvec{k}_{i}^{\rm out})\,
  \Psi_{N\beta}^\lambda(x_i^{\rm in}, \tvec{k}_{i}^{\rm in}) \, ,
\phantom{ \prod_{i=1}^N }
\end{eqnarray}
where the label $j$ denotes the struck parton and is summed over all
quarks with appropriate flavor in a given Fock state, and the labels
$(N,\beta)$ are summed over Fock states.  The wave function arguments
are given by momentum fractions, referring to the plus-momenta of
either the initial or the final state hadron,
\begin{eqnarray}
  \label{fractions-DGLAP}
x_i^{\rm in}  &=&     \frac{x_i}{1+\xi} , \hspace{3.7em}
x_i^{\rm out} \;=\;   \frac{x_i}{1-\xi} 
                            \qquad\qquad \mbox{for~} i\neq j ,
\nonumber \\
x_j^{\rm in} &=&  \frac{x_j+\xi}{1+\xi} , \hspace{3.5em}
x_j^{\rm out} =\; \frac{x_j-\xi}{1-\xi} ,
\end{eqnarray}
and by transverse parton momenta relative to the transverse momentum
of their parent hadron,
\begin{eqnarray}
  \label{k-perp-DGLAP}
\tvec{k}_{i}^{\rm in}  &=&  \tvec{k}_{i}^{\phantom{t}} 
        - \frac{x_i}{1+\xi}\, \tvec{p}^{\phantom{t}} , 
\hspace{3.7em}
\tvec{k}_{i}^{\rm out} \;=\; \tvec{k}_{i}^{\phantom{t}} 
        - \frac{x_i}{1-\xi}\, \tvec{p}^{\prime \phantom{t}} 
   \qquad\qquad \mbox{for~} i\neq j ,
\nonumber \\
\tvec{k}_{j}^{\rm in} &=&  \tvec{k}_{j}^{\phantom{t}} 
        + \frac{1-x_j}{1+\xi}\, \tvec{p}^{\phantom{t}}  , 
\hspace{3.5em}
\tvec{k}_{j}^{\rm out} =\; \tvec{k}_{j}^{\phantom{t}} 
        + \frac{1-x_j}{1-\xi}\, \tvec{p}^{\prime \phantom{t}} .
\end{eqnarray}
These transverse momentum arguments can readily be determined by
performing a transverse boost to a frame where the corresponding
hadron moves along the $z$ axis.  The representation in the region
$x\in [-1,-\xi]$ is obtained from~(\ref{DGLAP-imp}) by reversing the
overall sign according to (\ref{quark-helicities}), by changing
$\delta(x-x_j)$ into $\delta(x+x_j)$, and by summing $j$ over
antiquarks.  We thus find that in the DGLAP regions a GPD is
essentially given by the product $\Psi^{\phantom{*}}_{\mathrm{in}}
\Psi^*_{\mathrm{out}}$ of wave functions for the initial and final
state hadron, summed over all partonic configurations where the
emitted and the reabsorbed parton have specified flavor and
plus-momenta.  The parton configurations in the initial and final
state differ in their kinematics unless one has the forward case $p' =
p$.

In the ERBL region we have instead the overlap of wave functions where
the initial-state hadron has two partons more than the final-state
one,\footnote{The overlap formulae (\protect\ref{ERBL-mom}) and
(\protect\ref{ERBL-imp}) in the ERBL region differ by a global sign
from those in \protect\cite{Diehl:2000xz,Diehl:2002he}, due to the
phase convention (\protect\ref{spinors}) for antiquark spinors used
here.}
\begin{eqnarray}
  \label{ERBL-mom}
\lefteqn{
F^q_{\lambda'\lambda}(x,\xi,t) \;=\;
   - \sum_{N,\beta,\beta'} 
   \sqrt{\, 1-\xi}^{\, 3-N} \sqrt{1+\xi}^{\, 1-N}\, 
} \hspace{2em}
\nonumber \\
 &\times& 
   \sum_{j,j'} \frac{1}{\sqrt{\rule{0pt}{1.4ex} n_j n_{j'}}}\, 
   \int d x_j \prod_{i\neq j,j'}
     d x_i\; \delta\Big( 1- \xi - \sum_{i\neq j,j'} x_i \Big)
\nonumber \\[0.4em]
 &\times& (16\pi^3)^{-N+1}
  \int d^2 \tvec{k}_{j} \prod_{i\neq j,j'} d^2 \tvec{k}_{i}\;
     \delta^{(2)} \Big( \tvec{p}' 
       - \sum_{i\neq j,j'} \tvec{k}_{i} \Big)
\nonumber \\
 &\times& \delta(x-x_j)\;
  \Psi_{N-1,\, \beta'}^{*\, \lambda'}(x_i^{\rm out}, 
                  \tvec{k}_{i}^{\rm out})\,
  \Psi_{N+1,\, \beta}^\lambda(x_i^{\rm in}, \tvec{k}_{i}^{\rm in}) .
\phantom{ \prod_{i=1}^N }
\end{eqnarray}
The partons $j,j'$ are the ones emitted from the initial proton.  One
has to sum over all quarks $j$ and antiquarks $j'$ with opposite
helicities, opposite color, and appropriate flavor in the initial
state proton, over all Fock states $(N+1,\beta)$ containing such a
$q\bar{q}$ pair, and over all Fock states $(N-1,
\beta')$ of the outgoing proton whose quantum numbers match
those of the spectator partons $i\neq j,j'$.  The statistical factors
$n_j$ ($n_{j'}$) give the number of (anti)quarks in the Fock state
$(N+1,\beta)$ that have the same discrete quantum numbers as the
(anti)quark pulled out of the target.  The wave function arguments of
the spectator partons $i\neq j,j'$ are given as before in
(\ref{fractions-DGLAP}) and (\ref{k-perp-DGLAP}), whereas for the
partons removed from the target they now read
\begin{eqnarray}
  \label{fractions-ERBL}
x_j^{\rm in}    &=&   \frac{\xi+x_j}{1+\xi} , 
\qquad \qquad \qquad \qquad
x_{j'}^{\rm in} \;=\; \frac{\xi-x_j}{1+\xi} .
\nonumber \\
\tvec{k}_{j}^{\rm in} &=&  \tvec{k}_{j}^{\phantom{t}} 
        - \frac{\xi+x_j}{1+\xi}\, \tvec{p}
        - \frac{1}{2} \tvec{\Delta} , 
\hspace{2.3em}
\tvec{k}_{j'}^{\rm in} =\; {}- \tvec{k}_{j}^{\phantom{t}} 
        - \frac{\xi-x_j}{1+\xi}\, \tvec{p}
        - \frac{1}{2} \tvec{\Delta} .
\end{eqnarray}
In this region GPDs thus probe a color-singlet $q\bar{q}$ pair in the
initial-state hadron wave function, specifying the plus-momentum for
the $q$ and the $\bar{q}$, and the overall transverse momentum and
helicity of the pair. 

Going now to the impact parameter representation we introduce
operators describing quarks or antiquarks of definite transverse
position,
\begin{eqnarray}
  \label{def-creator}
\tilde{b}(k^+, \tvec{b}, \mu) &=& \int \frac{d^2 \tvec{k}}{16\pi^3}\,
    b(k^+, \tvec{k}, \mu)\, e^{i\tvec{k} \tvec{b}} ,
\nonumber \\
\tilde{d}(k^+, \tvec{b}, \mu) &=& \int \frac{d^2 \tvec{k}}{16\pi^3}\,
    d(k^+, \tvec{k}, \mu)\, e^{i\tvec{k} \tvec{b}} ,
\end{eqnarray}
with an analogous definition for gluons at definite transverse
position.  The corresponding Fock state expansion is
\begin{eqnarray}
  \label{Fock-imp}
\lefteqn{
|p^+, \tvec{b}, \lambda\rangle =  \sum_{N, \beta}\, 
  \frac{1}{\sqrt{f_{N\beta}}}
        \int [d x]_N \, [d^2 \tvec{b}]_{N} \;
  \widetilde{\Psi}^\lambda_{N\beta}(x_i, \tvec{b}_{i})\,
}
\nonumber \\
&\times&  
  \prod_{i=1}^{N_q} \,
   \frac{1}{\sqrt{x_i}}\, 
	\tilde{b}^{\,\dag}(x_i p^+, \tvec{b}_{i} + \tvec{b})
  \prod_{i=N_q+1}^{N_q+N_{\bar{q}}} 
   \frac{1}{\sqrt{x_i}}\, 
	\tilde{d}^{\,\dag}(x_i p^+, \tvec{b}_{i} + \tvec{b})
\nonumber \\
&& \hspace{8.1em} {}\times
  \prod_{i=N_q+N_{\bar{q}}+1}^{N}
   \frac{1}{\sqrt{x_i}}\, 
	\tilde{a}^\dag(x_i p^+, \tvec{b}_{i} + \tvec{b})
  \, |\, 0 \rangle 
\end{eqnarray}
with an integration element
\begin{eqnarray}
[d^2 \tvec{b}]_{N} &=& (4\pi)^{N-1} \prod_{i=1}^N 
   d^2\tvec{b}_i\; \delta^{(2)} \Big( \sum_{i=1}^N x_i \tvec{b}_i \Big) ,
\end{eqnarray}
and wave functions normalized as
\begin{eqnarray}
P^\lambda_{N\beta}
&=& \int [d x]_N \, [d^2 \tvec{b}]_{N} \;
 \Big| \widetilde\Psi^\lambda_{N\beta}(x_i, \tvec{b}_i) \Big|^2 .
\end{eqnarray}
The wave functions for definite transverse momentum or impact
parameter are related by Fourier transforms 
\begin{equation}
\Psi_{N\beta}^\lambda(x_i, \tvec{k}_{i} - x_i \tvec{p})
 = \int [d^2 \tvec{b}]_N\,
   \exp\!\Big[ {-i} \sum_{i=1}^N \tvec{k}_{i} \tvec{b}_{i} \Big]\, 
  \widetilde\Psi^\lambda_{N\beta}(x_i, \tvec{b}_{i})
\end{equation}
with $\tvec{p} = \sum_{i=1}^N \tvec{k}_{i}$ and
\begin{equation}
  \label{impact-lcwf}
\widetilde\Psi_{N\beta}^\lambda(x_i, \tvec{b}_{i} - \tvec{b})
 = \int [d^2 \tvec{k}]_N\,
   \exp\!\Big[ i \sum_{i=1}^N \tvec{k}_{i} \tvec{b}_{i} \Big]\, 
  \Psi^\lambda_{N\beta}(x_i, \tvec{k}_{i})
\end{equation}
with $\tvec{b} = \sum_{i=1}^N x_i \tvec{b}_{i}$.  The latter condition
confirms that the transverse position $\tvec{b}$ of the proton is the
center of plus-momentum of the partons (in each Fock state) as we
anticipated in Section~\ref{sec:impact}.

The overlap formulae are very similar to the one in transverse
momentum space.  For $x\in [\xi,1]$ we have
\begin{eqnarray}
  \label{DGLAP-imp}
\lefteqn{
\int \frac{d^2 \tvec{D}}{(2\pi)^2}\; e^{-i\, \tvec{D} \tvec{b}}\, 
F^q_{\lambda'\lambda}(x,\xi,t) \;=\; \sum_{N,\beta}
  \sqrt{1-\xi}^{\, 2-N} \sqrt{1+\xi}^{\, 2-N}\, 
	\sum_{j=q} [d x]_N \, [d^2 \tvec{b}]_{N} \;
}
\nonumber \\ 
 &\times&  \delta(x-x_j)\; 
        \delta^{(2)} \Big( \tvec{b}-\tvec{b}_{j} \Big)\;
  \widetilde\Psi_{N\beta}^{*\, \lambda'}(x_i^{\rm out}, 
                \tvec{b}_{i}^{\phantom{t}} - \tvec{b}_{0}^{\rm out})\,
  \widetilde\Psi_{N\beta}^\lambda(x_i^{\rm in}, 
                \tvec{b}_{i}^{\phantom{t}} - \tvec{b}_{0}^{\rm in})
\phantom{ \prod_{i=1}^N }
\end{eqnarray}
and for $x\in [-\xi,\xi]$
\begin{eqnarray}
  \label{ERBL-imp}
\lefteqn{
\int \frac{d^2 \tvec{D}}{(2\pi)^2}\; e^{-i\, \tvec{D} \tvec{b}}\, 
F^q_{\lambda'\lambda}(x,\xi,t) \;=\;
   - \sum_{N,\beta,\beta'} 
   \sqrt{1-\xi}^{\, 3-N}\sqrt{1+\xi}^{\, 1-N}\, 
}
\\ 
 &\times& 
   \sum_{j,j'} \frac{1}{\sqrt{\rule{0pt}{1.4ex} n_j n_{j'}}}\, 
   \int d x_j \prod_{i\neq j,j'}
     d x_i\; \delta\Big( 1- \xi - \sum_{i\neq j,j'} x_i \Big)
\nonumber \\[0.4em]
 &\times& (4\pi)^{N-1}\, 
  \int d^2 \tvec{b}_{j} \prod_{i\neq j,j'} d^2 \tvec{b}_{i}\;
     \delta^{(2)} \Big( \xi \tvec{b}_{j} 
       + \sum_{i\neq j,j'} x_i \tvec{b}_{i} \Big)
\nonumber \\
 &\times& \delta(x-x_j)\; \delta^{(2)} 
        \Big( \tvec{b}-\tvec{b}_{j} \Big)\:
  \widetilde\Psi_{N-1,\, \beta'}^{*\, \lambda'}(x_i^{\rm out}, 
             \tvec{b}_{i}^{\phantom{t}} - \tvec{b}_{0}^{\rm out})\,
  \widetilde\Psi_{N+1,\, \beta}^\lambda(x_i^{\rm in}, 
             \tvec{b}_{i}^{\phantom{t}} - \tvec{b}_{0}^{\rm in})
\phantom{ \prod_{i=1}^N }
\nonumber 
\end{eqnarray}
with $\tvec{b}_{j'} = \tvec{b}_{j}$.  The transverse positions of the
hadrons are given by
\begin{eqnarray}
  \label{positions-DGLAP}
\tvec{b}_{0}^{\rm in}  &=&    \frac{\xi}{1+\xi}\, \tvec{b}_{j} , 
\qquad
\tvec{b}_{0}^{\rm out} \;=\; -\frac{\xi}{1-\xi}\, \tvec{b}_{j} 
\end{eqnarray}
in both cases.  The interpretation of these expressions is exactly the
one we have previously given in Fig.~\ref{fig:impact}, where the blobs
represent the hadronic wave functions in the mixed representation of
plus-momentum and impact parameter.  In addition to the results
obtained in Section~\ref{sec:impact} we see that the transverse
locations of the spectator partons are conserved, as well as the
relative distance between the struck parton and the spectators in the
DGLAP region.  What \emph{does} change if $\xi \neq 0$ are the
relative positions of partons with respect to the center of
plus-momentum of their parent hadron.

Analogous representations can be obtained for all other twist-two GPDs
we have discussed so far.  The appropriate restrictions on the
helicities of the struck partons are readily determined from the
operators in Table~\ref{tab:helicities} and the correspondence between
left-handed incoming and right-handed outgoing partons.  The relevant
global signs and factors are given in
\cite{Diehl:2000xz}.\footnote{In \protect\cite{Diehl:2000xz} the
overall sign in the formula (68) for unpolarized gluon GPDs in the
ERBL region is incorrect and should be reversed.  The same correction
must be made in the corresponding formula for polarized gluon GPDs.}
An important difference between quark and gluon distributions is that
for gluons an extra factor of $(x+\xi)^{1/2} \, |x-\xi|^{1/2}$ appears
on the right-hand sides of the overlap formulae, both in the DGLAP and
the ERBL regions.  To understand this we note that the derivatives in
the gluon field strength operators $G^{+\mu}$ provide a factor of the
plus-momentum for both partons being probed, whereas the spinors from
the quark fields provide only a square root for each momentum.  Gluon
GPDs need however not vanish at $x = \xi$ due to the extra factor
$|x-\xi|^{1/2}$ since the relevant wave functions can diverge at these
points (see Section~\ref{sec:x-equal-xi}).  We also note that the
overlap formula for GPDs of scalar partons has a factor of
$(x+\xi)^{-1/2} \, |x-\xi|^{-1/2}$ on the right-hand side
\cite{Tiburzi:2002sx}.

Corresponding representations exist for targets with different spin.
Comparing (\ref{quark-gpd}) and (\ref{pion-gpd}) one finds that the
analogs of the overlap formulae (\ref{DGLAP-mom}), (\ref{ERBL-mom}),
(\ref{DGLAP-imp}), (\ref{ERBL-imp}) for a spin-zero target are simply
obtained by replacing $F^q_{\lambda'\lambda}(x,\xi,t)$ on the
left-hand sides with $H^q(x,\xi,t)$ and by omitting the helicity
labels $\lambda$, $\lambda'$ in the wave functions.  We finally remark
that a representation of twist-three GPDs (see
Section~\ref{sub:twist}) in terms of wave functions should also be
possible, along the lines of the studies of the forward twist-three
structure functions in~\cite{Mankiewicz:1991az,Mankiewicz:1991ji}.  A
perturbative study of the twist-three quark and gluon GPDs in a quark
target was performed in \cite{Mukherjee:2002xi}.

The overlap formulae in momentum or position space provide a concrete
interpretation of GPDs and related quantities:
\begin{itemize}
\item  Taking the forward limit $p=p'$ and equal hadron
helicities, our matrix elements represent the usual parton densities.
The momentum space wave functions in the DGLAP region then have the
same arguments, i.e., parton densities are given by
\emph{squared} wave functions $|\Psi(x_i,\tvec{k}_{i})|^2$ summed over
all parton configurations with a given flavor and momentum fraction
$x$ of the struck parton.  This justifies the interpretation as
classical probability densities in the parton model, as has long been
known \cite{Brodsky:1989pv}.
\item For $\xi=0$ and nonzero $t$ the momentum space wave functions
have a relative shift in their transverse momentum arguments. After
Fourier transforming to impact parameter space, the wave function
arguments of the initial and final state wave functions are however
identical. This makes the density interpretation discussed in
Section~\ref{sub:impact-forward} explicit in terms of squared impact
parameter wave functions $|\widetilde\Psi(x_i,\tvec{b}_{i})|^2$,
provided one has the same polarization for both hadron states.
\item Integrating over the momentum fraction $x$ and choosing a frame
with $\xi=0$ one obtains the wave function representations of elastic
form factors, going back to Drell and Yan \cite{Drell:1970km}, which
become again densities in impact parameter space~\cite{Soper:1977jc}.
\item For nonzero $\xi$ GPDs are \emph{not} expressed through squared
wave functions, neither in momentum space nor in the mixed
representation using impact parameter.  Instead, they express the
\emph{interference} or \emph{correlation} between wave functions for
different parton configurations in a hadron.  In the ERBL region, even
the parton content of the two wave functions is different.
\end{itemize}

The distributions $E$ and $\tilde{E}$ for quarks or gluons appear in
matrix elements between hadrons with opposite helicities.  On the
other hand, the overlap representations (\ref{DGLAP-mom}) or
(\ref{DGLAP-imp}) in the DGLAP region imply that the Fock states in
the initial and final hadron have the same quantum numbers, and in
particular the same helicity for each parton.  This means that in at
least one of the wave functions $\Psi_{\mathrm{in}}$ or
$\Psi_{\mathrm{out}}$ the helicity of the hadron is \emph{not} the sum
of helicities of its partons.  Conservation of angular momentum along
the $z$ axis (in a frame where the relevant parent hadron has zero
transverse momentum) implies that $E$ and $\tilde{E}$ are overlaps of
wave functions whose orbital angular momentum $L^3$ of the partons
differs by one unit.  This reflects from a different point of view the
finding of Ji's sum rule (Section~\ref{sec:spin}), where $E^q$ and
$E^g$ intervene.  $L^3 \neq 0$ wave functions cannot all be small in
the nucleon, because the Pauli form factors $F_2 = \int dx E$ at small
$t$ are large for both proton and neutron.  The role of $F_2(t)$ as an
indicator for light-cone wave functions with nonvanishing orbital
angular momentum has long been known \cite{Brodsky:1980zm}, for recent
discussions see for instance \cite{Jain:1993jf,Ralston:2002hu}.  A
dedicated study of the overlap representation for $F_2(t)$ and $\int
dx\, x E$ and its connection with orbital angular momentum was done by
Brodsky et al.~\cite{Brodsky:2000ii}.  Explicit overlap formulae for
the three-quark state of a proton target have recently been derived by
Ji et al.~\cite{Ji:2002xn}, and further work on light-cone wave
functions with orbital angular momentum can be found in
\cite{Ji:2003fw,Ji:2003yj}.

It is natural to ask about a wave function representation for GDAs.
They cannot be written as the simple overlap of the wave functions of
the individual hadrons.  Physically this is because the properties of
a multi-hadron state are not simply described by the individual
structure of each hadron because of the strong interaction between the
hadrons.  One can however write down a Fock state decomposition of the
interacting \emph{multi-hadron states} like $|\pi\pi\rangle$ or
$|p\bar{p}\rangle$ and understand the twist-two GDAs we have discussed
as the quark-antiquark wave functions of these states integrated over
the transverse parton momenta.

\subsubsection{Polynomiality and the covariant approach}
\label{sub:poly-wave}

Since the wave function representation we have derived are exact, the
resulting GPDs must satisfy the general properties we have discussed
in Section~\ref{sec:basic}.  For some of them this is easily verified:
in the forward limit one obtains the known wave function
representations of usual parton densities.  It is also straightforward
to show \cite{Diehl:2000xz} that the GPDs obtained by the overlap
formulae are even functions of $\xi$.  What is \emph{not} obvious to
see from the wave function representation is however the continuity of
GPDs at $x=\pm\xi$ (discussed in Section~\ref{sec:x-equal-xi}) and the
polynomiality condition.  In these cases both the DGLAP and the ERBL
regions must cooperate to lead to the required properties, and this
implies nontrivial relations between the wave functions for the
different Fock states relevant in the two regions.  An \textit{ad hoc}
ansatz for the wave functions would almost certainly lead to GPDs that
violate the above requirements.

It is instructive to see how the light-cone overlap representation we
have discussed so far comes about in a covariant framework, and how
polynomiality is achieved there.  Consider for simplicity a pion and
start with the Bethe-Salpeter wave function for the $q\bar{q}$ state.
This is the covariant vertex function between the pion and the two
quarks, including the full quark propagators,
\begin{equation}
  \label{BS}
\Psi^{\mathrm{BS}}_{\alpha\beta}(k) = \int d^4 x\, e^{i (k x)} \,
  \langle \pi(p) \,|\, T \bar{q}_\alpha(-\half x)\, q_\beta(\half x) 
  \,| 0\rangle
\end{equation}
where $k$ is half the difference between the quark and antiquark
four-momenta.  Choosing a particular frame and integrating over $k^-$
puts the relative light-cone time $x^+$ of the fields to zero, and
contracting with the Dirac matrix $\gamma^+\gamma_5$ selects the
``good'' field components which correspond to opposite helicities of
quark and antiquark.  Writing the quark operators at $x^+ =0$ in terms
of annihilation and creation operators for quarks and using the Fock
state decomposition of the meson one readily identifies the result
with the light-cone wave function $\Psi_{+-}$ given in
(\ref{wf-from-operator}),
\begin{eqnarray}
  \label{LC}
\lefteqn{
\int dk^-  \sum_{\alpha\beta} 
  (\gamma^+\gamma_5)_{\alpha\beta}^{\phantom{B}}
     \Psi^{\mathrm{BS}}_{\alpha\beta}(k) }
\nonumber \\
&=& 2\pi \int dx^- d^2 \tvec{x}\;  e^{i (k x)} \,
  \langle \pi(p) \,|\, \bar{q}(-\half x) \gamma^+\gamma_5\, q(\half x)\,
        | 0\rangle \Big|_{x^+ = 0}
\nonumber \\[0.2em]
&=& {}- 4\pi \sqrt{3}\; \Psi_{+-} .
	\phantom{\frac{1}{1}}
\end{eqnarray}
We have omitted the time ordering of the quark operators in the second
line of (\ref{LC}) following our discussion in Section
\ref{sec:light-cone}.  Further integration of $\Psi_{+-}$ over the
transverse parton momentum also puts $\tvec{x}$ to zero and gives the
meson distribution amplitude defined in (\ref{pseudo-da}).

\begin{figure}
\begin{center}
       \leavevmode
       \epsfxsize=0.88\textwidth
       \epsfbox{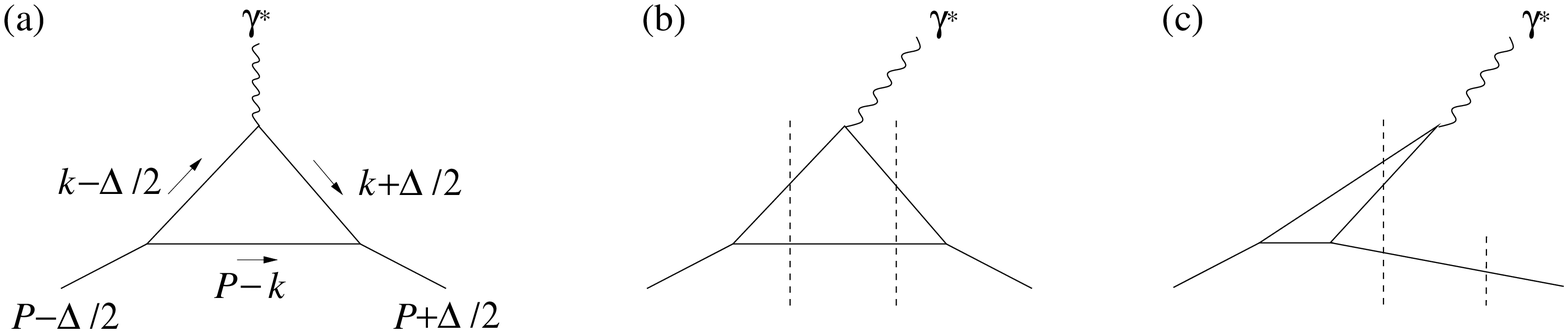}
\end{center}
\caption{\label{fig:triangle}  Triangle diagram in covariant
perturbation theory (a), and its light-cone representation in the
regions $\xi < x < 1$ (b) and $0< x<\xi$ (c).  Vertical lines show the
intermediate states in the wave function representation.}
\end{figure}

Let us now restrict ourselves to the overlap representation of the
elastic meson form factor, where polynomiality just means independence
of $\xi$.  Sawicki \cite{Sawicki:1992qj} has shown how this
independence comes about in a toy model, where scalar particles couple
to themselves by a $\phi^3$ interaction, and to a photon by the usual
derivative coupling of scalar QED.  The scalars play the role both of
quarks and of the meson.  In usual covariant perturbation theory at
leading order, the form factor is then given by the triangle diagram
of Fig.~\ref{fig:triangle}a.  Restricting oneself to the
plus-component of the current, which is enough to extract the form
factor, the diagram is proportional to
\begin{equation}
\int d^4 k\, k^+\, 
         D(P - k)\, D(k - \half \Delta)\, D(k + \half \Delta)
\end{equation}
with scalar quark propagators $D(l) = (l^2 - m^2 +i\epsilon)^{-1}$.
  
After choosing a particular frame, one can perform the integral over
$k^-$ by complex contour integration using Cauchy's theorem, with
poles originating from the quark propagators.  Defining $x = k^+/P^+$
and $\xi = - \Delta^+ /(2 P^+)$ as usual and taking a frame where $\xi
> 0$, one has to distinguish the following cases:
\begin{enumerate}
\item If $x<0$ or $x>1$ then the poles of all three propagators are on
the same side of the real $k^-$ axis.  Closing the integration contour
in the other half-plane one obtains a zero integral.
\item For $\xi<x<1$, corresponding to the kinematics of
Fig.~\ref{fig:triangle}b, one can close the contour so as to pick up
only the pole of $D(P-k)$.  The result is proportional to
\begin{equation}
  \label{triangle}
\int d k^+ d^2 \tvec{k}\; k^+ \, (P^+ - k^+)^{-1} \,
  D(k - \half \Delta) D(k + \half \Delta) 
        \Big|_{k^-_{\phantom{0}} = k^-_{\mathrm{os}}} \; ,
\end{equation}
where $k^-_{\mathrm{os}}$ is the value where $D(P-k) = 0$.  The
Bethe-Salpeter wave function of the incoming meson in this theory is
to lowest order in the coupling just given by a factor times the
product $D(k - \half \Delta)\, D(P - k)$ of quark propagators.  To
obtain the corresponding light-cone wave function one integrates over
$k^-$.  Using again complex contour integration one finds a nonzero
wave function only for $x \in [0,1]$, which can be written as a factor
times $D(k - \half\Delta)$ evaluated at $k^-_{\mathrm{os}}$.  In
analogy one can express the propagator $D(k + \half \Delta)$ at
$k^-_{\mathrm{os}}$ through the light-cone wave function of the final
state meson.  All together (\ref{triangle}) thus takes the form of an
overlap integral of the wave functions for the two mesons, as in the
overlap representation (\ref{DGLAP-mom}) for the DGLAP region.
\item For  $0<x<\xi$ one can either pick up the pole of $D(k -
\half\Delta)$ or the poles of the two other propagators.  The result
cannot be written in terms of the wave functions for the mesons.  This
is physically clear since the right-hand vertex in
Fig.~\ref{fig:triangle}c corresponds to a quark splitting into a meson
and a quark, which is not the kinematics of a wave function.  Brodsky
and Hwang \cite{Brodsky:1998hn} have realized that the result can
however be written as the light-cone wave function for three particles
in the meson (see the dotted line in the diagram) and the one-particle
wave function to find a meson in a meson, which is just proportional
to a $\delta$ function in momentum.  This is precisely the overlap of
wave functions with different particle content we have seen in the
ERBL region.
\end{enumerate}
Only when summing the contributions from the DGLAP and the ERBL
regions does one obtain a result for the meson form factor that is
independent of $\xi$, i.e.\ of the frame in which one has performed
the reduction from Bethe-Salpeter to light-cone wave functions.  The
wave functions for two and three partons in the meson are related in a
way so as to preserve Lorentz invariance: in fact they have been
identified from the same covariant Feynman diagrams in different
kinematical regimes.

It is not known which form corresponding relations between light-cone
wave functions take in general and in more realistic theories.  Such
relations can be provided by the equations of motion, since the Dirac
operator $i \Slash{D} - m$ contains terms which either conserve parton
number, or add or remove one gluon.

\subsubsection{Evolution}
\label{sub:wf-evol}

Because of the ultraviolet divergences of QCD, the quantities we have
manipulated in our derivation need to be renormalized.  This concerns
the fields $q_+$ and hence the parton operators and light-cone wave
functions, as well as the bilocal operators defining GPDs.  The
renormalization must ensure that the integrals one has to perform in
the overlap representations are finite, and the corresponding
renormalization scale $\mu$ provides the scale at which the wave
functions and parton distributions are defined.  It is intuitive to
think of $\mu$ as a cutoff in transverse momenta, so that the
available $\tvec{k}$ in the wave functions and their overlap are at
most of order $\mu$.  There are similar cutoff schemes, for instance
in the light-cone energy of the Fock states
\cite{Brodsky:1989pv}, which also cuts off small plus-momenta $k^+$.
For certain purposes, cutoff regularization will of course be
insufficient since it breaks Lorentz and gauge invariance of the
theory.

A study of the renormalization scale dependence for usual parton
densities in the context of the Fock state representation has been
performed in \cite{Burkardt:2002uc}, where evolution equations were
given for the contributions of individual Fock states to the parton
densities.  The evolution of an $N$ parton Fock state was seen to
depend also on the $(N-1)$ parton Fock states (which can radiate a
parton to become an $N$ parton state).  Summing over all Fock state
contributions the usual DGLAP equations were obtained.  In
\cite{Mukherjee:2002pq} the nonforward case was studied, and it
was shown how the splitting functions of the evolution are related
with the $k_T$ divergences of the perturbatively calculated wave
functions of a quark and gluon within a quark.

\subsubsection{A remark about the light-cone gauge}
\label{sub:light-cone-problems}

Brodsky et al.~\cite{Brodsky:2002ue} have claimed that the
representation of forward parton distributions as probabilities, which
follows from the wave function representation in
Section~\ref{sub:overlap-formulae} is not correct.  At the center of
the arguments in \cite{Brodsky:2002ue} are the effects of gluons
polarized in the plus-direction, which are described by a Wilson line
in the definition of parton distributions for a general gauge
(Section~\ref{sec:definitions}) but are absent in the light-cone gauge
$A^+=0$, where the parton picture and the overlap representation of
this section is derived.  Subsequent investigations of the theoretical
issues raised in \cite{Brodsky:2002ue} have been made by Collins
\cite{Collins:2002kn,Collins:2003fm} and by Ji and collaborators
\cite{Ji:2002aa,Belitsky:2002sm}.  We will not attempt a full
discussion of these problems, which are not finally settled in our
view, but nevertheless make some points relevant in the context of
this review.

\begin{figure}
\begin{center}
       \leavevmode
       \epsfxsize=0.4\textwidth
       \epsfbox{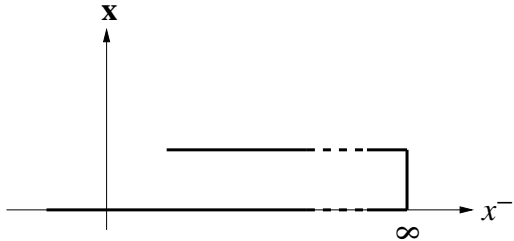}
\end{center}
\caption{\label{fig:gauge-path}  One of the possible paths for the
Wilson line in $k_T$ dependent parton distributions (see text).  The
part of the path going in the transverse direction is at $x^- \to
\infty$.}
\end{figure}

\begin{itemize}
\item The arguments in \cite{Brodsky:2002ue} do not affect the
leading-twist factorization theorems presented in
Section~\ref{sec:factor}, which connect GPDs to physical processes.
(In fact these theorems were proven in covariant gauges, where the
Wilson lines summing up the effects of $A^+$ gluons explicitly
appear.)
\item The general properties of GPDs discussed in
Sections~\ref{sec:definitions}, \ref{sec:basic} and
\ref{sec:spin} to \ref{sec:impact} are not affected either, since
they can be derived in covariant gauges, taking explicitly into
account the Wilson lines.  The discussion of helicity structure in
Section~\ref{sec:helicity} should not be affected either since $A^+$
gluons carry zero light-cone helicity.
\item Both the hard-scattering kernels and the matrix elements
defining GPDs are gauge invariant quantities, and it should in
particular be possible to evaluate them in $A^+=0$ gauge.  This was
indeed found in an explicit calculation of the quark density in scalar
QED \cite{Belitsky:2002sm} from the same diagrams that were considered
in \cite{Brodsky:2002ue}.  The regularization and renormalization
prescriptions needed to define parton distributions must of course be
gauge invariant to obtain this result, and simple cutoff schemes are
in general not suitable for this purpose.
\item The condition $A^+=0$ is not sufficient to fully fix the
gauge since it is preserved by gauge transformations which do not
depend on the coordinate $x^-$.  To specify the gauge completely one
may impose boundary conditions on the gauge field at $x^- \to \pm
\infty$, see \cite{Belitsky:2002sm} and references therein.  In
momentum space this is related with providing a prescription for the
$1 /k^+$ singularity of the gluon propagator
\begin{equation}
  \label{lc-gauge-prop}
\frac{1}{k^2 + i\epsilon} \left( g^{\alpha\beta} - 
	\frac{k^\alpha n_-^\beta + n_-^\alpha k^\beta}{k^+} \right) .
\end{equation}
Light-cone wave functions as defined in Section \ref{fock-space} are
not gauge invariant quantities and depend in particular on how the
residual gauge freedom is fixed.  In some of these prescriptions the
wave functions of stable particles have dynamical phases: regulators
like $1/ (k^+ + i\epsilon)$ for instance introduce imaginary parts in
addition to those from the Feynman prescription in the propagator $1
/(k^2 + i\epsilon)$.
\item Gauge issues are more prominent in distributions depending on
the parton $k_T$.  As remarked at the end of
Section~\ref{sub:impact-forward} the field operators in their
definition appear at finite transverse separation $\tvec{x}$ and hence
cannot be connected by a Wilson line along a purely light-like path.
A path often considered in the literature is shown in
Figure~\ref{fig:gauge-path}, and one may naively expect that in
light-cone gauge the Wilson line along this path reduces to unity.
This is however not true in general since in $A^+=0$ gauge the gauge
fields at $x^-\to \infty$ do not vanish, except for particular choices
of the residual gauge fixing.  The Wilson line along the transverse
direction at $x^-\to \infty$ then does appear in $k_T$ dependent
parton distributions, but reduces to unity in the $k_T$ integrated
distributions, where $\tvec{x} =\tvec{0}$ \cite{Belitsky:2002sm}.  In
$k_T$ dependent distributions the gluons described by the Wilson line
have important physical consequences for various spin and azimuthal
asymmetries in inclusive or semi-inclusive processes
\cite{Brodsky:2002cx,Brodsky:2002rv,Collins:2002kn,Ji:2002aa}.
\item The $1/k^+$ singularity of the light-cone gauge gluon propagator
(\ref{lc-gauge-prop}) also shows up as a singular term in the wave
function renormalization of the quark field, which directly affects
the definition of light-cone wave functions \cite{Brodsky:1989pv}.  A
cutoff in $k^+$ is sufficient to regulate the singularity in simple
situations as for instance in \cite{Burkardt:2002uc} but may not be in
general.\footnote{I thank J.~Collins for discussions about this
point.}
\end{itemize}
Little is known about how to appropriately define light-cone wave
functions in $A^+=0$ gauge taking into account the issues of
renormalization, of gauge fixing and of a proper regularization of
gauge artifacts.  It is not known how such a scheme should look like
in order to have physically appealing features such as the overlap
representation we have presented.  In this sense the overlap formulae
and their consequences presently do not have the status of all-order
results, and it remains to be seen which of these results will require
modification.  Ultimately these issues lead to the question of how to
properly implement the physical picture of the parton model in QCD.

\subsection{Positivity bounds}
\label{sec:positivity}

In the DGLAP regions GPDs are very similar to the usual parton
distributions, and one may expect quantitative relations between these
quantities.  Such relations indeed exist and provide upper bounds on
GPDs in terms of conventional parton densities.  That the wave
function representation of GPDs leads to such bounds was observed by
Martin and Ryskin \cite{Martin:1998wy}, who derived an upper bound on
GPDs in terms of the arithmetic mean of the usual parton densities at
two different momentum fractions.  Pire, Soffer and Teryaev
\cite{Pire:1998nw} later showed that stronger bounds exist which involve
the geometric mean instead.  A missing kinematical factor in these
bounds was corrected by Radyushkin \cite{Radyushkin:1998es}, and
\cite{Diehl:2000xz} corrected the incorrect treatment of the proton spin
structure in previous work, which had missed out the contribution from
the distributions $E$.  Bounds on GPDs in impact parameter space were
derived in \cite{Burkardt:2001ni,Diehl:2002he}.  An extensive
investigation of several aspects of positivity bounds was given by
Pobylitsa
\cite{Pobylitsa:2001nt,Pobylitsa:2002gw,Pobylitsa:2002iu,Pobylitsa:2002vi,Pobylitsa:2002vw,Pobylitsa:2002ru}.
All bounds discussed in the literature are restricted to the DGLAP
regions $x\in [\xi,1]$ and $x\in [-1,-\xi]$, and it is not known
whether anything similar holds in the ERBL region.

The positivity constraints we will discuss in the next subsection
provide upper bounds on GPDs in terms of the forward parton densities,
which are rather well known.  In contrast, the upper bounds on impact
parameter GPDs in Section~\ref{sub:imp-bounds} are given in terms of
parton densities in impact parameter space.  These are not known
phenomenologically, but the corresponding bounds provide constraints
on models or ans\"atze for the impact parameter dependence of GPDs
(and thus on their $t$-dependence after Fourier transform).  In fact,
not all current modeling strategies automatically respect the
positivity constraints, see Section~\ref{sub:t-depend}.

\subsubsection{Bounds in momentum representation}
\label{sub:mom-bounds}

A convenient method to derive the positivity constraints is to use the
wave function representation of GPDs given in
Section~\ref{sub:overlap-formulae}.  One readily sees that the overlap
formula (\ref{DGLAP-mom}) has the form of a scalar product in the
space of wave functions.  Taking a spin-zero target for simplicity,
one can then write $H_\pi^q(x,\xi,t)$ for $x\in [\xi,1]$ as a scalar
product $(\phi_{-\xi,\tvec{p'}} | \phi_{\xi,\tvec{p}})$ with
\begin{eqnarray}
\phi_{-\xi,\tvec{p'}}(x_i,\tvec{k}_i,N,\beta) &=& 
  \sqrt{1-\xi}^{\, 2-N}\,
  \Psi_{N\beta}(x_i^{\rm out}, \tvec{k}_{i}^{\rm out}) ,
\nonumber \\
\phi_{\xi,\tvec{p}}(x_i,\tvec{k}_i,N,\beta) &=&
  \sqrt{1+\xi}^{\, 2-N}\,
  \Psi_{N\beta}(x_i^{\rm in}, \tvec{k}_{i}^{\rm in}) .
\end{eqnarray}
The Schwartz inequality $| (\phi_{-\xi,\tvec{p'}} |
\phi_{\xi,\tvec{p}}) |^2 \le (\phi_{-\xi,\tvec{p'}} |
\phi_{-\xi,\tvec{p'}}) (\phi_{\xi,\tvec{p}} | \phi_{\xi,\tvec{p}})$
translates into
\begin{equation}
  \label{pion-quark-pos}
\Big| H^q_\pi(x,\xi,t) \Big|
	\le \sqrt{ q_\pi(x_{\rm in}) \, q_\pi(x_{\rm out}) }  
\qquad \qquad \mbox{for $x \in [\xi,1]$} ,
\end{equation}
where
\begin{equation}
x_{\rm in} = \frac{x+\xi}{1+\xi} , \hspace{3.5em}
x_{\rm out} =\; \frac{x-\xi}{1-\xi}
\end{equation}
are the momentum fractions of the struck parton before and after the
scattering, referring respectively to the initial and final hadron
momentum.  An analogous result holds for $x\in [-1,-\xi]$ and involves
antiquark instead of quark distributions.  For gluons one has
\begin{equation}
  \label{pion-glue-pos}
\Big| H^g_\pi(x,\xi,t) \Big| 
 \le    \sqrt{1-\xi^2}\, 
        \sqrt{x_{\rm in}\, g_\pi(x_{\rm in})} \,
	\sqrt{x_{\rm out}\, g_\pi(x_{\rm out})} 
\qquad  \mbox{for $x \in [\xi,1]$}
\end{equation}
instead.  

To extend these results to cases with nontrivial spin structure it is
useful to rewrite (\ref{pion-quark-pos}) as a positivity condition for
a matrix \cite{Pobylitsa:2002gw},
\begin{equation}
  \label{matrix condition}
\left( \begin{array}{cc}
q_\pi(x_{\rm in}) & H^q_\pi(x,\xi,t) \\
H^q_\pi(x,\xi,t)  & q_\pi(x_{\rm out})
\end{array} \right) 
\ge 0 ,
\end{equation}
again for $x \in [\xi,1]$, which we will always imply in the
following.  Analogs hold of course for the antiquark region and for
gluons.  For a spin $\half$ target the overlap representation of
$F^q_{\lambda'\lambda}$ in (\ref{DGLAP-mom}) is a scalar product
$(\phi^{\lambda'}_{-\xi,\tvec{p'}} | \phi_{\xi,\tvec{p}}^\lambda)$,
whose vectors are labeled by the proton helicity in addition to the
kinematical variables.  Positivity of the scalar product implies a
positive semidefinite $4\times 4$ matrix
\begin{equation}
  \label{pos-cond-proton}
\renewcommand{\arraystretch}{1.5}
\left( \begin{array}{cc}
F^q(x_{\rm in},0,0)         & F^q(x,\xi,t) \\ {}
\Big[F^q(x,\xi,t)\Big]^\dag & F^q(x_{\rm out},0,0)
\end{array} \right)
\ge 0 
\end{equation}
where $F^q(x,\xi,t)$ is the matrix
\begin{equation}
F^q =
  \label{proton-matrix}
\renewcommand{\arraystretch}{2.5}
\left( \begin{array}{cc}
  \displaystyle \sqrt{1-\xi^2} 
	\Big( H^q - \frac{\xi^2}{1-\xi^2}\, E^q \Big) &
  - e^{-i\varphi}\, \displaystyle \frac{\sqrt{t_0 - t}}{2m}\; E ^q\\
  e^{i\varphi}\, \displaystyle \frac{\sqrt{t_0 - t}}{2m}\; E^q &
  \displaystyle  \sqrt{1-\xi^2} 
	\Big( H^q - \frac{\xi^2}{1-\xi^2}\, E^q \Big)
\end{array} \right)
\end{equation}
obtained from (\ref{hel}).  The positivity condition
(\ref{pos-cond-proton}) can be written as
\cite{Pobylitsa:2001nt}
\begin{eqnarray}
  \label{proton-quark-pos}
(1-\xi^2) \Big( H^q - \frac{\xi^2}{1-\xi^2}\, E^q \Big)^2
+ \frac{t_0 - t}{4m^2}\; \Big( E^q \Big)^2
\le q(x_{\rm in}) \, q(x_{\rm out}) .
\end{eqnarray}
For better legibility we have omitted the arguments $x$, $\xi$, $t$ of
the GPDs in (\ref{proton-matrix}) and (\ref{proton-quark-pos}).

More involved bounds are obtained when also taking into account the
spin structure on the quark side.  The matrix $F^q$ in
(\ref{pos-cond-proton}) is then replaced with a $4\times 4$ matrix
formed by the helicity combinations $A^q_{\lambda'\mu', \lambda\mu}$
given in Section~\ref{sec:helicity-transitions}, where the rows of the
matrix are associated with the four helicity combinations
$(\lambda',\mu')$ and the columns with the four helicity combinations
$(\lambda,\mu)$.  The analog of (\ref{pos-cond-proton}) then states
the positivity of an $8\times 8$ matrix.  Using the parity constraints
(\ref{parity}) on $A^q_{\lambda'\mu', \lambda\mu}$ this can be
rewritten as the positivity of two $4\times 4$ matrices, which differ
by the relative signs between quark helicity conserving and quark
helicity changing GPDs \cite{Pobylitsa:2002gw}.  These positivity
conditions, which contain in particular the well-known bound $| \Delta
q(x) | \le q(x)$ and the Soffer bound \cite{Soffer:1995ww} on the
forward transversity distribution, can be translated into inequalities
involving GPDs by a general method given in \cite{Pobylitsa:2002gw}.
In the same reference it is shown how to obtain an upper bound on any
given linear combination of GPDs in terms of forward densities.
Examples are
\begin{eqnarray}
  \label{bound-non-flip}
\Bigg| \sqrt{1-\xi^2} 
	\Big( H^q - \frac{\xi^2}{1-\xi^2}\, E^q \Big) \Bigg|
&\le& \frac{1}{2}
  \Bigg( \sqrt{(q + \Delta q)_{\rm in} \, (q + \Delta q)_{\rm out}}
\nonumber \\
&& \hspace{0.6em} 
+ \sqrt{(q - \Delta q)_{\rm in} \, (q - \Delta q)_{\rm out}}
  \Bigg) ,
\\
  \label{bound-flip}
\Bigg| \frac{\sqrt{t_0 - t}}{2m}\; E^q \Bigg|
&\le& \frac{1}{2}
  \Bigg( \sqrt{(q + \Delta q)_{\rm in} \, (q - \Delta q)_{\rm out}}
\nonumber \\
&& \hspace{0.6em}
+ \sqrt{(q - \Delta q)_{\rm in} \, (q + \Delta q)_{\rm out}}
  \Bigg) ,
\end{eqnarray}
where we have used a shorthand notation $(q + \Delta q)_{\rm in} =
q(x_{\rm in}) + \Delta q(x_{\rm in})$ etc.  Stricter bounds involve in
addition the forward transversity distributions on the right-hand
sides.  Looser bounds are obtained by setting the polarized parton
densities to zero, compare also with (\ref{proton-quark-pos}).  We
remark that the bound involving only the distribution $H^q$ comes with
a factor on the left-hand side that vanishes like $\sqrt{t_0-t}$ in
the zero-angle limit, as happens in (\ref{bound-flip}).  Bounds for
the polarized quark GPDs are obtained by replacing $H\to
\tilde{H}$ and $E\to \tilde{E}$ in (\ref{bound-non-flip}) and $E \to
\xi \tilde{E}$ in (\ref{bound-flip}), with the same expressions on the
right-hand sides.

\subsubsection{Bounds in impact parameter representation}
\label{sub:imp-bounds}

Bounds on GPDs in the impact parameter representation are readily
derived from their overlap representation in (\ref{DGLAP-imp}).
Taking again a pion target for simplicity, the impact parameter GPD
has the form of a scalar product $(\tilde\phi_{-\xi} |
\tilde\phi_{\xi})$ with
\begin{eqnarray}
\widetilde\phi_{-\xi}(x_i,\tvec{b}_i,N,\beta) &=& 
  \sqrt{1-\xi}^{\, 2-N}\,
  \widetilde\Psi_{N\beta}(x_i^{\rm out},
	\tvec{b}_i - \tvec{b}_{0}^{\rm out}) ,
\nonumber \\
\widetilde\phi_{\xi}(x_i,\tvec{b}_i,N,\beta) &=& 
  \sqrt{1+\xi}^{\, 2-N}\,
  \widetilde\Psi_{N\beta}(x_i^{\rm in}, 
	\tvec{b}_i - \tvec{b}_{0}^{\rm in}) .
\end{eqnarray}
This leads to
\begin{equation}
  \label{pion-bound-impact}
(1-\xi^2)\, \Big|\,  \mathcal{F}\! H^q_\pi(x,\xi,\tvec{b}) \, \Big|
\le \sqrt{\mathcal{F}\! H^q_\pi
	\Big(x_{\rm in}, 0, \frac{\tvec{b}}{1+\xi} \Big)\,
          \mathcal{F}\! H^q_\pi
	\Big(x_{\rm out}, 0, \frac{\tvec{b}}{1-\xi} \Big)} ,
\end{equation}  
where $\mathcal{F}$ denotes the Fourier transform
\begin{equation}
\mathcal{F}\! f(x,\xi,\tvec{b}) = \int \frac{d^2\tvec{D}}{(2\pi)^2}\,
	 e^{-i \tvec{D} \tvec{b}} f(x,\xi,t)
\end{equation}
passing from momentum to impact parameter space, where it is
understood that $t$ depends on $\tvec{D}^2$ as given in
(\ref{t-by-D}).  Analogous expressions are obtained for proton GPDs by
replacing $H_\pi^q$ under the Fourier transform with the matrix
(\ref{proton-matrix}) or its generalization including the quark spin
dependent GPDs.  The impact parameter arguments on the right-hand side
of (\ref{pion-bound-impact}) are readily identified as the distance
between the probed quark and the center of the incoming or outgoing
proton in Figure~\ref{fig:impact}a.

A more general class of positivity conditions has been derived by
Pobylitsa \cite{Pobylitsa:2002iu,Pobylitsa:2002vi}.  For a spinless
target (and after translating Pobylitsa's convention for the impact
parameter into ours) they can be written as the requirement that
\begin{equation}
  \label{Poby-general}
(1-\xi^2)^{3/2-s}\, \mathcal{F}\! H_\pi (x,\xi,\tvec{b})
  = \sum_n Q_n^*\Big( x_{\rm out}, \frac{\tvec{b}}{1-x} \Big)\,
           Q_n\Big( x_{\rm in}, \frac{\tvec{b}}{1-x} \Big)
\end{equation}
with arbitrary functions $Q_n$, where the sum over $n$ may be discrete
or involve integration over continuous variables.  Here $s$ denotes
the spin of the partons, $s=\half$ for quarks, $s=1$ for gluons, and
$s=0$ for spinless partons in toy models.  The overlap representation
(\ref{DGLAP-imp}) can indeed be cast into the form
(\ref{Poby-general}).  To see this one changes variables to $x'{\!}_i
= x_i /(1-x)$ and $\tvec{b}'{\!}_i = \tvec{b}_i + \tvec{b}\, x /(1-x)$
for the spectators $i\neq j$.  The delta function constraints in the
overlap formula then become $\delta(1 - \sum_{i\neq j} x'_i) \,
\delta^{(2)}(\sum_{i\neq j} x'_i \tvec{b}'_i)$ and  hence
independent of $x_{\rm in}$, $x_{\rm out}$ and $\tvec{b}$.  The
functions $Q_n$ are just the wave functions $\widetilde\Psi$ times a
scale factor $(1-x_{\rm out})^{N-4}$ or $(1-x_{\rm in})^{N-4}$, and
the label $n$ includes the dependence on the spectator variables
$x'_i$ and $\tvec{b}'_i$.

Note that the distance $\tvec{b} /(1-x)$ between the probed quark and
the center of momentum of the \emph{spectator} system is the same
before and after the scattering, as was observed in
\cite{Burkardt:2002hr}.  If one rescales $\tvec{b}$ by $(1-x)$ then
the right-hand side of (\ref{Poby-general}) depends on $x$ and $\xi$
only through $x_{\rm out}$ and $x_{\rm in}$.  As a consequence one has
\begin{eqnarray}
\int_0^1 d x_{\rm in}\, \int_0^1 d x_{\rm out} \:
  p^*(x_{\rm out},\tvec{b})\, p(x_{\rm in},\tvec{b}) \,
  (1-\xi^2)^{3/2-s}\, 
  \mathcal{F}\! H_\pi \Big(x,\xi, (1-x)\tvec{b} \Big) \ge 0
\end{eqnarray}
for any function $p(z,\tvec{b})$ and for any fixed $\tvec{b}$.  By a
change of variables and a rescaling of the function $p$ this is
equivalent to the condition
\begin{equation}
  \label{Poby-final}
\int_{-1}^1 d\xi \, \int_{|\xi|}^1 dx \, (1-x)^{-2s}\,
	p^*(x_{\rm out},\tvec{b})\, p(x_{\rm in},\tvec{b}) \,
	\mathcal{F}\! H_\pi  \Big(x,\xi, (1-x)\tvec{b} \Big)
\ge 0 .
\end{equation}
Equivalent forms hold of course in the antiquark region $x < -|\xi|$.
The constraint (\ref{Poby-final}) is the most general one known so
far.  As shown in \cite{Pobylitsa:2002iu}, the positivity conditions
(\ref{pion-quark-pos}) and (\ref{pion-bound-impact}) are obtained from
(\ref{Poby-final}) by special choices of $p(z,\tvec{b})$, followed by
integration over $\tvec{b}$ in the case of (\ref{pion-quark-pos}).

For targets with spin the derivation goes in analogy, including the
matrix structure in the helicities of the target and the partons as
explained in the previous subsection.  The final result
(\ref{Poby-final}) then involves multiplication of the matrix
$\mathcal{F}\! A_{\lambda'\mu', \lambda\mu}$ with vector valued
functions $p^*_{\lambda'\mu'}$ and $p_{\lambda\mu}^{\phantom{*}}$.

\subsubsection{Derivation and validity}
\label{sub:pos-derive}

The constraints (\ref{Poby-final}) and their generalization to targets
with spin are stable under leading-logarithmic evolution to higher
evolution scales $\mu$ as shown in \cite{Pobylitsa:2002iu}.  The
essence of the argument is the interpretation of the LO evolution
kernels in the DGLAP region as the generalized parton distributions of
a parton target.  As is the case for the positivity constraints on
forward parton densities \cite{Altarelli:1998gn}, the same is however
not generally true for NLO evolution nor for backward evolution to
smaller scales, depending on the scheme used to renormalize the
distributions.  A model calculation \cite{Pobylitsa:2002vw} of
perturbative triangle diagrams as in Fig.~\ref{fig:triangle}, with a
Yukawa coupling between a scalar target and spin $\half$ partons, gave
for instance an additive term proportional to $\theta(x) \log \mu^2$
in the quark distribution $q(x)$, which will violate its positivity at
small renormalization scales.

The derivation of the positivity bounds we have sketched here relies
on the overlap representation and is hence subject to the caveats
concerning the proper definition and renormalization of light-cone
wave functions mentioned in Section~\ref{sub:light-cone-problems}.
This derivation is in essence equivalent to inserting a complete set
of intermediate states between the operators $\bar{q}(- \half z)$ and
$q(\half z)$ in the definition of quark densities \cite{Pire:1998nw},
or to using positivity of the norm in the Hilbert space of mixed
quark-hadron states $q(\half z) |p\rangle$, as done in Pobylitsa's
work.  All these derivation require the $A^+ = 0$ gauge in order to
have no Wilson line between the quark fields, and they all involve the
manipulation of ``states'' with net color (see our brief remark in
Section~\ref{sec:light-cone} in the context of light-cone
quantization).

An alternative derivation of the positivity bounds given in
\cite{Pobylitsa:2002ru} sheds further light onto the question of their
validity, and we will sketch the essence of the argument.  Starting
point are the positivity properties of the matrix element
\begin{eqnarray}
  \label{hadronic-tensor}
4\pi \epsilon_\alpha^{\phantom{*}}  W^{\alpha\beta} \epsilon^*_\beta
&=&  \epsilon_\alpha^{\phantom{*}} \left[
     \int d^4x\, e^{i (q+q') x/2} \langle p'| J^\beta(\half x) 
     J^\alpha(-\half x) | p\rangle \, \right] \epsilon^*_\beta
\\[0.3em]
&=&  \sum_X (2\pi)^4 \delta^{(4)}(p+q-p_X)\, 
	 \langle p'| \, \epsilon^* J(0) |X\rangle
	 \langle X | \, \epsilon\, J(0) |p \rangle
\nonumber 
\end{eqnarray}
with the local current $J^\alpha = \bar{q} \gamma^\alpha q$, where the
sum is over all \emph{physical} states with momentum $p_X$.  At this
point we anticipate the results of Sections~\ref{sec:factor} and
\ref{sec:compton-scatt}.  In the region where $q^2$ and $q'^2$ are
both spacelike $2\pi W^{\alpha\beta}$ is simply the absorptive part of
the hadronic tensor $T^{\alpha\beta}$ for doubly virtual Compton
scattering (up to the quark charges and the sum over flavors).  Notice
that in order to obtain this simple relation one must avoid the region
where one or both photons are timelike, since then the Compton
amplitude has additional cuts in the photon virtualities.  In the
generalized Bjorken limit (large $q^2$ and $q'^2$ and small $t$) the
absorptive part of the Compton amplitude at leading order in
$\alpha_s$ is simply proportional to $H^{q(+)}(\rho,\xi,t)$.  Having
both photons spacelike selects the DGLAP region $|\rho| \ge |\xi|$, as
seen in the definitions (\ref{xi-eta-def}).  To proceed one transforms
(\ref{hadronic-tensor}) to impact parameter space and obtains
$\mathcal{F}\!  H^{q(+)}(\rho,\xi,\tvec{b})$ represented as in
(\ref{Poby-general}).  The functions $Q_n$ are now matrix elements of
$J\cdot \epsilon$ between states of the target and the system $X$ at
definite plus-momentum and definite impact parameter, and $n$ labels
the different intermediate states $X$.  To separate quarks from
antiquarks, as well as the different parton helicity combinations, one
can use appropriate currents $J$ and polarization vectors $\epsilon$.
The current $J^\alpha = \bar{q} \gamma^\alpha (1-\gamma_5) q$ for
instance only couples to quarks with negative and antiquarks with
positive helicity.  To get the absorptive part of the amplitude one
has to cut the handbag diagrams in Fig.~\ref{fig:DDVCS}, so that the
parton is on shell before and after scattering on the current.  In the
Breit frame of this subprocess one sees that one can select the
positive or negative helicity parton by taking a polarization vector
$\epsilon$ corresponding to positive or negative helicity.

What this derivation requires is that $q^2$ and $q'^2$ be in a region
where the Compton amplitude is well approximated by the leading-twist
expression at leading order in $\alpha_s$.  This requires in
particular $\mu^2$ to be of the order of the hard scale of the
process.  The upshot of the argument is that the positivity bounds
should be valid at sufficiently large factorization scales $\mu^2$.
We note in passing that positivity constraints on the absorptive part
of the nonforward Compton amplitude have long ago been studied by De
R{\'u}jula and De Rafael~\cite{DeRujula:1973aa}.


\subsection{The point $x=\xi$}
\label{sec:x-equal-xi}

The point $x=\xi$ (or $x= -\xi$) where the DGLAP and ERBL regimes meet
is of particular interest and importance.  One one hand it is rather
directly accessible to experiment: at leading order in $\alpha_s$ the
imaginary parts of the amplitudes for DVCS and for meson
electroproduction just involve GPDs at $x = \pm\xi$, as we see from
the form (\ref{Born-convolution}) of the hard scattering kernels.
Physically the point $x=\xi$ corresponds to very peculiar parton
configurations, involving one parton with vanishing plus-momentum in
the initial or final state hadron (according to whether one approaches
the point from the ERBL or DGLAP region).  Intuitively one might fear
that such configurations are dangerous for the factorization of the
hard-scattering process, but one can show that this is not the case
for the processes we will consider (see Section~\ref{sec:factor}).

GPDs around $x= \xi$ are sensitive to the physics of partons with
small plus-momentum, but in a different way than the usual parton
distributions at small $x$.  In the wave function representation we
see that, as $x$ approaches $\xi$ from above, the kinematical mismatch
between the initial and final state wave functions becomes extreme,
with a parton momentum fraction going to zero in the final state but
staying finite (equal to $2\xi /(1+\xi)$ with respect to the parent
hadron) in the initial state.  As $x$ approaches $\xi$ from below, one
probes a parton pair with one momentum fraction finite and the other
going to zero, a configuration similar to the one of a meson
distribution amplitude at its endpoints $z=0$ or $z=1$.  We note that
the inequalities from positivity discussed in
Section~\ref{sub:mom-bounds} do not provide constraints in the limit
$x\to \xi$ since the parton distributions on their right-hand sides
diverge for $x_{\mathrm{out}} \to 0$ (unless the factorization scale
is very low).

The real parts of the amplitudes for DVCS and for meson
electroproduction involve principal value integrals as in
(\ref{Born-convolution}).  Consistency of factorization thus requires
leading-twist GPDs to be \emph{continuous} at $x = \pm\xi$, otherwise
the scattering amplitude would be logarithmically divergent.  An
analog to this is that leading-twist distribution amplitudes must
vanish at the endpoints $z=0$ and $z=1$ in order to give finite
convolutions with hard scattering kernels (see e.g.\
Section~\ref{sec:meson-lt}).

Using the connection between GPDs and parton-proton scattering
amplitudes discussed in Section~\ref{sec:light-cone}, Collins and
Freund \cite{Collins:1998be} have related the analyticity properties
in $x$ of GPDs with the analytic properties of Feynman graphs.  They
find that GPDs can be written as the sum of a term analytic at $x=\xi$
and a contribution which is nonanalytic but zero at this point.  This
ensures of course continuity at $x=\xi$.  A nonanalytic behavior at
this point is also suggested by the reduction formula (\ref{dd-red})
from double distributions to GPDs.  For fixed $\xi$ the length of the
integration line in the $\beta$-$\alpha$ plane is constant in the ERBL
region and proportional to $1-|x|$ in the DGLAP regions, i.e., the
change in this length has a jump at $x=\xi$.  Even with a smooth
double distribution one will then in general have a jump in some
derivative of the GPD at this point.

A discontinuity of the first derivative of GPDs has been found in
several model studies, for instance by Radyushkin
\cite{Radyushkin:1997ki} in scalar $\phi^3$ theory, where the GPD of
the scalar particle was evaluated perturbatively to first order in the
coupling.  Tiburzi and Miller \cite{Tiburzi:2002sx} and Theu{\ss}l et
al.~\cite{Theussl:2002xp} studied the GPDs of a bound state in
different toy field theories, using perturbation theory in a
Bethe-Salpeter approach (see Section~\ref{sub:covariant-models}).
Both groups obtain GPDs at $x=\xi$ that are continuous but have a jump
in the first derivative.  The same was found by Burkardt
\cite{Burkardt:2000uu} for the GPD of a meson state in the \mbox{'t
Hooft} model (QCD in $1+1$ dimensions and the large-$N_c$ limit
\cite{'tHooft:1974hx}), which is exactly solvable without recourse to
perturbation theory.  A discontinuous derivative is thus at least not
in contradiction with general principles of field theory.  The precise
behavior of the GPD at the transition point may depend on further
details such as the spin of the partons.  We note that the behavior of
gluon GPDs at $x=\xi$ has not much been investigated in the
literature.

In the overlap representation the continuity of GPDs at the boundary
between the DGLAP and the ERBL regimes is nontrivial and requires
relations between the wave functions for different Fock states when
the plus-momentum of one parton goes to zero.  Similarly to those
needed to guarantee polynomiality (see Section~\ref{sub:poly-wave})
such relations can be provided by the equations of motion.  {}From the
constraint equation (\ref{dirac-constraint}) of light-cone
quantization Antonuccio et al.~\cite{Antonuccio:1997tw} have in fact
obtained relations between wave functions for Fock states with $N$,
$N-1$ and $N+1$ partons when a quark or antiquark plus-momentum goes
to zero.  Explicit calculations in toy models do provide continuous
GPDs, which has been checked in the field theory examples mentioned
above
\cite{Radyushkin:1997ki,Tiburzi:2002sx,Theussl:2002xp,Burkardt:2000uu},
and also for the perturbatively calculable GPDs of an electron target
in QED \cite{Brodsky:2000xy}.  Not every approximation scheme in model
calculations will however correctly lead to continuous GPDs, as has
been seen in the Bethe-Salpeter framework, see
Section~\ref{sub:covariant-models}.

It is sometimes assumed that light-cone wave functions vanish when a
parton plus-momentum goes to zero.  For those wave functions that
reduce to leading-twist distribution amplitudes when integrated over
the parton $k_T$ this may be true, but in general it is most probably
not.  The work by Antonuccio et al.~\cite{Antonuccio:1997tw} suggests
that there are wave functions for higher Fock states that do not
vanish when a quark or antiquark momentum fraction $x$ tends to zero.
The usual argument \cite{Brodsky:1989pv} that such wave functions
would describe configurations with infinite light-cone energy is
invalidated by cancellations between the free (kinetic) and
interacting parts of the light-cone Hamiltonian.  The perturbative
wave function for an electron and a photon within an electron (given
e.g.\ in \cite{Brodsky:2000xy}) also remains finite for a vanishing
momentum fraction of the electron if the helicities of all particles
are aligned.  If the photon becomes slow, the same wave function
diverges like the inverse square root of the photon momentum fraction,
thus compensating the factor $|x-\xi|^{1/2}$ in the overlap
representation for spin-one partons (see
Section~\ref{sub:overlap-formulae}).  For scalar partons on the other
hand, a finite GPD at $x=\xi$ is already generated by wave functions
not vanishing faster than the square root of a parton momentum
fraction, given the factor $|x-\xi|^{-1/2}$ in the overlap formula.

The GPDs calculated from the lowest Fock states of the target were
indeed found to be nonzero at $x= \xi$ in $\phi^3$ theory and in QED
\cite{Radyushkin:1997ki,Brodsky:2000xy}, as well as in the bound state
calculations of~\cite{Tiburzi:2002sx,Theussl:2002xp} with scalar or
spin $\half$ partons.  We remark however that the meson GPDs in the
\mbox{'t Hooft} model do tend to zero at $x=\xi$
\cite{Burkardt:2000uu}, with the $q\bar{q}$ wave functions of the 
meson vanishing at the endpoints.

In QCD it is not plausible that GPDs should vanish at $x=\xi$.  Such a
situation is only stable under evolution if the GPD vanishes in the
entire DGLAP region (as it does in the limit $\mu\to\infty$),
otherwise evolution to a higher scale $\mu$ will immediately ``fill
up'' a zero at $x=\xi$.  We also recall the link between evolution and
the perturbative wave functions of two partons within a free parton
state (see Section~\ref{sub:wf-evol}), which do not vanish at the end
points.  A better understanding of these issues would further our
knowledge of how hadronic wave functions behave at their end points.


\subsection{Transition GPDs}
\label{sec:transition}

In hard exclusive processes where GPDs can be accessed one is not
limited to an elastic transition on the target side, say from a proton
to a proton.  One also has quasielastic transitions, say $p\to n$ or
$p \to \Delta$ to other baryons, and even to continuum states like
$p\to N \pi$ (where $N$ stands for a nucleon, $p$ or $n$).  As long as
the invariant mass of this hadronic system is small compared with the
hard scale $Q^2$ of the process, factorization works in the same way,
with the soft input being GPDs for the relevant hadron transition.

Transitions to final states like $\Delta$, $N^*$ or $N \pi$ compete
with the elastic transition and can thus present a background in
experiments.  To estimate its size, some phenomenological knowledge of
the relevant transition GPDs will be useful.  If on the other hand the
final state hadronic system can be sufficiently well detected, one can
think of studying transition GPDs to a given final state as a signal.
In fact they are quantities that allow one to study properties of
baryons not available as targets.  Moreover they are among the rare
quantities (apart from fragmentation functions) that specifically
contain information on their structure at \emph{parton} level.  In the
wave function representation they are interpreted as the overlap of
the final state hadron wave functions with the ``reference'' wave
functions of the target.

The definition of transition GPDs is a straightforward extension of
the usual one, e.g.
\begin{eqnarray}
  \label{trans-gpd}
\lefteqn{
\frac{1}{2} \int \frac{d z^-}{2\pi}\, e^{ix P^+ z^-}
  \langle n(p')|\, \bar{d}(-\half z)\, \gamma^+ u(\half z)\, 
  \,|p(p) \rangle \Big|_{z^+=0,\, \tvec{z}=0}
}
\\
&=& \frac{1}{2P^+} \left[
  H^{du}_{p\to n}(x,\xi,t)\, \bar{u}(p') \gamma^+ u(p) +
  E^{du}_{p\to n}(x,\xi,t)\, \bar{u}(p') 
                 \frac{i \sigma^{+\alpha} \Delta_\alpha}{2m} u(p)
  \, \right] .
\nonumber 
\end{eqnarray}
The symmetry relations (\ref{time-rev}) and (\ref{complex-conjug})
discussed in Section~\ref{sub:symm} change for transition GPDs,
because the actions of time reversal and of complex conjugation now
relate e.g.\ $H_{A\to B}$ with $H_{B\to A}$.  One thus no longer has
functions even in $\xi$, except if there is another symmetry relating
$H_{A\to B}$ with $H_{B\to A}$ such as isospin or SU(3) flavor
symmetry.

The combined action of time reversal and complex conjugation does
however constrain the transition GPDs to be real valued, provided that
they involve stable single hadrons.  Otherwise, a state like $|N
\pi\rangle_{\mathrm{out}}$ becomes $|N \pi\rangle_{\mathrm{in}}$ under
time reversal.  The transition GPDs to unstable states thus contain
dynamical phases, just as we saw in the case of GDAs in
Section~\ref{sec:gda}.

For the same reasons, transition form factors (related to transition
GPDs by sum rules) are real-valued only for transitions between stable
hadrons, and the number of independent form factors for the transition
${A\to B}$ is different from the number of form factors for elastic
transitions, unless flavor symmetry provides further relations between
${A\to B}$ and ${B\to A}$.  On the other hand the counting of
independent GPDs still proceeds according to the rules given in
Section~\ref{sub:counting}, because they do not make use of time
reversal or hermiticity.

\subsubsection{Transitions within the baryon octet}

Transition GPDs between members of the ground state baryon octet have
the same spin structure as the proton GPDs, namely there are four
quark helicity conserving GPDs $H$, $E$, $\tilde{H}$, $\tilde{E}$ for
each possible quark flavor transition.  There are no gluon GPDs for
transitions between different octet baryons because two gluons carry
zero isospin and hypercharge.  We remark that for determining quantum
number constraints one may either consider GPD matrix elements of the
type $\langle n | \bar{d} \gamma^+ u | p\rangle$ or the appropriate
one in the crossed channel $\langle n \bar{p}| \bar{d} \gamma^+ u |
0\rangle$.

GPDs for the proton-neutron transitions occur for instance in the
electroproduction processes $\gamma^* p\to \pi^+ n$ or $\gamma^* p\to
\rho^+ n$.  Scattering off neutrons in nuclei involves the reverse
transitions $\gamma^* n\to \pi^- p$ or $\gamma^* n\to \rho^- p$.  The
transition GPDs are related to the flavor diagonal ones by
\cite{Mankiewicz:1997aa} 
\begin{equation}
  \label{isospin-pn}
H^{du}_{p\to n} = H^{ud}_{n\to p} = H^{u}_p - H^{d}_p  ,
\end{equation}
at equal arguments $x$, $\xi$, $t$.  In the flavor diagonal sector one
has isospin relations $H^u_p = H^d_n$, $H^d_p = H^u_n$, $H^s_p =
H^s_n$ as for the usual parton densities.  Analogous relations hold
for the other GPDs $E$, $\tilde{H}$, $\tilde{E}$.

Among the other transitions within the ground state baryon octet,
those from a nucleon to a $\Sigma$ or $\Lambda$ are of practical
relevance, since they appear in kaon electroproduction processes, such
as $\gamma^* p \to K^+ \Lambda$, $\gamma^* p \to K^+ \Sigma^0$,
$\gamma^* p \to K^0 \Sigma^+$.  For a detailed account on the GPDs for
nucleon to hyperon transitions we refer to~\cite{Goeke:2001tz}.
Flavor SU(3) symmetry relates again these distributions to the flavor
diagonal ones in the nucleon,
namely~\cite{Frankfurt:1999fp,Goeke:2001tz}
\begin{eqnarray}
  \label{strange-relations}
H^{su}_{p\to \Lambda} &=&
        \frac{1}{\sqrt{6}} \Big[\, H^d_p + H^s_p - 2 H^u_p \,\Big] ,
\nonumber \\
H^{su}_{p\to \Sigma^0} &=&
        \frac{1}{\sqrt{2}} \Big[\, H^s_p - H^d_p \,\Big] ,
\nonumber \\
H^{sd}_{p\to \Sigma^+} &=& H^s_p - H^d_p  ,
\phantom{\frac{1}{\sqrt{2}}}
\end{eqnarray}
with similar relations for the transitions $n\to \Lambda, \Sigma^0,
\Sigma^-$.  Analogous relations hold for $E$ and $\tilde{H}$.  For
$\tilde{E}$ one expects substantial flavor SU(3) breaking effects, see
Section~\ref{sec:meson-lt}.

As discussed in \cite{Goeke:2001tz}, some quantities may be rather
sensitive to the SU(3) breaking.  The lowest $x$-moments of the four
GPDs $H$, $E$, $\tilde{H}$, $\tilde{E}$ for a given transition are now
related to six transition form factors, instead of the four elastic
form factors in (\ref{vector-ff}) and (\ref{axial-ff}).  The
additional form factors are zero in the SU(3) limit, where they are
related to elastic form factors that vanish due to time reversal.  One
of these additional form factors (which is unaccessible in weak
hyperon decays and thus experimentally unknown) contributes to sum
rule for $\tilde{E}$ with a prefactor $1/\xi$.  This does not imply a
steep rise of cross sections, where $\tilde{E}$ appears multiplied
with $\xi$, but it does provide an enhancement of this SU(3)
suppressed form factor compared with the SU(3) allowed one in the same
sum rule.  Information on the size of the $\tilde{E}$ transition GPDs
may be obtained from a transverse spin asymmetry in kaon
electroproduction, see Section~\ref{sec:meson-pheno}.

\subsubsection{Transitions to the $\Delta$}
\label{sub:delta-trans}

The nucleon-$\Delta$ transition is of particular practical importance
because it can contaminate the elastic $p\to p$ transition in
experiments.  $N \to \Delta$ GPDs have briefly been discussed in
\cite{Frankfurt:1998jq,Frankfurt:1999xe}; more detail can be
found in the review~\cite{Goeke:2001tz} and in the recent study
\cite{Guichon:2003ah}.

Isospin relates transition GPDs $f^{du}_{p\to \Delta^0}$ and
$f^{ud}_{p\to \Delta^{++}}$ to $f^u_{p\to \Delta^+} = - f^d_{p\to
\Delta^+}$ so that there is only one independent flavor structure.
For this a set of seven independent quark helicity conserving GPDs has
been introduced in \cite{Goeke:2001tz}, corresponding to the number of
independent $N\to \Delta$ transition form factors.  This set must be
incomplete, since the spin counting rules of
Section~\ref{sub:counting} give eight independent helicity
transitions.  On the other hand, phenomenological studies have made
use of the large-$N_c$ limit, where only three of the previous $N\to
\Delta$ GPDs are leading in $1/N_c$.  In this limit, the nucleon and
the $\Delta$ appear as different excitations of the same object,
namely of a soliton.  The three leading $N\to \Delta$ GPDs are then
respectively proportional to the isovector combinations
$\tilde{H}^{u-d}_p$, $E^{u-d}_p$, $\tilde{E}^{u-d}_p$.

When writing down hadronic matrix elements involving the $\Delta$ one
implicitly approximates it as a stable particle, in the same sense as
one assumes the $\rho$ to be stable when defining its distribution
amplitude.  In this formalism the $N\to \Delta$ GPDs are real valued,
and the dynamical phases associated with the decay of the resonance
must explicitly be taken into account when describing the transition
from $\Delta$ to $N \pi$, see e.g.~\cite{Guichon:2003ah}.  An
alternative treatment of the same process considers the transition
GPDs to an $N \pi$ state from the start, projecting the $N \pi$ system
on the partial wave and isospin combination corresponding to the
$\Delta$ resonance.  This formalism directly deals with the physically
observed $N \pi$ states with the appropriate quantum numbers, in the
same sense that one can trade a $\rho$ distribution amplitude for a
two-pion DA in a $P$-wave and the $C= -1$, $I=1$ projection.

\subsubsection{More transitions}
\label{sub:more-transitions}

GPDs for the $N \to N\pi$ transitions have been discussed in
\cite{Goeke:2001tz}.  The classification of their structure is more
involved than for single-particle states.  In addition to the usual
variables $x$, $\xi$, $t$ the transition GPDs depend on three
kinematical variables describing the internal degrees of freedom of
the $N\pi$ state (similar to the variables $\zeta$ and $s$ we have
encountered in GDAs).  There are now three independent momentum
fraction variables, both in the GPDs and in the analog of double
distributions, which have been studied by Bl\"umlein et
al.~\cite{Blumlein:2001sb}.  Apart from the $\Delta$ resonance region
just mentioned, the study of $N \to N\pi$ GPDs is particularly
interesting close to threshold, where again they are of practical
importance to experiment and where they are related to chiral dynamics
(see Section~\ref{sub:chiral}).

GPDs for other quantum number changing transitions have been discussed
in the literature.  Feldmann and Kroll have considered the $B\to \pi$
transition in~\cite{Feldmann:1999sm}.  The lowest $x$ moment of the
corresponding GPDs are the $B\to \pi$ transition form factors, which
are an essential ingredient to the study of various exclusive $B$
meson decays.  The GPD formalism has been used in order to classify
contributions to these form factors of different dynamical nature,
which can then be added without double counting.  The main tool in
this classification was the representation of GPDs in terms of
light-cone wave functions.  This representation had earlier been
considered for the $B\to \pi$ transition in \cite{Brodsky:1998hn},
whose main observation was that for this transition reference frames
with vanishing skewness $\xi$ are kinematically not possible if
$\Delta^2 >0$.  The ERBL regime with its emission of a $u\bar{b}$ or
$d\bar{b}$-pair from the initial $B$ meson then always appears in the
overlap representation of the form factors.

Frankfurt et al.~\cite{Frankfurt:1998jq} have pointed out that one can
even define GPDs for fermion number changing transitions such as $p\to
\pi$.  They involve operators with three quark fields (similarly to
the three-quark DA in the proton) and involve several partonic regimes
corresponding to the DGLAP and ERBL regions of ordinary GPDs.  They
are for instance sensitive to components of the nucleon wave functions
with three quarks at a small transverse distance, plus a number of
spectator partons that form the outgoing pion.  Such GPDs appear in
electroproduction processes like $\gamma^* p \to \pi p$ at large $Q^2$
and small Mandelstam variable $u$, i.e.\ in kinematics where in the
collision c.m.\ the $\pi$ moves in the direction of the initial
proton, and the final $p$ in the direction of the $\gamma^*$.  Note
however that the hard scattering process in this case is related to
the one for the elastic transition $\gamma^* p\to p$ at large $Q^2$;
experience with that reaction suggests that the leading-twist regime
may only be reached at extremely large values of $Q^2$ (see
Section~\ref{sec:large-s}).


\subsection{Spin 1 targets}
\label{sec:deuteron}

GPDs for hadrons with spin $J=1$ are of practical importance for
experiments on nuclear targets, in particular on the deuteron (see
Section~\ref{sec:nuclei}).  Their spin classification and basic
properties (in the parton helicity conserving sector) have been given
by Berger et al.~\cite{Berger:2001zb}.  The helicity counting rules of
Section~\ref{sub:counting} tell us that for each quark flavor and the
gluons there are 9 parton helicity conserving GPDs, 5 of them averaged
over parton helicities and 4 sensitive to parton polarization.  For
quarks one has\footnote{Note that the vector $P$ in
\protect\cite{Berger:2001zb} is defined with an extra factor of 2
compared with our convention (\protect\ref{basic-vectors}) here.}
\begin{eqnarray}
\lefteqn{
\frac{1}{2} \int \frac{d \lambda}{2\pi}\, e^{ix (P z)}
  \langle p'|\, \bar{q}(-\half z)\, \slash{n}_- \, q(\half z)\, 
  \,|p \rangle \Big|_{z = \lambda n_-}
= - (\epsilon'^*   \epsilon)\, H_1
}
\nonumber \\[0.2em]
&& {}+ \frac{(\epsilon n_-) (\epsilon'^* P)
       + (\epsilon'^* n_-) (\epsilon P)}{P n_-}\, H_2
  - \frac{2 (\epsilon P)(\epsilon'^* P)}{m^2}\, H_3
\nonumber \\
&& {}+ \frac{(\epsilon n_-) (\epsilon'^* P)
       - (\epsilon'^* n_-) (\epsilon P)}{P n_-}\, H_4
\nonumber \\
&& {}+ \Bigg[
m^2\, \frac{(\epsilon n_-)(\epsilon'^* n_-)}{(P n_-)^2}
 +\frac{1}{3} (\epsilon'^*   \epsilon) \Bigg] H_5 \; ,
\nonumber \\[0.3em]
\lefteqn{
\frac{1}{2} \int \frac{d \lambda}{2\pi}\, e^{ix (P z)}
  \langle p'|\, \bar{q}(-\half z)\, \slash{n}_- \gamma_5 \,
     q(\half z)\, \,|p \rangle \Big|_{z = \lambda n_-}
= {}- i \frac{\epsilon_{\mu \alpha \beta \gamma}
   n_-^\mu \epsilon'^{*\, \alpha}
  \epsilon^\beta P^\gamma}{P n_-}\,
\tilde{H}_1
}
\nonumber \\
&& {}+ 2i\, \frac{\epsilon_{\mu \alpha \beta \gamma}\, n_-^\mu
 \Delta^\alpha P^\beta}{P n_-}\,
 \frac{ \epsilon^\gamma (\epsilon'^* P) +
        \epsilon'^{* \,\gamma} (\epsilon P) }{m^2}\,
\tilde{H}_2
\nonumber \\ 
&& {}+ 2i\, \frac{\epsilon_{\mu \alpha \beta \gamma}\, n_-^\mu
 \Delta^\alpha P^\beta}{P n_-}\,
 \frac{ \epsilon^\gamma (\epsilon'^* P) -
        \epsilon'^{* \,\gamma} (\epsilon P) }{m^2}\,
\tilde{H}_3
\nonumber \\ 
&& {}+ \frac{i}{2}\, \frac{\epsilon_{\mu \alpha \beta \gamma}\, n_-^\mu
 \Delta^\alpha P^\beta}{P n_-}\,
 \frac{ \epsilon^\gamma (\epsilon'^* n_-) +
        \epsilon'^{* \,\gamma} (\epsilon  n_-) }{P n_-}\,
\tilde{H}_4\; ,
\end{eqnarray}
where $\epsilon$ and $\epsilon'$ are the respective polarization
vectors of the incoming and outgoing deuteron.  Time reversal
invariance gives that $H_4$ and $\tilde{H}_3$ are odd in $\xi$ and all
other GPDs even in $\xi$.  Taking the lowest moments in $x$ we recover
the vector and axial vector form factors for elastic deuteron
transitions,
\begin{eqnarray}
  \label{deut-sum-rules}
\int_{-1}^1 dx\, H_i(x,\xi,t) & = & G_i(t) \hspace{3em} (i=1,2,3) ,
\nonumber \\
\int_{-1}^1 dx\, \tilde{H}_i(x,\xi,t) & = &  \tilde{G}_i(t) 
\hspace{3em}     (i=1,2) ,
\end{eqnarray}
with
\begin{eqnarray}
\label{deut-ff}
\lefteqn{
  \langle p' |\,
    \bar{q}(0)\, \gamma^\mu\, q(0) \,| p \rangle
= - 2 (\epsilon'^*  \epsilon) P^\mu \; G_1(t)
}
\nonumber \\
&& {}+ 2 \Big[\epsilon^\mu (\epsilon'^*  P)
       + \epsilon'^{* \mu} (\epsilon  P)\Big] \, G_2(t)
     - 4 (\epsilon P)(\epsilon'^* P)\, 
	 \frac{P^\mu}{m^2} \; G_3(t) ,
\nonumber \\[0.3em]
\lefteqn{
  \langle p' |\,
    \bar{q}(0)\, \gamma^\mu \gamma_5\, q(0) \,| p \rangle
= - 2i \, \epsilon^\mu{}_{\!\alpha \beta \gamma}\,
    \epsilon'^{* \alpha} \epsilon^\beta P^\gamma\; \tilde{G}_1(t)
}
\nonumber \\[0.2em]
&& {}+ 4i \, \epsilon^\mu{}_{\!\alpha \beta \gamma}\,
    \Delta^\alpha P^\beta\, \frac{\epsilon^\gamma (\epsilon'^*  P)
    + \epsilon'^{* \gamma} (\epsilon P)}{m^2}\; \tilde{G}_2(t) .
\end{eqnarray}
One furthermore has
\begin{equation}
  \label{zero-sum-a}
\int_{-1}^1 dx\, H_4(x,\xi,t)  = 
\int_{-1}^1 dx\, \tilde{H}_3(x,\xi,t) = 0
\end{equation}
because of time reversal invariance, and
\begin{equation}
  \label{zero-sum-b}
\int_{-1}^1 dx\, H_5(x,\xi,t)  = 
\int_{-1}^1 dx\, \tilde{H}_4(x,\xi,t) = 0
\end{equation}
because the definitions of these GPDs involve the tensor $n^\mu
n^\nu$, whose analog is forbidden by Lorenz invariance in the
decomposition of the local vector and axial vector currents.

In the forward limit $H_1$, $H_5$, and $\tilde{H}_1$ tend to the
well-known quark densities in the deuteron, which at leading twist and
leading order in $\alpha_s$ are respectively proportional to the
structure functions $F_1(x)$, $b_1(x)$, and $g_1(x)$
\cite{Hoodbhoy:1989am}.  The vanishing of the integral of $H_5$ over
$x$ is directly related to the sum rule for $b_1(x)$ found by Close
and Kumano \cite{Close:1990zw}.


\section{Dynamics and models}
\label{sec:dynamics}

At present GPDs are still mostly unknown functions, apart from the
constraints from the forward limit and from sum rules relating them to
elastic form factors.  In this section we will review the current
strategies to make ans\"atze for these functions, which are needed for
two practical reasons.  One is that the projection of experiments
requires at least roughly correct predictions for cross sections and
event distributions.  On the other hand, the only known strategy at
present to ``extract'' GPDs from measurements is to assume a
functional form of GPDs with a number of adjustable parameters, and to
fit these by comparing the resulting cross sections and distributions
with experimental data, just as one does in the standard analyses of
usual parton densities.  For this to be reliable one needs some
confidence that the assumed functional dependence is adequate to
capture the actual shape of GPDs as functions of three kinematical
variables.  A better understanding of these issues finally means a
better understanding of hadron structure itself, which is the ultimate
goal of studying GPDs.

Among the features of GPDs one needs to understand is first the
interplay between the two momentum fractions, i.e.\ between the $x$
and $\xi$ dependence.
\begin{itemize}
\item In the DGLAP region, a way to put this question is how
strongly GPDs at a given $x$ and $\xi$ differ from the forward
distributions at the same $x$ or some combination of $x$ and $\xi$
like $x_{\mathrm{in}} = (x+\xi)/ (1+\xi)$.
\item The structure of GPDs in the ERBL region is yet more difficult
to guess from the usual parton densities.  Lorentz invariance (and its
realization by double distributions) indicates that there is cross
talk between the two regions, but also that there are contributions to
the ERBL region which are invisible in the DGLAP regions and hence in
the forward limit (see Section~\ref{sub:d-term}).
\item As discussed in Section~\ref{sec:x-equal-xi} a particularly
important point is the behavior of GPDs for $x$ in the vicinity of
$\xi$, both for the physics it reflects and for its prominent role in
physical scattering amplitudes.
\end{itemize}
A second feature is the correlation between the dependence on $t$ and
the longitudinal variables $x$ and $\xi$.  This concerns the interplay
between transverse and longitudinal momenta of partons, or in the
impact parameter representation the relation between the transverse
location of quarks or gluons and their longitudinal momenta.  In yet a
different form the question is how different the $t$ dependence is for
the various form factors one obtains by taking $x$-moments of GPDs for
the various parton species.

Some of these issues have been addressed in toy models.  These do not
claim to reflect the dynamics of QCD but allow one to explore which
structures can be generated in relativistic quantum field theories
where calculations can be pushed further than in QCD, since one may
use perturbation theory or can exactly solve the theory.  We have
already mentioned examples of such studies
\cite{Radyushkin:1997ki,Brodsky:2000xy,Tiburzi:2002sx,Theussl:2002xp,Burkardt:2000uu} 
when discussing the behavior of GPDs at the point $x=\xi$ in
Section~\ref{sec:x-equal-xi}.  Perturbative studies also provide a
glimpse into the relation between physics in the DGLAP and ERBL
regimes, which look very different in a partonic picture but in these
examples are generated by the same covariant Feynman diagrams in
different kinematical regions.


\subsection{Dynamics of GPDs}
\label{sub:dynamics}

There is also a growing number of studies using either dynamical
symmetries of QCD or models that aim to capture at least part of QCD
dynamics.  Whereas no dynamical model can presently give predictions
detailed enough to obtain, say, the full scattering amplitude of
deeply virtual Compton scattering, models can make predictions
concerning general features like those just mentioned.  Turning these
into predictions on observable quantities, one can hope to test
aspects of such models against experiment.  Ideally, this will allow
one to put to test the physical picture of the dynamics responsible
for the many nontrivial features of hadron structure encoded in GPDs.

\subsubsection{Chiral symmetry}
\label{sub:chiral}

Since GPDs are low-energy quantities in QCD it is natural to ask what
chiral symmetry has to say about them.  A number of properties can
indeed be inferred from chiral symmetry and its implementation in
chiral perturbation theory.  Notice however that, contrary to most of
the ``usual'' applications of these concepts, GPDs involve quark and
gluon fields.  Applying concepts of chiral physics therefore requires
a matching procedure to link low-energy degrees of freedom such as
pions to the parton degrees of freedom which manifest themselves in
processes at sufficiently high virtualities.

\begin{figure}
\begin{center}
	\leavevmode
	\epsfxsize=0.75\textwidth
	\epsfbox{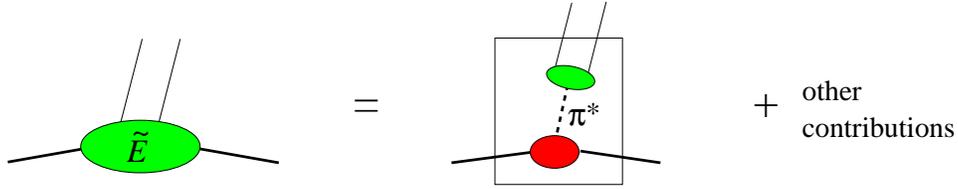}
\end{center}
\caption{\label{fig:pion-pole} Pion exchange contribution to 
$\tilde{E}^{u-d}$ at $t$ close to $m_\pi^2$.}
\end{figure}

Probably the most prominent manifestation of chiral symmetry is the
pion exchange contribution to the nucleon GPDs $\tilde{E}^q$, briefly
mentioned in \cite{Frankfurt:1998jq} and analyzed in
\cite{Mankiewicz:1998kg,Penttinen:1999th}.  Physically it corresponds
to the emission of a virtual pion by the initial nucleon, which for
small $|t|$ is only weakly off-shell.  The pion then annihilates into
the $q\bar{q}$ pair probed in the hard process, see
Fig.~\ref{fig:pion-pole}.  This contribution is only present in the
ERBL region.  Due to its quantum numbers it only contributes to the
$C$-even part of the isovector combination $\tilde{E}^{u-d}$ and to
the corresponding proton-neutron transition GPDs following the isospin
relation (\ref{isospin-pn}).  More precisely, one has a contribution
from an on-shell pion if one analytically continues to $t=m_\pi^2$,
where \cite{Mankiewicz:1998kg,Penttinen:1999th}
\begin{equation}
  \label{pion-pole}
\tilde{E}^{u-d}(x,\xi,t)  \stackrel{t\to m_\pi^2}{\to}
  \theta(|x| < |\xi|)\, \frac{1}{2|\xi|}\,
  \phi_\pi\Big( \frac{x+\xi}{2\xi} \Big) \,
  \frac{4 m^2 g_A(0)}{m_\pi^2 - t} .
\end{equation}
Here we have used the pion distribution amplitude $\phi_\pi$
normalized as
\begin{equation}
  \label{pion-da-normal}
\int_0^1 dz\, \phi_\pi(z) = 1 ,
\end{equation}
which is related to our definition in (\ref{pseudo-da}) by
$\phi_\pi(z) = \sqrt{2} f_\pi^{-1}\, \Phi_{\pi^0}^u(z)$ with $f_\pi
\approx 131$~MeV.  Note that both $\tilde{E}$ and $\phi_\pi$ depend on
the factorization scale $\mu$.  In (\ref{pion-pole}) use has been made
of the Goldberger-Treiman relation, which in the chiral limit relates
the axial charge $g_A(0) \approx 1.26$ of the nucleon to the
pion-nucleon coupling constant $g_{\pi N\!N}$ and is
phenomenologically valid with about $6\%$ accuracy
\cite{Weinberg:1996kr}.  The pion pole contribution to $\tilde{E}$ may
be seen as one particular instance of a resonance exchange
contribution to GPDs, but it is special in that for the pion it is
least questionable that one single resonance contribution may dominate
at small $t$ in the physical region, due to the smallness of
$m_\pi^2$.  Even for the pion there are corrections to the pole
contribution (\ref{pion-pole}) that are not negligible at $-t$ of a
few $0.1$~GeV$^2$, according to estimates in the chiral soliton model
(see Section~\ref{sub:chiral-soliton}).

Another application of chiral dynamics is to study nonanalytic terms
in the dependence of GPDs on $t$ and on the pion mass.  Whereas the
$t$ dependence is of obvious physical concern, the dependence on
$m_\pi$ is important for instance in lattice QCD studies, which
require extrapolation of results from unphysically large values of
$m_\pi$ to its physical value.  In particular, a logarithmic behavior
of GPDs in $t$ and $m_\pi^2$ is generated by loop corrections in
chiral perturbation theory.  Belitsky and Ji \cite{Belitsky:2002jp}
have studied these terms for the Mellin moments of the nucleon GPDs
$H^q$ and $E^q$, matching the relevant local quark-antiquark operators
to operators constructed from pion and nucleon fields, and then
calculating the corresponding one loop corrections in heavy-baryon
chiral perturbation theory.  Non-analytic terms in $t$ only occur for
the isoscalar combination of second moments $\int dx\, x H^{u+d}$ and
$\int dx\, x E^{u+d}$ and the isovector combination of first moments
$\int dx\,H^{u-d}$ and $\int dx\, E^{u-d}$.  For the other isospin
combinations and for all higher moments, the $t$ dependence is hence
at most linear at small $t$ to first order in chiral perturbation
theory.

For GPDs involving pions, constraints can be obtained for
configurations where a pion momentum becomes soft, more precisely when
its four-momentum goes to zero in the chiral limit.  Such soft-pion
theorems relate for instance the transition GPDs for $N\to N \pi$ in
the limit of a soft pion momentum to nucleon and pion GPDs.  Already
mentioned in \cite{Goeke:2001tz}, this has recently been investigated
in \cite{Guichon:2003ah}.

Not surprisingly, soft pion theorems also exist for GPDs of the pion.
Here the zero-momentum limit for the outgoing pion means $\xi\to 1$
and $t\to 0$, and one finds \cite{Polyakov:1998ze}
\begin{equation}
  \label{soft-pion-gpd}
\lim_{\xi\to 1} H_\pi^{u-d}(x,\xi,0) =
  \phi_\pi\Big( \frac{1+x}{2} \Big) , 
\qquad \qquad
\lim_{\xi\to 1} H_\pi^{u+d}(x,\xi,0) = 0
\end{equation}
in the chiral limit of QCD.  Note that $\xi=1$ is not in the physical
region in a world with massive pions, so that the above relations are
to be understood in the sense of an analytic continuation.  The
constraints (\ref{soft-pion-gpd}) have consequences for the GPDs at
$\xi$ away from 1.  They imply for instance that at $t=0$ the second
moment of $H^{u+d}_\pi$ is given by one instead of two parameters
\cite{Polyakov:1999gs}:
\begin{equation}
\int_{-1}^1 dx\, x H_\pi^{u+d}(x,\xi,0) = (1-\xi^2)\, A^{u+d}_\pi(0) ,
\end{equation}
where $A^{u+d}_\pi(0)$ is the momentum fraction carried by $u$ and $d$
quarks and antiquarks in the pion.  Taking the $\xi=1$ limit of
(\ref{dd-red}) also fixes the quark $D$-term at $t=0$ in terms of the
corresponding double distribution.  In particular its first Gegenbauer
moment defined in (\ref{D-Gegenbauer}) is fixed to be
$d_{1}^{u+d}(0,\mu^2) = - \frac{5}{4}\, A^{u+d}_\pi(0,\mu^2)$, which
is sizeable and of definite sign \cite{Kivel:2000fg}.

A systematic treatment of pion GPDs in the context of chiral
perturbation theory has been given in \cite{Kivel:2002ia}.  In lowest
order of the chiral expansion, the double distributions
$f(\beta,\alpha)$ of the pion coincide with generating functions for
the low-energy constants that match the local twist-two
quark-antiquark operators onto pion operators in the effective theory.
The soft pion theorems (\ref{soft-pion-gpd}) are readily obtained in
the chiral limit.  At one-loop level one obtains corrections to these
theorems, which involve logarithms of $t$ or $m_\pi^2$ and are
parametrically suppressed by $m_\pi^2 /(4 \pi f_\pi)^2$ as usual in
chiral perturbation theory.  An interesting finding is that to this
order a nonanalytic $t$-dependence is generated only in the ERBL
region; this predicts in particular a different $t$-behavior of the
real and imaginary parts of hard scattering amplitudes.  Taking
$\xi=0$ and converting to impact parameter space, one obtains the
transverse distributions of quarks at large impact parameter.  It is
found to come from low-momentum quarks, and by inspection of the
relevant diagrams it is naturally attributed to the long range pion
``cloud'' of the target.

\subsubsection{The large-$N_c$ limit}
\label{sub:large-Nc}

The limit of SU($N_c$) gauge theory with the number $N_c$ of colors
going to infinity has proven to be a valuable guide to the physics of
QCD with $N_c=3$, both for mesons \cite{'tHooft:1974jz} and for
baryons \cite{Witten:1979kh}.  We have already mentioned in
Section~\ref{sub:delta-trans} that the large-$N_c$ limit relates the
GPDs for the $N\to \Delta$ transitions with the GPDs of the nucleon.
The large-$N_c$ properties of nucleon GPDs for $u$- and $d$-quarks
have been investigated in \cite{Petrov:1998kf,Penttinen:1999th} and
are summarized in the review \cite{Goeke:2001tz}.  In the large-$N_c$
limit baryon masses scale as $m \sim N_c$, and to fix a power-counting
scheme the kinematical region with $t \sim N_c^0$ and $\xi \sim 1
/N_c$, $x \sim 1/N_c$ has been specified.  Note that the scaling of
the quark momentum fractions $x \pm \xi \sim 1 /N_c$ matches the
large-$N_c$ picture of a baryon as a bound state of $N_c$ quarks.

The large-$N_c$ limit then predicts the relative size of different
combinations of GPDs.  For leading isospin combinations one has
\cite{Goeke:2001tz}
\begin{eqnarray}
  \label{large-Nc-GPDs}
H^{u+d} &\sim& N_c^2 , \qquad 
E^{u-d}  \sim  N_c^3 ,
\nonumber \\
\tilde{H}^{u-d} &\sim& N_c^2 , \qquad
\tilde{E}^{u-d}  \sim  N_c^4 .
\end{eqnarray}
In each case, the other isospin combination is suppressed by one power
of $N_c$.  For the $D$-term one finds $D^{u+d} \sim N_c^2$ and
$D^{u-d} \sim N_c$, so that it contributes at leading order to $H^u$,
$H^d$ and the isosinglet combination $E^{u+d}$, but appears with a
relative suppression by $1/N_c^2$ in $E^{u-d}$. Taking into account
the power counting for the kinematical variables one obtains the
scaling behavior of the matrix elements $F^q_{\lambda'\lambda}$ and
$\tilde{F}^q_{\lambda'\lambda}$ in (\ref{hel}), which correspond to
definite nucleon helicities and occur in physical amplitudes.  For the
quark helicity averaged case one has
\begin{eqnarray}
F^{u+d}_{++} &\sim& N_c^2 , \qquad
F^{u-d}_{-+}  \sim  N_c^2
\nonumber \\[0.2em]
F^{u-d}_{++} &\sim& N_c , \qquad
F^{u+d}_{-+}  \sim  N_c ,
\end{eqnarray}
whereas the isovector combinations $\tilde{F}^{u-d}_{++}$ and
$\tilde{F}^{u-d}_{-+}$ scale like $N_c^2$, and the isoscalar
combinations $\tilde{F}^{u+d}_{++}$ and $\tilde{F}^{u+d}_{-+}$ scale
like $N_c$.  We note in passing that in physical applications one may
have a mix of isosinglet and isotriplet combinations, for instance
$\frac{4}{9} u + \frac{1}{9} d = \frac{5}{18} (u+d) + \frac{3}{18}
(u-d)$ in Compton scattering processes.

It may be useful to ``check'' the above relations in limits where the
GPDs are known.  Taking the forward limit, one finds the relations
$(u+ d) : (u- d) \sim N_c$ and $(\Delta u - \Delta d) : (\Delta u +
\Delta d) \sim N_c$ for the parton distributions.  At momentum
fractions $x\sim 1/ N_c$ in the valence region and for moderately
large factorization scale $\mu$ this is quite well satisfied in the
real world with $N_c=3$.  Another test case are the anomalous magnetic
moments $\kappa^q = F_2^q(0)$ for each separate quark flavor in the
proton, related to the anomalous magnetic moments of proton and
nucleon by $\kappa_p = \frac{2}{3} \kappa^u - \frac{1}{3} \kappa^d$
and $\kappa_n = \frac{2}{3} \kappa^d - \frac{1}{3} \kappa^u$ if one
neglects the contribution from strange quarks.  Numerically one has
$(\kappa^u - \kappa^d) : (\kappa^u + \kappa^d) \approx 3.7 : (-0.36)$,
whose size should be of order $N_c$ according to
(\ref{large-Nc-GPDs}).  These exercises indicate that (\textit{i})
terms suppressed by $1/N_c$ can indeed be as large as 30\% of the
leading terms, as they are entitled to be, and that (\textit{ii}) as
with any power counting argument one should allow for extra factors in
phenomenological applications, which may be of order 3 in one
direction or the other.  With these caveats in mind, one may still use
$1/N_c$ counting as a guide for the relative orders of magnitude
between the different flavor and spin combinations of GPDs.

\subsubsection{The chiral quark-soliton model}
\label{sub:chiral-soliton}

Quite detailed studies of GPDs have been performed in the chiral
quark-soliton model \cite{Petrov:1998kf,Penttinen:1999th}; a review
and references can be found in \cite{Goeke:2001tz}.  Essential
ingredients of this model are the spontaneously broken chiral symmetry
of QCD and the large-$N_c$ limit, which we have just discussed.  It is
based on an effective field theory of QCD, obtained from the instanton
model of the QCD vacuum.  Degrees of freedom are pions and quarks.
The quarks have a momentum dependent dynamical mass of order
$M=350$~MeV at low scale, which drops sharply at a momentum scale
$1/\rho \approx 600$~MeV given by the typical instanton size $\rho$ in
the vacuum of the model.  This scale provides an ultraviolet cutoff to
the effective theory when calculating parton distributions.  The
quarks and antiquarks in these distributions should then be seen as
``constituent quarks'' that have an unresolved substructure at length
scales smaller than $\rho$ \cite{Diakonov:1997vc}.  A simpler (but not
obviously equivalent) interpretation is to associate the distributions
obtained in the effective theory as the QCD quark and antiquark
distributions at a factorization scale $\mu \sim 1/\rho$.  Gluon GPDs
are parametrically suppressed by $\rho^2 M^2$ in this framework, and
have not been studied yet.

In this model the nucleon appears as a bound state of quarks in a
semi-classical pion field (the ``chiral soliton'').  The
single-particle spectrum of quarks in this field contains a bound
state and continuum states of positive or negative energy.  The
nucleon is obtained by occupying the bound state with $N_c$ quarks and
by filling the negative-continuum Dirac sea.  This leads to a
``two-component'' structure of nucleon GPDs with distinct
contributions from the ``valence'' bound state level and from the
Dirac sea.  Notice that the terms ``valence'' and ``sea'' have a
dynamical meaning in the context of this model, which is \emph{not}
equivalent to the usual meaning of ``valence'' distributions as the
difference $q- \bar{q}$ and of ``sea'' as $\bar{q}$.  In particular,
both the contributions from the ``valence'' and ``sea'' levels have
support in then entire regions $-1<x<1$ (although the valence
contribution is largest for moderate positive $x$).

So far, $u$ and $d$ quark GPDs in the nucleon have been studied in
this model to leading order in the large-$N_c$ expansion.  We note
that the model provides a fair description of the unpolarized and
polarized forward $u$ and $d$ quark distributions (including the
$1/N_c$ suppressed combination $u-d$ \cite{Diakonov:2000pa}) and of
the elastic nucleon form factors for $|t|$ up to about 1~GeV$^2$
(except for the electric form factor of the neutron)
\cite{Christov:1996vm}.  The analytic results in this model fulfill
various consistency requirements, namely the forward limit and the
reduction of the first Mellin moment to the elastic form factors
\cite{Petrov:1998kf,Penttinen:1999th}.  Schweitzer et al.\
\cite{Schweitzer:2002nm,Schweitzer:2003ms} have shown
that the higher Mellin moments satisfy polynomiality in $\xi$ when
analytically continued to $t=0$.  Several features of GPDs were found
in this model \cite{Petrov:1998kf,Penttinen:1999th,Goeke:2001tz}:
\begin{itemize}
\item The distribution $H^{u+d}$ exhibits rapid variations in $x$ around
$x=\pm \xi$, due to a rather smooth valence contribution and an
oscillating contribution from the Dirac sea.  Both $H^{u+d}$ and
$\tilde{H}^{u-d}$ have considerable structure in the ERBL region.  The
Dirac sea only contributes to the $C$-even combinations of $H^{u+d}$
and of $\tilde{H}^{u-d}$.
\item The pion pole contribution (\ref{pion-pole}) to
$\tilde{E}^{u-d}$ can be obtained analytically.  Perhaps even more
important, corrections to the pole term were also obtained and
parameterized for intermediate values of $t$.  These corrections were
found to have a non-negligible impact on observables in a
phenomenological study~\cite{Berger:2001zn}.
\item Study of both $H(x,\xi,t)$ and $\tilde{H}(x,\xi,t)$ shows that
the model does \emph{not} support a factorized $t$ dependence of the
form $F(t) f(x,\xi)$, which is often taken as an ansatz (see
Section~\ref{sub:t-depend}).
\item The isosinglet $D$-term for the nucleon is found to be rather
large.  The lowest Gegenbauer moments (\ref{D-Gegenbauer}) for $t=0$
were fitted numerically in \cite{Kivel:2000fg} and obtained
analytically in \cite{Schweitzer:2002nm}.
\end{itemize}

Since the pion appears as an explicit degree of freedom in the
effective theory, its quark GPDs can also readily be studied, which
has been done in \cite{Polyakov:1999gs,Praszalowicz:2003pr}.  Since
the model is constructed to respect the chiral symmetry breaking
pattern of QCD, it respects in particular the soft-pion relations
(\ref{soft-pion-gpd}) in the chiral limit.

\subsubsection{Quark models}
\label{sub:quark-models}

The very first dynamical study of quark GPDs was carried out in the
MIT bag model \cite{Chodos:1974je} by Ji et al.~\cite{Ji:1997gm}.  At
a very low factorization scale of order 450~MeV, quark GPDs were
obtained with support in the entire region $-1<x<1$, although strongly
concentrated at $x>0$.  The curves of all four distributions $H$, $E$,
$\tilde{H}$, $\tilde{E}$ as a function of $x$ were found to have an
extremely weak dependence on $\xi$.  The physics reason for this is
not understood, nor is the reliability of the results in the region
$x<\xi$, where antiquarks play an essential role.  Also using the MIT
bag model, Anikin et al.~\cite{Anikin:2001zv} have calculated the GPDs
in a pion, including the twist-three distributions to be discussed in
Section~\ref{sub:twist}.  The reliability of the model remains unclear
in the ERBL region, which requires nonvalence configurations in the
initial pion.

GPDs calculated in constituent quark models have support only in the
region $x>\xi$.  They are obtained as an overlap of Schr\"odinger wave
functions for three constituent quarks (corresponding to equal time
$z^0$ of the quark and antiquark operators, not to equal light-cone
time $z^+$).  Scopetta and Vento have given expressions for
nonrelativistic kinematics (requiring $|t| \ll m^2$ and hence small
$\xi$) and studied the quark distributions $H^u$, $H^d$ for different
quark potentials \cite{Scopetta:2002xq}.  The distributions $H$ and
$E$ in fully relativistic kinematics have been treated by Boffi et
al.~\cite{Boffi:2002yy}, including the effect of the Melosh
transformation for the quarks (which relates the usual states with
spin quantized along the $z$ axis with states of definite light-cone
helicity).  General features of the results of both groups are:
\begin{itemize}
\item GPDs that vanish at $x=\xi$ since the constituent wave functions
vanish for zero quark momentum.
\item As functions of $x$, the GPDs have a peak whose position shifts
to the right with increasing $\xi$ and also with increasing $|t|$.
\item To the right of this peak the shape in $x$ varies very little
with $\xi$.
\end{itemize}
Whereas the nonrelativistic approximation in
\cite{Scopetta:2002xq} leads to the artifact that GPDs
are small but nonvanishing at $x=1$, the relativistic treatment does
not have this drawback.  

An interesting feature of the results in \cite{Boffi:2002yy} is that
the distributions $E$ fall off faster at $x\to 1$ than the
distributions $H$.  We have seen in Section~\ref{sub:overlap-formulae}
that $E$ can be represented at the overlap of light-cone wave
functions where at least one wave function involves nonzero orbital
angular momentum $L^3$.  The Schr\"odinger wave functions of the
constituent quarks are in an $S$-wave, but the Melosh transform to
light-cone helicity states generates light-cone wave functions with
nonzero $L^3$.  The effect of this transform diminishes when the quark
momentum increases.  This may be the reason for the stronger decrease
of $E$ at large $x$, where the struck parton becomes fast (a more
careful analysis would have to take into account the orbital angular
momentum of the spectators).  The importance of the Melosh transform
for generating the proton helicity flip form factor $F_2(t)$ has also
been stressed by Miller and Frank \cite{Miller:2002qb}.

As $x$ approaches the boundary to the ERBL region, the results of
constituent quark models become more problematic.  A way to capture
the physics of higher Fock states and in particular of antiquarks in
the nucleon may be to assign a substructure to constituent quarks
\cite{Scopetta:2002xq}.  Simply applying the evolution equations
to constituent quark GPDs does of course also generate antiquarks and
gluons, fill up the zero at $x=\xi$, and populate the region $x<\xi$
when going to higher scales $\mu$.  It will be interesting to test
such scenarios against data for hard processes, where GPDs at $x \sim
\xi$ are of prime importance.

Restricted to the three-quark configurations in the proton,
constituent quark models will in general not respect the polynomiality
constraints.  In particular, elastic form factors can be obtained in
this framework by integrating over $x$, but only in frames with
$\xi=0$.

\subsubsection{Covariant approaches}
\label{sub:covariant-models}

Attempts for a consistent description of the physics in the ERBL
region have been made in the covariant approach of reducing
Bethe-Salpeter to light-cone wave functions, whose essentials we
explained in Section~\ref{sub:poly-wave}.  The different Fock states
required in the ERBL and the DGLAP regions are here generated by the
same covariant diagrams.  Remember that in the perturbative example of
Section~\ref{sub:poly-wave} the triangle diagram in the ERBL region
could be written in terms of the light-cone wave functions for three
particles or for one particle in a single-particle state.  The
situation is however more complicated if the external state is not an
elementary particle as in that example, but a bound state.
Interpreting the external lines in Fig.~\ref{fig:triangle}c as
``hadrons'' and the internal lines as ``partons'', one would obtain
the ``wave function'' for one hadron plus two partons in a single
hadron, which as it stands is not part of the Fock state expansion.

\begin{figure}
\begin{center}
	\leavevmode
	\epsfxsize=0.62\textwidth
	\epsfbox{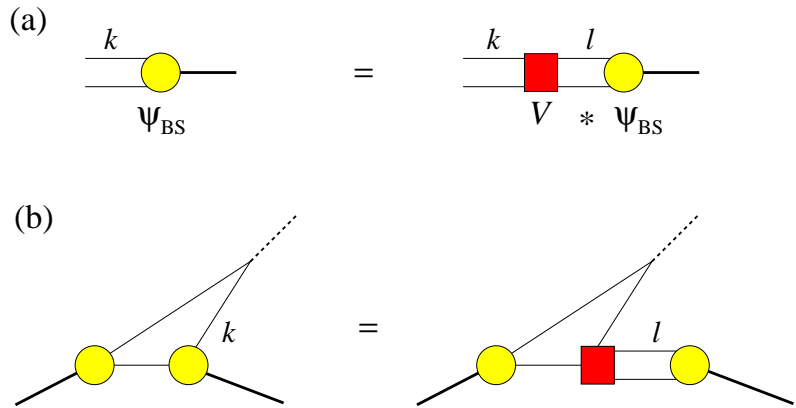}
\end{center}
\caption{\label{fig:BS-eq} (a) Schematic representation of the
Bethe-Salpeter equation. (b) The Bethe-Salpeter equation may be used
to trade a crossed wave function for the crossed interaction kernel.
The dashed line denotes the external (local or nonlocal) current.}
\end{figure}

\begin{figure}
\begin{center}
	\leavevmode
	\epsfxsize=0.66\textwidth
	\epsfbox{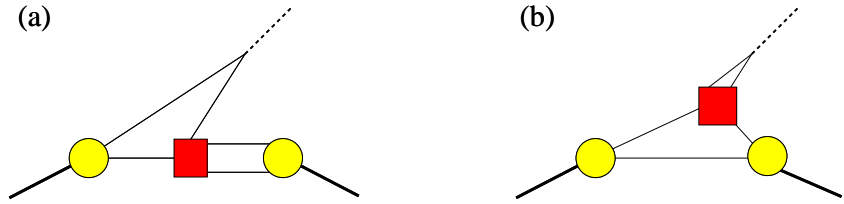}
\end{center}
\caption{\label{fig:BS-tri} Relevant configurations of the triangle
diagram in the ERBL region.}
\end{figure}

A strategy that has been pursued in the literature is to replace the
pointlike vertices in the triangle diagram Fig.~\ref{fig:triangle}
with the bound-state Bethe Salpeter wave functions
$\Psi_{\mathrm{BS}}$ and to make use of the Bethe-Salpeter equation,
depicted schematically in Fig.~\ref{fig:BS-eq}a.  The crossed vertex
function for the final state hadron in Fig.~\ref{fig:BS-eq}b is not in
the correct kinematics to reduce to a light-cone wave function after
integration over $k^-$.  Using the Bethe-Salpeter equation as in
Fig.~\ref{fig:BS-eq}b one obtains a form where crossing has to be
performed for the two-parton interaction kernel $V$ instead.
Integrating over the momentum $l^-$ and using Cauchy's theorem to pick
up appropriate singularities one may then hope to reduce
$\Psi_{\mathrm{BS}}$ to a light-cone wave function.  In general this
still is nontrivial since in addition to $\Psi_{\mathrm{BS}}(l)$ the
interaction kernel $V(k,l)$ can have a nontrivial dependence on $l^-$.
One also needs to take into account interactions at the vertex with
the external current operator, which involve configurations as in
Fig.~\ref{fig:BS-tri}b.  We remark that in the 't Hooft model the
interaction between quarks is exactly instantaneous in light-cone time
(i.e., $V$ is independent on $k^-$ and $l^-$), and the physical
picture of Fig.~\ref{fig:BS-tri} indeed appears in the evaluation of
GPDs in the ERBL region \cite{Burkardt:2000uu}.  For other
interactions the implementation of this program still requires
approximation schemes, which must be chosen with care in order to
obtain consistent results, in particular continuity of GPDs at
$x=\xi$.  This was clearly seen in the work of Choi et
al.~\cite{Choi:2002ic} compared with their earlier attempt
\cite{Choi:2001fc}.  Different approximations were studied by Tiburzi
and Miller \cite{Tiburzi:2001je}.  In subsequent work
\cite{Tiburzi:2002sw,Tiburzi:2002sx} these authors used a particular
expansion scheme in the reduction to the light cone and found that
configurations as in Fig.~\ref{fig:BS-tri} can be systematically
replaced by higher Fock states, thus recovering the overlap
representation.  So far such work has had to resort to the
weak-coupling limit, and it may be that the results obtained are more
immediately applicable not to bound states of quarks but rather to
bound states of nucleons (see Section~\ref{sec:nuclei}).  No scheme is
known where polynomiality of GPDs is exactly satisfied, but in
practice it may be sufficient to have approximate polynomiality in
kinematics where the scheme is valid, especially if there is a
systematic way of taking into account corrections as proposed in
\cite{Tiburzi:2002sw,Tiburzi:2002sx}.  The problems discussed here are
not only relevant for describing GPDs, but equally occur in form
factors for the transition from heavy to light hadrons at timelike
momentum transfer, see Section~\ref{sub:more-transitions}.

Theu{\ss}l et al.~\cite{Theussl:2002xp} have taken a different route
and calculated GPDs directly from the covariant diagrams with
four-dimensional Bethe-Salpeter vertices, without projecting on the
light-cone.  In this case full covariance is retained, and
polynomiality is exact.  To obtain the required wave functions one
must now solve the full Bethe-Salpeter equation, instead of its
light-cone projection.  The field theories considered in
\cite{Theussl:2002xp} are scalar electrodynamics and the Nambu
Jona-Lasinio model.

\subsubsection{Light-cone wave functions and partons}
\label{sub:soft-overlap-gpd}

The approaches discussed so far deal with \emph{effective} quark
degrees of freedom, which are not connected in a simple way to the
quark fields in the QCD Lagrangian.  A different line of work,
initiated in \cite{Diehl:1998kh}, deals directly with the overlap
representation of GPDs in terms of light-cone wave functions for
parton Fock states (Section~\ref{sec:overlap}), which directly
correspond to the QCD quark fields defining GPDs.  The rationale of
such efforts is to model the wave functions for the lowest Fock state
and to link a variety of processes where one can reasonably expect the
leading Fock state to provide the dominant contribution.  These are in
particular parton distributions at large $x$, and GPDs or form factors
at large values of $t$.  We make here the additional assumption that
in an important kinematical region parton configurations with low
internal virtualities provide the bulk contribution to GPDs, and that
contributions involving partons far off-shell become dominant only
very close to the limits $x\to 1$ or $|t|\to \infty$.  Such off-shell
configurations involve hard gluon exchange and lead to the well-known
dimensional counting rules of Brodsky and collaborators, see e.g.\
\cite{Brodsky:1989pv}.  In which kinematics such perturbative
configurations take over remains controversial (see our discussion in
Section~\ref{sec:hard-and-soft}).

Bolz and Kroll \cite{Bolz:1996sw} have developed a model wave function
for the three-quark Fock state in the nucleon, adjusting parameters to
describe the unpolarized $u$ and $d$ quark distributions at large $x$,
the Dirac form factor $F_1(t)$ at large $t$, and the decay $J/\Psi\to
p \bar{p}$.  The wave function is restricted to the configurations
with orbital angular momentum $L^3=0$, principally because theoretical
and phenomenological information on the $L^3 \neq 0$ wave functions is
even more scarce.  The wave function is understood to be \emph{soft}
(associated with a normalization scale $\mu_0=1$~GeV), excluding
configurations with large virtualities and in particular the
perturbative tail at large transverse momenta that is generated by
hard gluon exchange.  Its dependence on the $\tvec{k}_i$ is taken as
\begin{equation}
  \label{wf-ansatz}
\Psi(x_i,\tvec{k}_i) = N_3\, \phi(x_1,x_2,x_3) \;
   \frac{(4\pi a_3)^4}{x_1 x_2 x_3}\, 	
   \exp\left[ - a_3^2\, \sum_{i=1}^3 \frac{\tvec{k}_i^2}{x_i}
   \,\right] ,
\end{equation}
where $N_3$ is a normalization constant.  The probability for the
corresponding Fock state comes out as $P_3 \approx 0.17$.  This alone
makes it clear that one is not dealing with a constituent wave
function, which would be normalized to one.  The impact parameter form
of (\ref{wf-ansatz}) is
\begin{equation}
  \label{wf-ansatz-imp}
\tilde\Psi(x_i,\tvec{b}_i) = N_3\, \phi(x_1,x_2,x_3) \;
   \exp\left[ - \frac{1}{4 a_3^2}\, 
	\sum_{i=1}^3 x_i \tvec{b}_i^2 \,\right] ,
\end{equation}
and shows that the constant $a_3$ describes the average transverse
size of this particular Fock state.  A Gaussian ansatz for the
$\tvec{k}_i$ dependence has the virtue of allowing one to perform
integrals analytically.  A dependence on the combination
$\tvec{k}_i^2/ x_i^{\phantom{1}}$ naturally arises in light-cone
perturbation theory \cite{Lepage:1980fj,Brodsky:1989pv}, or
equivalently when integrating covariant Feynman diagrams over $k^-$ as
discussed in Section~\ref{sub:poly-wave}.

The Gaussian ansatz for the $\tvec{k}_i$ dependence in
(\ref{wf-ansatz}) goes back to Brodsky et al.~\cite{Brodsky:1981aa}.
Its analog for two partons instead of three has some theoretical
support from studies of the $q\bar{q}$ wave functions with $L_3=0$ in
a $\pi$ or a $\rho$.  Szczepaniak et al.\ \cite{Szczepaniak:1998sa}
found this form for the pion using QCD sum rules, although a different
dependence can be obtained in the same framework (namely by using
local parton hadron duality instead of the Borel transform).  Halperin
and Zhitnitsky \cite{Zhitnitsky:1995sk,Halperin:1997zk} strongly argue
in favor of a Gaussian dependence as in (\ref{wf-ansatz}) near the
endpoints $x_i\to 0$, based on a study of high moments $\int dx\,
d^2\tvec{k}\, (2x-1)^n\, \tvec{k}^{2m}\, \Psi(x,\tvec{k}^2)$ with QCD
sum rule techniques.\footnote{Note that the wave functions in
\protect\cite{Zhitnitsky:1995sk,Halperin:1997zk} are defined such that
these moments correspond to local operators containing covariant
derivatives $D^\mu$ in the transverse direction.  The nonlocal
operator associated with $\Psi(x,\tvec{k}^2)$ thus involves gluon
fields with transverse polarization and hence differs from our
definition using the Fock state expansion in
Section~\protect\ref{sec:overlap}.}
We caution that a discussion of the behavior of wave functions at
large $\tvec{k}_i^2 /x_i^{\phantom{1}}$, corresponding to large
off-shellness, requires careful consideration of the renormalization
and factorization scheme in which they are defined.  In the context
where we use them, wave functions at small scale $\mu^2$ are to
describe the region of small off-shellness, the hard region being
explicitly included in hard-scattering kernels.  {}From a pragmatic
point of view this is just what the above Gaussian ansatz ensures: in
practical applications there is little difference whether one includes
the region of high $\tvec{k}_i^2 /x_i^{\phantom{1}}$ in integrals or
not.

The function $\phi(x_1,x_2,x_3)$ in (\ref{wf-ansatz}) is the
leading-twist distribution amplitude in the proton at normalization
scale $\mu_0$.  In \cite{Bolz:1996sw} a simple polynomial form was
taken for $\phi(x_1,x_2,x_3)$, with a weak deviation from its
asymptotic form under evolution.  It has been argued that in the
exponent of (\ref{wf-ansatz}) the combination $\tvec{k}_i^2 +
m_{\mathrm{eff}}^2$ should appear, where $m_{\mathrm{eff}}$ is a quark
mass of some 100~MeV generated by nonperturbative physics
\cite{Lepage:1980fj,Stefanis:2000vd}.  This would generate a factor
\begin{equation}
  \label{mass-exponential}
\exp\left[ - a_3^2 \sum_{i=1}^3 \frac{m_{\mathrm{eff}}^2}{x_i} \right]
\end{equation}
in $\phi(x_1,x_2,x_3)$.  A polynomial form was taken in
\cite{Bolz:1996sw,Diehl:1998kh} to retain simplicity of expressions.
We emphasize that the corresponding wave function has been used to
describe the physics for configurations with parton momentum fractions
going down to order $0.1$.  In such a limited region a polynomial or a
polynomial times the exponential (\ref{mass-exponential}) should be
practically equivalent, given an adequate adjustment of the wave
function parameters.  The ansatz is \emph{not} intended for extremely
small values of $x_i$ or for the limit $x_i\to 0$, where there is a
qualitative difference between a polynomial and a polynomial times
(\ref{mass-exponential}).

An attempt has been made in \cite{Diehl:1998kh} to also model the wave
functions for the next highest Fock states with an extra gluon or with
an extra $q\bar{q}$ pair.  The principal aim was to obtain an
indication of how important higher Fock state contributions are for
various quantities.  The ansatz was tailored along the form
(\ref{wf-ansatz}) with simplicity as the main guide and free
parameters fitted to describe the unpolarized antiquark and gluon
distributions at large $x$.  A good description was obtained for
valence distributions $q-\bar{q}$ and $\Delta q- \Delta\bar{q}$ of $u$
and $d$ quarks at $x\gsim 0.5$, with the three-quark Fock component
dominating for $x\gsim 0.6$.  Notice that the polarized quark
densities were not used to constrain the model wave functions, so that
agreement in this sector is a nontrivial test of the model.  A fair
description of the Dirac form factors $F_1$ of proton and neutron was
also obtained for large $t$.

An important outcome of the study in \cite{Diehl:1998kh} is that the
nominal leading power behavior in $(1-x)$ of the soft contributions to
parton densities (which is easily identified in the analytical
expressions) gave a poor representation of the full result in a large
range of $x$.  The same holds for the nominal leading power behavior
in $t$ of elastic form factors.  This recommends care in using power
counting arguments for soft physics contributions of this particular
type.

Using the above ansatz to evaluate GPDs at large $x$ or large $t$, one
finds again that higher Fock states become gradually important as $x$
decreases, with the leading three-quark contribution dominating
roughly for $x\gsim 2\xi$.  An important feature is that the $t$
dependence is intimately related with $x$ and $\xi$, as suggested by
the generic structure of the overlap representation (\ref{DGLAP-mom}).
For $\xi=0$ the Gaussian ansatz for the wave functions leads to the
simple structure
\begin{equation}
  \label{wf-link}
H(x,0,t) = q(x) \, 
  \exp\left[ \frac{a^2}{2}\, \frac{1-x}{x}\, t \right] 
\end{equation}
for the separate contributions from each Fock state, which are summed
over in the end.  An analogous expression relates $\tilde{H}$ and
$\Delta q$.  Taking $x$-moments one obtains elastic form factors,
which at large $t$ are dominated by large $x$.  With the model
parameters in \cite{Diehl:1998kh} the dominant values of $x$ in the
integral giving the proton Dirac form factor range between $0.45$ and
$0.75$ for $|t|$ between 5~GeV$^2$ and 20~GeV$^2$, indicating that in
this region one is indeed far away from the $x\to 1$ limit.  The
expression (\ref{wf-link}) makes it clear why for this mechanism the
typical momentum scale of the $t$-dependence is not given by the
inverse of a typical transverse size $a^{-1}$, which is below 1~GeV,
but by this scale times a typical value of $(1-x)^{-1/2}$.  This
results in a momentum scale of several GeV.

The form (\ref{wf-link}) has also been used as a guide for an ansatz
with a larger range of validity and less dependence on model
parameters, assuming that the transverse size parameter $a$ is similar
for the lowest few Fock states.  In kinematics where these dominate,
one can then just sum over them and read (\ref{wf-link}) as a relation
between the full parton distributions, where $a$ is now an ``average''
transverse size of the relevant contributing Fock states.  {}From its
origin it is however clear that (\ref{wf-link}) is not sensible for
all $x$ and $t$.  If both variables are small, an increasing number of
Fock states becomes relevant for which the assumption of a common
dependence of the type (\ref{wf-ansatz}) with one average size
parameter $a$ is less and less plausible.  In particular,
(\ref{wf-link}) leads to an infinite value of $d F_1 /(dt)$ at $t=0$
since the relevant integrand $x^{-1} (q - \bar{q})$ is too singular at
$x=0$.  Restricting oneself to the contributions from the lowest Fock
states in the model, one obtains a quark distribution $q(x)$ vanishing
at $x=0$ and hence a finite value of $d F_1 /(dt)$ at $t=0$.  These
contributions have however no claim to dominate the actual charge
radius of the proton.

The ansatz (\ref{wf-ansatz}) refers to a low factorization scale of
$\mu_0=1$~GeV, and hence also the approximate relation
(\ref{wf-link}).  When going to higher scales both $H(x,0,t;\mu^2)$
and $q(x;\mu^2)$ evolve according to the usual DGLAP equations.  In a
study of the pion GPDs, Vogt \cite{Vogt:2000ku} has found that for
large $x$ and $t$ the relation (\ref{wf-link}) is approximately valid
over a finite range of $\mu$ if the parameter $a$ is taken to decrease
logarithmically with $\mu$ (without allowing it to depend on $x$).
This shows a certain robustness of the ansatz (\ref{wf-link}),
although its motivation from the overlap of soft wave function comes
from low scales $\mu$.

An ansatz analogous to (\ref{wf-link}) has also been proposed by
Radyushkin on the basis of the overlap of ``effective'' two-body wave
functions \cite{Radyushkin:1998rt}.  It has further been explored in
\cite{Afanasev:1998ym,Stoler:2001xa,Stoler:2003mx} for the nucleon,
and in \cite{Bakulev:2000eb} for the pion.

Clearly the Gaussian $\tvec{k}$ dependence in (\ref{wf-ansatz}) has
the status of an ansatz and may not capture all important features of
light-cone wave functions.  Tiburzi and Miller have explored the
overlap of wave functions for weakly bound two-body systems
\cite{Tiburzi:2001ta}.  They point out that, contrary to wave functions
with a power-law falloff such as $1/ \tvec{k}^4$, a Gaussian
dependence has special features (the product of two Gaussians with
shifted $\tvec{k}$ arguments, which appears in the overlap formula
(\ref{DGLAP-mom}), is again a Gaussian).  We also remark that in wave
functions with nonzero $L^3$ additional factors $\tvec{k}_i$ must
appear in order to have the correct properties under rotations about
the 3-axis.

\subsubsection{The limit of large $t$}
\label{sub:large-t-limit}

\begin{figure}
\begin{center}
	\leavevmode
	\epsfxsize=0.63\textwidth
	\epsfbox{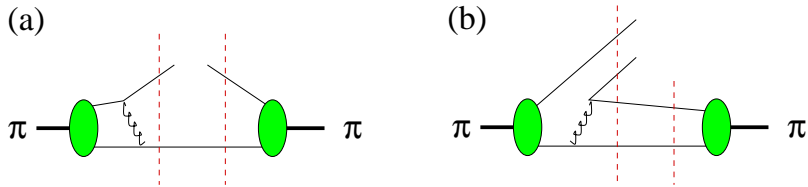}
\end{center}
\caption{\label{fig:large-t-pert} (a) One of the two diagrams for 
$H^q_\pi$ in the limit of large $t$ in the DGLAP region $x\in
[\xi,1]$.  In the second diagram the gluon is exchanged after the
struck quark returns.  (b) The diagram contributing in the ERBL
region.  Dashed lines indicate the representation as an overlap of
light-cone wave functions.}
\end{figure}

The dynamics discussed in the previous section involves low internal
virtualities.  At very large $t$, the perturbative contributions given
by the hard scattering mechanism will be dominant according to general
power counting arguments \cite{Lepage:1980fj,Brodsky:1989pv}.  This
limit may not be achievable in any realistic experiment, but its
features are of theoretical interest similar to those of the toy field
theory studies discussed earlier.  For the quark GPDs of the pion this
was investigated by Vogt \cite{Vogt:2001if}.  In the large-$t$ limit
$H^q_\pi$ can be expressed in terms of the leading-twist DAs of each
pion, convoluted with a hard-scattering kernel.  To leading order in
$\alpha_s$ the relevant diagrams are those given in
Fig.~\ref{fig:large-t-pert}, provided one uses light-cone gauge $A^+
=0$.  The scale of virtualities in these diagrams is set by $t$, which
thus provides the factorization scale of the obtained GPDs, and the
scale of $\alpha_s$.  Salient features of the results in
\cite{Vogt:2001if} are:
\begin{itemize}
\item The $t$ dependence is given as $H_\pi^q(x,\xi,t) = \alpha_s(t)\,
t^{-1} \, h^q(x,\xi)$.  This is compatible with dimensional counting
behavior, as for the pion form factor, which is obtained from the
lowest Mellin moment of $H_\pi^q$.  Corrections to the result are
power suppressed in $1/t$.
\item At $x=\xi$ the GPDs are finite and continuous, but their
derivative has a singular behavior like $\log|x - \xi|$.  This
confirms once more the suspected behavior discussed in
Section~\ref{sec:x-equal-xi}.
\item For $x\to 1$ the GPDs develop an unphysical logarithmic
singularity like $\log(1-x)$.  In this region the spectator quark in
one of the pions becomes slow and the internal virtualities of the
diagrams of Fig.~\ref{fig:large-t-pert} small, so that the collinear
hard scattering approximation is no longer justified.
\item {}From the diagrams in Fig.~\ref{fig:large-t-pert} one readily
sees that to this level of accuracy $H_\pi^s$ is zero.  Similarly,
$H^u_\pi$ is zero for $x<-\xi$ since the diagrams cannot generate
$\bar{u}$ antiquarks in a $\pi^+$.
\item The obtained GPDs exactly satisfy the polynomiality property.
\end{itemize}

The results of the calculation can also be cast into the form of a
wave function overlap.  The analysis is similar as for the pion form
factor, see Section 4 of \cite{Brodsky:1989pv} and also the discussion
in \cite{Tiburzi:2001ta}.  The perturbative tail at large $\tvec{k}$
generated by one-gluon exchange for a $q\bar{q}$ wave function is now
essential and easily identified in the diagram of
Fig.~\ref{fig:large-t-pert}a.  The integration over $\tvec{k}$ turns
one of the light-cone wave functions into a distribution amplitude.
In the ERBL region the hard-scattering diagrams generate the overlap
of a 4-parton wave function with the 2-parton one as shown in
Fig.~\ref{fig:large-t-pert}b.

\subsubsection{Lessons from impact parameter space}
\label{sub:lessons-impact}

The impact parameter representation of GPDs can be used to motivate
general features of the $t$-dependence in certain limits
\cite{Burkardt:2002hr}.  Consider first the case where $x\to 1$.  For
any parton configuration in a hadron localized at $\tvec{b}$ one has
the constraint $\tvec{b} = x_j \tvec{b}_j + \sum_{i\neq j} x_i
\tvec{b}_i$.  If $j$ denotes the struck parton, the momentum fraction
$x_j$ both in the initial and in the final state hadron goes to $1$,
and the spectator momentum fractions $x_i$ in both hadrons go to zero.
Barring the possibility that the transverse distribution of the
spectators becomes so wide that $\sum_{i\neq j} x_i \tvec{b}_i$ can
give a finite contribution even as all $x_i\to 0$, one finds that the
centers of plus-momentum of both initial and final hadron should
approximately coincide with the position $\tvec{b}_j$ of the fast
struck parton.  The $\tvec{b}$ profile of impact parameter GPDs should
hence become a narrow peak around $\tvec{0}$ for $x\to 1$ and any
$\xi$.  In momentum representation this translates into a flat $t$
dependence of the GPDs in this limit.  Note that the wave function
ansatz (\ref{wf-ansatz-imp}) and its consequence (\ref{wf-link}) do
satisfy this requirement.

As already discussed, the relation (\ref{wf-link}) is not acceptable
in the opposite limit $x\to 0$, where it would give a transverse
extension of partons growing as fast as $\langle \tvec{b}^2(x) \rangle
\sim 1/x$.  Gribov diffusion, i.e., the physical picture of repeated
parton branching at small $x$ as a diffusion process in impact
parameter space, leads to a growth as $\langle \tvec{b}^2(x) \rangle
\sim \log(x_0 /x)$ with a Gaussian profile both in impact parameter
and momentum space \cite{Gribov:1973jg}.  This gives a $t$-dependence
of the form
\begin{equation}
  \label{Gribov-diff}
H(x,0,t) \stackrel{x\to 0}{\sim} 
	\exp\left[ \alpha' \Big(\log\frac{x_0}{x}\Big) \, t \right]
\end{equation}
with constants $\alpha'$ and $x_0$.  One recognizes the interplay of
$t$ and $x$-dependence of Regge theory; we will come back to this
point in Sections \ref{sub:t-depend} and \ref{sec:small-x}.

Dedicated studies of impact parameter dependent GPDs at $\xi=0$ have
recently been performed for the pion within transverse lattice gauge
theory by Dalley \cite{Dalley:2003sz} and within quark models by
Broniowski and Ruiz Arriola \cite{Arriola:2003rp}.


\subsection{Moments}
\label{sub:moments}

Moments of GPDs are important in different respects.  They provide a
particular reduction of the rich information contained in functions of
three kinematical variables, which can be very useful for certain
aspects.  In Section~\ref{sub:impact-forward} we have seen that a
different $t$ dependence of different Mellin moments reflects how
partons are distributed in the transverse plane depending on their
momentum fraction.  The second moments of $H^q$ and $E^q$ (and the
first moments of $H^g$ and $E^g$) are of particular importance since
the associated operator of energy-momentum is of fundamental
importance in field theory and in particular provides access to the
angular momentum sum rule (Section~\ref{sec:spin}).  We have also seen
that conformal moments of GPDs have particularly simple properties
under evolution in leading logarithmic accuracy.

Among the dynamical investigations discussed above, dedicated results
on moments have been given in the MIT bag study \cite{Ji:1997gm}, in
the study of chiral loop contributions \cite{Belitsky:2002jp} and in
the chiral soliton model \cite{Penttinen:1999th}.  In the future one
can expect important contributions from lattice QCD, where moments are
the adequate quantities to be calculated, and which presents a
perspective to obtain results from first principles in QCD.  Lattice
calculations for moments of forward parton distributions have been
pursued for some time, see e.g.~\cite{Negele:2001rb,Gockeler:2002ek}
for recent reviews.  Evaluating matrix elements of local operators
between hadron states of different momentum is also possible using
lattice methods, as has been shown in first studies of the elastic
nucleon form factors $F_1(t)$, $F_2(t)$, $g_A(t)$ for values of $|t|$
up to about $2.5$~GeV$^2$ \cite{Gockeler:2001us,Gockeler:2003ay}.
First lattice studies of the energy-momentum form factors $A_q(t)$,
$B_q(t)$, $C_q(t)$ for $u$ and $d$ quarks in the proton have appeared
recently \cite{Gockeler:2003jf,Hagler:2003jd}.

The moments of GPDs are given as polynomials in $\xi$ whose
coefficients are form factors.  These are the quantities which have
simple analytic behavior in $t$ and are suitable for analytical
continuation to the crossed channel corresponding to GDAs.  Let us see
which quantum number combinations are allowed in the $p\bar{p}$
channel for the various moments of GPDs.  This is relevant for
determining which $t$-channel exchanges of resonances can contribute
to a given GPD, or for writing down dispersion relations for the
moments.  To identify the relevant quantum numbers one has to
analytically continue the form factor decompositions
(\ref{Ji-decomposition}) and (\ref{tilde-decomposition}) to
positive~$t$ (which we write again as $s$),
\begin{eqnarray}
  \label{crossed-decomposition}
\lefteqn{
\langle p(p)\bar{p}(p') |\, 
	\mathcal{O}_q^{\mu \mu_1  \ldots \mu_n} \,|0 \rangle 
}
\nonumber \\
&=& \mathbf{S}\, \bar{u}(p) \gamma^\mu v(p') 
  \sum_{i=0 \atop \scriptstyle{\rm even}}^n
   [A^q_{n+1,i}(s) + B^q_{n+1,i}(s)]\, P^{\mu_1} \ldots P^{\mu_i} \,
	\half\Delta^{\mu_{i+1}} \ldots \half\Delta^{\mu_n}
\nonumber \\
&-& \mathbf{S}\, \frac{\Delta^\mu}{2m}\, \bar{u}(p) v(p') 
  \sum_{i=0 \atop \scriptstyle{\rm even}}^n B^q_{n+1,i}(s)\, 
  P^{\mu_1} \ldots P^{\mu_i} \,
	\half\Delta^{\mu_{i+1}} \ldots \half\Delta^{\mu_n}
\nonumber \\
&+& \mathbf{S}\,\frac{P^{\mu}}{m} \bar{u}(p) v(p') \, 
  \mbox{mod}(n,2)\, C_{n+1}^q(s)\, 
  P^{\mu_1} \ldots P^{\mu_n} ,
\nonumber \\[0.9em]
\lefteqn{
\langle p(p)\bar{p}(p') |\, 
	\widetilde{\mathcal{O}}_q^{\mu \mu_1  \ldots \mu_n} \,|0 \rangle 
}
\nonumber \\
&=& \mathbf{S}\, \bar{u}(p) \gamma^\mu\gamma_5 v(p') 
  \sum_{i=0 \atop \scriptstyle{\rm even}}^n \tilde{A}^q_{n+1,i}(s)\, 
  P^{\mu_1} \ldots P^{\mu_i} \,
	\half\Delta^{\mu_{i+1}} \ldots \half\Delta^{\mu_n}
\nonumber \\
&+& \mathbf{S}\, \frac{P^\mu}{2m}\, \bar{u}(p) \gamma_5 v(p')
  \sum_{i=0 \atop \scriptstyle{\rm even}}^n \tilde{B}^q_{n+1,i}(s)\, 
  P^{\mu_1} \ldots P^{\mu_i} \,
	\half\Delta^{\mu_{i+1}} \ldots \half\Delta^{\mu_n} ,
\end{eqnarray}
where in the first equation we have used the Gordon identities to
trade the tensor current for the vector and scalar currents.  We write
$P = p+p'$ and $\Delta = p-p'$ in the crossed channel, in contrast to
our notation for GPDs.  To find which partial waves can contribute to
these matrix elements, one evaluates the spinor products on the
right-hand side for spinors of definite (usual) helicity in the
$p\bar{p}$ c.m.  Taking the plus-components of
(\ref{crossed-decomposition}) one then finds
\begin{eqnarray}
  \label{crossed-zero}
\lefteqn{
\langle p(p)\bar{p}(p') |\, 
	\mathcal{O}_q^{++ \ldots +} \,|0 \rangle 
= \eta_{\lambda'\lambda}\; (P^+)^{n+1}\, 
	2 \beta^{-1}\, (1-\beta^2)^{1/2}
}
\nonumber \\
&\times& \Bigg\{
  \sum_{i=0 \atop \scriptstyle{\rm even}}^n 
      \Big[A^q_{n+1,i}(s) + \frac{s}{4m^2} B^q_{n+1,i}(s)\Big]\, 
	(\half \beta \cos\theta)^{n-i+1}
\nonumber \\
&& {} + \mbox{mod}(n,2)\, 
  \Big[1 - \frac{s}{4m^2}\Big]\, C_{n+1}^q(s) \Bigg\} ,
\nonumber \\[0.3em]
\lefteqn{
\langle p(p)\bar{p}(p') |\, 
	\widetilde{\mathcal{O}}_q^{++ \ldots +} \,|0 \rangle 
= \tilde{\eta}_{\lambda'\lambda}\;  (P^+)^{n+1} (1-\beta^2)^{1/2}
}
\nonumber \\
&\times& 
  \sum_{i=0 \atop \scriptstyle{\rm even}}^n 
  \Big[\tilde{A}^q_{n+1,i}(s) + \frac{s}{4m^2} \tilde{B}^q_{n+1,i}(s)\Big]\,
	(\half \beta \cos\theta)^{n-i}
\end{eqnarray}
for the case where the helicities of proton and antiproton couple to
$\lambda-\lambda'=0$, and
\begin{eqnarray}
  \label{crossed-one}
\lefteqn{
\langle p(p)\bar{p}(p') |\, 
	\mathcal{O}_q^{++ \ldots +} \,|0 \rangle 
= \eta_{\lambda'\lambda}\; (P^+)^{n+1} 
}
\nonumber \\
&\times&
  \sum_{i=0 \atop \scriptstyle{\rm even}}^n 
      \Big[A^q_{n+1,i}(s) + B^q_{n+1,i}(s)\Big]\, 
	\sin\theta\, (\half \beta \cos\theta)^{n-i} ,
\nonumber \\[0.2em]
\lefteqn{
\langle p(p)\bar{p}(p') |\, 
	\widetilde{\mathcal{O}}_q^{++ \ldots +} \,|0 \rangle 
= \tilde{\eta}_{\lambda'\lambda}\;  (P^+)^{n+1}\, \beta
}
\nonumber \\
&\times& 
  \sum_{i=0 \atop \scriptstyle{\rm even}}^n 
  \tilde{A}^q_{n+1,i}(s)\, \sin\theta\, (\half \beta \cos\theta)^{n-i}
\end{eqnarray}
when they couple to $|\lambda-\lambda'|= 1$.  Here $\beta = (1 - 4
m^2/s)^{1/2}$ is the velocity and $\theta$ the polar angle of the
proton in the $p\bar{p}$ c.m., and $\eta$, $\tilde{\eta}$ are phase
factors which may depend on the azimuthal angle $\varphi$ of the
proton.  One can decompose (\ref{crossed-zero}) and
(\ref{crossed-one}) on partial wave states with total angular momentum
$J$.  They have $J^3=0$ because all Lorentz indices of the operators
were chosen as $+$.  The $\theta$ dependence of the partial waves is
given by the rotation functions $d^J_{J^3,\, |\lambda-\lambda'|}$ (see
e.g.\ \cite{Martin:1970aa}) and hence by
\begin{eqnarray}
d^J_{00} &=& P_J(\cos\theta)
\hspace{12em}  \mbox{for $\lambda-\lambda'=0$} ,
\nonumber \\
d^J_{01} &=& [J(J+1)]^{-1/2}\, \sin\theta\, P'_J(\cos\theta)
\hspace{3.5em} \mbox{for $|\lambda-\lambda'|= 1$} .
\end{eqnarray}
Comparing with the relations (\ref{general-moments}) and
(\ref{general-moments-tilde}) for the moments of the GPDs, one finds
that when continued to positive $t$ they receive contributions from
states with the quantum numbers
\begin{equation}
  \label{waves-even}
\begin{array}[b]{ll}
\displaystyle\int dx\, x^{n} 
	\Big(H^q + \frac{t}{4m^2} E^q\Big) \hspace{2em} & 
    J^{PC} = 1^{--}, 3^{--}, \ldots, (n+1)^{--}
\\
\displaystyle\int dx\, x^{n} (H^q + E^q) &
    J^{PC} = 1^{--}, 3^{--}, \ldots, (n+1)^{--}
\\
\displaystyle\int dx\, x^{n} 
	\Big(\tilde{H}^q + \frac{t}{4m^2} \tilde{E}^q\Big) &
    J^{PC} = 0^{-+}, 2^{-+}, \ldots, n^{-+}
\\
\displaystyle\int dx\, x^{n} \tilde{H}^q & 
    J^{PC} = 1^{++}, 3^{++}, \ldots, (n+1)^{++}
\\
\end{array}
\end{equation}
for $n$ even and 
\begin{equation}
  \label{waves-off}
\begin{array}[b]{ll}
\displaystyle\int dx\, x^{n} 
	\Big(H^q + \frac{t}{4m^2} E^q\Big) \hspace{2em} & 
    J^{PC} = 0^{++}, 2^{++}, \ldots, (n+1)^{++}
\\
\displaystyle\int dx\, x^{n} (H^q + E^q) &
    J^{PC} = 2^{++}, 4^{++}, \ldots, (n+1)^{++}
\\
\displaystyle\int dx\, x^{n} 
	\Big(\tilde{H}^q + \frac{t}{4m^2} \tilde{E}^q\Big) &
    J^{PC} = 1^{+-}, 3^{+-}, \ldots, n^{+-}
\\
\displaystyle\int dx\, x^{n} \tilde{H}^q & 
    J^{PC} = 2^{--}, 4^{--}, \ldots, (n+1)^{--}
\\
\end{array}
\end{equation}
for $n$ odd, where as usual the integrals are over $x\in [-1,1]$.
Corresponding relations hold for the conformal moments
$\mathcal{C}^q_n$, which are just linear combinations of the Mellin
moments with powers $x^{n}$, $x^{n-2}$, etc.  For gluon GPDs the
results are analogous with $x^n$ replaced by $x^{n-1}$.  These quantum
number assignments agree with the results of the general method of
\cite{Ji:2000id}, which does however not identify the relevant
combinations of distributions $H$, $E$ and of $\tilde{H}$,
$\tilde{E}$.  Notice that for $H+E$ and for $\tilde{H}$ no
contributions with $J=0$ appear since the proton helicities in
(\ref{crossed-one}) are coupled to $\pm 1$.  In particular we see that
for the moment $\int dx\, x (H^q+E^q)$ of Ji's sum rule only states
with the quantum numbers $J^{PC} = 2^{++}$ of $f_2$ mesons contribute.

In (\ref{crossed-zero}) and (\ref{crossed-one}) the factors $(\beta
\cos\theta)^{n-i}$ come from $n-i$ powers of $\Delta^+$ in the form
factor decompositions (\ref{crossed-decomposition}).  The partial wave
decomposition we have discussed is closely related to a decomposition
on polynomials in $2\zeta-1 = \beta \cos\theta$, which we have
encountered for the pion GDAs in Section~\ref{sub:gda-evolution}.  In
the proton case the appropriate decomposition is on Legendre
polynomials $P_l$ or their first derivatives $P'{\!}_l$, depending
on the proton helicity combinations.  Under crossing of matrix
elements $\langle p(p)\, \bar{p}(p') |\, \mathcal{O}\,|0
\rangle$ to $\langle p(p')|\, \mathcal{O}\,| p(p)\rangle$ one has to
change
\begin{equation}
(p^+ + p'^+ )^n\, Q_l \Bigg(\frac{p^+ - p'^+}{p^++p'^+}\Bigg) \
\to
(p'^+ - p^+ )^n \, Q_l \Bigg(\frac{p'^+ + p^+}{p'^+-p^+}\Bigg) \
\end{equation}
where $l\le n$ and $Q_l$ is either $P_l$ or $P'{\!}_l$.  An expansion
in $Q_l(2\zeta-1)$ for $t>0$ hence becomes an expansion in
$(-\xi)^{n}\, Q_l(-1/\xi)$ for $t<0$.  For GDAs and GPDs of the pion
more detail can be found in \cite{Polyakov:1999gs}.


\subsection{Ans\"atze for GPDs}
\label{sec:ansatz}

For full-fledged phenomenological studies the dynamical considerations
discussed in the previous section are not sufficient to calculate the
scattering amplitudes of hard processes, which require knowledge of
GPDs for several quark flavors and over the full range of $x$ (or at
least at $x=\pm \xi$).  For this purpose one has to develop general
ans\"atze for GPDs, whose basics we shall now present.  Before
starting we should say that the need to obtain a full description
presently forces one to make simplifications that are likely too
simple to capture important physics features.  Guidance and correction
from comparing with data, and further theoretical insight into the
dynamics of GPDs should lead to improvement.  The following can thus
only be a snapshot, focusing on basics and current problems, and some
of it will hopefully be outdated in the future.

The principal input for making an ansatz for GPDs is our knowledge of
the ``boundary conditions'' on GPDs provided by the forward parton
densities and also by the elastic form factors.  It is however clear
that important pieces of information about GPDs cannot be obtained
from knowing their forward limits.  In particular there are
contributions in the ERBL region that become ``invisible'' when $\xi$
is taken to zero.

\subsubsection{Exchange contributions}
\label{sub:exchange}

\begin{figure}[b]
\begin{center}
	\leavevmode
	\epsfxsize=0.5\textwidth
	\epsfbox{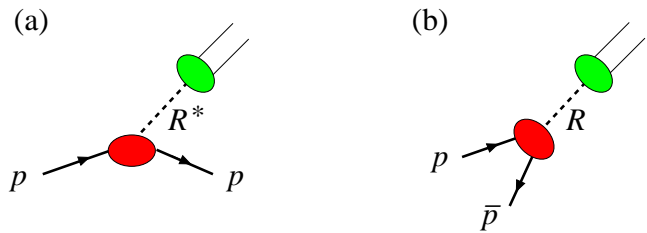}
\end{center}
\caption{\label{fig:resonance-xchange} (a) Resonance exchange
contribution to a GPD and (b) its analytic continuation to $t \sim
m_R^2$ in the GDA channel.}
\end{figure}

Among these contributions is for instance the pion pole contribution
to $\tilde{E}^{u-d}$ we have discussed in Section~\ref{sub:chiral}.
More generally one can think of the exchange of any resonance in the
$t$-channel, which then annihilates into a parton pair, see
Fig.~\ref{fig:resonance-xchange}a.  The corresponding contribution to
a generic GPD $f$ has the form
\begin{equation}
  \label{resonance-xchange}
f(x,\xi,t) = p_{n}\Big( \frac{1}{\xi} \Big)\, 
	\phi\Big( \frac{x}{\xi}, t \Big) ,
\end{equation}
where we have assumed $\xi>0$ for simplicity.  $\phi(u,t)$ has support
in $u\in [-1,1]$ and $p_{n}$ is a polynomial of order $n$.
Polynomiality requires that the moments $\int dx\, x^m f(x,\xi,t) =
\xi^{m+1} p_n(1/\xi) \int du\, u^m \phi(u,t)$ vanish for $m<n-1$.  The
identification of (\ref{resonance-xchange}) as the contribution from a
resonance with mass $m_R$ and width $\Gamma_R$ becomes clear when one
analytically continues $t$ to the vicinity of $m_R^2$, see
Fig.~\ref{fig:resonance-xchange}b.  According to our discussion in
Section~\ref{sub:resonance}, $\phi(u,t)$ should then be given by the
DA of the resonance times a Breit-Wigner propagator $(t - m_R^2 +
i\Gamma_{\!R}\, m_R)^{-1}$ times a factor describing the coupling of
the resonance to the nucleon.  The pion pole contribution
(\ref{pion-pole}) to $\tilde{E}$ has exactly this form (with zero
width of course).  The spin of the resonance determines the form of
the polynomial $p_n$ according to the partial wave decomposition
presented in Section~\ref{sub:moments}.

For describing GPDs in their physical region $t<0$ one has to keep in
mind that, except for the pion with its small mass, there is a large
interval between $t$ and squared resonance masses, so that one would
not expect any single resonance to dominate.  As we remarked in
Section~\ref{sub:chiral-soliton}, even the pion pole contribution
(\ref{pion-pole}) may receive non-negligible corrections at moderate
$t$.  The issues here are similar to those in modeling the Dirac and
Pauli form factors by vector meson dominance.

One readily sees from the hard-scattering kernel in
(\ref{Born-convolution}) that a single resonance term of the form
(\ref{resonance-xchange}) gives a contribution to the real part of a
scattering amplitude which grows like $\xi^{-n}$ at small $\xi$.
Unless $n$ is small enough, this is an unacceptably steep behavior.
This problem is reminiscent of Regge theory, where it is the
\emph{coherent sum} over an infinite number of $t$-channel exchange
terms that ensures a proper behavior of amplitudes in the high-energy
limit (corresponding to small $\xi$ in our context), see
e.g.~\cite{Collins:1977jy}.  Polyakov and Shuvaev
\cite{Polyakov:2002wz} have recently proposed a parameterization of
GPDs based on their representation as an infinite sum of $t$-channel
exchanges of the form (\ref{resonance-xchange}).  Taking for example a
pion target, one starts with the double expansion
(\ref{Polyakov-expansion}) of $\Phi^q$ in in a series of $\smash{
C^{3/2}_n(2z-1) }$ and $P^{\phantom{1}\!\!}_l(2\zeta-1)$ and obtains a
series representation of $H^q_\pi$ by analytic continuation, as was
already done in \cite{Polyakov:1999gs}.  The strategy in
\cite{Polyakov:2002wz} is to resum subsets of terms in this series in
order to obtain a simple approximation of GPDs for small $\xi$.  The
methods employed for this are similar to those in the Shuvaev
transform (Section~\ref{sub:Shuvaev}).  It will be interesting to see
the further development of this proposal.

Extending the ideas just discussed, one may also consider $t$-channel
exchange of continuum states instead of resonances.  An important
example is two-pion exchange, see the next subsection.

\subsubsection{The $D$-term}
\label{sub:model-d-term}

Another contribution which is invisible in the forward limit is the
$D$-term, whose theory we have presented in Section~\ref{sub:d-term}.
An input used in many recent phenomenological analyses is the
isoscalar combination $D^{u+d}$ obtained in the chiral soliton model
for the nucleon (Section~\ref{sub:chiral-soliton}) or an extension to
the flavor SU(3) singlet combination $D^{u+d+s}$ by Goeke et
al.~\cite{Goeke:2001tz}.  In accordance with the expansion scheme of
the model, isovector and gluon $D$-terms are neglected.  Kivel et
al.~\cite{Kivel:2000fg} have extrapolated the numerical result
obtained in \cite{Petrov:1998kf} to $t=0$ and fitted to Gegenbauer
polynomials as in (\ref{D-Gegenbauer}).  They provide the lowest
coefficients $d_1^{u+d} \approx -4.0$, $d_3^{u+d} \approx -1.2$,
$d_5^{u+d} \approx -0.4$, with higher coefficients being small.  We
remark that these values refer to the low factorization scale $\mu_0
\approx 600$~GeV of the chiral soliton model and become significantly
smaller at larger scales, given the large size of $\alpha_s$ at small
$\mu$.  Assuming a zero $D$-term of the gluon at the input scale and
evolving up to $\mu^2=5$~GeV$^2$ reduces for instance $d_1^{u+d}$ to
the value $-2.9$ in leading logarithmic approximation
\cite{Berger:2001xd} (see also~\cite{Belitsky:2001ns}).

Notice that according to the crossed-channel relations
(\ref{crossed-zero}) and (\ref{crossed-one}) the only $\theta$
independent terms for the parity even operators are the $C_{n}$.  They
are resummed in the $D$ term contribution to the distributions $H$ and
$E$.  Exchanges with quantum numbers $J^{PC} = 0^{++}$ are thus
exclusively contained in the $D$-term, but not in the double
distributions, which are related to the form factors $A_{n,i}$ and
$B_{n,i}$.  Exchanges with higher even spin contribute both to the
$D$-term and to double distributions, and odd-spin exchange
contributes only to the double distributions.  An explicit discussion
for the case of pion GPDs is given in \cite{Polyakov:1999gs}.

\begin{figure}[b]
\begin{center}
	\leavevmode
	\epsfxsize=0.2\textwidth
	\epsfbox{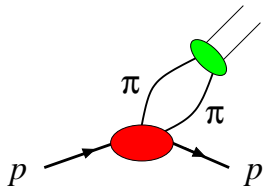}
\end{center}
\caption{\label{fig:d-term} A two-pion exchange graph contributing to
the $D$ term in the chiral quark-soliton model.}
\end{figure}

{}From a dynamical point of view, $D^{u+d}$ in the chiral soliton
model receives important contributions from diagrams as shown in
Fig.~\ref{fig:d-term}.  In this framework the $D$-term is mainly
generated from the ``pion cloud'' of nucleon.  In different terms one
may view the corresponding contributions as due to the $t$-channel
exchange of two pions in partial waves $0^{++}$, $2^{++}$, \ldots.

\subsubsection{Double distributions and connection to forward
densities} 
\label{sub:connect-forward}

Ans\"atze for GPDs used in the literature so far are typically taken
as an ``extrapolation'' of the forward densities to nonzero skewness
$\xi$ and nonzero $t$, possibly supplemented with the quark $D$-term
contribution just discussed.  This strategy is of course only possible
for the distributions $H$ and $\tilde{H}$, whose forward limits are
the known parton densities.  The modeling of $E$ and $\tilde{E}$ does
not have this guide and most often these distributions are neglected
in phenomenological studies, with the exceptions of the $D$-term
contribution to $E^{u+d}$ and the pion pole contribution
(\ref{pion-pole}) to $\tilde{E}^{u-d}$ (sometimes including the
corrections to the pion pole obtained in the chiral soliton model, see
Section~\ref{sub:chiral-soliton}).  It is clear that this is not
always appropriate: the $C$-odd part of $E^{u-d}$ for instance cannot
be small since at $t=0$ its integral over $x$ gives the large
isotriplet anomalous magnetic moment $\kappa^u - \kappa^d
\approx 3.7$.  For many observables, however, the contributions from
$E$ or $\tilde{E}$ are suppressed compared with those from $H$ or
$\tilde{H}$ by small kinematic prefactors $\xi^2$ or $t/ (4m^2)$, see
e.g.\ (\ref{gpd-compton-unpol}) and (\ref{gpd-inter-unpol}).  A first
attempt of more sophisticated modeling for $E^q$ has been made by
Goeke et al.~\cite{Goeke:2001tz}, with general features obtained in
the chiral soliton model as a guide.

A direct ansatz for GPDs has in general little chance to satisfy the
polynomiality constraints, so that the favored strategy in the
literature is to make instead an ansatz for double distributions,
which automatically leads to GPDs satisfying polynomiality.  Notice
that double distributions are indeed ``distributions'' in the
mathematical sense and need not be smooth functions.  An example of a
singular term in double distributions is $\delta^{(n)}(\beta)
\, h(\alpha,t)$ with $n\ge 0$.  It leads to GPDs of the form
(\ref{resonance-xchange}) with $p_n(1/\xi) = 1/ \xi^{n+1}$ and
$\phi(\alpha,t) = \partial^n /(\partial \alpha)^n\, h(\alpha,t)$,
which only has support in the ERBL region and is invisible in the
forward limit.  In models, such contributions concentrated at
$\beta=0$ may just be added to a double distribution with support in
the full region $|\beta| + |\alpha| \le 1$, which generates both the
forward parton densities and a contribution to GPDs in the DGLAP and
ERBL regions.

The connection with forward densities can of course only be made at
$t=0$, and in this section we will focus on the modeling of the
interplay between the $x$ and $\xi$ dependence, leaving the discussion
of the $t$-dependence for the next section.  We will thus suppress the
$t$ dependence of all functions, which may be understood as taken at
$t=0$.  Notice that for $\xi\neq 0$ this point is outside the physical
region due to the constraint $t \le - 4\xi^2 m^2 /(1-\xi^2)$, so that
the corresponding functions are to be understood in the sense of
analytic continuation.

To have a simple notation for the forward limit it is useful to extend
the definition of quark and gluon distributions to the full interval
$x\in [-1,1]$ by
\begin{eqnarray}
  \label{all-x}
q(-x)       &\stackrel{\mathrm{def}}{=}& -\bar{q}(x) , \qquad 
\Delta q(-x) \stackrel{\mathrm{def}}{=}  \Delta \bar{q}(x) ,
\nonumber \\
g(-x)        &\stackrel{\mathrm{def}}{=}&  -g(x) , \qquad 
\Delta g(-x)  \stackrel{\mathrm{def}}{=}   \Delta g(x)
\end{eqnarray}
for $x>0$.  Without assumption one can write the double distributions
as
\begin{equation}
  \label{profile-def}
f^q(\beta,\alpha) = q(\beta)\, h^q(\beta,\alpha) , \qquad
f^g(\beta,\alpha) = g(\beta)\, h^g(\beta,\alpha) ,
\end{equation}
with analogous forms for $\tilde{f}^{q,g}$, where the ``profile
functions'' $h^{q,g}$ are normalized to
\begin{equation}
  \label{profile-norm}
\int_{|\beta| -1}^{1- |\beta|} d\alpha\, h(\beta,\alpha) =1
\end{equation}
in order to produce the correct forward limit of the GPDs.  An ansatz
due to Radyushkin is to take the same $\alpha$ shape of the profile
function for all $\beta$, up to a rescaling by the length of the
integration interval in (\ref{profile-norm}),
\begin{equation}
  \label{special-profile}
h(\beta,\alpha) = \frac{1}{1-|\beta|} \, 	
   \rho\Big( \frac{\alpha}{1-|\beta|} \Big) 
\end{equation}
with $\rho$ normalized to $\int_{-1}^1 d\alpha\, \rho(\alpha) = 1$.  A
particular form for this function is~\cite{Musatov:1999xp}
\begin{equation}
  \label{profile}
\rho^{(b)}(\alpha) = \frac{\Gamma(2b+2)}{2^{2b+1}\, \Gamma^2(b+1)}\, 
	( 1 - \alpha^2 )^b ,
\end{equation}
where $b$ is a parameter that remains to be chosen.

One may refine this type of ansatz by separating the quark double
distributions into different components and choosing different profile
functions for each of them.  
\begin{enumerate}
\item A natural decomposition, used for instance in
\cite{Radyushkin:1998bz}, is into the $C$-even and $C$-odd
combinations $f^{q(+)}$, $\tilde{f}^{q(+)}$ and $f^{q(-)}$,
$\tilde{f}^{q(-)}$ defined in (\ref{dd-definite-charge}).
\item One can also separate the double distributions into
contributions from positive and negative $\beta$, i.e.\ $f^q = f^{>} +
f^{<}$ with $f^{>}(\beta,\alpha) = \theta(\beta) f^q(\beta,\alpha)$ and
$f^{<}(\beta,\alpha) = \theta(-\beta) f^q(\beta,\alpha)$.  Integration of
$f^{>}$ and $f^{<}$ over $\alpha$ respectively gives the quark and and
antiquark densities.  The respective contributions to GPDs have
support in the regions $x\in [-\xi,1]$ and $x\in [-1,\xi]$, both
including the ERBL region.  This corresponds to Radyushkin's
separation into ``quark'' and ``antiquark'' distributions discussed
briefly in Section~\ref{sec:double-d}, but has not been used for
modeling in the literature.
\item A hybrid of the previous schemes has been used in
\cite{Goeke:2001tz,Belitsky:2001ns}, decomposing $f^q =
f^q_{\mathrm{val}} + f^q_{\mathrm{sea}}$ with
\begin{eqnarray}
  \label{valence-sea}
f_{\mathrm{val}}(\beta,\alpha) &=& \Bigg\{
\begin{array}{cl}
f(\beta,\alpha) + f(-\beta,\alpha)  & \mbox{~for $\beta>0$} \\
0                  & \mbox{~for $\beta<0$}
\end{array}  
\nonumber \\
f_{\mathrm{sea}}(\beta,\alpha) &=& \Bigg\{
\begin{array}{cl}
-f(-\beta,\alpha) & \hspace{2em} \mbox{for $\beta>0$} \\
f(\beta,\alpha)   & \hspace{2em} \mbox{for $\beta<0$}
\end{array}
\end{eqnarray}
Integration of $f_{\mathrm{val}}$ and $f_{\mathrm{sea}}$ over $\alpha$
for $\beta > 0$ respectively gives the familiar valence and sea quark
distributions $q_{\mathrm{val}} = q - \bar{q}$ and $q_{\mathrm{sea}} =
\bar{q}$, which explains the naming scheme.  Notice however that
$f_{\mathrm{val}}$ generates GPDs with support $x\in [-\xi,1]$
including the ERBL region.  This clearly describes physics beyond the
simple idea of a distribution due to ``the three valence quarks'' in
the nucleon.  An analogous decomposition can be made for the polarized
double distributions $\tilde{f}^q$, replacing $f(\beta,\alpha) \to
\tilde{f}(\beta,\alpha)$ and $f(-\beta,\alpha)
\to -\tilde{f}(-\beta,\alpha)$ in (\ref{valence-sea}).
\end{enumerate}

Let us return to the choice of $b$ in the profile function
(\ref{profile}).  In the limit $b\to \infty$ one obtains a simple
profile $h^{(\infty)}(\beta,\alpha) = \delta(\alpha)$ and
corresponding GPDs
\begin{eqnarray}
  \label{super-simple}
H^q(x,\xi)        &=& q(x) , \qquad \phantom{x}
\tilde{H}^q(x,\xi) =  \Delta q(x) ,
\nonumber \\
H^g(x,\xi)        &=&  x g(x) , \qquad
\tilde{H}^g(x,\xi) =  x \Delta g(x) 
\end{eqnarray}
with no effect of skewness at all.  This may be the only consistent
ansatz for GPDs one can make without resort to double distributions.
It is however questionable from a physics point of view.  In the
global parton analyses the quark distributions $q(x)$ and $\Delta
q(x)$ come out singular at $x=0$.  For the gluon many
parameterizations have a singular behavior of $x g(x)$ at $x=0$,
unless one takes a rather low factorization scale.  A singular
behavior of a GPD at $x=0$ and nonzero $\xi$ is difficult to
understand physically, since this is the point in the ERBL region
where the two partons have equal and \emph{finite} plus-momenta.  This
is in contrast to the forward case $\xi=0$, where $x=0$ entails
\emph{zero} plus-momentum of both partons, which is indeed a singular
situation.  Note also that DGLAP evolution does not drive GPDs at
nonzero $\xi$ to become singular at $x=0$.

A choice made in many studies is to take $b=1$ for quark and $b=2$ for
gluon distributions~\cite{Radyushkin:1998bz}.  This is motivated by
the interpretation of $\alpha$ as a DA-like variable and the fact that
the asymptotic form of distribution amplitudes under evolution is like
$(1-\alpha^2)$ for quarks and $(1-\alpha^2)^2$ for gluons
(corresponding to $z(1-z)$ and $z^2 (1-z)^2$ in terms of the momentum
variable we use for distribution amplitudes, see
Section~\ref{sub:radon}).  Taking a simple analytic ansatz for the
forward quark distribution $q(x) = c x^a (1-x)^3$ allows analytic
evaluation of the integral necessary to obtain the GPD
\cite{Radyushkin:1998bz,Musatov:1999xp}.  Among the characteristic
features of the result is a continuous but nonanalytic behavior at the
points $x= \pm\xi$, both for $H^{q(+)}$ and $H^{q(-)}$.  In the model
defined by (\ref{profile-def}) to (\ref{profile}) the behavior of the
GPDs at these points is strongly influenced by the behavior of the
double distributions at the upper ``tip'' of the rhombus
$|\beta|+|\alpha|\le 1$ and thus by the behavior of the forward parton
densities at $x\to 0$.  We recall that in parameterizations of parton
densities the small-$x$ behavior is usually parameterized by a power
law,
\begin{equation}
  \label{small-x-power}
x f(x) \sim x^{-\lambda} .
\end{equation}
Typical values of $\lambda$ at factorization scale $\mu=1$~GeV are
slightly above zero for $f=\bar{q}$, and below zero for $f= q-\bar{q}$
and the polarized quark valence and sea distributions, see e.g.\
\cite{Martin:2002dr,Goto:1999by,Leader:2001kh}.  The behavior for
gluons at this scale is rather uncertain, with $\lambda$ of either
sign for $f=g$ and clearly below zero for $f=\Delta g$.  At the low
scales $\mu \sim 500$~MeV of the GRV and GRSV parameterizations
$\lambda$ is negative for all distributions, and smaller than $-1$ for
unpolarized and polarized gluons \cite{Gluck:1998xa,Gluck:2000dy}.

With a behavior (\ref{small-x-power}) the first derivative of GPDs in
the model (\ref{profile-def}) to (\ref{profile}) can become infinite.
Starting from (\ref{dd-red}) and trading $\partial/(\partial x)$ for
$\partial/(\partial \alpha)$ under the integral one finds
\begin{eqnarray}
\frac{\partial}{\partial x}  H^q(x,\xi) 
 \stackrel{x \to \xi}{\sim}
   - \int_{|x-\xi|} d\beta\, \beta^{b-2-\lambda} ,
\end{eqnarray}
which is divergent for $\lambda \ge b-1$ (see also
\cite{Berger:2001xd}).  This is in particular relevant for the
$C$-even combination $H^{q(+)}$ and a profile function with $b=1$.
Note that the principal-value integral (\ref{Born-convolution})
appearing in the real part of scattering amplitudes is very sensitive
to such a steep behavior.  For gluons, one has instead
\begin{equation}
\frac{\partial}{\partial x}  H^g(x,\xi)
\stackrel{x \to \xi}{\sim} 
   - \int_{|x-\xi|} d\beta\, \beta^{b-1-\lambda} ,
\end{equation}
which for practically relevant cases leads to a finite and continuous
first derivative of $H^g$ at $x=\xi$.

The sensitivity of the above model to forward parton densities at
small momentum fraction is not restricted to the points $x= \pm \xi$.
The integral in the $\beta$-$\alpha$ plane shown in
Fig.~\ref{fig:dd-sup} involves values of $|\beta|$ down to
$x_{\mathrm{out}} = |x-\xi| /(1-\xi)$ in the DGLAP regions.  In the
ERBL region the integration is over a region around $\beta = 0$ and
thus even involves parton densities down to $x=0$.  With $b=1$ for
quark distributions, the integral giving the GPD in the ERBL region
has an integrand behaving like $\beta^{-(1+\lambda)}$ for quark
densities going like (\ref{small-x-power}).  For the $C$-even
unpolarized combinations, where $\lambda>0$ at moderate scales $\mu$,
this is a non-integrable singularity, but since
$f^{q(+)}(\beta,\alpha)$ is odd in $\beta$ the corresponding integral
can be taken using the principal value prescription.  Equivalently one
can use the antisymmetry of $f^{q(+)}(\beta,\alpha)$ to rewrite the
integrand so as to have a small-$\beta$ behavior as
$\beta^{-\lambda}$, which is integrable
\cite{Lehmann-Dronke:1999ym,Berger:2001xd}.  The fact remains that in
this model GPDs at finite $\xi$ depend on forward densities at $x \ll
\xi$.  This was quantitatively investigated for $H^{q(+)}$ and a
profile parameter $b=1$ in \cite{Berger:2001xd,Freund:2002qf}.
Although differing in detail, both studies found a sensitivity of the
model GPDs to the sea quark density at $x$ values one or more orders
of magnitude below $\xi$, especially for the region $x\sim \xi$ of the
GPDs, which has strong impact on the corresponding hard-scattering
amplitudes.  For modeling GPDs at very small $\xi$ this drives one
into a region where the forward densities are unknown.  We note that
for the unpolarized gluon GPD, where the small-$x$ power $\lambda$ is
also positive, the corresponding sensitivity is much less since
$f^g(\beta,\alpha)$, and hence $g(\beta)$ in the model, enters the
reduction formula (\ref{dd-red-glue}) with an extra factor~$\beta$.

One may expect some sensitivity of GPDs to the physics of parton
momentum fractions $x$ below $\xi$.  Unlike forward parton
distributions at small $x$, GPDs around $x \approx \xi$ involve
however small momentum fractions in the initial but \emph{not} in the
final state wave function, so that a strong link between GPDs at
finite $\xi$ and parton densities at $x \ll \xi$ is not obvious to
understand physically.  {}From a different point of view, double
distributions in some narrow band of $\beta$ around zero on one hand
generate the small-$x$ behavior of forward densities.  They hence must
be large somewhere in this band, since the forward densities have a
strong rise towards $x=0$.  On the other hand the reduction of double
distributions to GPDs also involves this narrow band, even for $\xi$
much larger than the width of the band.  How important the influence
of this band is depends on the detailed shape of the profile function
$h(\beta,\alpha)$ in (\ref{profile-def}), and it is not clear whether
an $\alpha$ dependence only via the rescaled variable $\alpha /
(1-|\beta|)$ as in (\ref{special-profile}) is a viable representation
of the relevant physics.

The particular ansatz (\ref{profile-def}) to (\ref{profile}) must be
made at a specific factorization scale $\mu$.  A study of $H^g$ and
$H^{q(\pm)}$ for $\xi\approx 0.05$ in \cite{Musatov:1999xp} compared
the results obtained when either $(i)$ making the profile ansatz at
$\mu^2=1.5$~GeV$^2$ and evolving the obtained GPDs up to $\mu^2=
20$~GeV$^2$ or $(ii)$ directly making the profile ansatz with the same
forward densities evolved to $\mu^2= 20$~GeV$^2$.  For $b$ between 1
and 2 little deviation between the two versions was found.  We note
that the problem of sensitivity to the small-$x$ limit of the forward
densities is alleviated when making this ansatz at very small scale
$\mu$, where parton densities are less singular at small $x$
\cite{Weiss:2001pr}.  It would be interesting to know how evolution to
larger scale then changes the double distribution and its profile
function.

\subsubsection{Dependence on $t$}
\label{sub:t-depend}

An adequate description of the $t$ dependence of GPDs is essential in
all physical applications, because physical processes are measured for
$t \le t_0$ and the small-$t$ falloff of GPDs is in general expected
to be similarly steep as for elastic form factors.  The most widely
used ansatz of the $t$ dependence has so far been a factorized form
$F(t) f(x,\xi)$ for the GPD, with the form factors $F(t)$ chosen to
guarantee the correct sum rules (\ref{basic-sum-rules}).  In
particular cases one can make contact with experimentally known form
factors and has \cite{Guichon:1998xv}
\begin{eqnarray}
  \label{factorized-t}
H^u(x,\xi,t) &=& \frac{1}{2}F_1^u(t)\, H^u(x,\xi,0)  , \qquad
H^d(x,\xi,t)  =  F_1^d(t)\, H^d(x,\xi,0)  , 
\nonumber \\
\tilde{H}^q(x,\xi,t) &=& \frac{g_A(t)}{g_A(0)}\, \tilde{H}^q(x,\xi,0) , 
\qquad \mbox{for $q = u,d$} .
\end{eqnarray}
The GPDs continued to $t=0$ can then be modeled along the lines
described in the previous subsection.  Neglecting the small $s$ quark
contribution to the elastic nucleon form factors one can use isospin
invariance to obtain
\begin{equation}
  \label{nucleon-dirac}
F_1^u(t) = 2 F_1^p(t) + F_1^n(t) , \qquad 
F_1^d(t) = 2 F_1^n(t) + F_1^p(t) ,
\end{equation}
and use the electromagnetic form factors $F_1^p$, $F_1^n$ of proton
and neutron, which are quite well measured at small to intermediate
$t$ (see e.g.~\cite{Mergell:1996bf,Brash:2001qq}).  For the axial form
factor the above ansatz neglects again the $s$ quark contribution in
the nucleon and assumes that the isoscalar combination $g_A^{u+d}(t)$
has the same dependence on $t$ as the isovector one,
\begin{equation}
g^{\phantom{-}}_A(t) = g_A^{u-d}(t) .
\end{equation}
The latter is accessible in neutral current scattering, and also in
charged current nucleon transitions via the isospin relation
(\ref{isospin-pn}).  It is experimentally known up to $|t|$ of about
1~GeV$^2$ \cite{Bernard:2001rs}.  

An ansatz analogous to (\ref{factorized-t}) can be made for $E^q$,
with $F_1^q$ replaced by the Pauli form factor $F_2^q$.  In this case
one can of course not proceed modeling by using the forward limit.  A
form sometimes used is to set
\begin{eqnarray}
  \label{factorized-t-E}
E^u(x,\xi,t) &=& \frac{1}{2}F_2^u(t)\, H^u(x,\xi,0)  , \hspace{2em}
E^d(x,\xi,t)  =  F_2^d(t)\, H^d(x,\xi,0)  , 
\end{eqnarray}
which has not much motivation beyond simplicity.  In
(\ref{factorized-t}) and (\ref{factorized-t-E}) it is understood that
the factorized form does not include the $D$-term, which does not
appear in the Mellin moment giving the elastic form factors and has of
course the same $t$-dependence in $H$ and in $E$.  Here one typically
assumes a factorized form $D^{u+d}(x,\xi,t) = F(t)\,
D^{u+d}(x,\xi,0)$, with some choice for $F(t)$ and the term for $t=0$
taken as described in Section~\ref{sub:model-d-term}.  A factorized
ansatz as in (\ref{factorized-t}) can of course also be directly
written down at the level of double distributions.
 
Ans\"atze analogous to (\ref{factorized-t}) have also been made for
the gluon distributions $H^g$ and $\tilde{H}^g$, where one must model
the unmeasured form factors of the local currents $G^{\mu\alpha}\,
G_{\mu}{}^\beta$ and $G^{\mu\alpha}\,\tilde{G}_{\mu}{}^\beta$, see
e.g.~\cite{Lehmann-Dronke:1999ym,Belitsky:2001ns}.  We remark that in
processes at small $\xi$ the $t$ dependence is often described in a
different way, using an exponential form $\exp(b t)$ for the relevant
gluon or quark GPDs which will be briefly discussed in
Section~\ref{sec:small-x-t}.

The ansatz (\ref{factorized-t}) has the virtue of being simple and not
introducing additional free parameters.  It can probably describe the
overall drop in $t$ of scattering amplitudes at moderate $\xi$, but is
likely to fail in situations when different contributions to an
observable have different signs and can partially cancel, and in
observables sensitive to the interplay between $t$ and $\xi$.  {}From
a theoretical point of view it is for instance clear that the
electromagnetic and weak form factors in (\ref{factorized-t}) can only
constrain the $t$-dependence of $H^{q(-)}$ and $\tilde{H}^{q(+)}$, but
not of $H^{q(+)}$ and $\tilde{H}^{q(-)}$.  Along similar lines of
reasoning, Belitsky et al.\ \cite{Belitsky:2001ns} have taken separate
factorized expressions for the ``valence'' and ``sea'' components of
GPDs defined by (\ref{valence-sea}), with different form factors in
the valence and sea sectors.  Notice that for strange quarks an ansatz
like (\ref{factorized-t}) is clearly inadequate: since the proton has
no net strangeness the vector form factor $F_1^s(t)$ vanishes at
$t=0$, whereas the $C$-even combination $H^{s(+)}(x,0,0) = s(x) +
\bar{s}(x)$ in the forward limit is nonzero.  Not even for $H^{s(-)}$
would a factorized form be adequate beyond a certain accuracy, since
$s(x) - \bar{s}(x)$ has a vanishing lowest Mellin moment and is
overall small but does not vanish.

More generally one expects that the $t$ dependence of GPDs, or
equivalently the transverse spatial distributions of partons, will
depend on whether the partons are slow or fast.  This expectation is
borne out by explicit dynamical considerations, as we have discussed
for contributions from chiral logarithms (Section~\ref{sub:chiral}),
from the chiral soliton model (Section~\ref{sub:chiral-soliton}), from
the overlap of wave functions at large $x$
(Section~\ref{sub:soft-overlap-gpd}), or from the impact parameter
representation in the limits of small or large $x$
(Section~\ref{sub:lessons-impact}).  Factorization of the $t$
dependence from $x$ and $\xi$ does occur for particular contributions
or special limits: the pion pole contribution (\ref{pion-pole}) for
$t\to m_\pi^2$ has a global factor $(t - m_\pi^2)^{-1}$ and the
deviations from the pole form found in the chiral soliton model induce
little correlation between $t$ and $x / \xi$ \cite{Penttinen:1999th}.
Another example is the global power behavior in $t$ in the large-$t$
limit we saw in Section~\ref{sub:large-t-limit}, valid for $x$ not too
close to $1$.

First attempts to incorporate a correlation between $t$ and the
longitudinal variables in a practicable way have been made, with focus
on the simpler case $\xi=0$ so far.  As seen in (\ref{wf-link}) and
(\ref{Gribov-diff}), global factors
\begin{equation}
  \label{t-exponentials}
F_L(x,t) = \exp\left[ \frac{a^2}{2}\, \frac{1-x}{x}\, t \right]  , 
\qquad
F_R(x,t) = \exp\left[ \alpha' \Big(\log\frac{x_0}{x}\Big) \, t \right]
\end{equation}
for the $t$ dependence of the GPDs are motivated by dynamical
considerations in the respective limits of large and small $x$ (note
that neither form is stable under DGLAP evolution and must be taken at
a definite scale $\mu$).  Combined with a power-law dependence
(\ref{small-x-power}) of the forward quark or gluon densities at small
$x$, the form $F_R(x,t)$ can be seen as the extension to finite $t$ of
a behavior like $(x_0/x)^{\alpha(t)}$, with the form $\alpha(t) =
\alpha(0) + \alpha' t$ familiar from Regge theory (see
Sections~\ref{sec:small-x-gpd} and \ref{sec:small-x}).  In the chiral
soliton model such a behavior was reported to give a qualitative
description of $H^{u+d}$ and $\tilde{H}^{u-d}$
\cite{Goeke:2001tz,Penttinen:1999th}, with values $\alpha' \approx
0.8$~GeV$^{-2}$ and $x_0=1$ quoted for $H^{u+d}$ in
\cite{Goeke:2001tz}.  This value of $\alpha'$ is remarkably close to
the slope parameter for meson Regge trajectories in hadronic
collisions.  Notice that accidentally the large-$x$ behavior of
$F_R(x,t)$ with $x_0=1$ is $\exp[\, \alpha' (1-x)\, t]$ and hence the
same as for $F_L(x,t)$ \cite{Burkardt:2002hr}.  The constants
$\alpha'$ and $\half a^2$ in the respective forms have however very
different physical origins (the first being related to Fock states
with large parton number and the second to the lowest Fock states with
three quarks and possibly a few extra gluons or $q\bar{q}$ pairs).
The study \cite{Diehl:1998kh} found a value of $\half a^2 \approx
0.5$~GeV$^{-2}$ or smaller for the large-$x$ form $F_L(x,t)$ at scale
$\mu \approx 1$~GeV, which according to the considerations of Vogt
\cite{Vogt:2000ku} is expected to increase when evolving to the
lower scale $\mu \approx 600$~MeV of the chiral quark-soliton model,
so that the values of $\alpha'$ and $\half\, a^2$ may be accidentally
rather close to each other.  Further investigation would be required
to see whether the simple form $F_R(x,t)$ with $x_0=1$ can be used as
a quantitatively adequate ansatz for both small and large $x$.  Goeke
et al.~\cite{Goeke:2001tz} ``promoted'' this form to an ansatz
$f^q(\beta,\alpha,t) = h^q(\beta,\alpha)\, q(\beta)\,
|\beta|^{-\alpha' t}$ for a double distribution with a nontrivial
interplay between $t$ and the other variables.

A nontrivial $t$ dependence was obtained in a study for the pion by
Mukherjee et al.~\cite{Mukherjee:2002gb}.  Starting point was the
$q\bar{q}$ wave function obtained in a dynamical model.  This wave
function, having a power-law falloff as $|\tvec{k}|^{-2\kappa}$ at
large $|\tvec{k}|$ with $\kappa \approx 2$, was used to calculate the
quark GPD at $\xi=0$ from the overlap representation
(\ref{DGLAP-mom}).  The result was then rewritten in the form
$H_\pi^q(x,0,t) = \int d\alpha\, f^q(x,\alpha,t)$, where the integrand
had support for $|\alpha| \le 1-|x|$ and thus could be interpreted as
a double distribution.  At $t=0$ the form of $f^q$ is found to
coincide with the profile function ansatz (\ref{profile-def}) to
(\ref{profile}) with $b=\kappa - 1$.  The $t$ dependence does however
\emph{not} factorize from the dependence on $\beta$ and $\alpha$, and
at nonzero $t$ the shape in $\alpha$ depends on $\beta$ beyond the
simple rescaling in (\ref{special-profile}).  We note that the
procedure just described gives a result for $H_\pi^q(x,\xi,t)$ that
differs from the one obtained in the overlap representation at nonzero
$\xi$ with the same light-cone wave function.  As pointed out in
\cite{Tiburzi:2002kr} the result of \cite{Mukherjee:2002gb} actually
violates the positivity bound (\ref{pion-quark-pos}), which a GPD
obtained from the overlap formula respects by construction.  Tiburzi
and Miller \cite{Tiburzi:2002tq} have clarified the origin of this
mismatch: from $H_\pi^q(x,0,t)$ one can construct a double
distribution $f^q(\beta,\alpha,t)$ but not the second function
$g^q(\beta,\alpha,t)$ in the decomposition (\ref{general-pion-dd}),
which according to the reduction formula (\ref{general-red}) is
invisible in the GPD at $\xi=0$.  Without knowing both functions one
can however not calculate $H_\pi^q(x,\xi,t)$ unambiguously.  The
procedure in \cite{Mukherjee:2002gb} tacitly assumed $g^q = 0$ when
calculating $H_\pi^q(x,\xi,t)$ from $f^q$ alone, which no longer
guarantees consistent results even when starting from a consistent
dynamical model.  Lorentz invariance is not sufficient to reconstruct
a GPD from its knowledge at $\xi=0$.

\subsubsection{GPDs from triangle graphs}
\label{sub:triangle-again}

Ans\"atze for GPDs based on the double distribution representation
respect polynomiality by construction, but they do not necessarily
fulfill the positivity constraints unless the double distributions
possess certain properties, as already observed in
\cite{Radyushkin:1998es}.  An integral representation of double
distributions satisfying the most general positivity constraint
(\ref{Poby-final}) has been given in \cite{Pobylitsa:2002vi} but not
used in practice so far.  A different ansatz for double distributions
that respects positivity has very recently been suggested
in~\cite{Mukherjee:2002gb}.

Pobylitsa \cite{Pobylitsa:2002vw} has proposed a different way to
construct parameterizations of GPDs which fulfill the necessary
consistency requirements.  It is based on the calculation of GPDs in
perturbative toy models, which in lowest order of the coupling leads
to evaluating triangle graphs as we discussed in
Section~\ref{sub:poly-wave}.  Pobylitsa considered GPDs for a
spin-zero target, both in scalar $\phi^3$ theory and with spin $\half$
partons coupling to the target by a Yukawa interaction.  By
construction the GPDs calculated in covariant perturbation theory
respect polynomiality.  On the other hand the triangle graphs can be
rewritten in terms of light-cone wave functions and hence satisfy the
general positivity requirement (\ref{Poby-final}).  An exception are
ultraviolet divergences (occurring in the Yukawa case as mentioned in
Section~\ref{sub:pos-derive}), whose regularization can violate
positivity at low renormalization scale $\mu^2$.  The consistency
properties of the GPDs obtained in this way are retained if different
masses are taken for the different lines in the graphs.

Clearly one cannot expect GPDs obtained in perturbative models to be
close to those of hadrons in QCD.  The strategy pursued in
\cite{Pobylitsa:2002vw} is to take a superposition of GPDs obtained
with different parton masses $m_1$, $m_2$, $m_3$ in the graphs (and
also a superposition of the forms obtained in the $\phi^3$ and Yukawa
models).  The structure of the hadron GPDs is then encoded in the
weight functions $s(m_1, m_2, m_3)$ of this superposition.  Modeling
along these lines had previously been suggested by Brodsky et
al.~\cite{Brodsky:2000ii} for the nucleon, where the lines in the
triangle loop were associated with effective quark-diquark degrees of
freedom in the target.  Under suitable conditions on the weight
functions (which need not be positive definite) the positivity
properties of GPDs are preserved, as shown in \cite{Pobylitsa:2002vw}.
The ultraviolet divergence of the triangle graphs can be removed in
the superposition by an additional condition on the weights.  It will
be interesting to see whether the form of GPDs obtained by this
strategy is suitable in practice for parameterizing the structure of
QCD bound states.


\subsection{GPDs at small $x$}
\label{sec:small-x-gpd}

As we mentioned in the Introduction, an important context where GPDs
appear are exclusive processes at very high energy and hence very
small~$x$.  The small-$x$ regime has a number of specialties compared
with moderate or large $x$.  Here we focus on GPDs in the small-$x$
limit; broader aspects of the dynamics in this region will be
discussed in Section~\ref{sec:small-x}.  In the context of GPDs it is
more correct to speak of ``small $\xi$'', but we follow common usage,
where ``small $x$'' refers to the typical relevant scale of momentum
fractions in a process, which is indeed set by the external kinematic
variable $\xi$.

Most small-$x$ studies have focused on the gluon GPD $H^g$ and to a
lesser extent on the quark distributions $H^q$.  These are assumed to
dominate the experimentally most accessible observables as we will
explain in Section~\ref{sec:small-x}.  There we will also see why one
may concentrate on the imaginary part of the scattering amplitude,
which for DVCS and light meson production involves GPDs only in the
DGLAP region (see Section \ref{sec:compton-scatt}) and at Born level
just at $x= \pm\xi$.  Indicators for the effect of skewness in the
small-$x$ regime which have become rather common in the literature are
\begin{equation}
  \label{small-x-ratio}
R^g = \frac{H^g(\xi,\xi)}{2\xi\, g(2\xi)} , \qquad
R^q = \frac{H^q(\xi,\xi)}{q(2\xi)} ,
\end{equation}
where $t$ is set to zero in the spirit discussed in
Section~\ref{sub:connect-forward}.  Corresponding ratios defined for
other parameterizations of GPDs differ from those here by global
factors $(1+\xi)$, which can be approximated by $1$ to the precision
of interest here.

Early studies by Frankfurt et al.~\cite{Frankfurt:1998ha} and by
Martin and Ryskin \cite{Martin:1998wy} have approximated the skewed
gluon distribution $H^g(x,\xi)$ with the forward gluon distribution at
$x_{\mathrm{in}} = (x+\xi)/(1+\xi)$, i.e., with $1+\xi \approx 1$ they
set
\begin{equation}
  \label{old-diagonal-input}
H^g(x,\xi) = (x+\xi)\, g(x+\xi) ,
\end{equation}
which in particular implies $R^g=1$ (this explains the choice of
momentum fraction $2\xi$ instead of $\xi$ in the denominators of
(\ref{small-x-ratio})).  In terms of Radyushkin's conventions
(Sections~\ref{sec:conventions} and \ref{sec:definitions}) the ansatz
(\ref{old-diagonal-input}) reads more naturally
$\mathcal{F}^g_\zeta(X) = X g(X)$.  Making this ansatz at input scales
$\mu_0^2$ of order 1~GeV$^2$ and evolving to higher scales it was
found that the ratio $R^g$ increases with $\mu^2$, with typical values
up to 1.6 for $\mu^2$ around 100~GeV$^2$ and a weak $\xi$ dependence
for small $\xi < 10^{-2}$.  The ansatz (\ref{old-diagonal-input})
appears at odds with the symmetry of GPDs under $\xi\leftrightarrow
-\xi$ (although strictly speaking it does not violate general
principles when only used for positive $\xi$) and has in later work
mostly been supplanted by the ansatz
\begin{equation}
  \label{new-diagonal-input}
H^g(x,\xi) = x g(x) ,
\end{equation}
we already introduced in (\ref{super-simple}).  For $x\ge\xi$ and
small $\xi$ Freund and Guzey \cite{Freund:1998uf} found this relation
to be stable within 10\% to 20\% under evolution from $\mu^2$ around
$0.6$~GeV$^2$ to above 100~GeV$^2$.  Note that with the ansatz
(\ref{new-diagonal-input}) one has
\begin{equation}
  \label{super-simple-again}
R^g = \frac{\xi\, g(\xi)}{2\xi\, g(2\xi)} ,
\end{equation}
which increases with $\mu$ since at higher scales the gluon density at
small $x$ becomes steeper.  With a power-law behavior $x g(x) \sim
x^{-\lambda}$ or $x q(x) \sim x^{-\lambda}$ as in
(\ref{small-x-power}) one obtains
\begin{equation}
  \label{super-simple-ratio}
R^g = 2^{\lambda} , \qquad  R^q = 2^{\lambda+1} .
\end{equation}
from (\ref{new-diagonal-input}) and from its analog $H^q(x,\xi) =
q(x)$ for quarks.  Numerical values of $R^g$ up to 1.6 were found in
\cite{Freund:1998uf} for $\mu^2$ as quoted above
(\ref{super-simple-again}).  We remark that both
(\ref{old-diagonal-input}) and (\ref{new-diagonal-input}), as well as
simply setting $R^g = 1$ from the start, are equivalent in the leading
$\log \frac{1}{x}$ approximation, where small-$x$ arguments different
by factors of order 2 cannot be distinguished, see
Section~\ref{sub:beyond-collinear}.  The spread in values of $R^g$
just quoted indicates the precision of such an approximation in the
application at hand.

A more general connection between GPDs and their forward limit was
advocated in \cite{Frankfurt:1998ha} and \cite{Martin:1998wy}.  It
starts from the observation that for $x \gg \xi$ one should have
$H(x,\xi) \approx H(x,0)$ to a good precision.  In this region the
kinematical asymmetry between the two gluons is negligible, and one
may for instance expand $H(x,\xi) = \sum_{n} (\xi /x)^{2n} h_n(x)$ and
only keep the leading term.  The second point of the argument is to
calculate GPDs at high scale $\mu$ (where measurements are performed)
from input GPDs at low input scale $\mu_0$.  In the region $x\gg \xi$
the input distributions may be approximated with their diagonal
counterparts.  This is the region which controls the behavior of GPDs
at large $\mu$ and small momentum fraction $x \gsim \xi$ (which is
most relevant in the convolution with hard scattering kernels) since
under evolution to higher scales partons in the DGLAP region migrate
towards smaller momentum fractions.  Put differently, effects of
nonzero $\xi$ in the GPDs at the input scale $\mu_0$ are concentrated
in a region where $x \sim \xi$, and evolution to high $\mu$ shifts
these effects into the ERBL region.  At $x \gsim \xi$ and large $\mu$
the difference between $H$ and its forward counterpart is then of
perturbative origin, due to the effect of nonzero $\xi$ in the
nonforward evolution kernels.  Whereas all studies discussed so far
confined themselves to the DGLAP region, Shuvaev et
al.~\cite{Shuvaev:1999ce} argue that the ``washing out'' of skewness
effects in the input distributions is also effective in the ERBL
region.  In fact we have seen in Section~\ref{sub:solve-evolution}
that in this region evolution to large $\mu$ generates a universal
shape in the variable $x/\xi$.

A crucial question for using these observations in practice is how
long the evolution interval in $\mu$ must be to make this ``washing
out'' of the initial conditions effective.  Unless $\alpha_s(\mu)$ is
very large, evolution can indeed be rather slow, and its ``speed''
depends on the shape of the initial conditions.  We also note that in
many applications the relevant hard scale $\mu$ of the process is not
\emph{very} high, see Section~\ref{sub:meson-gluon}.  A numerical
study by Golec-Biernat et al.~\cite{Golec-Biernat:1999ib} evolved
different input distributions at $\mu_0^2 = 0.26$~GeV$^2$ up to higher
scales, both for small and for moderate $\xi$, and found indeed that
the shapes come closer to each other for $\mu^2=4$~GeV$^2$ (and
somewhat more for $\mu^2=100$~GeV$^2$).  It is difficult to asses the
general speed of this convergence: in the DGLAP regions the GPDs at
$\mu_0^2$ were already quite close to each other, whereas in the ERBL
region they differed widely at $\mu_0^2$ and still did considerably
after evolution (except for $H^g$).

A quantitative implementation of the above idea was developed in
\cite{Shuvaev:1999ce,Golec-Biernat:1999ib}, which makes use of the
Shuvaev transform (Section~\ref{sub:Shuvaev}).  It is based on
neglecting the $\xi$ dependence of the conformal moments
$\mathcal{C}_n(\xi)$ defined in (\ref{conf-mom}) and
(\ref{conf-mom-glue}).  This is equivalent to approximating Shuvaev's
effective forward distribution $f^{q}_\xi(x)$ or $f^{g}_\xi(x)$ by the
forward density $q(x)$ or $g(x)$.  One then has
\begin{equation}
H^{q,g}(x,\xi;\mu) \approx \int_{-1}^{1} dy\, 
  \mathcal{K}_{q,g}(x,\xi,y)\, f^{q,g}(y;\mu)
\label{shuv-approx}
\end{equation}
with $f^q(x) = q(x)$ and $f^g(x) = g(x)$, where the $\xi$ dependence
of the GPD is generated only by the integral kernel
$\mathcal{K}_q(x,\xi,y)$.  Such an ansatz is by construction stable
under evolution and thus relates GPDs with forward densities at any
scale $\mu$.  To justify this ansatz Golec-Biernat et
al.~\cite{Golec-Biernat:1999ib} argue that the $\xi$ dependence of
$H^{q,g}$ which comes from higher conformal moments can be neglected
at sufficiently large $\mu$, since all moments with $n > 1$
asymptotically evolve to zero.  In consequence, they claim that
(\ref{shuv-approx}) becomes more reliable as $\mu$ increases, but is
valid even if $\xi$ is not very small.  We find this argument
problematic since the $\xi$ independent parts of higher moments
$\mathcal{C}_n(\xi)$ evolve to zero with the same speed as their $\xi$
dependent parts.  Neglecting higher moments altogether would give the
asymptotic forms of GPDs (with $H(\xi,\xi) = 0$ in particular) and
clearly be insufficient at scales $\mu$ of practical relevance.

A different line of argument \cite{Shuvaev:1999ce} is to point out
that $\mathcal{C}_n(\xi) = \mathcal{C}_n(0) + \xi^2 \,
\mathcal{R}_n(\xi)$, so that neglecting the $\xi$ dependence of the
moments induces an error of $O(\xi^2)$.  This arguments seems in fact
independent on whether $\mu$ is small or large.  The crucial question
is then how small the induced corrections are in the GPDs at a given
value of $x$, since the inversion of conformal moments is a nontrivial
procedure (see Section~\ref{sub:solve-evolution}).

Given the form of the parton densities and of the kernels
$\mathcal{K}_{q,g}(x,\xi,y)$, the transformation (\ref{shuv-approx})
for $\xi\ll 1$ is dominated by $q(x)$ or $g(x)$ at small values of
$x$.  Approximating the forward densities by a power-law
(\ref{small-x-power}) as before, one can perform the integration
explicitly and obtains skewness ratios
\cite{Shuvaev:1999ce,Noritzsch:2000pr}
\begin{equation}
  \label{shuv-ratio}
R^g = \frac{2^{2\lambda+3}}{\sqrt{\pi}}\, 
	\frac{\Gamma(\lambda+\frac{5}{2})}{\Gamma(\lambda+4)} , \qquad
R^q = \frac{2^{2\lambda+3}}{\sqrt{\pi}}\, 
	\frac{\Gamma(\lambda+\frac{5}{2})}{\Gamma(\lambda+3)} .
\end{equation}

For modeling GPDs by double distributions, Musatov and Radyushkin have
given an approximate representation for small $\xi$
\cite{Musatov:1999xp},
\begin{eqnarray}
  \label{dd-approx}
H^g(x,\xi) &=& \int_{-1}^1 d\alpha\, \rho(\alpha) \, (x - \xi\alpha)\,
		g(x - \xi\alpha) , 
\nonumber \\
H^q(x,\xi) &=& \int_{-1}^1 d\alpha\,
		\rho(\alpha) \, q(x - \xi\alpha) ,
\end{eqnarray}
where one has neglected the $\beta$ dependence in the profile function
$h(\beta,\alpha)$ of (\ref{profile-def}) and thus obtains a function
of just one variable as in (\ref{special-profile}).  At small $x \sim
\xi$ this approximation has corrections of relative order $\xi$ for
sufficiently regular profile functions such as (\ref{special-profile})
with (\ref{profile}).  This follows from expanding $h(\beta,\alpha)$
around $\beta=0$, which is justified since for $x \sim \xi$ the
reduction formula (\ref{dd-red}) implies $\beta \sim \xi$. For $x\gg
\xi$ one can expand $x - \xi\alpha$ around $x$ in (\ref{dd-approx})
and thus has $H(x,\xi)$ approximated by the forward density up to
order $(\xi/x)^2$.

With this model a power law for the small-$x$ behavior of the forward
densities translates into a power behavior
\begin{equation}
  \label{xi-scaling}
H^g(x,\xi) = \xi^{-\lambda}\, \hat{H}^g(x/\xi)  , \qquad\qquad
H^q(x,\xi) = \xi^{-(1+\lambda)}\, \hat{H}_q(x/\xi) .
\end{equation}
With $\rho \propto (1-\alpha^2)^b$ as given in (\ref{profile}), the
integral in (\ref{dd-approx}) can explicitly be performed and yields
\cite{Musatov:1999xp}
\begin{equation}
  \label{dd-ratio}
R^g = \frac{\Gamma(2+2b)}{\Gamma(2+2b-\lambda)} \,
      \frac{\Gamma(1+b-\lambda)}{\Gamma(1+b)} ,
\hspace{1em}
R^q = \frac{\Gamma(2+2b)}{\Gamma(1+2b-\lambda)} \,
      \frac{\Gamma(b-\lambda)}{\Gamma(1+b)} ,
\end{equation}
where the different forms for gluons and quarks originate in the
different forward limits $H^g(x,0) = x g(x)$ and $H^q(x) = q(x)$ and
the corresponding extra factor $\beta$ in the relation
(\ref{dd-red-glue}) between $H^g$ and its double distribution.  As
long as $0<\lambda<1+b$ for gluons and $-1<\lambda<b$ for quarks, these
ratios decrease as $b$ increases, and tend to
(\ref{super-simple-ratio}) in the limit $b\to \infty$, which
corresponds to the ``forward model'' (\ref{super-simple}).  Other
special cases often used in the literature are
\begin{equation}
  \label{dd-special-ratio}
R^g_{(b=2)} = \frac{1}{(1-\frac{1}{3}\lambda) 
	(1 - \frac{1}{4}\lambda) (1 - \frac{1}{5}\lambda)} ,
\hspace{1em}
R^q_{(b=1)} = \frac{3}{(1-\lambda) (1 - \frac{1}{2}\lambda)} .
\end{equation}
We recall that the ansatz $b^q=1$ and $b^g=2$ is based on the analogy
with meson DAs and their asymptotic behavior.  For the asymptotic
behavior of double distributions under evolution we saw in
(\ref{dd-asy}) that all double distributions become concentrated at
$\beta=0$, with the shape in $\alpha$ being like $(1-\alpha^2)$ for
$f^{q(-)}$ and like $(1-\alpha^2)^2$ for $f^{g}$ , in accordance with
the above powers.  In contrast, the $\alpha$ shape for $f^{q(+)}$
tends to $(1-\alpha^2)^2$, which does not correspond to $b^q=1$.
Taking $b^q=2$ one obtains
\begin{equation}
  \label{dd-quark-ratio}
R^q_{(b=2)} = \frac{2.5}{(1-\frac{1}{2}\lambda) 
	(1 - \frac{1}{3}\lambda) (1 - \frac{1}{4}\lambda)} ,
\end{equation}
which for $|\lambda| < 1$ provides less enhancement than $R^q_{(b=1)}$
but still more than $R^g_{(b=2)}$.

\begin{figure}
\begin{center}
\leavevmode
\epsfxsize=0.48\textwidth
\epsfbox{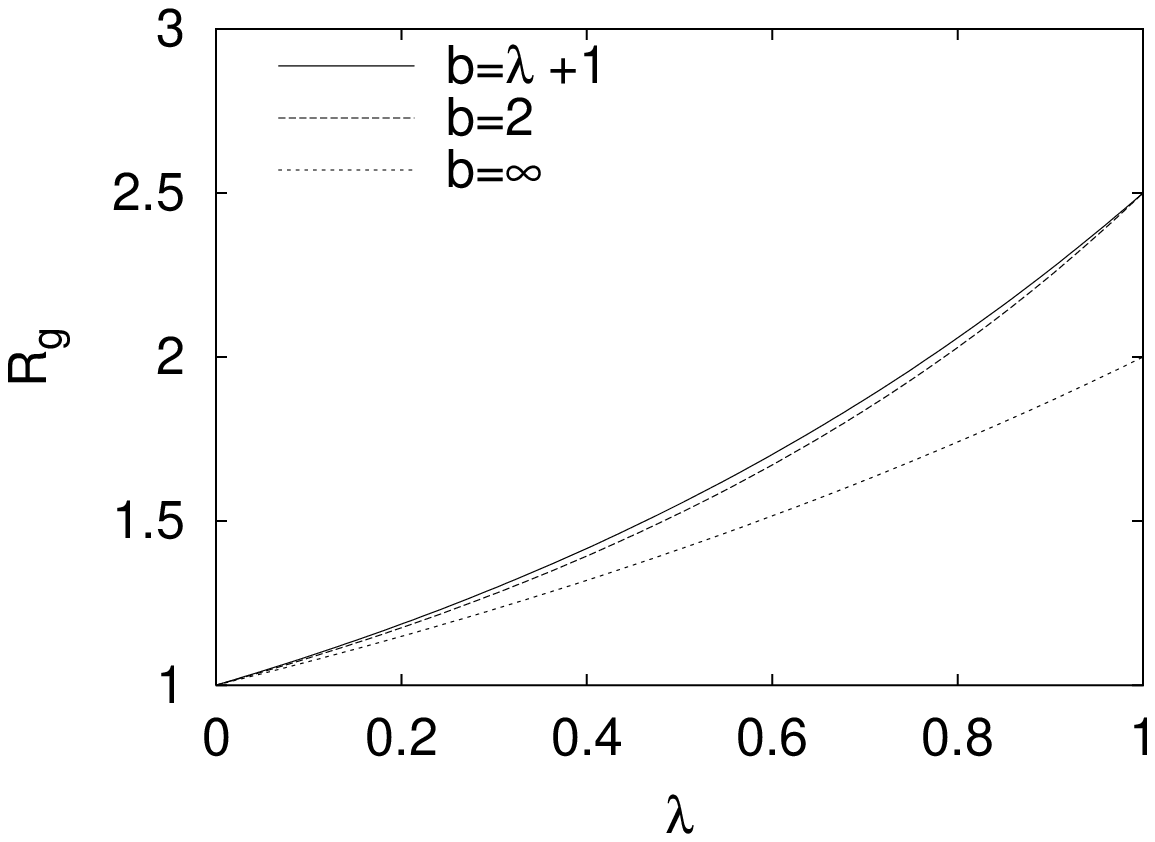} \\ 

\vspace{2em}

\epsfxsize=0.48\textwidth
\epsfbox{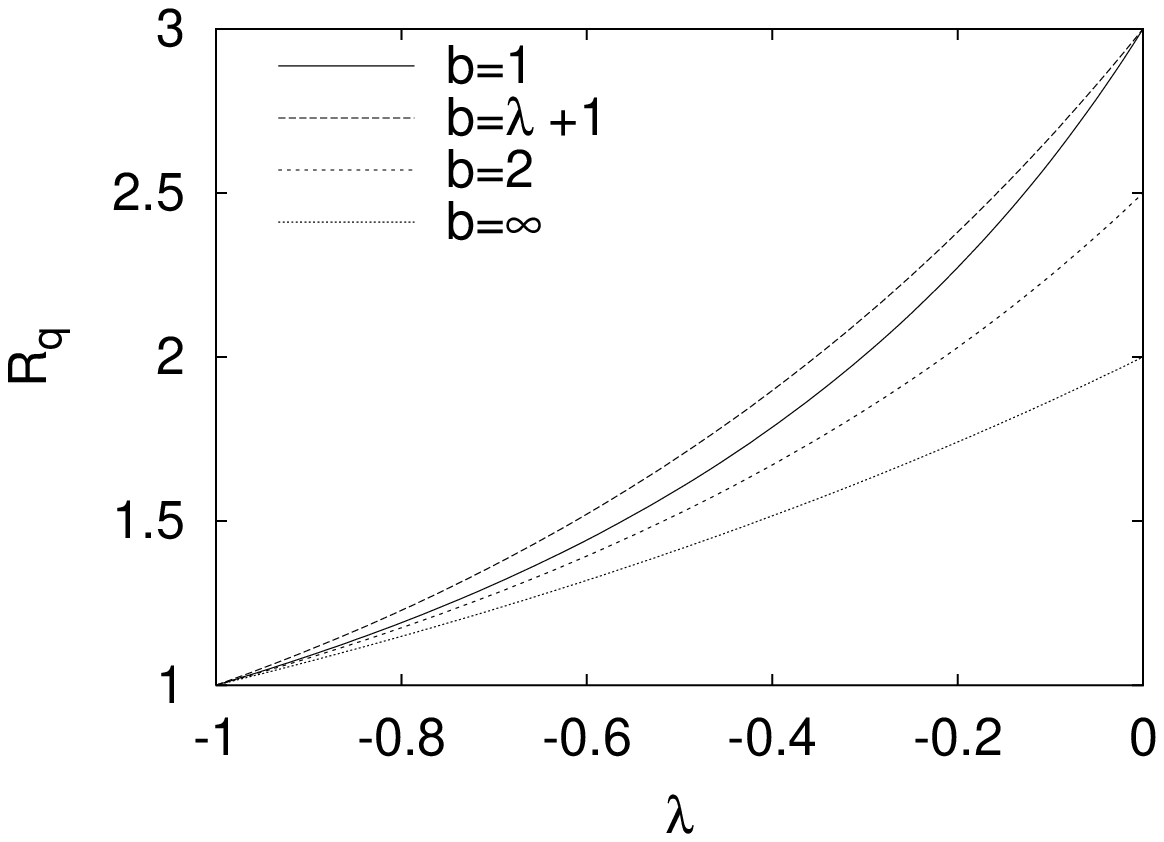} 
\epsfxsize=0.48\textwidth
\epsfbox{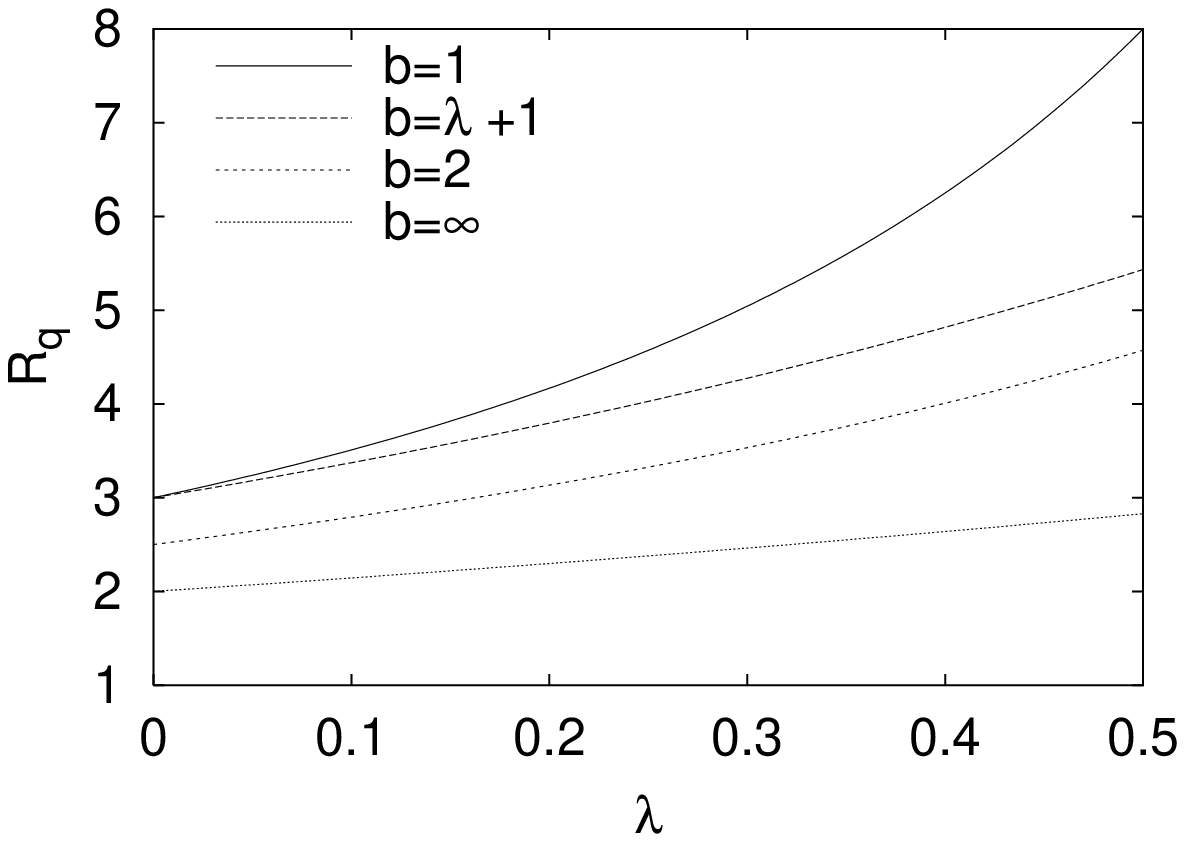} 
\end{center}
\caption{\label{fig:small-x-ratios} The skewness ratios
(\protect\ref{dd-ratio}) for gluons and quarks for forward densities
with a power behavior (\protect\ref{small-x-power}).  The choice
$b=\infty$ corresponds to the simple forward model in
(\protect\ref{super-simple}), and $b = \lambda + 1$ to the ansatz
(\protect\ref{shuv-approx}) based on the Shuvaev transform.}
\end{figure}

Musatov and Radyushkin \cite{Musatov:1999xp} have considered a model
based on a ``less asymptotic'' limit of the double distributions,
retaining for each separate Mellin moment $\int d\beta\, \beta^m
f(\beta,\alpha)$ in (\ref{dd-mellin}) the asymptotically dominant
term, which is proportional to $(1-\alpha^2)^{m+1}$ times the
corresponding Mellin moment of the forward quark or gluon density.
According to (\ref{dd-relate}) this is tantamount to retaining only
the $\xi$-independent term in the conformal moments $\mathcal{C}_n$ of
$H$ and thus to the model based on the Shuvaev transform discussed
above.  With a further approximation (which still allows one to
calculate $H$ at small $x$ and $\xi$ from the double distribution) one
can explicitly invert the Mellin moments of $f(\beta,\alpha)$.  For a
power-law behavior (\ref{small-x-power}) of parton densities the
result has the form (\ref{dd-approx}) with $\rho(\alpha)
\propto (1-\alpha^2)^b$ and $b=\lambda+1$ for both gluons and quarks.
The corresponding enhancement ratios are of course given by
(\ref{shuv-ratio}).  In Fig.~\ref{fig:small-x-ratios} we plot the
ratios $R^g$ and $R^q$ for the various models we have discussed in
this section.  We emphasize that $\lambda$ describes the forward
densities in the relevant range $x \sim \xi$, and as an effective
power depends weakly on the range of $x$ and rather strongly on the
scale $\mu$ (see Section~\ref{sec:small-x}).

In the phenomenologically relevant region of $\lambda$ the enhancement
factors for gluons are almost identical for the choice $b=2$ and for
the Shuvaev model.  For $\lambda \lsim 0.4$ there is in fact little
difference for all profile parameters between $b=1$ and $b=\infty$.
The ratio $R^g$ is relevant for vector meson production at small $x$,
where the gluon distribution dominates.  Compton scattering receives
important contributions from the $C$-even combination of quark
distributions even at small $x$ (see Section~\ref{sec:compton-scatt}).
For this combination one has $\lambda>0$ at scales relevant for the
process and obtains enhancement factors of appreciable size.

Freund et al.~\cite{Freund:2002qf} have made the strong claim that,
except when taking $b=\infty$, the underlying ansatz
(\ref{profile-def}) to (\ref{profile}) fails to describe the available
data on DVCS at LO or NLO in $\alpha_s$, overshooting both the H1
cross section data at small $\xB$ \cite{Adloff:2001cn} and the beam
spin asymmetries from HERMES and CLAS
\cite{Airapetian:2001yk,Stepanyan:2001sm} at moderate $\xB$.  They
propose an alternative ansatz using the forward model
(\ref{super-simple}) in the DGLAP region (corresponding to
$b=\infty$).  In the ERBL region they take a simple polynomial in $x$
with $\xi$-dependent coefficients chosen such that the resulting GPD
satisfies polynomiality for the lowest Mellin moments, arguing that
higher moments are practically irrelevant in the small $\xi$ region.
We caution that higher Mellin moments are essential for reconstructing
parton distributions (and the associated scattering amplitudes) at
achievable resolution scales $\mu$, and it remains to study how small
violations of polynomiality in higher Mellin moments propagate to
these quantities.

We finally remark that the general results presented in this section,
in particular the various expressions for the skewness ratios, can
readily be extended to the polarized GPDs, with the replacements $H^g
\to \tilde{H}^g$, $H^q \to \tilde{H}^q$ and their analogs for the
forward densities.


\subsection{Nuclei}
\label{sec:nuclei}

Some of the experimental facilities that can study hard exclusive
processes also run with nuclear targets, and it is natural to ask what
can be learned from the corresponding generalized parton
distributions.

Let us begin by discussing the deuteron, which in some respects is the
simplest case to be studied.  According to whether the deuteron stays
intact or breaks up one will deal with GPDs for the deuteron, or for
the breakup $d \to p+n$, or for the breakup to a more complicated
state (containing for instance additional soft pions).  Each case
offers characteristic physics information, whose extraction requires
of course that the final state can be experimentally identified.

The $d \to p+n$ transition may provide access to the GPDs of the
neutron (and thus to a flavor decomposition of the nucleon GPDs) in
kinematics where the neutron takes nearly all of the momentum transfer
$\Delta$.  In the impulse approximation the scattering then takes
place on a quasifree neutron in the deuteron, with the proton being a
spectator.  The $d \to p+n$ transition GPDs are thus given by the
neutron GPDs convoluted with the nuclear wave functions $\psi_{p+n}$
of the target.  Other nuclei might be used to the same end.  No
detailed studies of these issues have yet been performed in the
literature.

\begin{figure}[b]
\begin{center}
	\leavevmode
	\epsfxsize=0.8\textwidth
	\epsfbox{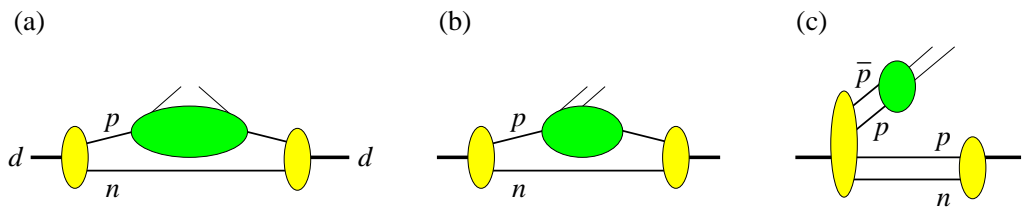}
\end{center}
\caption{\label{fig:deuteron} Different contributions to the
generalized parton distributions of the deuteron in a convolution
model.}
\end{figure}

The spin structure of the GPDs for the elastic $d \to d$ process has
been presented in Section~\ref{sec:deuteron}.  Consider again the
approximation of the deuteron as a weakly bound state of a proton and
a neutron, described by two-body wave functions $\psi_{p+n}$ with the
appropriate momentum and spin dependence.  In analogy to the
convolution model for the usual parton densities of the deuteron, the
deuteron GPDs are then a convolution of these nuclear light-cone wave
functions and the relevant nucleon GPDs, as shown in
Fig.~\ref{fig:deuteron}a and b.  As detailed in \cite{Berger:2001zb}
the GPDs $H_3$, $H_5$, $\tilde{H}_2$, $\tilde{H}_3$ require $D$-wave
configurations of the proton and neutron in the deuteron.  If
extracted experimentally, they may provide information on the
corresponding part of the nuclear wave function, in addition to what
can be inferred from the elastic deuteron form factors.

The GPDs of a nuclear system are interesting beyond their spin
structure.  For the mechanism of Fig.~\ref{fig:deuteron}a and b, we
observe that at nonzero $\xi$ the longitudinal momentum fraction of
the active nucleon in the bound state before and after the scattering
is different.  The $\xi$ dependence of the generalized parton
distribution reflects rather directly the width of the deuteron wave
function in longitudinal momentum: if the plus-momentum fraction of
the nucleon in the deuteron has width $w$, the deuteron GPDs will drop
at $\xi \sim w$ \cite{Berger:2001zb}.  Note that the dependence on the
longitudinal variable $\xi$ comes in correlation with the dependence
on transverse momentum, controlled by the variable $t$.  In kinematics
where the motion of the nucleons in the deuteron is dominantly
nonrelativistic the deuteron wave functions are rather well known, and
one may again use such processes to investigate the GPDs of the
neutron.  A similar statement holds for other nuclei.  A first
quantitative study of the deuteron GPDs in a convolution model has
been performed by Cano and Pire \cite{Cano:2003ju}.

The extension of the convolution model to GPDs of heavier nuclei has
been considered by Guzey and Strikman \cite{Guzey:2003jh}, who have
also estimated the relative importance of the case when the nucleus
breaks up.  In a phenomenological study by Kirchner and M\"uller
\cite{Kirchner:2003wt}, a simplification of the convolution picture
was used to obtain the $x$ and $\xi$ dependence, neglecting the
relative momenta of nucleons in the initial-state nucleus.  Note that
this implies nonzero relative nucleon momenta in the final-state
nucleus and hence violates the symmetry in $\xi$ of the nuclear GPDs.
By construction GPDs at $\xi\neq 0$ do not allow both the initial and
final nucleus to be treated as static systems of nucleons.  To neglect
the momentum mismatch between the initial and final state wave
functions is only consistent as an approximation for values of $\xi$
small compared with the typical momentum fractions over which the
nuclear wave functions vary significantly.

If follows from our discussion in Section \ref{sub:poly-wave} that the
convolution model described so far is not sufficient to guarantee the
polynomiality properties of GPDs.  In the ERBL region there is also
the possibility that a nucleon-antinucleon pair in the initial nucleus
annihilates into a $q\bar{q}$ system which is then emitted
(Fig.~\ref{fig:deuteron}c).  Experimental and theoretical studies of
nuclear GPDs at different $\xi$ may provide a glimpse on how important
such highly relativistic configurations are in a nuclear system.  For
deuteron GPDs at not too large $\xi$, only small to moderate
deviations from the sum rules (\ref{deut-sum-rules}),
(\ref{zero-sum-a}), (\ref{zero-sum-b}) were found for the convolution
model used in \cite{Cano:2003ju}.

Clearly the description of a nucleus in terms only of nucleon degrees
of freedom is an approximation.  The possible role of meson degrees of
freedom has briefly been discussed in \cite{Kirchner:2003wt}.  If
$\xi$ is so large that in the convolution picture one is forced into
the tail of the nuclear wave function, one will be sensitive to
quantum fluctuations of the deuteron that are more complicated than a
system of two almost free nucleons.  To study such configurations in
deeply inelastic scattering at $x_B>1$ has proven to be difficult
since one scatters on quarks with a very large momentum fraction in
the target.  For GPDs however one can have any plus-momentum fraction
of the struck quark, even for large~$\xi$.

An additional observable for nuclear targets is the dependence on the
atomic number $A$, which can distinguish different dynamical
mechanisms.  An example is a study of Polyakov \cite{Polyakov:2002yz},
which obtained a characteristic $A$ dependence of the $D$-term,
appealing to its connection with the spatial components of the
energy-momentum tensor described at the end of Section~\ref{sec:spin},
and using a simple model for a large nucleus.  Nuclear medium effects
in GPDs and their consequences in the scattering amplitude for DVCS
have recently been investigated by Freund and Strikman
\cite{Freund:2003wm}.

\begin{figure}[b]
\begin{center}
	\leavevmode
	\epsfxsize=0.33\textwidth
	\epsfbox{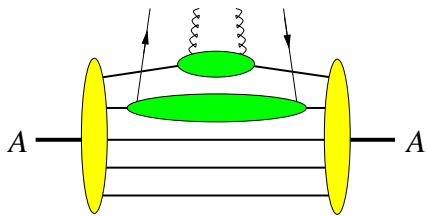}
\end{center}
\caption{\label{fig:nucleus-four} Contribution to a twist-four parton
distribution in a nucleus which involves two nucleon GPDs.}
\end{figure}

The occurrence of GPDs in a quite different context was studied by
Osborne and Wang \cite{Osborne:2002st}.  Applying the convolution
model to twist-four parton correlations in nuclei, contributions as
shown in Fig.~\ref{fig:nucleus-four} were identified as involving
nucleon GPDs, which are not close to the forward limit in all
kinematics.  These contributions scale as $A^{4/3}$ and can hence
become important for large nuclei since their twist-two parton
densities only grow like $A$ in the region where the convolution model
is applicable.


\subsection{Dynamics of GDAs}
\label{sec:gda-dynamics}

We have seen the intimate connection between GPDs and generalized
distribution amplitudes through analytic continuation in the
invariants $t$ or $s$.  Dynamical studies of the respective quantities
show similar connections but also important differences between the
physics in the $t$ and in the $s$ channel.  One important difference
already mentioned in Section~\ref{sub:overlap-formulae} is that GDAs
cannot be represented in terms of the wave functions of individual
hadrons they involve: they describe the formation of a hadronic system
including the interactions between hadrons.  A representation in terms
of the bound state wave function \emph{and} the corresponding
interaction kernel has been given by Tiburzi and Miller
\cite{Tiburzi:2001je,Tiburzi:2002sx} in the covariant framework we
discussed in Section~\ref{sub:covariant-models}.

Dynamical studies of GDAs have so far concentrated on the two-pion
system, both because of its relative theoretical simplicity and
phenomenological importance.  Moreover, most work has focused on the
quark GDAs.

\subsubsection{Very small $s$}
\label{sub:very-small-s}

At low invariant two-pion mass $\sqrt{s}$ the two-pion DAs have important
connections with the chiral symmetry of QCD and its spontaneous
breaking, and studies similar to those described in
Sections~\ref{sub:chiral} and \ref{sub:chiral-soliton} for the pion
GPDs have been performed.  In the chiral limit there are soft-pion
theorems for GDAs analogous to those for GPDs in
(\ref{soft-pion-gpd}).  At $s\to 0$ and $\zeta\to 0$ or $1$, where one
pion momentum becomes soft, one has~\cite{Polyakov:1998ze}
\begin{eqnarray}
  \label{soft-pion-gda}
\lim_{\zeta\to 1} \Phi^{u-d}(z,\zeta,0) &=& 
- \lim_{\zeta\to 0} \Phi^{u-d}(z,\zeta,0) \;=\; 2 \phi_\pi(z) , 
\nonumber \\
\lim_{\zeta\to 1} \Phi^{u+d}(z,\zeta,0) &=& 
\phantom{-} \lim_{\zeta\to 0} \Phi^{u+d}(z,\zeta,0) \;=\; 0 .
\end{eqnarray}
Similarly, one has $\lim_{\zeta\to 0,1} \Phi^{g}(z,\zeta,0) = 0$ for
the gluon GDA \cite{Kivel:1999sd}.  This type of relation can be
generalized to the three-pion DAs for the case where one or two pion
momenta become soft \cite{Pire:2000ky}.  One-loop contributions in
chiral perturbation theory generate nonanalytic terms in $s$ or in
$m_\pi^2$ for the two-pion DAs and have been calculated by Kivel and
Polyakov \cite{Kivel:2002ia}.  In particular, these terms provide
corrections to the soft-pion theorem (\ref{soft-pion-gda}).  They also
lead to an imaginary part of the two-pion DAs for $s > 4m_\pi^2$, due
to rescattering of the two pions.

The two-pion DAs have been evaluated in the chiral quark-soliton model
by Polyakov and Weiss \cite{Polyakov:1998td}.  Valid for $s \lsim 4
M^2 \approx (700~\mbox{MeV})^2$ and referring to a low factorization
scale $\mu \approx 600$~MeV, the GDAs obtained have a rather rich
structure in $z$, with sharp crossovers at the points $z = \zeta$ and
$z = 1-\zeta$ where one pion carries the same plus-momentum as the
quark.  Their shape also exhibits a clear dependence on $s$.  A
detailed study of both pion GDAs and pion GPDs in this model has
recently been performed by Prasza{\l}owicz and Rostworowski
\cite{Praszalowicz:2003pr}.

\subsubsection{Connection with resonance DAs}
\label{sub:resonance}

When $\sqrt{s}$ is in the vicinity of the mass of a resonance, a close
connection can be established between the GDA and the corresponding DA
of the resonance \cite{Polyakov:1998ze}.  This involves first a
projection on the appropriate partial wave of the two-pion state
through Legendre polynomials $P_l(\cos\theta)$ with $\beta\,
\cos\theta = 2\zeta -1$.  One can also directly work with the partial
wave Gegenbauer coefficients $\tilde{B}_{nl}$ discussed in
Section~\ref{sub:gda-evolution}.  Assuming that in some region $s$ the
resonance contribution dominates the appropriate partial wave in the
GDA, one can then match its behavior around the resonance peak onto
the DA of the resonance times a factor describing its decay to the
two-pion channel.  The latter is described by a Breit-Wigner
propagator $(s - m_R^2 + i \Gamma_{\! R} m_R)^{-1}$ and a constant
related to the relevant branching ratio.

{}From a conceptual point of view it is important that the GDA
describes both the ``resonance'' and ``continuum'' contribution in the
two-pion system, without the need to even separate the two unless one
explicitly wants to connect the GDA with resonance properties.  In
hard processes where a resonance DA appears, one can thus
alternatively work with the GDA of the resonance decay products.  This
can be especially useful when a separation of resonance ``signal'' and
continuum ``background'' is nontrivial, for instance for the very
broad $\rho$ resonance, or if one wants to extend study to invariant
masses $\sqrt{s}$ away from the resonance peak.  We will mention
practical applications of this in Sections
\ref{sub:meson-pairs-electro} and \ref{sub:meson-pair-pheno}.

\subsubsection{Very large $s$}
\label{sub:very-large-s}

\begin{figure}
\begin{center}
	\leavevmode
	\epsfxsize=0.22\textwidth
	\epsfbox{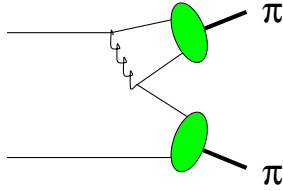}
\end{center}
\caption{\label{fig:large-s-pert} One of the diagrams for the pion
two-pion DA $\Phi^q$ in the limit of large $s$.}
\end{figure}

The behavior of the two-pion DAs in the limit $s\to \infty$ is fully
analogous to the large-$t$ limit of the pion GPDs discussed in
Section~\ref{sub:large-t-limit} and has been discussed in
\cite{Diehl:1999ek}.  The GDAs in this limit can be described by the
hard scattering mechanism, with diagrams as shown in
Fig.~\ref{fig:large-s-pert}.

The features of the result are analogous to those discussed in
Section~\ref{sub:large-t-limit}.  In particular one obtains a
power-law behavior $\Phi^q(z,\zeta,s) = \alpha_s(s)\, s^{-1} \,
\phi^q(z,\zeta)$ for the GDAs, whose factorization scale is
understood to be of order $s$.  As in the case of GPDs, one obtains
GDAs that exactly satisfy the polynomiality conditions.  In their
$z$-dependence a change of sign is observed at $z=\zeta$.  As $z$
approaches $\zeta$ one finds a logarithmic singularity like
$\log|z-\zeta|$, which signals the breakdown of the hard scattering
approximation in the vicinity of this point, where the momentum of the
quark or antiquark in one of the pions (the lower one in
Fig.~\ref{fig:large-s-pert}) becomes soft.  In an improved evaluation
one expects to find finite GDAs with a rapid change between positive
and negative values at this point.  Note that according to the
crossing rules~(\ref{gpd-gda-cross}) its location $z=\zeta$
corresponds to the point $x= 1$ of GPDs, where the hard scattering
formalism breaks down as well (see Section~\ref{sub:large-t-limit}).

\subsubsection{Modeling}
\label{sub:gda-models}

A modeling strategy for the two pion DAs $\Phi^u$ and $\Phi^d$ in the
mass region up to about $s \approx 1$~GeV$^2$ has been developed by
Polyakov \cite{Polyakov:1998ze}, with an extension to $\Phi^g$ in
\cite{Kivel:1999sd,Lehmann-Dronke:1999aq}.  It makes use of the
various connections of GDAs with other hadronic quantities we have
discussed so far.

Its first ingredient is the restriction to the asymptotic form in $z$
given in (\ref{asy-gda}).  We note that the only distribution
amplitudes which are strongly constrained from data are for a single
pion or a single $\eta$ or $\eta'$, where in particular the processes
$\gamma^* \gamma\to \pi^0$, $\eta$, $\eta'$ suggest a form rather
close to the asymptotic one, $\phi(z) = 6z(1-z)$, already at low
factorization scale $\mu \sim 1$~GeV (see
Section~\ref{sub:gamma-star-gamma}).  The same assumption for GDAs
should be taken with due care, given that the results of the chiral
quark-soliton model (Section~\ref{sub:very-small-s}) give shapes very
different from the asymptotic one at the very low scale $\mu\approx
600$~MeV, while giving a single-pion DA quite close to $\phi(z) =
6z(1-z)$ at the same scale \cite{Petrov:1998kg}.  Evolving the GDA up
to $\mu \sim 1$~GeV will bring its shape closer to the asymptotic form
since $\alpha_s$ in that region is quite large, but no quantitative
study of this has been performed.  Evolution from this scale upward is
rather slow because the anomalous dimensions are not very large, so
that strong differences from the asymptotic shape in $z$ will persist
over a long interval of~$\mu$.

When taking the asymptotic shape in $z$, the isovector GDA
$\Phi^{u-d}$ is expressed through the timelike pion form factor, which
is phenomenologically well known for small and moderate $s$.  This
leaves one to model the Gegenbauer coefficients for the isoscalar and
gluon GDAs.  More generally one may keep the $z$ dependence as in
(\ref{asy-gda}) but retain the scale dependent Gegenbauer coefficients
$B_{1l}^{u+d}(s,\mu^2)$ and $B_{1l}^g(s,\mu^2)$ instead of their
respective asymptotic limits, which depend only on $B^-_{1l}(s)$
\cite{Diehl:2000uv}.

The second ingredient in the method of \cite{Polyakov:1998ze} is to
use a dispersion relation in $s$ for these Gegenbauer coefficients.
As an input one needs the subtraction constants in this dispersion
relation, given by the $B_{1l}$ or their derivatives at the point
$s=0$.  At this point one has various pieces of information from
chiral symmetry and from the moments of quark densities in the pion,
which are phenomenologically known to some extent.  In particular, the
soft-pion theorem (\ref{soft-pion-gda}) gives $B_{10}^{u+d}(0) = -
B^{u+d}_{12}(0)$ and the analog for gluons, with corrections of
parametric suppression $m_\pi^2 /(4\pi f_\pi)^2$.  Furthermore,
$B^{u+d}_{12}(0)$ and $B^{g}_{12}(0)$ are respectively related to the
momentum fraction carried by $u$ and $d$ quarks or gluons in the pion,
see (\ref{pion-fraction}).  As remarked in \cite{Diehl:2000uv} these
fractions at moderate scales $\mu$ are still rather different from
their asymptotic values under evolution, and the corresponding
difference can have notable effects on physical amplitudes calculated
in this model \cite{Kivel:2000rq}.  The momentum fraction of $s$
quarks in the pion at moderate scale is presumably small, which
provides some justification for neglecting $\Phi^s$ in the pion at not
too large values of $s$ and $\mu^2$.  At $s$ of order $m_\pi^2$ the
coefficients $B_{1l}(s)$ of the two-pion DAs can also be related with
parameters in the Lagrangian of chiral perturbation theory, namely
with the couplings constants for the interaction of soft pions with a
gravitational field \cite{Lehmann-Dronke:2000xq}.  Polyakov
\cite{Polyakov:1998ze} has also studied these parameters within the
chiral soliton model.

A further ingredient in the dispersion relation are the phases of the
Gegenbauer coefficients $\tilde{B}_{1l}$ for definite partial waves.
As discussed in Section~\ref{sub:gda-evolution} they equal the
measured phase shifts $\delta_l(s)$ of elastic $\pi\pi$ scattering for
$s$ up to the region where inelastic scattering becomes important.
This limits the region of validity in $s$ of the simple dispersive
method described here.  Up to about $\sqrt{s} = 1.2$~GeV the $D$ wave
is approximately elastic and well described by a Breit-Wigner phase
from the $f_2(1270)$ resonance.  In the $S$ wave, a rather complicated
structure appears around $\sqrt{s} = 1$~GeV, where the narrow
$f_0(980)$ resonance and the opening of the two-kaon channel almost
coincide.  A first attempt to describe the behavior of the $S$ wave
beyond this energy has been made in \cite{Hagler:2002nf} by including
an inelasticity parameter.  A full treatment will likely require a
two-channel analysis, which has not been performed as yet.

As we saw in Section~\ref{sub:radon} not only GPDs but also GDAs can
be represented in terms of double distributions.  Only recently has
this been exploited for modeling purposes \cite{Mukherjee:2002gb}.

\subsubsection{GDAs and the hadronization process}
\label{sub:lund}

GDAs describe the hadronization  from quarks or gluons to a
specified system of hadrons.  This can be regarded as the
``exclusive'' counterpart of hadronization in the regime of large
multiplicities, where observables typically are taken to be more
inclusive.  Among the models used to describe this regime is the Lund
string model \cite{Andersson:1983ia}.  In this framework Maul
\cite{Maul:2000ky} has studied $\gamma^* \gamma$ annihilation into two
pions and other exclusive channels at $Q^2 \gg s$, thus taking the
Lund model to its ``exclusive limit'', where the Lund string breaks
only once before the desired final state is obtained.  As the Lund
model is semiclassical, this study was performed at the cross section
level, in contrast to the description of the same process by GPDs,
which enter at amplitude level.  At $s = 1$~GeV$^2$ the cross section
obtained in \cite{Maul:2000ky} differs by a factor of about 3.5 from
the cross section calculated with GPDs modeled along the lines of the
previous subsection \cite{Diehl:2000uv}.  This is remarkably close
given considerable uncertainties in both model studies and above all
the very different dynamical pictures they represent, the chosen value
of $s$ being at the lower limit of applicability for one model and at
the upper limit for the other.


\section{Exclusive processes to leading power accuracy}
\label{sec:leading-power}

\subsection{Factorization}
\label{sec:factor}

The possibility to study GPDs in suitable exclusive scattering
processes rests on factorization theorems, as does the program to
extract usual parton densities from inclusive and semi-inclusive
measurements.  Collins, Frankfurt and Strikman \cite{Collins:1997fb}
have given a detailed proof for factorization in light meson
production, and Collins and Freund \cite{Collins:1998be} for Compton
scattering, see \cite{Collins:1999yw} for a brief account of both
works.  These proofs are based on the properties of Feynman diagrams
and very similar to the factorization proofs for inclusive DIS or
Drell-Yan pair production \cite{Collins:1988gx}.  Radyushkin has
analyzed Compton scattering and meson production using the
$\alpha$-representation of Feynman graphs \cite{Radyushkin:1997ki}.
Another analysis of Compton scattering has been given by Ji and
Osborne \cite{Ji:1998xh}.  Recently, Bauer et al.~\cite{Bauer:2002nz}
have investigated Compton scattering in an effective field theory
framework.  An effective field theory formulation had earlier been
used by Derkachov and Kirschner in a study focusing on
evolution~\cite{Derkachov:2001km}.

Consider first the amplitude for virtual Compton scattering
\begin{equation}
  \label{compton-process}
\gamma^*(q) + p(p) \to \gamma^*(q') + p(p') .
\end{equation}
So far we have discussed DVCS, where $q^2$ is large and spacelike,
whereas the final photon is on shell.  Another relevant case is when
the outgoing photon is timelike and decays into a lepton pair
$\ell^+\ell^-$.  The factorization proof \cite{Collins:1998be}
requires that at least one of the photons be far off-shell.  This
includes the case where the initial photon is real and $q'^2$ is large
and timelike \cite{Collins:2001pr}, which was called ``timelike
Compton scattering'' (TCS) in \cite{Berger:2001xd}.  The case where
both photons are off shell has been dubbed ``double deeply virtual
Compton scattering'' (DDVCS) by Guidal and
Vanderhaeghen~\cite{Guidal:2002kt} and appears in the process $ep\to
ep\, \ell^+\ell^-$.  The factorization theorem for Compton scattering
is valid in the generalized Bjorken limit
\begin{equation}
  \label{compton-limit}
|q^2| + |q'^2| \to \infty \qquad \mbox{at fixed~}
q^2/W^2,\,  q'^2/W^2,\,  t .
\end{equation}
With the case of a timelike final-state photon in mind we define $Q^2
= - q^2$ and $Q'^2 = q'^2$.  The amplitude can be written as
\begin{equation}
  \label{factorize-compton}
\mathcal{A}(\gamma^*p\to \gamma^* p) =
\sum_i \int_{-1}^1 dx \, 
  T^i(x,\rho,\xi, Q^2-Q'^2)  F^i(x,\xi,t) ,
\end{equation}
up to terms suppressed by inverse powers of the large momentum scale
$Q$ or $Q'$.  Here $F^i$ stands for the matrix elements $F^q$, $F^g$
and $\tilde{F}^q$, $\tilde{F}^g$.  The corresponding hard scattering
amplitudes $T^i$ further depend on the photon helicities, as will be
discussed in Section~\ref{sub:selection}.  There are now two scaling
variables,\footnote{A variety of notations is used for these variables
in the literature.  In particular, ``$\xi$'' sometimes has the meaning
of $\rho$ in (\protect\ref{xi-eta-def}).}
\begin{equation} 
  \label{xi-eta-def}
\rho = - \frac{(q+q')^2}{2 (p+p') \cdot (q+q')} , \qquad
\xi = - \frac{(q-q') \cdot (q+q')}{(p+p') \cdot (q+q')} ,
\end{equation}
which up to terms of order $t/W^2$ and $m^2/W^2$ equal
\begin{eqnarray}
  \label{xi-eta-prop}
\rho &\approx& \frac{Q^2 - Q'^2}{2 W^2 + Q^2 - Q'^2} \approx
             - \frac{(q+q')^+}{(p+p')^+} ,
\nonumber \\
\xi &\approx& \frac{Q^2 + Q'^2}{2 W^2 + Q^2 - Q'^2} \approx
               \frac{(p-p')^+}{(p+p')^+} .
\end{eqnarray}
Special cases are DVCS with $Q'^2=0$ and $\rho = \xi$, and TCS with
$Q^2=0$ and $\rho = - \xi$.  For simplicity we will in the following
refer to the large scale as $Q$, keeping in mind the exception of TCS.
In (\ref{factorize-compton}) we have not written out the dependence on
the factorization scale $\mu_F$ of the GPDs and on the renormalization
scale $\mu_R$ of the running coupling.  Explicit logarithms $\log
(Q^2-Q'^2) /\mu_F^2$ appear in the hard scattering kernel starting at
$O(\alpha_s)$, and logarithms $\log (Q^2-Q'^2) /\mu_R^2$ appear
starting at $O(\alpha_s^2)$.  The dependence of the amplitude on
$\mu_F$ and $\mu_R$ cancels of course order by order in the strong
coupling.  In choosing $Q^2-Q'^2$ as argument of the logarithms we
have assumed that one does not have $Q'^2 = Q^2$, in which case one
would take the argument $Q^2+Q'^2$ instead.  We note that in part of
the literature the generalized Bjorken limit is defined not by
$|q^2|+|q'^2|\to \infty$ but by $|q^2+q'^2|\to \infty$, which excludes
the point $Q'^2 = Q^2$.  It is not clear why factorization should not
hold in this case, but there is no dedicated investigation of this
issue in the literature.

The plus-momenta in (\ref{xi-eta-prop}) refer to a frame where the
external momenta of the Compton amplitude have small transverse
components of order $\sqrt{-t}$.  Various particular choices have been
made in the literature, for instance $\tvec{p}=0$ or $\tvec{p} = -
\tvec{p}' = - \half \tvec{\Delta}$ on the proton side, and $\tvec{q} =
0$ or $\tvec{q'} = 0$ or $\tvec{q} = - \tvec{q}' = \half
\tvec{\Delta}$ on the photon side.  At leading order in the large
scale all these frames are equivalent.

The relevant limit for the production of a meson $M$
\begin{equation}
  \label{meson-process}
\gamma^*(q) + p(p) \to M(q') + p(p') 
\end{equation}
is
\begin{equation}
  \label{meson-limit}
Q^2  \to \infty \qquad \mbox{at fixed~} Q^2/W^2,\,  t .
\end{equation}
In this limit, the amplitude for longitudinal polarization of photon
and meson reads
\begin{eqnarray}
\mathcal{A}(\gamma^*_L\, p\to M_L^{\phantom{*}}\, p) 
&=& \frac{1}{Q}
\sum_{ij} \int_{-1}^1 dx \, \int_0^1 dz\,
  T^{ij}(x,\xi,z, Q^2) F^{i}(x,\xi,t)\, \Phi^j(z)
        \label{factorize-meson}
\end{eqnarray}
up to power corrections in $1/Q$.  All other helicity transitions are
of order $1/Q^2$ or higher.  Again we have not displayed the
dependence on the factorization scale $\mu_F$ of the GPDs and the
meson DAs and on the renormalization scale $\mu_R$ of $\alpha_s$.  The
dependence of $T^{ij}$ on $Q^2$ is through logarithms $\log
Q^2/\mu_F^2$ and $\log Q^2/\mu_R^2$, both starting at $O(\alpha_s^2)$.
In a suitable reference frame the external momenta in
(\ref{meson-process}) have small transverse components, the incoming
and outgoing protons have large plus-momentum, and the outgoing meson
has large minus-momentum.  $x$ and $\xi$ then represent plus-momentum
fractions as before.  $z$ is now the minus-momentum fraction of a
quark or gluon in the meson and thus corresponds to reversing plus-
and minus-components in the definitions (\ref{vector-da}) and
(\ref{pseudo-da}) of distribution amplitudes $\Phi^j$ for the relevant
quark flavors or gluons.

The factorization theorems apply to a larger class of processes than
(\ref{compton-process}) and (\ref{meson-process}).  One may replace
the outgoing proton $p(p')$ by any single- or multiparticle state $Y$
with appropriate quantum numbers; transition GPDs then appear in the
factorization formula.  Similarly one may replace the outgoing meson
$M$ by a system $X$ of particles and the meson DA by the appropriate
GDA, see \cite{Freund:1999xg} for a more detailed discussion.  What
counts for factorization is that the invariant masses $M_Y$ and $M_X$
of these systems remain fixed in the limits (\ref{compton-limit}) or
(\ref{meson-limit}).  In practice this means $M_Y^2, M_X^2 \ll W^2$ so
that $Y$ is well separated in rapidity from the outgoing $\gamma^*$ or
$X$.

\begin{figure}
\begin{center}
	\leavevmode
	\epsfxsize=0.68\textwidth
	\epsfbox{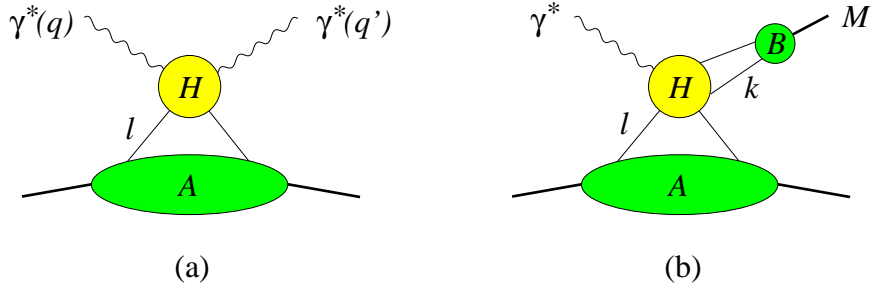}
\end{center}
\caption{\label{fig:factorize} Graphs for the factorization formulae of
Compton scattering (a) and meson production (b).  An arbitrary number
of additional gluons with suitable polarizations can connect $H$ with
$A$ and with $B$.  Diagrams with soft subgraphs as in
Fig.~\protect\ref{fig:soft} are discussed in the text.}
\end{figure}

Let us discuss some important points in the construction of the
factorization theorem, following the approach of
\cite{Collins:1998be,Collins:1999yw}.  Although it technically uses
Feynman diagrams, the key elements are general principles (gauge
invariance, dimensional analysis, boost properties and analyticity of
scattering amplitudes) and should hold beyond perturbation theory.
\begin{itemize}
\item  The  factorization formulae are associated with the graphs
shown in Fig.~\ref{fig:factorize}, which provide the leading power
behavior of the limit (\ref{compton-limit}) or (\ref{meson-limit}).
These ``reduced graphs'' specify both the topology of the relevant
Feynman graphs and regions of loop momentum space.  Lines in subgraph
$A$ are collinear to the incoming and outgoing proton, lines in
subgraph $B$ are collinear to the outgoing meson, while lines in $H$
have large components in both the plus- and minus directions and are
thus far off-shell.
\item To leading power accuracy one can Taylor expand the hard
subgraph $H$ in those components of its external momenta that are
small on the scale of $Q$.  For Fig.~\ref{fig:factorize}a we have
\begin{equation}
  \label{coll-approx}
\int d^4 l\, H(l) A(l) \approx 
     \int d l^+\, H(l) \Big|_{l^-=0, \tvec{l}=0}\,
     \int d l^-\, d^2 \tvec{l}\, A(l) .
\end{equation}
Writing $A(l)$ as the Fourier transform of a position space operator
sandwiched between proton states, we then obtain a light-cone
operator, as shown in (\ref{off-shell}).  On the other hand, the hard
scattering $H$ is now evaluated with external partons on shell and
strictly collinear.  For meson production one also expands $H$ around
$k^+=0$ and $\tvec{k}=0$.  The collinear expansion has important
consequences, in particular the helicity selection rules we will
shortly discuss.
\item The operators for the external partons in the subgraphs $A$
and $B$ are time ordered, whereas those in the definition of parton
distributions and distribution amplitudes are not.  It is therefore
crucial that the time ordering of the operators can be dropped in the
integral $\int d l^- A(l)$ and analogously in $\int d k^+ B(k)$, as we
discussed in Section~\ref{sec:light-cone}.
\item In addition to the lines shown in Fig.~\ref{fig:factorize} there
can be an arbitrary number of collinear gluons with polarization along
the plus-direction between $H$ and $A$, and an arbitrary number of
collinear gluons with polarization along the minus-direction between
$H$ and $B$.  Using Ward identities, these graphs can be summed so
that Wilson lines appear in the definitions of GPDs and DAs, whereas
the hard scattering $H$ is evaluated with the minimum number of
external lines shown in Fig.~\ref{fig:factorize}.  At this stage one
also obtains that the operators associated with the subgraphs $A$ and
$B$ are gauge invariant.
\item The integrals over momentum fractions in the factorization
formulae (\ref{factorize-compton}) and (\ref{factorize-meson}) involve
configurations where lines in $H$ are not far off-shell.  A simple
example is the point $x = \rho$ in the Born level diagram of the
Compton amplitude in Fig.~\ref{fig:DDVCS}a (which corresponds to
$x=\xB$ in the forward case describing inclusive DIS).  The quark
struck by the initial photon has zero plus-momentum and (in the
collinear approximation) is on shell for these configurations.  The
corresponding pole provides the imaginary part of the scattering
amplitude via Feynman's $i\epsilon$ in the propagator.  The important
point is that one can avoid these poles by deforming the $x$
integration in the complex plane into a region where the plus-momenta
in $H$ are of order $Q$.  In contrast this is not possible for the
lines joining $H$ to the other subgraphs: the integration path in
complex loop momenta becomes pinched between different singularities
in the limit $Q\to \infty$ and hence cannot be deformed.  In this
sense the lines going into the collinear subgraphs represent partons
with ``truly'' small virtualities, whereas on-shell configurations
inside the hard scattering graph $H$ (obtained in particular when
calculating its imaginary part by cutting lines) do not.  A physically
intuitive representation of reduced diagrams as in
Fig.~\ref{fig:factorize} is obtained from the Coleman-Norton theorem
\cite{Coleman:1965aa}, which states that lines pinched in this manner
correspond to the space-time trajectories of a particle in a
classically allowed scattering process.

\begin{figure}
\begin{center}
	\leavevmode
	\epsfxsize=0.75\textwidth
	\epsfbox{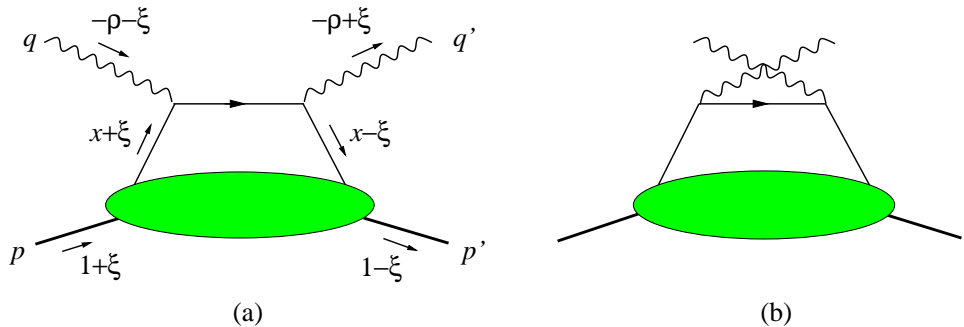}
\end{center}
\caption{\label{fig:DDVCS} Born graphs for the Compton amplitude with
independent photon virtualities, with plus-momentum fractions
referring to the average proton momentum $\half (p+p')$.}
\end{figure}

\item In addition to the graphs in Fig.~\ref{fig:factorize} there are
  reduced graphs with a further soft subgraph (whose momenta have all
  components small compared with $Q$) connected to any other subgraph.
  Important examples are shown in Fig.~\ref{fig:soft}.  Other examples
  have the form of Fig.~\ref{fig:factorize}b with an additional soft
  subgraph connected to $A$ and $B$ by soft gluons.  For a detailed
  classification see \cite{Collins:1997fb,Collins:1998be}.  Such
  configurations correspond to nonperturbative cross talk between the
  target and the produced photon and meson, and an important piece
  of the factorization proofs is to show that they do not occur in the
  final result.
  
  Note that configurations of this type occur in the momentum
  integrals of the factorization formulae at the transition $x \approx
  \pm\xi$ between ERBL and DGLAP regions and in the endpoint regions
  $z\approx 0$ or $1$ of the meson DA.  Examples are shown in
  Fig.~\ref{fig:soft-examples}.  Consider for instance the region in
  the DVCS graph of Fig.~\ref{fig:soft-examples}a where the momentum
  $l$ becomes soft.  The horizontal quark line in the diagram is then
  of low virtuality and no longer belongs to a hard subgraph (and
  neglecting its transverse momentum as is done in the hard scattering
  subprocess is no longer justified).  The reduced graph of this
  configuration is then of the form of Fig.~\ref{fig:soft} rather than
  of Fig.~\ref{fig:factorize}.  As shown in the factorization
  proof~\cite{Collins:1998be}, the contribution from the region where
  $l^+ l^- \sim \tvec{l}^2 \sim \Lambda^2$ is power suppressed and may
  hence be included in the integral over $x$ in the factorization
  formula to leading power accuracy.  Here $\Lambda$ stands for a
  typical hadronic scale, say 1~GeV.  The region where $\tvec{l}^2
  \sim \Lambda^2$ while $l^+$ and $l^-$ are of order $\Lambda^2 /Q$
  does contribute to leading power, but in this case one can deform
  the integration over $l^+$ in the complex plane to a region where
  $l^+ \sim Q$, as mentioned in the previous point.

\begin{figure}
\begin{center}
	\leavevmode
	\epsfxsize=0.68\textwidth
	\epsfbox{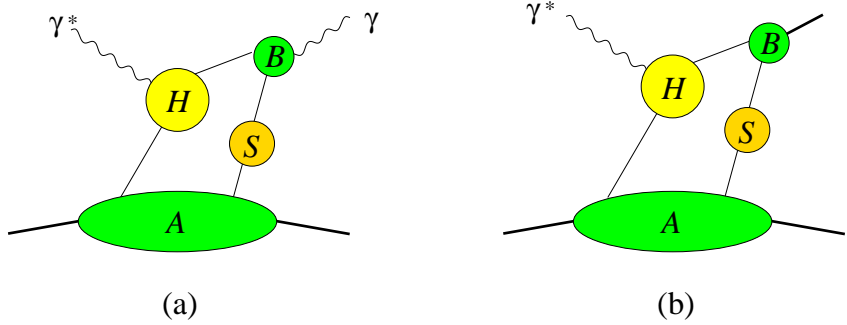}
\end{center}
\caption{\label{fig:soft} Examples of graphs with extra soft subgraphs
  for DVCS (a) and meson production (b).}
\end{figure}

\begin{figure}
\begin{center}
	\leavevmode
	\epsfxsize=0.85\textwidth
	\epsfbox{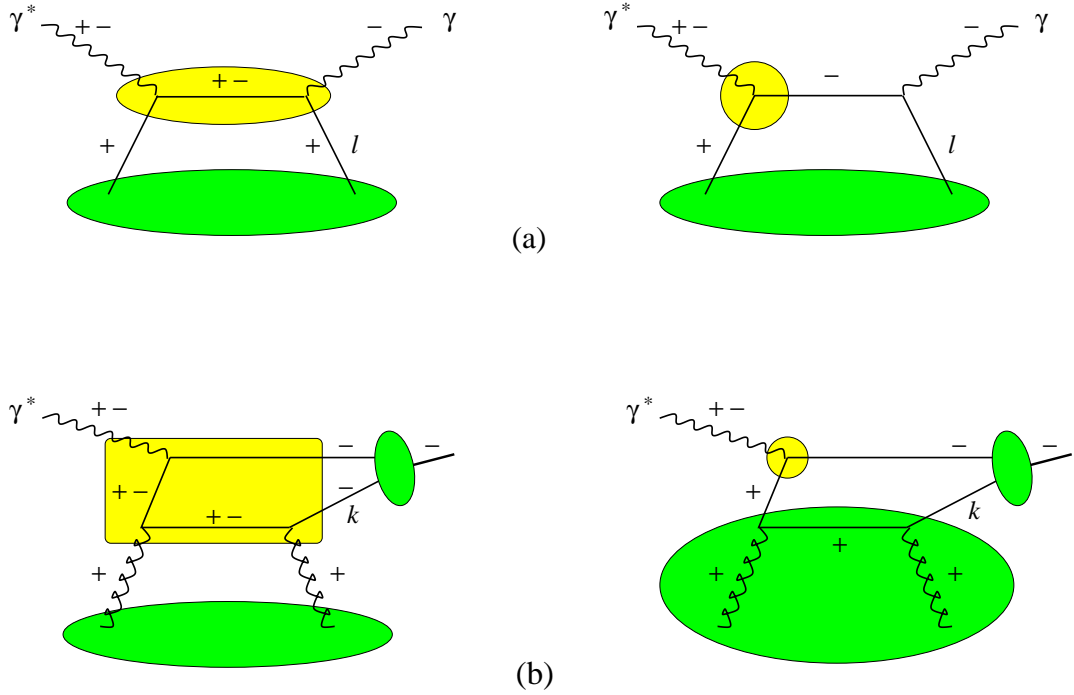}
\end{center}
\caption{\label{fig:soft-examples} Feynman diagrams whose reduced
graphs are either as in Fig.~\protect\ref{fig:factorize} or as in
Fig.~\protect\ref{fig:soft}.  A $+$ or $-$ next to a line indicates
large plus- or minus-momenta.  Lines with both $+$ and $-$ are hard,
and lines without labels are soft and constitute the soft subgraphs
here.  The external proton lines are omitted for simplicity.}
\end{figure}

For meson production one finds \cite{Collins:1997fb} that with
longitudinal $\gamma^*$ the soft region contributes to the amplitude
only as $1/Q^2$ and is hence power suppressed compared with the
dominant behavior $1/Q$.  Gauge invariance is important to establish
this: the soft region can provide leading $1/Q$ contributions for
individual Feynman diagrams, but these cancel in the sum over all
diagrams.  A similar finding in the small-$x$ approximation was made
by Hebecker and Landshoff \cite{Hebecker:1998rv}.  Another important
ingredient of the argument is the possibility to deform the
integration contour in the region $k^+ k^- \ll \tvec{k}^2 \sim
\Lambda^2$ and relies heavily on the fact that in the initial state
there is no hadron moving in the direction of the final meson.  This
would fail in processes where there are oppositely moving hadrons in
both the initial and the final state (see
Section~\ref{sec:hadron-hadron}).  If the $\gamma^*$ in meson
production is transverse, the amplitude is power suppressed.
Contributions at level $1/Q^2$ come from \emph{both} the collinear and
the soft region, i.e.\ from both $k^- \sim Q$ and $k^- \sim \Lambda$.
\item For DVCS and TCS only one of the two photons is far off-shell in
the limit (\ref{compton-limit}) and one may wonder whether using the
pointlike photon-quark coupling is adequate.  We have already remarked
that configurations where one of the quarks coupling to the photon is
soft (so that the other has small virtuality) are power suppressed or
can be avoided by complex contour deformation.  Configurations where
both quarks coupling to the photon are approximately collinear to it
(and will thus be subject to strong interactions among themselves) are
described by reduced graphs as in meson production
(Fig.~\ref{fig:factorize}b).  These are power suppressed compared with
the leading diagrams for Compton scattering
(Fig.~\ref{fig:factorize}a).
\end{itemize}

The hard scattering kernels in (\ref{factorize-compton}) and
(\ref{factorize-meson}) have important general properties which we now
discuss.  The hard scattering diagrams depend on the photon and parton
momenta, but not on the average proton momentum $P$, to which the
momentum fractions $x$, $\xi$, $\rho$ refer.  Thus one can write the
Compton amplitude as
\begin{eqnarray}
  \label{scaled-compton}
\mathcal{A}(\gamma^*p\to \gamma^* p)
 &=& \sum_q \int_{-1}^1 \frac{dx}{\xi}\,
  t^q\left(\frac{x}{\xi}, \frac{\rho}{\xi}, 
  \log(Q^2-Q'^2) \right) F^q(x,\xi,t) 
 + \Big( t^q \to \tilde{t}^q,\, F^q \to \tilde{F}^q \Big)
\nonumber \\
&& \hspace{-0.5em} {}+ \int_{-1}^1 \frac{dx}{\xi}\,
  t^g\left(\frac{x}{\xi}, \frac{\rho}{\xi}, 
  \log(Q^2-Q'^2) \right) 
  \frac{1}{\xi} F^g(x,\xi,t)
 + \Big( t^g \to \tilde{t}^g,\, F^g \to \tilde{F}^g \Big)  .
\nonumber \\
\end{eqnarray}
It is understood that the scale restoring the proper dimension in the
logarithms is $\mu_F^2$ or $\mu_R^2$.  The overall factor of $1/\xi$
can be traced back to rewriting $d l^+ = P^+ d x$, and for gluons
there is a further factor of $1/\xi$ originating from the explicit
$1/P^+$ in the definition (\ref{gluon-gpd}) of gluon GPDs.  For meson
production one similarly has
\begin{eqnarray}
\mathcal{A}(\gamma^*_L\, p\to M_L^{\phantom{*}}\, p) 
&=& \frac{1}{Q}
\int_{-1}^1 \frac{dx}{\xi} \, \int_0^1 dz\, \left[ \sum_{qj}
  t^{qj}\left(\frac{x}{\xi},z, \log Q^2 \right)  
  F^q(x,\xi,t)\, \Phi^j(z) \right.
\nonumber \\
&& \left. \hspace{5.9em} {}+ \sum_{j}
  t^{gj}\left(\frac{x}{\xi},z, \log Q^2 \right)  
  \frac{1}{\xi} F^g(x,\xi,t)\, \Phi^j(z)  \right]
  \label{scaled-meson}
\end{eqnarray}
for mesons with natural parity, and an analogous relation with
$\tilde{F}^q$ and $\tilde{F}^g$ for mesons with unnatural parity.  The
factorization formulae predict the scaling of the Compton amplitude
for large $Q^{(\prime)}$ at fixed $\xi$, $\rho$ and $t$, up to
logarithmic corrections.  This is the exact analog of Bjorken scaling
in DIS.  For meson production, the dominant amplitude is predicted to
decrease as $1/Q$ for large $Q$ at fixed $\xi$ and $t$, again up to
logarithms.

The forms (\ref{scaled-compton}) and (\ref{scaled-meson}) have
immediate consequences on the $\xi$ or $(\xi, \rho)$ dependence of the
amplitude.  If the quark or gluon GPDs at small $\xi$ scale as in
(\ref{xi-scaling}), $H^g(x,\xi) = \xi^{-\lambda}\, \hat{H}^g(x/\xi)$
or $H^q(x,\xi) = \xi^{-(1+\lambda)}\, \hat{H}^q(x/\xi)$, their
contribution to the Compton amplitude gives a power law
\begin{equation}
  \label{compton-scaling}
\mathcal{A}_{q,g} =
\xi^{-(1+\lambda)} C_{q,g}(\rho / \xi)
\end{equation}
for $\xi \ll 1$.  At fixed $Q^2$ and $Q'^2$ this translates into a
high-energy behavior as $W^{2 (1+\lambda)}$.  For meson production
amplitudes the result is analogous, with $C_{q,g}(\rho/\xi)$
replaced by constants.  Such a power behavior was explicitly found in
numerical studies, see
\cite{Mankiewicz:1998uy,Freund:2001rk,Belitsky:2001ns}. 

Resonance exchange contributions (\ref{resonance-xchange}) to GPDs
also lead to a power behavior in $\xi$ times a function of $\rho/
\xi$.  Replacing in particular $\tilde{F}^{q}$ in (\ref{scaled-compton})
with the pion pole contribution (\ref{pion-pole}) to $\tilde{E}^{q}$
gives a term $\xi^{-1} C_{\pi}(\rho / \xi)$.  According to (\ref{hel})
the distribution $\tilde{E}^q$ appears in amplitudes with at least one
factor of $\xi$, which leads to at most a constant high-energy
behavior of the amplitude at fixed $Q^2$ and $Q'^2$, as it must be for
spin-zero exchange.  Finally, replacing $F^q$ or $F^g$ in
(\ref{scaled-compton}) with the $D$-term contributions (\ref{dd-red})
or (\ref{dd-red-glue}) to $H^q$, $E^q$ or $H^g$, $E^g$ gives a result
depending only on $\rho/\xi$ for both quarks and gluons.  Further
$\xi$ dependence comes from the appropriate prefactors in the proton
matrix elements (\ref{hel}).

\subsubsection{Selection rules}
\label{sub:selection}

Because the hard scattering kernels express collinear scattering, the
helicities of incoming and outgoing particles must balance to ensure
conservation of angular momentum $J^3$.  Together with chirality
conservation in the hard scattering (where light quark masses are set
to zero) this leads to selection rules for the photon (or meson)
helicities.  It can be shown that the axial anomaly of QCD does not
invalidate chirality conservation in hard scattering coefficients
\cite{Collins:1999un}.  For Compton scattering this allows at leading
power the transitions summarized in Table~\ref{tab:compton}.  The
transition $\gamma^*_L \to \gamma^*_L$ is of course absent if one of
the photons is real.  In DDVCS this transition only starts at one-loop
level, as does its forward analog, the longitudinal structure function
$F_L$ of inclusive DIS.  For the hadronic tensor
(\ref{compton-tensor}) at leading order in $1/Q$ and in $\alpha_s$
this implies a nonforward analog of the Callan-Gross relation $F_2 =
2x F_1$ between deep inelastic structure functions
\cite{Blumlein:1999sc,Blumlein:2000cx}.  Note that in the $\gamma^*_L
\to \gamma^*_L$ transition $\tilde{H}^{q,g}$ and $\tilde{E}^{q,g}$ do
not appear because of parity invariance.  Photon helicity flip by two
units can be balanced by gluon helicity flip GPDs, and thus also
occurs at one-loop level.  Amplitudes with one longitudinal and one
transverse photon are suppressed by a power $1/Q^{(\prime)}$ and
explicitly appear at twist-three level (see
Section~\ref{sub:dvcs-three}).

\begin{table}
\caption{\label{tab:compton}Allowed photon helicity transitions in
Compton scattering at leading power in $1/Q$ or $1/Q'$, together with
the relevant GPDs and the order of the hard scattering where they
occur.  LO corresponds to Born level and NLO to $O(\alpha_s)$.
Further allowed transitions are obtained by reversing both helicities.
For each parton species $H$ stands for $H$, $E$ and $\tilde{H}$ for
$\tilde{H}$, $\tilde{E}$, whereas $H_T$ represents all four
transversity distributions.}
\begin{center}
\vspace{1ex}
\leavevmode
\renewcommand{\arraystretch}{1.3}
\begin{tabular}{llcc} \hline
$\gamma^*\to$  & $\gamma^*$ &
  LO           &        NLO      \\ \hline
$+$ & $+$ & $H^{q(+)}, \tilde{H}^{q(+)}$ & 
            $H^{q(+)}, \tilde{H}^{q(+)},\, H^g,\, \tilde{H}^g$ \\
$0$ & $0$ & & $H^{q(+)},\, H^g$ \\
$+$ & $-$ & & $H_T^g$ \\ \hline
\end{tabular}
\renewcommand{\arraystretch}{1}
\end{center}
\end{table}

The selection rules for meson production are more intricate.  To make
the symmetry between the proton and meson side in the factorization
formula explicit, it is useful to consider the GPDs in the ERBL
region, where they have the parton kinematics of a distribution
amplitude.  The hard scattering is then the same as for an elastic
meson form factor or a meson transition form factor, with the incoming
partons moving to the right and the outgoing ones to the left in the
Breit frame.

The leading-twist operators for DAs and GPDs describe coupling of the
parton pair to helicity $0$, helicity $\pm 1$ (with quark
transversity), or helicity $\pm 2$ (with gluon transversity).  Because
of chirality conservation in the hard scattering, chiral-odd DAs and
GPDs must occur in pairs.  This was originally thought to offer a way
for measuring quark transversity GPDs in the production of a
transversely polarized meson \cite{Collins:1998be}.  The hard
scattering kernel is however zero, as was explicitly checked at
leading \cite{Mankiewicz:1998uy} and next-to-leading
\cite{Hoodbhoy:2001da} order in $\alpha_s$.  It holds in effect to all
orders \cite{Diehl:1998pd,Collins:1999un}: quark helicity conservation
would force the initial and final quark pair to have opposite total
helicity in the Breit frame, and a photon cannot mediate a change of
$J^3$ by two units.  

The matrix elements of the tensor operator $\mathbf{S} G^{+i} G^{j+}$
cannot appear either.  Helicity conservation in the hard scattering
would only allow them in pairs, but one cannot have a gluon GPD and a
gluon GDA in the same diagram: two gluons couple only to $C=+$ and a
photon does not couple to the transition between two $C$-even states.
One is thus left with transitions $q\bar{q}\to q\bar{q}$, $gg\to
q\bar{q}$, or $q\bar{q}\to gg$ from helicity 0 to helicity 0.  This
also establishes that only $\gamma^*_L$ amplitudes are leading in
$1/Q$.  The corresponding statement for meson form factors is the
well-known hadron helicity conservation in hard exclusive scattering
processes
\cite{Brodsky:1981kj}.

The parity constraints for the hard scattering with a $\gamma_L^*$
require the operators of the DAs and the GPDs to have the same parity,
so that natural parity mesons only go with $H$, $E$ and unnatural
parity ones only with $\tilde{H}$, $\tilde{E}$.  In Table
\ref{tab:mesons} we list the relevant combinations for several
mesons.  This compilation is not complete.\footnote{Detailed tables
including quark flavor combinations are given by Goeke et
al.~\cite{Goeke:2001tz}.  In their Table~3 the entries in the first
column correctly read $(2u + \bar{u}) - (2d + \bar{d})$, $(2u -
\bar{u}) - (2d - \bar{d})$, $(2\Delta u + \Delta\bar{u}) - (2\Delta d
+ \Delta\bar{d})$, $(2\Delta u - \Delta\bar{u}) - (2\Delta d -
\Delta\bar{d})$ \protect\cite{Polyakov:2001pr}.}
The selection of GPDs and DAs for $a_1$ mesons is the same as for
pions, and for $f_1$ the same as for the $\eta$, $\eta'$ system.  The
combinations for $a_0$ and $a_2$ coincide with those for $f_0$ and
$f_2$ except that there is no $\Phi^g$.  For neutral $b_1$ and $h_1$
mesons one has access to $\tilde{H}^{q(+)}$, and for charged $b_1$ to
$\tilde{H}^{q(+)} \pm 3 \tilde{H}^{q(-)}$.  One can however expect
experimental study of these mesons to be difficult because they have
many-body decay modes or poorly known branching ratios.

\begin{table}
\caption{\label{tab:mesons}Combinations of GPDs and DAs entering in
selected meson electroproduction channels at leading order in $1/Q$
and in $\alpha_s$.  The same results hold with appropriate GDAs if the
meson is replaced by a many-body system with appropriate quantum
numbers.  For each parton species $H$ stands for $H$, $E$ and
$\tilde{H}$ for $\tilde{H}$, $\tilde{E}$.  In the charged meson
channels the isospin relations (\protect\ref{isospin-pn}) have been
used to relate the $p\to n$ and $p\to n$ GPDs to the flavor diagonal
ones in the proton.}
\begin{center}
\vspace{1ex}
\leavevmode
\renewcommand{\arraystretch}{1.3}
\begin{tabular}{llcc} \hline
$J^{PC}$ or $J^{P}$ & meson & GPD & DA \\ \hline
$1^{--}$ & $\rho^0$, $\omega$, $\phi$ & $H^{q(+)}, H^g$ & $\Phi^q$ \\
$0^{++}, 2^{++}$ & $f_0, f_2$ & $H^{q(-)}$ & $\Phi^q, \Phi^g$\\
$0^{-+}$ & $\pi^0$, $\eta$, $\eta'$ & $\tilde{H}^{q(-)}$ & $\Phi^q$ \\
\hline
$1^-$ & $\rho^\pm$ & $H^{q(+)} \pm  3 H^{q(-)}$ & $\Phi^q$ \\ 
$0^-$ & $\pi^\pm$  & 
        $\tilde{H}^{q(-)} \pm 3 \tilde{H}^{q(+)}$ & $\Phi^q$ \\ \hline
\end{tabular}
\renewcommand{\arraystretch}{1}
\end{center}
\end{table}

One may expect that for isoscalar mesons with unnatural parity the
gluon DA, going with the operator $ G^{+\mu}\, \tilde{G}_{\mu}{}^{+}$
also enters.  At leading order in $\alpha_s$ the corresponding
collinear scattering kernel is however zero, as was found in a study
of $\eta$ and $\eta'$ production \cite{Kroll:2002nt}.  At one-loop
level the gluon DA should appear explicitly in the hard scattering,
since it mixes with the quark DA under evolution.  The symmetry
between the meson and proton side in the hard scattering implies that
likewise the gluon GPD $\tilde{H}^g$ appears in $h_1$ production only
at $O(\alpha_s^2)$.
 
{}From Tables~\ref{tab:compton} and \ref{tab:mesons} we see that
access to $H^g$ and $E^g$ is provided in Compton scattering at NLO and
in the production of neutral vector mesons.  $\tilde{H}^g$ and
$\tilde{E}^g$ can only be accessed Compton scattering at NLO (assuming
that a NLO analysis of $h_1$ production is not practical).
Possibilities to access the polarized gluon GPDs in processes beyond
leading twist have been proposed and will be discussed in
Section~\ref{sec:beyond-twist-two}.  Access to the $C$-even
combinations of quark GPDs is given in Compton scattering, with each
quark flavor weighted by its squared charge $e_q^2$.  {}From
\begin{equation}
  \label{DVCS-iso}
\sum_q e_q^2 H^q = \frac{5}{18} H^{u+d} + \frac{3}{18} H^{u-d} +
\frac{1}{9} H^s
\end{equation}
and the analog for the other GPDs one sees in particular that the
isoscalar combination is slightly enhanced over the isotriplet one by
the charge weighting.

Information for flavor separation of $H^{q(+)}$ and $E^{q(+)}$ may be
obtained from vector meson production, and for $\tilde{H}^{q(+)}$,
$\tilde{E}^{q(+)}$ from charged pseudoscalar production at small
$\xi$, where one may expect that $|\tilde{H}^{q(+)}| \gg
|\tilde{H}^{q(-)}|$ in analogy with the forward densities.  The
$C$-odd combinations of quark GPDs may be studied in the relevant
neutral meson channels.  A different possibility to access
combinations of GPDs not occurring in Compton scattering is in
principle given by replacing photons in the Compton process by weak
gauge bosons, but this has not been studied so far.  Finally, flavor
separation of $u$ and $d$ quark GPDs could be obtained with nuclear
targets by scattering on a neutron instead of a proton (see
Section~\ref{sec:nuclei}).  In analogy to the situation for flavor
separation of inclusive parton densities this requires sufficient
knowledge of the nuclear wave functions.


\subsection{Compton scattering}
\label{sec:compton-scatt}

Compton scattering is the most thoroughly studied process involving
GPDs, with most work focused on DVCS so far.  The process is
conveniently analyzed in terms of the hadronic tensor
\begin{equation}
  \label{compton-tensor}
T^{\alpha\beta} = i \int d^4x\, 
  e^{i (q+q') x/2} \langle p'| T J^\alpha_{\mathrm{em}}(-\half x) \,
    J^\beta_{\mathrm{em}}(\half x) | p\rangle ,
\end{equation}
where $e J^\alpha_{\mathrm{em}}(x)$ is the electromagnetic current.
The helicity amplitudes for $\gamma^* p\to \gamma^* p$ are then given
by
\begin{equation}
  \label{hel-amp-compton}
e^2 M_{\lambda'\mu', \lambda\mu} = 
e^2\, \epsilon_\alpha^{\phantom{*}} 
    T^{\alpha\beta} \epsilon'^*_\beta
\end{equation}
where $\epsilon$ ($\epsilon'$) and $\mu$ ($\mu'$) denote the
polarization and helicity of the initial (final) photon, and $\lambda$
($\lambda'$) the helicity of the initial (final) proton.  {}From
parity invariance one has
\begin{equation}
  \label{Compton-parity}
M_{-\lambda'-\mu', -\lambda-\mu} =
(-1)^{\lambda'-\mu' - \lambda+\mu}\, M_{\lambda'\mu', \lambda\mu}
\; ,
\end{equation}
provided one takes a frame where the initial and final proton momenta
are in the $x$-$z$ plane, which we will assume from now on.  We
further choose a frame where both protons move fast into the positive
$z$ direction.

To leading order in $1/Q$ one can write
\begin{equation}
  \label{leading-order-compton}
T^{\alpha\beta}
 = - g_T^{\alpha\beta} \mathcal{F}
   - i \epsilon_T^{\alpha\beta} \tilde{\mathcal{F}} + \ldots ,
\end{equation}
where the $\ldots$ indicate terms which appear at $O(\alpha_s)$ and go
with the photon helicity combinations $(\mu'\mu) = (00)$, $(+-)$ or
$(-+)$.  The handbag diagrams in Fig.~\ref{fig:DDVCS} give
\begin{eqnarray}
  \label{compton-integrals}
\mathcal{F}(\rho,\xi,t) &=& \sum_q e_q^2
\int_{-1}^1 dx\, F^q(x,\xi,t)\, 
    \left( \frac{1}{\rho - x - i\epsilon}
           - \frac{1}{\rho + x - i\epsilon} \right) + O(\alpha_s),
\nonumber \\
\tilde{\mathcal{F}}(\rho,\xi,t) &=& \sum_q e_q^2
\int_{-1}^1 dx\, \tilde{F}^q(x,\xi,t)\, 
    \left( \frac{1}{\rho - x - i\epsilon}
           + \frac{1}{\rho + x - i\epsilon} \right) + O(\alpha_s) ,
\end{eqnarray}
where at $O(\alpha_s)$ one has also contributions from $F^g$ in
$\mathcal{F}$ and from $\tilde{F}^g$ in $\tilde{\mathcal{F}}$.
Following the convention of Belitsky et
al.~\cite{Belitsky:2000gz,Belitsky:2001ns} we further define
$\mathcal{H}$, $\mathcal{E}$ by replacing $F^{q,g}$ with $H^{q,g}$ or
$E^{q,g}$, and $\tilde{\mathcal{H}}$, $\tilde{\mathcal{E}}$ by
replacing $\tilde{F}^{q,g}$ with $\tilde{H}^{q,g}$ or
$\tilde{E}^{q,g}$ in (\ref{compton-integrals}).  

To obtain explicit amplitudes we choose polarization vectors
\begin{equation}
  \label{photon-pol}
\epsilon(\pm) = \frac{1}{\sqrt{2}} (0,\, \mp 1,\, i,\, 0)
\end{equation}
for both photons when their momentum points in the negative $z$
direction in the photon-proton c.m., and correspondingly if they have
a small momentum in the $x$ direction.  Since deviations from strict
collinearity are suppressed by $1/Q$ in the hard scattering, one then
has
\begin{equation}
  \label{polaris-products}
- \epsilon_\alpha\, g_T^{\alpha\beta} \epsilon'^*_\beta = 1 ,
\qquad
-i \epsilon_\alpha\, \epsilon_T^{\alpha\beta} \epsilon'^*_\beta 
	= \mu
\end{equation}
for $\mu=\mu'=\pm 1$, up to corrections in $1/Q$.  Explicitly one
obtains from (\ref{hel}) and (\ref{leading-order-compton})
\begin{eqnarray}
  \label{compton-two-gpd}
M_{++,++} &=& \sqrt{1-\xi^2} \left( \mathcal{H} + \tilde{\mathcal{H}}
    - \frac{\xi^2}{1-\xi^2} (\mathcal{E} + \tilde{\mathcal{E}})
  \right) ,
\nonumber \\
M_{-+,-+} &=& \sqrt{1-\xi^2} \left( \mathcal{H} - \tilde{\mathcal{H}}
    - \frac{\xi^2}{1-\xi^2} (\mathcal{E} - \tilde{\mathcal{E}})
  \right) ,
\nonumber \\
M_{++,-+} &=& \frac{\sqrt{t_0-t}}{2m}\,
        (\mathcal{E} - \xi \tilde{\mathcal{E}}) ,
\phantom{\Bigg( \Big)}
\nonumber \\
M_{-+,++} &=& {}- \frac{\sqrt{t_0-t}}{2m}\,
        (\mathcal{E} + \xi \tilde{\mathcal{E}}) .
\end{eqnarray}
These amplitudes correspond to the phase conventions of the light-cone
helicity spinors (\ref{spinors}) in a frame where the incoming proton
momentum $p$ points along the positive $z$ axis, whereas the
components of the outgoing proton momentum satisfy $p'^1 \le 0$, $p'^2
=0$, $p'^3 >0$.  Up to $1/Q$ suppressed terms, the amplitudes are the
same for the usual helicity spinors in the $\gamma^* p$ c.m., where
both protons move fast (see Appendix~\ref{app:spinors}).

In the case where the final or initial state photon is real, the
imaginary part of the amplitudes is given by GPDs at the boundaries $x
= \pm \xi$ of the ERBL and DGLAP regions.  The same is true for vector
meson production.  In contrast, with two nonzero photon virtualities
one has access to the quark GPDs evaluated at fixed $x = \pm \rho$ in
the ERBL region, since kinematics enforces $|\rho| \le \xi$.  This
unique possibility of DDVCS has been emphasized already in
\cite{Radyushkin:1997ki}.

The hard scattering kernels in (\ref{compton-integrals}) provide a
simple relation between DVCS and TCS at leading order in $\alpha_s$.
{}From
\begin{equation}
  \label{convolution-rel}
\mathcal{H}(-\xi,\xi,t) = \Big[ \mathcal{H}(\xi,\xi,t) \Big]^* ,
\qquad \qquad
\tilde{\mathcal{H}}(-\xi,\xi,t) 
	= - \Big[ \tilde{\mathcal{H}}(\xi,\xi,t) \Big]^* ,
\end{equation}
and the analogs for $\mathcal{E}$ and $\tilde{\mathcal{E}}$ one finds
that at leading-twist accuracy
\begin{equation}
   \label{dvcs-tcs-rel}
M_{\lambda'\mu, \lambda\mu} \,\Big|_{\mathrm{TCS}} =
  \Big[ M_{\lambda' -\mu, \lambda -\mu} \,\Big]^*_{\mathrm{DVCS}} 
+ O(\alpha_s) 
\end{equation}
with $\mu=\pm 1$, where it is understood that the amplitudes
are evaluated at the same $\xi$ and $t$ and at equal values of $Q^2$
and $Q'^2$, which set the factorization scale in the GPDs.  In this
approximation the amplitudes for DVCS and for TCS thus carry the same
information.  Relations similar to (\ref{convolution-rel}) and
(\ref{dvcs-tcs-rel}) hold for $|\rho| < \xi$ and connect the
amplitudes for DDVCS at equal $(Q^2 + Q'^2)$ and opposite $(Q^2 - Q'^2)$.
Verification of the relations (\ref{dvcs-tcs-rel}) would constitute a
model independent test of leading-twist dominance and the Born level
approximation for the Compton amplitude.  At $O(\alpha_s)$ accuracy
these relations no longer hold, neither for the NLO corrections to the
quark handbag diagrams nor for the diagrams involving gluon GPDs.

The hard scattering kernels at NLO have been calculated by several
groups.  They can be found in
\cite{Belitsky:1998rh,Mankiewicz:1998bk,Ji:1998nk,Ji:1998xh}\footnote{
The journal version of \protect\cite{Mankiewicz:1998bk} contains a
typo in the coefficient of $\tilde{F}^g$, which is corrected in
version 3 in the hep-ph archive.}
for the photon helicity transition $(\mu'\mu) = (++)$, and in
\cite{Mankiewicz:1998bk} for $(\mu'\mu) = (00)$.  These results are
given for independent $\rho$ and $\xi$, explicit kernels for DVCS can
be found in~\cite{Belitsky:1999sg}.\footnote{The global factor
$1/|\eta|$ in eq.~(6) of \protect\cite{Belitsky:1999sg} should be
omitted, and for gluon kernels it must read $\mp (t\to -t)$ instead of
$\pm (t\to -t)$ in eq.~(9) \protect\cite{Muller:2000pr}.}
The helicity flip case $(\mu'\mu) = (+-)$ has been calculated in
\cite{Hoodbhoy:1998vm,Belitsky:2000jk}.  For DVCS one finds a rather
simple result \cite{Diehl:2001pm}
\begin{eqnarray}
M_{\lambda'\mu', \lambda\mu} &=& 
- \frac{\alpha_s}{4\pi} \sum_q e_q^2 \int_{-1}^1 
  \frac{dx}{x}\, A^g_{\lambda'\mu', \lambda\mu}(x,\xi,t)\,
      \left( \frac{1}{\xi - x - i\epsilon}
           - \frac{1}{\xi + x - i\epsilon} \right) 
\end{eqnarray}
for $(\mu'\mu) = (+-)$ and $(-+)$, with the phase conventions
(\ref{photon-pol}).  Here $A^g$ stands for the combinations of gluon
transversity GPDs given in Section~\ref{sec:helicity-transitions}.
The term in brackets is proportional to $x$, so that the integrand has
no $1/x$ singularity.

Beyond leading order in $\alpha_s$, both real and imaginary parts of
the Compton amplitude involve GPDs in integrals over $x$.  For DVCS
the imaginary part of the amplitude involves only the DGLAP regions,
with $x\ge \xi$ corresponding to the $s$-channel cut and $x\le -\xi$
to the $u$-channel cut of the parton-photon subprocess.  The same
holds for light meson production amplitudes.  If the final state
photon in the Compton process is timelike the situation is more
complicated since there are additional cuts in $q'^2$, and the
imaginary part of the amplitude is no longer related in a simple way
to the discontinuities in the external invariants.

Numerical studies of DVCS amplitudes at NLO accuracy have been
performed by Belitsky et al.\ in~\cite{Belitsky:1999sg} and later in
\cite{Belitsky:2001ns}, and by Freund and McDermott 
\cite{Freund:2001hm,Freund:2001rk,Freund:2001hd}.  Both groups find
a significant dependence of the size and pattern of NLO corrections on
the input GPDs, so that general conclusions are to be taken with some
caution.  The input dependence in
\cite{Freund:2001hm,Freund:2001rk,Freund:2001hd} is generally larger
in the small-$\xB$ region and larger for the real than for the
imaginary part of the amplitude.  With identical input GPDs at a given
scale, NLO corrections tend to decrease the unpolarized DVCS cross
section.  For a factorization scale $\mu = Q$, both groups find that
even at moderate $\xB \sim 0.1$ the contributions from the gluon GPDs
are not always negligible.  On the other hand even at small $\xB
\sim 10^{-3}$ the contribution from quark GPDs cannot be ignored.
This is in contrast to vector meson production, where quarks and
gluons enter at the same order in $\alpha_s$, so that at small $\xB$
the gluon contribution should strongly dominate over quarks.  Note
that for DVCS at NLO the distinction between quark and gluon
contributions depends on the factorization scale and scheme (the
$\overline{\mathrm{MS}}$ scheme has been used in all studies).
Belitsky et al.~\cite{Belitsky:2001ns} have pointed out that the
contribution of gluon GPDs at NLO can be made very small by choosing a
scale which for their model GPDs was about $\mu \approx 3.8\, Q$.  A
physics reason for taking such a high $\mu$ is however not obvious.
Comparing on the other hand the NLO results obtained with $\mu^2 =
Q^2$, $\mu^2 = 2Q^2$ and $\mu^2 = Q^2/2$ the studies
\cite{Belitsky:1999sg} and
\cite{Freund:2001rk} found rather moderate effects.

We remark that the one-loop kernels going with gluon GPDs are so far
only available for massless quarks.  This leads to an uncertainty in
the range of $\xB$ and $Q^2$ where the collision energy is large
enough for charm to be important but $Q^2$ is not so large that the
charm quark mass can be neglected in the loop.  This is presumably the
case for the kinematical region of the DVCS measurements of H1 and
ZEUS \cite{Adloff:2001cn,Zeus:2003ip}.  Notice for comparison that the
charm contribution to the inclusive proton structure function $F_2$ is
in the 10\% to 20\% range for the same values of $\xB$ and $Q^2$
\cite{Breitweg:1999ad,Adloff:2001zj}.


\subsection{Meson electroproduction}
\label{sec:meson-lt}

Meson production processes offer the possibility to study various
aspects of GPDs that are unaccessible in Compton scattering, in
addition to being sensitive to the structure of the produced meson
itself.  Dedicated phenomenological studies of meson production at
leading $1/Q$ accuracy have been performed by Mankiewicz et al.\ for
neutral~\cite{Mankiewicz:1998uy} and charged
\cite{Mankiewicz:1997aa} mesons, by Frankfurt et
al.\ for pseudoscalar mesons~\cite{Frankfurt:1999fp} and for processes
with a $\Delta$ or a strange baryon in the final
state~\cite{Frankfurt:1999xe}, by Eides et al.~\cite{Eides:1998ch} and
by Mankiewicz et al.~\cite{Mankiewicz:1998kg} for neutral
pseudoscalars, and by Vanderhaeghen et
al.~\cite{Vanderhaeghen:1998uc,Vanderhaeghen:1999xj}.

To leading order in $\alpha_s$ the diagrams involving quark
GPDs and quark DAs give a contribution
\cite{Mankiewicz:1998uy,Mankiewicz:1997aa,Vanderhaeghen:1998uc,Frankfurt:1999fp}  
\begin{eqnarray}
  \label{meson-1}
\mathcal{A}_1 &=& e\, \frac{16\pi \alpha_s}{9} \frac{1}{Q}
\int_0^1 dz\, \sum_{q q'} \Phi^{qq'}(z)\,
\int_{-1}^1 dx\, F^{q'\!q}(x,\xi,t)
\left[ \frac{1}{\bar{z}}\, \frac{e_q}{\xi -x -i\epsilon} -
       \frac{1}{z} \frac{e_{q'}}{\xi + x -i\epsilon} \right]
\end{eqnarray}
to the amplitude for a longitudinal $\gamma^*$ and a longitudinal
meson with natural parity.  Here $\Phi^{qq'}(z)$ is defined as in
(\ref{vector-da}) with the operator $\bar{q} \gamma^+ q'$, so that $z$
is the momentum fraction of the quark $q$ and $\bar{z}= 1-z$ the
momentum fraction of the antiquark $\bar{q}^{\,\prime}$.  For
$F^{q'\!q}$ the relevant operator is $\bar{q}^{\,\prime} \gamma^+ q$,
so that the momentum fraction $x+\xi$ belongs to flavor $q$ and
$x-\xi$ to flavor $q'$.  The expression for mesons with unnatural
parity is analogous with the appropriate DA and $\tilde{F}^{q'\!q}$
instead of $F^{q'\!q}$.  If the meson quantum numbers admit there is
in addition a contribution with either gluon GPDs
\cite{Mankiewicz:1998uy,Lehmann-Dronke:1999aq,Lehmann-Dronke:2000xq}, 
\begin{eqnarray}
  \label{meson-2}
\mathcal{A}_2 &=& e\, \frac{\pi \alpha_s}{3} \frac{1}{Q}
\int_0^1 dz\, \sum_{q} e_q \frac{\Phi^q(z)}{z\bar{z}} \,
\int_{-1}^1 dx\, \frac{F^g(x,\xi,t)}{x}  
\left[ \frac{1}{\xi - x -i\epsilon} -
                        \frac{1}{\xi + x -i\epsilon} \right]
\end{eqnarray}
or with the gluon DA of the meson
\cite{Lehmann-Dronke:1999aq,Lehmann-Dronke:2000xq},
\begin{eqnarray}
  \label{meson-3}
\mathcal{A}_3 &=& e\, \frac{4\pi \alpha_s}{3} \frac{1}{Q}
\int_0^1 dz\, \frac{\Phi^g(z)}{z\bar{z}} \,
\int_{-1}^1 dx\, \sum_{q} e_q F^q(x,\xi,t)  
\left[ \frac{1}{\xi - x -i\epsilon} +
                     \frac{1}{\xi + x -i\epsilon} \right] .
\end{eqnarray}
Note that to leading order in $1/Q$ and in $\alpha_s$ the meson
structure only enters through a few constants.  For each quark flavor
combination one needs $\int dz\, (z\bar{z})^{-1}\, \Phi$ and $\int
dz\, (z\bar{z})^{-1}\, (z-\bar{z})\, \Phi$, one of which is zero if
the meson is a $C$ eigenstate.  If applicable there is a further
constant $\int dz\, (z\bar{z})^{-1}\, \Phi^g$ from the gluon DA.
Relations between DAs for different quark flavors follow from flavor
symmetry.  Isospin invariance gives for instance $\Phi_{\pi^+}^{ud} =
\Phi_{\pi^-}^{du} = \sqrt{2}\, \Phi_{\pi^0}^{u} = - \sqrt{2}\,
\Phi_{\pi^0}^{d}$ and corresponding relations for the $\rho$
mesons.\footnote{Note that the relative sign between DAs for different
isospin states is convention dependent, see Appendix~A of
\protect\cite{Diehl:2000uv}.}
In analogy to GPDs we use the shorthand notation $\Phi^{u-d} =
\Phi^u - \Phi^d$ etc.

Let us discuss some aspects of the phenomenology following from these
expressions, and of the possibilities to extract information on GPDs
and meson structure from various channels.  We caution that in
experimental observables one has no direct access to the amplitudes
discussed here, but rather to their squares or interference terms,
summed over appropriate proton spin combinations (see
Section~\ref{sec:meson-pheno}).

For the neutral vector mesons $\rho^0$, $\omega$, $\phi$ both quark
and gluon GPDs contribute.  The ratio of their production amplitudes
is
\begin{eqnarray}
  \label{vector-ratios}
\mathcal{A}_{\rho^0} : \mathcal{A}_{\omega} : \mathcal{A}_{\phi} 
&=& \int_{-1}^1 \frac{dx}{\xi - x - i\epsilon}
        \left( \frac{2 F^{u(+)} + F^{d(+)}}{\sqrt{2}} 
        + \frac{9}{8\sqrt{2}}\, \frac{F^g}{x} \right)
\nonumber \\
&:&  \int_{-1}^1 \frac{dx}{\xi - x - i\epsilon}
        \left( \frac{2 F^{u(+)} - F^{d(+)}}{\sqrt{2}} 
        + \frac{3}{8\sqrt{2}}\, \frac{F^g}{x} \right)
\nonumber \\
&:&  \int_{-1}^1 \frac{dx}{\xi - x - i\epsilon}
     \left( - F^{s(+)} - \frac{3}{8}\, \frac{F^g}{x} \right) ,
\end{eqnarray}  
if one assumes that their respective $q\bar{q}$ content is
$\frac{1}{\sqrt{2}} ( |u\bar{u}\rangle - |d\bar{d}\rangle )$,
$\frac{1}{\sqrt{2}} ( |u\bar{u}\rangle + |d\bar{d}\rangle )$ and
$|s\bar{s}\rangle$, and that their distribution amplitudes are related
as $\Phi_{\rho}^{u-d}(z) = \Phi_\omega^{u+d}(z) = \sqrt{2}\,
\Phi_\phi^s(z)$.\footnote{These relations do not simply follow from
SU(3) flavor symmetry.  They neglect in addition the mixing of
$q\bar{q}$ with gluons, which induces e.g.\ differences between
$\Phi_{\rho}^{u-d}$ and $\Phi_\omega^{u+d}$.}
This hypothesis reproduces rather well the combinations $(M_{V
\rule{0pt}{1.3ex}} \Gamma_{V\to e^+e^-})^{1/2}$ of mass and partial
leptonic width for these mesons, which are proportional to $\sum_q
\int dz\, e_q \Phi^q(z)$.  Measurement gives $(M_{\rho
\rule{0pt}{1.3ex}} \Gamma_{\rho\to e^+e^-}) : (M_{\omega
\rule{0pt}{1.3ex}} \Gamma_{\omega\to e^+e^-}) : (M_{\phi
\rule{0pt}{1.3ex}} \Gamma_{\phi\to e^+e^-}) \approx 9 : 0.8 : 2.2$,
compared with the ratios $9 : 1 : 2$ obtained under the above
assumptions.  With (\ref{vector-ratios}) the same ratios are obtained
for the electroproduction cross sections $\sigma_{\rho^0}$,
$\sigma_{\omega}$, $\sigma_{\phi}$ in the region where the gluon GPDs
dominate over those for quarks.  This is expected and found in the
data at small $\xB$, see Section~\ref{sec:small-x-mesons}.  At larger
values of $\xB$ one expects the quark contributions to become
important, at least for the $\rho^0$ and the $\omega$.  How important
the respective contributions are at intermediate $\xB$ and from which
point quarks dominate is not clear at present.  Existing studies by
Vanderhaeghen et al.~\cite{Vanderhaeghen:1998uc,Vanderhaeghen:1999xj}
(see also \cite{Goeke:2001tz}) have added the cross sections modeled
for quark and gluon GPDs but not taken into account their
interference.  For the gluon contributions they furthermore used a
model by Frankfurt et al.~\cite{Frankfurt:1996jw}, which specifically
used approximations for the region of very small $\xB$ (see
Section~\ref{sec:small-x-mesons}).

In a region of $\xB$ and $Q^2$ where one can argue for the dominance
of quark GPDs for $\rho$ and $\omega$ production (possibly by
comparison with the $\phi$ channel) the cross section ratio
$d\sigma_{\omega}/ d\sigma_{\rho^0}$ in the approximation of
(\ref{vector-ratios}) gives rather direct information about the
relative size of unpolarized $u$ and $d$ quark GPDs.  Note that quark
dominance and the simple assumption $F^u : F^d = 2:1$ yield a ratio
$d\sigma_{\rho^0} : d\sigma_{\omega} = 25 : 9$, which is significantly
smaller than in the gluon dominated region, as already observed in
\cite{Collins:1997fb}.  

The production of charged mesons involves a combination of $C$-even
and $C$-odd GPDs as seen in Table~\ref{tab:mesons}.  In analogy to
forward parton distributions one expects that at small $\xB$ the
$C$-even combinations will become dominant, so that for instance the
cross sections for $\gamma^* p\to \rho^+ n$ and $\gamma^* n\to \rho^-
p$ or their analogs for $\pi^+$ and $\pi^-$ production will become
equal.  In a model study \cite{Mankiewicz:1997aa} for $\rho^\pm$
production this was found to hold for $\xB \lsim 0.1$.  Using the
isospin relations $\Phi_{\rho^+}^{ud} = \Phi_{\rho^-}^{du} =
\sqrt{2}\, \Phi_{\rho^0}^{u}$ and those in (\ref{isospin-pn}) relating
the nucleon transition GPDs to the flavor diagonal ones, one finds the
three $\rho$ channels connected by
\begin{eqnarray}
  \label{rho-ratios}
\mathcal{A}_{\rho^0} : \mathcal{A}_{\rho^+} : \mathcal{A}_{\rho^-} 
&=& \int_{-1}^1 \frac{dx}{\xi-x-i\epsilon}
        \left( \frac{2 F^{u(+)} + F^{d(+)}}{\sqrt{2}} 
        + \frac{9}{8\sqrt{2}}\, \frac{F^g}{x} \right)
\\
&:&  \int_{-1}^1 \frac{dx}{\xi-x-i\epsilon}
        \left( \frac{F^{u(+)} - F^{d(+)}}{2} 
            + \frac{3F^{u(-)} - 3F^{d(-)}}{2} \right)
\nonumber \\
&:&  \int_{-1}^1 \frac{dx}{\xi-x-i\epsilon}
        \left( \frac{F^{u(+)} - F^{d(+)}}{2} 
            - \frac{3F^{u(-)} - 3F^{d(-)}}{2} \right) .
\nonumber 
\end{eqnarray}  
Together with the relation (\ref{vector-ratios}) between $\rho^0$ and
$\omega$ production we see that these four channels offer in principle
enough observables to separate $F^g$, $F^{u(+)}$, $F^{d(+)}$, and
$F^{u(-)}- F^{d(-)}$.  The simple estimate $F^u : F^d = 2:1$ also
suggests that the cross section for charged $\rho$ production is
significantly smaller than for $\rho^0$ production, even in a region
where the latter is not enhanced by gluon exchange.

Note that the vector meson channels channels offer the possibility to
obtain flavor information beyond the combination (\ref{DVCS-iso}) from
Compton scattering for the distributions $H^{q(+)}$ and $E^{q(+)}$,
which enter in Ji's sum rule (\ref{Ji-sum}).  $\phi$ production is one
of the very few known processes where one could separately access
$F^{s(+)}$, although in direct competition with $F^{g}$, apart
possibly from $K^*$ production (see below) and neutrino induced
processes (Section~\ref{sub:heavy-light}).

The production of $\pi^0$ involves the $C$-odd combination $2
\tilde{F}^{u(-)} + \tilde{F}^{d(-)}$ of polarized quark GPDs.  Note
that the pion pole contribution to $\tilde{E}$ is $C$-even and does
not appear in this process.  In the $\eta$, $\eta'$ system there is a
nontrivial pattern of mixing and flavor SU(3) breaking, related to the
U(1) axial anomaly of QCD, for reviews we refer to
\cite{Feldmann:1999uf,Feldmann:2002kz}.  Studies of electroproduction
focusing on this issue have been performed in \cite{Eides:1998ch} and
\cite{Kroll:2002nt}.  Neglecting isospin violation, one has a set
$\Phi_{\eta}^{u+d}$, $\Phi_{\eta}^{s}$ and $\Phi_{\eta'}^{u+d}$,
$\Phi_{\eta'}^{s}$ of independent quark DAs.  In the naive SU(3) limit
one has $\Phi_{\eta}^{u+d} = - \Phi_{\eta}^{s}$ and
$\Phi_{\eta'}^{u+d} = 2 \Phi_{\eta'}^{s}$.  As we mentioned in
Section~\ref{sub:selection} the gluon DAs of $\eta$ and $\eta'$ enter
in electroproduction only through mixing in the LO evolution and at
order $\alpha_s^2$ in the hard scattering.  The relative weight of the
quark DAs is given by
\begin{equation}
  \label{eta-electro}
\mathcal{A}_{\eta} \propto 
\int_{-1}^1 \frac{dx}{\xi-x -i\epsilon} \int_0^1 \frac{dz}{z\bar{z}}\, 
\left[ \Big( 2 \tilde{F}^{u(-)} - \tilde{F}^{d(-)} \Big)  
        \Phi_\eta^{u+d} 
     - 2 \tilde{F}^{s(-)} \Phi_\eta^s \right] 
\end{equation}
for the $\eta$.  This involves the same $z$ integral as the process
$\gamma^* \gamma \to \eta$ at leading order in $1/Q$ and $\alpha_s$
(see Section~\ref{sub:gamma-star-gamma}):
\begin{equation}
  \label{eta-photo}
\mathcal{A}(\gamma^* \gamma \to \eta) \propto
\int_0^1 \frac{dz}{z\bar{z}}\, 
\left[ 5 \Phi_\eta^{u+d} + 2 \Phi_\eta^s \right] ,
\end{equation}
Relations analogous to (\ref{eta-electro}) and (\ref{eta-photo}) hold
for the $\eta'$.  We see that electroproduction and two-photon
annihilation probe different quark flavor combinations of the DAs.
The forward limit of $\tilde{F}^{s(-)}$ is $\Delta s - \Delta
\bar{s}$, so that one may plausibly neglect $\tilde{F}^{s(-)}$ in
(\ref{eta-electro}), especially at larger values of $\xi$
\cite{Eides:1998ch}.  The electroproduction ratio $d\sigma_\eta /
d\sigma_{\eta'}$ then gives rather direct information about the
relative size of $\Phi^{u+d}_\eta$ and $\Phi^{u+d}_{\eta'}$, which is
independent from what is measured in $\gamma^*\gamma$ annihilation.
Conversely, particular scenarios for $\eta$, $\eta'$ mixing and for
the shape of their DAs can be tested in $d\sigma_\eta /
d\sigma_{\eta'}$ without uncertainties from the nucleon GPDs if the
above approximations are valid.  With knowledge about the relative
size of $\eta$, $\eta'$ and $\pi^0$ quark DAs as input one could
finally obtain information about the relative size of
$2\tilde{F}^{u(-)} - \tilde{F}^{d(-)}$ and $2\tilde{F}^{u(-)} +
\tilde{F}^{d(-)}$.

\begin{figure}[b]
\begin{center}
	\leavevmode
	\epsfxsize=0.82\textwidth
	\epsfbox{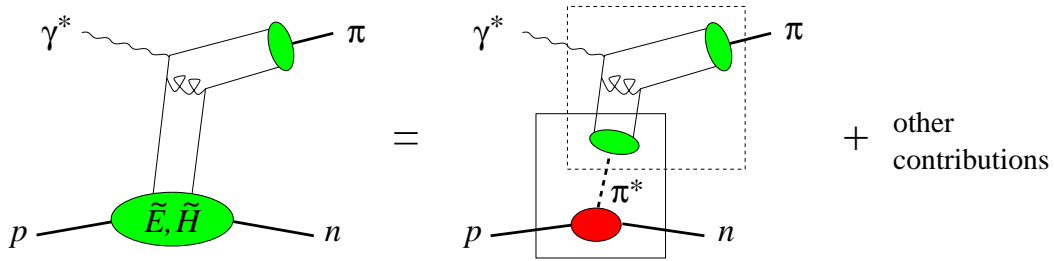}
\end{center}
\caption{\label{fig:pion-pole-process} Pion exchange contribution to
$\gamma^* p \to \pi^+ n$.  The dashed box gives the pion form factor
at leading twist when $t$ is analytically continued to $m_\pi^2$.}
\end{figure}

Charged pion production receives a contribution from the pion pole
part (\ref{pion-pole}) of $\tilde{E}^{u-d}$, which according to
phenomenological estimates
\cite{Mankiewicz:1998kg,Vanderhaeghen:1999xj} strongly dominates the
cross section for sufficiently large $\xB$ and small $t$.  This pole
contribution to the amplitude can be written as
\begin{equation}
  \label{pion-pole-process}
\mathcal{A}_{\pi^+}^{\mathrm{pole}} =
- \mathcal{A}_{\pi^-}^{\mathrm{pole}} =
-e\, \frac{Q F_\pi(Q^2)}{f_\pi}\, \frac{2 m\, g_A(0)}{m_\pi^2-t}\,
   \bar{u}(p') \gamma_5 u(p) 
\end{equation}
with $f_\pi \approx 131$~MeV, where 
\begin{equation}
   \label{pion-ff}
F_\pi(Q^2) = \frac{2\pi \alpha_s}{9}\, \frac{f_\pi^2}{Q^2}\, 
  \left[ \, \int_0^1 dz\, \frac{\phi_\pi(z)}{z\bar{z}} \right]^2
\end{equation}
is the electromagnetic pion form factor to leading order in $1/Q$ and
$\alpha_s$ in the hard scattering picture \cite{Brodsky:1989pv}, which
is the exact analog for form factors of the factorization scheme
discussed in Section~\ref{sec:factor}.  The relation
(\ref{pion-pole-process}) readily generalizes to higher orders in
$\alpha_s$.  As in (\ref{pion-da-normal}) we have used the pion
distribution amplitude $\phi_\pi(z)$ normalized to unit integral.  The
physical picture behind (\ref{pion-pole-process}) is that the nucleon
emits a pion (whose off-shellness is neglected as one approaches the
pion pole) on which the photon scatters, see
Fig.~\ref{fig:pion-pole-process}.  This is the basis of the attempt to
measure the electromagnetic pion form factor in $\gamma^* p\to \pi^+\,
n$.  In the large $Q^2$ limit the GPD formalism offers ways to study
the extent to which the pion pole dominates the amplitude in given
kinematics.  As we have emphasized in Section~\ref{sub:chiral-soliton}
the chiral soliton model estimates important non-pole contributions to
$\tilde{E}^{u-d}$ if $t$ is not small enough.  The minimal
kinematically allowed value of $|t|$ is $m^2 \xB^2 /(1-\xB)$ at large
$Q^2$, which favors measurement at smaller $\xB$.  On the other hand,
as $\xB$ decreases the contributions from $\tilde{H}^{u-d}$ are
expected to become increasingly important, in analogy with the
increase of the forward polarized quark densities with $\xB$.  In
contrast the pion pole contribution (\ref{pion-pole-process}) does not
grow with decreasing $\xB$.  A model estimate of the relative size of
the two contributions has been given by Mankiewicz et
al.~\cite{Mankiewicz:1998kg}, who in particular point out the
importance of an adequate description of the $t$-dependence of
$\tilde{H}$.  We note that the physical process corresponding to the
contribution from $\tilde{H}$ and $\tilde{E}$ in the DGLAP region,
including the boundary $x=\xi$ which provides the imaginary part of
the amplitude at leading $O(\alpha_s)$, has been discussed long ago by
Carlson and Milana \cite{Carlson:1990zn}.  To obtain information on
the size of non-pole contributions in $\pi^\pm$ production from the
$\pi^0$ channel is unfortunately difficult, given the different
combinations of quark flavor and $C$-parity for the GPDs.  On the
other hand, information on the relative size of the contributions from
$\tilde{H}$ and $\tilde{E}$ in the $\pi^\pm$ channels can be obtained
with transverse target polarization \cite{Frankfurt:1999fp}, as we
will discuss in Section~\ref{sec:meson-pheno}.  We remark in passing
that the pion pole term in $\tilde{E}^{u-d}$ also contributes to DVCS.
In this case it has however to compete not only with contributions
from $\tilde{H}$ but above all from $H$, which is expected to be
larger than $\tilde{H}$ in analogy with the situation for the forward
densities (where this follows from positivity).  In addition, the
squared quark charges favor the isosinglet over the isotriplet GPDs in
DVCS, see (\ref{DVCS-iso}).

In addition to the production of pions from spacelike photons, one may
consider the production of a timelike photon with large virtuality
$Q'^2$ from charged pion beams, with the $\gamma^*$ decaying into an
$e^+e^-$ or $\mu^+\mu^-$ pair \cite{Berger:2001zn}.  This process
factorizes in analogy to its spacelike counterpart if the photon is
longitudinally polarized, which can be studied using the angular
distribution of the lepton pair.  To leading order in $1/Q^{(\prime)}$
and in $\alpha_s$ the amplitudes of space- and timelike processes are
closely related~\cite{Berger:2001zn},
\begin{eqnarray}
\mathcal{A}(\pi^- p\to \gamma^* n) &=& 
  \Big[ \mathcal{A}(\gamma^* p \to \pi^+ n) \Big]^* \, , 
\nonumber \\
\mathcal{A}(\pi^+ n\to \gamma^* p) &=& 
  \Big[ \mathcal{A}(\gamma^* n \to \pi^- p) \Big]^* \, , 
\end{eqnarray}
where both sides are evaluated for equal proton and neutron
helicities, for equal values of $\xi$ and $t$, and with $Q'^2$ in the
timelike processes equal to $Q^2$ in the spacelike ones.  These
relations will be modified by corrections in $\alpha_s$, in a similar
way as in Compton scattering, and also by higher twist effects.
Comparison of the channels may thus provide indications for the size
of such corrections, and more generally on the relation between
analogous space- and timelike processes.  The timelike channels
receive a pion pole contribution, which reads like
(\ref{pion-pole-process}) with the pion form factor in the timelike
region.  This can directly be measured in $e^+e^-\to \pi^+\pi^-$
(although the available data \cite{Bollini:1975pn} at high $Q'^2$ has
large experimental errors).  Good data for both processes would
provide a rather unique chance to directly test pion pole dominance,
which does not seem experimentally feasible in the spacelike case for
large $Q^2$.

Let us finally point out some specialties of strangeness production,
studied in \cite{Frankfurt:1999xe} and reviewed in detail in
\cite{Goeke:2001tz}.  Due to flavor SU(3) breaking, the distribution
amplitudes for neutral and charged kaons cannot be expected to be
symmetric in $z\leftrightarrow \bar{z}$, in contrast to pions and
$\rho$ mesons.  The relative contribution of $C$-even and $C$-odd GPDs
then depends on the relative size of the integrals $\int dz\,
(z\bar{z})^{-1}\, \Phi$ and $\int dz\, (z\bar{z})^{-1}\, (z-\bar{z})\,
\Phi$.  As a consequence one finds kaon pole contributions 
not only for $\gamma^* p\to K^+ \Sigma^0$ and $\gamma^* p\to K^+
\Lambda$ but also for $\gamma^* p\to K^0 \Sigma^+$.  This pole
contribution in the relevant flavor transition GPDs $\tilde{E}$ has a
form similar to the pion pole contribution (\ref{pion-pole}) in the
nucleon sector.  Due to the large mass splitting between $\pi$ and $K$
one expects a substantial violation of the flavor SU(3) relations for
$\tilde{E}$, which as in (\ref{strange-relations}) relate the
strangeness transition GPDs with the flavor diagonal ones for the
proton.  We finally note that when using these relations for the $H$
and $E$ distributions, independent flavor information about $s$ quarks
in the nucleon may in principle be obtained in the production of
$K^*(892)$ mesons, without the presence of gluon distributions.

\subsubsection{Beyond leading order in $\alpha_s$}
\label{sub:NLO-pions}

An important question is how well the Born level expressions we have
discussed so far approximate the physical scattering amplitudes.  The
only NLO study to date has been performed by Belitsky and M\"uller
\cite{Belitsky:2001nq} for $\gamma^* p\to \pi^+ n$.  The
hard-scattering kernel can be taken over from the known results for the
electromagnetic pion form factor by appropriate analytic continuation.
In fact, for the pion pole contribution to $\tilde{E}$, which is
believed to dominate the cross section at larger $\xB$, the NLO
corrections are the same in both cases, as expressed in
(\ref{pion-pole-process}).  Since the meson production amplitude is
proportional to $\alpha_s$ already at leading order, the choice of
renormalization scale $\mu_R$ is of particular importance for the
result.  One particular way of setting this scale is the BLM procedure
\cite{Brodsky:1983gc}, which in the case at hand can be implemented by
choosing $\mu_R$ such that the term multiplying $\beta_0 = 11 -
\frac{2}{3} n_f$ in the NLO expressions vanishes.  With the model they
used for $\tilde{H}$, Belitsky and M\"uller found a small scale
$\mu_{\mathrm{BLM}}^2$ below $0.1\, Q^2$ in a wide range of $\xB$.
The BLM scale is particularly small for the pion form factor and hence
for the pion pole part of $\tilde{E}$.  For the asymptotic pion DA one
finds $\mu_{\mathrm{BLM}}^2 \approx 0.01\, Q^2$.  At such low scales
one can of course not evaluate the perturbative running coupling for
practically relevant values of $Q^2$, and the study
\cite{Belitsky:2001nq} froze $\alpha_s /\pi$ to a value of $0.1$ for
$\mu_R \le 1$~GeV.  The result was a considerable size of the NLO
corrections, and a huge spread of the NLO result when comparing the
results for the BLM scale and the choice $\mu_R^2 = Q^2$.  We do
however not think that, as it stands, this implies the process to be
beyond perturbative control at any $Q^2$ available in experiment.  The
intimate connection between the $\log(Q/\mu_R)$ dependence of the
one-loop kernels and the running of $\alpha_s(\mu_R)$ in the tree
level term is destroyed when modifying $\alpha_s$ for
non-perturbative effects.  Another point is that, as discussed in
\cite{Belitsky:2001nq,Bakulev:2000uh},  the NLO kernel contains large
terms which do not go with $\beta_0$, so that BLM scale setting cannot
render the size of one-loop corrections small.  We remark that in a
study of the pion form factor, Meli\'c et al.~\cite{Melic:1998qr} have
found that for $\mu_R^2 \approx 0.05\, Q^2$ and the asymptotic pion DA,
the NLO corrections are small and show little variation when $\mu_R^2$
is changed.\footnote{The choice $\mu_R^2 = e^{-5/3}\, Q^2 /4
\approx 0.05\, Q^2$ is referred to as BLM scale in
\protect\cite{Melic:1998qr}, but the scale that renders the $\beta_0$
dependent term in the NLO amplitude zero for the asymptotic DA is
$\mu_R^2 = e^{-14/3}\, Q^2 \approx 0.01\, Q^2$.}
For moderate $Q^2$ this choice, which has also been discussed in
\cite{Bakulev:2000uh}, still leads to scales where $\alpha_s$ is large
and one may doubt the applicability of perturbation theory.  For $Q^2
\gsim 10$~GeV the total NLO result for $F_\pi(Q^2)$ in
\cite{Melic:1998qr} was found to have important but not extremely
large variation when $\mu_R^2$ was chosen between $Q^2$ and $0.05\,
Q^2$.  NLO corrections not only affect the overall size of the cross
section but also its dependence on kinematical variables and spin.
The study by Belitsky and M\"uller reports that the transverse target
spin asymmetry in $\gamma^* p\to \pi^+ n$ (see
Section~\ref{sec:meson-pheno}) is rather stable under NLO corrections.

Notice that the Born level expressions (\ref{meson-1}) to
(\ref{meson-3}) can be written as an integral over $x$ involving the
GPDs times an integral over $z$ involving the DAs, or as a sum of two
such terms if the DA is neither symmetric nor antisymmetric under $z
\leftrightarrow \bar{z}$.  This simple property is lost at the level
of $\alpha_s^2$ corrections, where there is a nontrivial interplay
between $x$ and $z$ in the hard scattering kernels.  In other words,
the meson structure determines how the nucleon GPDs are probed, and
the process amplitude can have different $\xi$ and $t$ dependence for
one and the same set of GPDs but different choices of the meson DA.
How important this cross talk is has however not been studied so far.
Whether and how a trustworthy perturbative description of $F_\pi(Q^2)$
and meson electroproduction can be obtained for achievable $Q^2$, and
how well different observables can be theoretically controlled,
remains to be further investigated.


\subsection{Meson pair production and GDAs}
\label{sub:meson-pair-prod}

GDAs appear in a variety of hard scattering processes.  We will
discuss in some detail their occurrence in two-photon annihilation and
in hard electroproduction.  They also appear in decay processes where
a large scale is set by a heavy quark mass.  Studies have been
performed for the decays of $J/\Psi$ or $\Upsilon$ into $\gamma\,
\pi^+\pi^-$ \cite{Ma:2001qq} and $\ell^+\ell^-\, \pi^+\pi^-$
\cite{Ma:2001ar}, which at leading order in $\alpha_S$ are sensitive
only to the gluon GDA $\Phi^g$, and for exclusive semileptonic
\cite{Maul:2001zn,Ma:2001ar} and hadronic
\cite{Chen:2002th} decays of $B$ mesons.

\subsubsection{Two-photon annihilation}
\label{sub:gamma-star-gamma}

The analog of Compton scattering in the crossed channel is two-photon
annihilation
\begin{equation}
  \label{gamma-annihilation}
\gamma^*(q) + \gamma^*(q') \to h(p) + \bar{h}(p') ,
\end{equation}
where $h$ can be either a baryon or a meson.  In analogy to the
Compton case one can show factorization of this process
\cite{Freund:1999xg} in the limit
\begin{equation}
|q^2| + |q'^2| \to \infty  \qquad \mbox{at fixed~} q'^2 /q^2,\, W^2,\,
\cos\theta , 
\end{equation}
where $W^2 = (p+p')^2$ is the squared invariant mass of the hadronic
system and $\theta$ the polar angle of the hadron $h$ in the c.m.  If
both photons are spacelike this process can be observed in $e^+e^-$
collisions (Section~\ref{sec:gaga}), and we define $Q^2 = -q^2$ and
$Q'^2 = -q'^2$.  As in the Compton channel one may have both photons
off shell or only one of them.  The scaling behavior and helicity
selection rules are analogous to those of Compton scattering.
Amplitudes with both photons transverse or longitudinal are leading,
and those with one transverse and one longitudinal photon are power
suppressed by $1/Q^{(\prime)}$.

Studies in the literature have so far focused on the channels
$\gamma^* \gamma \to \pi^+\pi^-$ and $\gamma^* \gamma \to
\pi^0\pi^0$.  To lowest order in $1/Q$ and $\alpha_s$ the hadronic
tensor is given by
\begin{eqnarray}
  \label{gamma-gamma-two-pi}
T^{\alpha\beta} &=& i \int d^4 x\, e^{i (q-q') x /2} \, 
\langle \pi(p) \pi(p')| \, T J^\alpha_{\mathrm{em}}(-\half x) \,
  J^\beta_{\mathrm{em}}(\half x) \, | 0\rangle
\nonumber \\
 &=& {}- \frac{1}{2}\, g_T^{\alpha\beta}\, \sum_q e_q^2\int_0^1 dz\,
    \frac{2z-1}{z(1-z)}\, \Phi^q(z,\zeta,W^2)
\end{eqnarray}
in terms of the quark GDAs for the pion pair.  For pairs of hadrons
with spin like $p\bar{p}$ or $\rho\rho$ there are also terms with the
tensor $\epsilon_T^{\alpha\beta}$.  Both $g_T^{\alpha\beta}$ and
$\epsilon_T^{\alpha\beta}$ filter out the photon helicity combinations
$(\mu'\mu) = (++)$ and $(\mu'\mu) = (--)$ in the collision c.m., so
that the hadrons are produced in a state with total $J^3=0$ along the
photon axis.  The hard scattering kernels at $O(\alpha_s)$ can be
obtained from those of Compton scattering by crossing and are
explicitly given in \cite{Kivel:1999sd} for the $\pi\pi$ channel.
Modeling the two-pion DAs along the lines described in
Section~\ref{sub:gda-models} Kivel et
al.~\cite{Kivel:1999sd,Kivel:2000rq} found that the contribution from
the gluon GDA at NLO depends rather strongly on the Gegenbauer
coefficient $B^g_{12}$ in the model, which is related to the momentum
fraction taken by gluons in the pion.  The process is thus rather
sensitive to the relative strength of $q\bar{q}$ and $gg$ coupling to
a pion pair.

A different probe of the gluon coupling is given by the helicity
amplitudes with $(\mu'\mu) = (+-)$ or $(-+)$, corresponding to pion
pairs with $J^3 = \pm 2$, which to leading order in $1/Q$ receive only
contributions going with the tensor gluon GDA $\Phi_T^g$.  The
corresponding amplitudes at $O(\alpha_s)$ have been discussed in
\cite{Kivel:1999sd}.  As pointed out by Braun and Kivel
\cite{Braun:2000cs} this may also be used to investigate the gluon
content of meson resonances, for instance in $\gamma^* \gamma \to
f_2$.  Notice that in hard processes one is sensitive to $q\bar{q}$ or
$gg$ configurations at small transverse separation, which is quite
different from and complementary to the quark-antiquark or two-gluon
picture in constituent models and meson spectroscopy.

The annihilation of $\gamma^* \gamma$ into a single light pseudoscalar
meson factorizes in an analogous way for large $Q^2$, with the
hadronic tensor being proportional to
\begin{equation}
  \label{gamma-gamma-pi}
\sum_q e_q^2\int_0^1 dz\, \frac{1}{z(1-z)}\, \Phi^q(z) .
\end{equation}
Measurements by CLEO \cite{Gronberg:1998fj} for $\pi^0$, $\eta$,
$\eta'$ and by L3 \cite{Acciarri:1998yx} for $\eta'$ indicate that the
scaling behavior of the leading twist-mechanism sets in already at
moderate $Q^2$.  The data for this process gives probably the cleanest
available constraints on the $q\bar{q}$ distribution amplitudes of
these mesons and is consistent with a shape close to the asymptotic
one, $\Phi^q \propto z(1-z)$, at low factorization scale $\mu \sim
1$~GeV.  For a recent analysis and references for $\pi^0$ we refer
to~\cite{Diehl:2001dg}, and for $\eta$ and $\eta'$ to
\cite{Feldmann:1999uf,Kroll:2002nt}.  We will discuss limitations of
the leading-twist description in Section~\ref{sub:two-photon}.

Using the model two-pion DAs we discussed in
Section~\ref{sub:gda-models} it was estimated in \cite{Diehl:2000uv}
that the cross section for $\gamma^*\gamma \to \pi^0\pi^0$ integrated
over $W$ from threshold to 1~GeV is only about 10\% of the one for
$\gamma^*\gamma \to \pi^0$.  In the leading-twist mechanism the pion
pair is produced by an intermediate $q\bar{q}$ or $gg$ state, so that
it can only have total isospin $I=0$ but not $I=2$.  As a consequence
the production amplitudes for $\pi^0\pi^0$ and for $\pi^+\pi^-$ are
equal (up to a convention dependent phase).

It is instructive to compare the exclusive annihilation of $\gamma^*
\gamma$ with the inclusive annihilation  into hadrons, both in the
limit of large $Q^2$ at fixed $W^2$.  The inclusive channel is then
described by the transverse and longitudinal photon structure
functions $F^\gamma_T$ and $F^\gamma_L$ in the limit $\xB = Q^2
/(W^2+Q^2) \to 1$.  It can be calculated at partonic level as
$\gamma^* \gamma\to q\bar{q}$ to leading order in $\alpha_s$, provided
that one explicitly regulates the region where either the quark or
antiquark carry the entire momentum of the hadronic system in the
Breit frame.  In the leading-twist description of the exclusive
channels this corresponds to the endpoint regions $z\to 0,1$ and is
regulated by the behavior of the distribution amplitudes, i.e.\ by the
hadronization process.  Taking the limit of large $Q^2$ at fixed $W^2$
one finds a scaling behavior~\cite{Diehl:2000uv}
\begin{equation}
F_T^\gamma \sim c(W^2) , \qquad
F_L^\gamma \sim W^2 / Q^2
\end{equation}
with some function $c$.  This coincides indeed with the $Q^2$
dependence predicted for the exclusive channels discussed above.  The
study \cite{Diehl:2000uv} also compared the total cross section for
$\gamma^*\gamma$ calculated $(i)$ as open $q\bar{q}$ production with a
quark mass around 300~MeV as regulator (or equivalently a transverse
momentum cutoff of the same size) and $(ii)$ as the sum of
leading-twist cross sections for the exclusive channels.  Amusingly,
the inclusive $q\bar{q}$ cross section integrated up to $W=1$~GeV is
comparable in size with the sum of the exclusive channels $\pi^0$,
$\eta$, $\eta'$ and the (small) estimated $\pi\pi$ continuum.

\subsubsection{Electroproduction}
\label{sub:meson-pairs-electro}

The concept of GDAs allows one to extend meson production studies to a
wider class of processes.  Of particular practical importance is the
case of pion pairs.  When the invariant mass $m_{\pi\pi}$ of the pair
is in the vicinity of the $\rho(770)$ or $f_2(1270)$ resonances, an
analysis in terms of GPDs can be complementary to a more conventional
one in terms of single-meson states and provide additional physics
insights.

Clerbaux and Polyakov have used $\pi^+\pi^-$ electroproduction data
from H1 and ZEUS to analyze the spectrum in the two-pion invariant
mass $m_{\pi\pi}$.  As we saw in Section~\ref{sub:gda-evolution}, the
$m_{\pi\pi}$ dependence of $\Phi^{q(-)}$ is given by the timelike pion
form factor $F_\pi(m^2_{\pi\pi})$ if the $z$ dependence has its
asymptotic shape $z (1-z)$.  This predicts the invariant mass
distribution in electroproduction at asymptotically large $Q^2$ to be
\begin{equation}
  \label{rho-peak}
\frac{1}{N}\, \frac{dN}{d m^2_{\pi\pi}} 
\stackrel{\log Q^2 \to \infty}{\to}
\beta^3 \, | F_\pi(m^2_{\pi\pi}) |^2 ,
\end{equation}
where $\beta = (1 - 4 m_\pi^2 /m^2_{\pi\pi})^{1/2}$.  Note that
$|F_\pi(m^2_{\pi\pi})|$ around the $\rho$ mass is well measured from
$e^+e^-$ annihilation data.  The form (\ref{rho-peak}) shows very
clearly that leading-twist dominance does \emph{not} predict the
nonresonant ``background'' under the $\rho$ peak to vanish at large
photon virtuality, since $F_\pi$ contains a contribution from the
continuum in addition to the resonance.  Clerbaux and Polyakov find
that a fit to (\ref{rho-peak}) of the measured $m_{\pi\pi}$ spectrum
between about 550 and 1100~MeV becomes increasingly good with larger
$Q^2$ (with the highest bin being around 20~GeV$^2$).  This is
consistent with $\Phi^{q(-)}$ in this mass range being close to its
asymptotic $z (1-z)$ shape, although it might also be that the
coefficients of higher terms in the Gegenbauer expansion
(\ref{Polyakov-expansion}) have an $m_{\pi\pi}$ dependence similar to
that of $F_\pi(m^2_{\pi\pi})$.  In
\cite{Clerbaux:2000hb} an attempt was made to extract some of the
parameters describing the deviation from the asymptotic form of the
GDA, but the statistical errors in the data were too large to permit
strong conclusions.

This study was performed for the mass spectrum integrated over the
angular distribution of the pion pair.  It is known that at moderate
values of $Q^2$ the contribution from $J^3 = \pm 1$ states (such as a
transverse $\rho$) is not negligible, whereas the leading-twist
description only holds for the amplitudes with $J^3 = 0$.  The authors
of \cite{Clerbaux:2000hb} point out that the $J^3 = \pm 1$
contributions can be suppressed by taking appropriate moments in the
polar angle $\theta$ of the $\pi^+$ in the two-pion c.m.

The production of pion pairs in the $C$-even projection has been
studied by Lehmann-Dronke et
al.~\cite{Lehmann-Dronke:1999aq,Lehmann-Dronke:2000xq}, for an earlier
discussion see~\cite{Diehl:1999cg}.  The invariant mass regions with
the clearest signal for this channel, compared with the overwhelming
dominance of the $\rho$ resonance, were estimated to be close to the
$\pi\pi$ threshold and around the $f_2$ resonance.  In addition it is
advantageous to go to larger values of $\xB$, since the $C$-even
two-pion state involves the GPD combinations $F^{q(-)}$, which at
small $\xB$ are well below $F^{q(+)}$ and $F^g$ occurring in the
$C$-odd $\pi\pi$ channel.  One can however make use of the strong
$C$-odd contribution by measuring its interference with the $C$-even
channel, which can be filtered out from the $\theta$ distribution, see
Section~\ref{sub:meson-pair-pheno}.

The electroproduction of $C$-even states involves both their quark and
gluon GDAs.  It was reported in \cite{Lehmann-Dronke:2000xq} that for
the model of the GDAs described in Section~\ref{sub:gda-models} the
total amplitude is only weakly sensitive to the relative size of
$\Phi^{u+d}$ and $\Phi^g$.  It would require further study to assess
whether this process alone or together with two-photon annihilation
could be used to estimate the relative quark and gluon coupling to
pion pairs or in particular to the $f$ resonances.


\section{Beyond leading power}
\label{sec:beyond}

In order to extract twist-two GPDs and GDAs from data one must have
sufficient control over power corrections to the leading-twist
description reviewed in Section~\ref{sec:factor}.  Moreover, in
certain situations it may be possible to treat power suppressed
effects as a signal which by itself carries information about hadron
structure and dynamics at the quark-gluon level.  

In the leading-twist approximation the amplitude is factorized into a
hard and one or several collinear subgraphs.  The subgraphs are
connected by the minimal necessary number of partons, which is two in
the cases we deal with.  The hard scattering is evaluated in the
collinear approximation, and the masses of hadrons and quarks are
neglected.  The study of power corrections thus involves hadronic
matrix elements with more than two external partons, corrections to
evaluating the hard scattering for collinear on-shell partons, and
target mass corrections.  Further contributions can arise from soft
subgraphs as in Fig.~\ref{fig:soft}.  As discussed in
Section~\ref{sec:factor} they are intimately related with
configurations where partons entering the hard scattering become soft.

A systematic approach to power corrections in Compton scattering and
two-photon annihilation is the operator product expansion (OPE) of the
product of two electromagnetic currents.  Here information about
hadron structure is encoded in matrix elements of higher-twist
operators.  These operators are not all independent since they satisfy
the QCD equations of motion when physical matrix elements are taken.
In order not to over-parameterize nonperturbative physics one needs to
choose an appropriate operator basis.  Different choices are possible,
as discussed in the analysis of deep inelastic scattering by Ellis,
Furmanski and Petronzio~\cite{Ellis:1983cd}.  Since the equations of
motion involve operators with different parton content, effects due to
parton transverse momentum or off-shellness and effects due to extra
partons are related.  One hence needs to specify a particular
framework in order to state which is the ``dynamical origin'' of a
particular contribution.  Note also that gauge invariance relates
different terms.  The operators $\bar{q} \gamma^+ (i\lrpartial_{\!T})
\, q$ and $\bar{q} \gamma^+ (g A_T)  \, q$ for instance, respectively
describing quark transverse momentum and an additional gluon, mix
under gauge transformations, and only their sum is gauge invariant.

Complications arise for real photons.  A real photon can split into a
quark and antiquark that are both of low virtuality and almost
collinear to each other.  Such configurations are prone to long-range
interactions, so that the pointlike quark-photon vertex is insufficient
to represent them.  In analogy to quark-antiquark configurations in a
meson they can be described by distribution amplitudes for the photon,
which have been discussed in detail by Ball et al.~\cite{Ball:2002ps}.
Furthermore, a real photon can split into one fast near-shell quark
and a soft one.  Such end-point configurations are associated with
soft subgraphs and hence with dynamics that cannot obviously be
described by the hadronic matrix elements of the OPE.  At the level of
some subleading power one may thus expect the OPE description to break
down when one photon is on shell.  A signal of this would be the
occurrence of divergences in the convolution of the appropriate hard
scattering kernels with the appropriate parton distributions or
distribution amplitudes.

In this case, as well as for processes like exclusive meson
production, one cannot use the OPE and there is to date no systematic
way of describing the dynamics beyond leading-power accuracy.  A
suitable framework to achieve this may be an effective field theory
formulation.  First steps in this direction (with main focus on $B$
meson decays) have been made but this field is still in an early stage
of development, see e.g.~\cite{Chay:2002vy},
\cite{Beneke:2002ph,Beneke:2002ni} and
\cite{Pirjol:2002km,Bauer:2002aj,Rothstein:2003wh}.  Existing methods
attempt to estimate or model power suppressed effects, typically
focusing on particular sources of power corrections and making
assumptions going beyond first principles of QCD.  A detailed account
of these methods is well beyond the scope of this review.  We rather
aim to give an overview of what has been done for the processes where
GPDs or GDAs occur, and point out related work on processes that are
very similar.

We note that there are processes that cannot be described within the
OPE but where nevertheless power corrections up to a certain accuracy
can be expressed in terms of hard scattering coefficients and hadronic
matrix elements of higher-twist operators.  An example are the $1/Q^2$
power corrections in the unpolarized Drell-Yan pair cross section
\cite{Qiu:1991xx,Qiu:1991xy}.  What has to be shown is that the
reduced diagrams contributing to that accuracy still have the form of
a hard subgraph connected to several collinear ones, without soft
interactions between the collinear graphs that would imply
non-perturbative cross talk between the two incoming hadrons.


\subsection{Compton scattering at $1/Q$ accuracy}
\label{sec:twist-three}

The $1/Q$ suppressed power corrections to virtual Compton scattering
have been studied at leading order in $\alpha_s$ by a number of
groups.  To this level of accuracy the structure of the process is
well understood and can be written again as the convolution of hard
scattering kernels with hadronic matrix elements.  The theoretical
analysis closely resembles that of the $1/Q$ suppressed terms in deep
inelastic scattering on a polarized target, but there are important
particularities due to the nonforward kinematics in our case.  Just
recently, part of the $\alpha_s$ corrections to the $1/Q$ suppressed
DVCS amplitudes was obtained by Kivel and Mankiewicz
\cite{Kivel:2003jt}.

\subsubsection{Operator product expansion and twist}
\label{sub:twist}

A framework to treat factorization in Compton scattering is given by
the OPE.  If at least one of the photons is far off-shell, the time
ordered product of currents in the hadronic tensor
(\ref{compton-tensor}) is dominated by distances $z^2$ close to the
light-cone and one can expand
\begin{eqnarray}
  \label{light-cone-ope}
\lefteqn{
i T J^\alpha_{\mathrm{em}}(-\half z)\, J^\beta_{\mathrm{em}}(\half z) 
}
\\[0.2em]
 &=& \bar{q}(-\half z) \frac{\gamma^\alpha\, \slash{z}\, 
	\gamma^\beta}{2\pi^2 z^4}\, W[-\half z,\half z]\, q(\half z)
  \,+\,\Big \{ z \to -z,\, \alpha \leftrightarrow \beta \Big\}
  \,+\, O(z^{-2})
\nonumber \\[0.4em]
 &=& \frac{z_\rho}{2 \pi^2 z^4}\, \Bigg(
     s^{\alpha\rho\beta\sigma}\, \Big\{
	\bar{q}(-\half z)\, \gamma_\sigma W[-\half z,\half z]\, 
		q(\half z) -  \{z \to -z\} \Big\}
\nonumber \\
 && \hspace{2.1em} {} - i\epsilon^{\alpha\rho\beta\sigma}\, \Big\{
	\bar{q}(-\half z)\, \gamma_\sigma \gamma_5\, 
            W[-\half z,\half z]\, q(\half z) 
	+ \{z \to -z\} \Big\} \Bigg)
    + O(z^{-2})
\nonumber 
\end{eqnarray}
up to higher-order corrections in $\alpha_s$.  Here
$s^{\alpha\rho\beta\sigma} = g^{\alpha\rho} g^{\beta\sigma} +
g^{\alpha\sigma} g^{\beta\rho} - g^{\alpha\beta} g^{\rho\sigma}$, and
$W[a,b]$ denotes a Wilson line along the straight-line path from
$a^\mu$ to $b^\mu$, generalizing our notation of
Section~\ref{sec:definitions}.  The expression (\ref{light-cone-ope})
directly corresponds to the handbag diagrams for Compton scattering in
Fig.~\ref{fig:DDVCS}, supplemented by the gluons summed in the Wilson
lines.

In the \emph{local} OPE one expands the product of currents in terms
of local operators.  This can be obtained from (\ref{light-cone-ope})
using
\begin{eqnarray}
  \label{local-expansion}
\bar{q}(-\half z)\, \gamma^\sigma W[-\half z,\half z]\, q(\half z) 
&=& \sum_{n=0}^\infty \frac{1}{n!}\, z_{\alpha_1} \ldots z_{\alpha_n}
    \Big[\bar{q} \gamma^\sigma \lrD^{\alpha_1}
		\ldots \lrD^{\alpha_n} q \, \Big]
\end{eqnarray}
and its analog for the axial vector operator, where all fields on the
right-hand side are evaluated at $z=0$.  An early treatment of the
nonforward virtual Compton amplitude in this framework has been
given by Watanabe \cite{Watanabe:1981ce,Watanabe:1982ue}, and more
recent ones by Chen \cite{Chen:1998rc} and by Ji and Osborne
\cite{Ji:1998xh}.  These works are confined to leading power accuracy,
whereas a study by White \cite{White:2001pu} includes $1/Q$ power
corrections.  In close analogy to the classical analysis of inclusive
DIS, the local OPE provides an expansion of the Compton tensor
$T^{\alpha\beta}$ in powers of $1 /\rho$, i.e.\ for $|\rho| \to
\infty$ in the unphysical region.  The coefficients of this expansion
can be identified with the Mellin moments of the discontinuity of
$T^{\alpha\beta}$ in $\rho$, and the full amplitude is recovered by a
dispersion relation in this variable \cite{White:2001pu}.  At leading
$1/Q$ accuracy the $1/\rho$ expansion involves the Mellin moments of
the twist-two GPDs we discussed in Section~\ref{sub:polynom}; the
corresponding Wilson coefficients at $O(\alpha_s)$ can be found in
\cite{Ji:1998xh}.  We must remark that the considerations of
\cite{White:2001pu} cannot be taken as a proof for factorization as
claimed in that work, since the arguments crucially depend on
properties of the Wilson coefficients which are not proven (but do
hold in the explicit Born level and one-loop calculations).

An alternative approach, developed independently by Anikin and
Zavyalov \cite{Anikin:1978tj} and by Balitsky and Braun
\cite{Balitsky:1989bk}, is to expand directly in terms of ``string''
operators as in (\ref{light-cone-ope}), where fields occur at
positions along a straight line from $-\kappa z$ to $\kappa z$.  For
$z^2=0$ these operators are also called ``light-ray operators''.  The
local OPE expansion is organized in terms of operators with definite
``twist'', which is defined as the spin of an operator minus its
canonical mass dimension.  The twist-two projections of the operators
on the right-hand side of (\ref{local-expansion}) are obtained by
symmetrization in $\sigma$, $\alpha_1$, \ldots, $\alpha_n$ and by
subtractions making the operators traceless in any two indices;
higher-twist operators are antisymmetric in some of their indices.  A
string operator of definite twist is defined by resumming the
corresponding twist projections of the local operators in
(\ref{local-expansion}).  The construction of the string operators
with definite twist relevant for the Compton amplitude has been
performed in a series of works by Geyer et
al.~\cite{Geyer:1999uq,Geyer:2000ig,Geyer:2001qf}.  One finds in
particular that for light-like distance $z$ the decomposition of these
operators involves only a finite number of twists.

The notion of twist defined in this way is also called ``geometrical
twist''.  It differs from the ``dynamical twist'', which refers to
matrix elements of light-ray operators and specifies at which power in
$1/Q$ they first appear in hard processes.  Dynamical twist is closely
related to the decomposition of operators in terms of ``good'' and
``bad'' field components in light-cone quantization (see
Section~\ref{sec:light-cone}); for a discussion see
\cite{Jaffe:1996zw}.  For twist two the geometrical and dynamical
twist definitions coincide, but not for higher twist.  We will
continue to call the GPDs and GDAs introduced in
Section~\ref{sec:proper} ``twist-two'' or ``leading-twist''
quantities, even when they appear in power suppressed amplitudes.
When referring to the ``twist'' of a scattering amplitude, we will
imply the notion of dynamical twist.

An important finding is that the projection of operators on
geometrical twist two gives a Compton amplitude which in nonforward
kinematics respects electromagnetic gauge invariance at leading power
in $1/Q$ but not beyond, as can be seen by explicit calculation
\cite{Blumlein:2000cx}.   Radyushkin and
Weiss~\cite{Radyushkin:2000ap} have emphasized that the twist-two
projection of the operator expanded product $T
J^\alpha_{\mathrm{em}}(z_1) J^\beta_{\mathrm{em}}(z_2)$ does not
satisfy electromagnetic current conservation.  The terms which break
current conservation are ``total derivatives'' of operators, i.e.\
derivatives with respect to an overall translation such as
\begin{equation}
\frac{\partial}{\partial X_\mu}\, \Bigg[\,
    \bar{q}(X-\half z)\, \gamma^\sigma
	W[X-\half z, X+\half z]\, q(X+\half z) \,\Bigg]_{X=0} \: .
\end{equation}
In matrix elements these derivatives turn into factors of $i
\Delta^\mu$, so that their forward matrix elements are zero but not
their nonforward ones.  A consistent evaluation of nonforward Compton
scattering at $1/Q$ accuracy requires the inclusion of operators with
geometrical twist three.  As already mentioned, the QCD equations of
motion are needed at this level in order to obtain a basis of
independent operators.  Belitsky and M\"uller \cite{Belitsky:2000vx}
have given an explicit representation for the geometrical twist-three
parts of the string operators in (\ref{light-cone-ope}) at $z^2=0$.  It
involves total derivatives of their geometrical twist-two projections,
and quark-antiquark-gluon operators $\bar{q} \gamma^+ G^{+\rho} q$ and
$\bar{q} \gamma^+ \gamma_5\, \tilde{G}^{+\rho} q$ with all fields
along the light-cone (for simplicity we have omitted Wilson line
factors).  A representation away from the light-cone has recently been
given by Kiptily and Polyakov \cite{Kiptily:2002nx}.

The contribution to the nonforward Compton amplitude from the
geometrical twist-two operators and their total derivatives is
separately gauge invariant.  This provides a minimal set of operators
to restore current conservation of the handbag amplitude to order
$1/Q$ \cite{Kivel:2000rb,Radyushkin:2000jy,Radyushkin:2000ap}.  The
resulting $1/Q$ terms are often called ``kinematical'' twist-three
corrections.  Radyushkin and Weiss have pointed out that to $1/Q$
accuracy in the Compton amplitude the effect of the operators with
total derivatives can be represented as a ``spin rotation'' of the
twist-two string operators, which acts on the Dirac indices of the
quark and antiquark fields \cite{Radyushkin:2001fc}.  The
contributions from the quark-antiquark-gluon operators are commonly
referred to as ``dynamical'' or ``genuine'' twist-three terms.

As a result of the operator decomposition just mentioned the hadronic
matrix elements
\begin{eqnarray}
  \label{vector-gpd-def} F^\mu &=& P^+
\int \frac{d z^-}{2\pi}\, e^{ix P^+ z^-}
  \langle p'|\, \bar{q}(-\half z)\, \gamma^\mu q(\half z)\, 
  \,|p \rangle \Big|_{z^+=0,\, \tvec{z}=0} \: , 
\nonumber \\
\tilde{F}^\mu &=& P^+
\int \frac{d z^-}{2\pi}\, e^{ix P^+ z^-}
  \langle p'|\, \bar{q}(-\half z)\, \gamma^\mu \gamma_5\, q(\half z)\, 
  \,|p \rangle \Big|_{z^+=0,\, \tvec{z}=0} \: , 
\end{eqnarray}
can be represented in terms of functions parameterizing the
quark-antiquark-gluon contributions, and of a ``Wandzura-Wilczek
part'' involving the twist-two GPDs or double distributions discussed
so far in this review.  This representation is limited to $1/Q$
accuracy in the Compton amplitude, which will be implied in the
remainder of this section.  Only the plus and transverse components of
(\ref{vector-gpd-def}) are relevant then, since the minus components
only appear at order $1/Q^2$.  Most studies in the literature have
used the ``Wandzura-Wilczek approximation'', where the
quark-antiquark-gluon matrix elements are neglected.  For the forward
Compton amplitude this leads to the well-known relation between the
polarized structure functions $g_1$ and $g_2$ of DIS
\cite{Wandzura:1977qf}, which agrees rather well with recent
measurements and a number of theoretical calculations
\cite{Anthony:2002hy}.  Little has been done so far to explore whether
one may expect this approximation to be valid in nonforward
kinematics.  A recent study in the chiral quark-soliton model
\cite{Kiptily:2002nx} has estimated that the first nonvanishing
moments of the quark-antiquark-gluon distributions, which are matrix
elements of the local operators $\bar{q} \gamma^+ G^{+\rho} q$ and
$\bar{q} \gamma^+ \gamma_5\, \tilde{G}^{+\rho} q$, are parametrically
suppressed at low normalization scale and small $t$ (as are the
twist-two gluon distributions in the same model, see
Section~\ref{sub:chiral-soliton}).

The Wandzura-Wilczek part of the above matrix elements on the proton
reads~\cite{Kivel:2000cn}
\begin{eqnarray}
  \label{WW-relations}
F^\mu_{W} &=& \left( P^\mu - \frac{\Delta_T^\mu}{2 \xi} \right)
  \frac{1}{P^+} \left[ H(x,\xi)\, \bar{u} \gamma^+ u + 
  E(x,\xi)\, \bar{u} 
	\frac{i \sigma^{+\alpha} \Delta_\alpha}{2m} u \right]
\nonumber \\[0.1em]
&& {}+ \frac{1}{2} \int_{-1}^1 dy\,
       \Big[ W(x,y,\xi) + W(-x,-y,\xi) \Big]\,  G^\mu(y,\xi)
\nonumber \\[0.2em]
&& {}+ \frac{1}{2} \int_{-1}^1 dy\,
       \Big[ W(x,y,\xi) - W(-x,-y,\xi) \Big]\,
       i \epsilon_T^{\mu\nu}\, \tilde{G}_\nu(y,\xi) ,
\nonumber \\[0.2em]
\tilde{F}^\mu_{W} &=& 
          \left( P^\mu - \frac{\Delta_T^\mu}{2 \xi} \right)
  \frac{1}{P^+} \left[ \tilde{H}(x,\xi)\, \bar{u} \gamma^+\gamma_5 u +
         \tilde{E}(x,\xi)\, \bar{u} \frac{\gamma_5 \Delta^+}{2m} u
  \right]
\nonumber \\[0.1em]
&& {}+ \frac{1}{2} \int_{-1}^1 dy\,
	 \Big[ W(x,y,\xi) + W(-x,-y,\xi) \Big]\, \tilde{G}^\mu(y,\xi)
\nonumber \\[0.2em]
&& {}+ \frac{1}{2} \int_{-1}^1 dy\,
       \Big[ W(x,y,\xi) - W(-x,-y,\xi) \Big]\,
       i \epsilon_T^{\mu\nu}\, G_\nu(y,\xi),
\end{eqnarray}
where
\begin{equation}
W(x,y,\xi) =   \frac{1}{\xi-y} \Big[ \,
	\theta(\xi > x > y) - \theta(\xi < x < y) \, \Big]
\end{equation}
is the Wandzura-Wilczek kernel, and 
\begin{eqnarray}
  \label{G-defs}
G^\mu(x,\xi) &=& \bar{u} \gamma_T^\mu u \,
			\Big[ H(x,\xi) + E(x,\xi) \Big]
  - \frac{\Delta_T^\mu}{2 \xi} \Bigg[ \,
    \frac{\bar{u} \gamma^+ u}{P^+} \,
    \Big( x \frac{\partial}{\partial x} 
	+ \xi \frac{\partial}{\partial \xi} \Big) 
              \Big[ H(x,\xi) + E(x,\xi) \Big]
\nonumber \\
 && \hspace{13.7em} {} 
    - \frac{\bar{u} u}{m} \,
    \Big( x \frac{\partial}{\partial x} 
	+ \xi \frac{\partial}{\partial \xi} \Big) E(x,\xi) \Bigg] ,
\nonumber \\
\tilde{G}^\mu(x,\xi) &=& \bar{u} \gamma_T^\mu\gamma_5 u \;
				\tilde{H}(x,\xi)
  - \frac{\Delta_T^\mu}{2 \xi} \Bigg[ \,
    \frac{\bar{u} \gamma^+\gamma_5 u}{P^+} \, 
    \Big( x \frac{\partial}{\partial x} 
	+ \xi \frac{\partial}{\partial \xi} \Big) \tilde{H}(x,\xi)
\nonumber \\
 && \hspace{9.5em} {} 
    - \frac{\bar{u} \gamma_5 u}{m} \,
    \Big( x \frac{\partial}{\partial x} 
	+ \xi \frac{\partial}{\partial \xi} \Big) \,
              \xi \tilde{E}(x,\xi)
  \Bigg] .
\end{eqnarray}    
For legibility we have not displayed the $t$ dependence of GPDs, and
omitted spinor arguments and quark flavor labels.  The plus-components
of (\ref{WW-relations}) reduce to $2 P^+$ times the matrix elements
$F^q$ and $\tilde{F}^q$ defining twist-two GPDs, as it must be.  As
noted in \cite{Kivel:2000cn} the $D$-term contribution to $H$ and $E$,
as well as the pion pole part of $\tilde{E}$ cancel in $G^\mu$ and
$\tilde{G}^\mu$.  These terms hence appear in $F^\mu$ and
$\tilde{F}^\mu$ in the same form as in twist-two matrix elements, in
particular without being convoluted with the Wandzura-Wilczek kernels.
For a spin-zero target there are expressions similar to the above
\cite{Kivel:2000rb,Radyushkin:2000ap,Anikin:2001ge}.  They  are
simpler since there are no twist-two distributions for the axial
vector operator due to parity invariance.  Expressions of the matrix
elements $F^\mu$ and $\tilde{F}^\mu$ for targets with spin zero or
$\half$ including the quark-antiquark-gluon contributions are given in
\cite{Belitsky:2000vx}, and a general form factor decomposition
of $F^\mu$ and $\tilde{F}^\mu$ for spin $\half$ targets can be found
in \cite{Kiptily:2002nx}.  Notice that the evolution of the
Wandzura-Wilczek parts of $F^\mu$ and $\tilde{F}^\mu$ directly follows
from the known evolution at twist two-level.  The evolution of the
genuine twist-three operators is discussed in \cite{Belitsky:2000vx}.

The structure of (\ref{WW-relations}) to (\ref{G-defs}) is similar to
the Wandzura-Wilczek relations for meson distribution amplitudes
\cite{Ball:1998fj,Ball:1998je} or GDAs \cite{Anikin:2001ge}, which
originate from the same decompositions of string operators.  We recall
that these relations involve the inclusion of operators with
geometrical twist three, the use of the QCD equations of motion, and
the neglect of quark-antiquark-gluon matrix elements.  They are hence
different in form and content from the Wandzura-Wilczek relations
between matrix elements of geometrical twist-two operators in
\cite{Blumlein:1999sc,Blumlein:2000cx,Blumlein:2001sb}.  A
discussion of these differences in the case of distribution amplitudes
is given in \cite{Ball:2001uk}.

Let us finally mention sum rules for the matrix elements $F^\mu$ and
$\tilde{F}^\mu$, which were discussed in
\cite{Penttinen:2000dg,Kivel:2000cn,Goeke:2001tz,Kiptily:2002nx} and
in particular give the Burk\-hardt-Cotting\-ham
\cite{Burkhardt:1970ti} and Efremov-Leader-Teryaev
\cite{Efremov:1997hd} sum rules  in the forward limit.  The first and
second moments ${\smash \int_{-1}^1 dx\, F^\mu }$ and ${\smash
\int_{-1}^1 dx\, x F^\mu }$ as well as their analogs for
$\tilde{F}^\mu$ receive no contributions from the
quark-antiquark-gluon operators and are hence readily obtained from
the Wandzura-Wilczek relations (\ref{WW-relations}).  The first
moments simply lead to the form factors $F_1$, $F_2$ and $g_A$, $g_P$
for each quark flavor, whereas the second moments involve these form
factors and the second moments of the GPDs $H$, $E$ and $\tilde{H}$,
$\tilde{E}$.

\subsubsection{The DVCS amplitude at $1/Q$ accuracy}
\label{sub:dvcs-three}

The DVCS amplitude at $1/Q$ accuracy has been studied to leading order
in $\alpha_s$ by several groups with different methods.  The
investigation of Anikin et al.~\cite{Anikin:2000em} is for spinless
targets and closely follows the techniques of Ellis, Furmanski and
Petronzio, using in particular the ``transverse'' basis of operators
introduced in \cite{Ellis:1983cd}.  Belitsky and M\"uller give a
complete treatment for targets of spin 0 or $\half$ in
\cite{Belitsky:2000vx}.  Other studies have used the Wandzura-Wilczek
approximation.  Penttinen et al.~\cite{Penttinen:2000dg} consider a
nucleon target and proceed in a manner motivated by parton model
ideas.  The work of Kivel et al.\ for targets of spin 0
\cite{Kivel:2000rb} and spin $\half$ \cite{Kivel:2000cn}, and the work
of Radyushkin and Weiss
\cite{Radyushkin:2000jy,Radyushkin:2000ap,Radyushkin:2001fc} for
spin-zero targets closely follows the string operator formalism of
Balitsky and Braun \cite{Balitsky:1989bk}.  Studies concerning the
phenomenology of DVCS at twist-three level are reviewed in
Section~\ref{sec:dvcs-pheno}.

At $1/Q$ accuracy the calculation of DVCS involves the handbag
diagrams of Fig.~\ref{fig:DDVCS}, to be evaluated beyond the collinear
approximation by performing the Taylor expansion in
(\ref{coll-approx}) one order further, and in addition diagrams with
an extra (transversely polarized) gluon connecting the quark line of
the hard scattering with the collinear subgraph of the proton.
Applying the equations of motion for the quark field operator one
finds \cite{Anikin:2000em,Belitsky:2000vx} that the complete amplitude
can be expressed in terms of the matrix elements $F^\mu$ and
$\tilde{F}^\mu$ defined in (\ref{vector-gpd-def}).  It can hence be
parameterized by twist-two quark-antiquark GPDs and
quark-antiquark-gluon GPDs of twist three.  These steps ensure that
the result respects electromagnetic current conservation and involves
only QCD operators that are gauge invariant.  The hadronic tensor for
DVCS then reads~\cite{Kivel:2000cn}
\begin{eqnarray}
  \label{compton-tensor-three}
T^{\alpha\beta} &=& 
\sum_q e_q^2 \int_{-1}^{1} dx\, \Bigg( 
  - g_{T}^{\alpha\gamma}\, \frac{F^{q +}}{2P^+} \, C
  - i \epsilon_{T}^{\alpha\gamma}\, 
		\frac{\tilde{F}^{q +}}{2P^+} \, \tilde{C} 
\nonumber \\
 && + \frac{(q + 4\xi P)^\alpha}{2\, q P}
   \left\{ g_{T}^{\gamma\delta}\, F^{q}_{\,\delta}\, C
    - i \epsilon_{T}^{\gamma\delta}\, 
		\tilde{F}^{q}_{\,\delta}\, \tilde{C}
   \right\}  \Bigg)
   \Bigg( g_{\gamma}{}^{\beta} + \frac{\Delta_{T \gamma}\,
		P^\beta}{q P} \Bigg)
\end{eqnarray}
up to $1/Q^2$ suppressed corrections, with the familiar
hard-scattering kernels
\begin{eqnarray}
C(x,\xi) &=& \frac{1}{\xi-x -i\epsilon} - \frac{1}{\xi+x -i\epsilon} ,
\nonumber \\
\tilde{C}(x,\xi) &=&
  \frac{1}{\xi-x -i\epsilon} + \frac{1}{\xi+x -i\epsilon} .
\end{eqnarray}
The tensor is given for a coordinate system where the transverse
directions are defined with respect to the initial photon momentum $q$
and to the average hadron momentum $P$, and where $p$ and $p'$ have
large components along the positive $z$ axis.  For a spinless target
one has in particular $\tilde{F}^{q +} = 0$ from parity invariance,
and the parameterization of the amplitude requires one GPD for the
transverse components of each matrix element $F^{q \mu}$ and
$\tilde{F}^{q \mu}$ in addition to the GPDs $H^q$ parameterizing $F^{q
+}$ \cite{Anikin:2000em}.  The Compton amplitude for two off-shell
photons is given in \cite{Belitsky:2000vx}.  It involves the same
hadronic matrix elements, with hard-scattering kernels $C(x,\rho)$ and
$\tilde{C}(x,\rho)$ and appropriately modified tensor structures.

Let us discuss the electromagnetic gauge invariance of the Compton
amplitude, which requires $q_\alpha T^{\alpha\beta} = T^{\alpha\beta}
q'_\beta = 0$.  The Compton tensor (\ref{leading-order-compton}) in
leading-twist approximation is gauge invariant except for terms of
order $\Delta_T /Q$ and is hence consistent to the accuracy at which
it is calculated.  Vanderhaeghen et al.\
\cite{Vanderhaeghen:1998uc,Guichon:1998xv} have proposed to use an
exactly gauge invariant amplitude by the replacement
\begin{equation}
  \label{vdh-presc}
T^{\alpha\beta} \to T^{\alpha\gamma} \Bigg(  
	g_{\gamma}{}^{\beta} 
	- \frac{q'_\gamma\, v^\beta}{q' v} \Bigg) ,
\end{equation}
where $v$ is an appropriately chosen auxiliary vector with $q' v$ of
order $Q^2$.  If the Compton tensor calculated up to corrections of
order $1 /Q^{n+1}$ satisfies $T^{\alpha\beta} q'_\beta = O(1 /Q^n)$
then this replacement is allowed within the accuracy of the
calculation.  An analogous prescription can of course be used for the
initial photon, but with the coordinate system chosen here the tensor
already satisfies $q_\alpha T^{\alpha\beta} = 0$.  We remark that the
prescription (\ref{vdh-presc}) is equivalent to using the axial gauge
$v A_{\mathrm{em}} =0$ for the electromagnetic field, where the
polarization vectors of the final state photon in Feynman gauge are to
be replaced according to
\begin{equation}
  \epsilon_\beta \to \epsilon_\beta
	 - \frac{\epsilon v}{q' v}\, q'_\beta .
\end{equation}
Taking $v = P$ and using that $q' P = q P$ and $q'_T = -
\Delta_T^{\phantom{*}}$ one finds that the terms going with $F^{q +}$
and $\tilde{F}^{q +}$ in the Compton tensor
(\ref{compton-tensor-three}) are precisely those anticipated by the
prescription (\ref{vdh-presc}).  Note that the terms in
(\ref{compton-tensor-three}) where $F^q_{\,\delta}$ or
$\tilde{F}^q_{\,\delta}$ is multiplied with $\Delta_{T \gamma}$
contribute to the amplitude only at order $1/Q^2$. They have been
added by hand following the above rationale in order to obtain an
exactly gauge invariant tensor.

No general analysis comparable to the leading-twist factorization
theorem guarantees that at $1/Q$ accuracy the Compton amplitude
\emph{can} actually be represented in terms of hadronic matrix elements
and hard-scattering kernels.  According to our discussion at the
beginning of this section this is particularly problematic when one
photon is on shell.  A consistency check of the factorized result
(\ref{compton-tensor-three}) is the absence of collinear or soft
singularities which would signal a breakdown of the factorization
scheme at this order.  As has been discussed in \cite{Kivel:2000rb}
and \cite{Radyushkin:2000jy,Radyushkin:2000ap} for the pion and in
\cite{Kivel:2000cn} for the nucleon, the convolution of the Wandzura
Wilczek kernels with smooth GPDs leads to functions with a
discontinuity.  Namely, the convolution
\begin{eqnarray}
f_W(x,\xi) &=& \int_{-1}^1 dy\, W(x,y,\xi) f(y,\xi) 
\nonumber \\
&=& \theta(x > \xi) \int_{x}^1 dy\, \frac{f(y,\xi)}{y - \xi} - 
    \theta(x < \xi) \int_{-1}^x dy\, \frac{f(y,\xi)}{y - \xi} ,
\end{eqnarray}
where $f(y,\xi)$ stands for a generic GPD, has a discontinuity
\begin{equation}
  \label{WW-disc}
f_W(\xi + \delta,\xi) - f_W(\xi - \delta,\xi) 
\stackrel{\delta\to 0}{\to}
  \pv{2}\int_{-1}^1 dy\, \frac{f(y,\xi)}{y - \xi}
\end{equation}
at $x=\xi$, which involves the same integral as the real part of the
twist-two amplitude and has little chance to be zero.  The convolution
with $W(-x,-y,\xi)$ is discontinuous at $x= -\xi$ in turn.  The
Compton tensor (\ref{compton-tensor-three}) involves however only the
combinations $(\xi + x - i\epsilon)^{-1}\, W(x,y,\xi)$ and $(\xi - x -
i\epsilon)^{-1}\, W(-x,-y,\xi)$ where these discontinuities are
harmless.  In the calculation of $T^{\alpha\beta}$ one also finds a
contribution
\begin{equation}
  \label{dumb-tensor}
\sum_q e_q^2 \int_{-1}^{1} dx\,
  \left\{ g_{T}^{\alpha\delta}\, F^{q}_{\,\delta}\, C + i
    \epsilon_{T}^{\alpha\delta}\, \tilde{F}^{q}_{\,\delta}\, \tilde{C}
    \right\} \frac{(q + 2\xi P)^\beta}{2\, q P}
\end{equation}
which involves $(\xi - x - i\epsilon)^{-1}\, W(x,y,\xi)$ and $(\xi + x
- i\epsilon)^{-1}\, W(-x,-y,\xi)$ and thus has logarithmic
singularities.  As emphasized in
\cite{Kivel:2000rb,Radyushkin:2000jy,Radyushkin:2000ap,Kivel:2000cn}
this tensor gives however a $1/Q^2$ suppressed term when contracted
with any \emph{physical} polarization vector of the final photon.
This is beyond the accuracy of the calculation and therefore has been
omitted in (\ref{compton-tensor-three}).  Indeed, one has $(q + 2\xi
P)^\beta \approx (q' + \Delta_T)^\beta$ up to yet higher terms in
$1/Q$, and physical polarization vectors are of course orthogonal to
$q'^\beta$.  Anikin and Teryaev \cite{Anikin:2001ge} have pointed out
the analog of this situation in the crossed channel, where the
twist-three GDAs for pion pairs have finite values at the end points
$z=0$ or $z=1$.  This leads to logarithmic singularities in the
unphysical sector when convoluted with $z^{-1}$ or $(1-z)^{-1}$ from
the hard-scattering kernels, but no singularities appear in the
physical part of the hadronic tensor.  Note that the discontinuities
of the matrix elements $F^\mu$ and $\tilde{F}^\mu$ at $x= \pm \xi$ do
not lead to problems in the doubly virtual Compton amplitude, where
the hard-scattering kernels have their poles at $x= \pm \rho$, as
remarked in \cite{Kivel:2000rb}.

Radyushkin and Weiss have explicitly shown that to order $1/Q^2$
accuracy the total derivative operators discussed in the previous
subsection provide terms which result in replacing $(q + 2\xi
P)^\beta$ by $q^{\prime \beta}$ in the contribution
(\ref{dumb-tensor}), so that its factorization breaking singularities
do not appear in the physical sector of the Compton tensor
\cite{Radyushkin:2000ap}.  A result free of singularities was also
obtained in a recent evaluation of $\alpha_s$ corrections to DVCS at
order $1/Q$, which was restricted to the Wandzura Wilczek
approximation and to the contribution involving quark GPDs
\cite{Kivel:2003jt}.  In general it remains however unknown whether
factorization still holds for DVCS at $1/Q^2$ or at $\alpha_s /Q$
accuracy.

Calculating amplitudes for definite photon helicity in the $\gamma^*
p$ c.m.\ from the Compton tensor (\ref{compton-tensor-three}), one
finds that the $1/Q$ corrections do not affect the transverse photon
amplitudes (\ref{compton-two-gpd}) calculated at leading-twist
accuracy.  Instead they give nonzero transitions for a longitudinal
photon, which are absent at leading power.  The new amplitudes can be
expressed in terms of $(i)$ the Compton form factors $\mathcal{H}$,
$\mathcal{E}$, $\tilde\mathcal{H}$, $\tilde\mathcal{E}$ already
appearing at leading twist, $(ii)$ form factors given by
\begin{eqnarray}
  \label{twist-three-ffs}
\mathcal{H}_{W} &=& \sum_q e_q^2 \int_{-1}^1 dx\,
  \frac{1}{\xi-x -i\epsilon} \, \int_{-1}^1 dy\, W(-x,-y,\xi)\;
  H^{q(+)}(y,\xi,t) 
\nonumber \\
 &=& \sum_q e_q^2 \int_{-1}^1 dx\, \frac{1}{\xi +x}\,
     \ln \frac{2\xi}{\xi -x - i\epsilon} \; H^{q(+)}(x,\xi,t)
\end{eqnarray}
and its analogs with $E^{q(+)}$, $\tilde{H}^{q(+)}$,
$\tilde{E}^{q(+)}$, and $(iii)$ four independent form factors
involving quark-antiquark-gluon distributions \cite{Belitsky:2001ns}.
The derivatives in (\ref{G-defs}) can be written according to 
\begin{equation}
\sum_q e_q^2 \int_{-1}^1 dx\, \frac{1}{\xi +x}\,
     \ln \frac{2\xi}{\xi -x - i\epsilon} \,
  \Big( x \frac{\partial}{\partial x} 
	+ \xi \frac{\partial}{\partial \xi} \Big) H^{q(+)}(x,\xi,t) 
= \xi \frac{\partial}{\partial \xi}\, \mathcal{H}_{W} .
\end{equation}
The form factors $\mathcal{H}$, $\mathcal{E}$, $\tilde\mathcal{H}$,
$\tilde\mathcal{E}$ can in principle be extracted from the four
independent leading-twist amplitudes $M_{\lambda'+,\lambda+}$.  The
four independent twist-three amplitudes $M_{\lambda'+,\lambda\, 0}$
then depend on another eight unknowns, namely on $\mathcal{H}_{W}$,
$\mathcal{E}_{W}$, $\tilde\mathcal{H}_{W}$, $\tilde\mathcal{E}_{W}$
and on the genuine twist-three form factors.  To decide from measured
Compton amplitudes whether the Wandzura-Wilczek approximation is
adequate would hence require sufficient knowledge of the
$x$-dependence of GPDs at given $\xi$, so that both the $\mathcal{H}$
and the $\mathcal{H}_{W}$ type convolutions can be evaluated.  If in
turn this approximation can be sufficiently motivated from other
sources, measurement of the twist-three amplitudes will provide
information about $H^q$, $E^q$, $\tilde{H}^q$, $\tilde{E}^q$ beyond
what can be deduced from the leading amplitudes
$M_{\lambda'+,\lambda+}$.


\subsection{Two-photon annihilation}
\label{sub:two-photon}

Among all hard exclusive processes $\gamma^* \gamma \to \pi^0$ plays a
special role.  On one hand rather good experimental data exist for
$Q^2$ up to about 8~GeV$^2$.  On the other hand this is one of the
simplest processes from a theoretical point of view, and several
investigations of power corrections have been carried out.  The
methods they employ are of direct relevance for processes where GDAs
or GPDs can be measured.

For two-photon annihilation into a meson one can write down the
operator product expansion.  The first power corrections to the
amplitude appear at order $1/Q^2$ and can be found in
\cite{Khodjamirian:1997tk}.  Taking the asymptotic forms of the
relevant twist-two and twist-four distribution amplitudes one finds a
ratio $\mathcal{A}_4 /\mathcal{A}_{2} = - \frac{8}{9} \delta^2 /Q^2$
of the twist-four and twist-two contributions to the amplitude.  Here
$\delta$ parameterizes the local matrix element $\langle \pi(p) |
\bar{d}\, g \tilde{G}^{\mu\nu} \gamma_\nu\, u | 0\rangle$.  Light-cone
sum rule estimates give $\delta^2 \approx 0.2$~GeV$^2$ at a scale
$\mu=1$~GeV, with an error of about 30\% being quoted
in~\cite{Khodjamirian:2000mi}.

Twist-four corrections of similar size have been estimated in
\cite{Gosdzinsky:1998fs} using the renormalon technique (see
\cite{Beneke:1998ui}).  This method starts with a resummation to all
orders in perturbation theory of fermion loop insertions in the gluon
lines of the relevant hard scattering subgraphs (which are those at
NLO in the present case).  Due to the infrared behavior of the running
coupling this resummation contains an ambiguity which is power
suppressed by $1/Q^2$, and which must be compensated in the process
amplitude by a corresponding ambiguity in the twist-four
contributions.  One obtains an estimate for the power corrections
under the assumption that the size of this ambiguity approximates the
size of the twist-four corrections themselves.

As explained in the beginning of this section, the process $\gamma^*
\gamma \to \pi$ differs from $\gamma^* \gamma^* \to \pi$ in that the
real photon can lead to additional power corrections, associated with
soft regions of phase space and signaled by endpoint divergences in
the amplitude calculated from hard scattering graphs.  Such
corrections do indeed show up in the renormalon calculation of
\cite{Gosdzinsky:1998fs}.  A different method to estimate them is
based on light-cone sum rules, whose principal ingredients are
parton-hadron duality and the use of dispersion relations.  In the
analysis by Khodjamirian~\cite{Khodjamirian:1997tk} the amplitude for
$\gamma^* \gamma \to \pi$ is expressed in terms of the one for
$\gamma^* \gamma^* \to \pi$ in the region where both photons are far
off-shell, so that soft end-point contributions for the photon are not
kinematically possible.  In \cite{Khodjamirian:1997tk} the twist-two
and twist-four terms of the $\gamma^* \gamma \to \pi$ amplitude were
evaluated to lowest order in $\alpha_s$, whereas the one-loop
corrections to the leading-twist term were included in
\cite{Schmedding:1999ap,Bakulev:2002uc}.  A different way of using
light-cone sum rules for $\gamma^* \gamma \to \pi$ has been taken by
Radyushkin and collaborators
\cite{Radyushkin:1996pm,Radyushkin:1996tb,Musatov:1997pu}.  Here the
$\gamma^*\gamma \pi$ vertex was related to the Green function between
two photons and the axial current, in kinematics where the momentum
transfer by the axial current is sufficiently far off-shell to use
perturbation theory.  No input of pion distribution amplitudes is
required in this case.

Building on the work of Sterman and collaborators
\cite{Botts:1989kf,Li:1992nu}, Jakob et al.\ have developed a
different approach to estimating power corrections
\cite{Jakob:1993iw,Jakob:1996hd}.  Instead of expanding hard
scattering graphs in the parton momenta around the point where they
are on-shell and collinear to each other, one explicitly keeps the
dependence of the hard scattering on transverse parton momenta.  This
is then convoluted not with the distribution amplitudes of the
external hadrons, but with their light-cone wave functions, thus
keeping the full nonperturbative $k_T$ dependence.  This formalism
incorporates the Sudakov factors of the ``modified hard scattering
approach'' of Sterman et al.\ \cite{Botts:1989kf,Li:1992nu}.  Going
beyond the ``standard hard scattering approach''
\cite{Brodsky:1989pv}, these Sudakov form factors resum effects of
soft gluons in the hard scattering to all orders in perturbation
theory.  Calculations in this framework are conveniently performed in
impact parameter space rather than transverse momentum.  Both the
behavior of the impact parameter wave function and the Sudakov factors
tend to suppress the region of large impact parameters, where the use
of perturbation theory is problematic.  In particular this suppresses
the contribution from the soft endpoint regions of the wave
functions.\footnote{There are cases where the suppression from the
Sudakov factor is not efficient enough to suppress large impact
parameters, as pointed out by Descotes-Genon and Sachrajda for the
$B\to \pi$ transition form factor
\protect\cite{Descotes-Genon:2001hm}.}
In the momentum space formulation, it is precisely in the endpoint
regions when the neglect of parton $k_T$ in the hard scattering is not
justified, since parton virtualities become small there.  Note that
this way of including transverse momentum in the hard scattering (as
well as the light-cone sum rule approaches discussed above) does not
include power suppressed effects order by order in $1/Q$, but rather
sums certain types of power corrections to all orders in a $1/Q^{n}$
expansion.  The relation between the inclusion of $k_T$ in the hard
scattering kernel and the operator product expansion is not clear, as
has been pointed out by Musatov and Radyushkin \cite{Musatov:1997pu}.
We also remark that the question of gauge invariance in this approach
has not much been discussed in the literature.

In a phenomenological study, Jakob et al.~\cite{Jakob:1996hd} used a
Gaussian form as in (\ref{wf-ansatz}) for the $\tvec{k}$ dependence of
the $q\bar{q}$ wave function of the pion,
\begin{equation}
  \label{pion-wf-ansatz}
\Psi_{+-}(z,\tvec{k}) = N\, \frac{(4\pi a)^2}{z(1-z)}\, 
  \exp\left[ - a^2 \frac{\tvec{k}^2}{z(1-z)} \right] \,
  \Phi_\pi(z)
\end{equation}
with a normalization factor $N$ fixed by (\ref{pion-wf-da}).  Taking
the asymptotic form for the distribution amplitude $\Phi_\pi(z)$ the
authors obtain a $\gamma^*\gamma\to \pi$ scattering amplitude smaller
than the leading-twist one by about $33\%$ at $Q^2=2$~GeV$^2$ and by
about $15\%$ at $Q^2=8$~GeV$^2$.  In this kinematics, the effect of
the Sudakov form factors turns out to be negligible compared to the
intrinsic $k_T$ dependence of the pion wave function and of the quark
propagator in the hard scattering \cite{Diehl:2001dg}.

Given the precision of experimental data, an analysis of the
leading-twist contribution calls for the inclusion of the
$O(\alpha_s)$ corrections to the hard scattering.  Contrary to the
elastic pion form factor discussed in Section~\ref{sub:NLO-pions},
these are relatively moderate in $\gamma^*\gamma\to \pi$ for $Q^2$
larger than about 2~GeV$^2$.  A recent evaluation of the full
$O(\alpha_s^2)$ corrections \cite{Melic:2002ij} also suggests that the
perturbation series is well behaved in this range (whereas the
$O(\alpha_s^2)$ terms become of the same size as the $O(\alpha_s)$
ones when going down to $Q^2=1$~GeV$^2$, given the increase of the
strong coupling constant).  The evaluation of the $O(\alpha_s^2)$
terms with fermion loops also allows one to determine the BLM scale
for the running coupling in this process, which for the asymptotic
pion DA and in the $\overline{\mathrm{MS}}$ scheme is $\mu_R^2 \approx
0.11\, Q^2$ when taken to be independent of parton momentum fractions
\cite{Melic:2001wb}.  This leads to rather small scales for most
relevant $Q^2$.  To use the BLM scale one must hence modify the
running of $\alpha_s$ in the infrared region, and thus go beyond the
usual leading-twist formalism.  Such an approach has in particular
been pursued in \cite{Brodsky:1998dh} and
\cite{Stefanis:2000vd}.\footnote{Notice that in
\protect\cite{Brodsky:1998dh} the BLM scale for $\gamma^*\gamma\to
\pi$ is assumed to be the one for the pion form factor
(Section~\protect\ref{sub:NLO-pions}) and is hence too small.}

The CLEO data \cite{Gronberg:1998fj} for the $\gamma^*\gamma\to \pi$
amplitude show an approximate leading-twist scaling behavior already
at moderate $Q^2$ above 2~GeV$^2$.  The normalization of the amplitude
is rather close to its asymptotic value at $\log Q^2\to \infty$ (given
by the leading-twist expression with the asymptotic pion DA and
without $\alpha_s$ corrections).  It is difficult to assess whether
the remaining difference for $Q^2$ between 2 and 8 GeV$^2$ is due to
the shape of $\Phi_\pi(z)$ or to power suppressed contributions.  A
pure leading-twist analysis at $O(\alpha_s)$ can describe the data
with $\Phi_\pi$ mildly deviating from its asymptotic form
\cite{Diehl:2001dg}, and so can several descriptions including the
power corrections discussed above.  Analyses agree that very
asymmetric shapes of $\Phi_\pi$ are strongly disfavored by the
$\gamma^*\gamma\to \pi$ data
\cite{Kroll:1996jx,Schmedding:1999ap,Bakulev:2002uc}, but by just how
much $\Phi_\pi$ deviates from its asymptotic form strongly depends on
the theory assumptions concerning power suppressed terms.  In other
words, values and errors on the leading-twist pion DA obtained in
existing analyses crucially depend on the values and errors assumed
for the power corrections.  Attempts to determine the latter from
present data as well will likely give results with poor statistical
significance.

We also note that the amplitude for $\gamma^*\gamma\to \pi$ depends on
$\Phi_\pi$ through the integral (\ref{gamma-gamma-pi}) over $z$ and
does not allow one to reconstruct the function $\Phi_\pi(z)$ without
further theoretical assumptions. The leading-twist logarithmic scaling
violations provide only little further constraints in the $Q^2$ range
available \cite{Diehl:2001dg}.  In scenarios where power suppressed
terms also depend on $\Phi_\pi$, this situation is somewhat better
\cite{Schmedding:1999ap,Bakulev:2002uc}.  An often made assumption is
that one can truncate the Gegenbauer expansion of $\Phi_\pi$ after a
few terms.  As a theoretical ansatz this is corroborated by several
attempts to calculate or model $\Phi_\pi$ within QCD, where the
coefficients of higher Gegenbauer polynomials tend to decrease with
the degree of the polynomial.

Summarizing this discussion, the annihilation process $\gamma^*\gamma
\to \pi$ is a case where a leading-twist description with a plausible
shape for the nonperturbative input describes the data rather well,
even at moderate $Q^2$.  Compared with data from other processes it
provides the cleanest constraints on $\Phi_\pi$ we have.  Attempts to
determine the integral (\ref{gamma-gamma-pi}) of $\Phi_\pi(z)$ with
high precision are limited by theoretical control over power
corrections.  Precise data at higher $Q^2$ may help to overcome these
limitations.  Further and similarly clean information on the shape of
$\Phi_\pi(z)$ could be obtained from $\gamma^*\gamma^* \to \pi$ in the
region where one photon virtuality is much smaller than the other, as
has been discussed in \cite{Diehl:2001dg,Melic:2002ij}.

The production of pion pairs, $\gamma^*\gamma \to \pi\pi$, which gives
access to generalized distribution amplitudes, is very similar to
$\gamma^*\gamma \to \pi$, and one can expect a similar situation
concerning power corrections.  The analysis in \cite{Diehl:2000uv}
observed that the production of pion pairs may be more sensitive to
the soft physics at the endpoints $z=0$ and $z=1$.  This is because
compared with single pion production the $z$ region around $\half$ is
suppressed by an extra factor $(2z-1)^2$, with one factor $(2z-1)$
coming from the hard-scattering kernel (\ref{gamma-gamma-two-pi}) and
one from the Gegenbauer expansion (\ref{Polyakov-expansion}) of the
two-pion DA.  A light-cone sum rule analysis by Kivel and Mankiewicz
\cite{Kivel:2000rq}, following closely the treatment of
$\gamma^*\gamma\to \pi$ by Khodjamirian \cite{Khodjamirian:1997tk},
obtains rather moderate power corrections.  At $Q^2=2$~GeV$^2$ they
reduce the leading-twist result by about $30\%$ and are completely
negligible at $Q^2=8$~GeV$^2$.

The study just mentioned focused on the helicity amplitude where the
$\pi\pi$ pair is produced in the $L^3=0$ partial wave.  The situation
for the case $L^3=\pm 1$ and $L^3=\pm 2$ can be inferred from work by
Braun and Kivel \cite{Braun:2000cs} on $\gamma^* \gamma$ annihilation
into $f_2(1270)$ (which in fact predominantly decays into $\pi\pi$).
The transition amplitudes to the various meson helicity states were
calculated using the operator product expansion on the light-cone,
i.e., the same formalism described in Section~\ref{sec:twist-three}
for DVCS.  The transition to $L^3=\pm 1$ is suppressed by $1/Q$ and in
the Wandzura-Wilczek approximation depends on the twist-two quark DA.
The production of the $L^3=\pm 2$ state has a leading-twist part
involving the tensor gluon DA briefly discussed in
Section~\ref{sec:gda}.  This competes with a $1/Q^2$ suppressed
contribution, which in the Wandzura-Wilczek approximation again
involves the twist-two quark DA.  Taking the asymptotic forms for the
$z$ dependence of the respective distribution amplitudes and assuming
that the decay constants describing their normalization are of similar
size, one finds that the relative size of leading and subleading power
contributions is $\alpha_s /\pi$ versus $m^2 /Q^2$, where $m$ is the
meson mass.  Unless the two-gluon DA has a significantly larger
normalization than the quark DA one will hence need rather large $Q^2$
to have clean access to the two-gluon component in this process, given
its suppression by $\alpha_s /\pi$.


\subsection{Compton scattering beyond $1/Q$ accuracy} 
\label{sub:compton-corrections}

Whereas the structure of the virtual Compton amplitudes is well
understood including $1/Q$ contributions, no systematic treatment yet
exists for higher power suppressed terms.  This concerns in particular
the corrections to the leading helicity amplitudes conserving the
photon polarization.

First attempts have been made to resum target mass corrections to all
orders in $m^2/Q^2$ in DDVCS.  Belitsky and M\"uller
\cite{Belitsky:2001hz} have performed this resummation for the vector
and axial vector quark operators of geometrical twist two.  The
resulting hadronic tensor is conveniently expressed in terms of double
distributions, convoluted with a modified hard scattering kernel,
where the target mass and invariant momentum transfer appear via the
variable
\begin{equation}
  \label{target-mass-variable}
\mathcal{M}^2 = - 4\rho^2\, 
  \frac{(4m^2 -t) \beta^2 + t \alpha^2}{(q+q')^2\,
  (\beta+\xi\alpha)^2} ,
\end{equation}
where $\rho$ is defined in (\ref{xi-eta-def}).  In the forward limit
$t=0$, $\xi=0$, this reduces to the well-known combination $4 \xB^2
m^2/Q^2$ in the target mass corrections of deep inelastic scattering.
An important feature of this result is that the target mass
corrections are proportional to $\rho^2$ and should thus be rather
small in much of the experimentally relevant kinematics.  As it
stands, the result can however not be used for phenomenology: as in
the case of $1/Q$ corrections, the hadronic tensor obtained from
geometrical twist-two operators does not respect electromagnetic gauge
invariance.  A full treatment will therefore require the inclusion of
operators with higher geometrical twist.  Building on results of
\cite{Geyer:2001qf}, Eilers and Geyer \cite{Eilers:2002yn} have done
this for scalar quark operators (but not for the vector and axial
vector ones) and found again that the target mass appears through the
variable (\ref{target-mass-variable}).

In a similar fashion as for $\gamma^*\gamma\to \pi$ one may estimate
twist-four corrections to DVCS using the renormalon technique
discussed in the previous subsection.  Belitsky and Sch\"afer
\cite{Belitsky:1998wz} have given a detailed presentation of the
necessary resummation of fermion loops in the NLO hard-scattering
diagrams.  A numerical estimate has been reported by V\"anttinen et
al.~\cite{Vanttinen:1998pp}.  At $Q^2=4$~GeV$^2$ the authors found an
enhancement of the squared DVCS amplitude growing from about 20\% at
$\xB=0.1$ to about 60\% at $\xB=0.8$, with some dependence on the
particular ansatz taken for the twist-two GPDs.

Vanderhaeghen et al.~\cite{Vanderhaeghen:1999xj} have estimated power
corrections due to parton $k_T$ by extending to DVCS the framework of
Jakob et al.\ presented in the previous subsection.  In analogy to the
ansatz (\ref{pion-wf-ansatz}), which extends a distribution amplitude
to a wave function, the authors introduced a $k_T$ dependence in the
proton GPDs by multiplying the usual double distributions with
\begin{equation}
  \frac{a^2}{(1-|\beta|)^2 -\alpha^2}\, 
  \exp\left[ - a^2 
	\frac{4 \tvec{k}^2}{(1-|\beta|)^2 -\alpha^2} \right] 
\end{equation}
times an appropriate normalization factor, taking a transverse size
parameter $a \approx 0.86$~GeV$^{-1}$ as in the pion wave function.
The $k_T$ dependent forward quark densities obtained with the ansatz
in \cite{Vanderhaeghen:1999xj} give an average parton transverse
momentum of $\langle k_T^2 \rangle^{1/2} \approx (1-x) \times 520$~MeV
at fixed parton plus-momentum fraction $x$.  With $\xB=0.3$ the
authors find a reduction of the squared leading-twist amplitude by
about 40\% for $Q^2=2$~GeV$^2$, going down to about 20\% reduction for
$Q^2=6$~GeV$^2$.

To estimate the effect of the hadron-like component of the real photon
in DVCS it is suggestive to compare this process with the
electroproduction of a $\rho$ meson.  Donnachie and Gravelis
\cite{Donnachie:2000rz} have used a model that gives a fair
description of the H1 and ZEUS data on $\rho$ production, and
converted $\gamma^*_T\, p\to \rho^{\phantom{*}}_T\, p$ into a
contribution to the $\gamma^*_T\, p\to \gamma p$ amplitude using the
standard $\rho - \gamma$ conversion factor of the vector dominance
model.  Comparing with the measured DVCS cross section from H1
\cite{Adloff:2001cn}, they found that for $Q^2 = 2$ to 4~GeV$^2$ the
so defined $\rho$ contribution to the DVCS amplitude is about 18\%.
For the range $Q^2=11$ to 20~GeV$^2$ this ratio decreased to about
12\%.  In a similar study, Cano and Laget \cite{Cano:2002zw} used a
model which describes well the HERMES data for transverse $\rho$
electroproduction \cite{Airapetian:2000ni}.  The corresponding $\rho$
contribution to the DVCS amplitude accounts for about 50\% of the
electron beam spin asymmetries measured at CLAS (for $Q^2 =
1.25$~GeV$^2$) \cite{Stepanyan:2001sm} and at HERMES (for $Q^2 =
2.6$~GeV$^2$) \cite{Airapetian:2001yk}.  In both studies the
dependence on the particular model used for $\rho$-production should
be rather weak since $\rho$-production and DVCS are compared in
similar kinematics (one needs however assumptions about the proton
spin dependence in both processes and about the yet unmeasured $t$
dependence of DVCS).  We remark that the above conversion between
amplitudes with a real photon and amplitudes with a $\rho(770)$
corresponds to the simplest form of vector dominance, which is by
itself a model.  It may be used as an estimate of how important the
hadronic part of the photon is in DVCS, but should be taken as an
indicator of power-suppressed contributions only with great care.
Nevertheless, the studies cited above suggest some caution concerning
a pure leading-twist interpretation of DVCS at rather low $Q^2$.  The
situation for timelike photons with virtuality around the $\rho$ mass
will be discussed in Section~\ref{sub:dual-time}.

We conclude this subsection by reporting a study by Kivel and
Mankiewicz \cite{Kivel:2001rw} on $1/Q^2$ corrections to the Compton
amplitude with double photon helicity flip.  These were calculated in
the Wandzura-Wilczek approximation, with the result expressed in terms
of the twist-two GPDs $H^q$, $E^q$, $\tilde{H}^q$, $\tilde{E}^q$.
Their convolution with hard-scattering kernels was found to have only
integrable logarithmic singularities at $x=\pm \xi$, so that
factorization does not manifestly break down in this amplitude at
$O(\alpha_s^0 /Q^2)$.  In close analogy to the study
\cite{Braun:2000cs} of $\gamma^*\gamma \to f_2$ mentioned in
Section~\ref{sub:two-photon}, the ratio of the leading-twist amplitude
to the $1/Q^2$ corrections was found to go like $\alpha_s /\pi$ to
$m^2/Q^2$, where $m$ is the proton mass.  There is however a rather
complicated dependence of the hard-scattering kernels on $\xi$ and
$x$, and the tensor gluon GPDs in the leading-twist term are
completely unknown.  Without further study it is hence difficult to
say whether in a suitable range of $\xi$ and $Q^2$ the leading-twist
contribution may be favored.


\subsection{Meson form factors and meson electroproduction} 
\label{sub:form-factors}

The elastic pion form factor has been the object of rather detailed
studies of power corrections.  This observable may be a case where the
leading-twist contribution is not a good approximation of the total
amplitude, even for $Q^2$ values of 10~GeV$^2$ and higher.  Jakob et
al.~\cite{Jakob:1993iw} have included the transverse parton momentum
in the hard scattering in the way discussed in
Section~\ref{sub:two-photon}, and found a substantial suppression of
the leading-twist result in this $Q^2$ range.  Most of the suppression
is not due to the Sudakov form factors but rather to the $k_T$
dependence of the pion wave function and of the gluon propagator in
the hard scattering.  Note that transverse momentum enters a typical
gluon virtuality as $z_1 \bar{z}_2\, Q^2 + (\tvec{k}_1 -
\tvec{k}_2)^2$, where the indices refer to one or the other pion,
whereas in a quark virtuality it appears in the form $z_1 Q^2 +
\tvec{k}_1^2$ (see Fig.~\ref{fig:hadron-hadron}c).  The dominant
effect of $k_T$ is hence in the gluon propagator.  The $k_T$ in the
quark propagator is neglected in most investigations, including the
one just cited.  The integration over one of the impact parameters
conjugate to $\tvec{k}_1$ and $\tvec{k}_2$ then conveniently
simplifies to a delta function.  A study by Li found only a moderate
further suppression when taking into account $k_T$ in the quark
propagator as well \cite{Li:1993ce}.

\begin{figure}[b]
\begin{center}
	\leavevmode
	\epsfxsize=0.7\textwidth
	\epsfbox{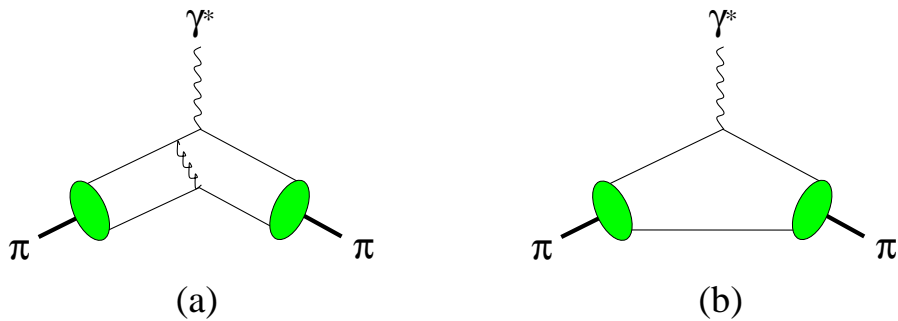}
\end{center}
\caption{\label{fig:pion-ff} (a) Hard-scattering diagram for the pion
form factor.  (b) Soft overlap contribution to the pion form factor.}
\end{figure}

The strong suppression found in~\cite{Jakob:1993iw} indicates that
even for $Q^2$ of several GeV$^2$ a substantial part of the
leading-twist expression (\ref{pion-ff}) is due to the regions where a
quark or antiquark in one or both pions becomes soft.  A power
suppressed contribution to $F_\pi$ not included in the hard-scattering
diagram of Fig.~\ref{fig:pion-ff}a is the soft overlap of the
$q\bar{q}$ Fock states for both pions, shown in
Fig.~\ref{fig:pion-ff}b.  Evaluating both contributions with the same
wave functions Jakob et al.~\cite{Jakob:1996hd} found the soft overlap
to be dominant over the graphs with one-gluon exchange.  Notice that
one has here a partial cancellation of negative power corrections from
parton $k_T$ in the $O(\alpha_s)$ graphs and of the positive soft
overlap contribution at $O(\alpha_s^0)$.  How large each contribution
is depends not only on the pion wave function but also on the details
of the perturbative treatment, compare e.g.\ the results of
\cite{Jakob:1996hd} and of \cite{Stefanis:2000vd}.  The sum of all
contributions in \cite{Jakob:1993iw} is close to the data if
$\Phi_\pi$ is close to its asymptotic form, whereas strongly
asymmetric pion DAs strongly overshoot the data.  We caution however
that the data for $F_\pi(Q^2)$ at large $Q^2$ (see
\protect\cite{Volmer:2000ek} and references therein) are obtained from
$\gamma^* p\to \pi^+ n$ by an extrapolation to the pion pole, which is
subject to criticism (see Section~\protect\ref{sec:meson-lt}).

A model-dependent way to treat both hard-scattering and soft
contributions in the same framework is provided by the light-cone sum
rule technique.  A detailed investigation has been given by Braun et
al.~\cite{Braun:1999uj}, where one of the pions was replaced with the
local axial current and the other was kept and described by its
distribution amplitudes up to twist four.  Calculating the graphs of
$O(\alpha_s^0)$ and $O(\alpha_s)$, the authors find a large positive
contribution to $F_\pi(Q^2)$ from regions where a quark or antiquark
in the pion is soft.  These power suppressed contributions are in part
canceled by negative power suppressed terms from the hard-scattering
region, due in this case to higher-twist pion DAs.

Given the model dependence of any existing treatment it is difficult
to asses how large power corrections are to $F_\pi$ in the
experimentally relevant $Q^2$ range.  The cancellations found in the
above studies also show the dangers of discussing only selected
sources of power corrections.  Moreover, the separation of different
contributions, in particular of ``hard'' versus ``soft'' ones,
requires definition within a given framework, with care taken not to
double count.

Comparing this situation for $F_\pi$ with the one for $\gamma^*\gamma
\to \pi$ one can identify why $F_\pi$ is much more sensitive to power
corrections at the same value of $Q^2$.  A typical virtuality inside
the leading-twist graphs for $\gamma^*\gamma \to \pi$ is $z_1 Q^2$.
The leading-twist graphs for $F_\pi$ start with an additional gluon
line, and the hard momentum of the external photon is diluted more
strongly inside a graph: in particular the typical virtuality of the
gluon is $z_1 \bar{z}_2\, Q^2$ and hence tends to be significantly
smaller than in $\gamma^*\gamma \to \pi$.  This leads to greater
sensitivity to corrections due for instance to the parton $k_T$, and
ultimately to the soft end-point regions.  Furthermore the soft
overlap contribution to $F_\pi$ in Fig.~\ref{fig:pion-ff}b comes with
a factor of $\alpha_s$ less than the leading power contribution, which
starts at $O(\alpha_s)$.  In contrast, the soft overlap contribution
between the real photon and the pion in $\gamma^*\gamma \to \pi$
(corresponding to the endpoint region of $z_1$) comes with the same
power of $\alpha_s$ as the leading-twist hard scattering piece.

Vanderhaeghen et al.~\cite{Vanderhaeghen:1999xj} have estimated the
effect of parton $k_T$ in meson electroproduction, using meson
light-cone wave function with a Gaussian $k_T$ dependence and a
similar $k_T$ dependence for GPDs, as described in
Section~\ref{sub:compton-corrections}.  They considered the quark GPD
contributions to both vector and pseudoscalar production.  The
inclusion of $k_T$ dependence in the relevant diagrams shifts the
point giving the imaginary part of the scattering amplitude from
$x=\xi$ to $x>\xi$, which reduces the result for current models of
GPDs.  Note that the imaginary part of the amplitude is typically
larger than the real part, except for charged pion production, which
is dominated by the pion pole contribution.  In the case of
$\gamma^*_L\, p \to \rho_L^{\phantom{*}}\, p$, the
study~\cite{Vanderhaeghen:1999xj} found indeed a rather strong
reduction of the leading-twist cross section, by a factor of about 3.5
at $\xB=0.3$ and $Q^2=4$~GeV$^2$ and a factor of about 2.5 at
$\xB=0.3$ and $Q^2=8$~GeV$^2$, with yet bigger effects at larger
$\xB$.  A strong suppression for the same process in the small-$x$
regime of the H1 and ZEUS data has been reported by Postler
\cite{Postler:2001zf} in a study including the $k_T$ dependence for
the $\rho$ wave function, but staying collinear for the gluon GPDs in
the proton.

\begin{figure}
\begin{center}
	\leavevmode
	\epsfxsize=0.67\textwidth
	\epsfbox{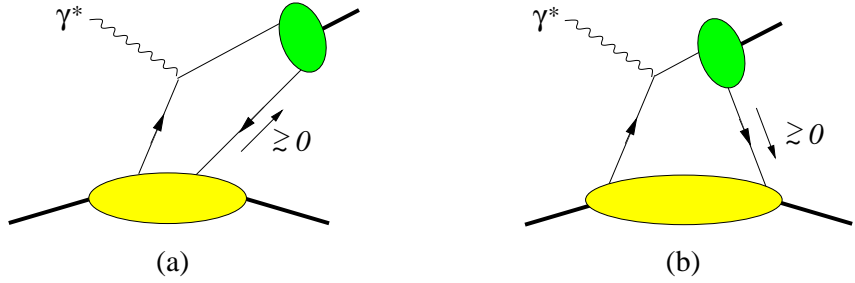}
\end{center}
\caption{\label{fig:meson-overlap} Soft overlap contributions to meson
electroproduction in (a) the ERBL and (b) the DGLAP regime.
Plus-momentum fractions $\protect\gsim 0$ refer to the average nucleon
momentum.}
\end{figure}

Like the elastic pion form factor, meson electroproduction admits a
soft overlap regime at $O(\alpha_s^0)$, shown in
Fig.~\ref{fig:meson-overlap}.  Vanderhaeghen et
al.~\cite{Vanderhaeghen:1999xj} have modeled this in the ERBL region
of Fig.~\ref{fig:meson-overlap}a, where the pion can be described by
its light-cone wave function.  How to estimate the contribution from
the DGLAP configurations of Fig.~\ref{fig:meson-overlap}b is not
known.  Note that the soft overlap (at least as calculated in this
model) contributes only to the real part of the amplitude, so that its
effect was found negligible for $\rho$ production in
\cite{Vanderhaeghen:1999xj}.  For the production of charged pions,
however, the pion pole contribution described by
(\ref{pion-pole-process}) was strongly enhanced and led to a strong
enhancement of the cross section for $\gamma^*_L\, p\to \pi^+ n$,
clearly outweighing the suppression from parton $k_T$ in the hard
scattering contributions.

The renormalon approach described in Section~\ref{sub:two-photon} has
been applied to meson production by V\"anttinen et
al.~\cite{Vanttinen:1998pp}.  For $\gamma^* p\to \pi^0 p$ at
$Q^2=4$~GeV$^2$ a substantial enhancement of the cross section was
found, with a strong dependence on the GPD model used.  Substantial
power corrections estimated from renormalons have also been reported
for $\gamma^* p\to \pi^+ n$ by Belitsky \cite{Belitsky:2003tm}, who
finds in addition that they tend to cancel in the transverse target
spin asymmetry (see Section~\ref{sec:meson-pheno}) for the kinematics
considered in his study.

\begin{figure}[b]
\begin{center}
	\leavevmode
	\epsfxsize=0.27\textwidth
	\epsfbox{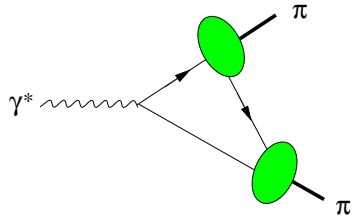}
\end{center}
\caption{\label{fig:pion-time-overlap} Soft overlap for the timelike
pion form factor.}
\end{figure}

The pion form factor at large \emph{timelike} photon virtuality is
measured to be significantly larger than at the corresponding
\emph{spacelike} virtualities.  The leading-twist formula gives
$F_\pi(q^2) = F_\pi(-q^2)$ at $O(\alpha_s)$; whether the NLO
corrections would significantly enhance the timelike region has not
been studied.  Gousset and Pire \cite{Gousset:1995yh} have studied the
effect of finite parton $k_T$ in the same formalism as Jakob et al.,
and found an \emph{enhancement} over the leading-twist $O(\alpha_s)$
result (although not quite large enough to describe the data).  This
trend is readily understood from the gluon virtuality in the hard
scattering, which is $(z_1 \bar{z}_2\, q^2 - \tvec{k}^2)$ when
transverse parton momentum is included.  For $q^2>0$ this inclusion
can provide an enhancement of the propagator, whereas for $q^2<0$ it
always gives suppression.  The soft overlap contribution to the
timelike form factor cannot be described in terms of pion light-cone
wave functions, as is seen in Fig.~\ref{fig:pion-time-overlap}.  A
study using light-cone sum rule techniques has been performed by
Bakulev et al.~\cite{Bakulev:2000uh}, with both pions replaced by
local axial currents.  The authors also calculated Sudakov-type
radiative corrections to the quark-photon vertex in this mechanism.
These are found to decrease $F_\pi$ in the spacelike region, whereas
for timelike photons they increase $F_\pi$ and also provide an
imaginary part.  Power corrections to the timelike process $\pi N \to
\gamma^* N$ discussed in Section~\ref{sec:meson-lt} have not been
studied in the literature.  Here one may expect a similar pattern
compared to pion electroproduction $\gamma^* N \to \pi N$ as between
the time- and spacelike pion form factors.

A number of studies of power corrections to the leading meson
production amplitude have been performed in the specific context of
small $\xB$.  These will be reviewed in
Section~\ref{sec:small-x-mesons}.


\subsection{Parton-hadron duality}
\label{sec:duality}

Parton-hadron duality appears in several contexts in the reactions of
interest for us.  We have already mentioned the technique of
light-cone sum rules, where duality is used to relate amplitudes
involving hadrons and Green functions involving local currents, which
are more readily accessible for perturbative evaluation in suitable
kinematics.  In Section~\ref{sub:meson-wave} we will report on the use
of parton-hadron duality to relate the diffractive amplitude $\gamma^*
p\to q\bar{q} + p$ with vector meson production.

\subsubsection{Timelike photons}
\label{sub:dual-time}

The application of the leading-twist formalism to processes with
timelike photons, in particular to $\gamma^* p \to \gamma^* p$,
$\gamma p\to \gamma^* p$ and to $\pi N\to \gamma^* N$, requires care
in kinematics where the produced photon has an invariant mass $Q'$ in
the vicinity of meson resonances.  Note that the limit for which
factorization is derived implies $Q'^2 \to \infty$ and hence does not
address this issue.  This also holds for DDVCS, where the $Q^2$ of the
incident photon provides a second hard scale: here the limit
(\ref{compton-limit}) sends both $Q^2$ and $Q'^2$ to infinity.

\begin{figure}[b]
\begin{center}
	\leavevmode
	\epsfxsize=0.7\textwidth
	\epsfbox{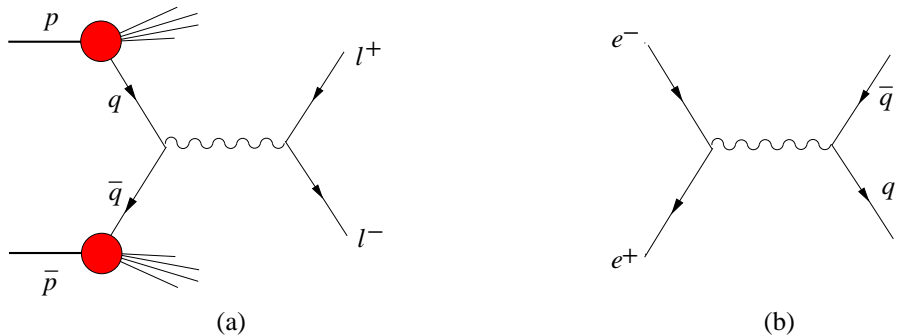}
\end{center}
\caption{\label{fig:timelike-friends} Born-level diagrams for (a)
Drell-Yan lepton pair production and (b) $e^+ e^-$ annihilation into
hadrons.}
\end{figure}

It is suggestive to compare the above cases to other processes
involving the coupling of a quark line to a timelike photon with large
virtuality, such as Drell-Yan pair production and $e^+e^-$
annihilation into hadrons (Fig.~\ref{fig:timelike-friends}).  As
discussed in \cite{Berger:2001xd} one must however be careful when
using such analogies, since the relevant kinematic configurations are
qualitatively different.  In the $\gamma^*$ production processes where
we want to study GPDs, one of the quark lines attached to the timelike
photon is part of a hard scattering kernel and thus to be regarded as
far off shell, whereas the other quark line is connected to the proton
GPD and corresponds to propagation over long distances.  In contrast,
both $q$ and $\bar{q}$ have small virtualities in the Drell-Yan
process, and both are hard in $e^+e^- \to \mathit{hadrons}$.  (In the
tree-level graph $e^+e^- \to q \bar{q}$ the final quarks are on shell,
but such singularities can be avoided by analytic continuation as
explained in Section~\ref{sec:factor}.)  Moreover, the space-time
structure is not the same in these cases: in Compton scattering at
tree level the final $\gamma^*$ is formed by $q\bar{q}$ in the ERBL
region, but in the DGLAP region the corresponding vertex is $q\to
\gamma^* q$ or $\bar{q}\to \gamma^* \bar{q}$, which does not suggest
resonant behavior.

With these caveats in mind one may nevertheless use the analogy with
Drell-Yan and $e^+e^-$ annihilation to estimate the importance of
resonance effects in exclusive $\gamma^*$ production processes.
Drell-Yan production data \cite{Badier:1984ik,Freudenreich:1990mu}
permit a leading-twist description for heavy-photon masses $Q'$ above
4 or 5~GeV, excluding the region of the $\Upsilon$ resonances.  Below
this mass range, problems arise due to decay leptons from charm
production, a background of course absent in the exclusive processes
we are interested in.  As to $e^+e^- \to \mathit{hadrons}$, the data
\cite{Bai:2001ct,Swartz:1996hc} show little resonance structure and
good agreement with the leading-twist perturbative result in the $Q'$
range from 2 to 3~GeV and above 5~GeV, excluding again the $\Upsilon$
resonances.  One may thus regard these ranges as hopeful candidates
for a leading-twist analysis of $\gamma p\to \gamma^* p$ and $\pi N\to
\gamma^* N$.

For $\gamma^* p\to \gamma^* p$ one may extend this range if the
initial photon virtuality $Q^2$ provides a hard scale to the process,
and in particular also include the low $Q'$ region down to the limit
$Q'=0$ of DVCS.  Care is nevertheless advised, especially in a $Q'$
band of several 100~MeV around 770~MeV, given the strong $\rho$ peak
in many similar situations.  The importance of the resonance effects
in this region can directly be assessed by converting data on
$\gamma^* p \to \rho p \to (\pi^+\pi^-)p$ to $\gamma^* p \to \rho p
\to (\ell^+\ell^-) p$ using the known $\rho$ branching ratios.  The
latter contributes directly to the amplitude $\gamma^* p\to \gamma^* p
\to (\ell^+\ell^-) p$, where DDVCS can be observed.  This contribution
is of the same order in the electromagnetic coupling since the
$\rho\to \ell^+\ell^-$ decay proceeds via a $\gamma^*$.  At moderately
large $Q^2$ the amplitudes for $\gamma^*_L\, p \to
\rho_L^{\phantom{*}}\, p$ and $\gamma^*_T\, p \to
\rho_T^{\phantom{*}}\, p$ are experimentally known to be  of
comparable size, whereas for very large $Q^2$ the longitudinal
contribution dominates.  Guidal and Vanderhaeghen \cite{Guidal:2002kt}
have estimated the ``$\rho$ contribution'' to DDVCS by calculating
$\gamma^*_L\, p \to \rho_L^{\phantom{*}}\, p$ in the leading-twist
approximation, and report rather little effect on DDVCS observables in
the kinematics they considered.

Given the above theoretical caveats it may be best from a
phenomenological point of view to assess the importance of resonance
effects directly from the data of the processes under study, since
such effects will likely show up in the $Q'$ distribution.  Such a
study and its comparison to other processes may itself provide
information on the mechanism of parton-hadron duality in different
environments.  In $Q'$ regions where resonance structures do appear, a
leading-twist analysis of the data will need to invoke parton-hadron
duality after averaging over a suitable $Q'$ range, as can be done for
$e^+e^- \to \mathit{hadrons}$.

\subsubsection{Bloom-Gilman duality}
\label{sub:Bloom-Gilman}

This particular type of parton-hadron duality has recently been
studied with increasing interest in inclusive DIS, both experimentally
and theoretically, see e.g.~\cite{Jeschonnek:2002nm} for a recent
overview.  An extension of such studies to DVCS and DDVCS for $W^2$ in
the resonance region seems natural.  It is amusing to note that
virtual Compton scattering with a timelike final $\gamma^*$ can
simultaneously involve parton-hadron duality in two channels,
corresponding to the variables $Q'$ and $W$.

Close and Zhao \cite{Close:2002tm} have investigated a toy model for
the proton, consisting of two scalar constituents with electromagnetic
charges $e_1$ and $e_2$ bound by a potential.  They have shown how the
superposition of resonances reproduces features of the leading-twist
mechanism, namely Bjorken scaling and the absence of terms with $e_1
e_2$, where the photons couple to different constituents.  These
features are found both for the imaginary part of the forward Compton
amplitude and for $\im \mathcal{A}(\gamma^* p \to \gamma^* p)$ in
nonforward kinematics at large $Q^2$ and small $t$.  The authors
proceed by discussing the imaginary part of the DVCS amplitude in
terms of two independent scaling variables, $Q^2 /(2pq)$ and $-t
/(2pq)$, and draw conclusions on generalized parton distributions as
functions of the independent variables $x$, $\xi$, and $t$.  Since the
leading-twist expression of $\im \mathcal{A}(\gamma^* p \to \gamma p)$
at Born level only gives access to GPDs at $x=\pm\xi$, these
conclusions are unfortunately incorrect.  It may be possible to apply
the considerations of \cite{Close:2002tm} to $\im \mathcal{A}(\gamma^*
p \to \gamma^* p)$ for small $t$ and different spacelike $q^2$ and
$q'^2$, which involves the GPDs in the DGLAP region at leading-twist
accuracy (although no practicable way is known to measure it).


\section{Beyond the factorization theorems}
\label{sec:beyond-twist-two}

A number of investigations have been concerned with processes or
observables to which the leading-twist factorization theorems
discussed in Section~\ref{sec:factor} cannot be applied, with the
prime motivation to see whether information about GPDs or GDAs can be
obtained nevertheless.  The most advanced example is Compton
scattering at the $1/Q$ level (Section~\ref{sec:twist-three}).  The
hope of such studies is often to probe GPDs or GDAs to which
leading-twist observables give only little or no access, for instance
the polarized gluon GPD $\tilde{H}^g$ or transversity GPDs.


\subsection{Transverse vector mesons}
\label{sec:transverse-mesons}

The electroproduction of transverse vector mesons from transverse
photons is power suppressed by $1/Q$ compared with the dominant
longitudinal amplitude, and there is no factorization theorem which
allows one to calculate it in terms of meson distribution amplitudes
and 
nucleon GPDs.  It is nevertheless instructive to see what an attempt
at such a calculation gives.  This has been done by Mankiewicz and
Piller \cite{Mankiewicz:1999tt}, and later by Anikin and Teryaev
\cite{Anikin:2002wg} with a different method.  The results of both
studies agree.  Only the chiral-even meson wave functions were
considered in both cases.  To make a transverse meson from a
$q\bar{q}$ pair with helicities coupled to zero, one needs one unit of
orbital angular momentum or one extra gluon, and the relevant meson
distribution amplitudes start at dynamical twist three.  To obtain the
dominant part of $\gamma^*_T\, p \to \rho_T^{\phantom{*}}\, p$ one is
then restricted to dynamical twist two on the GPD side.  The
calculation of Mankiewicz and Piller uses the distribution amplitudes
defined in the QCD string operator approach of Braun and collaborators
\cite{Ball:1998fj} and stays within the Wandzura-Wilczek
approximation.  Anikin and Teryaev use the ``transverse'' operator
basis of Ellis et al.~\cite{Ellis:1983cd} and give results both with
and without the Wandzura-Wilczek approximation, restricting themselves
to a spinless target.

Calculation of the graphs in Fig.~\ref{fig:mesons} (plus those with an
additional gluon line attached to the meson in the case of
\cite{Anikin:2002wg}) gives logarithmic singularities, which affect
both the meson DAs and the target GPDs.  For the meson they take the
form
\begin{equation}
\int_0^1 \frac{dz}{z} \int_z^1 \frac{du}{u}\, \Phi^q(u) ,
  \label{bad-meson-convolution}
\end{equation}
where $\Phi^q$ is the usual twist-two DA.  This integral is only
finite if $\int_0^1 du \, u^{-1} \Phi^q(u)$ is accidentally zero,
which can at most occur at a particular factorization scale.  On the
GPD side the result involves the familiar convolutions with $(\xi \mp
x - i\epsilon)^{-1}$, but in addition the integrals
\begin{eqnarray}
\int_{-1}^1 dx\, \frac{1}{(\xi \mp x - i\epsilon)^2}\,
   H^{q,g}(x,\xi,t) 
&=& \mp \int_{-1}^1 dx\, \frac{1}{\xi \mp x - i\epsilon}\,
   \frac{\partial}{\partial x} H^{q,g}(x,\xi,t)
  \label{bad-proton-convolution}
\end{eqnarray}
and analogous ones with $E$, $\tilde{H}$, $\tilde{E}$ for quarks and
gluons.  To be well defined and finite, they require the
$x$-derivative of the GPDs to be finite at $x=\pm \xi$.  As we
discussed in Section~\ref{sec:x-equal-xi}, this is not the case for
the quark GPDs in a large number of field theoretical models, where
this derivative has a discontinuity at these points.  We have also
seen that the double distribution formalism suggests that such
discontinuities can occur.  The situation for the gluon GPDs is less
clear.  One may observe that the asymptotic form of a gluon $D$-term
goes like $\theta(|x| < |\xi|)\, [1 - (x/\xi)^2]^2$ and has a
vanishing first derivative at $x=\pm \xi$, whereas the asymptotic form
$\theta(|x| < |\xi|)\, [1 - (x/\xi)^2]$ of a quark $D$-term has a
first derivative jumping from a finite value to zero at that point.
No further investigation of this point has been made.

These findings show that for the amplitude at hand an expansion of the
partonic scattering amplitude around zero $k_T$ of the partons (or
equivalently an expansion of nonlocal operators around $z_T=0$ in
position space) is not adequate.  The occurrence of divergences in the
calculation may be cured by keeping finite parton $k_T$ in the
partonic scattering.  This will regulate end point divergences by
replacing for instance an inverse propagator $z Q^2$ by $z Q^2 +
\tvec{k}^2$.  An even simpler phenomenological prescription is to cut
off $z$-integrals like (\ref{bad-meson-convolution}) at a value
$\langle\tvec{k}^2\rangle /Q^2$, with $\langle\tvec{k}^2\rangle$ being
of typical hadronic size.  The integral will then have a logarithmic
enhancement by $\log \langle\tvec{k}^2\rangle /Q^2$.  As remarked in
\cite{Anikin:2002wg}, this may explain why the observed ratio $R =
\sigma_L /\sigma_T$ of longitudinal to transverse $\rho$ production
seems to grow more slowly with $Q^2$ than expected from simple power
counting arguments (see Section~\ref{sub:mesons-twist}).  On the GPD
side, the inclusion of transverse momentum would change the location
of the singularity in (\ref{bad-proton-convolution}) from $x=\xi$ to
$x>\xi$, where the GPDs are expected to be regular.

As a matter of principle, the above divergences signal however the
presence of soft contributions to the amplitude, in full accordance
with the general analysis of the factorization theorems we presented
in Section~\ref{sec:factor}.  Such soft exchanges between the meson
and target side break the factorization into soft matrix elements
depending only on either the meson or the target.  It is not known how
serious such violations of universality are, and it remains
controversial in the literature if phenomenological prescriptions as
discussed above are adequate to describe the essential dynamics.  We
will again come across this issue in Sections~\ref{sec:aligned-jet}
and \ref{sec:small-x-mesons}.


\subsection{Meson pairs with large invariant mass}
\label{sec:meson-pairs}

Lehmann-Dronke et al.~\cite{Lehmann-Dronke:1999ym} have studied the
electroproduction production of $\pi^+\pi^-$ and $K^+K^-$ pairs with
an invariant pair mass $M$ of a few GeV.  In contrast to the case $M^2
\ll Q^2$, where the formation of the meson pair can be described by
the appropriate GDA, the process was modeled by calculating the
production of a $q\bar{q}$ pair at the cross section level and
describing the subsequent hadronization to a meson pair using the Lund
string fragmentation model \cite{Andersson:1983ia}.  This is
reminiscent of the study of exclusive $\gamma^*\gamma$ annihilation
into hadrons we reported on in Section~\ref{sub:lund}.  The authors of
\cite{Lehmann-Dronke:1999ym} estimated cross sections for kinematics
achievable at HERMES and found them to be too small for the luminosity
one may expect there.

The description of open $q\bar{q}$ production in the GPD framework
(which we will again discuss in Section~\ref{sec:open-diff}) is of
theoretical interest by itself.  Lehmann-Dronke et al.\ use the
collinear factorization formalism to calculate the amplitudes for both
transverse and longitudinal photons.  The result for gluon exchange
diagrams involves the convolution (\ref{bad-proton-convolution}) for
the GPDs $H^{g}$, $E^{g}$, $\tilde{H}^{g}$, $\tilde{E}^{g}$.  In
contrast, the scattering kernels for the quark distributions $H^{q}$,
$E^{q}$, $\tilde{H}^{q}$, $\tilde{E}^{q}$ are found to have only
simple poles at $x=\pm \xi$ and $x= \pm \rho$, with $\rho$ defined in
(\ref{xi-eta-def}) if we set $Q'= M$.  Dangerous double poles would
thus only occur for $M \to 0$, where $\xi$ and $\rho$ coincide.  The
momentum fractions $z$ and $1-z$ of the final $q$ and $\bar{q}$ are
external kinematical variables in this case.  One may hence avoid the
dangerous soft regions $z\to 0,1$ by imposing a minimum transverse
momentum $\tvec{k}$ for the $q\bar{q}$ pair, given that $\tvec{k}^2 =
z(1-z) M^2$ .  In physical applications one must of course assume that
the quark momentum is sufficiently correlated with a measured momentum
in the final state.


\subsection{Access to transversity distributions}
\label{sec:transversity-access}

We have seen in Sections~\ref{sec:compton-scatt} and
\ref{sub:gamma-star-gamma} that one may access the gluon transversity
GPDs at leading-twist level in Compton scattering with photon helicity
flip, and the corresponding tensor gluon DA or GDA in two-photon
annihilation with $L^3=\pm 2$.  Kivel \cite{Kivel:2001qw} has
investigated the occurrence of these quantities in power suppressed
meson electroproduction amplitudes with helicity transfer between the
photon and the meson.  He calculated the diagrams of
Fig.~\ref{fig:mesons} in the light-cone string operator formalism of
Braun et al., giving results for a spinless target.  Two cases were
found where tensor gluon GPDs or DAs provide the only contribution to
the considered amplitude at level $1/Q^2$.  Both involve one unit of
orbital angular momentum in the hard scattering kernel, which gives
the $1/Q$ suppression compared with the leading-twist amplitude, where
photon and meson are longitudinal.  The corresponding quark GPDs or
DAs of dynamical twist three were reduced to the usual twist-two ones
using the Wandzura-Wilczek approximation.
\begin{itemize}
\item  The transition $\gamma^*(+1)\, h \to f_2(+2)\, h$, where
helicities are given in parentheses and $h$ denotes the target,
involves the usual quark GPDs $H^q$ for the target, and the tensor
gluon DA $\Phi^g_T(z)$ of the $f_2$ in the form
\begin{equation}
\int_0^1 dz\, \frac{\Phi^g_T(z)}{z^2 (1-z)^2} .
\end{equation}
The expansion of $\Phi^g_T(z)$ in Gegenbauer polynomials motivated by
the evolution equations contains a factor $z^2 (1-z)^2$, as does its
analog (\ref{Polyakov-expansion}) for $\Phi^g$, so that this integral
does not have an endpoint divergence.
\item The transition $\gamma^*(+1)\, h \to \rho(-1)\, h$ goes with the
usual meson DAs $\Phi^q$, and with the tensor gluon GPD in the
convolution
\begin{equation}
\int_{-1}^1 dx\, \frac{1}{(\xi - x - i\epsilon)^2 
                          (\xi + x - i\epsilon)^2}\, H^g_T(x,\xi,t)  .
\end{equation}
As for the usual gluon GPDs, it is not known whether $H^g_T$ has a
continuous first derivative at $x=\pm \xi$, which is required to make
this integral well-defined.  We remark that in a small-$x$ study to be
discussed in Section~\ref{sub:meson-wave}, Ivanov and Kirschner
\cite{Ivanov:1998gk} obtained a $1/Q^2$ contribution to the same
amplitude which does not involve the gluon transversity distribution
in an obvious way.  Why this contribution seems does not appear in the
formalism used by Kivel is presently not understood.
\end{itemize}

\begin{figure}[b]
\begin{center}
	\leavevmode
	\epsfxsize=0.32\textwidth
	\epsfbox{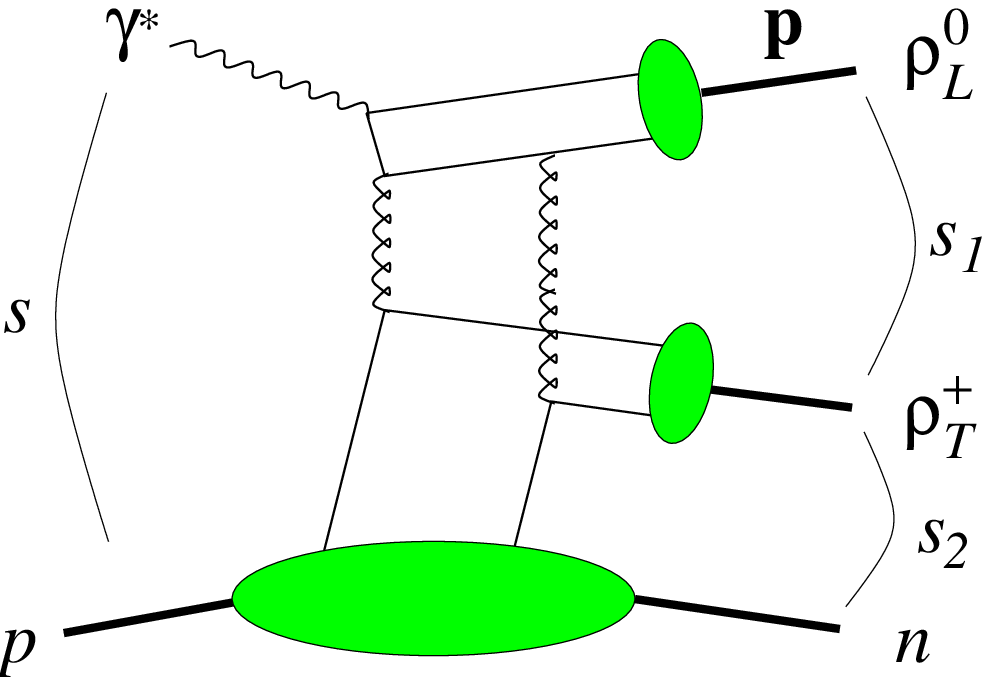}
\end{center}
\caption{\label{fig:quark-transversity} Production of two mesons with
a large gap in rapidity between them as a candidate process for
measuring quark transversity GPDs.}
\end{figure}

A candidate mechanism to access \emph{quark} transversity GPDs has
been proposed by Ivanov et al.~\cite{Ivanov:2002jj}.  (Remember that
the early proposal to use processes like $\gamma^*_L\, p \to
\rho_T^{\phantom{*}}\, p$ to this end cannot be realized as explained
in Section~\ref{sub:selection}.)  The authors of \cite{Ivanov:2002jj}
considered the process $\gamma^* p \to \rho_L^0\, \rho_T^+\, n$ in
kinematics with a large rapidity gap between the two mesons.  In terms
of the variables defined in Fig.~\ref{fig:quark-transversity} the
chosen kinematics is $s \sim s_1 \gg s_2 \gg -t \sim
\Lambda_{\mathrm{QCD}}^2$ and $s_1 \gg Q^2 \gg
\Lambda_{\mathrm{QCD}}^2$.  This implies that the transverse momentum
of the $\rho_L^0$ is also large, $\tvec{p}^2 \sim s_2$.  Factorization
as shown in in Fig.~\ref{fig:quark-transversity} has not been proven
for this process, but the authors could show that the off-shell gluon
propagators in the relevant diagrams do not have pinched
singularities.  Large $Q^2$ was imposed in order to suppress the
hadron-like component of the initial (transverse or longitudinal)
photon, which is more prone to factorization breaking soft exchanges.
This requirement may be avoided by replacing the $\rho_L^0$ with a
$J/\Psi$.  A large subenergy $s_2$ was taken in order to select gluon
exchange in the upper part of the diagram, whereas the charge of the
$\rho_T^+$ ensures quark exchange in the lower part.  The
leading-twist collinear approximation, which should be relevant for
large $\tvec{p}^2$, selects the chiral-odd distribution amplitudes
$\Phi^q_T$ for the transverse meson.  This in turn selects the
chiral-odd nucleon GPDs, which appear in the amplitude through a
convolution in $x$ restricted to the ERBL region, at skewness $\xi
\approx \zeta / (2-\zeta)$ with $\zeta \approx (Q^2 + s_1) /s \approx
\tvec{p}^2 /(s_2 + \tvec{p}^2)$.  Further investigation will have to
show whether this proposal can be used in practice to provide access
to chiral-odd GPDs, about which nothing is presently known.


\subsection{Heavy meson production}
\label{sec:heavy-mesons}

\subsubsection{Charmonium}

Exclusive production of heavy quarkonium states, in particular of the
$J/\Psi$, has repeatedly been investigated as a process to access gluon
GPDs.  Some of these studies are dedicated to the small $x$ regime and
will be reviewed in Section~\ref{sub:meson-wave}.

The relevant graphs for heavy meson production are as for light mesons
in Fig.~\ref{fig:mesons}b.  In the limit where $Q^2 \gg M_{J/\Psi}^2$,
the charm quarks in the loop may be treated as massless, so that one
recovers the predictions of the leading-twist factorization theorem:
the dominant $\gamma^* p\to J/\Psi\, p$ transition has both photon and
meson longitudinally polarized, and this transition is described in
terms of nucleon GPDs and the leading-twist distribution amplitude
$\Phi_{J/\Psi}(z)$ of the $J/\Psi$.

The opposite of this situation is photoproduction, $Q^2=0$.  In this
situation it is the charm quark mass that provides a hard scale to the
process and protects the off-shell propagators in the graphs of
Fig.~\ref{fig:mesons}b from going close to their mass shell.  Of
course, the charm quark mass is not negligible if $Q^2$ is comparable
in size to $M_{J/\Psi}^2$.  The finite quark mass also cuts off the
endpoint regions $z=0,1$ of the meson wave function, where
factorization breaks down, so that both amplitudes with $\gamma_T^*$
and $\gamma_L^*$ can be investigated in a GPD framework.  A further
characteristic of the process is that charmonium bound states are
approximately nonrelativistic.  The simplest approximation of their
$c\bar{c}$ wave function corresponds to the static case, where $c$ and
$\bar{c}$ equally share the meson four-momentum, with zero relative
momentum and $m_c$ approximated by half the bound state mass.  This
static approximation is widely used in the literature.  As a
refinement, one may expand the wave function in powers of the relative
velocity $v$ of the quarks in the meson rest frame, as is done in the
non-relativistic QCD approach \cite{Bodwin:1995jh}.  So far, no
systematic analysis has been given for issues of factorization and
power counting in $\gamma^{(*)} p \to J/\Psi\, p$ when the quark mass
cannot be neglected.

Hoodbhoy~\cite{Hoodbhoy:1997zg} has studied the deviation from the
static wave function approximation to order $v^2$ in the
non-relativistic expansion \emph{and} to leading accuracy in the $1/Q$
expansion of $J/\Psi$ electroproduction.  A very small reduction of
the amplitude was found compared with the static wave function.  We
note that the static approximation implies a distribution amplitude
$\Phi(z) \propto \delta(z-\half)$, and that the integral $\int dz\,
z^{-1} \Phi(z)$ \emph{increases} when this delta-peak becomes wider,
unless there are cancellations between regions with positive and
negative $\Phi(z)$.  It is hence not clear how the departure from the
static approximation could lead to \emph{suppression} of the
electroproduction amplitude in the large $Q$ limit.  We also observe
that for $Q^2 \gg m_c^2$ it is appropriate to use the meson DA at a
renormalization scale of order $Q^2$ to prevent large logarithms $\log
Q/m_c$ appearing in loop corrections to the hard-scattering process.
The physics of modes with virtualities between $m_c$ and $Q$ is
however not included in a wave function obtained in a non-relativistic
approximation.

A study by Ryskin \cite{Ryskin:1997fm} of $J/\Psi$ electro- and
photoproduction suggested that one may access the polarized gluon
distributions ($\tilde{H}^g$ and $\tilde{E}^g$ in our notation) in
suitable polarization observables.  Unfortunately, this hope was the
consequence of a mistake in the calculation, as pointed out by
V\"anttinen and Mankiewicz \cite{Vanttinen:1998en}.  Calculating the
graphs of Fig.~\ref{fig:mesons}b with the collinear approximation for
the gluons, keeping the charm quark mass in the loop, one finds that
$\tilde{H}^g$ and $\tilde{E}^g$ do not contribute to $\gamma^{(*)}
p\to J/\Psi\, p$ in the static approximation for the meson wave
function.  Taking into account the corrections of $O(v^2)$ in a
non-relativistic expansion, V\"anttinen and Mankiewicz
\cite{Vanttinen:1998zd} found only very small effects.  The double
polarization asymmetry of $J/\Psi$ photoproduction with the helicities
of photon and target proton aligned or antialigned in the c.m.\ is
proportional to the ratio of polarized to unpolarized gluon GPDs, but
a value of at most 1\% was estimated for this observable.

One should finally remark that the charm quark mass is not very large,
and that the treatment of the charmonium as a non-relativistic system
is not undisputed, see Section~\ref{sub:meson-wave}.  This is less
problematic for the $\Upsilon$ resonances.

\subsubsection{Heavy-light mesons}
\label{sub:heavy-light}

With the prospect of future high-intensity neutrino beam facilities,
charged and neutral current production process may be studied at the
same footing as the photon induced reactions we have discussed so far.
Due to the quantum numbers of $W$ and $Z$ bosons, these currents give
access to different combinations of GPDs than those one can probe in
electromagnetic reactions.

Lehmann-Dronke and Sch\"afer \cite{Lehmann-Dronke:2001wu} have studied
the process $\bar{\nu}_\mu\, p \to \mu^+ p \, D_s^-$ for large
virtuality $Q^2$ and longitudinal polarization of the exchanged $W^-$
boson.  The graphs of Fig.~\ref{fig:mesons}a and
Fig.~\ref{fig:mesons}b were calculated in the collinear hard
scattering approximation.  Both the mass of the charm quark and of the
meson were kept finite, neglecting terms of order $M_{D_S}^4/ (Q^2 +
M_{D_S}^2)^2$.  The result involves the $s\bar{c}$ distribution
amplitude of the meson and the proton GPDs $H^g$, $H^s - \tilde{H}^s$,
$\tilde{H}^g$ together with their $E$-counterparts.  The polarized
gluon GPDs $\tilde{H}^g$, $\tilde{E}^g$ appeared with a factor
$M_{D_S}^2/ (Q^2 + M_{D_S}^2)$ and were completely negligible in a
numerical study of the amplitude for $Q^2 > 12$~GeV$^2$.  In this
region, effects from the finite masses $m_c$ and $M_{D_S}$ were found
to be quite small, so that the results of the massless leading-twist
approach gave a good approximation of the result.


\subsection{Hadron-hadron collisions}
\label{sec:hadron-hadron}

The reactions we have discussed so far are all in photon or lepton
induced scattering on a hadronic target.\footnote{An exception are
cases like $\pi^- p \to \gamma^* n$ where an off-shell photon is in
the final instead of the initial state.  When referring to
hadron-hadron scattering here, we imply more than one hadron in both
the initial and the final state.}
It is natural to ask if one cannot also use hadron-hadron collisions
to study GPDs, as one does for the extraction of the usual parton
densities.  This turns out to be difficult, since the factorization
theorems which provide a sound basis to extract GPDs do not hold for
hadron-hadron scattering processes.

\begin{figure}
\begin{center}
	\leavevmode
	\epsfxsize=0.82\textwidth
	\epsfbox{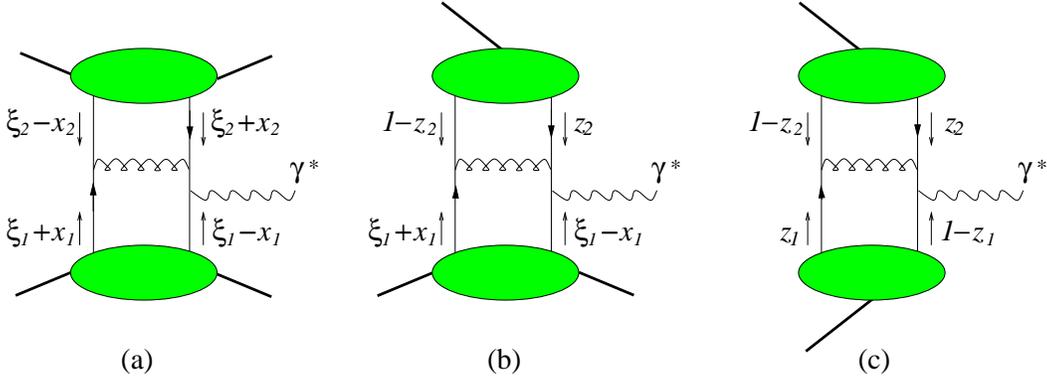}
\end{center}
\caption{\label{fig:hadron-hadron} Example graphs of the hard
scattering mechanism for $p p \to \gamma^* p p$ (a), for $\pi^- p \to
\gamma^* n$ (b), and for the timelike meson form factor (c).  Momentum
fractions refer to the meson for DAs and to the average hadron
momentum for GPDs.}
\end{figure}

A simple example of the problems which occur is the exclusive analog
of Drell-Yan pair production, i.e., $p p \to \gamma^* p p$ or $p
\bar{p} \to \gamma^* n \bar{n}$ with the timelike photon decaying into
a lepton pair.\footnote{I thank B.~Pire and O.~Teryaev for discussions
about this issue.}
Naively employing leading-twist factorization one obtains the diagram
of Fig.~\ref{fig:hadron-hadron}a and further diagrams with different
attachments of the photon to a quark line.  The hard-scattering
subprocess is the same as for $\pi^- p \to \gamma^* n$ in
Fig.~\ref{fig:hadron-hadron}b and for the timelike pion form factor
$\pi^- \pi^+ \to \gamma^*$ in Fig.~\ref{fig:hadron-hadron}c, where
leading-twist factorization does hold.  The corresponding parton
scattering amplitudes are related by the correspondence between $z$ in
a meson DA and $(\xi + x) /(2 \xi)$ for a GPD, both giving the
momentum fraction of the quark with respect to the total momentum
carried by the two partons.  This gives hard scattering kernels
\begin{eqnarray}
K_a &\propto& 
  \frac{e_{q'}}{(\xi_1 + x_1) (\xi_2 - x_2) + i\epsilon} -
  \frac{e_{q}}{(\xi_1 - x_1) (\xi_2 + x_2) + i\epsilon},
\nonumber \\
K_b &\propto& 
  \frac{e_{q'}}{(\xi_1 + x_1) (1 - z_2) + i\epsilon} -
  \frac{e_{q}}{(\xi_1 - x_1) z_2 + i\epsilon} ,
\nonumber \\
K_c &\propto& 
  \frac{e_{q'}}{z_1 (1 - z_2) + i\epsilon} -
  \frac{e_{q}}{(1 - z_1) z_2 + i\epsilon} ,
\end{eqnarray}
for the three cases of Fig.~\ref{fig:hadron-hadron}, where the
$i\epsilon$ prescription follows directly from the Feynman
prescription for the gluon propagator in the diagrams (the
denominators of the quark propagators cancel against the numerators in
the sum of the relevant diagrams).  $e_q$ and $e_{q'}$ respectively
are the charges of the quark on the left- and right-hand sides of the
graphs.  The crucial difference between the case of a meson DA and a
GPD is that the poles of the gluon propagators in $z_{1,2}$ are at the
endpoints of the $z_{1,2}$ integration and will be canceled by the
zeroes of the meson DA.  In contrast, the poles in $x_{1,2}$ are at
$\pm \xi_{1,2}$, where the corresponding GPDs do in general not
vanish.  For $K_b$ this poses no problem, since the $i\epsilon$
prescription makes the integral well defined.  For $K_a$ however, this
prescription is not sufficient to regulate the simultaneous poles in
$x_1$ and $x_2$ in the convolution with the corresponding GPDs.
Explicitly one finds that the imaginary part of this convolution
diverges as $\log (1/\epsilon)$.  The divergence cannot be canceled
between the terms with $e_q$ and $e_{q'}$ in the hard scattering,
since the GPDs are in general different at $x=\xi$ and $x= -\xi$.
This shows that the ``hard scattering kernel'' naively calculated in
the collinear expansion for $p p \to \gamma^* p p$ is not actually
hard: the configuration in Fig.~\ref{fig:hadron-hadron}a where the
gluon and both quark lines on the left have soft momenta is not
suppressed.

As we mentioned in Section~\ref{sec:factor}, the factorization proof
for meson electroproduction had to resort to complex contour
deformation of loop momenta in order to show that nonfactorizable soft
interactions do not contribute to the scattering amplitude, and that
this was only possible because there were not two hadrons both in the
initial and in the final state.  Nonfactorizable soft interactions in
hadron-hadron scattering can be shown to interfere destructively only
in \emph{inclusive} processes like Drell-Yan production, where one
sums over all remnant states of the initial hadrons.  For a brief
discussion and references see~\cite{Collins:2001ga}.

An example where factorization breaking expected for the above reasons
seems to be observed in data are diffractive processes.  For
diffraction in $ep$ collisions one can establish factorization in
terms of hard scattering coefficients and diffractive parton densities
\cite{Collins:1998sr}.  Taking densities fitted to diffractive $ep$
scattering at HERA and calculating processes like diffractive $W$
production, $p \bar{p} \to W + p + X$ with a large gap in rapidity
between the scattered $p$ and all other final-state particles, one
finds cross sections significantly larger than measured at the
Tevatron \cite{Alvero:1998ta,Jung:2002mx}.  Suggestive of the picture
that factorization is broken by \emph{soft} interactions is the
observation that the momentum distribution of the particles produced
in the hard subprocess does seem to agree with predictions obtained by
assuming factorization \cite{Jung:2002mx}.  There are attempts in the
literature to model the effect of such soft interactions, which is
often called ``rapidity gap survival'' with the idea in mind that a
potential rapidity gap left behind from a hard subprocess can
subsequently be filled by particles from soft interactions.

\begin{figure}
\begin{center}
	\leavevmode
	\epsfxsize=0.22\textwidth
	\epsfbox{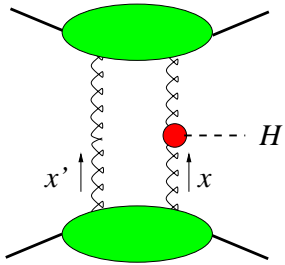}
\end{center}
\caption{\label{fig:higgs} Diagram for exclusive Higgs production in
hadron-hadron collisions.  The coupling between gluons and Higgs is
via a top quark loop.}
\end{figure}

This was for instance done by Khoze et
al.~\cite{Khoze:1997dr,Khoze:2000cy,DeRoeck:2002hk} in an
investigation of exclusive Higgs production, $p p\to p + H + p$, which
has long been proposed as a very clean process to detect and study the
Higgs at the LHC.  In this work the survival probability multiplies
the result obtained from calculating the diagram in
Fig.~\ref{fig:higgs}.  Note that, unlike in
Fig.~\ref{fig:hadron-hadron}a, one parton exchanged between the two
protons is now disconnected from the hard scattering subgraph $g g \to
H$.  Khoze et al.\ calculate this graph using $k_T$ dependent GPDs
(see Section~\ref{sub:beyond-collinear}), including a Sudakov form
factor which suppresses the region of small parton $k_T$.  The authors
find the typical transverse gluon momentum to satisfy
$\Lambda^2_{\mathrm{QCD}} \ll k_T^2 \ll M_H^2$, and typical
longitudinal momentum fractions
\begin{equation}
\frac{\Lambda_{\mathrm{QCD}}}{\sqrt{s}} 
      \:\:\ll\:\:  |x'| \sim \frac{k_T}{\sqrt{s}} 
      \:\:\ll\:\:  x \sim \frac{M_H}{\sqrt{s}} ,
\end{equation}
so that the gluon which ``bypasses'' the hard scattering is not
actually soft.  We remark that the theoretical description of $p p\to
p + H + p$ is not uncontroversial, with a wide spread in size of cross
section estimates in the literature, see
e.g.~\cite{Khoze:2002py,DeRoeck:2002pr}.

\section{GPDs and small-$x$ physics}
\label{sec:small-x}

Hard exclusive processes at high energy and hence at small $x$ have
been investigated in theory and experiment for some time, in
particular since the operation of the HERA $ep$ collider at DESY.  In
this section we will discuss some of the dynamical issues which render
the small-$x$ regime special compared with moderate and large $x$, and
put into context the role played by GPDs.  As in
Section~\ref{sec:small-x-gpd} we will use the term ``small $x$''
having in mind that the small external variable is $\xi$ or $\xB$ in
the electroproduction processes we focus on.

To start with let us recall some general results for high-energy
reactions and parton distributions at small $x$.
\begin{itemize}
\item The dominant parton densities at small $x$ are the unpolarized
gluon and next the flavor singlet combination of unpolarized quarks
and antiquarks, for scales $\mu^2=Q^2$ large enough to make contact
with physical observables.  Evolution to higher scales enhances this
trend.  It is natural to expect the same situation for GPDs.  The
small-$x$ behavior of parton helicity dependent distributions is
interesting in itself, but these distributions will presumably be hard
to access experimentally.
\item Helicity flip amplitudes in hadron-hadron scattering at high
energy are measured to be relatively small, although nonzero, so that
at small $\xi$ one may expect $H^g$ and $\sum_q H^q$ to dominate over
$E^g$ and $\sum_q E^q$.  This is in line with the comparison between
the isoscalar combinations of the Dirac and Pauli form factors
$(F_1^p+F_1^n)$ and $(F_2^p+F_2^n)$, with the caveats spelled out in
Section~\ref{sub:t-depend}.
\item The real and imaginary parts of scattering amplitudes are
related by dispersion relations, which involve the imaginary part in
the entire physical region of squared c.m.\ energy $s$.\footnote{To be
precise, one should replace $\im \mathcal{A}$ in this discussion by
the energy discontinuity $\mbox{disc}_s\,\mathcal{A}(s,t) =
[\mathcal{A}(s+i\epsilon,t) - \mathcal{A}(s-i\epsilon,t)] /(2i)$.
This makes for instance a difference when $\mathcal{A}$ involves
phases due to spinors.}
As an example consider a once subtracted dispersion relation at fixed
$t$ for an amplitude even under crossing $s \leftrightarrow u$, see
e.g.\ \cite{Martin:1970aa,Collins:1977jy},
\begin{equation}
\re \frac{\mathcal{A}_{+}(s,t)}{s} \approx
  \frac{2 s}{\pi} 
  \pv{3}\int_{s_{\mathrm{min}}}^\infty
  \frac{ds'}{s'^2 - s^2}\,  \im \frac{\mathcal{A}_{+}(s',t)}{s'} ,
\end{equation}
where we have taken the limit of large $s$ and small $t$ and neglected
subtraction and pole terms (which would decrease as $s^{-1}$ on the
right-hand side).  This relation can be simplified by Taylor expanding
$\im \mathcal{A}_{+}(s',t) /s'$ in $\log(s'/s)$ and approximating
$s_{\mathrm{min}} \approx 0$.  The Taylor series can be summed
explicitly, and one obtains a so-called derivative analyticity
relation~\cite{Bronzan:1974jh}
\begin{eqnarray}
  \label{large-s-dispersion}
\re \frac{\mathcal{A}_{+}(s,t)}{s} &\approx &
  \tan\left( \frac{\pi}{2}\, \frac{d}{d \log{s}} \right) 
  \frac{\im \mathcal{A}_{+}(s,t)}{s} 
 \approx \frac{\pi}{2}\, \frac{d}{d \log{s}} \,
        \im \frac{\mathcal{A}_{+}(s,t)}{s} .
\end{eqnarray}
For amplitudes odd under crossing one starts with 
\begin{equation}
\re \mathcal{A}_{-}(s,t) \approx
  \frac{2 s}{\pi} 
  \pv{3}\int_{s_{\mathrm{min}}}^\infty
   \frac{ds'}{s'^2 - s^2}\,  \im \mathcal{A}_{-}(s',t) ,
\end{equation}
and finds
\begin{eqnarray}
  \label{odd-dispersion}
\re \mathcal{A}_{-}(s,t) &\approx &
  \tan\left( \frac{\pi}{2}\, \frac{d}{d \log{s}} \right) 
  \im \mathcal{A}_{-}(s,t)
 \approx \frac{\pi}{2}\, \frac{d}{d \log{s}} \,
        \im \mathcal{A}_{-}(s,t) .
\end{eqnarray}
A more sophisticated version of these relations is also given in
\cite{Bronzan:1974jh}.  Most small-$x$ studies calculate the imaginary
part of a scattering amplitude and recover the real part by a relation
of the type (\ref{large-s-dispersion}).
\item The parameterization  (\ref{small-x-power}) of small-$x$ parton
densities as a power of $x$ is motivated by Regge theory, which
describes strong interactions at high $s$ and small $t$ starting from
general principles like analyticity of scattering amplitudes
\cite{Martin:1970aa,Collins:1977jy}.  In accordance with these
principles, Regge behavior is also found for scattering amplitudes in
QCD \cite{Forshaw:1997dc,Donnachie:2001zz}.  The simplest energy
dependence obtained in Regge theory is a sum of ``Regge poles''
\begin{equation}
\label{Regge-amplitude}
\mathcal{A}(s,t) \sim \sum_j c_j(t)\, 
   \left( \frac{s}{s_0} \right)^{\alpha_j(t)}\, 
        (1+ S_j\, e^{- i\pi \alpha_j(t)}) ,
\end{equation}
where the real-valued coefficients $c_j(t)$ depend on the scattering
particles, but the ``Regge trajectories'' $\alpha_j(t)$ do not.
Phenomenologically, trajectories are well described by a linear form
$\alpha_j^{\phantom{'}}(t) = \alpha_j^{\phantom{'}}(0) + \alpha'_j \,
t$ for $-t$ not too large.  Each term $j$ is associated with a
$t$-channel exchange of definite quantum numbers, including the
``signature'' $S_j=\pm 1$ which distinguishes even and odd terms under
$s \leftrightarrow u$ crossing (equivalent to $s \leftrightarrow -s$
at high energies).  Notice that the relations
\begin{equation}
\label{Regge-phase}
\re \mathcal{A}_+ = 
	\tan\Big[ {\textstyle\frac{\pi}{2}} (\alpha-1) \Big]\,
			\im \mathcal{A}_+ ,
\qquad
\re \mathcal{A}_- = 
	\tan\Big[ {\textstyle\frac{\pi}{2}} \alpha \,\Big]\, 
			\im \mathcal{A}_-
\end{equation}
for a single term with positive or negative signature are in agreement
with (\ref{large-s-dispersion}) and (\ref{odd-dispersion}) as it must
be.  More complicated dependence on $s$ involves logarithms from
``Regge cuts'' in addition to the powers in (\ref{Regge-amplitude}).

The form (\ref{Regge-amplitude}) translates into a representation of
quark and gluon densities as a sum over powers $x^{-\alpha_j(0)}$
\cite{Landshoff:1971ff}.  This can be seen by representing the parton
distributions as parton-hadron scattering amplitudes (see
Section~\ref{sec:light-cone}), or by using (\ref{Regge-amplitude}) for
the forward Compton amplitude $\gamma^* p\to \gamma^* p$ and comparing
with its expression through parton densities.  An extension of these
arguments underlies the ansatz (\ref{t-exponentials}) for the
$t$-dependence of GPDs at small $x$ and $\xi=0$, where the trajectory
$\alpha_j(t)$ appears.
\item A special role is played by exchanges with vacuum quantum
numbers ($P=C=1$ and isospin $I=0$ and positive signature), which
dominate elastic amplitudes in the high-energy limit.  In QCD these
quantum numbers correspond to two-gluon exchange in the $t$-channel.
Data for hadron-hadron scattering and for DIS down to $\xB \approx
10^{-2}$ can be well described by Regge poles corresponding to meson
exchange in addition to a ``Pomeron'' pole with vacuum quantum numbers
and a trajectory with $\alpha_{I\!\!P}^{\phantom{'}}(0) \approx 1.08$
and $\alpha'_{I\!\!P} \approx 0.25$~GeV$^{-2}$
\cite{Donnachie:1986iz,Donnachie:1992ny}. The HERA data for deep
inelastic processes at $\xB$ down to about $10^{-4}$ have however
shown that the situation is not as simple, with the rise in energy
becoming considerably steeper as $Q^2$ increases.  The nature of this
phenomenon and its description in QCD remains a major unsettled and
controversial issue.  Cudell et al.~\cite{Cudell:1999kf} have
emphasized that in Regge theory the powers of energy in
(\ref{Regge-amplitude}) depend on $t$ but not on the virtuality $Q^2$
of a scattering particle.  When parameterizing the high-energy
behavior of cross sections by a single power-law with a $Q^2$
dependent exponent, one should hence keep in mind that this is an
``effective power'', which approximates a sum of powers or a more
complicated form involving logarithms in a certain region of energy.
Correspondingly, a single power-law term $x f(x) \sim x^{-\lambda}$ in
a parton distribution can provide an approximate description over a
finite $x$-range and for a given factorization scale $\mu$.
\end{itemize}

We already observed in Section~\ref{sec:factor} that a power-law
behavior of GPDs as in (\ref{xi-scaling}) leads to a power-law
behavior in $\xi$ of the scattering amplitude, due to the general
structure of hard-scattering kernels.  The relations
(\ref{Regge-phase}) between real and imaginary part of the amplitude
have also been reproduced, both in numerical \cite{Freund:2001rk} and
analytical \cite{Noritzsch:2000pr,Belitsky:2001ns} studies with
appropriate input GPDs.  It is believed that the leading-twist
factorization formulae satisfy in fact the dispersion relations
underlying (\ref{Regge-phase}).  We remark however that the proof
given for this in \cite{Collins:1997fb} is incorrect as it stands,
since the argument does not properly take into account the $\xi$
dependence of the GPDs .

We finally note that for electroproduction processes the Regge limit
means $\xB\to 0$ at fixed $Q^2$, whereas the limit underlying the
collinear factorization theorems is $Q^2\to \infty$ at fixed $\xB$.
Whether for finite values of $Q^2$ and $\xB$ one approximation scheme
of the other is appropriate (or both) can at present not be decided
from theory and is at times controversial in practice.


\subsection{Collinear factorization and beyond}
\label{sub:beyond-collinear}

\begin{figure}
\begin{center}
	\leavevmode
	\epsfxsize=0.3\textwidth
	\epsfbox{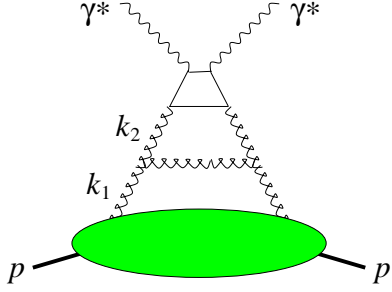}
\end{center}
\caption{\label{fig:k-perp-dglap} Two-loop diagram for the forward
Compton amplitude.}
\end{figure}

The description of hard scattering processes we have used so far is
based on collinear factorization, hard scattering kernels evaluated at
fixed order in $\alpha_s$, and the resummation of logarithms
$\alpha_s^m (\alpha_s^{\phantom{m}\!\!\!} \log Q^2)^n$ to all orders
in $n$ by DGLAP evolution or its nonforward generalization.  In the
small-$x$ limit one has however a two-scale problem, $s\gg Q^2 \gg
\Lambda^2_{\mathrm{QCD}}$, and large logarithms $\log \frac{1}{x}$
appear in loops.  Their resummation to all orders of perturbation
theory can be performed using different schemes with different
accuracy of the resulting scattering amplitudes, such as leading power
in $1/Q$, resummed leading logarithms $(\alpha_s \log \frac{1}{x})^n$,
resummed leading double logarithms $(\alpha_s \log Q^2\, \log
\frac{1}{x})^n$, etc.  An important ingredient in this context are
$k_T$ dependent, or unintegrated parton distributions and
corresponding hard scattering kernels, where $k_T$ denotes the
transverse momentum of the partons entering the hard scattering
\cite{Catani:1990xk,Collins:1991ty}.  The simplest definition of the
$k_T$ dependent gluon density is through
\begin{equation}
  \label{k-t-gluon}
x g(x,\mu^2) = \int^{\mu^2} \frac{d \tvec{k}^2}{\tvec{k}^2}\, 
               f(x,\tvec{k}^2) ,
\end{equation}
but there are more sophisticated versions taking into account $k_T$
effects to different accuracy, which may include an additional
dependence of the unintegrated density on a renormalization scale
$\mu^2$.  For a recent overview we refer to \cite{Andersson:2002cf}.
$k_T$ dependent GPDs have been discussed in a particular scheme by
Martin and Ryskin \cite{Martin:2001ms}.  Notice that the usual
collinear factorization framework also partly takes into account
finite $k_T$ of partons within hadrons: this is the case for the gluon
with momentum $k_2$ in the two-loop diagram of
Fig.~\ref{fig:k-perp-dglap}, where the transverse momentum resulting
from the splitting of the gluon with momentum $k_1$ explicitly appears
in the hard scattering kernel.  Conversely, using only the one-loop
hard scattering coefficient together with the unintegrated gluon
density takes into account part of the two-loop corrections of the
collinear framework.

\begin{figure}
\begin{center}
	\leavevmode
	\epsfxsize=0.8\textwidth
	\epsfbox{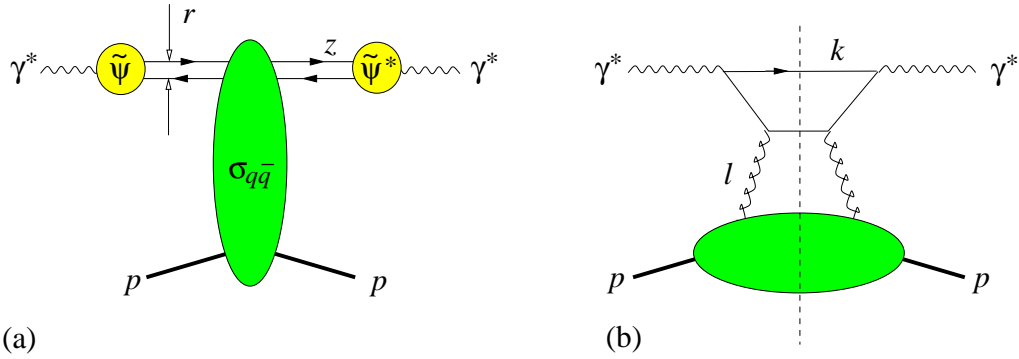}
\end{center}
\caption{\label{fig:color-dipole} Imaginary part of the forward
Compton amplitude in the dipole picture (a) and in collinear
factorization (b).}
\end{figure}

A particular factorization scheme for amplitudes like $\gamma^* p \to
\gamma^* p$ at high energy originates in the BFKL framework
\cite{Lipatov:1976zz,Kuraev:1977fs,Balitsky:1978ic,Forshaw:1997dc} at
leading $\log \frac{1}{x}$ accuracy.  It can be formulated in terms of
color dipoles
\cite{Nikolaev:1991ja,Mueller:1994rr,Mueller:1994jq}, as shown in
Fig.~\ref{fig:color-dipole}a for $\gamma^* p \to \gamma^* p$ with
longitudinal $\gamma^*$ polarization.  The imaginary part of the
forward scattering amplitude then takes the form
\begin{equation}
  \label{dipole-formula}
\im \mathcal{A}(\gamma^*_L\, p\to \gamma^*_L\, p) 
\propto \int d^2\tvec{r} \int dz\, 
  \widetilde{\Psi}^*(\tvec{r},z)\, 
     \sigma_{q\bar{q}}(\tvec{r},\xB)\,
  \widetilde{\Psi}(\tvec{r},z) ,
\end{equation}
where $\widetilde{\Psi}$ denotes the impact parameter wave function
for a $q\bar{q}$ pair in a photon of virtuality $Q^2$, see Section
\ref{sub:overlap-formulae}.  Because of translation invariance
$\widetilde{\Psi}$ only depends on the difference $\tvec{r} =
\tvec{b}_q - \tvec{b}_{\bar{q}}$ of quark and antiquark positions.
$\sigma_{q\bar{q}}$ is the total cross section for scattering a
$q\bar{q}$ dipole with transverse size $\tvec{r}$ on the target
(related to the corresponding forward elastic amplitude by the optical
theorem).  This scattering conserves the quark helicities, which are
to be summed over in (\ref{dipole-formula}) and have not been
displayed.  The factorization formula has a natural interpretation in
the target rest frame, where it corresponds to a sequence of
processes: the $\gamma^*$ splitting into $q\bar{q}$ far ahead of the
target by a distance of order $1 /(\xB m)$, the interaction of this
$q\bar{q}$ pair with the target, and its recombination into a
$\gamma^*$.

This dipole factorization scheme can be derived in the leading $\log
\frac{1}{x}$ approximation, requiring $Q^2$ only to be sufficiently
large to use perturbation theory.  Its result coincides with the one
of collinear factorization if one takes the limit of leading power in
$1/Q$ and of the leading double logarithm $\log Q^2\, \log
\frac{1}{x}$ in each scheme.  Identifying the two results one obtains
the relation
\begin{equation}
  \label{dipole-dglap}
\sigma_{q\bar{q}}(\tvec{r},x) \sim \tvec{r}^2 \, \alpha_s\,
      x g(x)
\end{equation}
at small dipole separation $\tvec{r}$, which is selected in the
process by taking large $Q^2$.  The factorization scale of the gluon
density here is of order $\mu^2 \sim 1 /\tvec{r}^2$, corresponding to
$\mu^2 \sim Q^2$ after convoluting with the $\gamma^*$ wave functions
in (\ref{dipole-formula}), and the same holds for the scale of
$\alpha_s$.  A more precise statement cannot be made at leading $\log
Q^2$ accuracy.  Similarly, since one has taken the leading $\log
\frac{1}{x}$ limit one obtains the gluon distribution evaluated at
$\xB$ and not as an integral over gluon momentum fractions $x$ as in
the full calculation in the collinear factorization diagram shown in
Fig.~\ref{fig:color-dipole}b.

The BFKL dipole formalism can be extended to scattering at finite $t$,
or equivalently at definite impact parameter between the dipole and
the target state, with a dipole-target scattering amplitude depending
in addition on $t$ or on the impact parameter \cite{Navelet:1998tx}.
In the analog of (\ref{dipole-dglap}) the GPD $H^g$ or its impact
parameter equivalent will then appear instead of $x g(x)$.  We also
note that BFKL dynamics is not restricted to the unpolarized sector,
and there are studies of the polarized proton structure function $g_1$
\cite{Bartels:1996wc} and of the double helicity-flip structure
function of a photon target \cite{Ermolaev:1998jv}.

The simple physical picture expressed in the dipole formalism makes it
appealing to use it beyond the limits where its validity has so far
been firmly established.  Strictly speaking, we already did this in
applying (\protect\ref{dipole-formula}) to a proton target, since the
derivation of the dipole representation requires a target whose
structure is amenable to perturbation theory, such as a heavy
quarkonium state \cite{Mueller:1994rr,Mueller:1994jq}.  A further step
is to use dipole factorization also for processes like $\gamma^* p \to
\rho p$ or $\gamma p\to \gamma p$, where one or both wave functions in
(\ref{dipole-formula}) cease to be perturbatively calculable.  As
$\tvec{r}$ becomes comparable to a hadronic scale, the relation
between the dipole cross section and the gluon distribution no longer
holds, and one can currently only model the behavior of
$\sigma_{q\bar{q}}$ in this region.

There are other limits of extending the dipole formalism in its simple
form, namely when going beyond the leading $\log \frac{1}{x}$.  Bialas
et al.~\cite{Bialas:2000xs} have shown that exactly taking into
account the kinematical correlation between the plus-momentum
fractions and the $k_T$ of the gluons in the diagram of
Fig.~\ref{fig:color-dipole}b is incompatible with the conservation of
the dipole size in Fig.~\ref{fig:color-dipole}a.  BFKL calculations at
the next-to-leading $\log \frac{1}{x}$ level further indicate that not
only $q\bar{q}$ dipoles but also higher Fock states in the photon like
$q\bar{q}\, g$ will need to be explicitly taken into account
\cite{Bartels:2001mv}.

Finally, the effect of nonzero skewness $\xi$ in nonforward processes
is beyond the leading $\log \frac{1}{x}$ approximation, which cannot
distinguish between different small momentum fractions of the same
order.  This was already observed in the early study of Bartels and
Loewe \cite{Bartels:1982jh}, who considered BFKL dynamics in processes
like $\gamma^* p \to Z p$ and found that the relevant variable for the
energy logarithm is the larger of $\xi$ and $\rho$ defined in
(\ref{xi-eta-def}).  By the same token, $H^g(x,\xi,0)$ for $x \gsim
\xi$ and the usual gluon density $x g(x)$ taken at momentum fraction
$\xi$ cannot be distinguished to this precision, as emphasized by
Brodsky et al.~\cite{Brodsky:1994kf}.  It remains to be seen whether
and how the effect of different parton momentum fractions can be taken
into account beyond leading $\log \frac{1}{x}$ in the dipole picture
or in an extension of it.

There is on the other hand physics at small $x$ which goes beyond the
description in terms of parton distributions and collinear
factorization.  An example of topical interest is the phenomenon of
parton saturation (see \cite{Mueller:2001fv} for a recent overview).
Under DGLAP evolution to high scales $\mu$, the density of partons at
small $x$ continues to increase, but one can expect that the density
of partons in a given phase space cannot become arbitrarily large and
at some point will ``saturate''.  This phenomenon can be implemented
phenomenologically into the dipole formalism through a corresponding
behavior of the dipole cross section
$\sigma_{q\bar{q}}(\tvec{r},\xB)$, and even a simple model ansatz as
the one of Golec-Biernat and W\"usthoff \cite{Golec-Biernat:1998js} is
remarkably successful in describing data for a large class of
small-$x$ processes in $ep$ scattering.  For presently available
energies and observables the situation remains however ambiguous as to
whether the leading-twist description breaks down and saturation sets
in.

Another example of physics that cannot be described by leading-twist
GPDs (but can within the BFKL approach) is given by processes where
due to quantum numbers the minimal number of gluons exchanged in the
$t$-channel is three instead of two.  In the framework of Regge theory
this corresponds to the odderon, which is the $P=C=-1$ counterpart of
the pomeron.  Compared with quark-antiquark exchange in the
$t$-channel this does not correspond to the minimal number of external
partons in the hard scattering subprocess and is hence power
suppressed in the hard scale at given $x$.  On the other hand,
three-gluon exchange increases faster than $q\bar{q}$ exchange as $x$
decreases at given $Q^2$ and should hence dominate at high energies.
No firm evidence for odderon exchange in hard or soft processes so far
exists.  Suggestions to search for three-gluon exchange in hard
processes have been made for $\gamma^* p \to \eta_c\, p$ in
\cite{Czyzewski:1997bv,Bartels:2001hw}, for electroproduction of
various mesons with positive charge parity in \cite{Engel:1998cg}, and
for $\gamma^* p \to \pi^+\pi^-\, p$ in
\cite{Hagler:2002nh,Hagler:2002nf}.  In the latter case the
interference between pion pairs in the $C$-odd and $C$-even state
(respectively produced by the exchange of two and three gluons) can be
accessed in $C$-odd angular asymmetries (see
Section~\ref{sub:meson-pair-pheno}).

Whether GPDs and collinear factorization, or an appropriate
generalization to $k_T$ dependent GPDs as suggested in
\cite{Martin:2001ms}, provide an appropriate framework to describe
physics in a given range of $Q^2$ and small $x$ remains to be seen (as
it does for the description of DIS by collinear factorization).  On
the other hand, GPDs so far present the only framework where effects
of skewness, which are beyond a description at leading $\log
\frac{1}{x}$ precision, are explicitly taken into account.


\subsection{Longitudinal vs.~transverse photons}
\label{sec:aligned-jet}

In processes involving light quarks longitudinal and transverse
virtual photons play distinct roles in the context of factorization.
We have already seen this in Section~\ref{sec:transverse-mesons} and
shall now investigate this issue in the specific context of small $x$.
To this end, consider the forward Compton amplitude as shown in
Fig.~\ref{fig:color-dipole}b.  Calculating the relevant four cut
diagrams in the small-$x$ limit and expressing the lower blob by the
unintegrated gluon distribution (\ref{k-t-gluon}) one obtains
\begin{equation}
  \label{k-t-start}
\im \mathcal{A} \propto  \int d^2\tvec{l}\; d^2\tvec{k}\; dz\,
   K(\tvec{l},\tvec{k},z) \, \frac{f(\xB,\tvec{l}^2)}{\tvec{l}^4}  .
\end{equation}
The plus-momentum fraction of the gluon in the cut diagrams is $x =
\xB$ up to terms of order $\tvec{k}^2 /Q^2$ or $\tvec{l}^2 /Q^2$.  To
simplify the discussion we have taken the leading $\log\frac{1}{x}$
approximation and set $x=\xB$.  The quark loop kernel can be written
as the overlap of photon wave functions,
\begin{equation}
  \label{quark-loop}
K(\tvec{l},\tvec{k},z) =  \Psi^*(\tvec{k},z)\, \Big[ 
  \Psi(\tvec{k+l},z) + \Psi(\tvec{k-l},z) - 2 \Psi(\tvec{k},z) 
\Big] .
\end{equation}
The crucial difference between transverse and longitudinal photons can
be traced back to their wave functions.  For massless quarks they
read\footnote{The normalization of the photon wave function used in
the context of small-$x$ physics differs in general from the one we
have introduced in Section~\protect\ref{sec:overlap}.}
\begin{eqnarray}
  \label{photon-wf}
\Psi^{0}_{\lambda\lambda'}(\tvec{k},z) 
  &=& 2 e e_q\, \frac{z\bar{z}\, Q}{z\bar{z}\ Q^2 + \tvec{k}^2}\,
                       \delta_{\lambda,\, -\lambda'} ,
\nonumber \\
\Psi^{+}_{+-}(\tvec{k},z) &=& - \sqrt{2} e e_q\,
        \frac{z\, (k^1 + i k^2)}{z\bar{z}\, Q^2 + \tvec{k}^2} ,
\nonumber \\
\Psi^{+}_{-+}(\tvec{k},z) &=& \phantom{-} \sqrt{2} e e_q\,
        \frac{\bar{z}\, (k^1 + i k^2)}{z\bar{z}\, Q^2 + \tvec{k}^2} ,
\nonumber \\
\Psi^{+}_{++}  &=& \Psi^{+}_{--} = 0 , \phantom{\frac{1}{2}}
\end{eqnarray}
where the upper index refers to the photon helicity and the lower
indices to the quark and antiquark, respectively.  The wave functions
$\Psi^-_{\lambda\lambda'}$ are obtained from
$\Psi^+_{-\lambda\,-\lambda'}$ via parity invariance.  The
denominators in (\ref{photon-wf}) correspond to the off-shell quark
propagators in the cut diagrams of Fig.~\ref{fig:color-dipole}b.  The
quark or antiquark is guaranteed to be off-shell by an order $Q^2$,
except for asymmetric configurations where $z$ or $\bar{z}$ is close
to 0.  Such configurations are suppressed by the numerator factors of
the wave functions for longitudinal photons, but not for transverse
ones.

The typical scale for the dependence of the quark loop kernel
(\ref{quark-loop}) on $\tvec{l}^2$ is $\mu^2 \sim z\bar{z}\, Q^2 +
\tvec{k}^2$.  For $z$ away from the endpoints this is much larger than
the hadronic scale governing the $\tvec{l}^2$ dependence of
$f(\xB,\tvec{l}^2)$.  The leading logarithmic behavior in $Q^2$ is
then obtained by Taylor expanding $K(\tvec{l},\tvec{k},z)$ around
$\tvec{l}=0$, corresponding to the region $\tvec{l}^2 \ll \mu^2$, and
taking $\mu^2$ as the upper integration limit.  The lowest
nonvanishing term in this expansion is proportional to $\tvec{l}^2$
after integration over the the azimuth of $\tvec{l}$, and one obtains
\begin{equation}
  \label{collin-1}
\im \mathcal{A} \propto \int d^2\tvec{k}\; dz\; \Psi^*(\tvec{k},z) 
  \left[ \frac{\partial}{\partial k^i} 
     \frac{\partial}{\partial k^i}\, \Psi(\tvec{k},z) \right] \,
  \int^{\mu^2} \frac{d \tvec{l}^2}{\tvec{l}^2}\, f(\xB,\tvec{l}^2) .
\end{equation}
With the help of (\ref{k-t-gluon}) one readily identifies the last
integral as the usual gluon density at a factorization scale $\mu^2$.
For longitudinal photon polarization the integral over $z$ in
(\ref{collin-1}) is dominated by values of $z\sim \half$ due to the
factor $z\bar{z}$ in the photon wave functions, and one has recovered
the collinear factorization at the leading $\log Q^2$ level.

For transverse photons, however, the end-point regions in $z$ are not
suppressed.  Taking the scale $\mu^2$ of the gluon density fixed one
can explicitly perform the integration over $\tvec{k}$ and obtains a
logarithmically divergent integral $\int_0^1 dz\, (z\bar{z})^{-1}$.
One may formally render this finite by taking literally the dependence
$\mu^2 \sim z\bar{z}\, Q^2$ in the gluon density (\ref{k-t-gluon}),
which for $z\bar{z}\to 0$ will then also vanish and make the $z$
integration finite \cite{Martin:1997bp,Ivanov:1998gk}.  In the region
of small $z\bar{z}$ the initial approximation $\tvec{l}^2 \ll
z\bar{z}\, Q^2 + \tvec{k}^2$ leading to (\ref{collin-1}) was however
not justified from the start, and such a procedure can only be
sensible if the endpoint regions do not give a substantial
contribution to the result.  Even without approximating
(\ref{k-t-start}) by (\ref{collin-1}) one has an amplitude with
contributions from the region where $\tvec{l}^2$, $\tvec{k}^2$, and
$z\bar{z}\, Q^2$ are of hadronic size.  In this region the starting
expression (\ref{k-t-start}) does not reflect a separation of soft and
hard physics, since the quark virtualities in the kernel $K$ are not
perturbatively large.  For these ``aligned jet'' configurations it is
then appropriate to factorize the diagrams into the quark or antiquark
distributions and a hard scattering $\gamma^* q \to \gamma^* q$ or
$\gamma^* \bar{q} \to \gamma^* \bar{q}$, where the parton entering the
hard scattering is the one with small $z$ or $\bar{z}$.\footnote{Note
that the calculation discussed here uses Feynman gauge for the gluons.
A gluon attached to the parton with large $z$ or $\bar{z}$ in
Fig.~\protect\ref{fig:color-dipole}b then contributes to the Wilson
line in the definition of the quark distribution.}
The same discussion holds not only for the forward Compton amplitude
$\gamma^*_T\, p \to \gamma^*_T\, p$ but also for DVCS, $\gamma^*_T\, p
\to \gamma p$, where the aligned jet region leads to the quark handbag
diagram in Fig.~\ref{fig:handbag}b.

For light meson production such as $\gamma^* p \to \rho\, p$ the
discussion goes along similar lines, with $\Psi^*(\tvec{k},z)$ in
(\ref{quark-loop}) replaced by the relevant meson light-cone wave
function, and $f(\xB,\tvec{l}^2)$ with the corresponding unintegrated
generalized gluon distribution, which at $t=0$ is equal to the forward
distribution in the leading $\log \frac{1}{x}$ limit.  With
longitudinal photons one obtains the collinear approximation for the
gluons in the leading $\log Q^2$ approximation as above.  A definite
statement about the situation for transverse photon or meson
polarization can only be made when specifying the behavior of the
meson wave function.  Unless one takes this to be strongly suppressed
for $z\bar{z}\to 0$ one finds again the above problem of infrared
sensitivity, see e.g.~\cite{Ivanov:1998gk}.  How important this region
is and whether a perturbative factorization as in (\ref{k-t-start})
can be justified for the production of light mesons from transverse
photons remains controversial in the literature, compare
e.g.~\cite{Martin:1997bp} and \cite{Abramowicz:1997hb}.  Contrary to
Compton scattering one cannot factorize the diagram in terms of a
quark distribution and a hard scattering subprocess, since the
coupling of the meson to the quark lines is itself nonperturbative.

In the case where both the $\gamma^*$ and the meson have longitudinal
polarization, one can proceed and take the collinear approximation on
the meson side by neglecting $\tvec{k}^2$ compared with $z\bar{z}\,
Q^2$ in the photon wave function.  The integration over $\tvec{k}$
then concerns only the meson wave function (when taking a $\tvec{k}$
independent scale $\mu^2$ in the gluon distribution) and gives the
meson distribution amplitude at a factorization scale of order $Q^2$.
Together with the approximation leading to the $k_T$ integrated gluon
distribution, one then has obtained the collinear factorization
formula of Section \ref{sec:factor} at leading double $\log
\frac{1}{x}\, \log Q^2$.

For massive quarks the starting expression (\ref{k-t-start})
remains valid, with the appropriate photon wave functions given e.g.\
in \cite{Bartels:2003yj}.  The quark helicities in the $\gamma^* p$
c.m.\ are not changed by the interaction with the gluons; this is a
feature of the coupling of a vector particle to a fast-moving fermion.
For heavy quarks the end-point region in $z$ is not associated with
low virtualities in the quark loop, since the denominator of the
photon wave functions then reads $(z\bar{z}\, Q^2 + \tvec{k}^2 +
m_q^2)$.  Assuming that $m_c$ is large enough to warrant a
perturbative treatment one can hence use the factorized expressions
(\ref{k-t-start}) or (\ref{collin-1}) for the charm quark loop in the
Compton amplitude with either longitudinal or transverse photons , and
for any polarization in charmonium production from a real or virtual
photon.  For the production of heavy mesons, the appropriate argument
of the gluon density in the leading $\log \frac{1}{x}$ approximation
is
\begin{equation}
  \label{leading-log-x}
x = \frac{Q^2 + M_V^2}{W^2}
\end{equation}
instead of $\xB \approx Q^2 /W^2$.  This corresponds to $\xi \approx
x/2$ as given in $(\ref{xi-eta-prop})$ for DDVCS kinematics, with $Q'$
replaced by $M_V$.

It is easy to translate the preceding discussion into the impact
parameter representation, which is used in the dipole picture.
Starting from (\ref{quark-loop}) one rewrites
\begin{equation}
  \label{impact-overlap}
\int d^2\tvec{k}\, K(\tvec{l},\tvec{k},z) = - 64\pi^4
  \int d^2\tvec{r}\, \widetilde{\Psi}^*(\tvec{r},z) \,
  (1 - e^{i \tvec{l} \tvec{r}}) \,(1 - e^{-i \tvec{l} \tvec{r}}) \,
  \widetilde{\Psi}(\tvec{r},z) .
\end{equation}
Inserting this into the starting expression (\ref{k-t-start}) and
comparing with (\ref{dipole-formula}) one finds that
\begin{equation}
  \label{dipole-uninteg}
\sigma_{q\bar{q}}(\tvec{r},\xB)  \propto \int \frac{d^2
  \tvec{l}}{\tvec{l}^4}\, f(\xB,\tvec{l}^2)\,  
  (1 - e^{i \tvec{l} \tvec{r}}) \,(1 - e^{-i \tvec{l} \tvec{r}}) .
\end{equation}
Under the same requirements as discussed above one now obtains the
leading $\log Q^2$ approximation by Taylor expanding $(1 - e^{i
\tvec{l} \tvec{r}}) \,(1 - e^{-i \tvec{l} \tvec{r}})$ around
$\tvec{l}=0$ for the region $\tvec{l}^2\, \tvec{r}^2 \ll 1$ and
integrating over $\tvec{l}^2$ up to $\mu^2 \sim 1/\tvec{r}^2$.  The
leading term is proportional to $\tvec{l}^2\, \tvec{r}^2$ after
integration over the azimuth of $\tvec{l}$ and gives the double
leading logarithmic relation (\ref{dipole-dglap}) between the dipole
cross section and the gluon density.

For $\gamma^*_L\, p \to \rho_L^{\phantom{*}}\, p$ one obtains
(\ref{impact-overlap}) with $\widetilde{\Psi}^*$ replaced by the
impact parameter wave function of the meson.  The photon wave function
has a logarithmic divergence at $\tvec{r}=0$ and varies with
$\tvec{r}$ over typical distances of order $1/Q$, whereas the
$\tvec{r}$ dependence of the meson wave function is controlled by the
much larger meson radius.  Approximating $\tvec{r}^2$ with a value of
order $1/Q^2$ in the meson wave function one obtains the impact
parameter analog of the collinear approximation on the meson side.


\subsection{The $t$ dependence}
\label{sec:small-x-t}

Our discussion so far has focused on the scattering amplitude at $t=0$
(or more precisely at $t=t_0\approx -4 \xi^2\, m^2$, which is
negligible in the kinematics of interest).  Important physics resides
however in the dependence on $t$ or on the impact parameter, and on
its interplay with the other variables.  Within the dipole formalism
one can identify the impact parameter $\tvec{b}$ as the distance
between the transverse centers of momentum of the dipole and of the
target \cite{Bartels:2003yj}.  The dipole cross section then depends
on two transverse vectors $\tvec{r}$ and $\tvec{b}$ with a nontrivial
interplay.  In the limit where one can connect the dipole cross
section with the nonforward gluon distribution, $\tvec{b}$ becomes the
distance between the gluon on which one scatters and the transverse
center of the target, as we already have seen in
Section~\ref{sec:impact}.

The $t$ dependence is correlated with the longitudinal variable $x$.
In a Regge theory description, a single pole with a linear trajectory
gives for instance a factor $x^{-\alpha' t} = \exp[ \alpha' t
\log\frac{1}{x} ]$.  Translated to impact parameter this states that
the transverse extension of the target increases logarithmically with
energy.  Bartels and Kowalski \cite{Bartels:2000ze} have argued that
in processes with a hard scale this phenomenon is intimately related
with the dynamics of confinement.

On a practical level, most small-$x$ studies of meson production or
DVCS describe the $t$-dependence by an exponential ansatz
\begin{equation}
  \label{b-slope}
\frac{d\sigma(\gamma^* p)}{dt} \propto e^{B t} ,
\end{equation}
where $B$ may depend on $\xB$ and on $Q^2$ in the sense of an
``effective power''.  We remark that the validity of such an
exponential form may be restricted to rather small $t$ and should not
be taken for granted over a wide $t$ range.


\subsection{Vector meson production}
\label{sec:small-x-mesons}

The production of vector mesons $\rho^0$, $\omega$, $\phi$ and
$J/\Psi$, $\Psi(2S)$, $\Upsilon$ at small $\xB$ and small $t$ has been
studied quite extensively in theory.  Measurements in the $\xB$ range
from $10^{-4}$ to $10^{-2}$ have been made by H1 and ZEUS at HERA.  We
do not attempt here a complete review of the subject but concentrate
on issues regarding the description in terms of GPDs and the wider
question of how adequate a leading-twist description is in
experimentally accessible kinematics.  An extensive review of the
experimental situation up to 1997 has been given by
Crittenden~\cite{Crittenden:1997yz}, for a recent overview see
\cite{Kreisel:2002vt}.  Original references are
\cite{Aid:1996ee,Adloff:1997jd,Adloff:1999kg,Adloff:2000nx,Adloff:2002tb}
and
\cite{Derrick:1995yd,Derrick:1996nb,Breitweg:1998nh,Breitweg:1999fm,Breitweg:2000mu} for electroproduction of light mesons, and
\cite{Aid:1996ee,Aid:1996dn,Adloff:1998yv,Adloff:1999zs,Adloff:2000vm,Adloff:2002re}
and
\cite{Breitweg:1998nh,Breitweg:1997rg,Breitweg:1998ki,Chekanov:2002xi}
for photo- and electroproduction of heavy mesons.  Salient features of
the data are \cite{Kreisel:2002vt}:
\begin{itemize}
\item The dependence on $W$ for photoproduction of light mesons is
comparable to the energy dependence of elastic hadron-hadron
scattering, whereas it becomes significantly steeper if either $Q^2$
or the meson mass are large.  This is in line with the change in
energy dependence of $\sigma_{\mathrm{tot}}(\gamma^* p)$ as one goes
from the photoproduction limit to the deeply inelastic region.
\item In $\rho$ production the slope parameter $B$ defined by
(\ref{b-slope}) decreases as $Q^2$ increases from zero to several
GeV$^2$, down to values comparable to those measured in $J/\Psi$
photoproduction \cite{Adloff:2000vm,Chekanov:2002xi}.
\item The production cross sections of $\rho$, $\omega$, $\phi$
plotted as a function of $Q^2+M_V^2$ closely follow a common curve
when rescaled by the factors $\rho : \omega : \phi = 9 : 1 : 2$
explained in Section~\ref{sec:meson-lt}.  To a lesser degree of
accuracy the same holds for $J/\Psi$ production when rescaling
according to $\rho : J/\Psi = 9 : 8$ (see our comment in
Section~\ref{sub:meson-wave}).
\item Preliminary data are consistent with  the ratio of $d\sigma
/dt$ for elastic and proton dissociative $\rho$ electroproduction
being independent of $Q^2$.
\end{itemize}
These features support the qualitative picture of $\gamma^* p
\to V p$ at large $Q^2$ or large $M_V^2$ which follows from
Fig.~\ref{fig:mesons}b, with
\begin{itemize}
\item a quark loop at the top, which has little $t$-dependence and
satisfies flavor SU(3) symmetry, up to SU(3) breaking effects due to
the meson wave functions (see~\cite{Frankfurt:1996jw} for a
discussion) and due to quark masses in kinematics where they are
important,
\item a quantity describing gluon exchange with the proton, whose
behavior at fixed small $t$ is similar to the $x$ behavior of the
usual gluon density at scale $\mu^2 \sim Q^2 + M_V^2$.
\end{itemize}

The $\gamma^* p$ cross section just discussed corresponds to a mixture
of transverse and longitudinal photons, and is to a good approximation
$\sigma(\gamma^* p) \approx \sigma_T + \sigma_L$ in the kinematics of
the experiments.  Detailed studies of the helicity structure in vector
meson production have been carried out using angular distributions,
with the most precise data available for $\rho$ production
\cite{Adloff:1999kg,Adloff:2002tb,Breitweg:1999fm}:
\begin{itemize}
\item  The ratio $R = \sigma_L / \sigma_T$ increases as a function of
$Q^2$, but is significantly smaller than $R = Q^2 /M_\rho^2$ in
electroproduction, taking a value of $R=1$ for $Q^2 \approx
2$~GeV$^2$.  The behavior of $R$ above $Q^2 = 10$~GeV$^2$ is not easy
to assess because statistical errors are still rather large.
Preliminary data suggest that for fixed $Q^2$ there is little
dependence of $R$ on $W$ in the high-energy region $W > 40$~GeV
\cite{Kreisel:2002vt}.
\item  $s$-channel helicity conservation (SCHC) is approximately
observed in the data but broken at the $10\%$ level, with the dominant
helicity changing transition at large $Q^2$ being $\gamma^*_T
\to \rho_L^{\phantom{*}}$.
\end{itemize}

\subsubsection{The gluon distribution}
\label{sub:meson-gluon}

The description of $\rho$ electroproduction by two-gluon exchange and
a quark loop has a long history, starting with work by Donnachie and
Landshoff \cite{Donnachie:1987pu}.  Brodsky et
al.~\cite{Brodsky:1994kf} have analyzed the process in the framework
of factorization: the scattering amplitude for $\gamma^*_L\, p\to
\rho_L^{\phantom{*}}\, p$ at $t=0$ was obtained in terms of the meson
distribution amplitude, and of the forward gluon density evaluated at
$x= \xB$ and $\mu^2 =Q^2$ since the analysis was done to leading
double $\log \frac{1}{x}\, \log Q^2$ accuracy.

The choice of factorization scale in the gluon distribution entails a
considerable uncertainty at small $\xB$, where the scaling violation
in $g(\xB,\mu^2)$ is particularly strong.  This affects both the
overall normalization of the cross section and its dependence on $\xB$
and $Q^2$.  Although the choice of scale remains principally
undetermined at leading $\log Q^2$ accuracy, an ``educated guess'' of
appropriate scale is phenomenologically of great importance in the
absence of a full NLO calculation of the process.  We remark that
commonly the renormalization scale in $\alpha_s$ is taken equal to the
scale of the gluon distribution, or equal to $\tvec{l}^2$ if the
unintegrated gluon density (\ref{k-t-gluon}) is used.  Frankfurt et
al.\
\cite{Frankfurt:1996jw,Frankfurt:1998fj,Frankfurt:1998yf,Frankfurt:2000ez}
estimate the scale in $g(\xB,\mu^2)$ by a procedure based on the
dipole representation (\ref{dipole-formula}) and (\ref{dipole-dglap})
for $\gamma^*_L\, p\to \gamma^*_L\, p$ and for $\gamma^*_L\, p\to
\rho_L^{\phantom{*}}\, p$.  The estimated scale for $\rho$ production
is found to be lower than $Q^2$, corresponding to larger average
dipole sizes $\tvec{r}^2$ in $\rho$ production than in the forward
longitudinal Compton amplitude.  The same procedure was applied by the
authors to heavy quarkonium production.  Martin et
al.~\cite{Ryskin:1997hz,Martin:1997bp,Martin:1999rn,Martin:1999wb}
advocate the scale $\mu^2 = \frac{1}{4} (Q^2 + M_V^2)$ as argument of
$g(x,\mu^2)$.  In numerical studies they use the unintegrated density.
For $\tvec{l}^2$ above a value $\tvec{l}_0^2$ where $g(x,\tvec{l}^2)$
is available from global parton analyses $f(x,\tvec{l}^2)$ is
calculated from (\ref{k-t-gluon}), supplemented by a Sudakov form
factor in \cite{Martin:1999wb}, whereas for $\tvec{l}^2$ below
$\tvec{l}_0^2$ different procedures are used, see~\cite{Levin:1997vf}.
Compared with using the collinear gluon density, an increase of the
cross section was found when including the gluon $k_T$.

Efforts have also been undertaken to improve on the leading $\log
\frac{1}{x}$ approximation and take into account the finite skewness
in meson production.  Kinematics tells us that $x$ in
(\ref{leading-log-x}) is the difference $x_1 - x_2$ of plus-momentum
fractions for the emitted and the reabsorbed gluon, where both
fractions refer to the incoming proton.  In the DGLAP region, which
gives the imaginary part of the amplitude, one has $x_1 \ge x$.  In
studies of $\Upsilon$ and $J/\Psi$ production, Frankfurt et
al.~\cite{Frankfurt:1998yf,Frankfurt:2000ez} have estimated typical
values for $x_1$, identified the generalized and forward gluon at a
starting scale and then evolved up to the hard scale of the process.
In \cite{Frankfurt:2000ez} the relation (\ref{dipole-dglap}) between
the dipole cross section and the gluon density was only used at small
$\tvec{r}$, whereas an ansatz was made for the behavior of
$\sigma_{q\bar{q}}$ at larger $\tvec{r}$, thereby going beyond a
leading-twist description.  Martin et
al.~\cite{Martin:1999rn,Martin:1999wb} estimated skewness effects with
the relation (\ref{shuv-ratio}) obtained in the Shuvaev model for the
ratio $R^g$ of $H^g(\xi,\xi)$ and $2\xi \, g(2\xi)$, with $2\xi
\approx x$ given by (\ref{leading-log-x}).  The authors report
significant enhancement factors from skewness for the cross sections
of $\rho$ and $J/\Psi$ electroproduction at large $Q^2$.  These
factors also influence the $Q^2$ dependence, since the effective power
$\lambda$ in the gluon density depends on $Q^2$.  Put in more general
terms, the scale dependence of the nonforward and the forward gluon
distribution is not the same.

For photoproduction of $\Upsilon$, both groups
\cite{Frankfurt:1998yf,Martin:1999rn} find significant
enhancement from their estimates of skewness, since at the
corresponding large scales $\mu^2$ the gluon density is very steep.
They emphasize that the enhancement factor for the cross section is
essential for describing data \cite{Breitweg:1998ki,Adloff:2000vm},
being about 2 with $\mu^2 = \frac{1}{4} M_\Upsilon^2$ in
\cite{Martin:1999rn} and about 2.3 with $\mu^2$ between $40$ and
$75$~GeV$^2$ in \cite{Frankfurt:1998yf}.  Ma and Xu \cite{Ma:2001yf}
obtain a description of the data using the forward model
(\ref{super-simple}) for the gluon GPD, with an enhancement factor
$R^g = 2^\lambda$ and a scale $\mu^2 = m_b^2$.  In all three studies
the contribution from the real part of the scattering amplitude,
obtained from the imaginary part using (\ref{large-s-dispersion}), is
significant given the steepness of the gluon density.

We remark that vector meson production has been studied in the dipole
formalism by several groups, using the analog of
(\ref{dipole-formula}) together with a model for the dipole cross
section, see e.g.\
\cite{Dosch:1997ss,Hufner:2000jb,Caldwell:2001ky,Dosch:2002ig}.
Munier et al.~\cite{Munier:2001nr} have taken the ``reverse'' approach
and attempted to extract the dipole cross section at given impact
parameter using the measured cross section for $\gamma^*_L\, p\to
\rho_L^{\phantom{*}}\, p$ and different models for the $\rho$ wave
function.  No definite conclusion concerning the onset of parton
saturation (which implies the breakdown of the leading-twist
description) in current data could be drawn in this study.  The same
holds for $J/\Psi$ production: Gotsman et al.~\cite{Gotsman:2001ic}
find better agreement with the data when including saturation effects,
whereas Frankfurt et al.~\cite{Frankfurt:2000ez} do not require
saturation while making a particular choice of factorization scale.
Dosch and Ferreira \cite{Dosch:2002ig} have modeled saturation effects
in $J/\Psi$ photoproduction in the dipole approach and found them too
small to be identified in present data, given other theoretical
uncertainties and experimental errors.

\subsubsection{The meson wave function}
\label{sub:meson-wave}

A substantial source of theoretical uncertainty in the normalization
and the $Q^2$ dependence of light meson electroproduction is due to
the $\tvec{k}$ dependence of the meson light-cone wave functions.
Frankfurt et al.~\cite{Frankfurt:1996jw,Frankfurt:1998fj} have
compared the convolution (\ref{collin-1}) of wave functions for
$\gamma^*_L\, p\to \rho_L^{\phantom{*}}\, p$ with its collinear
approximation, where $\tvec{k}$ in the photon wave function is set to
zero, and found large suppression factors in the cross section due to
the meson $\tvec{k}$.  Values at $Q^2=10$~GeV$^2$ are between about
$0.64$ and $0.06$, with variations due to the parameters taken in the
wave functions but above all to its assumed large-$\tvec{k}$ behavior,
which ranged from a Gaussian falloff at large $\tvec{k}^2$ to a
power-law decrease like $1/\tvec{k}^2$.  The $1/\tvec{k}^2$ behavior
is generated by hard gluon exchange between the quark and antiquark,
which gives rise to the ERBL evolution of the distribution amplitude.
Rather than an estimate of higher-twist effects, its explicit
inclusion in the hard scattering thus is an estimate of higher-loop
corrections, similar to the use of unintegrated parton distributions
we discussed after (\ref{k-t-gluon}).  Comparison with experiment
\cite{Breitweg:1998nh} indicates that the very strong suppression from
a $1/\tvec{k}^2$ falloff, together with the choice made in
\cite{Frankfurt:1996jw} for the scale of the gluon distribution
underestimates the data.  The suppression factor of $0.64$ quoted
above corresponds to a wave function proposed by Halperin and
Zhitnitsky \cite{Halperin:1997zk}, who report large cancellations
between the $1/Q^2$ and the $1/Q^4$ corrections to the collinear
approximation when expanding (\ref{collin-1}) in moments $\int
d^2\tvec{k} \, \tvec{k}^{2n} \Psi_\rho(\tvec{k}^2,z) / Q^{2n}$.  In
such a situation, successive improvement of the leading-twist
expression by higher-power terms in $1/Q^2$ would be an unfortunate
strategy.

Martin et al.~\cite{Martin:1997bp,Martin:1999wb} have proposed to
describe the transition from $q\bar{q}$ to the meson not by a wave
function but by using parton-hadron duality, based on the picture that
a $q\bar{q}$ pair with invariant mass $M$ in the vicinity of $m_\rho$
has few other channels to hadronize into (an isospin factor of $9/10$
allowing for $\omega$ production is taken into account).  They project
the amplitude for open $q\bar{q}$ production onto the relevant partial
wave and integrate over a certain range of $M^2 = \tvec{k}^2/
(z\bar{z})$ at the level of the cross section.  This approach is taken
for both $\gamma^*_L\, p\to \rho_L^{\phantom{*}}\, p$ and
$\gamma^*_T\, p\to \rho_T^{\phantom{*}}\, p$.  In the latter case
there are non-negligible contributions from soft regions of phase
space, but the integrals are finite and the authors report little
variation of the results when changing the parameter $\tvec{l}_0^2$
mentioned above, which separates the hard and the soft regime in the
unintegrated gluon density.  The overall normalization is difficult to
predict in this framework (due to the choice of mass window in $M^2$),
and a substantial $K$-factor estimating higher-order corrections in
$\alpha_S$ was used in \cite{Martin:1997bp}.  In particular the
measured $Q^2$ dependence of the ratio $R = \sigma_L / \sigma_T$ is
however well described.  That its growth is slower than a linear
behavior like $Q^2 / M_\rho^2$ can be traced back to the fact that
$\sigma_T$ is sensitive to the gluon density at lower scale $\mu^2$
than $\sigma_L$ (as emphasized earlier by Nemchik et
al.~\cite{Nemchik:1994fp}), and that the increase of $g(x,\mu^2)$ with
$\mu^2$ is faster at low values of $\mu^2$.  This result is not
special to using parton-hadron duality.  In fact, Ivanov and Kirschner
\cite{Ivanov:1998gk} have obtained the same expression for $R$ as in
\cite{Martin:1997bp}, using a particular ansatz for the structure of
the meson wave function.  The helicity dependence of the $\rho$ wave
function in this ansatz is taken as for a photon, given by $\bar{u}
\slash{\epsilon}_\rho v$, where $u$ and $v$ are massless on-shell
spinors for the quark and antiquark and $\epsilon_\rho$ is the meson
polarization vector.  This is proportional to the corresponding
rotation functions $e^{i\mu \varphi}\, d^1_{1 \mu}(\theta)$, where
$\mu$ denotes the meson helicity and $\varphi$, $\theta$ the azimuthal
and polar angles of the quark in the meson rest frame.  The meson
production cross section in \cite{Ivanov:1998gk} has the structure
\begin{equation}
	\label{ivanov-ansatz}
\frac{d\sigma}{dt} \propto \Bigg| \int d M^2\, 
  \int d(\cos{\theta})\, d\varphi\;
  \mathcal{A}_{q\bar{q}}(M,\theta,\varphi)\,
  w(M)\, e^{-i\mu \varphi}\, d^1_{1 \mu}(\theta) \, 
  \Bigg|^2 ,
\end{equation}
where $\mathcal{A}_{q\bar{q}}$ describes the production of a
$q\bar{q}$-pair with squared invariant mass $M^2 = \tvec{k}^2
/(z\bar{z})$, and $w(M)$ comes from the meson wave function.  This is
similar to the structure of the result using parton-hadron duality,
\begin{equation}
\frac{d\sigma}{dt} \propto \int_{M_1^2}^{M_2^2} d M^2\,
\Bigg| \int d(\cos{\theta})\, d\varphi\;
  \mathcal{A}_{q\bar{q}}(M,\theta,\varphi)\,
  e^{-i\mu \varphi}\, d^1_{1 \mu}(\theta)\,
\Bigg|^2 ,
\end{equation}
in particular in the dependence on $\cos\theta = 2z-1$, which is
sensitive to the polarizations of meson and photon and which
determines the typical scale of hardness in the quark propagators
$z\bar{z}\, Q^2 + \tvec{k}^2 = z\bar{z}\, (Q^2 + M^2)$.  Ivanov and
Kirschner have used the ansatz (\ref{ivanov-ansatz}) to calculate not
only the helicity conserving transitions with longitudinal and
transverse photons, but also the helicity violating ones.  Their
leading power behavior was found to be
\begin{eqnarray}
\mathcal{A}_{0,0} &\sim& 1/Q , \qquad\:\, \mathcal{A}_{1,1} \sim 1/Q^2, 
\nonumber \\
\mathcal{A}_{0,1} &\sim& 1/Q^2 , \qquad \mathcal{A}_{1,0} \sim 1/Q^3 , 
\qquad 
\mathcal{A}_{1,-1} \sim 1/Q^2 ,
\end{eqnarray}
where the first subscript refers to the meson and the second to the
photon helicity.  A sensitivity like $(z\bar{z})^{-1}$ to aligned-jet
configurations was found in $\mathcal{A}_{1,1}$, $\mathcal{A}_{0,1}$,
$\mathcal{A}_{1,0}$, as well as in a $1/Q^4$ contribution to
$\mathcal{A}_{1,-1}$.  A rather good description of the HERA data on
the helicity structure of $\rho$
\cite{Adloff:1999kg,Adloff:2002tb,Breitweg:1999fm} and $\phi$
\cite{Adloff:2000nx} production can be obtained in the model.  Similar
results were obtained by Kuraev et al.~\cite{Kuraev:1998ht}, who used
constituent wave functions for the $\rho$.

For heavy-quark systems like the $J/\Psi$ or the $\Upsilon$ the
simplest approximation to the light-cone wave function is the static
ansatz $\Psi(\tvec{k},z) \propto \delta^{(2)}(\tvec{k})\,
\delta(z-\half)$ together with a quark mass $m_q = \half M_V$.  Ryskin
et al.~\cite{Ryskin:1997hz} have investigated corrections to this,
including a Gaussian $\tvec{k}$ dependence of the wave function in the
hard scattering while keeping $z=\half$.  They find that the ensuing
suppression tends to be canceled by taking a current quark mass $m_c$
instead of $\half M_{J/\Psi}$ in the hard scattering (especially for
photoproduction the cross section depends on a high power of
$m_c^{-1}$).  Frankfurt et al.~\cite{Frankfurt:1998fj} have reached
different conclusions, in particular including in $\Psi(\tvec{k},z)$ a
nontrivial $z$ dependence of and taking a perturbative falloff like $1
/\tvec{k}^2$ at large $\tvec{k}$.  They obtain suppression factors for
the photoproduction cross section as small as 0.25 relative to taking
the static wave function and $m_q = \half M_{J/\Psi}$.  This
suppression becomes weaker for large $Q^2$ and eventually turns into
an enhancement, in accordance with our arguments in
Section~\ref{sec:heavy-mesons}.

A commonly used way to obtain the meson light-cone wave function is by
a kinematical transformation of the Schr\"odinger wave function
calculated e.g.\ in a potential model.  This corresponds to the
calculation of GPDs in constituent quark models discussed in
Section~\ref{sub:quark-models}.  H\"ufner et al.~\cite{Hufner:2000jb}
have pointed out the importance in charmonium production of including
the Melosh transform between the different spin bases used in
rest-frame Schr\"odinger wave functions and in light-cone wave
functions.  They further caution that the purely kinematic transform
between the two wave functions has the status of a model since in
quantum field theory the connection between matrix elements at equal
time $x^0$ and equal light-cone time $x^+$ includes dynamical effects.

Frankfurt et al.~\cite{Frankfurt:1998fj} have pointed out that a
non-relativistic description of charmonium is problematic since it
typically yields wave functions with a significant part of
high-momentum modes not satisfying $v \ll c$.  For bottomonium the
situation is less critical, and the suppression compared with the
static approximation found in~\cite{Frankfurt:1998fj} for $\Upsilon$
is rather mild.

An expectation often encountered in the literature is that at small
$x$ the electroproduction cross sections for $\rho$, $\omega$, $\phi$,
$J/\Psi$ should behave as $9 : 1 : 2 : 8$ in the limit where $Q^2 \gg
M_{J/\Psi}$.  This holds if these ratios are followed by the squared
integrals $[\, \sum_q e_q\, \int dz\, z^{-1} \Phi^q(z) \,]^2$ of the
respective distribution amplitudes.  The squared integrals $[\, \sum_q
e_q\, \int dz\, \Phi^q(z) \,]^2$ are proportional to $M_{V
\rule{0pt}{1.3ex}} \Gamma_{V\to e^+e^-}$, 
whose measured values for the \emph{light} mesons (see
Section~\ref{sec:meson-lt}) fit rather well with the above ratios,
which follow from flavor SU(3) and the neglect of mixing between
$q\bar{q}$ and gluon configurations.  As pointed out in
\cite{Frankfurt:1996jw} the extension to flavor SU(4) fails however
quite badly, with experimental values $(M_{\rho \rule{0pt}{1.3ex}}
\Gamma_{\rho\to e^+e^-}) : (M_{J/\Psi \rule{0pt}{1.3ex}}
\Gamma_{J/\Psi \to e^+e^-}) \approx 9 : (3.5 
\times 8)$.  This is not surprising for quantities determined by
physics at scales not larger than the respective meson masses.
Furthermore, the respective $z$ dependence of the distribution
amplitudes may well differ between light and heavy mesons, unless
$\log Q^2$ is so large that even $\Phi_{J/\Psi}(z)$ has evolved to the
asymptotic shape.  If the quantities $[\, \sum_q e_q\, \int dz\,
z^{-1} \Phi^q(z) \,]^2$ relevant for electroproduction were to follow
the naive SU(4) expectation $9 : 8$ for $\rho$ versus $J/\Psi$ at some
factorization scale, this would be a rather accidental conspiracy of
two SU(4) breaking effects.

\subsubsection{Dynamics and observables}
\label{sub:meson-observ}        

The description of $\gamma^*_L\, p \to V_L\, p$ at small $x$ to
leading-twist and leading $O(\alpha_s)$ accuracy (implying the leading
$\log Q^2$ approximation) allows very direct access to the gluon
structure of the proton.  In particular, the meson structure only
appears through a global factor $\sum_q e_q\, \int dz\, z^{-1}
\Phi^q(z)$ in the amplitude.  Let us try and assess how particular
observables are affected by the various corrections we have discussed.
\begin{itemize}
\item \textit{The overall normalization} of the cross section may be
severely affected by finite meson $\tvec{k}$ in the hard scattering
subprocess according to estimates.  It is hence most sensitive to the
details of the meson wave function.
\item \textit{The $Q^2$ dependence} is influenced by several sources.
It may be strongly affected by the finite $\tvec{k}$ effects just
mentioned.  Its theoretical description is subject to the uncertainty
in relating the factorization scale of the gluon distribution with
$Q^2$.

We note that in many experimental studies the $Q^2$ dependence of the
$\gamma^* p$ cross section is given and parameterized at fixed $W^2$,
whereas the predictions of the factorization theorems are for fixed
$\xB$.
\item \textit{The dependence on $\xB$} is again strongly influenced by
the choice of scale in the gluon distribution.  Physics beyond leading
twist such as saturation will affect the $\xB$ behavior, but the data
does at present not allow one to decide whether such effects are at
work in HERA kinematics.

One may expect less impact on the $\xB$ dependence from the details of
the meson structure.  Such a cross-talk does exist at some level: in
the approach of Frankfurt et al.\ the choice of scale in the gluon
distribution is for instance connected with the relevant dipole sizes
in the convolution $\int dz \, \widetilde{\Psi}^*(\tvec{r},z) \,
\widetilde{\Psi}(\tvec{r},z)$ of photon and meson wave functions.  How
important such cross-talk may be has however not been studied
quantitatively.
\item \textit{The $t$ dependence} may also be expected to be
rather independent of the detailed meson structure.  Again there is a
caveat of cross-talk via the resolution scale of the process: in the
leading-twist approximation the behavior of $H^g(x,\xi,t)$ with $t$
depends on the scale $\mu^2$ where the gluons are probed.  Indeed
evolution in $\mu^2$ changes the $t$ dependence unless it is
completely uncorrelated with the dependence on $x$.
\end{itemize}
In light of this discussion a better control of the relevant
resolution scale in meson production would be of great importance.  To
put the theoretical discussion on a firmer footing, a full
next-to-leading order calculation is called for, which would also
include effects of the meson structure at small distance (i.e.\ at
large~$\tvec{k}$).


\subsection{DVCS}
\label{sec:small-x-dvcs}

Let us now turn to DVCS in the small-$x$ regime of the H1 and ZEUS
measurements \cite{Adloff:2001cn,Zeus:2003ip}.  In
Section~\ref{sec:compton-scatt} we have already discussed the NLO
analyses in the GPD framework by McDermott and Freund
\cite{Freund:2001hm,Freund:2001rk,Freund:2001hd} and by Belitsky et
al.~\cite{Belitsky:2001ns}.  Contrary to vector meson production, the
forward and nonforward Compton amplitudes for transverse photons
receive important contributions from the quark distributions (at least
for ``standard'' choices of the factorization scale).  In the dipole
representation this is related to the importance of aligned-jet
configurations for transverse photons.  Apart from the leading-twist
analyses, a number of other approaches have been applied to DVCS at
small $x$, which we shall briefly present.

An early study by Frankfurt et al.~\cite{Frankfurt:1998at} starts by
modeling the ratio 
\begin{equation}
  \label{dvcs-ratio}
R(Q^2,\xB) = 
  \frac{\im [ \mathcal{A}(\gamma^*_T p \to \gamma^*_T p) ] \;\:\, }{
        \im [ \mathcal{A}(\gamma^*_T p \to \gamma p) ]_{t=0}}
\end{equation}
at $Q_0^2 = 2.6$~GeV$^2$, using an ansatz based on the aligned-jet
model.  The change of the amplitudes with $Q^2$ is then investigated
using nonforward evolution, which at small $x$ is dominated by the
gluon distribution.  $H^g$ at the starting scale was fixed by the
ansatz (\ref{old-diagonal-input}).  The authors find little variation
of the ratio $R$ for $Q^2$ between $Q_0^2$ and 45~GeV$^2$, with $R$
between $0.5$ and $0.58$ for $\xB$ of order $10^{-2}$ to $10^{-4}$.
The real part of the DVCS amplitude is recovered by a derivative
analyticity relation and the $t$-dependence as described in
Section~\ref{sec:small-x-t}.  The numerator of (\ref{dvcs-ratio}) is
proportional to the transverse inclusive structure function $F_T = F_2
- F_L$.  For estimating the DVCS cross section in
\cite{Frankfurt:1998at}, $F_L$ was assumed to be negligible in the
relevant kinematics, and $F_T$ was approximated with the measured
$F_2$ \cite{Freund:2003pr}.

DVCS has been analyzed in the dipole formalism by Donnachie and
Dosch~\cite{Donnachie:2000px}, by McDermott et
al.~\cite{McDermott:2001pt}, and by Favart and Machado
\cite{Favart:2003cu} with different models for the dipole cross
section.  We remark that, even though $\sigma_{q\bar{q}}(\tvec{r},
\xB)$ does not take skewness into account, the amplitudes for
$\gamma^*_T p\to \gamma^*_T p$ and for DVCS differ in the dipole
formulation because the analog of the master formula
(\ref{dipole-formula}) involves the overlap of different wave
functions for DIS and for DVCS.  The studies
\cite{McDermott:2001pt,Favart:2003cu} compared the $\tvec{r}$
dependence of $\sigma_{q\bar{q}}(\tvec{r},\xB) \int dz\,
\widetilde{\Psi}^*(\tvec{r},z)\, \widetilde{\Psi}(\tvec{r},z)$ in both
cases.  As expected, the typical dipole size is larger in DVCS than in
$F_T$ at equal $Q^2$ (except in the limit $Q^2\to 0$ where both agree
by definition); by exactly how much is somewhat model dependent.  The
results of all the studies mentioned here
\cite{Frankfurt:1998at,Donnachie:2000px,McDermott:2001pt,Favart:2003cu}
are in agreement with the H1 data within errors.

A special feature of DVCS is that through interference with the
Bethe-Heitler process in $ep\to ep \gamma$ it allows measurement of
both imaginary and real part of the scattering amplitude (see
Section~\ref{sec:dvcs-pheno}).  The derivative analyticity relation
(\ref{large-s-dispersion}) suggests that $\re \mathcal{A}$ does not
carry any new information compared to $\im \mathcal{A}$ in the
high-energy region.  This does not seem to hold in practice.
Frankfurt et al.~\cite{Frankfurt:1998et} report that different
parameterizations of the small-$x$ data for $F_2(\xB,Q^2)$ lead to
quite distinct predictions for $\re \mathcal{A}_{\mathrm{DVCS}}$, in
particular when comparing power-law and logarithmic parameterizations
of the $\xB$-dependence.  A similar finding has been made in the study
by McDermott et al.~\cite{McDermott:2001pt}, where the predictions of
two different models for the dipole cross section are much more
similar for $\im \mathcal{A}_{\mathrm{DVCS}}$ than for $\re
\mathcal{A}_{\mathrm{DVCS}}$.  This indicates that in a finite range
of $\xB$ different curves may describe the cross section data within
errors, but have rather different slopes in $\log \xB$.  Note that
assessing the importance of higher derivatives with respect to $\log
\xB$ in  (\ref{large-s-dispersion}) requires even higher quality of
data for $\im \mathcal{A}$, so that direct measurement of $\re
\mathcal{A}$ may indeed contribute important extra information.

We finally note that DVCS gives direct access to the Compton amplitude
with a transverse $\gamma^*$ (the transition $\gamma^*_L\, p
\to \gamma p$ being suppressed at large $Q^2$ and small $t$), in
contrast to inclusive DIS, where the separation of $F_T$ and $F_L$
requires measurement at different beam energies.  The inclusive
structure function $F_2$ and the DVCS amplitude may thus show
important differences beyond the effects of skewness.  We remark in
particular that an estimate by Bartels et al.~\cite{Bartels:2000hv}
found in certain regions of small $\xB$ and moderate $Q^2$ important
twist-four effects in $F_T$ and in $F_L$, which largely canceled in
their sum $F_2$.


\subsection{Other diffractive channels}
\label{sec:open-diff}

A different type of small-$x$ processes which can be studied at HERA
and where gluon GPDs are relevant are diffractive reactions
$\gamma^{(*)} p \to X p$ with suitable non-exclusive final states $X$.
For a general review we refer to \cite{Wusthoff:1999cr}.
Golec-Biernat et al.~\cite{Golec-Biernat:1998vf} have investigated the
effect of skewness in the ``exclusive'' production of dijets,
described by $\gamma^{(*)} p\to q \bar{q} + p$ at parton level, and
pointed out interesting effects due to the kinematical correlation
between the longitudinal momenta of the gluons and the large
transverse momentum $\tvec{k}$ of the jets.  This reaction is not easy
to access: states corresponding to $q\bar{q} g$ or more complicated
configurations at parton level dominate in a large part of phase
space, and a dedicated experimental study is not yet available for
events in a region where a sole $q\bar{q}$ pair should adequately
describe the diffractive final state.

Another case of interest is the diffractive production of charm, where
the channel $\gamma^{(*)} p\to c \bar{c} + p$ allows description in
terms of gluons GPDs.  Here the charm quark mass protects against the
infrared sensitive region of aligned-jet configurations, as in the
case of $J/\Psi$ production.  Again, this channel has to be separated
from states with additional gluons at parton level.  Dominance of
$c\bar{c}$ over $c\bar{c} g$ and higher states can be expected at
large values of $\beta = Q^2 /(Q^2 + M^2)$, i.e., at small invariant
mass $M$ of the diffractive system \cite{Levin:1997vf}.

For inclusive diffraction $\gamma^{*} p\to X + p$ the situation is
similar.  In order to have $X$ adequately represented by a $q\bar{q}$
pair one needs to go to large $\beta$ \cite{Wusthoff:1999cr}.  For
light quarks, however, the aligned-jet region is part of the phase
space.  With the collinear approximation for the gluons in the hard
scattering one finds integrals behaving as $\sigma_L \sim \int dz\,
(z\bar{z})^{-1}$ and $\sigma_T \sim \int dz\, (z\bar{z})^{-2}$,
depending on the photon polarization, so that one has to go back to
the unintegrated gluon distribution or introduce some infrared
regularization.  A detailed study of this process in the context of
GPDs has been performed by Hebecker and Teubner
\cite{Hebecker:2000xs}.  The authors estimate in particular that a
region where a GPD based description is viable and where the cross
section is dominated by $\sigma_L$ (which is less infrared sensitive
than $\sigma_T$) is given by $\Lambda^2 /Q^2 \ll 1-\beta \ll
(\Lambda^2 /Q^2)^{1/2} $, where $\Lambda$ is a hadronic scale.


\section{Phenomenology}
\label{sec:pheno}

In this section we present the basic phenomenology of the reactions
where GPDs and GDAs can be accessed.  The phenomenological analysis
can be separated into two distinct steps, as was discussed in detail
for DVCS in \cite{Diehl:1997bu}.  The first step is to go from
differential cross sections of processes like $ep\to ep \gamma$ or
$ep\to ep \rho$ to amplitudes for $\gamma^* p\to \gamma p$ or
$\gamma^* p\to \rho\, p$.  These can then be analyzed making use of the
factorization theorems.  This first means to test whether the general
scaling and helicity structure is as predicted for the regime of
leading-twist dominance (see Section~\ref{sec:factor}).  If this is
the case, the task remains to extract GPDs from the appropriate
process amplitudes, where they appear in convolutions with the hard
scattering kernels.  How to tackle this ``deconvolution'' problem
remains an outstanding task for theory, and its status will be
discussed in Section~\ref{sec:deconvolute}.

The principles of analyzing meson electroproduction have long been
established, see for instance the detailed work on vector meson
production by Schilling and Wolf \cite{Schilling:1973ag}.  The
phenomenology of two-photon reactions like DVCS is special because of
competing Bethe-Heitler processes.  On one hand this makes their
analysis more involved, but on the other hand it provides much more
detailed access to the amplitudes at two-photon level which one aims
to study.  We will report on the detailed work that has been done for
DVCS.  The situation for the cross-channel counterparts $\gamma p \to
\gamma^* p$ and the annihilation of $\gamma^* \gamma$ into hadron pairs
is very similar, and we shall point out important special aspects of
these reactions.  For processes involving two off-shell photons the
corresponding theory is in the process of development.  First brief
results have been given for DDVCS in
\cite{Guidal:2002kt,Belitsky:2002tf}, whose phenomenology combines the
cases of DVCS and of TCS.


\subsection{DVCS}
\label{sec:dvcs-pheno}

\begin{figure}
\begin{center}
	\leavevmode
	\epsfxsize=0.83\textwidth
	\epsfbox{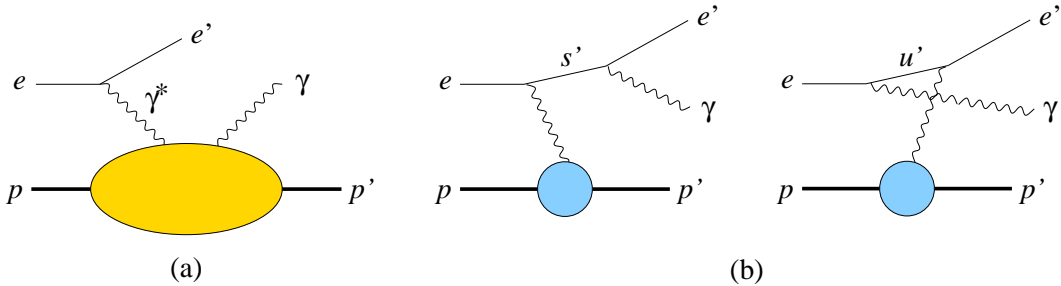}
\end{center}
\caption{\label{fig:vcs-bh} (a) Compton scattering and (b)
Bethe-Heitler contributions to $ep\to ep\gamma$.}
\end{figure}

DVCS is measured in electro- or muoproduction of real photons, $ep\to
ep\gamma$ or $\mu p\to \mu p\gamma$.  In the kinematics of relevance
for the factorization theorems the lepton mass is negligible and will
be set to zero in the following.  The two cases can then be considered
on equal footing.  For definiteness we will refer to electroproduction
in the following and also specialize to targets with spin~$\half$.
Aspects of targets with spin zero or one will be discussed in
Section~\ref{sub:spin-0-1}.  Apart from Compton scattering, the
Bethe-Heitler process contributes to the same final state (see
Fig.~\ref{fig:vcs-bh}), and both mechanisms have to be added in the
amplitude.  In particular situations, the interference term between
both mechanisms can be used to study $\gamma^* p \to \gamma p$ at
amplitude level, including its phase.  Belitsky and M\"uller
\cite{Belitsky:2002ep} have compared this situation with holography,
in the sense that the phase of the Compton amplitude is measured
against the known ``reference phase'' of the Bethe-Heitler process.
With lepton beams of both charges one can filter out the interference
term, as remarked long ago by Brodsky et al.~\cite{Brodsky:1972vv}.
In a similar way one can use the beam polarization asymmetry, as was
observed by Kroll et al.~\cite{Kroll:1996pv} in the context of virtual
Compton scattering at large $t$.  The general structure of $ep\to
ep\gamma$ at twist-two level was first discussed by
Ji~\cite{Ji:1997nm}.  A strategy to analyze DVCS in the context of the
$1/Q$ expansion was proposed in \cite{Diehl:1997bu} and forms the
basis of our presentation here.  Belitsky et al.\ have investigated in
detail how the cross section of $ep\to ep\gamma$ is related to
integrals involving GPDs at twist-two \cite{Belitsky:2000gz} and
twist-three \cite{Belitsky:2001yp,Belitsky:2001ns} accuracy.  Detailed
numerical predictions for DVCS observables using models for twist-two
GPDs were made by Vanderhaeghen et
al.~\cite{Vanderhaeghen:1998uc,Guichon:1998xv,Vanderhaeghen:1999xj},
by Belitsky et al.~\cite{Belitsky:2000gz}, and by Freund and McDermott
\cite{Freund:2001hm,Freund:2001hd}.  Studies for particular
fixed-target experiments can be found in \cite{Bertin:1998hj} for
JLAB, in \cite{Korotkov:2001zn,Korotkov:2002ym} for HERMES, and in
\cite{d'Hose:2002ia} for COMPASS.  Predictions at twist-three level
were made in \cite{Kivel:2000fg,Goeke:2001tz}, \cite{Belitsky:2001ns},
and \cite{Freund:2003qs}.  DVCS with the $p\to \Delta$ transition, $e
p\to e \Delta^+ \gamma$, is briefly discussed in
\cite{Frankfurt:1998jq,Goeke:2001tz}, and the more general case of the
$p\to p\pi^0$ and $p\to n\pi^+$ transitions has recently been
investigated in \cite{Guichon:2003ah}.

\begin{figure}[ht]
\begin{center}
	\leavevmode
	\epsfxsize=0.57\textwidth
	\epsfbox{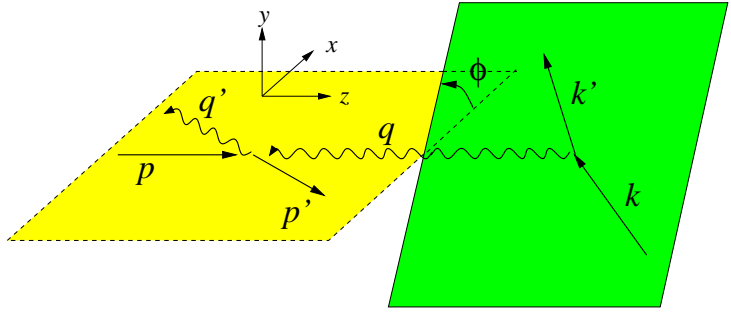}
\end{center}
\caption{\label{fig:dvcs-kin} Kinematics of $ep\to ep\gamma$ in the
c.m.\ of the final-state proton and photon.}
\end{figure}

The dynamics of the $\gamma^* p\to \gamma p$ subprocess can be
parameterized in different ways.  A form factor decomposition of the
hadronic tensor $T^{\alpha\beta}$ defined in (\ref{compton-tensor})
has for instance been used in \cite{Belitsky:2001ns}.  Alternatively
one may use the helicity amplitudes $e^2 M_{\lambda'\mu', \lambda\mu}$
defined in (\ref{hel-amp-compton}), which we find particularly
suitable given their immediate physical interpretation.  With the
constraints from parity invariance there are 12 such amplitudes for
DVCS, of which four conserve the photon helicities, four describe
single helicity flip and four describe double helicity flip of the
photon.  Both the Compton form factors and the helicity amplitudes
depend only on the kinematical variables $Q^2$, $t$, and $\xB$ (which
may be traded for $\xi$).  Additional variables needed to describe the
electroproduction process are the usual inelasticity parameter $y$ or
the ratio $\epsilon$ of longitudinal and transverse polarization of
the virtual photon in DVCS,
\begin{equation}
  \label{y-eps-def}
y = \frac{k \cdot p}{q \cdot p} , \qquad
\epsilon = 
\frac{1 - y - \delta}{1 - y + y^2/2 + \delta}
\end{equation}
with $\delta = y^2 \xB^2 m^2/Q^2$, and the azimuthal angle $\phi$
between hadron and lepton planes defined in
Fig.~\ref{fig:dvcs-kin}.\footnote{This definition follows the
convention used by HERMES \protect\cite{Ely:2002th} and by CLAS
\protect\cite{Stepanyan:2003pc}.  It is related to the angle used in
\cite{Diehl:1997bu} by $\phi_{\mathrm{here}} =
-\varphi_{\smallcite{Diehl:1997bu}}$ and to the one in
\protect\cite{Belitsky:2001ns} by $\phi_{\mathrm{here}} =
\pi - \phi_{\smallcite{Belitsky:2001ns}}$.}

The Bethe-Heitler process can be calculated given the knowledge of the
electromagnetic proton form factors, which are well measured at small
$t$ (see e.g.\ \cite{Mergell:1996bf}).  One can hence give exact
expressions for the differential $ep\to ep\gamma$ cross section in
terms of the Compton amplitudes.  Since however the theoretical
description of the Compton process is based on an expansion in $1/Q$
at fixed $\xB$ and $t$, it is meaningful to apply the same expansion
to the Bethe-Heitler process.  This expansion reveals a simple
structure of the electroproduction cross section in the kinematics of
DVCS \cite{Diehl:1997bu,Belitsky:2000gz,Belitsky:2001ns}.  Treating
the Bethe-Heitler process exactly (while approximating the Compton
process to twist-two or three accuracy) is of course possible at the
level of a numerical analysis, as has for instance been done in
\cite{Vanderhaeghen:1998uc,Kivel:2000fg}.

The differential electroproduction cross section on an unpolarized
target is given by
\begin{eqnarray}
\frac{d\sigma(ep\to ep\gamma)}{d\phi\, dt\, dQ^2\, d\xB} =
  \frac{\alpha_{\mathrm{em}}^3}{8\pi}\, \frac{\xB y^2}{Q^4}\, 
  \frac{1}{\sqrt{1 + 4 \xB^2 m^2 /Q^2}}\,  \frac{1}{e^6}\,
 {\sum_{\mathrm{spins}}}' 
 | \mathcal{T}_{\mathrm{VCS}} + \mathcal{T}_{\mathrm{BH}} |^2 ,
\end{eqnarray}
where $\sum_{\mathrm{spins}}'$ denotes the sum over the spins of the
final-state proton and photon and the average over the initial proton
polarization.  The contribution from Compton scattering is
\begin{eqnarray}
  \label{VCS}
\lefteqn{
\frac{1}{e^6}\, {\sum_{\mathrm{spins}}}'
	| \mathcal{T}_{\mathrm{VCS}} |^2 =
\frac{1}{Q^2}\, \frac{2}{1-\epsilon}\,
\sum_{\lambda'\lambda} \Bigg[ \,
   \frac{|M_{\lambda'+,\lambda+}|^2 + |M_{\lambda'+,\lambda-}|^2}{2}
 + \epsilon |M_{\lambda'+,\lambda\, 0}|^2
}
\nonumber \\[0.2em]
 & & {}- \cos\phi \, \sqrt{\epsilon (1+\epsilon) }\, 
  \re\left\{ 
     M_{\lambda'+,\lambda+}^*\, M_{\lambda'+,\lambda\, 0}^{\phantom{*}}
   - M_{\lambda'+,\lambda-}^*\, M_{\lambda'+,\lambda\, 0}^{\phantom{*}}
  \right\} 
{}- \cos 2\phi\, \epsilon\, \re\left\{ 
    M_{\lambda'+,\lambda+}^*\, M_{\lambda'+,\lambda-}^{\phantom{*}}
                                    \right\}
    \phantom{\frac{1}{2}}
\nonumber \\[0.4em]
 & & {}- P_\ell \, \sin\phi \, \sqrt{\epsilon (1-\epsilon) }\, 
  \im\left\{ 
     M_{\lambda'+,\lambda+}^*\, M_{\lambda'+,\lambda\, 0}^{\phantom{*}}
   - M_{\lambda'+,\lambda-}^*\, M_{\lambda'+,\lambda\, 0}^{\phantom{*}}
  \right\}
\Bigg]
\end{eqnarray}
where $P_\ell$ is the lepton beam polarization, $-1\le P_\ell \le 1$.
The full expression of the Bethe-Heitler contribution can be found in
\cite{Belitsky:2001ns}.  To make its behavior in DVCS kinematics
apparent we take the limit of large $Q$ at fixed $t$, $\xB$, $y$ and
$\phi$, and require in particular that $1-\xB$ and $1-y$ be large
compared with $m^2/Q^2$.  Expanding in $1/Q$ we get
\begin{eqnarray}
  \label{BH}
\frac{1}{e^6}\, {\sum_{\mathrm{spins}}}'
| \mathcal{T}_{\mathrm{BH}} |^2 &=&
- \frac{1}{t}\, \frac{4}{\epsilon}\, 
  \frac{1}{P}
  \Bigg[ \frac{1-\xi^2}{\xi^2}\, \frac{t - t_0}{t}\, 
         \Big( F_1^2 - \frac{t}{4m^2} F_2^2 \Big) + 2 (F_1+F_2)^2 
{}+ O\Big(\frac{1}{Q}\Big) \Bigg] 
\end{eqnarray}
with the Dirac and Pauli form factors $F_1$, $F_2$ evaluated at
momentum transfer $t$.  The factor $P = 1 + \cos\phi\; O(1/Q)$ will be
discussed shortly.  In the same expansion the interference term reads
\begin{eqnarray}
  \label{VCS-BH}
\lefteqn{
\frac{1}{e^6}\, {\sum_{\mathrm{spins}}}'
( \mathcal{T}_{\mathrm{BH}}^* \mathcal{T}_{\mathrm{VCS}}^{\phantom{*}}  
+ \mathcal{T}_{\mathrm{VCS}}^* \mathcal{T}_{\mathrm{BH}}^{\phantom{*}} )
 = e_\ell\; \frac{1}{t}\, \frac{m}{Q}\,
         \frac{8 \sqrt{2}}{\xi}\, \frac{1}{P}
}
\nonumber \\
 &\times&  \Bigg[ 
   \cos\phi \, \frac{1}{\sqrt{\epsilon (1-\epsilon)}}\, 
           \re \widehat{M}_{++}
 - \cos 2\phi \, \sqrt{\frac{1+\epsilon}{1-\epsilon}}\, 
           \re \widehat{M}_{+0}
{}- \cos 3\phi \, \sqrt{\frac{\epsilon}{1-\epsilon}}\, 
           \re \widehat{M}_{+-}
\nonumber \\
 & & {}+ P_\ell \sin\phi \, \sqrt{\frac{1+\epsilon}{\epsilon}}\, 
           \im \widehat{M}_{++}
     - P_\ell \sin 2\phi \, \im \widehat{M}_{+0} 
     + O\Big(\frac{1}{Q}\Big) \Bigg] ,
\end{eqnarray}
where $e_\ell = \pm 1$ is the lepton beam charge.  We have not yet
used the $Q^2$ behavior of the Compton amplitudes
$M_{\lambda'\mu',\lambda\mu}$ at this point, i.e., the terms denoted
by $O(1/Q)$ in (\ref{VCS-BH}) involve Compton amplitudes multiplied
with kinematical factors going at least like $1/Q$.  We have
introduced linear combinations
\begin{equation}
  \label{eff-amplitudes}
\widehat{M}_{\mu'\mu} = \frac{1}{2}
	\sum_{\lambda'\lambda} g_{\lambda'\lambda}\,
	 M_{\lambda'\mu', \lambda\mu}
\end{equation}
with coefficients
\begin{eqnarray}
  \label{inter-coeffs}
g_{++} &=& \frac{\sqrt{t_0-t}}{2m} \Big[ F_1 + \xi (F_1+F_2) \Big] ,
\nonumber \\
g_{--} &=& \frac{\sqrt{t_0-t}}{2m} \Big[ F_1 - \xi (F_1+F_2) \Big] ,
\nonumber \\
g_{+-} &=& \sqrt{1-\xi^2}\; \frac{t_0-t}{4m^2}\, F_2 ,
\nonumber \\
g_{-+} &=& {}- \sqrt{1-\xi^2}\; \frac{t_0-t}{4m^2}\, F_2 
           - \frac{2\xi^2}{\sqrt{1-\xi^2}}\, (F_1+F_2) .
\end{eqnarray}
Notice that $\widehat{M}_{+\mu} \propto \sqrt{t_0-t}$ since $g_{-+}$
is multiplied with $M_{-+,+\mu}$, which due to angular momentum
conservation vanishes at $t=t_0$ at least like $\sqrt{t_0-t}$.  Our
phase convention for the Compton amplitudes is to take the (usual)
helicity spinors of Appendix~\ref{app:spinors} for the proton and the
photon polarizations specified in Section~\ref{sec:compton-scatt},
using the coordinate system shown in
Fig.~\ref{fig:dvcs-kin}.\footnote{Different phase conventions were
used in \protect\cite{Diehl:1997bu}, and the proton helicity
conserving DVCS amplitudes there differ from the ones used here by a
global sign.}

The relative weight of the Compton and Bethe-Heitler cross sections is
parametrically given by
\begin{equation}
 | \mathcal{T}_{\mathrm{VCS}} |^2 : | \mathcal{T}_{\mathrm{BH}} |^2
\sim \left( \frac{1}{Q^2}\, \frac{1}{1-\epsilon} \right) :
     \left( -\frac{1}{t}\, \frac{1}{\epsilon} \right)  ,
\end{equation}
with the interference term being of order of their geometric mean.
The condition $Q^2 \gg -t$ of DVCS favors the Bethe-Heitler
contribution to start with.  This can be counteracted by having
$\epsilon$ close to 1, i.e., small $y$.  This means a large $ep$ c.m.\
energy $\sqrt{s}$ at given $Q^2$ and $\xB$, or large $\xB$ at given
$Q^2$ and $\sqrt{s}$, given the relation $y \xB = Q^2 /(s-m^2)$.  We
notice in (\ref{BH}) that the Bethe-Heitler cross section grows like
$\xi^{-2}$ at small $\xi$, except in the very forward region where
$|t_0 - t| \ll |t_0|$.  According to our discussion in
Section~\ref{sec:small-x-gpd} one expects a small-$x$ behavior like
$\xi^{-(1+\lambda)}$ with some $\lambda>0$ for the leading-twist
Compton amplitudes, so that the Compton cross section should increase
somewhat faster than the Bethe-Heitler one at small $\xi$.

The lepton virtualities in the Bethe-Heitler graphs of
Fig.~\ref{fig:vcs-bh} are $s' = (q'+k')^2$ and $u' = (q'-k)^2$.  They
are given by
\begin{eqnarray}
   \label{lepton-props}
s' &=& \frac{Q^2}{y} \left[ 1 - 
	\frac{2 \sqrt{1-y}\, \sqrt{t_0-t}}{Q}\, 
	\sqrt{\frac{1-\xi}{1+\xi}}\, \cos\phi \right] + O(t,m^2) ,
\nonumber \\
-u' &=& \frac{Q^2}{y} \left[ 1-y +
	\frac{2 \sqrt{1-y}\, \sqrt{t_0-t}}{Q}\, 
	\sqrt{\frac{1-\xi}{1+\xi}}\, \cos\phi \right] + O(t,m^2)
\end{eqnarray}
and hence $\phi$ independent to leading order in $1/Q$.  It turns
however out that for experimentally relevant kinematics, this
dependence is numerically not always negligible, as has been
emphasized in \cite{Belitsky:2000gz,Belitsky:2001ns}.  Because the
leading term in $u'$ is proportional to $1-y$, this holds in
particular for the kinematics of the HERMES and CLAS measurements
\cite{Airapetian:2001yk,Stepanyan:2001sm}, where $y$ is not small.
Neglecting the $\cos\phi$ terms is an even worse approximation in
$1/(s' u')$, which appears in the cross section, because the geometric
series $(1-x)^{-1} = 1 + x + \ldots$ converges rather slowly.  This is
why we have kept the ratio of exact and approximated propagators
\begin{equation}
  P = \frac{- s' u'}{(1-y) y^{-2}\, Q^4}
\end{equation}
in the expressions of the Bethe-Heitler and interference terms above
(the exact expressions for the propagators can be found in eqs.~(28)
to (32) of \cite{Belitsky:2001ns}).  Taking this ratio into account is
particularly important in an angular analysis of the Compton process,
which we now discuss.

\subsubsection{Angular structure}
\label{sub:Compton-angular}

Let us take a closer look on how the simple angular structure in
(\ref{BH}) and (\ref{VCS-BH}) comes about and how general it is.  To
this end we write both $\mathcal{T}_{\mathrm{VCS}}$ and
$\mathcal{T}_{\mathrm{BH}}$ as a product of a leptonic and a hadronic
tensor, the latter being the Compton tensor for VCS and the
proton-photon vertex for Bethe-Heitler.  For Compton scattering we
have introduced polarization vectors (\ref{photon-pol}) of the
intermediate photon with momentum $q$, and in analogy we introduce
polarizations for the photon of momentum $\Delta$ in the Bethe-Heitler
process.  We choose these vectors, as well as the polarizations of the
external particles, with reference only to the hadronic plane (i.e.,
define them in terms of the momenta $p$, $p'$, $q$) and contract them
with the hadronic and leptonic tensors.  The hadronic parts of the
amplitudes are then $\phi$ independent and all $\phi$ dependence is
explicit in the leptonic parts.  Further $\phi$ dependence arises if
the target has a definite polarization longitudinal or transverse to
the lepton beam (see Section~\ref{sub:Compton-polar}).  We obtain the
expression (\ref{VCS}) for the unpolarized Compton cross section and a
similar one with target polarization.  For the Bethe-Heitler part we
get
\begin{equation}
\sum_{\mathrm{spins}} \rho_{\tilde{\lambda} \lambda}\, 
| \mathcal{T}_{\mathrm{BH}} |^2 =
\sum_{\lambda', \tilde{\lambda}, \lambda  \atop 
      \scriptstyle{\tilde{\nu}, \nu}}
   L_{\mathrm{BH}}(P_\ell, \tilde{\nu}, \nu) \;
   \Gamma^*_{\tilde{\nu} \lambda' \tilde{\lambda}}\, 
   \Gamma_{\nu \lambda' \lambda \phantom{\tilde{l}} }^{\phantom{*}} \,
   \rho_{\tilde{\lambda} \lambda}^{\phantom{*}} ,
\end{equation}
where $\sum_{\mathrm{spins}}$ denotes summation over the spins of
incoming and outgoing particles.  $\rho_{\tilde{\lambda} \lambda}$ is
the spin density matrix of the target, with helicities $\lambda$ and
$\tilde{\lambda}$ in the amplitude and its complex conjugate. The
leptonic part $L_{\mathrm{BH}}$ of the squared amplitude depends on
the lepton beam polarization (taken as longitudinal since transverse
polarization effects are suppressed by the lepton mass) and on the
helicities $\nu$ and $\tilde{\nu}$ of the photon with momentum
$\Delta$ in the amplitude and its complex conjugate.  $\Gamma$ is the
proton-photon vertex for fixed helicities.  One finds that the
denominators of the lepton propagators appear as $(s' - m_\ell^2)^{-1}
(u' - m_\ell^2)^{-1}$ and $m_\ell^2\, (s' - m_\ell^2)^{-2} (u' -
m_\ell^2)^{-2}$.  If we neglect the lepton mass, $L_{\mathrm{BH}}$ can
thus be written as $1/P$ times a trigonometric polynomial in $\phi$.
The $P_\ell$ independent part of this polynomial is of order 2, and
the $P_\ell$ dependent part of order 1.  Both parts are of order 0 to
leading order in $1/Q$ as we see in (\ref{BH}).

Similarly, the interference term is
\begin{eqnarray}
  \label{structure-interf}
\lefteqn{
\sum_{\mathrm{spins}} \rho_{\tilde{\lambda} \lambda}\, \Big(
  \mathcal{T}_{\mathrm{BH}}^* \mathcal{T}_{\mathrm{VCS}}^{\phantom{*}}  
+ \mathcal{T}_{\mathrm{VCS}}^* \mathcal{T}_{\mathrm{BH}}^{\phantom{*}}
\Big)
}
\nonumber \\
&=& \sum_{\lambda', \tilde{\lambda}, \lambda \atop 
      \scriptstyle{\tilde{\nu}, \mu', \mu}}
  L_{\mathrm{INT}}(P_\ell, \mu', \mu, \tilde{\nu}) \;
    \Gamma^*_{\tilde{\nu} \lambda' \tilde{\lambda}} \, 
  M_{\lambda'\mu', \lambda\mu \phantom{\tilde{l}} }^{\phantom{*}} \,
    \rho_{\tilde{\lambda} \lambda}^{\phantom{*}} \, 
+ \mathrm{c.c.} 
\end{eqnarray}
The leptonic part now has the form $1/P$ times a trigonometric
polynomial in $\phi$, with a $P_\ell$ independent part of order 3, and
the $P_\ell$ dependent part of order 2.  Its dependence on $\phi$ and
the photon polarizations is
\begin{equation}
L_{\mathrm{INT}} \propto \frac{1}{P(\cos\phi)}\,  
     e^{i (\mu - 2\mu') \phi} \, 
     \delta_{\tilde{\nu} \mu'} + O\Big( \frac{1}{Q} \Big) ,
\end{equation}
where the coefficient of $e^{\pm 3i \phi}$ does not depend on $P_\ell$.
This simple structure readily translates into the angular dependence
in (\ref{VCS-BH}), where for fixed $\mu'$ each helicity $\mu$ comes
with a different angular dependence, and where a $\sin 3\phi$ term is
missing in the $P_\ell$ dependent part.  The constraint $\tilde{\nu} =
\mu'$ in the leading order part of $L_{\mathrm{INT}}$ has important
consequences: in the interference term the Compton amplitudes
$M_{\lambda'\mu', \lambda\mu}$ are summed over the proton helicities
with weights $\Gamma^*_{\mu' \lambda' \tilde{\lambda}}\,
\rho_{\tilde{\lambda} \lambda}^{\phantom{*}}$ that are
\begin{itemize}
\item dependent on $\xi$ and $t$ but not on $\epsilon$,
\item the same for the $P_\ell$ independent and dependent part,
\item the same for all $\mu$ at fixed $\mu'$.
\end{itemize}
These properties are of course realized in (\ref{VCS-BH}) to
(\ref{inter-coeffs}), with the coefficients $g_{\lambda'\lambda}$
being just the proton-photon vertices $\Gamma^*_{+ \lambda'
\lambda}$ up to a global kinematical factor.  They are also found in
the explicit calculation for a polarized target
\cite{Belitsky:2001ns}.  One finds in particular that with different
target polarizations one can obtain enough different weighting factors
$\Gamma^*_{\mu' \lambda' \tilde{\lambda}}\,
\rho_{\tilde{\lambda} \lambda}^{\phantom{*}}$ to separate the Compton
amplitudes for different $\lambda'$, $\lambda$ (see
Section~\ref{sub:Compton-GPDs}).

Going beyond the leading order in $1/Q$ one finds that for given $\mu$
and $\mu'$ other frequencies $e^{i n \phi}$ appear in
$L_{\mathrm{INT}}$, suppressed by powers of $1/Q$.  The interference
term for an unpolarized target has the general $\phi$ dependence
\begin{eqnarray}
  \label{inter-angular}
P(\cos\phi)\, \frac{d\sigma_{\mathrm{INT}}(ep\to ep\gamma)}{d\phi\, 
	dt\, dQ^2\, d\xB} &=& 
c_0 + c_1  \cos\phi + c_2  \cos 2\phi + c_3  \cos 3\phi
\nonumber \\
 &+& P_\ell\, ( s_1  \sin\phi + s_2  \sin 2\phi ) ,
\end{eqnarray}
where the $c_i$, $s_i$ are linear combinations of the Compton
amplitudes with prefactors depending on $Q^2$, $t$, $\xi$ and
$\epsilon$.  Closer inspection reveals in particular that the
amplitudes appear in the hierarchy
\begin{itemize}
\item $Q^{-1} M_{\lambda'+,\lambda+}$,  $\, Q^{-2}
M_{\lambda'+,\lambda\, 0}$, $\, Q^{-3} M_{\lambda'+,\lambda-}$ in
$c_0$,
\item  $M_{\lambda'+,\lambda+}$, $\, Q^{-1} M_{\lambda'+,\lambda\,
0}$, $\, Q^{-2} M_{\lambda'+,\lambda-}$ in $c_1$ and $s_1$,
\rule{0pt}{1.2em}
\item $M_{\lambda'+,\lambda\, 0}$,  $\, Q^{-1} M_{\lambda'+,\lambda-}$,
$\, Q^{-3} M_{\lambda'+,\lambda+}$ in $c_2$ and $s_2$.
\rule{0pt}{1.2em} 
\end{itemize}
The linear combination of the different twist-two amplitudes
$M_{\lambda'+,\lambda+}$ in the subleading coefficient $c_0$ differs
from the one in $c_1$.  We will come back to this in
Section~\ref{sub:Compton-GPDs}.

If one takes into account the power behavior predicted by the
dynamical analysis of the factorization theorem (where
$M_{\lambda'+,\lambda\, 0}$ is $1/Q$ suppressed compared with
$M_{\lambda'+,\lambda+}$) one altogether finds that each Fourier
coefficient in (\ref{inter-angular}) is given by the leading-order
formula (\ref{VCS-BH}) up to corrections of order $1/Q^2$ and not only
$1/Q$.  An analysis based on (\ref{VCS-BH}) should therefore be
feasible even at moderately large $Q^2$.  In the analogous case of
timelike Compton scattering (see Section~\ref{sec:gaga}) this was
corroborated by numerical comparison of the interference term obtained
with and without expanding the Bethe-Heitler amplitude in the inverse
large scale \cite{Berger:2001xd}.

An important corollary to the preceding discussion is that the $\phi$
dependence is a consequence of the \emph{leptonic} part of the
Bethe-Heitler and interference terms.  The results of this subsection
therefore hold not only for the process $ep\to ep\gamma$, but for
targets of \emph{any} spin, and for the case where the target is
scattered into a different hadron or hadronic system, whose details
(including polarization) are unobserved in the measured cross section.
This implies in particular that the principles of the angular analysis
of $ep\to ep\gamma$ using the above results are not invalidated if
there is a background of proton dissociation.

\subsubsection{Beam and target polarization}
\label{sub:Compton-polar}

Let us again specialize to a spin $\half$ target, but allow for the
possibility that the proton dissociates into a system of any spin.
The angular analysis described in the previous subsection becomes
modified in the presence of target polarization, since this
polarization is experimentally fixed in the laboratory and not with
respect to the scattering plane of the Compton process.  Let us recall
the essentials of the transformation from definite polarization in the
laboratory (defined to be the target rest frame or a frame where
electron and proton collide head on) to polarization with respect to
the ``hadronic'' coordinate system of Fig.~\ref{fig:dvcs-kin}, where
we have also defined the helicity states for the initial proton.
\begin{itemize}
\item Longitudinal polarization in the lab induces in the hadronic
system a longitudinal polarization, plus a transverse polarization
which is suppressed by a factor of order $2 \xB m/Q$ and whose
direction in the $x$-$y$ plane depends on $\phi$.  The spin density
matrix of the target is hence $\phi$ dependent to $1/Q$ accuracy, and
an analysis of $1/Q$ suppressed terms in the interference term or the
Compton part of the $ep$ cross section will have the $\phi$ dependence
modified compared with the unpolarized case.
\item Transverse polarization in the lab induces in the hadronic
system a transverse polarization whose direction depends on $\phi$ and
on the azimuthal angle $\Psi$ between the target polarization and the
lepton scattering plane in the laboratory.  It further induces a
longitudinal polarization whose size depends on $\Psi$ and is again
suppressed by a factor of order $2 \xB m/Q$.  In this case the
spin-density matrix of the target hence depends on $\phi$ already to
leading order and accordingly modifies the angular structure compared
with the unpolarized case.
\end{itemize}

Important constraints on the structure of the $ep$ cross section follow
from parity and time reversal invariance:
\begin{itemize}
\item Since the helicity and the covariant spin vector of a spin
$\half$ particle are parity odd, a single (beam or target) spin
asymmetry must be proportional to the antisymmetric tensor
$\epsilon_{\alpha\beta\gamma\delta}$, which is the only other parity
odd structure available.  This leads to an azimuthal dependence via
$\sin(n \phi)$ (and $\sin(n \phi \pm \Psi)$ for transverse target
polarization) with integer $n$.  Conversely, the unpolarized cross
section or a double (beam and target) spin asymmetry cannot be
proportional to $\epsilon_{\alpha\beta\gamma\delta}$ and thus has an
azimuthal dependence via $\cos(n \phi)$ and $\cos(n \phi \pm \Psi)$.
\item The tensor $\epsilon_{\alpha\beta\gamma\delta}$ is odd under
time reversal.  Time reversal invariance implies that terms
proportional to this tensor must involve absorptive parts of the
scattering amplitude, i.e., correspond to possible on-shell
intermediate states in the process.  As we are working to leading
order in $\alpha_{\mathrm{em}}$, the leptonic parts of the Compton and
Bethe-Heitler amplitudes do not have absorptive parts.  Together with
the above consequence of parity invariance one thus finds that single
spin asymmetries must go with $\im (M^* M)$, $\im (\Gamma^* M)$ or
$\im (\Gamma^* \Gamma)$, whereas the unpolarized cross section and
double spin asymmetries are proportional to the real parts of these
products, where for simplicity we have omitted the helicity
labels.\footnote{The identification of the ``imaginary'' with the
``absorptive'' parts for $M$ and $\Gamma$ only holds if no further
phases come from the proton or photon polarizations.  This is the case
for our phase convention.}
The elastic proton-photon vertex $\Gamma$ does not have an imaginary
part, so that $\im (\Gamma^* M)= \Gamma \, \im M$ and $\im (\Gamma^*
\Gamma) =0$.

An important consequence is that the Bethe-Heitler contribution is
zero for \emph{any} single-spin asymmetry on a spin $\half$ target.
This is an exact result to leading order in
$\alpha_{\mathrm{em}}$,\footnote{The result even holds for the QED
radiative corrections which are enhanced by logarithms of the lepton
mass.  These correspond to the additional emission of soft or
collinear photons from the lepton and to the associated virtual
corrections, and do not generate absorptive parts.}
confirmed by the explicit calculation in \cite{Belitsky:2001ns} and
generalizing the observation for the beam spin asymmetry of
\cite{Kroll:1996pv}.  Notice that it no longer holds for proton
dissociation, where the Bethe-Heitler contribution to single spin
asymmetries must be estimated from the phases of the electromagnetic
transition $\Gamma$ from the proton to the final state.  If however
the cross section for $ep\to e X \gamma$ is summed over all hadronic
final states $X$ of given invariant mass (as is done when the proton
remnant is not experimentally detected) one has again a zero
Bethe-Heitler contribution to single spin asymmetries.  This is
readily seen by writing the Bethe-Heitler cross section as the
contraction of a leptonic tensor with the imaginary part of the
hadronic tensor for the forward Compton amplitude $\gamma^*(\Delta) +
p\to
\gamma^*(\Delta) + p$.  Due to parity and time reversal invariance the
respective lepton and proton polarization dependent parts of these
tensors are antisymmetric, whereas the polarization independent parts
are symmetric.
\end{itemize}

\subsubsection{Cross sections and asymmetries}
\label{sub:cross-asy}

There are different observables from which information on the Compton
amplitudes can be extracted.  The pure Bethe-Heitler part of the
differential $ep$ cross section is calculable and can be subtracted,
provided it does not dominate so strongly that the remainder would
have unacceptably large errors.  Such a subtraction has been performed
in the cross section measurements at small $x$ by H1 and ZEUS
\cite{Adloff:2001cn,Zeus:2003ip}.  As discussed in the previous
subsection, a different possibility to eliminate the Bethe-Heitler
contribution is to take the difference of cross sections for opposite
beam or target polarization (the latter either longitudinal or
transverse).  In both cases, the cross section difference receives
contributions from both Compton scattering and the
Compton-Bethe-Heitler interference.  The cleanest separation of these
pieces can be achieved in experiments with beams of either charge.
Since the Compton contribution to the $ep$ amplitude is linear and the
Bethe-Heitler contribution quadratic in the lepton charge, the
interference term is projected out in the difference $d
\sigma(e^+ p) - d \sigma(e^- p)$ of cross sections, whereas it is
absent in their sum.

To understand the relative importance of Compton and interference
contributions in different observables, we schematically write
\begin{eqnarray}
\lefteqn{
\frac{d\sigma_{\mathrm{VCS}}}{d\phi\, dt\, dQ^2\, d\xB} =
  \frac{1}{Q^2}  \sum_{\lambda'\lambda} \Big[
       \ldots |M_{\lambda'+,\lambda+}|^2
   + \ldots \cos\phi \;
          \re ( M_{\lambda'+,\lambda+}^*\, 
	        M_{\lambda'+,\lambda\, 0}^{\phantom{*}} )
}^{\phantom{*}}
\nonumber \label{struct-compt}
\\[0.3em]
  && {}+ \ldots P_\ell\, \sin\phi \;
          \im ( M_{\lambda'+,\lambda+}^*\, 
                M_{\lambda'+,\lambda\, 0}^{\phantom{*}} )
     + O(Q^{-2}) + O(\alpha_s) \Big]
\\[0.5em]
\lefteqn{
\frac{d\sigma_{\mathrm{INT}}}{d\phi\, dt\, dQ^2\, d\xB} = 
  \frac{1}{Q}\,
   \frac{e_\ell}{P(\cos\phi)}\,  \sum_{\lambda'\lambda} 
   \Big[ \ldots Q^{-1} \, \re M_{\lambda'+,\lambda+}
}
\label{struct-int}
\nonumber \\[0.3em]
 && {}+ \ldots \cos\phi \, \re M_{\lambda'+,\lambda+}
      + \ldots \cos 2\phi\ \re M_{\lambda'+,\lambda\, 0}
\nonumber
\\[0.5em]
 && {}+ \ldots P_\ell\, \sin\phi\, \im M_{\lambda'+,\lambda+} 
      + \ldots P_\ell\, \sin 2\phi\, \im M_{\lambda'+,\lambda\, 0} 
    + O(Q^{-2}) + O(\alpha_s)  \Big] ,
\end{eqnarray}
with 
\begin{equation}
P(\cos\phi) = 1 + \ldots Q^{-1} \cos\phi + O(Q^{-2}) ,
\end{equation}
where we have made explicit the dependence on $\phi$ and on the lepton
charge and polarization.  For the power counting we have now also used
$M_{\lambda'+,\lambda\, 0} \sim 1/Q$ and $M_{\lambda'+,\lambda-} \sim
\alpha_s$.  We see that when integrated over $\phi$, the subtracted
cross section $d \sigma /(dt\, dQ^2\, d\xB) - d \sigma_{\mathrm{BH}}
/(dt\, dQ^2\, d\xB)$ receives a contribution
\begin{itemize}
\item from Compton scattering, proportional to the squared
leading twist amplitudes $|M_{\lambda'+,\lambda+}|^2$,
\item and from the interference term, at the level of  $1/Q$
suppressed terms.  Part of this contribution comes from the $\phi$
independent contribution in brackets and part from the $\cos\phi$
term, with both pieces involving the leading-twist amplitudes $\re
M_{\lambda'+,\lambda+}$.  We remark that with the model GPDs used in
\cite{Belitsky:2001ns}, the total interference contribution to the
$\phi$ integrated cross section was estimated to be at the percent
level in the kinematics of the H1 measurement.
\end{itemize}

Another important observable is the cross section difference for
different lepton helicities
\begin{equation}
   \label{good-spin}
P(\cos\phi)\, \left[
  \frac{d\sigma(e^\uparrow p)}{d\phi\, dt\, dQ^2\, d\xB} -
  \frac{d\sigma(e^\downarrow p)}{d\phi\, dt\, dQ^2\, d\xB} 
\right] ,
\end{equation}
where the prefactor removes the $\phi$ dependence due to the lepton
propagators.  This observable has a $\sin\phi$ term, which receives a
contribution from $d\sigma_{\mathrm{INT}}$ going with $\im
M_{\lambda'+,\lambda+}$, and a contribution from
$d\sigma_{\mathrm{VCS}}$ involving the product of twist-two and
twist-three amplitudes.  Similarly, the subtracted unpolarized cross
section
\begin{equation}
  \label{good-sub}
P(\cos\phi)\, \left[
  \frac{d\sigma(e p)}{d\phi\, dt\, dQ^2\, d\xB} -
  \frac{d\sigma_{\mathrm{BH}}(e p)}{d\phi\, dt\, dQ^2\, d\xB} 
\right]
\end{equation}
has a $\cos\phi$ term with an interference contribution involving $\re
M_{\lambda'+,\lambda+}$ and a contribution from $1/Q$ suppressed terms
in $P(\cos\phi)\, d\sigma_{\mathrm{VCS}} /(d\phi\, dt\, dQ^2\, d\xB)$.
It was estimated in \cite{Belitsky:2001ns} that despite being power
suppressed, this Compton contribution can be comparable to the
interference one, especially in the small $x$ regime.  To understand
this, notice that $\re ( M_{\lambda'+,\lambda+}^*
M_{\lambda'+,\lambda\, 0}^{\phantom{*}} )$ contains the product of the
imaginary parts of two Compton amplitudes, which at small $x$ tend to
be larger than the real parts.  This recommends care when trying to
extract $\re M_{\lambda'+,\lambda+}$ from an angular analysis of the
cross section alone, without the benefit of the beam charge asymmetry.

In cases where information on the Compton process is sought from the
interference term, the analysis of the $\phi$ dependence becomes
considerably cleaner if the measured cross section is re-weighted by
the propagator factor $P(\cos\phi)$ as in (\ref{good-spin}) and
(\ref{good-sub}), provided one has sufficient experimental resolution
for the kinematic quantities $(q'+k')^2$ and $(q'-k)^2$.  The weighted
cross section is then a trigonometric polynomial, whose coefficients
have a simple relation to the fundamental quantities describing the
Compton process, at least for sufficiently large $Q^2$ (see
Section~\ref{sub:Compton-angular}).  These coefficients are readily
extracted by fitting the $\phi$ dependence or by additional weighting
with $\cos(n\phi)$ and $\sin(n\phi)$, i.e., by measuring quantities
like
\begin{equation}
  \label{good-boy}
\int d\phi\, \sin(\phi)\, P(\cos\phi)\, 
     \frac{d\sigma(e p)}{d\phi\, dt\, dQ^2\, d\xB}
\end{equation}
and its analog for appropriate cross section differences.  The
weighted cross sections including the factor $P(\cos\phi)$ may be
smaller than the ones without.  This does \emph{not} determine which
of the two is more sensitive to the quantities to be extracted, since
not only the size of the observable but also its error depends on the
weight chosen.  The same holds for the comparison of other weighting
factors, for instance of $\mbox{sgn}(\sin\phi)$ instead of $\sin\phi$,
which leads to an ``up-down'' asymmetry $\int_0^\pi d\phi\, (d\sigma
/d\phi) - \int_\pi^{2\pi} d\phi\, (d\sigma /d\phi)$.

As discussed in Section~\ref{sub:Compton-angular} the coefficients of
$\cos\phi$ or $\sin\phi$ in the re-weighted interference term are
proportional to $\widehat{M}_{++}$ and those of $\cos 2\phi$ or $\sin
2\phi$ are proportional to $\widehat{M}_{+0}$, up to corrections in
$1/Q^2$ if one is in a region where the Compton amplitudes have the
power behavior predicted by the large $Q^2$ analysis.  In contrast,
angular distributions without re-weighting by $P(\cos\phi)$ have the
dynamical information about Compton scattering entangled with $1/Q$
effects from the Bethe-Heitler propagators, which can be numerically
important.  These propagators depend strongly on $t$, see
(\ref{lepton-props}), and their effect cannot readily be separated in
$\phi$ spectra integrated over a certain $t$ range.  For a particular
parameterization of the relevant GPDs one can of course calculate the
corresponding twist-two and twist-three Compton amplitudes and
directly evaluate observables from them, as was for instance done in
the study by Kivel et al.~\cite{Kivel:2000fg}.  Given however the
considerable freedom in modeling the dependence of the different GPDs
on their three variables, the connection between the shape of GPDs and
the differential $ep$ cross section is highly nontrivial.  The studies
\cite{Kivel:2000fg} and \cite{Belitsky:2001ns} found that twist-three
amplitudes can have quite visible effects on observables in the
kinematics of current experiments, for instance on the beam charge and
to a lesser extent on the beam spin asymmetry.  Only if one assumes
twist-three quark-gluon correlation to be negligible are these
twist-three amplitudes expressed in terms of the leading-twist input
GPDs.  A separation of Compton amplitudes with different dynamical
twist, or similarly of Compton form factors as proposed in
\cite{Belitsky:2001ns}, should help towards keeping the analysis
transparent and assessing possible ambiguities in attempting to
extract GPDs from the data.

As discussed in \cite{Belitsky:2001ns}, differences of absolute cross
sections (for different polarizations or beam charges) are simpler to
analyze theoretically than the corresponding asymmetries (normalized
to the sum of relevant cross sections).  This is because the
denominator of these asymmetries also depends on unknown Compton
amplitudes, which are in general different from those in the
numerator.  Notice that there are different versions of normalized
observables.  The complications in the analysis introduced by the
normalization factor are for instance different for
\begin{equation}
  \label{exp-SSA}
A(\phi) = 
\frac{d\sigma(e^\uparrow p) - d\sigma(e^\downarrow p)}{
      d\sigma(e^\uparrow p) + d\sigma(e^\downarrow p)} ,
\end{equation}
and for
\begin{equation}
A^{\sin\phi} = 
\frac{2 \displaystyle \int d\phi\, \sin\phi\, \Bigg[ 
	\frac{d\sigma(e^\uparrow p)}{d\phi}
      - \frac{d\sigma(e^\downarrow p)}{d\phi} \Bigg]}{
      \displaystyle \int d\phi\, \Bigg[ 
	\frac{d\sigma(e^\uparrow p)}{d\phi}
      + \frac{d\sigma(e^\downarrow p)}{d\phi} \Bigg]}
\end{equation}
or the analog with additional weighting factors $P(\cos\phi)$ under
the integrals.\footnote{The factor $2$ in the numerator of
$A^{\sin\phi}$ follows the HERMES convention
\protect\cite{Airapetian:2001yk} and has the effect that $A^{\sin\phi}
= s/c$ for $d\sigma /d\phi = c + s\, P_\ell\, \sin\phi$.}
Note that the factor $1/P$ in the Bethe-Heitler and the interference
term cancels only approximately in (\ref{exp-SSA}) when numerator and
denominator are separately integrated over a range in $t$.  Which
observable presents the best compromise between theoretical and
experimental uncertainties can only be decided case by case.

We finally remark on the specific situation for experiments with muon
beams, which can e.g.\ be performed by the COMPASS experiment at the
CERN SPS \cite{d'Hose:2002ia}.  The natural polarization of muon beams
produced from pion decays changes with the beam charge, and one may
wonder what the separation power is for the combined reversal of beam
charge and polarization.  {}From (\ref{struct-compt}) and
(\ref{struct-int}) one readily finds that the corresponding difference
of cross sections (where the Bethe-Heitler contribution drops out)
contains the $\sin\phi$ term of $d\sigma_{\mathrm{VCS}}$ and the
$\cos(n\phi)$ terms of $d\sigma_{\mathrm{INT}}$.  The sum of cross
sections involves the $\cos(n\phi)$ terms in $d\sigma_{\mathrm{VCS}}$
and the $\sin(n\phi)$ terms in $d\sigma_{\mathrm{INT}}$.  This is a
consequence of parity invariance as discussed in
Section~\ref{sub:Compton-polar}.  Unless one of them strongly
dominates, the Compton and interference contributions can hence be
separated in such an experiment by the angular distribution, requiring
in fact only a separation between terms even and odd under $\phi\to
-\phi$.

\subsubsection{Access to GPDs}
\label{sub:Compton-GPDs}

The ultimate goal of studying Compton scattering in our context is the
extraction of amplitudes which carry information about GPDs.  Using
the polarization and $\phi$ dependence of the $ep$ cross section
discussed in the previous sections one can first test in a model
independent way whether the behavior in $Q^2$ is as predicted by the
power counting arguments underlying the factorization theorem, and
whether at given $Q^2$ power suppressed amplitudes like
$M_{\lambda'+,\lambda\, 0}$ are indeed smaller than the leading-twist
ones $M_{\lambda'+,\lambda+}$.  To be sure, the detailed dynamics
contributing to the $1/Q$ amplitudes $M_{\lambda'+,\lambda\, 0}$ and
to the $1/Q^2$ corrections in $M_{\lambda'+,\lambda+}$ is different,
and either of them may be untypically small or large.  With this
caveat in mind one can take the smallness of twist-three amplitudes as
a generic consistency check for the relevance of the $1/Q$ expansion.
If this check is passed one may proceed and analyze the relevant
Compton amplitudes in terms of GPDs.

The detailed analysis of Belitsky et al.~\cite{Belitsky:2001ns} shows
that one can in principle separate all Compton form factors
corresponding to the amplitudes $M_{\lambda'+,\lambda+}$ and
$M_{\lambda'+,\lambda\, 0}$ from the $\phi$ dependent terms in the
weighted interference term (\ref{inter-angular}) and its analogs for a
longitudinally and for a transversely polarized target.  One further
finds that the same reconstruction is in principle possible without
transverse target polarization if one has the full information from
the Compton cross section and from the interference term, including
the $\phi$ independent term $c_0$ in (\ref{inter-angular}) and its
analog for longitudinal target polarization.  We see however from
(\ref{gpd-compton-unpol}) to (\ref{gpd-inter-lpol}) that in the
combinations of twist-two amplitudes $M_{\lambda'+,\lambda+}$ relevant
for an unpolarized or longitudinally polarized target, $E$ always
comes with a kinematical suppression factor compared with other GPDs,
which should make its separation difficult.  To access $E$, which
carries crucial information about orbital angular momentum, transverse
target polarization hence seems inevitable.  To disentangle the
different double flip amplitudes $M_{\lambda'+,\lambda-}$ would
require both the Compton and interference terms for longitudinal and
transverse targets, given the absence of the highest harmonics in the
lepton helicity dependent part of the cross section (see
Section~\ref{sub:Compton-angular}).

Using the representation (\ref{compton-two-gpd}) of the twist-two
Compton amplitudes in terms of the Compton form factors of Belitsky et
al., which are convolutions of the corresponding GPDs with
hard-scattering kernels, one can readily see which combinations of
GPDs particular observables are sensitive to.  For an unpolarized
target the Compton cross section involves
\begin{eqnarray}
  \label{gpd-compton-unpol}
\frac{1}{2}\, \sum_{\lambda'\lambda} |M_{\lambda'+,\lambda+}|^2
 &=& (1-\xi^2) \Big( |\mathcal{H}|^2  + |\tilde\mathcal{H}|^2 \Big)
   - \left( \xi^2 + \frac{t}{4m^2} \right) |\mathcal{E}|^2 
   - \xi^2 \frac{t}{4m^2}\, |\tilde\mathcal{E}|^2 
\nonumber \\
 && {} - 2 \xi^2\, \re (\mathcal{H}^* \mathcal{E} 
                  + \tilde\mathcal{H}^* \tilde\mathcal{E} ) .
\end{eqnarray}
The $\cos\phi$ and $\sin\phi$ terms in the interference go with
\begin{equation}
  \label{gpd-inter-unpol}
\widehat{M}_{++} = \sqrt{1-\xi^2}\, \frac{\sqrt{t_0-t}}{2m}\, 
	\left[ F_1\, \mathcal{H} + 
               \xi (F_1 + F_2) \tilde{\mathcal{H}}
               - \frac{t}{4m^2}\, F_2\, \mathcal{E} \right] .
\end{equation}
Note that $\tilde\mathcal{E}$ is absent in this term.  A numerical
study by Korotkov and Nowak \cite{Korotkov:2002ym} found
(\ref{gpd-compton-unpol}) strongly dominated by $(1-\xi^2)
|\im\mathcal{H}|^2$ in typical HERMES kinematics, and
$\im\widehat{M}_{++}$ strongly dominated by the term with
$\im\mathcal{H}$.  Such observations are however dependent on the
specific GPD models and kinematics considered and should be taken with
due care: Belitsky et al.~\cite{Belitsky:2001ns} report for instance
that for certain models the $\mathcal{E}$ contribution to
$\widehat{M}_{++}$ can be significant in small-$x$ kinematics.  While
being a useful guide for a rough phenomenological analysis, the
neglect of particular GPDs in a particular observable should if at all
possible be cross checked with estimates of their size from different
observables, or if these are not available by scanning a suitable
range of ans\"atze for the GPDs.

The $\phi$ independent term in $P(\cos\phi)\, d\sigma_{\mathrm{INT}}
/d\phi$ has been calculated in \cite{Belitsky:2001ns}.  To leading
order in $1/Q$ its contribution to $e^{-6}\, \sum_{\mathrm{spins}}' (
\mathcal{T}_{\mathrm{BH}}^* \mathcal{T}_{\mathrm{VCS}}^{\phantom{*}} +
\mathcal{T}_{\mathrm{VCS}}^* \mathcal{T}_{\mathrm{BH}}^{\phantom{*}} )
$ in (\ref{VCS-BH}) can be written as
\begin{eqnarray}
  \label{inter-const}
&& {}- e_\ell\; \frac{1}{Q^2}\, \frac{8}{\xi}\, \frac{1}{P} \,
	\sqrt{\frac{1+\epsilon}{1-\epsilon}}\,
  \re \Bigg\{ \Bigg[ F_1\, \mathcal{H}
        - \xi^2 (F_1 + F_2) (\mathcal{H} + \mathcal{E}) 
	- \frac{t}{4m^2}\, F_2\, \mathcal{E} \Bigg]
\nonumber \\
&& \phantom{- e_\ell\;\;}
{}- \frac{1+\epsilon}{\epsilon}\, \frac{t-t_0}{t}\,
    (1-\xi) \Bigg[ F_1\, \mathcal{H} + 
               \xi (F_1 + F_2) \tilde{\mathcal{H}}
               - \frac{t}{4m^2}\, F_2\, \mathcal{E} \Bigg] \Bigg\} .
\end{eqnarray}
Numerical estimates in \cite{Belitsky:2001ns} show that this term need
not be small compared with the $\cos\phi$ term.  The second line of
(\ref{inter-const}) involves the same combination of GPDs as the
$\cos\phi$ term, whereas the first line, which survives the zero angle
limit $t=t_0$, involves a different combination.  The difference
between the two terms in square brackets is however proportional to
$\xi \tilde\mathcal{H} + \xi^2 (\mathcal{H} + \mathcal{E})$ and should
be rather small unless $\xi$ is large, as was confirmed by numerical
estimates in \cite{Belitsky:2001ns}.  At sufficiently small $\xi$ the
constant and the $\cos\phi$ term in the interference are hence
approximately sensitive to the \emph{same} linear combination of
Compton form factors or helicity amplitudes.

For longitudinal target polarization the relevant combinations in the
Compton cross section and interference term respectively read
\begin{eqnarray}
  \label{gpd-compton-lpol}
\lefteqn{
\frac{1}{2}\, \sum_{\lambda'} 
      \left( |M_{\lambda'+,++}|^2 - |M_{\lambda'+,-+}|^2 \right)
}
\nonumber \\
 && = 2 \re \Bigg[ (1-\xi^2) \mathcal{H}^* \tilde\mathcal{H}
     - \Big( \frac{\xi^2}{1+\xi} + \frac{t}{4m^2} \Big)\,
         \xi \mathcal{E}^* \tilde\mathcal{E}
     - \xi^2 \Big( \tilde\mathcal{H}^* \mathcal{E} 
                       + \mathcal{H}^* \tilde\mathcal{E} \Big) \Bigg]
\end{eqnarray}
and
\begin{eqnarray}
  \label{gpd-inter-lpol}
\lefteqn{
\frac{1}{2} \sum_{\lambda'}
  \left( g_{\lambda'+}\, M_{\lambda'+,++} 
       - g_{\lambda'-}\, M_{\lambda'+,-+} \right)
= \sqrt{1-\xi^2}\, \frac{\sqrt{t_0-t}}{2m}
}
\nonumber \\
&& {}\times \Bigg[ F_1\, \tilde\mathcal{H}
         + \xi (F_1 + F_2) {\mathcal{H}} 
         + \frac{\xi^2}{1+\xi}\, (F_1+F_2)\, \mathcal{E}
         - \Big( \frac{\xi^2}{1+\xi}\, F_1 
               + \xi \frac{t}{4m^2}\, F_2 \Big)\,
             \tilde\mathcal{E} \Bigg] .
\end{eqnarray}
Given the kinematical prefactors they should be most sensitive to
$\mathcal{H}$ and $\tilde\mathcal{H}$, so that together with the
unpolarized terms (\ref{gpd-compton-unpol}), (\ref{gpd-inter-unpol})
one should be able to separate these two contributions, with the
caveats spelled out below (\ref{gpd-inter-unpol}).  Note that the
Compton term (\ref{gpd-compton-lpol}) is only accessible in the double
asymmetry for beam and target polarization, in accordance with the
rules of Sect.~\ref{sub:Compton-polar}.  The imaginary part of
(\ref{gpd-inter-lpol}) appears in the single target spin asymmetry,
and its real part in the double spin asymmetry for beam and target.

Observables where $\mathcal{E}$ and $\tilde\mathcal{E}$ are not
kinematically suppressed compared with $\mathcal{H}$ and
$\tilde\mathcal{H}$ require \emph{transverse} target polarization,
which involves the combinations
\begin{eqnarray}
  \label{gpd-compton-tm}
\lefteqn{ \frac{1}{2i}\, \sum_{\lambda'} 
      \left( M^*_{\lambda'+,-+}\, M_{\lambda'+,++}^{\phantom{*}} -
	     M^*_{\lambda'+,++}\, M_{\lambda'+,-+}^{\phantom{*}} \right)
}
\nonumber \\
 &=& 2 \im \Bigg[ \sqrt{1-\xi^2}\, \frac{\sqrt{t_0-t}}{2m}\,
     \Big( \mathcal{E}^* \mathcal{H}
                - \xi \tilde\mathcal{E}^* \tilde\mathcal{H} \Big)
     \Bigg] ,
\\[0.3em]
  \label{gpd-compton-tp}
\lefteqn{
\frac{1}{2}\, \sum_{\lambda'} 
      \left( M^*_{\lambda'+,-+}\, M_{\lambda'+,++}^{\phantom{*}} +
	     M^*_{\lambda'+,++}\, M_{\lambda'+,-+}^{\phantom{*}} \right)
}
\nonumber \\
 &=& 2\re \Bigg[ \sqrt{1-\xi^2}\, \frac{\sqrt{t_0-t}}{2m}\,
    \Big( \mathcal{E}^* \tilde\mathcal{H}
       - \xi \tilde\mathcal{E}^* \mathcal{H} 
       - \frac{\xi^2}{1+\xi}\, \mathcal{E}^* \tilde\mathcal{E} \Big)
    \Bigg]
\end{eqnarray}
in the Compton cross section, with (\ref{gpd-compton-tm}) appearing
in the single target spin asymmetry and (\ref{gpd-compton-tp}) in the
double spin asymmetry for target and beam.  In the interference term
one has the combinations
\begin{eqnarray}
  \label{gpd-inter-tm}
\lefteqn{
\frac{1}{2} \sum_{\lambda'}
  \left( g_{\lambda'-}\, M_{\lambda'+,++}
       - g_{\lambda'+}\, M_{\lambda'+,-+} \right)
}
\nonumber \\
 &=& \left[ \xi^2 F_1 - \frac{t}{4m^2}\, (1-\xi^2) F_2 \right] 
	\mathcal{H}
   + \left[ \xi^2 F_1 + \frac{t}{4m^2}\, (F_1 + \xi^2 F_2) \right]
	\mathcal{E}
\nonumber \\
 && {} - \xi^2 (F_1+F_2)\, \tilde\mathcal{H} 
   - \frac{t}{4m^2}\, \xi^2 (F_1 + F_2)\, \tilde\mathcal{E} ,
\\[0.3em]
  \label{gpd-inter-tp}
\lefteqn{
\frac{1}{2} \sum_{\lambda'}
  \left( g_{\lambda'-}\, M_{\lambda'+,++}
       + g_{\lambda'+}\, M_{\lambda'+,-+} \right)
}
\nonumber \\
 &=& \Bigg[ \xi^2 F_1 - \frac{t}{4m^2}\, (1-\xi^2) F_2 \Bigg] 
	\tilde\mathcal{H}
   + \Bigg[ \frac{\xi^2}{1+\xi}\, F_1 
	+ \frac{t}{4m^2}\, (F_1 + \xi F_2) \Bigg]
	\xi\, \tilde\mathcal{E}
\nonumber \\
 && {} - \xi^2 (F_1+F_2)\, \mathcal{H} 
   - \left( \frac{\xi^2}{1+\xi} + \frac{t}{4m^2} \right)
     \xi (F_1 + F_2)\, \mathcal{E} ,
\end{eqnarray}
which can be separated by the $\phi$ dependence in the cross section.
Their imaginary parts are accessible in the single target spin
asymmetry, and their real parts in the double spin asymmetry for
target and beam.  Only in the combination (\ref{gpd-inter-tm}) is
$\mathcal{E}$ not kinematically suppressed compared with
$\mathcal{H}$.  In \cite{Belitsky:2000gz} it was observed that for
$t=t_0$, both (\ref{gpd-inter-tm}) and (\ref{gpd-inter-tp}) become
proportional to $\xi^2 (F_1+F_2)\, (\mathcal{H} - \tilde\mathcal{H})$
up to terms down by further factors of at least $\xi^2$.  This extra
suppression of $\mathcal{E}$ is however only relevant for very forward
angles and no longer effective as soon as $|t-t_0| \sim |t_0|$.

Various asymmetries have been studied using models for GPDs.  For the
small-$x$ kinematics of the H1 and ZEUS measurements, it was estimated
that the beam charge asymmetry may be in the 10\% range
\cite{Belitsky:2000gz,Belitsky:2001ns,Freund:2001hd}.  The beam spin
asymmetry was found sizeable provided that $y$ is not too small.  This
is because due to the polarization transfer from the lepton to the
virtual photon of DVCS, lepton polarization effects at small $y$ come
with a suppression factor of $y \sim \sqrt{1-\epsilon}$ relative to
unpolarized contributions, as can be seen in (\ref{VCS}) and
(\ref{VCS-BH}).  Estimates of similar size have been obtained in
\cite{Frankfurt:1998at,Freund:1999hy} using the simple model of
Frankfurt et al.\ for DVCS at small $x$ (see
Section~\ref{sec:small-x-dvcs}).

For the fixed-target kinematics of the HERMES or Jefferson Lab
experiments the studies
\cite{Belitsky:2000gz,Belitsky:2001ns,Freund:2001hd} found sizeable
effects possible for the beam charge asymmetry, as well as for the
single spin asymmetries for the lepton and for longitudinal or
transverse target polarization.  In contrast, the double polarization
asymmetry for longitudinal beam and target was estimated to be rather
small \cite{Belitsky:2000gz,Freund:2001hd}. 

For some observables, different models for the GPDs give distinctly
different results.  Kivel et al.~\cite{Kivel:2000fg} observed that the
beam charge asymmetry in HERMES kinematics can change sign when adding
the model $D$-term mentioned in Section~\ref{sub:model-d-term} to a
double distribution based ansatz.  In contrast, the study of Belitsky
et al.~\cite{Belitsky:2001ns} found that when taking the same $D$-term
the asymmetry becomes close to zero.  This was traced back to
important differences in the double distribution part of both studies.
Note that although the forward quark densities (at not too small $x$)
are rather well constrained by experiment, this is not the case for
the nonforward behavior of GPDs, in particular for their $t$
dependence.  Given these uncertainties it seems premature to claim
evidence e.g.\ of a sizeable $D$-term from a single observable.

The importance of different contributions to the cross section changes
significantly when going from a proton to a neutron target.  The
electromagnetic Dirac form factor of the neutron is small at low $t$,
but its Pauli form factor is large.  This suppresses the $1/\xi^2$
enhanced term in the Bethe-Heitler cross section (\ref{BH}) so that
its contribution is less important compared with the proton case.  We
see in (\ref{gpd-inter-unpol}) that there is also a suppression for
the $\cos\phi$ and $\sin\phi$ part of the interference term with an
unpolarized target, where the normally dominant contribution from
$\mathcal{H}$ is accompanied by the Dirac form factor.  Furthermore
$\tilde\mathcal{H}$ for the neutron may be small due to a cancellation
between $u$ and $d$ quark contributions, assuming that their relative
sign and size is as for the polarized forward
densities~\cite{Berger:2001xd}.

\subsubsection{First data}
\label{sub:dvcs-data}

\begin{figure}
\begin{center}
	\leavevmode
	\epsfxsize=0.43\textwidth
	\epsfbox{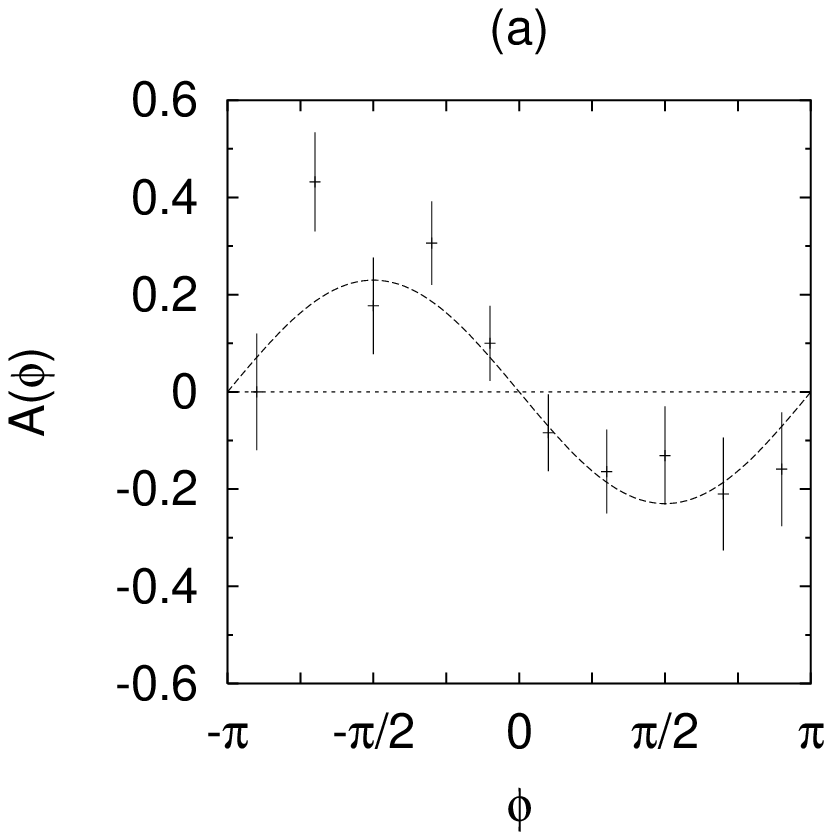}
\hspace{0.04\textwidth}
	\epsfxsize=0.43\textwidth
	\epsfbox{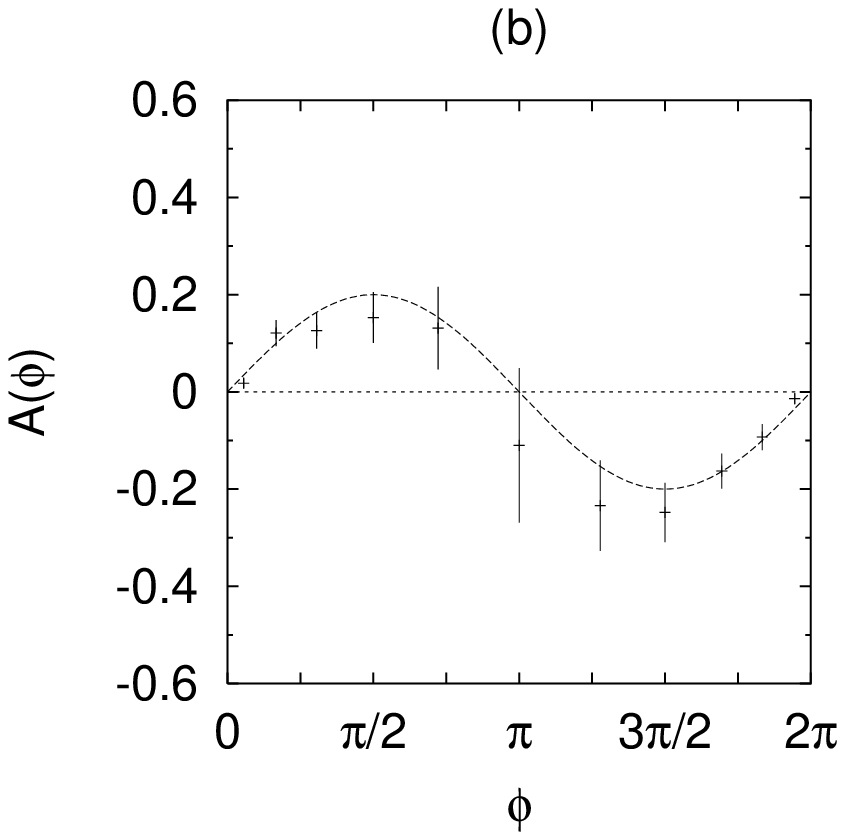}
\end{center}
\caption{\label{fig:DVCS-SSA} Data for the beam spin asymmetry
$A(\phi)$ in DVCS.  In both cases the error bars show the statistical
errors, which dominate over the systematic ones.  (a) The HERMES
measurement \protect\cite{Airapetian:2001yk} for $e^+ p\to e^+ \gamma
X$ with an experimentally reconstructed missing mass of $M_X^2$
between $- (1.5~\mbox{GeV})^2$ and $(1.7~\mbox{GeV})^2$.  The curve
corresponds to $A(\phi) = -0.23 \sin\phi$. (b) The CLAS data for $e^-
p\to e^- \gamma p$ \protect\cite{Stepanyan:2001sm} together with the
curve $A(\phi) = 0.2 \sin\phi$.}
\end{figure}

Measurements of DVCS at fixed target energies have been reported by
HERMES \cite{Airapetian:2001yk} and CLAS \cite{Stepanyan:2001sm}.
Both measure the single beam spin asymmetry $A(\phi)$ defined in
(\ref{exp-SSA}).  The results are shown in Fig.~\ref{fig:DVCS-SSA}.
With respective lepton beam energies of $E=27.6$~GeV and $E=4.25$~GeV
the Bethe-Heitler process is prominent in the cross section, except
for larger values of $\xB$ at HERMES.  The average kinematical values
of the HERMES measurement are $Q^2= 2.6$~GeV$^2$, $\xB= 0.11$, $-t=
0.27$~GeV$^2$, and $y= 0.46$, whereas the CLAS Collaboration compared
their data with theory predictions for fixed $Q^2= 1.25$~GeV$^2$,
$\xB= 0.19$, $-t= 0.19$~GeV$^2$, and $y= 0.82$.  We note that HERMES
has detected the final $e^+$ and $\gamma$ but not the scattered
proton, whereas CLAS measured the $e^-$ and $p$ but not the
final-state $\gamma$.  Important experimental backgrounds are
therefore $ep\to ep\, \pi^0$ for CLAS and proton dissociation for
HERMES.

Both experiments find a sizeable single spin asymmetry $A(\phi)$
compatible with a $\sin\phi$ behavior; a fit by CLAS to $A(\phi) =
\alpha\, \sin\phi + \beta\, \sin 2\phi$ found $\beta$ compatible with
zero within errors.  With the caveats concerning the Bethe-Heitler
propagators discussed in Sections~\ref{sub:Compton-angular} and
\ref{sub:cross-asy} this provides evidence that the leading-twist
combination $\im \widehat{M}_{++}$ is sizeable whereas $\im
\widehat{M}_{+0}$ is small.  To be sure, by itself this is not a
\emph{proof} that one is in the region where the large $Q^2$ analysis
applies.  In particular, $\widehat{M}_{+0}$ also has to vanish in the
photoproduction limit $Q^2\to 0$.  More precise data will hopefully
allow one to establish the size of the longitudinal photon amplitudes
and their behavior with $Q^2$.

The sign of the measured $A(\phi)$ translates into $\im
\widehat{M}_{++} > 0$.  According to (\ref{gpd-inter-unpol}) this is
expected by a twist-two analysis if one assumes that the observable is
dominated by $H^q(\xi,\xi,t)$ and that these GPDs have the same
positive sign as their forward limits.  The current GPD models
discussed in Section~\ref{sec:ansatz} provide values of $A(\phi)$ with
the correct size.  This holds for the leading-twist LO analysis of
\cite{Korotkov:2001zn}, for the leading-twist NLO analysis of
\cite{Freund:2001hd}, and for the LO analyses including twist-three
effects in the Wandzura-Wilczek approximation of
\cite{Belitsky:2001ns} and of \cite{Kivel:2000fg} (the latter being
shown in the HERMES publication).  In all cases the theory predictions
are however somewhat above the data.  Given the existing uncertainties
concerning power corrections and the rather large number of assumptions
and parameters in GPD parameterizations it is not simple to identify
the ``culprit'' for this situation.  Freund and McDermott
\cite{Freund:2001hd} argue that the enhancement factor of nonforward
to forward distributions (see Section~\ref{sec:small-x-gpd}) is not
large according to the data and further conclude that modeling based
on double distributions with a profile as in (\ref{profile-def}) to
(\ref{profile}) is problematic.

The DVCS cross section at small $\xB$ has been measured by the H1 and
ZEUS Collaborations \cite{Adloff:2001cn,Zeus:2003ip}, subtracting the
Bethe-Heitler contribution and assuming that after integration over
$\phi$ the interference term can be neglected (see
Section~\ref{sub:cross-asy}).  The kinematical region was $Q^2=2$ to
20~GeV$^2$ and $W=30$ to 120~GeV for H1, and $Q^2=5$ to 100~GeV$^2$
and $W=40$ to 140~GeV for ZEUS.  As in meson electroproduction or
photoproduction of $J/\Psi$, a rather steep increase of the cross
section with energy was observed.  With the GPD models just mentioned,
the study \cite{Freund:2001hd} found a cross section too large in this
kinematics.  The analysis in \cite{Belitsky:2001ns} could achieve
agreement with the H1 data by taking a large profile parameter $b$
(see Section~\ref{sub:connect-forward}) and a very steep falloff in
$t$ for the sea quarks.\footnote{Notice that the $t$-slope parameter
$B_{\mathrm{sea}}$ used in \protect\cite{Belitsky:2001ns} refers to
the amplitude and not to the cross section level, unlike the common
parameterization $d\sigma /dt \propto e^{B t}$.}  As in the case of
the fixed-target data it may be too early to draw definite conclusion,
but it is clear that theoretical predictions are sufficiently
different for some of them to be ruled out by data.  Both for the
fixed-target and the collider experiments, measurement of the $t$
dependence would be of great help to reduce the freedom and
uncertainty in modeling the input GPDs.

\subsubsection{Targets with spin zero or one}
\label{sub:spin-0-1}

Many theoretical investigations have focused on DVCS on a spinless
target, mainly because of its relative simplicity compared with the
spin $\half$ case.  The case of a pion target has often been studied
as a concrete example, where furthermore rather detailed dynamical
predictions are available (see Section~\ref{sec:dynamics}).  One may
think of studying DVCS on a virtual pion target in the process $ep
\to e \gamma\, \pi^+n$ at small invariant momentum transfer from the
proton to the neutron.\footnote{I thank M. Amarian for discussions on
this issue.}
To obtain the pion GPDs $H_\pi^q(x,\xi,t)$ would then require a
careful extrapolation to the pion pole.  The analogy with elastic form
factors suggests that the situation in this respect should resemble
the one for extracting $F_\pi(t)$ at small $t$ from $ep\to e\, \pi^+
n$.

Another context where targets with different spin $J$ appear is
scattering on nuclei.  Preliminary data for the beam spin asymmetry
$A(\phi)$ on neon ($J=0$) and on the deuteron ($J=1$) have for
instance been reported by HERMES \cite{Ellinghaus:2002zw}.  Model
estimates for the cross section and various asymmetries have been
given in \cite{Cano:2003ju,Kirchner:2003wt} for a deuteron target and
in \cite{Guzey:2003jh,Kirchner:2003wt} for neon.  The cross section
dependence on the azimuthal angle $\phi$ and on the lepton beam
polarization follows the principles spelled out in
Sections~\ref{sub:Compton-angular} and \ref{sub:Compton-polar} for an
unpolarized target of \emph{any} spin.  Analogous observables hence
project out the real or imaginary parts of Compton amplitudes in an
analogous way.  In contrast, the structure of observables with target
polarization is different.  A general analysis in the framework of
Belitsky et al.~\cite{Belitsky:2001ns} has been presented in
\cite{Kirchner:2003wt} for a spin-one target.  The case of a spin zero
target was studied along the same lines to twist-three accuracy in
\cite{Belitsky:2000vk}.

Note that for DVCS on a spinless target there are only three
independent helicity amplitudes, $M_{++}$, $M_{+0}$, $M_{+-}$, where
the first index refers to the outgoing and the second to the incoming
photon.  This makes the analysis rather simple.  The beam spin
independent part of the interference term depends on $\re M_{++}$,
$\re M_{+0}$, $\re M_{+-}$ and offers four observable quantities,
given as Fourier coefficients of $P(\cos\phi)\,
d\sigma_{\mathrm{INT}}/ d\phi$, whereas the beam spin dependent part
depends on $\im M_{++}$, $\im M_{+0}$ and contains two independent
Fourier coefficients.  One can hence extract all independent dynamical
quantities from the $\phi$ distribution of the interference, using
either its $1/Q$ expanded or its exact expression in terms of the
Compton amplitudes.  A spin zero target hence presents a clean
environment for testing the large $Q^2$ expansion underlying a GPD
based analysis.


\subsection{Timelike Compton scattering}
\label{sec:tcs}

\begin{figure}[b]
\begin{center}
	\leavevmode
	\epsfxsize=0.62\textwidth
	\epsfbox{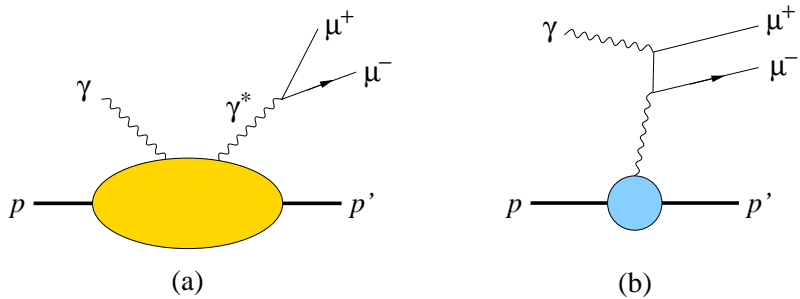}
\end{center}
\caption{\label{fig:tcs-bh} (a) Compton scattering and (b)
Bethe-Heitler contribution to $\gamma p\to \mu^+\mu^-\, p$.  A second
Bethe-Heitler diagram is obtained by reversing the charge flow of the
muon line.}
\end{figure}

The phenomenology of TCS is very similar to the one of DVCS and has
been presented in \cite{Berger:2001xd}, closely following the DVCS
treatment we have just discussed.  The Compton subprocess in
\begin{equation}
  \label{TCS-def}
\gamma(q) + p(p) \to \ell^-(k) + \ell^+(k') + p(p')
\end{equation}
interferes with a Bethe-Heitler process as shown in
Fig.~\ref{fig:tcs-bh}.  As we have seen in
Section~\ref{sec:compton-scatt} it is natural to compare TCS and DVCS
at the same $\xi$ and $t$ and at corresponding values of $Q'^2$ and
$Q^2$.  According to the kinematical relation (\ref{xi-eta-prop}) the
analog of $\xB \approx 2\xi /(1+\xi) \approx Q^2 /(Q^2 + W^2)$ in DVCS
is thus $2\xi /(1+\xi) \approx Q'^2 /W^2$ in the timelike process.
The full kinematics of lepton pair production is suitably described by
two further variables $\theta$ and $\phi$, the polar angles of
$\ell^-$ in the c.m.\ of the lepton pair, with the $z$-axis pointing
opposite to the scattered proton.  The azimuth $\phi$ plays the same
role as its analog in DVCS and is the key to obtain information on
Compton scattering from its interference with the Bethe-Heitler
process.  The analogous roles played in the cross section formulae by
$\theta$ in TCS and the inelasticity $y$ in DVCS are elucidated by the
correspondence
\begin{equation}
\frac{1+\cos\theta}{2} \approx \frac{k\cdot p'}{(k+k')\cdot p'}
	\:\:\leftrightarrow\:\:
\frac{k\cdot p}{(k-k')\cdot p} = \frac{1}{y} ,
\end{equation}
where in the first relation we have neglected the masses of the lepton
and the target.  This relation reveals a crucial difference in the
phenomenology of the two processes: whereas at small $y$ a factor
$1/y^2 \sim 1/(1-\epsilon)$ enhances the DVCS cross section over
Bethe-Heitler (see Section~\ref{sec:dvcs-pheno}) the corresponding
factor in TCS is bounded between $-1$ and 1.  Given the relative
factors $1/t$ and $1/Q'^2$ from the photon propagators, the
Bethe-Heitler process hence tends to dominate over TCS in deeply
virtual kinematics, so that information about the Compton amplitudes
must be gained through the interference term.

To filter out the interference term one can use that the
$\ell^+\ell^-$ pair is produced in a $C$ odd state by Compton
scattering and in a $C$ even state by the Bethe-Heitler process.  Any
observable that changes sign under the exchange of $\ell^-$ and
$\ell^+$ momenta hence projects out the interference.  This is the
analog of the beam charge asymmetry in DVCS, and it is readily
available from the angular distribution of the leptons instead of
requiring beams of opposite charge.  The structure of the interference
term is analogous to (\ref{VCS-BH}) in DVCS: the linear combinations
$\re \widehat{M}_{--}$, $\re \widehat{M}_{0-}$, $\re \widehat{M}_{+-}$
of Compton amplitudes appear multiplied with $\cos\phi$, $\cos 2\phi$,
$\cos 3\phi$ and with a known lepton propagator factor $1/
P(\cos\phi)$.  A suitable observable to extract the leading-twist
amplitudes $\re \widehat{M}_{--}$ is for instance the cross section
weighted by $\cos\phi$ or by $\cos\phi\, P(\cos\phi)$.  Note that this
requires one to reconstruct both the hadronic and leptonic scattering
planes in order to measure $\phi$.  In principle one may also access
the amplitudes $\re M_{\lambda'-,\lambda-}$ from $\phi$ integrated
observables.  An example is the cross section weighted by
$\cos\theta$.  This comes however with an additional kinematic
suppression factor $1/Q'$ and was estimated to be numerically very
small in kinematics of fixed-target experiments \cite{Berger:2001xd}.

The imaginary parts of the Compton amplitudes can be obtained with a
polarized photon beam.  With circular photon polarization (obtained
for instance by brems\-strah\-lung from a longitudinally polarized
lepton beam) one finds $\im \widehat{M}_{--}$, $\im \widehat{M}_{0-}$,
$\im \widehat{M}_{+-}$ multiplied with $\sin\phi$, $\sin 2\phi$, $\sin
3\phi$ times $1/ P(\cos\phi)$ in the interference term.  Note that
unlike in the \emph{lepton} spin asymmetry of DVCS, the contribution
with $\sin 3\phi$ is not absent here and provides access to the double
helicity flip amplitudes of the photon.

No experimental results on TCS have been reported so far.  A numerical
estimate of the $\cos\phi$ asymmetry in \cite{Berger:2001xd} found
rather small effects for fixed-target kinematics and $Q'^2$ around
5~GeV$^2$, but the estimate strongly depended on the model for GPDs in
the ERBL region and should be taken with care.  This is typical of
observables involving the real parts of Compton amplitudes, which
crucially depend on GPDs in the region where least is known about
them.

The yet unexplored phenomenology of DDVCS, to be measured in $ep\to
ep\, \ell^+\ell^-$, will combine the elements of DVCS and of TCS, with
Bethe-Heitler processes of both types in Figs.~\ref{fig:vcs-bh} and
\ref{fig:tcs-bh} contributing to the cross section.  In the limit
where one of the two photon virtualities becomes small one expects to
find the features of either DVCS or TCS in the pattern of observable
quantities.


\subsection{Two-photon annihilation}
\label{sec:gaga}

\begin{figure}
\begin{center}
	\leavevmode
	\epsfxsize=0.68\textwidth
	\epsfbox{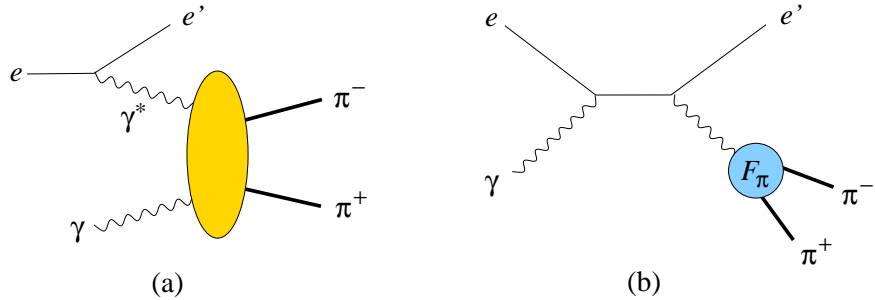}
\end{center}
\caption{\label{fig:brems} (a) Two-photon annihilation and (b)
bremsstrahlung contribution to $e\gamma \to e\, \pi^+\pi^-$.  A second
bremsstrahlung diagram is obtained by interchanging the photon
vertices.}
\end{figure}

The analysis strategy we have presented for Compton scattering carries
over to the crossed-channel process of exclusive $\gamma^* \gamma$
annihilation and has been developed in detail in \cite{Diehl:2000uv}
for the case of $e \gamma \to e\, \pi\pi$.  The analog of the
Bethe-Heitler process is now bremsstrahlung of a timelike $\gamma^*$
which decays into the final state hadrons, as shown in
Fig.~\ref{fig:brems}.  It can be calculated given the timelike pion
form factor $F_\pi(s)$ at low to moderate $s$, whose square is
measured in $e^+e^-\to \pi^+\pi^-$ and whose phase can be taken from
the phase shift analysis of elastic $\pi\pi$ scattering using Watson's
theorem (see Section~\ref{sub:gda-evolution}).  Modeling the two-pion
DA along the lines of Section~\ref{sub:gda-models} one finds that the
bremsstrahlung process strongly enhances and dominates the $e \gamma
\to e\, \pi^+\pi^-$ cross section in a broad region of invariant
two-pion masses $m_{\pi\pi}$ around the $\rho(770)$ mass.  Angular
analysis in the azimuth $\phi$ between the leptonic and hadronic
planes of the process can then be used to filter out the interference
with two-photon annihilation, which depends on the helicity amplitudes
$M_{\mu'\mu}$ for $\gamma^*\gamma \to \pi\pi$.  This is achieved by
using the charge conjugation properties of the hadronic system, which
is produced $C$-even by $\gamma^*\gamma$ annihilation and $C$-odd by
bremsstrahlung.  One of the simplest observables for this purpose is
the cross section weighted with $\cos\phi$.  More refined quantities
to separate the three independent quantities $\re (F_\pi^* M_{++})$,
$\re (F_\pi^* M_{0+})$ and $\re (F_\pi^* M_{-+})$ appearing in the
interference term are discussed in \cite{Diehl:2000uv}, as well as the
possibilities to simultaneously perform a partial-wave analysis of the
two-pion system.

In the approximation of leading twist and leading $O(\alpha_s)$ the
phase of $M_{++}$ is equal to the dynamical phase of the isosinglet
GDA $\Phi^{u+d}$, so that the interference with bremsstrahlung gives
rather clean access to this quantity.  Consider $m_{\pi\pi}$ in a
region where the two pions are predominantly in the $S$ or $P$ wave
and where Watson's theorem can be applied (see
Section~\ref{sub:gda-evolution}).  Then $\re (F_\pi^* M_{++})$ is
proportional to $\cos(\delta_0 - \delta_1)$, where
$\delta_l(m_{\pi\pi})$ is the phase shift of the $l$th partial wave.
Significant structure can thus be expected in the $m_{\pi\pi}$
dependence of interference observables as shown in
\cite{Diehl:2000uv}.  Studies at $m_{\pi\pi} \sim 1$~GeV may in
particular shed light on the nature of the $f_0(980)$ resonance, in a
context where its structure is probed at \emph{short distance}.

$e \gamma$ scattering at large $Q^2$ can be studied at $e^+ e^-$
colliders in events with one lepton scattered at large and the other
at small invariant momentum transfer.  The estimates in
\cite{Diehl:2000uv} find that the event rate at high-luminosity
experiments like BABAR and BELLE should be sufficient to study the
interference signal in $e^+e^-\to e^+e^-\, \pi^+\pi^-$, as well as the
cross section of $e^+e^-\to e^+e^-\, \pi^0\pi^0$, where bremsstrahlung
does not contribute since the $\pi^0\pi^0$ system is charge
conjugation even.

The principles of phenomenological analysis spelled out in
\cite{Diehl:2000uv} generalize to channels with different produced
hadrons like $\bar{p} p$ or $\rho\rho$, with the additional
complication of having more unknown amplitudes (as in Compton
scattering on targets with nonzero spin).  Anikin et
al.~\cite{Anikin:2003fr} have recently studied $e^+e^- \to e^+e^-
\rho^0\rho^0$ at large $Q^2$, where bremsstrahlung is absent for the
same reason as in $\pi^0\pi^0$ production.  First data on this process
has been presented by the L3 Collaboration \cite{Achard:2003qa} and is
in fair agreement with the $Q^2$ behavior predicted by the
leading-twist mechanism.


\subsection{Meson production}
\label{sec:meson-pheno}

In this section we present basics of analyzing meson electroproduction
in the framework of GPDs, and point out some salient features of
relevant data at fixed-target energies.  For a discussion of vector
meson production at small $x$ see Section~\ref{sec:small-x-mesons}.  A
number of numerical estimates of production cross sections in the GPD
framework have been made in the literature.  Mankiewicz et al.\ have
considered $\rho^0$ and $\pi^0$ production in
\cite{Mankiewicz:1998uy}, charged $\rho$ production in
\cite{Mankiewicz:1997aa}, and charged pion production and its relation
to the pion form factor in \cite{Mankiewicz:1998kg}.  Cross sections
for $\rho^0$ and $\pi^0$ in the leading-twist framework were given by
Vanderhaeghen et al.~\cite{Vanderhaeghen:1998uc}, whereas the study
\cite{Vanderhaeghen:1999xj} also estimated the size of power
corrections (see Section~\ref{sub:form-factors}).  Frankfurt et al.\
have studied channels where the initial proton is scattered into a
$\Delta$ or a strange baryon \cite{Frankfurt:1999xe}, and the
production ratio of $\eta$ and $\pi^0$ production was considered by
Eides et al.~\cite{Eides:1998ch}.  Several of these predictions are
collected in the review by Goeke et al.~\cite{Goeke:2001tz}.

With Hand's convention \cite{Hand:1963bb} for the virtual photon flux
one has the electroproduction cross section for a meson $M$ on an
unpolarized target given by
\begin{equation}
\frac{d\sigma(ep\to ep M)}{dt\, dQ^2\, d\xB} =
  \frac{\alpha_{\mathrm{em}}}{2\pi}\, \frac{y^2}{1-\epsilon}\, 
  \frac{1-\xB}{\xB}\, \frac{1}{Q^2}\, 
                  \left[ \frac{d\sigma_T}{dt} + 
                         \epsilon \frac{d\sigma_L}{dt} \right]
\end{equation}
with $\gamma^* p$ cross sections
\begin{equation}
\frac{d\sigma_{i}}{dt} = \frac{1}{16\pi}\, 
	\frac{\xB^2}{1-\xB}\, \frac{1}{Q^4}\,
        \frac{1}{\sqrt{1 + 4 \xB^2 m^2 /Q^2}}\, 
 {\sum_{\mathrm{spins}}}' \, 
      \Big| \mathcal{A}(\gamma^*_{i} p\to M p) \Big|^2 ,
\qquad \qquad i=T,L
\end{equation}
for transverse and longitudinal photon polarization.  $y$ and
$\epsilon$ are defined in (\ref{y-eps-def}).  We have not explicitly
labeled the dependence of the scattering amplitude $\mathcal{A}$ on
the hadron polarizations, which are summed or averaged over in
$\sum_{\mathrm{spins}}'$.

The leading-twist expression for $d \sigma_L /dt$ is readily obtained
from the amplitudes given in Section~\ref{sec:meson-lt}.  The relevant
combinations of GPDs are obtained as
\begin{eqnarray}
  \label{unpol-meson-nat}
\frac{1}{2} \sum_{\lambda'\lambda} 	
	|\mathcal{F}_{\lambda'\lambda}|^2
  &=& (1-\xi^2)\, |\mathcal{H}|^2
   - \left( \xi^2 + \frac{t}{4m^2} \right) |\mathcal{E}|^2 
   - 2 \xi^2\, \re (\mathcal{E}^* \mathcal{H} )
\end{eqnarray}
for mesons with natural parity, and 
\begin{eqnarray}
  \label{unpol-meson-unnat}
\frac{1}{2} \sum_{\lambda'\lambda} 
	|\tilde\mathcal{F}_{\lambda'\lambda}|^2
  &=& (1-\xi^2)\, |\tilde\mathcal{H}|^2
   - \xi^2 \frac{t}{4m^2}\, |\tilde\mathcal{E}|^2 
   - 2\xi^2\, \re (\tilde\mathcal{E}^* \tilde\mathcal{H} ) .
\end{eqnarray}
for mesons with unnatural parity.  Here $\mathcal{F}$ and
$\tilde\mathcal{F}$ are appropriate convolutions of the proton matrix
elements $F$ and $\tilde{F}$, summed over quark flavors and gluons as
specified by the scattering amplitudes (\ref{meson-1}) to
(\ref{meson-3}).  $\mathcal{H}$, $\mathcal{E}$, $\tilde\mathcal{H}$,
$\tilde\mathcal{E}$ are defined in analogy from the corresponding
quark or gluon GPDs.  Note that the flavor structure and the
hard-scattering kernels in the convolution differ from those in
Compton scattering, where we have used the same notation.  Since they
only reflect the decomposition of $F_{\lambda'\lambda}$ and
$\tilde{F}_{\lambda'\lambda}$ in terms of proton spinors, the
structures (\ref{unpol-meson-nat}) and (\ref{unpol-meson-unnat})
readily generalize to higher order corrections in $\alpha_s$.

In addition to the cross section for a longitudinal photon and
unpolarized target there is one more observable which involves only
leading-twist amplitudes and can be expressed using the factorization
theorem, namely the single spin asymmetry for a transversely polarized
target (again with a longitudinal photon).  This was first discussed
for pseudoscalar production by Frankfurt et
al.~\cite{Frankfurt:1999fp}, and the analog for vector meson
production was given in \cite{Goeke:2001tz}.  The cross section
difference for target polarization along the positive and negative $y$
axis (with the coordinate system defined as for Compton scattering in
Fig.~\ref{fig:dvcs-kin}) involves the GPD combinations
\begin{equation}
  \label{tr-meson-nat}
\frac{1}{2i}\, \sum_{\lambda'} 
      \left( \mathcal{F}^*_{\lambda'-} 
             \mathcal{F}_{\lambda'+}^{\phantom{*}} -
	     \mathcal{F}^*_{\lambda'+} 
             \mathcal{F}_{\lambda'-}^{\phantom{*}} \right)
 = \sqrt{1-\xi^2}\, \frac{\sqrt{t_0-t}}{m}\,
     \im ( \mathcal{E}^* \mathcal{H} )
\end{equation}
for mesons with natural parity, and
\begin{equation}
  \label{tr-meson-unnat}
\frac{1}{2i}\, \sum_{\lambda'} 
\left( \tilde\mathcal{F}^*_{\lambda'-} 
       \tilde\mathcal{F}_{\lambda'+}^{\phantom{*}} -
       \tilde\mathcal{F}^*_{\lambda'+} 
       \tilde\mathcal{F}_{\lambda'-}^{\phantom{*}} \right)
 = {}- \sqrt{1-\xi^2}\, \frac{\sqrt{t_0-t}}{m}\,
     \xi\, \im ( \tilde\mathcal{E}^* \tilde\mathcal{H} )
\end{equation}
for mesons with unnatural parity.  The corresponding normalized
polarization asymmetry goes with the ratio of (\ref{tr-meson-nat}) and
(\ref{unpol-meson-nat}) or of (\ref{tr-meson-unnat}) and
(\ref{unpol-meson-unnat}), and thus depends on $\mathcal{E}
/\mathcal{H}$ or on $\tilde\mathcal{E} /\tilde\mathcal{H}$.
Restricting ourselves to $O(\alpha_s)$ accuracy we find from the
discussion of Section~\ref{sec:meson-lt} that for suitable mesons like
$\pi$ or $\rho$ the amplitude depends on the meson structure only
through a global factor, which cancels in these normalized
asymmetries.  As emphasized in \cite{Frankfurt:1999fp} the difference
of polarized cross sections for pseudoscalar production gives access
to the product of the GPDs $\tilde{H}$ and $\tilde{E}$, whereas in
large kinematical domains either $\tilde{H}$ or $\tilde{E}$ is
estimated to dominate the unpolarized cross section.  In analogy, the
transverse asymmetry in vector meson production is sensitive to $E$
even in kinematics where $H$ dominates the cross section
(\ref{unpol-meson-unnat}).  Numerical estimates
\cite{Frankfurt:1999fp,Goeke:2001tz} found that both for pseudoscalar
and for vector meson production the transverse asymmetry may be
sizeable.

\subsubsection{Leading twist and beyond}
\label{sub:mesons-twist}

The above predictions refer to the $\gamma^* p$ cross section with
longitudinal polarization of the photon and of the meson if
applicable.  In the case where the final meson has spin zero or where
its polarization is unobserved, one needs to extract the longitudinal
cross section from $\sigma_T + \epsilon \sigma_L$ by a Rosenbluth
separation: measuring at different $\epsilon$ but equal values of
$Q^2$, $\xB$, $t$ (on which $\sigma_{T,L}$ depend) requires different
$ep$ collision energies.  There are however indirect ways of
constraining the size of power suppressed amplitudes.  For a target
with no polarization or longitudinal polarization in the $\gamma^*p$
c.m.\ a $\phi$ dependence of $d\sigma(ep\to ep M) /(d\phi\, dt\,
dQ^2\, d\xB)$ is due to the interference of different $\gamma^*$
polarizations.  $\cos\phi$ or $\sin\phi$ terms go with the product of
$\gamma^*_L$ and $\gamma^*_T$ amplitudes and are predicted to be power
suppressed as $1/Q$ relative to $\sigma_L$.  A $\cos 2\phi$ or $\sin
2\phi$ dependence comes from the interference of amplitudes with
positive and negative photon helicity, resulting in the same $1/Q^2$
suppression as for $\sigma_T$.  HERMES has observed a nonzero
$\sin\phi$ dependence of the single spin asymmetry for $ep\to e\,
\pi^+ n$ on a longitudinally polarized target
\cite{Airapetian:2001iy}.  Since the longitudinal polarization is with
respect to the beam direction it induces a small transverse
polarization in the hadronic process $\gamma^* p\to M p$ (see
Section~\ref{sub:Compton-polar}), but this effect alone cannot account
for the asymmetry observed.  A quantitative analysis will require data
with transverse target polarization so that both effects can be
separated, but the above observation already implies that the
$\gamma^*_T$ amplitude cannot be negligible in the kinematics of the
measurement, with average values of $Q^2 = 2.2$~GeV$^2$, $\xB = 0.15$,
$-t = 0.46$~GeV$^2$.

For the production of $\rho$ mesons, the decay angular distribution of
$\rho\to \pi^+\pi^-$ contains information on the $\rho$ helicity and
in combination with the $\phi$ dependence can be used for a detailed
polarization analysis, worked out in the classical work of Schilling
and Wolf \cite{Schilling:1973ag}.  Data for $\rho$ production in the
large-$Q^2$ region has in particular been taken by HERMES
\cite{Ackerstaff:2000bz} for $W=3.8$~GeV to 6.5~GeV and $Q^2 \le
4$~GeV$^2$ and by E665 at Fermilab \cite{Adams:1997bh} for $W$ from
about 10~GeV to 25~GeV and $Q^2 \le 20$~GeV$^2$.  In both experiments
angular analysis found approximate helicity conservation between the
photon and meson in the $\gamma^* p$ c.m.  This can then be used to
translate a measurement of the cross sections for transverse and
longitudinal $\rho$ mesons (separated via the decay angular
distribution) into a measurement of $\sigma_T$ and $\sigma_L$ for
transverse and longitudinal photons, without Rosenbluth separation.
The ratio $R = \sigma_L /\sigma_T$ reaches a value of 1 at $Q^2$
around 2 to 3~GeV$^2$, providing a clear indication of nonnegligible
power suppressed amplitudes.  HERMES has also observed a nonzero
double spin asymmetry for longitudinal beam and target polarization
\cite{Airapetian:2001hq,Airapetian:2003yv}.  The effect cannot be
explained alone by the induced transverse target polarization with
respect to the $\gamma^* p$ axis, and is evidence for an asymmetry
between the $\gamma^*_T$ cross sections with the target spin in the
same or in the opposite direction as the spin of the photon.  As to
the longitudinal cross section itself, the HERMES data
\cite{Airapetian:2000ni} for $\gamma^*_L\, p\to \rho_L^{\phantom{*}}\,
p$ with $W=4$~GeV to 6~GeV and $Q^2 \le 5$~GeV$^2$ and the data from
E665 \cite{Adams:1997bh} are reasonably well reproduced by the
calculation of \cite{Vanderhaeghen:1999xj}, which contains a
substantial amount of power suppression from parton $k_T$ in this
kinematics.

\subsubsection{Meson pair production}
\label{sub:meson-pair-pheno}

As we have discussed in Section~\ref{sub:meson-pairs-electro},
measurement of $ep\to ep\, (\pi^+\pi^-)$ with the pion pair off the
$\rho$ mass peak offers a possibility to access the two-pion system
both in the $C$-even and in the $C$-odd state.  At large enough $Q^2$
the process contains information about the two-pion distribution
amplitude.  The angular distribution of the pion pair in its rest
frame carries again detailed information about the angular momentum
state of the two-pion system, as it does on the $\rho$ peak.  The
general framework of angular analysis including $J=0,1,2$ partial
waves has been given by Sekulin \cite{Sekulin:1973mk}, and results
specific to leading twist dynamics were presented by Lehmann-Dronke et
al.\ \cite{Lehmann-Dronke:2000xq}.  Access to the $C$-even component,
i.e., to $S$ and $D$ waves can in particular be obtained by
interference with the $C$-odd component in the tail of the $\rho$
peak.  Any angular observable reversing sign under exchange of the
pion momenta projects out the interference of the production
amplitudes for $C$-even and $C$-odd two-pion states, as we discussed
for $e \gamma\to e\, \pi^+\pi^-$ in Section~\ref{sec:gaga}.  This
interference is very sensitive to the dynamical phases of the GDAs,
with phase differences $\delta_J(m_{\pi\pi}) -
\delta_{J'}(m_{\pi\pi})$ resulting in a characteristic structure in
the $m_{\pi\pi}$ distribution.

A simple observable projecting out the interference is the
$\cos\theta$ moment of the cross section, where $\theta$ is the polar
angle of the $\pi^+$ in the two-pion c.m.  Preliminary data for a
nonzero signal has been reported by HERMES \cite{diNezza:2002vp}.
Leading-twist factorization only applies to amplitudes with $J^3=0$.
As pointed out in \cite{Lehmann-Dronke:2000xq} these can be selected
by taking the moment\footnote{The prefactor $\sqrt{7/3}$ of
$P_3(\cos\theta)$ in \protect\cite{Lehmann-Dronke:2000xq} is mistaken
and should be replaced with $7/3$ \protect\cite{Polyakov:2003pr}.}
\begin{equation}
  \label{Poly-moment}
\Big\langle P_1(\cos\theta) 
	+ \frac{7}{3} P_3(\cos\theta) \Big\rangle 
= \frac{2}{\sqrt{3}} \,
  \re \Big[ \rho_{00}^{10} + \sqrt{5}\, \rho_{00}^{12} \, \Big] 
\end{equation}
with $P_1(x) = x$ and $P_3(x) = \frac{1}{2} (5 x^3 - 3 x)$.  Partial
waves with $J>2$ have been neglected in (\ref{Poly-moment}).  The
moment $\langle f(\cos\theta) \rangle$ is defined through
\begin{equation}
\int_{-1}^1 d(\cos\theta)\, f(\cos\theta) \,
	\frac{d\sigma}{d(\cos\theta)} = \langle f(\cos\theta) \rangle
\, \int_{-1}^1 d(\cos\theta)\, \frac{d\sigma}{d(\cos\theta)} ,
\end{equation}
and the $\pi\pi$ system is described by its spin density matrix
$\rho_{\lambda\, \lambda'}^{J J'}$, whose diagonal entries
$\rho_{\lambda\, \lambda}^{J J}$ give the probability of producing the
pion pair with quantum numbers $J$ and $J^3 = \lambda$ and whose
off-diagonals describe the corresponding interference terms.  In
contrast, the moment
\begin{equation}
\Big\langle P_1(\cos\theta) 
	- \frac{14}{9} P_3(\cos\theta) \Big\rangle 
= \frac{2}{\sqrt{3}} \,
  \re \Big[ \rho_{00}^{10} 
	+ 2 \sqrt{\frac{5}{3}}\, \rho_{11}^{12} \, \Big]
\end{equation}
is sensitive to the interference of $S$ and $P$ waves with $J^3 = 0$
and to the interference of $D$ and $P$ waves with $J^3 = \pm 1$.
Again partial waves with $J>2$ have been neglected.


\subsection{The deconvolution problem}
\label{sec:deconvolute}

So far we have been concerned with extracting helicity amplitudes from
differential cross sections.  A different task is to retrieve GPDs or
GDAs from the appropriate amplitudes, where they appear convoluted
with known hard-scattering kernels as discussed in
Section~\ref{sec:factor}.  For GDAs the situation is analogous to the
one for single-meson DAs (see our detailed discussion of the pion case
in Section~\ref{sub:two-photon}).  One can access integrals of the
distribution amplitudes over $z$, with $\zeta$ and $s$ just appearing
as additional variables in the GDAs.  Notice that at leading
$O(\alpha_s)$ the same $z$ integrals appear in two-photon annihilation
and in electroproduction of states with even $C$ parity, see
Sections~\ref{sec:meson-lt} and \ref{sub:meson-pair-prod}.

Let us turn to the extraction of GPDs.  For the leading-twist DVCS
amplitudes one has to deconvolute
\begin{eqnarray}
  \label{dvcs-decon}
\mathcal{A}(\xi,Q^2,t) &=& \sum_{i} \, \int_{-1}^1 dx\, 
   T^i\Big(x,\xi, \log(Q^2/\mu_F^2) \Big)\,  F^i(x,\xi,t; \mu_F^2) ,
\end{eqnarray}
where $F^i$ stands for $F^q$, $F^g$, $\tilde{F}^q$, $\tilde{F}^g$.  We
have explicitly given the dependence on the factorization scale but
suppressed the renormalization scale $\mu_R$.  The structure for
electroproduction of light mesons is similar, with the kernels $T^i$
replaced by the convolution of hard-scattering kernels and meson
distribution amplitudes.  Since the variable $t$ appears in the GPDs
but not in the hard-scattering kernels, it does not pose particular
problems, and one may consider the deconvolution problem at a given
value of $t$.  The interplay between the variables $x$ and $\xi$ is
however nontrivial.

It is instructive to compare this problem with the extraction of
forward parton densities from appropriate inclusive cross sections,
where the experimental observables are convolutions of the type
\begin{equation}
  \label{inclusive-decon}
\sum_{i=q,g} \, \int_{-1}^1 dx\, 
	t^i\Big(x,\xB,  \log(Q^2/\mu_F^2) \Big)\, f^i(x; \mu_F^2) ,
\end{equation}
where the $f^i$ denote the forward limits of the $F^i$.  In this case
the external scaling variable $\xB$ (or its analog in processes like
Drell-Yan or jet production) appears in the hard-scattering kernels
but not in the distributions.  At leading order in $\alpha_s$ the
kernels reduce to delta functions in the momentum fraction and one has
a one-to-one correspondence between the measured $\xB$ and the
argument of the parton densities.  At higher orders in the coupling
one has nontrivial convolutions, and the evolution of the parton
densities in $\mu_F$ is used in an essential way to extract for
instance the gluon density from the $Q^2$ and $\xB$ dependence of the
DIS cross sections, where gluons do not appear at Born level.
	
What makes the convolution in (\ref{dvcs-decon}) more complicated is
the appearance of $x$ and $\xi$ in both the hard-scattering kernels
and the distributions.  At leading order in $\alpha_S$ the imaginary
part of the amplitude involves still a delta function and one has
direct access to $F^i$ at the points $x=\pm \xi$ separating DGLAP and
ERBL regions.  At NLO accuracy the convolution in $\im \mathcal{A}$
involves only the DGLAP regions, whereas the ones in $\re \mathcal{A}$
involve the full interval $x\in [-1,1]$ already starting at Born
level.  It is not known whether as a mathematical problem the system
(\ref{dvcs-decon}) can be solved for the GPDs, given the nontrivial
properties of GPDs as functions of $x$ and $\xi$ and given their known
evolution behavior in $\mu_F^2$.  That this might be the case is
suggested by a study of Freund \cite{Freund:1999xf}.  The particular
method considered there is based on the expansion of GPDs in
polynomials discussed at the end of Section~\ref{sub:solve-evolution},
and can hardly be used in practice unless one finds a set of
polynomials where such an expansion can be truncated after a
sufficiently small number of terms.

The only other known way to constrain GPDs is to choose a particular
parameterization for them, calculate process amplitudes (or directly
the relevant observables) and to fit the free parameters to data.
This is just the analog of what is done in conventional extractions of
forward parton distributions.  To make it reliable in our context
requires sufficient understanding of the functional dependence of GPDs
on $x$, $\xi$, and $t$.  To make use of the information from the $Q^2$
dependence of the observables and the evolution of the GPDs requires
of course to be in a region where the $Q^2$ variation of amplitudes is
adequately described by the leading-twist approximation, as in the
case of forward distributions.

A possibility to gain more direct information on the shape of GPDs in
their two momentum variables is provided by DDVCS, where one has two
measured scaling variables
\begin{equation}
  \label{ddvcs-decon}
\mathcal{A}(\xi,\rho,Q^2,t) = \sum_{i} \int_{-1}^1 dx\, 
   T^i\Big(x,\xi,\rho, \log(Q^2/\mu_F^2) \Big)\,  F^i(x,\xi,t; \mu_F^2)
\end{equation}
and hence an inversion problem more akin to the forward case.  At Born
level the imaginary part of the amplitude gives GPDs at $x=\pm \rho$
and hence in the ERBL region as discussed in
Section~\ref{sec:compton-scatt}.  We note that at NLO accuracy $\im
\mathcal{A}$ involves now both ERBL and DGLAP regions, unlike in the
case of DVCS.  Especially since the ERBL region is the region where
GPDs are most different from their forward counterparts, DDVCS has a
unique potential.  Data and phenomenological analysis will have to
show to what extent this potential can be realized.


\section{Processes at large $s$, $t$ and $u$}
\label{sec:large-t}

This section deals with exclusive processes with a large momentum
transfer to a hadron, such as elastic form factors and Compton
scattering at large $t$, and with their counterparts in the crossed
channel, such as timelike form factors and two-photon annihilation
into hadrons at large $s$.  These processes receive contributions from
handbag-type diagrams, which can be expressed in terms of GPDs at
large $t$ or of GDAs at large $s$.  Following the historical
development, we will first concentrate on large-$t$ scattering
processes and discuss large-$s$ annihilation processes at the end of
this section.

\subsection{Hard scattering and soft overlap contributions}
\label{sec:hard-and-soft}

\begin{figure}
\begin{center}
	\leavevmode
	\epsfxsize=0.6\textwidth
	\epsfbox{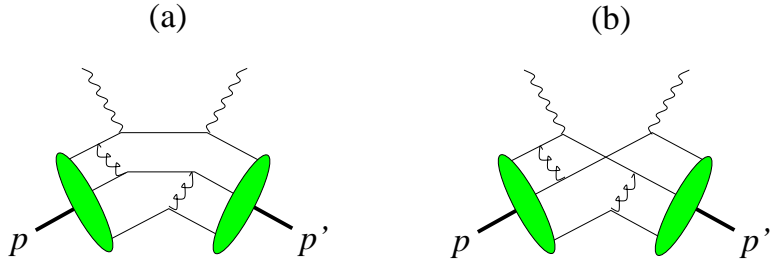}
\end{center}
\caption{\label{fig:hard-scattering} Hard-scattering diagrams for
wide-angle Compton scattering.  Removing one photon one obtains
diagrams for the elastic form factor.}
\end{figure}

Consider the spacelike Dirac form factor $F_1(t)$ of the nucleon at
large $t$, or Compton scattering $\gamma p\to \gamma p$ at large $s
\sim -t \sim  -u$.  In a frame where both initial and final particles
are fast, the nucleon is deviated by a large angle, and so is each of
its partons.  An exception are partons with soft momenta, which
``fit'' into both the initial and final wave function without
undergoing a large momentum transfer.  As shown by Brodsky and Lepage
\cite{Lepage:1980fj,Brodsky:1989pv}, the dominant reaction mechanism
in the large-$t$ limit (at fixed ratios $t/s$, $u/s$ of the large
invariants) involves for each nucleon the Fock state with minimal
number of partons, which undergo a hard scattering process described
by a connected graph, as shown in Fig.~\ref{fig:hard-scattering} for
Compton scattering.  The amplitude is given as a convolution of the
leading-twist distribution amplitude for each nucleon with a hard
scattering kernel evaluated in the collinear approximation.  This is
the exact analog of the hard-scattering mechanism described in
Section~\ref{sec:leading-power} when applied to meson form factors.
It leads in particular to ``dimensional counting rules''
\cite{Brodsky:1973kr,Matveev:1973ra}, which predict the power-law
dependence in the hard scale up to logarithmic corrections.

Compared with the meson case, scattering on baryons involves more
internal lines in the hard scattering, where hence the external hard
scale becomes more ``diluted''.  In such processes one may thus expect
power suppressed effects to be more prominent at moderate values of
the hard scale.  Evaluation of the proton form factor $F_1(t)$, where
data exist up to just over $-t= 30$~GeV$^2$, with the leading-twist
formula leads to values well below the data unless one takes
distribution amplitudes strongly concentrated in the endpoint regions.
Configurations with soft partons in a hadron wave function, for which
the approximations of the hard scattering approach break down, then
provide a substantial part of the leading-twist result, although they
are power suppressed in the asymptotic analysis.  This has been
pointed out long ago by Isgur and Llewellyn Smith and by Radyushkin
\cite{Isgur:1989cy,Isgur:1989iw,Radyushkin:1991te}.  Such regions of
phase space are additionally suppressed when taking into account the
Sudakov form factors in the ``modified hard scattering'' approach of
Sterman and collaborators~\cite{Botts:1989kf,Li:1992nu}.  Yet stronger
suppression results when one includes the $k_T$ dependence of the
hadronic wave functions in the hard scattering subprocess.  This led
to results well below the data for all distribution amplitudes
considered in a study by Bolz et al.~\cite{Bolz:1995hb}.  More
recently, Kundu et al.~\cite{Kundu:1998gv} claimed agreement between
the data for $F_1(t)$ at $-t \gsim 10$~GeV$^2$ and a calculation using
the modified hard scattering approach.  Bolz et al.~\cite{Bolz:1998gq}
have however remarked that the main numerical contributions to this
result are still from regions, where it is not justified to neglect
the $k_T$ dependence of the hadronic wave functions when evaluating
the hard scattering.

The analysis of wide-angle Compton scattering in the collinear hard
scattering framework leads to results below the available data, even
when endpoint concentrated DAs are used \cite{Brooks:2000nb}.  In the
same analysis the ratio of the Compton cross section and the square of
$F_1$ at the same $t$ came out too small.  {}From the analogy with the
pion form factor (Section~\ref{sub:NLO-pions}) one may expect
enhancement of the leading-order result from radiative corrections to
the hard scattering kernel, but their calculation for baryon
scattering processes has not been attempted so far.

\begin{figure}[b]
\begin{center}
	\leavevmode
	\epsfxsize=0.9\textwidth
	\epsfbox{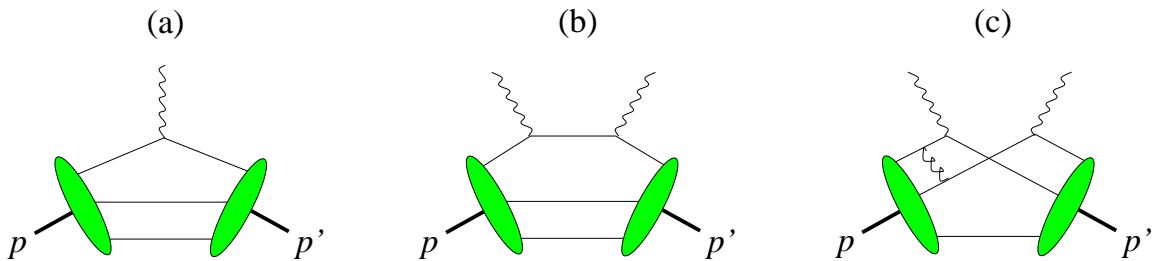}
\end{center}
\caption{\label{fig:Feynman-mechanism} Soft overlap mechanism for 
(a) the elastic proton form factor and (b) wide-angle Compton
scattering.  (c) A soft overlap contribution with ``cat's ears''
topology.}
\end{figure}

A contribution to elastic form factors that involves only small
virtualities is the Feynman mechanism, where almost all momentum
transfer to the nucleon is absorbed by a single quark
(Fig.~\ref{fig:Feynman-mechanism}a).  This quark carries nearly the
full momentum of the nucleon before and after the scattering, whereas
all other partons in the hadron have soft momenta and hence need not
be ``turned around''.  This corresponds to quite rare configurations
in the nucleon wave function, but the configurations required for the
hard scattering mechanism are rare as well, namely three quarks in a
small transverse area of size $1/Q^2$ as encoded in the distribution
amplitude at factorization scale $Q$.  To decide which of these rare
configurations dominates is difficult to judge from intuition.
Evaluating the soft overlap contribution to $F_1(t)$ using light-cone
wave functions with a Gaussian falloff (\ref{wf-ansatz}) in
$\tvec{k}_i$ and an $x_i$ dependence rather close to the asymptotic
form, a rather good description of the large-$t$ data can be achieved
\cite{Bolz:1996sw}, whereas strongly endpoint concentrated wave
functions strongly overshoot the data.  Modeling the wave functions
for the next higher Fock states with an additional gluon or $q\bar{q}$
pair, the study \cite{Diehl:1998kh} found that their importance in the
overlap decreases with increasing $t$.  This is in line with naive
expectation: larger $t$ requires a smaller momentum fraction of the
spectator system, which should be more difficult for states with more
partons.  One expects that the soft overlap contribution of
Fig.~\ref{fig:Feynman-mechanism} receives radiative corrections from
the quark-photon vertex, which can be resummed into a Sudakov form
factor and will lead to a suppression for spacelike momentum transfer
$t$ \cite{Lepage:1979zb}.  How to evaluate these effects for the
processes of interest here is however not known.

In agreement with the leading-twist analysis one finds that the soft
overlap mechanism evaluated with the above wave function models falls
off faster with $t$ than the hard-scattering mechanism.  In
particular, the soft overlap contribution to the scaled Dirac form
factor $t^2 F_1(t)$ decreases with $t$.  This happens however only at
$t$ of several 10~GeV$^2$.  As explained in
Section~\ref{sub:soft-overlap-gpd} and emphasized in
\cite{Radyushkin:1998rt,Diehl:1999tr}, one finds that the relevant
scale of the $t$-dependence for the soft overlap mechanism is
\emph{several} GeV$^2$, although the basic nonperturbative 
parameters in the wave functions are around 1~GeV$^2$.  This warns us
that asymptotic power counting cannot simply be used for soft overlap
contributions to processes involving baryons unless the hard scale is
\emph{very} large.  One can explicitly show how the form
(\ref{wf-link}) of the soft overlap contributions to form factors can
mimic the scaling behavior of dimensional counting rules over a rather
wide but finite range of $t$ in the multi-GeV region
\cite{Diehl:1999tr}.  For various form factors differing by their
flavor or spin structure this happens in approximately the same region
of $t$ and may be seen as a rather universal property of the hadron in
question.  Apparent dimensional counting behavior can thus be found in
several processes where the physics of the soft overlap mechanism can
be parameterized by such form factors.  This shows that, although the
mimicking of dimensional counting behavior by the soft overlap may be
seen as a numerical ``accident'', it can occur on a systematic basis
in several places.

The situation for wide-angle Compton scattering is similar to the
elastic form factors but more complex.  There is a soft overlap
contribution to this process, shown in
Fig.~\ref{fig:Feynman-mechanism}b, which involves the same
low-virtuality parton configurations in the nucleon as the form
factors, and Compton scattering $\gamma q\to \gamma q$ on one quark.
The Compton subprocess is hard and suggests the factorized description
of this mechanism we will discuss in the following.  As discussed in
\cite{Radyushkin:1998rt,Diehl:1998kh} there are other overlap
contributions where the photons couple to two different quarks, as
shown in Fig.~\ref{fig:Feynman-mechanism}c, whereas the remaining
spectator partons have soft momenta.  For large $s$, $t$, $u$
kinematics does not allow such a process to be entirely soft, and if
the wave functions in the figure are understood as soft wave
functions, at least one hard gluon exchange is required.  This type of
contribution is intermediate between the soft overlap in
Fig.~\ref{fig:Feynman-mechanism}b and the hard-scattering mechanism of
Fig.~\ref{fig:hard-scattering} with respect to the number of hard
internal propagators and the restrictions on the nucleon wave
functions, and one may expect that its size is intermediate between
the two extreme mechanisms.  There are further power-suppressed
mechanisms (see \cite{Radyushkin:1998rt}), for instance those with
hadron-like configurations of the photons, which one might model in
terms of vector dominance and wide-angle meson-nucleon scattering.  To
date there is no formalism allowing a systematic treatment of all
relevant contributions to the amplitude in a situation where the
leading-twist mechanism is not dominant, and phenomenological analysis
is required to identify situations where one particular mechanism
turns out to dominate.

Generalized parton distributions provide a unified description of the
soft phy\-sics involved in soft overlap contributions.  The Feynman
mechanism for the elastic form factor just corresponds to the overlap
of soft wave functions in the relevant GPDs we discussed in
Section~\ref{sub:soft-overlap-gpd}.  As we describe in the next
section, one may also evaluate the soft overlap contributions of
Fig.~\ref{fig:Feynman-mechanism}b in terms of GPDs.  This leads to a
description of the Compton amplitude involving a calculable
short-distance process and quantities encoding the relevant soft
physics, which one may extract from measurement and confront with
efforts to calculate or model them theoretically.


\subsection{The soft overlap for Compton scattering and meson
production}
\label{sec:soft-handbag}

\subsubsection{Calculating the handbag diagrams}
\label{sub:calc-handbag}

The soft overlap diagrams in Fig.~\ref{fig:Feynman-mechanism}b can be
rearranged in the form of the familiar handbag diagrams of
Fig.~\ref{fig:handbag}b (with external photons on shell).  The details
of the wave function overlap (including summation over the number of
soft spectators) are lumped together in the matrix element of the
bilocal operator $\bar{q}_\alpha(-\half z) q_\beta(\half z)$ between
the external hadrons.  To be more precise, only the \emph{soft} part
of this matrix element is implied here.  Contributions involving large
virtualities are not part of what is to be calculated.  They are
implied to be subtracted, unless they are sufficiently small to be
simply neglected.  Part of such hard contributions within the handbag
are in fact included in leading-twist hard scattering diagrams as in
Fig.~\ref{fig:hard-scattering}a.  This situation precludes one from
using certain familiar arguments based on dimensional analysis, since
the matrix element $\langle p'| \bar{q}_\alpha(-\half z) q_\beta(\half
z) | p\rangle_\mathrm{soft}$ depends on a large invariant $t$ but by
definition contains no internal virtualities of that size.

The calculation of the soft handbag has been carried out in
\cite{Radyushkin:1998rt} and in \cite{Diehl:1998kh} using different
techniques, whose results agree if suitable approximations are made
within each scheme.  The calculation of \cite{Diehl:1998kh} is in
momentum space and organizes the diagrams into Compton scattering on
an on-shell quark and Fourier transformed matrix elements of
$\bar{q}(-\half z) \gamma^+ q(\half z)$ and $\bar{q}(-\half z)
\gamma^+ \gamma_5\, q(\half z)$ at $z^2=0$.  They are given as
$1/x$ moments of GPDs,
\begin{eqnarray}
   \label{compton-form-factors}
R_V(t) = \sum_q e_q^2 \int_{-1}^1 \frac{dx}{x}\, H^q(x,0,t) ,
&\qquad&
R_T(t) = \sum_q e_q^2 \int_{-1}^1 \frac{dx}{x}\, E^q(x,0,t) ,
\nonumber \\
R_A(t) = \sum_q e_q^2 
	\int_{-1}^1 \frac{dx}{|x|}\, \tilde{H}^q(x,0,t) ,
\end{eqnarray}
and referred to as ``Compton form factors''.  The calculation
explicitly chooses a frame where $\xi=0$, and where all external
particles have energies and momenta of order $\sqrt{-t}$.  (The wave
function representation of the matrix elements is simplest in such a
frame because it avoids the parton number changing overlap in the ERBL
region.)  At $\xi=0$ no Compton form factor related to $\tilde{E}^q$
appears, since these distributions are multiplied with $\Delta^+$ in
their definition (\ref{quark-gpd}).  An attempt was made to quantify
the approximations of the calculation parametrically, assuming that
the soft wave functions ensure $\tvec{k}_i^2 /x_i^{\phantom{2}} \lsim
\Lambda^2$ for all partons at the wave function vertices in
Fig.~\ref{fig:Feynman-mechanism}b.  This ensures that all virtualities
at these vertices are bounded by a soft physics scale $\Lambda^2 \sim
1$~GeV$^2$.  The final result for the Compton amplitude was then found
to be accurate up to corrections of $O(\Lambda^2 /t)$.

The calculation of Radyushkin \cite{Radyushkin:1998rt} starts with the
handbag diagrams in position space and expands the operator
$\bar{q}_\alpha(-\half z) q_\beta(\half z)$ around light-like
distances $z^2=0$.  Representing the resulting matrix elements by
double distributions according to (\ref{dd-def}) and Fourier
transforming to momentum space, one obtains quark propagators with
denominators
\begin{eqnarray}
  \label{rad-prop}
\hat{s} &=& \beta (s-m^2)
  - {\textstyle\frac{1}{4}}\Big[ (1-\beta)^2 - \alpha^2 \Big]\, t
  + \beta^2 m^2 ,
\nonumber \\
\hat{u} &=& \beta (u-m^2)
  - {\textstyle\frac{1}{4}}\Big[ (1-\beta)^2 - \alpha^2 \Big]\, t
  + \beta^2 m^2 
\end{eqnarray}
for $\beta > 0$, respectively corresponding to the $s$ and $u$ channel
handbag graphs.  The calculation proceeds by approximating $\hat{s}
\approx \beta (s-m^2)$ and $\hat{u} \approx \beta (u-m^2)$, which in
\cite{Radyushkin:1998rt} is justified on numerical rather than
parametrical grounds.  After further approximations the integration
over the double distribution parameters leads to
\begin{equation}
  \label{rad-ff}
\sum_q e_q^2 
\int d\beta\, d\alpha\, \frac{1}{\beta}\, f^q(\beta,\alpha,t)
\end{equation}
and its analog for the other double distributions, and hence to the
same $1/x$ moments of GPDs as above.  The form factors corresponding
to $E^q$ and $\tilde{E}^q$ were explicitly neglected in
\cite{Radyushkin:1998rt}, and as a further simplification $R_A$ was
approximated by $R_V$.

The result of both calculations, whose phenomenology we shall discuss
in the next subsection, quite clearly involves approximations which
one would like to improve on, especially when faced with data at
values of $s$, $t$ and $u$ that are not significantly above 1~GeV$^2$.
How this can be done consistently is not clear at the present stage.
Electromagnetic gauge invariance for Compton scattering on a quark is
ensured if the quarks are approximated to be on shell.  To be
consistent with kinematical constraints, this approximation requires
one to set the quark momentum equal to the momentum of its parent
hadron and to neglect the target mass.  Since the quark handbag
diagrams in Fig.~\ref{fig:handbag} are not gauge invariant by
themselves, improvement over these approximations will require careful
study, as the case of $1/Q$ corrections to DVCS has shown (see
Section~\ref{sec:twist-three}).  Whether one obtains a gauge invariant
Compton amplitude when keeping the full propagators (\ref{rad-prop})
in the approach of Radyushkin is for instance not known.  The same
caveat concerns the inclusion of the $t$-dependent terms in
(\ref{rad-prop}) up to linear accuracy in $t/s$, considered by Bakulev
et al.~\cite{Bakulev:2000eb}.

Identifying the momenta of the struck parton and its parent hadron
implies in particular setting the quark light-cone momentum fraction
$x$ to 1 in the hard scattering.  One of the theoretical uncertainties
of this approximation concerns the factors $1/x$ and $1/|x|$ in the
GPD moments (\ref{compton-form-factors}), since the way they appear
in the calculation of \cite{Diehl:1998kh} does not unambiguously
assign them to either the hard scattering or the soft matrix element.
Within the present accuracy of the approach, one might also choose to
replace them respectively with $\mbox{sgn}(x)$ and $1$.  For an
individual quark flavor, the Compton form factor $R_A^q$ is then
replaced by the axial form factor $g_A^q$.  In the case of $R_V^q$ and
$R_T^q$ one does not obtain form factors of local currents because the
region $x<0$ is explicitly weighted with a factor of $-1$.  This
region corresponds to scattering on an antiquark in the proton, which
is much less likely than a quark to carry most of the hadron momentum.
Neglecting these configurations one may count the integral over this
region with the ``wrong sign'', and then finds the Dirac and Pauli
form factors $F_1^q$ and $F_2^q$ of the local vector current.  We
remark however that with the model wave functions used in
\cite{Diehl:1998kh} the omission of the $1/x$ factors in the Compton
form factors is of some numerical importance, since the $x$ values
dominating the integral for experimentally relevant $t$ are not very
close to $1$ (see Section~\ref{sub:soft-overlap-gpd}).  This calls for
an improvement of the theoretical description beyond the $x=1$
approximation in the hard scattering.

Different possibilities of approximately taking into account the
nucleon mass when relating the subprocess kinematics to the external
Mandelstam variables have been explored in \cite{Diehl:2002ee}.  Their
comparison may be taken as an estimate of theoretical uncertainties in
kinematics where the target mass is not negligible.  They were found
to significantly depend on the observable and on the scattering angle,
and to decrease substantially when going from the energies of present
experiments at Jefferson Lab \cite{Chen:2000wa} to those of the
proposed 12~GeV upgrade \cite{Cardman:2001jl}.

\subsubsection{Phenomenology of Compton scattering}
\label{sub:soft-compton-pheno}

With the approximations we have discussed, the helicity amplitudes
$e^2 M_{\lambda'\mu', \lambda\mu}$ for Compton scattering at large
$s\sim -t\sim -u$ are given by
\cite{Diehl:1998kh,Huang:2001ej}
\begin{eqnarray}
  \label{compton-handbag}
M_{\lambda'+, \lambda+} =
  \Bigg( \delta_{\lambda'\lambda}\, R_V(t) 
      - 2\lambda\, \delta_{\lambda'\, -\lambda}\,
  	 	         \frac{\sqrt{-t}}{2m} R_T(t) \Bigg) \,
  \frac{s-u}{\sqrt{-su}} 
+ 2\lambda\, \delta_{\lambda'\lambda}\, R_A(t) \,
  \frac{s+u}{\sqrt{-su}} ,
\end{eqnarray}
$M_{\lambda'+, \lambda-} = 0$, and the parity relations
(\ref{Compton-parity}).  The light-cone helicities $\lambda$,
$\lambda'$ of the initial and final state proton are normalized to
$\half$ and refer to a frame with momentum components $(p-p')^+ = 0$,
$(p-p')^1 > 0$, $(p-p')^2=0$.  Their explicit transformation to usual
helicities in the $\gamma p$ c.m.\ is given in \cite{Huang:2001ej}.
The amplitudes for virtual Compton scattering $\gamma^* p\to \gamma p$
with $Q^2$ at most of the same order as the other large invariants can
readily be calculated in the same formalism and are given in
\cite{Diehl:1998kh}, where various observables have been estimated.
Amplitudes which flip the photon helicity by one or two units are
respectively found to be proportional to $Q$ and $Q^2$.

The result (\ref{compton-handbag}) corresponds to the hard parton
subprocess at Born level.  Radiative corrections of $O(\alpha_s)$ have
been calculated by Huang et al.~\cite{Huang:2001ej} for the case of
real photons.  It was found that the collinear and soft divergences
obtained with on-shell external partons are multiplied with the
leading-order scattering amplitudes and hence can be absorbed into the
Compton form factors.  It is suggestive to assume that this can be
implemented by an appropriate renormalization prescription, although
this has not been studied in detail.  The $O(\alpha_s)$ corrections
also induce nonzero amplitudes that flip the photon helicity by two
units.  They further involve handbag diagrams where the scattering is
on a fast gluon instead of a quark.  In line with expectations, the
corresponding moments of gluon GPDs estimated in \cite{Huang:2001ej}
are smaller compared with their quark analogs, although not
negligible.  The $O(\alpha_s)$ kernels for virtual Compton scattering
have recently been evaluated by Huang and Morii \cite{Huang:2003uy}.

{}From (\ref{compton-handbag}) one obtains for the unpolarized cross
section
\begin{eqnarray}
  \label{handbag-cross-section}
\frac{d\sigma}{dt}
= \frac{\pi \alpha_{\rm em}^2}{(s-m^2)^2}\,
\left[ \frac{(s - u)^2}{|s u|} \Big( R_V^2(t)
                        - \frac{t}{4m^2} R_T^2(t) \Big)
                 + \frac{(s + u)^2}{|s u|} R_A^2(t) \right]
\end{eqnarray}
where here and in the following we keep the target mass in the phase
space and flux factor while neglecting it in the squared process
amplitudes.  The dependence on the scattering angle in the collision
c.m.\ is readily obtained from the relations $t \approx -\half s
(1-\cos\theta)$, $u \approx -\half s (1+\cos\theta)$, which are valid
up to $O(m^2/s)$ corrections.  The $O(\alpha_s)$ corrections to the
cross section have been estimated to increase the cross section by
about 10\% for backward angles and by about 30\% in the forward
direction \cite{Huang:2001ej}.  In that study the range of $\theta$
was restricted to $|\cos\theta| <0.6$ in order to ensure that $-t$ or
$-u$ do not become significantly smaller than $s$, and this
restriction will be implied in the following.

Whereas the form factors $R_V$ and $R_A$ may be obtained from the
overlap of soft model wave functions
\cite{Radyushkin:1998rt,Diehl:1998kh}, modeling of $R_T$ along the
same lines would require wave functions with nonzero orbital angular
momentum between the partons and has not been attempted.  Instead one
may assume that $R_T/R_V$ behaves approximately like the ratio
$F_2/F_1$ of Pauli and Dirac form factors, given the similarity of
these quantities.  Measurements from Jefferson Lab using the recoil
polarization of the scattered proton find a behavior $F_2(t) /F_1(t)
\approx 0.37 \times 2m /\sqrt{-t}$ for $t$ between 1 and 5.6~GeV$^2$
\cite{Gayou:2001qd}, whereas older measurements using Rosenbluth
separation found a steeper falloff $F_2(t) /F_1(t) \propto 4m^2
/(-t)$.  Which result is correct awaits clarification, for recent
discussions see \cite{Arrington:2002cr,Guichon:2003qm}.  If $R_T(t)
/R_V(t) \sim 2m /\sqrt{-t}$ then $R_T$ contributes to the cross
section at the same level as $R_V$ and $R_A$, and proton helicity
conserving and helicity changing amplitudes in (\ref{compton-handbag})
are of comparable size.

The wave function model in \cite{Radyushkin:1998rt,Diehl:1998kh} gave
form factors $R_V(t)$ and $R_A(t)$ which for $-t$ between about 5 and
15~GeV$^2$ behave approximately like $t^{-2}$ and thus mimic
dimensional counting behavior in this range.  The cross section
(\ref{handbag-cross-section}) of the soft handbag then mimics the
dimensional counting behavior $d\sigma /(dt) \sim s^{-6}$ at fixed
$t/s$.  It is only at larger values of $t$ that the soft part of the
form factors implied in the handbag formulas is power suppressed
compared to $t^{-2}$, with a corresponding power suppression of the
soft overlap contribution to the Compton cross section as required by
the asymptotic analysis.

Using these model form factors, the handbag result
(\ref{compton-handbag}) gives a fair description of the available data
for $d\sigma/(dt)$ from experiments with photon beam energies
$E_\gamma$ from 3~GeV to 6~GeV in the target rest frame
\cite{Diehl:1998kh,Huang:2001ej}.  As discussed in
Section~\ref{sec:hard-and-soft}, the hard scattering mechanism has
difficulties to reproduce the cross section.  A satisfactory
description can however be obtained in the diquark model
\cite{Kroll:1996pv}.  This treats the proton as an effective
quark-diquark system, thus implying nonperturbative correlations among
the quarks, and evaluates hard exclusive processes as hard scattering
on the quark-diquark system, with phenomenological diquark form
factors describing again effects beyond perturbation theory.

To establish which dynamics is actually at work in the process it is
important to find features of the different mechanisms that are
independent of the nonperturbative physics input, i.e., of the Compton
form factors in our case.  As we have seen, the $s$ behavior of the
cross section at fixed $t/s$ cannot be used to distinguish
experimentally between different reaction mechanisms in Compton
scattering, given that values of $t$ much above 15~GeV$^2$ will hardly
be realizable in experiment.  Among the generic features of the soft
handbag mechanism is that photon helicity flip for $\gamma p\to \gamma
p$ only occurs at $O(\alpha_s)$, which directly reflects a feature of
real Compton scattering on massless fermions.  This can be tested with
suitable polarization observables.  A example is the asymmetry
$\Sigma$ of cross sections for linear photon polarization either
normal to or in the scattering plane, for which the soft handbag
generically predicts $\Sigma = O(\alpha_s)$.  Under weak assumptions
on the gluon Compton form factors one further obtains that $\Sigma
<0$, and numerical estimates suggest that this quantity may indeed be
a good discriminator between different mechanism \cite{Huang:2001ej}.

As another generic feature, the soft overlap mechanism predicts small
phases of the amplitudes for real and virtual Compton scattering:
since the Compton form factors are real valued, imaginary parts of the
amplitude are only generated by the partonic subprocess at
$O(\alpha_s)$.  In the hard-scattering mechanism and its diquark
variant, imaginary parts arise at leading order in the strong coupling
and phases are generically large.  As emphasized by Kroll et
al.~\cite{Kroll:1996pv}, the imaginary part of the virtual Compton
amplitude can be accessed in the lepton polarization asymmetry of
$ep\to ep\gamma$ using the interference between Compton and
Bethe-Heitler amplitudes, i.e., the same principle we discussed for
DVCS in Section~\ref{sec:dvcs-pheno}.

If follows from (\ref{compton-handbag}) that the soft handbag
mechanism makes specific predictions about the $s$-dependence of
amplitudes at fixed $t$: each helicity amplitude can be written as a
finite sum of known functions of $s$, whose coefficients are
determined by the Compton form factors.  Including $O(\alpha_s)$
corrections, the $s$-dependent functions are modified, and additional
terms due to gluon form factors appear.  This structure can be tested,
as observed by Nathan \cite{Nathan:1999sr}.  Dividing for convenience
the Compton cross section (\ref{handbag-cross-section}) by the
Klein-Nishina cross section $d\sigma_{\mathrm{KN}} /dt$ on a pointlike
target, one obtains
\begin{equation}
  \label{Nathan-scaling}
\frac{d\sigma /dt}{d\sigma_{\mathrm{KN}} /dt} =
  f_V \Bigg( R_V^2(t) - \frac{t}{4m^2} R_T^2(t) \Bigg) +
  (1-f_V) R_A^2(t) 
\end{equation}
with
\begin{equation}
f_V = \frac{(s-u)^2}{2(s^2 + u^2)} ,
\end{equation}
where we neglect the proton mass throughout.  Note that
$d\sigma_{\mathrm{KN}} /dt$ is obtained from
(\ref{handbag-cross-section}) by setting $R_A = R_V$ and $R_T=0$,
which fixes the coefficient of $R_A^2$ in (\ref{Nathan-scaling}).  The
$O(\alpha_s)$ corrections to the quark handbag diagrams are readily
taken into account using the results of \cite{Huang:2001ej}: if one
divides $d\sigma /dt$ by $d\sigma /dt|_{R_A = R_V, R_T=0}$, the
structure of (\ref{Nathan-scaling}) is preserved with a modified
function $f_V$.  The remaining $O(\alpha_s)$ corrections due to the
gluon form factors were estimated to be at most 10\% in
\cite{Huang:2001ej}.

As observed in \cite{Nathan:1999sr}, confronting
(\ref{Nathan-scaling}) with data at equal $t$ and different $s$ would
not only provide a distinctive test of the mechanism but may in
principle be used to separate $R_A$ from the other form factors.  The
lever arm for this is however rather small since $f_V$ is rather close
to 1 in most kinematics, so that sensitivity to $R_A$ will be smaller
than to the combination of $R_V$ and $R_T$ in (\ref{Nathan-scaling}).
An alternative way to separate the different form factors is given by
spin asymmetries, which were studied in detail in
\cite{Diehl:1999tr,Huang:2001ej}.  An example are the correlation
parameters between the helicity of the incoming photon and the
helicity of the incoming ($A_{LL}$) or outgoing ($K_{LL}$) proton,
which in the handbag are equal and given by
\begin{eqnarray}
  \label{all-kll}
A_{LL}\, \frac{d\sigma}{dt} &=& K_{LL}\, \frac{d\sigma}{dt}
 \;=\; \frac{2\pi \alpha_{\rm em}^2}{(s-m^2)^2} 
     \frac{s^2 - u^2}{|su|}\, R_A(t)\, \Bigg( R_V(t)
               - \frac{t}{s + \sqrt{-su}}\, R_T(t) \Bigg)
\end{eqnarray}
to lowest order in $\alpha_s$.  Dividing by the Klein-Nishina value
$A_{LL}^{\mathrm{KN}} = K_{LL}^{\mathrm{KN}} = (s^2-u^2)/(s^2+u^2)$
for a target without structure and mass one obtains
\begin{equation}
\frac{A_{LL}}{A_{LL}^{\mathrm{KN}}} =
\frac{K_{LL}}{K_{LL}^{\mathrm{KN}}} \simeq \frac{R_A}{R_V} \,
\left[ 1 - \frac{t^2}{2(s^2 + u^2)} \Big( 1 - \frac{R_A^2}{R_V^2}
                                        \Big) \right]^{-1}  ,
\end{equation}
where we have neglected terms involving $R_T$.  The kinematical
prefactor in brackets is small in typical kinematics, so that this
ratio is approximately a measure of $R_A /R_V$.

An observable sensitive to the tensor form factor $R_T$ is for
instance the correlation $K_{LS}$ between the helicity of the initial
photon and the transverse (``sideways'') polarization of the recoil
proton in the scattering plane.  With the sign convention of
\cite{Huang:2001ej} this parameter is 
\begin{eqnarray}
  \label{kls}
K_{LS}\, \frac{d\sigma}{dt} =
\frac{2\pi \alpha_{\rm em}^2}{(s-m^2)^2} 
\frac{s^2 - u^2}{|su|}\, \frac{\sqrt{-t}}{2m}\, 
  \Bigg( \frac{4m^2}{s + \sqrt{-su}}\, R_V(t) - R_T(t) \Bigg) \,R_A(t)
\end{eqnarray}
to leading $O(\alpha_s)$.  A useful variable is the ratio
$K_{LS}/K_{LL}$, where the form factor $R_A$ drops out
\cite{Huang:2001ej} and where the target mass corrections were
estimated to be small \cite{Diehl:2002ee}.

\subsubsection{Meson production}

The soft overlap mechanism gives contributions to a number of
exclusive wide-angle processes.  Huang et al.~\cite{Huang:2000kd} have
investigated exclusive photo- and electroproduction of mesons at large
scattering angle, generalizing the approach to Compton scattering
described in the previous two subsections.  The photon-meson
transition was described within the collinear hard scattering
mechanism, so that the resulting diagrams had the form of
Fig.~\ref{fig:mesons}.  The scattering amplitudes for the production
of longitudinally polarized mesons can then be expressed in terms of
$(i)$ form factors which apart from the flavor structure are the same
as for wide-angle Compton scattering and of $(ii)$ the leading-twist
meson distribution amplitudes $\Phi^q(z)$ convoluted with hard
scattering kernels depending on $z$ and on the external invariants
$s$, $t$ and $Q^2$.  As already remarked in
Section~\ref{sec:hard-and-soft}, dimensional counting behavior for
these processes is again mimicked by the mechanism in the $t$ range
where the corresponding form factors approximately fall of as
$t^{-2}$.

Compared with available data for large angle photoproduction of
$\pi^0$, $\rho^0$ and $\omega$ at $s \sim 10$~GeV$^2$, Huang et al.\
find that the soft overlap cross sections are too small by orders of
magnitude.  The authors suggest that these data may be dominated by
the hadronic component of the photon, arguing that the $s$ dependence
of available cross section data is more compatible with that of
meson-hadron scattering than with the dimensional counting behavior of
$\gamma p\to M p$ in the hard-scattering approach or its soft overlap
analog.  This hadronic component may be modeled in terms of vector
meson dominance as was for instance done in \cite{Cano:2001sb}.  Its
relevance should decrease when the incident photon is virtual, but
data in the relevant kinematics are not yet available.  Kroll and
Passek-Kumeri\v{c}ki \cite{Kroll:2002nt} have pointed out that
large-angle electroproduction of $\eta'$ has some sensitivity to the
two-gluon component of the $\eta'$, unlike its counterpart in the
small-$t$ region.  For large-angle real Compton scattering, Huang et
et al.\ found that the hadronic component of the photon is not
dominant for $s$ around 10~GeV$^2$, using available data and vector
meson dominance to estimate the corresponding contribution to $\gamma
p\to \gamma p$.


\subsection{Two-photon annihilation}
\label{sec:large-s}

Let us now discuss processes related to Compton scattering by crossing
symmetry.  Exclusive two-photon annihilation offers the possibility to
investigate a variety of final states.  Data at high energies
is available from VENUS \cite{Hamasaki:1997cy}, CLEO
\cite{Dominick:1994bw,Artuso:1994xk,Anderson:1997ak}, LEP
\cite{Achard:2002ez,Abbiendi:2002bx,Achard:2003jc,Anulli:2004}, and
results with yet higher statistics can be expected from Belle at KEK
\cite{Anulli:2004}.  The time reversed process of exclusive $p\bar{p}$
annihilation into photon pairs would likely be accessible at the
planned HESR facility at GSI \cite{Gutbrod:2001cd}.

\subsubsection{Baryon pair production}
\label{sub:baryon-ann}

The dynamics of $\gamma\gamma \to p\bar{p}$ at large $s \sim -t \sim
-u$ shares important features with wide-angle Compton scattering on
the proton.  The problems of the hard-scattering approach at $s$ of
order 10~GeV$^2$ are the same as those discussed in
Section~\ref{sec:hard-and-soft}, with the leading-twist calculation of
\cite{Farrar:1985gv} being far below the available cross section data
for baryon pair production.  The diquark model gives a rather good
description of the data \cite{Berger:2002vc}, as it does for
wide-angle Compton scattering.

\begin{figure}
\begin{center}
	\leavevmode
	\epsfxsize=0.82\textwidth
	\epsfbox{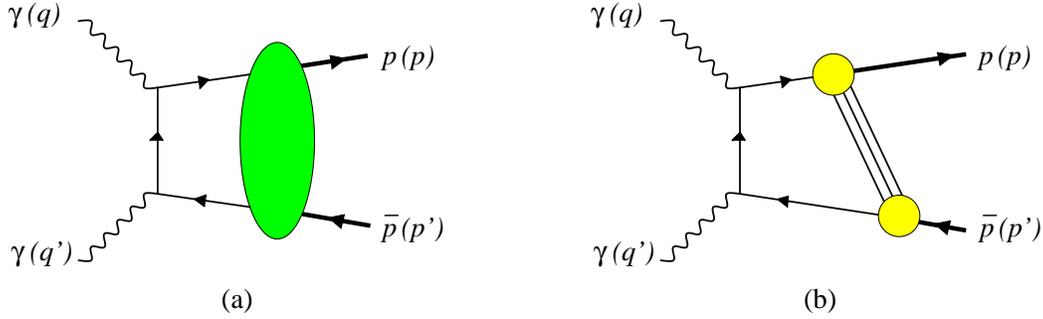}
\end{center}
\caption{\label{fig:soft-timelike} (a) Handbag factorization of 
$\gamma\gamma\to p \bar{p}$ at large $s$, $t$, $u$.  A second graph is
obtained by interchanging the photon vertices.  (b) The physical
mechanism of handbag factorization.  Any number of soft partons
connect the two parton-hadron vertices.}
\end{figure}

The analog of the soft overlap mechanism for this process is shown in
Fig.~\ref{fig:soft-timelike}: a hard photon-photon collision produces
a quark-antiquark pair, which then hadronizes into $p\bar{p}$ in a
soft process.  For the hadronization to involve no large virtualities,
the additional partons created must have soft momenta, so that the
initial quark and antiquark respectively carry almost the full
momentum of one of the final-state baryons.  This process cannot be
represented in terms of a light-cone wave function overlap since one
of the soft vertices in Fig.~\ref{fig:soft-timelike}b has both
incoming and outgoing partons.  It can however be treated within a
covariant framework starting with Bethe-Salpeter wave functions, as
remarked in Section \ref{sec:gda-dynamics}.  In any case, the soft
dynamics can be parameterized in terms of crossed versions of the
Compton form factors introduced above.  Our discussion of Compton
scattering at the end of Section~\ref{sec:hard-and-soft} has its
equivalent in the annihilation channel: whether the soft handbag
provides the dominant mechanism in given kinematics must be
investigated phenomenologically.  Since the detailed dynamics is
different in the time- and spacelike domains, the ranges in $s$ and
$-t$ where the soft handbag is relevant need not be the same.

In \cite{Diehl:2002yh} the soft handbag diagrams for $\gamma\gamma \to
p\bar{p}$ have been calculated within the same momentum-space
formalism as wide-angle Compton scattering \cite{Diehl:1998kh}.  This
was done in a reference frame where the baryons are produced back to
back and have zero three-momentum along the $z$ axis.  The skewness
parameter of GDAs is then $\zeta=\half$, and the baryon and antibaryon
appear in a symmetrical fashion.  The approximation of setting $x=1$
in wide-angle Compton scattering now corresponds to approximating the
quark plus-momentum fraction $z$ with $\half$.  In order to obtain a
representation of the amplitude in terms of GDAs it turned out
necessary to neglect configurations where the \emph{antiquark}
hadronizes into the proton, in a similar way as discussed at the end
of Section~\ref{sub:calc-handbag}.  The scattering amplitude is then
parameterized in terms of three annihilation form factors,
\begin{eqnarray}
R_V(s) = \sum_q e_q^2 \int_0^1 dz\, \Phi^q_V(z,\half, s) , 
\nonumber
\\
R_A(s) = \sum_q e_q^2 \int_0^1 dz\, \Phi^q_A(z,\half, s) , 
&\qquad&
R_P(s) = \sum_q e_q^2 \int_0^1 dz\, \Phi^q_P(z,\half, s) ,
\end{eqnarray}
defined from the proton GDAs $\Phi^q(z,\zeta,s)$ we introduced in
(\ref{GDA-proton}).  Up to the different weighting of quark flavors
these form factors respectively correspond to the magnetic form factor
$G_M(s) = F_1(s) + F_2(s)$ and the axial and pseudoscalar form factors
$g_A(s)$, $g_P(s)$ in the timelike region.  Because of time reversal
invariance, the helicity amplitudes of $\gamma\gamma \to p\bar{p}$ and
$p\bar{p}\to \gamma\gamma$ are of course equal.

Freund et al.~\cite{Freund:2002cq} have considered the same process
using the position space method of Radyushkin
\cite{Radyushkin:1998rt}.  In this approach, the amplitudes are
expressed in terms of integrals over double distributions of the type
\begin{equation}
\sum_q e_q^2
\int d\beta\, d\alpha\, \frac{1}{\beta}\, f^q(\beta,\alpha,s) .
\end{equation}
{}From the reduction formula (\ref{cross-red}) one sees that this
cannot be written as a moment of a GDA because of the factor
$1/\beta$.  The approximations made in \cite{Diehl:2002yh} correspond
to replacing this factor with 1, after which the results of both
approaches agree.

For the unpolarized cross section one obtains \cite{Diehl:2002yh}
\begin{eqnarray}
  \label{gaga-cross-sect}
\frac{d\sigma}{dt} (\gamma\gamma\to p\bar{p}) &=&
  \frac{\pi \alpha_{\mathrm{em}}^2}{s^2} \Bigg[
  \frac{s^2}{tu}\, \Big( \, |R_A(s) + R_P(s)|^2 
	        + \frac{s}{4m^2} |R_P(s)|^2 \, \Big)
   {}+ \frac{(t-u)^2}{tu}  |R_V(s)|^2
\Bigg] ,
\end{eqnarray}
where in the calculation of the amplitude we have neglected hadron
masses as we did for Compton scattering.  For $p\bar{p}\to
\gamma\gamma$ one obtains the same cross section times a factor $(1 -
4m^2/s)^{-1}$ due to the different phase space and flux.

Many phenomenological aspects in the annihilation channel are quite
similar to the case of Compton scattering.  As one difference we point
out that in the annihilation channel it is a combination of the form
factors $R_A$, $R_P$ associated with the axial current, rather than the
one associated with the vector current, which tends to dominate the
cross section.  This is due to the relative kinematical factor
$(t-u)^2/s^2 \approx \cos^2\theta$, which is small except in the
forward and backward regions, where the approach is not valid.  In
\cite{Diehl:2002yh} the size of $|R_A(s)|$ was estimated using
$\gamma\gamma \to p\bar{p}$ data between $s=6.5$~GeV$^2$ and
11~GeV$^2$, finding $s^2 |R_A(s)| \approx 5 \div 8$~GeV$^4$ in
this region.  This is somewhat larger than the estimate $s^2 |R_V(s)|
\approx 2.5 \div 5$~GeV$^4$ obtained from the data on the
magnetic form factor $G_M(s)$, but there is no known constraint that
would require $|R_A(s)| \le |R_V(s)|$.  Note that, as in the spacelike
region, the $s$-dependence of these form factors is compatible with
dimensional scaling for these values of $s$.  The $\cos\theta$
distribution obtained with these form factor estimates is well
approximated by $d\sigma/dt \propto 1/\sin^2\theta$ and was found in
reasonable agreement with available data.

A particularity of the soft handbag mechanism is that the final hadron
pair originates from an intermediate $q\bar{q}$ state and can
therefore only carry isospin $I=0$ or $I=1$, but not $I=2$.  Together
with flavor SU(3) symmetry this allows one to express the annihilation
form factors for all ground state octet baryons in terms of the
individual flavor contributions $R_i^u$, $R_i^d$, $R_i^s$ ($i=V,A,P$)
in the proton \cite{Diehl:2002yh}.  One may expect that for large
values of $s$ the strange quark contributions $R_i^s$ are negligible
compared with $R_i^u$ and $R_i^d$, since the quark is required to
carry almost all the proton momentum.  To the extent that the form
factor combination $R_A + R_P$ dominates the cross section
(\ref{gaga-cross-sect}) one then obtains a simple approximate relation
between the cross sections for all octet channels in terms of the
ratio of the relevant form factors for $u\bar{u}$ and $d\bar{d}$
fragmenting into $p\bar{p}$.  This relation was found to be in
reasonable agreement with data for $\Sigma^0
\bar{\Sigma}{}^0$ and $\Lambda \bar{\Lambda}$ pair production in
two-photon annihilation, and predicts in particular that the cross
section for the mixed $\Lambda \bar{\Sigma}^0 + \Sigma^0
\bar{\Lambda}$ channel should be much lower than for $p\bar{p}$ pairs.

\subsubsection{Meson pair production}
\label{sub:meson-ann}

Two-photon annihilation into pairs of pseudoscalar mesons is in some
respects simpler than annihilation into baryons.  The hard-scattering
mechanism involves a partonic subprocess with fewer internal off-shell
lines, and one might expect it to be relevant at lower values of $s$
than for baryons.  Comparing with the data for $\gamma\gamma\to
\pi^+\pi^-$ at $s$ of order 10~GeV$^2$, one finds however
\cite{Diehl:2001fv} that the leading-twist cross section is well below
the observed one, except if one takes a strongly endpoint concentrated
pion DA, which is in conflict with the measurements of
$\gamma^*\gamma\to \pi^0$ (see Section~\ref{sub:two-photon}).  The
situation does not improve when the effect of parton $k_T$ is included
in the hard scattering \cite{Vogt:2000bz}.

The soft handbag contribution to meson pair production was studied in
\cite{Diehl:2001fv} using the same methods as for baryon pair
production \cite{Diehl:2002yh}.  Evaluating the hard scattering
subprocess with the parton momenta approximated by the momenta of the
final-state mesons, one obtains however a zero result, due to a
conspiracy of invariance under rotation and charge conjugation (which
is not effective in the case of baryons because of their spin degree
of freedom).  The calculation in \cite{Diehl:2001fv} hence included
the first subleading term of the Taylor expansion in the quark
plus-momentum fraction $z$ around $\half$.  The scattering amplitude
is then expressed in terms of the $(2z-1)$ moment of the two-pion
distribution amplitude,
\begin{equation}
  R_{\pi\pi}(s) = \frac{1}{2} \sum_q e_q^2 \int_0^1 dz\,  (2z-1)
	\Phi^q_{\pi\pi}(z,\half, s) .
\end{equation}
According to Section~\ref{sub:gda-evolution} this form factor belongs
to the quark part of the energy-momentum tensor, except for the
weighting by the quark charges.  The unpolarized annihilation cross
section is found to be
\begin{equation}
\frac{d\sigma}{dt}(\gamma\gamma\to \pi^+\pi^-) =
  \frac{\pi \alpha_{\mathrm{em}}^2}{2s^2} 
	\left( \frac{s^2}{tu} \right)^2
	|  R_{\pi\pi}(s) |^2 .
\end{equation}
Notice that by expanding the hard scattering subprocess to first order
in $(2z-1)$ one obtains a cross section going like $s^4/(4tu)^2
\approx 1/\sin^4\theta$, instead of $s^2/(4tu) \approx 1/\sin^2\theta$
in the baryon case.  This gives a rather good description of the
angular dependence seen in preliminary data on $\gamma\gamma\to
\pi^+\pi^-$ and $\gamma\gamma\to K^+K^-$ \cite{Diehl:2001fv}.  Fitting
to the normalization of preliminary cross section data, the same study
obtained an annihilation form factor $|R_{\pi\pi}(s)|$ compatible with
dimensional scaling behavior $s^{-1}$ for $s$ between about 6~GeV$^2$
and 30~GeV$^2$, with a size comparable to the timelike pion form
factor (where data is available for $s \lsim 9$~GeV$^2$).

As in the baryon channels, the absence of isospin $I=2$ states in the
handbag mechanism allows one to relate the cross section for different
final states.  Using in addition SU(3) flavor symmetry (which is
approximately seen in the comparison of $\pi^+\pi^-$ and $K^+K^-$
production data) and neglecting the strange quark contribution to
$R_{\pi\pi}(s)$, one predicts a cross section ratio $\sigma(K^0
\bar{K}{}^0) : \sigma(K^+K^-) \simeq 4 : 25$ for kaons.  In
the two-pion channel only $I=0$ is allowed in the handbag mechanism,
so that one obtains
\begin{equation}
  \label{pion-test}
 \frac{d\sigma}{dt}(\gamma\gamma\to \pi^0\pi^0) =
 \frac{d\sigma}{dt}(\gamma\gamma\to \pi^+\pi^-)  .
\end{equation}
This is in stark contrast to the hard-scattering mechanism, where due
to destructive interference between $I=0$ and $I=2$ amplitudes the
cross section for neutral pions is about an order of magnitude smaller
when a pion DA close to the asymptotic form is used
\cite{Brodsky:1981rp}.   The relation (\ref{pion-test}) is very
robust as it follows only from isospin invariance and from the
assumption that the relevant diagrams have handbag topology, i.e.,
both photons coupling to the same quark line.  Its test at large $s$
may thus be one of the cleanest discriminators between the two
reaction mechanisms.  On the basis of parton-hadron duality Close and
Zhang \cite{Close:2002sc} have suggested that this relation may be
valid at smaller $s$ values, where the detailed mechanism we have
discussed may not yet be applicable.  We finally note that
$\rho^0\rho^0$ and $\rho^+\rho^-$ production are connected as in
(\ref{pion-test}) by the handbag mechanism.


\section{Conclusions}
\label{sec:con}

To conclude, let us return to the introduction and take a second look
at the different motivations for studying generalized parton
distributions and generalized distribution amplitudes.

The link which GPDs provide between the scale evolution of forward
parton densities and of distribution amplitudes gives insight into the
inner working of evolution itself.  It points to a similarity between
GPDs in the ERBL region and meson distribution amplitudes which
persists beyond the formal level: in both cases one probes
quark-antiquark or gluon pairs of small transverse size in a hadron,
and dynamical considerations have shown intimate relations between the
two types of quantities.  A further extension is given by GDAs, which
may be seen as probing small-size parton configurations in the hadron
continuum, or as describing hadronization in a very specific setting.
Both the interplay between GPDs in the DGLAP and the ERBL regions
imposed by Lorentz invariance and the crossing symmetry between GPDs
and GDAs relate ``spacelike'' and ``timelike'' dynamics at the
interface between partons and hadrons.  To understand these relations
at a dynamical level remains a challenge.

Distributions like $E^q$ and $E^g$ are only nonzero because partons
carry orbital angular momentum, as can be seen in the helicity
balance, in the wave function representation, and in Ji's sum rule.
The state of the art does not yet allow us to quantify how precisely
one may evaluate this sum rule from experimental data.  Ji's sum rule
provides a very concise way of quantifying the role of orbital angular
momentum, but one should not forget that comparing for instance the
$\xi$ and $t$ dependence of $E^q(\xi,\xi,t)$ and $H^q(\xi,\xi,t)$
(which is more directly accessible in measurements) would already give
information about this poorly known degree of freedom.  The wave
function and impact parameter representations provide heuristic tools
to interpret this information.

Whereas the evaluation of Ji's sum rule puts a focus on the region of
small $t$ (to be eventually continued to $t=0$), the $t$ dependence by
itself opens up a new dimension.  Going beyond the ``one dimensional
projection'' inherent in conventional parton densities, it provides
access to the transverse structure of hadrons.  The impact parameter
representation formulates this in a manner well adapted to the parton
picture and to relativistic field theory.  Elastic form factors like
the electromagnetic ones specify the size of hadrons in a ``global''
way, independent of whether the hadron is probed at a resolution of
$0.5$~fm or $0.05$~fm.  Higher moments of GPDs, or the combined
dependence of GPDs on longitudinal and transverse variables, will give
specific information about spatial structure at parton level,
distinguishing slow from fast partons and quarks from gluons, which
play very different roles in the dynamics.

The impact parameter picture of GPDs is most simple for $\xi=0$, where
it extends the density interpretation which makes the parton model so
intuitive.  The relevant processes are however measured at finite
$\xi$.  We do not see this as a terrible shortcoming.  On one hand the
transverse ``shift'' induced by nonzero $\xi$ is not large in an
important part of experimental phase space.  Having $\xi$ not exactly
zero is on the other hand what gives access to the rather unique
points $x=\pm \xi$ and to the region $-\xi<x<\xi$.  To adequately
describe them is a test for major approaches to understanding hadron
structure: examples we have discussed are chiral dynamics and the
large-$N_c$ limit, the concept of constituent quarks, and efforts
starting from light-cone quantization and wave functions.  The points
$x=\pm \xi$ are rather directly accessible in DVCS and light meson
production, and if DDVCS can be experimentally realized the region
$-\xi<x<\xi$ will be as well.

Even in small-$x$ physics, where one initially expected the difference
between forward and nonforward distributions to be least important,
theoretical studies and comparison with data have revealed important
effects of nonzero skewness $\xi$.  Whether for experimentally
relevant kinematics these effects can be ascribed to perturbative
evolution from a low starting scale (which would allow one to
constrain the ordinary gluon distribution in exclusive processes)
remains a matter of appreciation.  Given our arguments in
Section~\ref{sec:small-x-gpd} we feel that results like the Shuvaev
formula (\ref{shuv-approx}) should rather be tested than used.  Better
understanding GPDs in this region is ultimately related with
understanding the dynamics of small-$x$ partons beyond the leading
$\log\frac{1}{x}$ approximation.

Other fields of study are still rather unexplored, like transition
GPDs and GPDs of nuclei.  To the extent that these can be studied in
experiment, theoretical investigation will hopefully reveal their
physics potential in more detail.

The dynamics of exclusive processes at large $s$ and large $t$ remains
a poorly understood area of QCD.  The concept of GPDs and GDAs allows
one to describe soft Feynman-type contributions in a two-step
strategy, separating the dynamical mechanism from quantitative aspects
of the hadronic structure.  Applying this strategy to processes like
wide-angle Compton scattering or two-photon annihilation into meson
pairs has identified a number of key observables which should help to
pin down the underlying dynamics.

Where do we stand?  The wealth of physics information in the functions
we have discussed comes with a considerable degree of complexity,
which must be managed in practice.  The theory for calculating
two-photon processes like DVCS from given GPDs is well advanced, at
the level of NLO for leading twist and for part of the leading
power-suppressed terms.  Likewise, theoretical tools are in place to
attempt an extraction of information at amplitude level from
experimental observables.  To take the final step and reconstruct GPDs
from amplitudes will require some prior understanding of the interplay
between $x$ and $\xi$ in these functions, and the most promising
strategy at present is to combine theory studies of dynamics and
phenomenological analysis of data in a learning process.

A quantitative description of hard meson production in the framework
of GPDs remains a challenge to theory.  In a sense these processes
have ``inherited'' the dynamical complexity of meson form factors at
large momentum transfer concerning the importance of radiative and
power corrections.  It should however not be overlooked that meson
production provides a wealth of observables from which different
aspects of dynamics can be studied phenomenologically.  A minimum
requirement is to substantiate which observables are least prone to
theory uncertainties.  Then one may gain quantitative information,
both about GPDs and nucleon structure, and about the small-scale
structure of mesons and continuum states from their distribution
amplitudes.  The latter can be studied independently in exclusive
reactions at high-luminosity $e^+ e^-$ colliders, which would be most
valuable.

Experiment has seen impressive progress in the study of hard exclusive
processes in the last years, with important measurements at facilities
that have not been specifically designed to this end.  Further
progress in the immediate future can be expected in a wide range of
kinematics, from collider energies at H1 and ZEUS to fixed-target
energies of HERMES (who already have preliminary results on the beam
charge asymmetry in DVCS \cite{Ellinghaus:2002bq}) and of the
Jefferson Lab experiments \cite{Chen:2000jl,Sabatie:2002he}.
Measurements in a region between these extremes will hopefully come
from COMPASS at CERN \cite{d'Hose:2002ia}.  On an intermediate time
scale there is the proposed energy upgrade to 12~GeV at JLAB
\cite{Cardman:2001jl} and the project of an Electron-Ion-Collider
EIC \cite{Holt:2002bn}.  At a certain level of detail and precision,
studying the physics described in the review will likely require a
dedicated experimental facility.


\section*{Acknowledgments}

I have very much benefited from discussions with many experimental and
theoretical colleagues over the years.  Space does not permit to list
them all.  Thanks go in particular to E.~R.~Berger, S.~J.~Brodsky,
F. Cano, J.~C.~Collins, Th.~Feldmann, T.~Gousset, H.~W.~Huang,
D.~S.~Hwang, R. Jakob, P.~Kroll, B. Pire, J.~P.~Ralston, O. Teryaev,
and C. Vogt, with whom I had the pleasure to collaborate on subjects
reviewed here.  Special thanks go to D.~M\"uller and M.~V.~Polyakov
for patiently answering innumerable questions concerning their work.
I am indebted to M.~Amarian, J.~C.~Collins, Th.~Feldmann,
G. Iacobucci, R. Jakob, X. Janssen, A.~D.~Martin, B. Pire, and
M.~V.~Polyakov for valuable remarks on parts of the manuscript.

I thank the Institute for Nuclear Theory at the University of
Washington for its hospitality and the Department of Energy for
partial support during the completion of this work.


\section*{Note added}

Concerning the behavior of GPDs at $x=\xi$ (Sections
\ref{sec:x-equal-xi} and \ref{sec:transverse-mesons}), Braun et
al.~\cite{Braun:2002wu} have observed that for the perturbatively
calculated gluon GPD of a free quark target not only $H^g(x,\xi,t)$
but also its first derivative in $x$ is continuous at this point.

Chen and Savage \cite{Chen:2003jm} have investigated the GPDs for the
transition from a nucleon to a nucleon and a soft pion
(Section~\ref{sub:chiral}).  They argue that chiral symmetry is
insufficient to reduce these quantities to the elastic nucleon GPDs
and claim that the results by Guichon et al.~\cite{Guichon:2003ah} are
incorrect.

Hoodbhoy et al.~\cite{Hoodbhoy:2003uu} have recalculated the GPDs of
the pion in the large $t$ limit (Section~\ref{sub:large-t-limit}) and
revealed an error in the result by Vogt \cite{Vogt:2001if}.


\appendix

\section{Acronyms}
\label{app:acro}

In the following we list the acronyms used in the text and refer to
sections where the corresponding notions are defined or discussed in
detail.

\bigskip

\begin{tabular}{ll}
DA    & Distribution amplitude (Section~\ref{sec:gda}) \\
DIS   & Deeply inelastic scattering \\
DGLAP & Dokshitser, Gribov, Lipatov, Altarelli, Parisi
        (Sections~\ref{sec:definitions}, \ref{sec:evolution}) \\
DVCS  & Deeply virtual Compton scattering (Sections~\ref{sec:factor},
        \ref{sec:compton-scatt}, \ref{sec:dvcs-pheno}) \\
DDVCS & Double deeply virtual Compton scattering
        (Sections~\ref{sec:factor}, \ref{sec:compton-scatt}) \\
ERBL  & Efremov, Radyushkin, Brodsky, Lepage (Sections
        \ref{sec:definitions}, \ref{sec:gda}, \ref{sec:evolution}) \\
GDA   & Generalized distribution amplitude (Section \ref{sec:gda}) \\
GPD   & Generalized parton distribution \\
LO    & Leading order \\
NLO   & Next-to-leading order \\
OPE   & Operator product expansion (Section \ref{sub:twist}) \\
QCD   & Quantum Chromodynamics \\
TCS   & Timelike Compton scattering (Sections \ref{sec:factor},
        \ref{sec:compton-scatt}, \ref{sec:tcs})
\end{tabular}


\section{Light-cone helicity spinors}
\label{app:spinors}

Here we give the explicit spinors we have used in calculations for
fermions with definite light-cone helicity~\cite{Soper:1972xc}.  In
the usual Dirac representation they read
\begin{eqnarray}
  \label{spinors}
u(p,+) &=& N
     \left( \begin{array}{c} 
      p^0 + p^3 + m \\ p^1 + i p^2 \\ p^0 + p^3 - m \\ p^1 + i p^2
     \end{array} \right) ,
\hspace{2.5em}
u(p,-) = N
     \left( \begin{array}{c} 
      - p^1 + i p^2 \\ p^0 + p^3 + m \\ p^1 - i p^2 \\ - p^0 - p^3 + m
     \end{array} \right) ,
\nonumber \\
v(p,+) &=& - N
     \left( \begin{array}{c} 
      - p^1 + i p^2 \\ p^0 + p^3 - m \\ p^1 - i p^2 \\ - p^0 - p^3 - m
     \end{array} \right) ,
\hspace{0.8em}
v(p,-) = - N
     \left( \begin{array}{c} 
     p^0 + p^3 - m  \\ p^1 + i p^2 \\ p^0 + p^3 + m \\ p^1 + i p^2
     \end{array} \right) ,
\end{eqnarray}
where $N^{-1} = \sqrt{2(p^0 + p^3)}$.  For quark spinors this
corresponds to the phase conventions of Brodsky and Lepage
\cite{Brodsky:1998de}, whereas for antiquark spinors we differ from
\cite{Brodsky:1998de} by the overall sign.  The spinors in
(\ref{spinors}) satisfy the charge conjugation relations $v(p,\nu) =
S(C)\, \bar{u}^T(p,\nu)$ with $S(C) = i\gamma^2\gamma^0$.  For
massless spinors one simply has $v(p,\nu) = -u(p,-\nu)$.  The
covariant spin vectors belonging to our spinors are given by
\begin{equation}
s = \frac{2\nu}{m} \left( p - \frac{m^2}{p n_-}\, n_- \right) 
\end{equation}
for helicity $\nu=\pm \half$.  This vector is normalized as $s^2 =
-1$.  In the frame where the particle is at rest, $s = (0,\tvec{s})$
points into the spin direction.  It appears in the projector relations
\begin{eqnarray}
u(p,\nu)\, \bar{u}(p,\nu) &=& (\slash{p} + m)\,
	 \frac{1 + \gamma_5 \slash{s}}{2} ,
\nonumber \\
v(p,\nu)\, \bar{v}(p,\nu) &=& (\slash{p} - m)\,
	 \frac{1 + \gamma_5 \slash{s}}{2} .
\end{eqnarray}

For states with ordinary helicity we use the convention
\renewcommand{\arraystretch}{1.2}
\begin{eqnarray}
  \label{hel-spinors}
u(p,+) &=& 
     \left( \begin{array}{r} 
       \sqrt{p^0 + m}\,\; \chi_+(p) \\
       \sqrt{p^0 - m}\; \chi_+(p)
     \end{array} \right) ,
\hspace{2.5em}
u(p,-) =
     \left( \begin{array}{r} 
       \sqrt{p^0 + m}\,\; \chi_-(p) \\
       - \sqrt{p^0 - m}\; \chi_-(p) 
     \end{array} \right) ,
\nonumber \\
v(p,+) &=& {}-
     \left( \begin{array}{r} 
       \sqrt{p^0 - m}\; \chi_-(p) \\
       - \sqrt{p^0 + m}\,\; \chi_-(p)
     \end{array} \right) ,
\hspace{0.5em}
v(p,-) = {}-
     \left( \begin{array}{r} 
       \sqrt{p^0 - m}\; \chi_+(p) \\
       \sqrt{p^0 + m}\,\; \chi_+(p)
     \end{array} \right)
\nonumber \\
\end{eqnarray}
with two-spinors
\begin{eqnarray}
\chi_+(p) &=& \frac{1}{\sqrt{2 |\vec{p}\,|\, (|\vec{p}\,| + p^3)}}\,
     \left( \begin{array}{c} 
        |\vec{p}\,| + p^3 \\
	\phantom{-} p^1 + i p^2
     \end{array} \right)
 =  \left( \begin{array}{c} 
        \cos\half\vartheta \\
	\phantom{-} e^{i\varphi}\, \sin\half\vartheta
     \end{array} \right) ,
\nonumber \\
\chi_-(p) &=& \frac{1}{\sqrt{2 |\vec{p}\,|\, (|\vec{p}\,| + p^3)}}\,
     \left( \begin{array}{c} 
	- p^1 + i p^2 \\
        |\vec{p}\,| + p^3
     \end{array} \right)
 =  \left( \begin{array}{c} 
	- e^{-i\varphi}\, \sin\half\vartheta  \\
        \cos\half\vartheta
     \end{array} \right) ,
\end{eqnarray}
where $\vec{p} = (p^1, p^2, p^3)$ is the three-momentum vector and
$\vartheta$, $\varphi$ are its polar angles.  The transformation from
the light-cone spinors $u_{\mathrm{LC}}$, $v_{\mathrm{LC}}$ in
(\ref{spinors}) to the spinors $u_{\mathrm{H}}$, $v_{\mathrm{H}}$ in
(\ref{hel-spinors}) reads
\renewcommand{\arraystretch}{1.1}
\begin{equation}
\left( \begin{array}{r} u_{\mathrm{H}}(+) \\ 
                        u_{\mathrm{H}}(-) \end{array} \right) = \\
U
\left( \begin{array}{r} u_{\mathrm{LC}}(+) \\ 
                        u_{\mathrm{LC}}(-) 
       \end{array} \right) , \qquad
\left( \begin{array}{r} v_{\mathrm{H}}(+) \\ 
                        v_{\mathrm{H}}(-) \end{array} \right) = \\
U^*
\left( \begin{array}{r} v_{\mathrm{LC}}(+) \\ 
                        v_{\mathrm{LC}}(-) 
       \end{array} \right)
\end{equation}
with a unitary matrix
\begin{eqnarray}
U &=&  \frac{\sqrt{p^0+|\vec{p}\,|}}{\sqrt{2 
             |\vec{p}\,|\, (|\vec{p}\,| + p^3)\, (p^0 + p^3)}}
\renewcommand{\arraystretch}{2}
\left( \begin{array}{cc} 
   |\vec{p}\,| + p^3 \, , \ \ &
   (p^1 + i p^2)\, \displaystyle\frac{m}{p^0 + |\vec{p}\,|} \\
 - (p^1 - i p^2)\, \displaystyle\frac{m}{p^0 + |\vec{p}\,|} \, , &
   |\vec{p}\,| + p^3
       \end{array} \right) 
\nonumber \\[0.5em]
 &=& \frac{\sqrt{p^0+|\vec{p}\,|}}{
	\sqrt{p^0 + p^3 \phantom{|}}} \,
\left( \begin{array}{cc}
   \cos\half\vartheta \, , & e^{i\varphi}\, \sin\half\vartheta\;
                        \displaystyle\frac{m}{p^0 + |\vec{p}\,|} \\
   - e^{-i\varphi}\, \sin\half\vartheta\;
   \displaystyle\frac{m}{p^0 + |\vec{p}\,|} \, , & \cos\half\vartheta
       \end{array} \right) .
\renewcommand{\arraystretch}{1}
\end{eqnarray}
We see that the ratio $|U_{+-} /U_{++}| = |U_{-+} /U_{--}|$ between
off-diagonal and diagonal elements in this matrix is not greater than
$(\tan\half\vartheta)\, m /(2 |\vec{p}\,|)$. This means that for a
particle moving fast at an angle $\vartheta \neq \pi$ the difference
between usual and light-cone helicity is small.



\newpage

\begin{thebibliography}{100}

\bibitem{Altarelli:1972sw}
G.~Altarelli and G.~Preparata,
\newblock Phys. Lett. {\bf B39}, 371 (1972).

\bibitem{Gatto:1972sy}
R.~Gatto and G.~Preparata,
\newblock Nucl. Phys. {\bf B47}, 313 (1972).

\bibitem{DeRujula:1973aa}
A.~De~R{\'u}jula and E.~De~Rafael,
\newblock Annals Phys. {\bf 78}, 132 (1973).

\bibitem{Wieczorek:1974zu}
E.~Wieczorek, V.~A. Matveev, and D.~Robaschik,
\newblock Theor. Math. Phys. {\bf 19}, 315 (1974).

\bibitem{Watanabe:1981ce}
K.~Watanabe,
\newblock Prog. Theor. Phys. {\bf 66}, 1003 (1981).

\bibitem{Watanabe:1982ue}
K.~Watanabe,
\newblock Prog. Theor. Phys. {\bf 67}, 1834 (1982).

\bibitem{Geyer:1985vw}
B.~Geyer, D.~Robaschik, M.~Bordag, and J.~Ho\v{r}ej\v{s}i,
\newblock Z. Phys. {\bf C26}, 591 (1985).

\bibitem{Braunschweig:1986nr}
T.~Braunschweig, B.~Geyer, J.~Ho\v{r}ej\v{s}i, and D.~Robaschik,
\newblock Z. Phys. {\bf C33}, 275 (1986).

\bibitem{Dittes:1988xz}
F.~M. Dittes, D.~M{\"u}ller, D.~Robaschik, B.~Geyer, and J.~Ho\v{r}ej\v{s}i,
\newblock Phys. Lett. {\bf B209}, 325 (1988).

\bibitem{Muller:1994fv}
D.~M{\"u}ller, D.~Robaschik, B.~Geyer, F.~M. Dittes, and J.~Ho\v{r}ej\v{s}i,
\newblock Fortschr. Phys. {\bf 42}, 101 (1994), hep-ph/9812448.

\bibitem{Gribov:1972ri}
V.~N. Gribov and L.~N. Lipatov,
\newblock Sov. J. Nucl. Phys. {\bf 15}, 438 (1972).

\bibitem{Lipatov:1975qm}
L.~N. Lipatov,
\newblock Sov. J. Nucl. Phys. {\bf 20}, 94 (1975).

\bibitem{Altarelli:1977zs}
G.~Altarelli and G.~Parisi,
\newblock Nucl. Phys. {\bf B126}, 298 (1977).

\bibitem{Dokshitzer:1977sg}
Y.~L. Dokshitzer,
\newblock Sov. Phys. JETP {\bf 46}, 641 (1977).

\bibitem{Efremov:1980qk}
A.~V. Efremov and A.~V. Radyushkin,
\newblock Phys. Lett. {\bf B94}, 245 (1980).

\bibitem{Lepage:1979zb}
G.~P. Lepage and S.~J. Brodsky,
\newblock Phys. Lett. {\bf B87}, 359 (1979).

\bibitem{Bartels:1982jh}
J.~Bartels and M.~Loewe,
\newblock Zeit. Phys. {\bf C12}, 263 (1982).

\bibitem{Ryskin:1993ui}
M.~G. Ryskin,
\newblock Z. Phys. {\bf C57}, 89 (1993).

\bibitem{Brodsky:1994kf}
S.~J. Brodsky, L.~Frankfurt, J.~F. Gunion, A.~H. Mueller, and M.~Strikman,
\newblock Phys. Rev. {\bf D50}, 3134 (1994), hep-ph/9402283.

\bibitem{Ji:1997ek}
X.-D. Ji,
\newblock Phys. Rev. Lett. {\bf 78}, 610 (1997), hep-ph/9603249.

\bibitem{Radyushkin:1996nd}
A.~V. Radyushkin,
\newblock Phys. Lett. {\bf B380}, 417 (1996), hep-ph/9604317.

\bibitem{Ji:1997nm}
X.-D. Ji,
\newblock Phys. Rev. {\bf D55}, 7114 (1997), hep-ph/9609381.

\bibitem{Radyushkin:1996ru}
A.~V. Radyushkin,
\newblock Phys. Lett. {\bf B385}, 333 (1996), hep-ph/9605431.

\bibitem{Collins:1997fb}
J.~C. Collins, L.~Frankfurt, and M.~Strikman,
\newblock Phys. Rev. {\bf D56}, 2982 (1997), hep-ph/9611433.

\bibitem{Jain:1993jf}
P.~Jain and J.~P. Ralston,
\newblock hep-ph/9305250.

\bibitem{Burkardt:2000za}
M.~Burkardt,
\newblock Phys. Rev. {\bf D62}, 071503 (2000), hep-ph/0005108,
\newblock Erratum-ibid.\ D {\bf 66}, 119903 (2002).

\bibitem{Grozin:1983aa}
A.~G. Grozin,
\newblock Sov. J. Nucl. Phys. {\bf 38}, 289 (1983).

\bibitem{Baier:1985aa}
V.~A. Baier and A.~G. Grozin,
\newblock Sov. J. Part. Nucl. {\bf 16}, 1 (1985).

\bibitem{Diehl:1998dk}
M.~Diehl, T.~Gousset, B.~Pire, and O.~Teryaev,
\newblock Phys. Rev. Lett. {\bf 81}, 1782 (1998), hep-ph/9805380.

\bibitem{Radyushkin:1998rt}
A.~V. Radyushkin,
\newblock Phys. Rev. {\bf D58}, 114008 (1998), hep-ph/9803316.

\bibitem{Diehl:1998kh}
M.~Diehl, T.~Feldmann, R.~Jakob, and P.~Kroll,
\newblock Eur. Phys. J. {\bf C8}, 409 (1999), hep-ph/9811253.

\bibitem{Diehl:2001fv}
M.~Diehl, P.~Kroll, and C.~Vogt,
\newblock Phys. Lett. {\bf B532}, 99 (2002), hep-ph/0112274.

\bibitem{Diehl:2002yh}
M.~Diehl, P.~Kroll, and C.~Vogt,
\newblock Eur. Phys. J. {\bf C26}, 567 (2003), hep-ph/0206288.

\bibitem{Freund:2002cq}
A.~Freund, A.~V. Radyushkin, A.~Sch{\"a}fer, and C.~Weiss,
\newblock Phys. Rev. Lett. {\bf 90}, 092001 (2003), hep-ph/0208061.

\bibitem{Ji:1998pc}
X.-D. Ji,
\newblock J. Phys. {\bf G24}, 1181 (1998), hep-ph/9807358.

\bibitem{Radyushkin:2000uy}
A.~V. Radyushkin,
\newblock hep-ph/0101225.

\bibitem{Goeke:2001tz}
K.~Goeke, M.~V. Polyakov, and M.~Vanderhaeghen,
\newblock Prog. Part. Nucl. Phys. {\bf 47}, 401 (2001), hep-ph/0106012.

\bibitem{Lepage:1980fj}
G.~P. Lepage and S.~J. Brodsky,
\newblock Phys. Rev. {\bf D22}, 2157 (1980).

\bibitem{Radyushkin:1997ki}
A.~V. Radyushkin,
\newblock Phys. Rev. {\bf D56}, 5524 (1997), hep-ph/9704207.

\bibitem{Golec-Biernat:1998ja}
K.~J. Golec-Biernat and A.~D. Martin,
\newblock Phys. Rev. {\bf D59}, 014029 (1999), hep-ph/9807497.

\bibitem{Diehl:2000xz}
M.~Diehl, T.~Feldmann, R.~Jakob, and P.~Kroll,
\newblock Nucl. Phys. {\bf B596}, 33 (2001), hep-ph/0009255,
\newblock Erratum-ibid.\ B {\bf 605}, 647 (2001).

\bibitem{Polyakov:2003pr}
M.~V. Polyakov,
\newblock private communication (2003).

\bibitem{Pobylitsa:2002vi}
P.~V. Pobylitsa,
\newblock Phys. Rev. {\bf D67}, 034009 (2003), hep-ph/0210150.

\bibitem{Tiburzi:2002kr}
B.~C. Tiburzi and G.~A. Miller,
\newblock Phys. Rev. {\bf D67}, 013010 (2003), hep-ph/0209178.

\bibitem{Landshoff:1971ff}
P.~V. Landshoff, J.~C. Polkinghorne, and R.~D. Short,
\newblock Nucl. Phys. {\bf B28}, 225 (1971).

\bibitem{Frankfurt:1998ha}
L.~Frankfurt, A.~Freund, V.~Guzey, and M.~Strikman,
\newblock Phys. Lett. {\bf B418}, 345 (1998), hep-ph/9703449.

\bibitem{Diehl:1998sm}
M.~Diehl and T.~Gousset,
\newblock Phys. Lett. {\bf B428}, 359 (1998), hep-ph/9801233.

\bibitem{Jaffe:1983hp}
R.~L. Jaffe,
\newblock Nucl. Phys. {\bf B229}, 205 (1983).

\bibitem{Kogut:1970xa}
J.~B. Kogut and D.~E. Soper,
\newblock Phys. Rev. {\bf D1}, 2901 (1970).

\bibitem{Brodsky:1989pv}
S.~J. Brodsky and G.~P. Lepage,
\newblock in: A. H. Mueller (Ed.), \emph{Perturbative {Q}uantum
  {C}hromodynamics} (World Scientific, Singapore, 1989) p.~93.

\bibitem{Brodsky:1998de}
S.~J. Brodsky, H.-C. Pauli, and S.~S. Pinsky,
\newblock Phys. Rept. {\bf 301}, 299 (1998), hep-ph/9705477.

\bibitem{Jaffe:1996zw}
R.~L. Jaffe,
\newblock hep-ph/9602236.

\bibitem{Yamawaki:1998cy}
K.~Yamawaki,
\newblock hep-th/9802037.

\bibitem{Soper:1972xc}
D.~E. Soper,
\newblock Phys. Rev. {\bf D5}, 1956 (1972).

\bibitem{Diehl:2001pm}
M.~Diehl,
\newblock Eur. Phys. J. {\bf C19}, 485 (2001), hep-ph/0101335.

\bibitem{Belitsky:2000jk}
A.~V. Belitsky and D.~M{\"u}ller,
\newblock Phys. Lett. {\bf B486}, 369 (2000), hep-ph/0005028.

\bibitem{Hoodbhoy:1998vm}
P.~Hoodbhoy and X.-D. Ji,
\newblock Phys. Rev. {\bf D58}, 054006 (1998), hep-ph/9801369.

\bibitem{Ralston:1979ys}
J.~P. Ralston and D.~E. Soper,
\newblock Nucl. Phys. {\bf B152}, 109 (1979).

\bibitem{Jaffe:1997yz}
R.~L. Jaffe,
\newblock hep-ph/9710465.

\bibitem{Barone:2001sp}
V.~Barone, A.~Drago, and P.~G. Ratcliffe,
\newblock Phys. Rept. {\bf 359}, 1 (2002), hep-ph/0104283.

\bibitem{Jaffe:1989xy}
R.~L. Jaffe and A.~Manohar,
\newblock Phys. Lett. {\bf B223}, 218 (1989).

\bibitem{Artru:1990zv}
X.~Artru and M.~Mekhfi,
\newblock Z. Phys. {\bf C45}, 669 (1990).

\bibitem{Ji:2000id}
X.-D. Ji and R.~F. Lebed,
\newblock Phys. Rev. {\bf D63}, 076005 (2001), hep-ph/0012160.

\bibitem{Windmolders:1999ra}
R.~Windmolders,
\newblock Nucl. Phys. Proc. Suppl. {\bf 79}, 51 (1999), hep-ph/9905505.

\bibitem{Filippone:2001ux}
B.~W. Filippone and X.-D. Ji,
\newblock Adv. Nucl. Phys. {\bf 26}, 1 (2001), hep-ph/0101224.

\bibitem{Sehgal:1974rz}
L.~M. Sehgal,
\newblock Phys. Rev. {\bf D10}, 1663 (1974),
\newblock Erratum-ibid.\ D {\bf 11}, 2016 (1975).

\bibitem{Ratcliffe:1987dp}
P.~G. Ratcliffe,
\newblock Phys. Lett. {\bf B192}, 180 (1987).

\bibitem{Ji:1998pf}
X.-D. Ji,
\newblock Phys. Rev. {\bf D58}, 056003 (1998), hep-ph/9710290.

\bibitem{Brodsky:2000ii}
S.~J. Brodsky, D.~S. Hwang, B.-Q. Ma, and I.~Schmidt,
\newblock Nucl. Phys. {\bf B593}, 311 (2001), hep-th/0003082.

\bibitem{Teryaev:1999su}
O.~V. Teryaev,
\newblock hep-ph/9904376.

\bibitem{Gockeler:2003jf}
QCDSF Collaboration, M.~G{\"o}ckeler {\em et~al.},
\newblock hep-ph/0304249.

\bibitem{Hagler:2003jd}
LHPC and SESAM Collaborations, P.~H{\"a}gler {\em et~al.},
\newblock hep-lat/0304018.

\bibitem{Hoodbhoy:1998bt}
P.~Hoodbhoy, X.-D. Ji, and W.~Lu,
\newblock Phys. Rev. {\bf D59}, 074010 (1999), hep-ph/9808305.

\bibitem{Jaffe:1990jz}
R.~L. Jaffe and A.~Manohar,
\newblock Nucl. Phys. {\bf B337}, 509 (1990).

\bibitem{Hoodbhoy:1998yb}
P.~Hoodbhoy, X.-D. Ji, and W.~Lu,
\newblock Phys. Rev. {\bf D59}, 014013 (1999), hep-ph/9804337.

\bibitem{Jaffe:2000kr}
R.~L. Jaffe,
\newblock Phil. Trans. Roy. Soc. Lond. {\bf A359}, 391 (2001), hep-ph/0008038.

\bibitem{Balitsky:1997rs}
I.~Balitsky and X.-D. Ji,
\newblock Phys. Rev. Lett. {\bf 79}, 1225 (1997), hep-ph/9702277.

\bibitem{Barone:1998dx}
V.~Barone, T.~Calarco, and A.~Drago,
\newblock Phys. Lett. {\bf B431}, 405 (1998), hep-ph/9801281.

\bibitem{Mathur:1999uf}
N.~Mathur, S.~J. Dong, K.~F. Liu, L.~Mankiewicz, and N.~C. Mukhopadhyay,
\newblock Phys. Rev. {\bf D62}, 114504 (2000), hep-ph/9912289.

\bibitem{Gadiyak:2001fe}
V.~Gadiyak, X.-D. Ji, and C.-W. Jung,
\newblock Phys. Rev. {\bf D65}, 094510 (2002), hep-lat/0112040.

\bibitem{Bass:2001dg}
S.~D. Bass,
\newblock Phys. Rev. {\bf D65}, 074025 (2002), hep-ph/0102036.

\bibitem{Polyakov:2002wz}
M.~V. Polyakov and A.~G. Shuvaev,
\newblock hep-ph/0207153.

\bibitem{Polyakov:2002yz}
M.~V. Polyakov,
\newblock Phys. Lett. {\bf B555}, 57 (2003), hep-ph/0210165.

\bibitem{Sachs:1962aa}
R.~G. Sachs,
\newblock Phys. Rev. {\bf 126}, 2256 (1962).

\bibitem{Diehl:2000uv}
M.~Diehl, T.~Gousset, and B.~Pire,
\newblock Phys. Rev. {\bf D62}, 073014 (2000), hep-ph/0003233.

\bibitem{Kivel:1999sd}
N.~Kivel, L.~Mankiewicz, and M.~V. Polyakov,
\newblock Phys. Lett. {\bf B467}, 263 (1999), hep-ph/9908334.

\bibitem{Polyakov:1998ze}
M.~V. Polyakov,
\newblock Nucl. Phys. {\bf B555}, 231 (1999), hep-ph/9809483.

\bibitem{Pire:2000ky}
B.~Pire and O.~V. Teryaev,
\newblock Phys. Lett. {\bf B496}, 76 (2000), hep-ph/0007014.

\bibitem{Anikin:2003fr}
I.~V. Anikin, B.~Pire, and O.~V. Teryaev,
\newblock hep-ph/0307059.

\bibitem{Polyakov:1999gs}
M.~V. Polyakov and C.~Weiss,
\newblock Phys. Rev. {\bf D60}, 114017 (1999), hep-ph/9902451.

\bibitem{Watson:1954uc}
K.~M. Watson,
\newblock Phys. Rev. {\bf 95}, 228 (1954).

\bibitem{Chase:1980hj}
M.~K. Chase,
\newblock Nucl. Phys. {\bf B174}, 109 (1980).

\bibitem{Baier:1982aa}
A.~G. Baier, V. A.~Grozin,
\newblock Sov. J. Nucl. Phys. {\bf 35}, 596 (1982).

\bibitem{Kroll:2002nt}
P.~Kroll and K.~Passek-Kumeri\v{c}ki,
\newblock Phys. Rev. {\bf D67}, 054017 (2003), hep-ph/0210045.

\bibitem{Gribov:1983tu}
L.~V. Gribov, E.~M. Levin, and M.~G. Ryskin,
\newblock Phys. Rept. {\bf 100}, 1 (1983).

\bibitem{Martin:1998wy}
A.~D. Martin and M.~G. Ryskin,
\newblock Phys. Rev. {\bf D57}, 6692 (1998), hep-ph/9711371.

\bibitem{Balitsky:1997mj}
I.~I. Balitsky and A.~V. Radyushkin,
\newblock Phys. Lett. {\bf B413}, 114 (1997), hep-ph/9706410.

\bibitem{Radyushkin:1998es}
A.~V. Radyushkin,
\newblock Phys. Rev. {\bf D59}, 014030 (1999), hep-ph/9805342.

\bibitem{Blumlein:1997pi}
J.~Bl{\"u}mlein, B.~Geyer, and D.~Robaschik,
\newblock Phys. Lett. {\bf B406}, 161 (1997), hep-ph/9705264.

\bibitem{Blumlein:1999sc}
J.~Bl{\"u}mlein, B.~Geyer, and D.~Robaschik,
\newblock Nucl. Phys. {\bf B560}, 283 (1999), hep-ph/9903520.

\bibitem{Belitsky:1998vj}
A.~V. Belitsky and D.~M{\"u}ller,
\newblock Nucl. Phys. {\bf B527}, 207 (1998), hep-ph/9802411.

\bibitem{Belitsky:1998gc}
A.~V. Belitsky and D.~M{\"u}ller,
\newblock Nucl. Phys. {\bf B537}, 397 (1999), hep-ph/9804379.

\bibitem{Belitsky:1999gu}
A.~V. Belitsky, D.~M{\"u}ller, and A.~Freund,
\newblock Phys. Lett. {\bf B461}, 270 (1999), hep-ph/9904477.

\bibitem{Belitsky:1999fu}
A.~V. Belitsky and D.~M{\"u}ller,
\newblock Phys. Lett. {\bf B464}, 249 (1999), hep-ph/9906409.

\bibitem{Belitsky:1999hf}
A.~V. Belitsky, A.~Freund, and D.~M{\"u}ller,
\newblock Nucl. Phys. {\bf B574}, 347 (2000), hep-ph/9912379.

\bibitem{Belitsky:2000yn}
A.~V. Belitsky, A.~Freund, and D.~M{\"u}ller,
\newblock Phys. Lett. {\bf B493}, 341 (2000), hep-ph/0008005.

\bibitem{Bukhvostov:1985rn}
A.~P. Bukhvostov, G.~V. Frolov, L.~N. Lipatov, and E.~A. Kuraev,
\newblock Nucl. Phys. {\bf B258}, 601 (1985).

\bibitem{Balitsky:1989bk}
I.~I. Balitsky and V.~M. Braun,
\newblock Nucl. Phys. {\bf B311}, 541 (1989).

\bibitem{Belitsky:1998uk}
A.~V. Belitsky, D.~M{\"u}ller, L.~Niedermeier, and A.~Sch{\"a}fer,
\newblock Nucl. Phys. {\bf B546}, 279 (1999), hep-ph/9810275.

\bibitem{Freund:2001hd}
A.~Freund and M.~McDermott,
\newblock Eur. Phys. J. {\bf C23}, 651 (2002), hep-ph/0111472.

\bibitem{Freund:1998uf}
A.~Freund and V.~Guzey,
\newblock Phys. Lett. {\bf B462}, 178 (1999), hep-ph/9806267.

\bibitem{Golec-Biernat:1998vf}
K.~Golec-Biernat, J.~Kwieci{\'n}ski, and A.~D. Martin,
\newblock Phys. Rev. {\bf D58}, 094001 (1998), hep-ph/9803464.

\bibitem{Golec-Biernat:1999ib}
K.~J. Golec-Biernat, A.~D. Martin, and M.~G. Ryskin,
\newblock Phys. Lett. {\bf B456}, 232 (1999), hep-ph/9903327.

\bibitem{Belitsky:1998pc}
A.~V. Belitsky, B.~Geyer, D.~M{\"u}ller, and A.~Sch{\"a}fer,
\newblock Phys. Lett. {\bf B421}, 312 (1998), hep-ph/9710427.

\bibitem{Musatov:1999xp}
I.~V. Musatov and A.~V. Radyushkin,
\newblock Phys. Rev. {\bf D61}, 074027 (2000), hep-ph/9905376.

\bibitem{Belitsky:1998pq}
A.~V. Belitsky, D.~M{\"u}ller, L.~Niedermeier, and A.~Sch{\"a}fer,
\newblock Phys. Lett. {\bf B437}, 160 (1998), hep-ph/9806232.

\bibitem{Freund:2001bf}
A.~Freund and M.~F. McDermott,
\newblock Phys. Rev. {\bf D65}, 056012 (2002), hep-ph/0106115.

\bibitem{Efremov:1980rn}
A.~V. Efremov and A.~V. Radyushkin,
\newblock Theor. Math. Phys. {\bf 42}, 97 (1980).

\bibitem{Makeenko:1981bh}
Y.~M. Makeenko,
\newblock Sov. J. Nucl. Phys. {\bf 33}, 440 (1981).

\bibitem{Ohrndorf:1982qv}
T.~Ohrndorf,
\newblock Nucl. Phys. {\bf B198}, 26 (1982).

\bibitem{Shuvaev:1999fm}
A.~Shuvaev,
\newblock Phys. Rev. {\bf D60}, 116005 (1999), hep-ph/9902318.

\bibitem{Kivel:1999wa}
N.~Kivel and L.~Mankiewicz,
\newblock Nucl. Phys. {\bf B557}, 271 (1999), hep-ph/9903531.

\bibitem{Mankiewicz:1998uy}
L.~Mankiewicz, G.~Piller, and T.~Weigl,
\newblock Eur. Phys. J. {\bf C5}, 119 (1998), hep-ph/9711227.

\bibitem{Belitsky:1999sg}
A.~V. Belitsky, D.~M{\"u}ller, L.~Niedermeier, and A.~Sch{\"a}fer,
\newblock Phys. Lett. {\bf B474}, 163 (2000), hep-ph/9908337.

\bibitem{Noritzsch:2000pr}
J.~D. Noritzsch,
\newblock Phys. Rev. {\bf D62}, 054015 (2000), hep-ph/0004012.

\bibitem{Shuvaev:1999ce}
A.~G. Shuvaev, K.~J. Golec-Biernat, A.~D. Martin, and M.~G. Ryskin,
\newblock Phys. Rev. {\bf D60}, 014015 (1999), hep-ph/9902410.

\bibitem{Radyushkin:1998bz}
A.~V. Radyushkin,
\newblock Phys. Lett. {\bf B449}, 81 (1999), hep-ph/9810466.

\bibitem{Radyushkin:1983wh}
A.~V. Radyushkin,
\newblock Phys. Lett. {\bf B131}, 179 (1983).

\bibitem{Teryaev:2001qm}
O.~V. Teryaev,
\newblock Phys. Lett. {\bf B510}, 125 (2001), hep-ph/0102303.

\bibitem{Belitsky:2000vk}
A.~V. Belitsky, D.~M{\"u}ller, A.~Kirchner, and A.~Sch{\"a}fer,
\newblock Phys. Rev. {\bf D64}, 116002 (2001), hep-ph/0011314.

\bibitem{Kivel:2002ia}
N.~Kivel and M.~V. Polyakov,
\newblock hep-ph/0203264.

\bibitem{Collins:1981uk}
J.~C. Collins and D.~E. Soper,
\newblock Nucl. Phys. {\bf B193}, 381 (1981),
\newblock Erratum-ibid.\ B {\bf 213}, 545 (1983).

\bibitem{Burkardt:2002hr}
M.~Burkardt,
\newblock Int. J. Mod. Phys. {\bf A18}, 173 (2003), hep-ph/0207047.

\bibitem{Diehl:2002he}
M.~Diehl,
\newblock Eur. Phys. J. {\bf C25}, 223 (2002), hep-ph/0205208,
\newblock Erratum-ibid.\ to appear.

\bibitem{Ralston:2001xs}
J.~P. Ralston and B.~Pire,
\newblock Phys. Rev. {\bf D66}, 111501 (2002), hep-ph/0110075.

\bibitem{Soper:1977jc}
D.~E. Soper,
\newblock Phys. Rev. {\bf D15}, 1141 (1977).

\bibitem{Burkardt:2000wq}
M.~Burkardt,
\newblock hep-ph/0008051.

\bibitem{Kogut:1974ub}
J.~B. Kogut and L.~Susskind,
\newblock Phys. Rev. {\bf D9}, 3391 (1974).

\bibitem{Pobylitsa:2002iu}
P.~V. Pobylitsa,
\newblock Phys. Rev. {\bf D66}, 094002 (2002), hep-ph/0204337.

\bibitem{Belitsky:2002ep}
A.~V. Belitsky and D.~M{\"u}ller,
\newblock Nucl. Phys. {\bf A711}, 118 (2002), hep-ph/0206306.

\bibitem{Belitsky:2003tm}
A.~V. Belitsky,
\newblock hep-ph/0307256.

\bibitem{Ji:2003ak}
X.-D. Ji,
\newblock hep-ph/0304037.

\bibitem{Mulders:1996dh}
P.~J. Mulders and R.~D. Tangerman,
\newblock Nucl. Phys. {\bf B461}, 197 (1996), hep-ph/9510301.

\bibitem{Mulders:1999mc}
P.~J. Mulders,
\newblock hep-ph/9912493.

\bibitem{Burkardt:2002ks}
M.~Burkardt,
\newblock Phys. Rev. {\bf D66}, 114005 (2002), hep-ph/0209179.

\bibitem{Burkardt:2003uw}
M.~Burkardt,
\newblock hep-ph/0302144.

\bibitem{Pire:2002ut}
B.~Pire and L.~Szymanowski,
\newblock Phys. Lett. {\bf B556}, 129 (2003), hep-ph/0212296.

\bibitem{Brodsky:2000xy}
S.~J. Brodsky, M.~Diehl, and D.~S. Hwang,
\newblock Nucl. Phys. {\bf B596}, 99 (2001), hep-ph/0009254.

\bibitem{Burkardt:2002uc}
M.~Burkardt, X.-D. Ji, and F.~Yuan,
\newblock Phys. Lett. {\bf B545}, 345 (2002), hep-ph/0205272.

\bibitem{Ji:2002xn}
X.-D. Ji, J.-P. Ma, and F.~Yuan,
\newblock Nucl. Phys. {\bf B652}, 383 (2003), hep-ph/0210430.

\bibitem{Ball:1998fj}
P.~Ball and V.~M. Braun,
\newblock hep-ph/9808229.

\bibitem{Ball:1998je}
P.~Ball,
\newblock JHEP {\bf 01}, 010 (1999), hep-ph/9812375.

\bibitem{Braun:2000kw}
V.~Braun, R.~J. Fries, N.~Mahnke, and E.~Stein,
\newblock Nucl. Phys. {\bf B589}, 381 (2000), hep-ph/0007279.

\bibitem{Tiburzi:2002sx}
B.~C. Tiburzi and G.~A. Miller,
\newblock Phys. Rev. {\bf D67}, 054015 (2003), hep-ph/0210305.

\bibitem{Mankiewicz:1991az}
L.~Mankiewicz and A.~Sch{\"a}fer,
\newblock Phys. Lett. {\bf B265}, 167 (1991).

\bibitem{Mankiewicz:1991ji}
L.~Mankiewicz and Z.~Ryzak,
\newblock Phys. Rev. {\bf D43}, 733 (1991).

\bibitem{Mukherjee:2002xi}
A.~Mukherjee and M.~Vanderhaeghen,
\newblock Phys. Rev. {\bf D67}, 085020 (2003), hep-ph/0211386.

\bibitem{Drell:1970km}
S.~D. Drell and T.-M. Yan,
\newblock Phys. Rev. Lett. {\bf 24}, 181 (1970).

\bibitem{Brodsky:1980zm}
S.~J. Brodsky and S.~D. Drell,
\newblock Phys. Rev. {\bf D22}, 2236 (1980).

\bibitem{Ralston:2002hu}
J.~P. Ralston, P.~Jain, and R.~V. Buniy,
\newblock AIP Conf. Proc. {\bf 549}, 302 (2000), hep-ph/0206074.

\bibitem{Ji:2003fw}
X.-D. Ji, J.-P. Ma, and F.~Yuan,
\newblock hep-ph/0301141.

\bibitem{Ji:2003yj}
X.-D. Ji, J.-P. Ma, and F.~Yuan,
\newblock hep-ph/0304107.

\bibitem{Sawicki:1992qj}
M.~Sawicki,
\newblock Phys. Rev. {\bf D46}, 474 (1992).

\bibitem{Brodsky:1998hn}
S.~J. Brodsky and D.~S. Hwang,
\newblock Nucl. Phys. {\bf B543}, 239 (1999), hep-ph/9806358.

\bibitem{Mukherjee:2002pq}
A.~Mukherjee and M.~Vanderhaeghen,
\newblock Phys. Lett. {\bf B542}, 245 (2002), hep-ph/0206159.

\bibitem{Brodsky:2002ue}
S.~J. Brodsky, P.~Hoyer, N.~Marchal, S.~Peign{\'e}, and F.~Sannino,
\newblock Phys. Rev. {\bf D65}, 114025 (2002), hep-ph/0104291.

\bibitem{Collins:2002kn}
J.~C. Collins,
\newblock Phys. Lett. {\bf B536}, 43 (2002), hep-ph/0204004.

\bibitem{Collins:2003fm}
J.~C. Collins,
\newblock Acta Phys. Polon. {\bf B34}, 3103 (2003), hep-ph/0304122.

\bibitem{Ji:2002aa}
X.-D. Ji and F.~Yuan,
\newblock Phys. Lett. {\bf B543}, 66 (2002), hep-ph/0206057.

\bibitem{Belitsky:2002sm}
A.~V. Belitsky, X.-D. Ji, and F.~Yuan,
\newblock Nucl. Phys. {\bf B656}, 165 (2003), hep-ph/0208038.

\bibitem{Brodsky:2002cx}
S.~J. Brodsky, D.~S. Hwang, and I.~Schmidt,
\newblock Phys. Lett. {\bf B530}, 99 (2002), hep-ph/0201296.

\bibitem{Brodsky:2002rv}
S.~J. Brodsky, D.~S. Hwang, and I.~Schmidt,
\newblock Nucl. Phys. {\bf B642}, 344 (2002), hep-ph/0206259.

\bibitem{Pire:1998nw}
B.~Pire, J.~Soffer, and O.~Teryaev,
\newblock Eur. Phys. J. {\bf C8}, 103 (1999), hep-ph/9804284.

\bibitem{Burkardt:2001ni}
M.~Burkardt,
\newblock hep-ph/0105324.

\bibitem{Pobylitsa:2001nt}
P.~V. Pobylitsa,
\newblock Phys. Rev. {\bf D65}, 077504 (2002), hep-ph/0112322.

\bibitem{Pobylitsa:2002gw}
P.~V. Pobylitsa,
\newblock Phys. Rev. {\bf D65}, 114015 (2002), hep-ph/0201030.

\bibitem{Pobylitsa:2002vw}
P.~V. Pobylitsa,
\newblock Phys. Rev. {\bf D67}, 094012 (2003), hep-ph/0210238.

\bibitem{Pobylitsa:2002ru}
P.~V. Pobylitsa,
\newblock hep-ph/0211160.

\bibitem{Soffer:1995ww}
J.~Soffer,
\newblock Phys. Rev. Lett. {\bf 74}, 1292 (1995), hep-ph/9409254.

\bibitem{Altarelli:1998gn}
G.~Altarelli, S.~Forte, and G.~Ridolfi,
\newblock Nucl. Phys. {\bf B534}, 277 (1998), hep-ph/9806345.

\bibitem{Collins:1998be}
J.~C. Collins and A.~Freund,
\newblock Phys. Rev. {\bf D59}, 074009 (1999), hep-ph/9801262.

\bibitem{Theussl:2002xp}
L.~Theussl, S.~Noguera, and V.~Vento,
\newblock nucl-th/0211036.

\bibitem{Burkardt:2000uu}
M.~Burkardt,
\newblock Phys. Rev. {\bf D62}, 094003 (2000), hep-ph/0005209.

\bibitem{'tHooft:1974hx}
G.~'t~Hooft,
\newblock Nucl. Phys. {\bf B75}, 461 (1974).

\bibitem{Antonuccio:1997tw}
F.~Antonuccio, S.~J. Brodsky, and S.~Dalley,
\newblock Phys. Lett. {\bf B412}, 104 (1997), hep-ph/9705413.

\bibitem{Mankiewicz:1997aa}
L.~Mankiewicz, G.~Piller, and T.~Weigl,
\newblock Phys. Rev. {\bf D59}, 017501 (1999), hep-ph/9712508.

\bibitem{Frankfurt:1999fp}
L.~L. Frankfurt, P.~V. Pobylitsa, M.~V. Polyakov, and M.~Strikman,
\newblock Phys. Rev. {\bf D60}, 014010 (1999), hep-ph/9901429.

\bibitem{Frankfurt:1998jq}
L.~L. Frankfurt, M.~V. Polyakov, and M.~Strikman,
\newblock hep-ph/9808449.

\bibitem{Frankfurt:1999xe}
L.~L. Frankfurt, M.~V. Polyakov, M.~Strikman, and M.~Vanderhaeghen,
\newblock Phys. Rev. Lett. {\bf 84}, 2589 (2000), hep-ph/9911381.

\bibitem{Guichon:2003ah}
P.~A.~M. Guichon, L.~Moss{\'e}, and M.~Vanderhaeghen,
\newblock hep-ph/0305231.

\bibitem{Blumlein:2001sb}
J.~Bl{\"u}mlein, J.~Eilers, B.~Geyer, and D.~Robaschik,
\newblock Phys. Rev. {\bf D65}, 054029 (2002), hep-ph/0108095.

\bibitem{Feldmann:1999sm}
T.~Feldmann and P.~Kroll,
\newblock Eur. Phys. J. {\bf C12}, 99 (2000), hep-ph/9905343.

\bibitem{Berger:2001zb}
E.~R. Berger, F.~Cano, M.~Diehl, and B.~Pire,
\newblock Phys. Rev. Lett. {\bf 87}, 142302 (2001), hep-ph/0106192.

\bibitem{Hoodbhoy:1989am}
P.~Hoodbhoy, R.~L. Jaffe, and A.~Manohar,
\newblock Nucl. Phys. {\bf B312}, 571 (1989).

\bibitem{Close:1990zw}
F.~E. Close and S.~Kumano,
\newblock Phys. Rev. {\bf D42}, 2377 (1990).

\bibitem{Mankiewicz:1998kg}
L.~Mankiewicz, G.~Piller, and A.~Radyushkin,
\newblock Eur. Phys. J. {\bf C10}, 307 (1999), hep-ph/9812467.

\bibitem{Penttinen:1999th}
M.~Penttinen, M.~V. Polyakov, and K.~Goeke,
\newblock Phys. Rev. {\bf D62}, 014024 (2000), hep-ph/9909489.

\bibitem{Weinberg:1996kr}
S.~Weinberg,
\newblock {\em The quantum theory of fields. {V}ol.~{II}: Modern applications}
  (Cambridge University Press, 1996).

\bibitem{Belitsky:2002jp}
A.~V. Belitsky and X.-D. Ji,
\newblock Phys. Lett. {\bf B538}, 289 (2002), hep-ph/0203276.

\bibitem{Kivel:2000fg}
N.~Kivel, M.~V. Polyakov, and M.~Vanderhaeghen,
\newblock Phys. Rev. {\bf D63}, 114014 (2001), hep-ph/0012136.

\bibitem{'tHooft:1974jz}
G.~'t~Hooft,
\newblock Nucl. Phys. {\bf B72}, 461 (1974).

\bibitem{Witten:1979kh}
E.~Witten,
\newblock Nucl. Phys. {\bf B160}, 57 (1979).

\bibitem{Petrov:1998kf}
V.~Y. Petrov {\em et~al.},
\newblock Phys. Rev. {\bf D57}, 4325 (1998), hep-ph/9710270.

\bibitem{Diakonov:1997vc}
D.~Diakonov, V.~Y. Petrov, P.~V. Pobylitsa, M.~V. Polyakov, and C.~Weiss,
\newblock Phys. Rev. {\bf D56}, 4069 (1997), hep-ph/9703420.

\bibitem{Diakonov:2000pa}
D.~Diakonov and V.~Y. Petrov,
\newblock hep-ph/0009006.

\bibitem{Christov:1996vm}
C.~V. Christov {\em et~al.},
\newblock Prog. Part. Nucl. Phys. {\bf 37}, 91 (1996), hep-ph/9604441.

\bibitem{Schweitzer:2002nm}
P.~Schweitzer, S.~Boffi, and M.~Radici,
\newblock Phys. Rev. {\bf D66}, 114004 (2002), hep-ph/0207230.

\bibitem{Schweitzer:2003ms}
P.~Schweitzer, M.~Colli, and S.~Boffi,
\newblock Phys. Rev. {\bf D67}, 114022 (2003), hep-ph/0303166.

\bibitem{Berger:2001zn}
E.~R. Berger, M.~Diehl, and B.~Pire,
\newblock Phys. Lett. {\bf B523}, 265 (2001), hep-ph/0110080.

\bibitem{Praszalowicz:2003pr}
M.~Prasza{\l}owicz and A.~Rostworowski,
\newblock Acta Phys. Polon. {\bf B34}, 2699 (2003), hep-ph/0302269.

\bibitem{Chodos:1974je}
A.~Chodos, R.~L. Jaffe, K.~Johnson, C.~B. Thorn, and V.~F. Weisskopf,
\newblock Phys. Rev. {\bf D9}, 3471 (1974).

\bibitem{Ji:1997gm}
X.-D. Ji, W.~Melnitchouk, and X.~Song,
\newblock Phys. Rev. {\bf D56}, 5511 (1997), hep-ph/9702379.

\bibitem{Anikin:2001zv}
I.~V. Anikin, D.~Binosi, R.~Medrano, S.~Noguera, and V.~Vento,
\newblock Eur. Phys. J. {\bf A14}, 95 (2002), hep-ph/0109139.

\bibitem{Scopetta:2002xq}
S.~Scopetta and V.~Vento,
\newblock Eur. Phys. J. {\bf A16}, 527 (2003), hep-ph/0201265.

\bibitem{Boffi:2002yy}
S.~Boffi, B.~Pasquini, and M.~Traini,
\newblock Nucl. Phys. {\bf B649}, 243 (2003), hep-ph/0207340.

\bibitem{Miller:2002qb}
G.~A. Miller and M.~R. Frank,
\newblock Phys. Rev. {\bf C65}, 065205 (2002), nucl-th/0201021.

\bibitem{Choi:2002ic}
H.-M. Choi, C.-R. Ji, and L.~S. Kisslinger,
\newblock Phys. Rev. {\bf D66}, 053011 (2002), hep-ph/0204321.

\bibitem{Choi:2001fc}
H.-M. Choi, C.-R. Ji, and L.~S. Kisslinger,
\newblock Phys. Rev. {\bf D64}, 093006 (2001), hep-ph/0104117.

\bibitem{Tiburzi:2001je}
B.~C. Tiburzi and G.~A. Miller,
\newblock Phys. Rev. {\bf D65}, 074009 (2002), hep-ph/0109174.

\bibitem{Tiburzi:2002sw}
B.~C. Tiburzi and G.~A. Miller,
\newblock Phys. Rev. {\bf D67}, 054014 (2003), hep-ph/0210304.

\bibitem{Bolz:1996sw}
J.~Bolz and P.~Kroll,
\newblock Z. Phys. {\bf A356}, 327 (1996), hep-ph/9603289.

\bibitem{Brodsky:1981aa}
S.~J. Brodsky, T.~Huang, and G.~P. Lepage,
\newblock in: \emph{Particles and Fields 2}, Procs.\ of the Banff Summer
  Institute, Banff, Alberta, 1981, Eds.\ A.~Z.~Capri and A.~N.~Kamal (Plenum,
  {N}ew {Y}ork, 1981) p.~143.

\bibitem{Szczepaniak:1998sa}
A.~Szczepaniak, A.~Radyushkin, and C.-R. Ji,
\newblock Phys. Rev. {\bf D57}, 2813 (1998), hep-ph/9708237.

\bibitem{Zhitnitsky:1995sk}
A.~R. Zhitnitsky,
\newblock Phys. Lett. {\bf B357}, 211 (1995), hep-ph/9410228.

\bibitem{Halperin:1997zk}
I.~E. Halperin and A.~Zhitnitsky,
\newblock Phys. Rev. {\bf D56}, 184 (1997), hep-ph/9612425.

\bibitem{Stefanis:2000vd}
N.~G. Stefanis, W.~Schroers, and H.-C. Kim,
\newblock Eur. Phys. J. {\bf C18}, 137 (2000), hep-ph/0005218.

\bibitem{Vogt:2000ku}
C.~Vogt,
\newblock Phys. Rev. {\bf D63}, 034013 (2001), hep-ph/0007277.

\bibitem{Afanasev:1998ym}
A.~V. Afanasev,
\newblock hep-ph/9808291.

\bibitem{Stoler:2001xa}
P.~Stoler,
\newblock Phys. Rev. {\bf D65}, 053013 (2002), hep-ph/0108257.

\bibitem{Stoler:2003mx}
P.~Stoler,
\newblock hep-ph/0307162.

\bibitem{Bakulev:2000eb}
A.~P. Bakulev, R.~Ruskov, K.~Goeke, and N.~G. Stefanis,
\newblock Phys. Rev. {\bf D62}, 054018 (2000), hep-ph/0004111.

\bibitem{Tiburzi:2001ta}
B.~C. Tiburzi and G.~A. Miller,
\newblock Phys. Rev. {\bf C64}, 065204 (2001), hep-ph/0104198.

\bibitem{Vogt:2001if}
C.~Vogt,
\newblock Phys. Rev. {\bf D64}, 057501 (2001), hep-ph/0101059.

\bibitem{Gribov:1973jg}
V.~N. Gribov,
\newblock hep-ph/0006158.

\bibitem{Dalley:2003sz}
S.~Dalley,
\newblock hep-ph/0306121.

\bibitem{Arriola:2003rp}
E.~Ruiz Arriola and W.~Broniowski,
\newblock hep-ph/0307198.

\bibitem{Negele:2001rb}
J.~W. Negele,
\newblock Nucl. Phys. {\bf A699}, 18 (2002), hep-lat/0107010.

\bibitem{Gockeler:2002ek}
M.~G{\"o}ckeler {\em et~al.},
\newblock hep-lat/0209160.

\bibitem{Gockeler:2001us}
M.~G{\"o}ckeler {\em et~al.},
\newblock hep-ph/0108105.

\bibitem{Gockeler:2003ay}
M.~G{\"o}ckeler {\em et~al.},
\newblock hep-lat/0303019.

\bibitem{Martin:1970aa}
A.~D. Martin and T.~D. Spearman,
\newblock {\em Elementary Particle Theory} (North-Holland, Amsterdam, 1970).

\bibitem{Collins:1977jy}
P.~D.~B. Collins,
\newblock {\em An Introduction to {R}egge Theory and High-Energy Physics}
  (Cambridge University Press, Cambridge, 1977).

\bibitem{Berger:2001xd}
E.~R. Berger, M.~Diehl, and B.~Pire,
\newblock Eur. Phys. J. {\bf C23}, 675 (2002), hep-ph/0110062.

\bibitem{Belitsky:2001ns}
A.~V. Belitsky, D.~M{\"u}ller, and A.~Kirchner,
\newblock Nucl. Phys. {\bf B629}, 323 (2002), hep-ph/0112108.

\bibitem{Martin:2002dr}
A.~D. Martin, R.~G. Roberts, W.~J. Stirling, and R.~S. Thorne,
\newblock Phys. Lett. {\bf B531}, 216 (2002), hep-ph/0201127.

\bibitem{Goto:1999by}
Asymmetry Analysis Collaboration, Y.~Goto {\em et~al.},
\newblock Phys. Rev. {\bf D62}, 034017 (2000), hep-ph/0001046.

\bibitem{Leader:2001kh}
E.~Leader, A.~V. Sidorov, and D.~B. Stamenov,
\newblock Eur. Phys. J. {\bf C23}, 479 (2002), hep-ph/0111267.

\bibitem{Gluck:1998xa}
M.~Gl{\"u}ck, E.~Reya, and A.~Vogt,
\newblock Eur. Phys. J. {\bf C5}, 461 (1998), hep-ph/9806404.

\bibitem{Gluck:2000dy}
M.~Gl{\"u}ck, E.~Reya, M.~Stratmann, and W.~Vogelsang,
\newblock Phys. Rev. {\bf D63}, 094005 (2001), hep-ph/0011215.

\bibitem{Lehmann-Dronke:1999ym}
B.~Lehmann-Dronke, M.~Maul, S.~Schaefer, E.~Stein, and A.~Sch{\"a}fer,
\newblock Phys. Lett. {\bf B457}, 207 (1999), hep-ph/9901283.

\bibitem{Freund:2002qf}
A.~Freund, M.~McDermott, and M.~Strikman,
\newblock Phys. Rev. {\bf D67}, 036001 (2003), hep-ph/0208160.

\bibitem{Weiss:2001pr}
C.~Weiss,
\newblock private communication (2001).

\bibitem{Guichon:1998xv}
P.~A.~M. Guichon and M.~Vanderhaeghen,
\newblock Prog. Part. Nucl. Phys. {\bf 41}, 125 (1998), hep-ph/9806305.

\bibitem{Mergell:1996bf}
P.~Mergell, U.~G. Meissner, and D.~Drechsel,
\newblock Nucl. Phys. {\bf A596}, 367 (1996), hep-ph/9506375.

\bibitem{Brash:2001qq}
E.~J. Brash, A.~Kozlov, S.~Li, and G.~M. Huber,
\newblock Phys. Rev. {\bf C65}, 051001 (2002), hep-ex/0111038.

\bibitem{Bernard:2001rs}
V.~Bernard, L.~Elouadrhiri, and U.~G. Meissner,
\newblock J. Phys. {\bf G28}, R1 (2002), hep-ph/0107088.

\bibitem{Mukherjee:2002gb}
A.~Mukherjee, I.~V. Musatov, H.~C. Pauli, and A.~V. Radyushkin,
\newblock Phys. Rev. {\bf D67}, 073014 (2003), hep-ph/0205315.

\bibitem{Tiburzi:2002tq}
B.~C. Tiburzi and G.~A. Miller,
\newblock Phys. Rev. {\bf D67}, 113004 (2003), hep-ph/0212238.

\bibitem{Adloff:2001cn}
H1 Collaboration, C.~Adloff {\em et~al.},
\newblock Phys. Lett. {\bf B517}, 47 (2001), hep-ex/0107005.

\bibitem{Airapetian:2001yk}
HERMES Collaboration, A.~Airapetian {\em et~al.},
\newblock Phys. Rev. Lett. {\bf 87}, 182001 (2001), hep-ex/0106068.

\bibitem{Stepanyan:2001sm}
CLAS Collaboration, S.~Stepanyan {\em et~al.},
\newblock Phys. Rev. Lett. {\bf 87}, 182002 (2001), hep-ex/0107043.

\bibitem{Cano:2003ju}
F.~Cano and B.~Pire,
\newblock hep-ph/0307231.

\bibitem{Guzey:2003jh}
V.~Guzey and M.~Strikman,
\newblock hep-ph/0301216.

\bibitem{Kirchner:2003wt}
A.~Kirchner and D.~M{\"u}ller,
\newblock hep-ph/0302007.

\bibitem{Freund:2003wm}
A.~Freund and M.~Strikman,
\newblock hep-ph/0307211.

\bibitem{Osborne:2002st}
J.~Osborne and X.-N. Wang,
\newblock Nucl. Phys. {\bf A710}, 281 (2002), hep-ph/0204046.

\bibitem{Polyakov:1998td}
M.~V. Polyakov and C.~Weiss,
\newblock Phys. Rev. {\bf D59}, 091502 (1999), hep-ph/9806390.

\bibitem{Diehl:1999ek}
M.~Diehl, T.~Feldmann, P.~Kroll, and C.~Vogt,
\newblock Phys. Rev. {\bf D61}, 074029 (2000), hep-ph/9912364.

\bibitem{Lehmann-Dronke:1999aq}
B.~Lehmann-Dronke, P.~V. Pobylitsa, M.~V. Polyakov, A.~Sch{\"a}fer, and
  K.~Goeke,
\newblock Phys. Lett. {\bf B475}, 147 (2000), hep-ph/9910310.

\bibitem{Petrov:1998kg}
V.~Y. Petrov, M.~V. Polyakov, R.~Ruskov, C.~Weiss, and K.~Goeke,
\newblock Phys. Rev. {\bf D59}, 114018 (1999), hep-ph/9807229.

\bibitem{Kivel:2000rq}
N.~Kivel and L.~Mankiewicz,
\newblock Eur. Phys. J. {\bf C18}, 107 (2000), hep-ph/0008168.

\bibitem{Lehmann-Dronke:2000xq}
B.~Lehmann-Dronke, A.~Sch{\"a}fer, M.~V. Polyakov, and K.~Goeke,
\newblock Phys. Rev. {\bf D63}, 114001 (2001), hep-ph/0012108.

\bibitem{Hagler:2002nf}
P.~H{\"a}gler, B.~Pire, L.~Szymanowski, and O.~V. Teryaev,
\newblock Eur. Phys. J. {\bf C26}, 261 (2002), hep-ph/0207224.

\bibitem{Andersson:1983ia}
B.~Andersson, G.~Gustafson, G.~Ingelman, and T.~Sj{\"o}strand,
\newblock Phys. Rept. {\bf 97}, 31 (1983).

\bibitem{Maul:2000ky}
M.~Maul,
\newblock Phys. Rev. {\bf D63}, 036003 (2001), hep-ph/0003254.

\bibitem{Collins:1999yw}
J.~C. Collins,
\newblock hep-ph/9907513.

\bibitem{Collins:1988gx}
J.~C. Collins, D.~E. Soper, and G.~Sterman,
\newblock in: A. H. Mueller (Ed.), \emph{Perturbative {Q}uantum
  {C}hromodynamics} (World Scientific, Singapore, 1989) p.~1.

\bibitem{Ji:1998xh}
X.-D. Ji and J.~Osborne,
\newblock Phys. Rev. {\bf D58}, 094018 (1998), hep-ph/9801260.

\bibitem{Bauer:2002nz}
C.~W. Bauer, S.~Fleming, D.~Pirjol, I.~Z. Rothstein, and I.~W. Stewart,
\newblock Phys. Rev. {\bf D66}, 014017 (2002), hep-ph/0202088.

\bibitem{Derkachov:2001km}
S.~E. Derkachov and R.~Kirschner,
\newblock Phys. Rev. {\bf D64}, 074013 (2001), hep-ph/0101174.

\bibitem{Collins:2001pr}
J.~C. Collins,
\newblock private communication (2001).

\bibitem{Guidal:2002kt}
M.~Guidal and M.~Vanderhaeghen,
\newblock Phys. Rev. Lett. {\bf 90}, 012001 (2003), hep-ph/0208275.

\bibitem{Freund:1999xg}
A.~Freund,
\newblock Phys. Rev. {\bf D61}, 074010 (2000), hep-ph/9903489.

\bibitem{Coleman:1965aa}
S.~Coleman and R.~E. Norton,
\newblock Nuovo Cim. {\bf 38}, 438 (1965).

\bibitem{Hebecker:1998rv}
A.~Hebecker and P.~V. Landshoff,
\newblock Phys. Lett. {\bf B419}, 393 (1998), hep-ph/9710296.

\bibitem{Freund:2001rk}
A.~Freund and M.~F. McDermott,
\newblock Phys. Rev. {\bf D65}, 074008 (2002), hep-ph/0106319.

\bibitem{Collins:1999un}
J.~C. Collins and M.~Diehl,
\newblock Phys. Rev. {\bf D61}, 114015 (2000), hep-ph/9907498.

\bibitem{Blumlein:2000cx}
J.~Bl{\"u}mlein and D.~Robaschik,
\newblock Nucl. Phys. {\bf B581}, 449 (2000), hep-ph/0002071.

\bibitem{Hoodbhoy:2001da}
P.~Hoodbhoy,
\newblock Phys. Rev. {\bf D65}, 077501 (2002), hep-ph/0108214.

\bibitem{Diehl:1998pd}
M.~Diehl, T.~Gousset, and B.~Pire,
\newblock Phys. Rev. {\bf D59}, 034023 (1999), hep-ph/9808479.

\bibitem{Brodsky:1981kj}
S.~J. Brodsky and G.~P. Lepage,
\newblock Phys. Rev. {\bf D24}, 2848 (1981).

\bibitem{Polyakov:2001pr}
M.~V. Polyakov,
\newblock private communication (2001).

\bibitem{Belitsky:2000gz}
A.~V. Belitsky, D.~M{\"u}ller, L.~Niedermeier, and A.~Sch{\"a}fer,
\newblock Nucl. Phys. {\bf B593}, 289 (2001), hep-ph/0004059.

\bibitem{Belitsky:1998rh}
A.~V. Belitsky and D.~M{\"u}ller,
\newblock Phys. Lett. {\bf B417}, 129 (1998), hep-ph/9709379.

\bibitem{Mankiewicz:1998bk}
L.~Mankiewicz, G.~Piller, E.~Stein, M.~V{\"a}nttinen, and T.~Weigl,
\newblock Phys. Lett. {\bf B425}, 186 (1998), hep-ph/9712251.

\bibitem{Ji:1998nk}
X.-D. Ji and J.~Osborne,
\newblock Phys. Rev. {\bf D57}, 1337 (1998), hep-ph/9707254.

\bibitem{Muller:2000pr}
D.~M{\"u}ller,
\newblock private communication (2000).

\bibitem{Freund:2001hm}
A.~Freund and M.~F. McDermott,
\newblock Phys. Rev. {\bf D65}, 091901 (2002), hep-ph/0106124.

\bibitem{Zeus:2003ip}
ZEUS Collaboration, S.~Chekanov {\em et~al.},
\newblock hep-ex/0305028.

\bibitem{Breitweg:1999ad}
ZEUS Collaboration, J.~Breitweg {\em et~al.},
\newblock Eur. Phys. J. {\bf C12}, 35 (2000), hep-ex/9908012.

\bibitem{Adloff:2001zj}
H1 Collaboration, C.~Adloff {\em et~al.},
\newblock Phys. Lett. {\bf B528}, 199 (2002), hep-ex/0108039.

\bibitem{Eides:1998ch}
M.~I. Eides, L.~L. Frankfurt, and M.~I. Strikman,
\newblock Phys. Rev. {\bf D59}, 114025 (1999), hep-ph/9809277.

\bibitem{Vanderhaeghen:1998uc}
M.~Vanderhaeghen, P.~A.~M. Guichon, and M.~Guidal,
\newblock Phys. Rev. Lett. {\bf 80}, 5064 (1998).

\bibitem{Vanderhaeghen:1999xj}
M.~Vanderhaeghen, P.~A.~M. Guichon, and M.~Guidal,
\newblock Phys. Rev. {\bf D60}, 094017 (1999), hep-ph/9905372.

\bibitem{Frankfurt:1996jw}
L.~Frankfurt, W.~Koepf, and M.~Strikman,
\newblock Phys. Rev. {\bf D54}, 3194 (1996), hep-ph/9509311.

\bibitem{Feldmann:1999uf}
T.~Feldmann,
\newblock Int. J. Mod. Phys. {\bf A15}, 159 (2000), hep-ph/9907491.

\bibitem{Feldmann:2002kz}
T.~Feldmann and P.~Kroll,
\newblock Phys. Scripta {\bf T99}, 13 (2002), hep-ph/0201044.

\bibitem{Carlson:1990zn}
C.~E. Carlson and J.~Milana,
\newblock Phys. Rev. Lett. {\bf 65}, 1717 (1990).

\bibitem{Bollini:1975pn}
D.~Bollini {\em et~al.},
\newblock Nuovo Cim. Lett. {\bf 14}, 418 (1975).

\bibitem{Belitsky:2001nq}
A.~V. Belitsky and D.~M{\"u}ller,
\newblock Phys. Lett. {\bf B513}, 349 (2001), hep-ph/0105046.

\bibitem{Brodsky:1983gc}
S.~J. Brodsky, G.~P. Lepage, and P.~B. Mackenzie,
\newblock Phys. Rev. {\bf D28}, 228 (1983).

\bibitem{Bakulev:2000uh}
A.~P. Bakulev, A.~V. Radyushkin, and N.~G. Stefanis,
\newblock Phys. Rev. {\bf D62}, 113001 (2000), hep-ph/0005085.

\bibitem{Melic:1998qr}
B.~Meli{\'c}, B.~Ni\v{z}i{\'c}, and K.~Passek,
\newblock Phys. Rev. {\bf D60}, 074004 (1999), hep-ph/9802204.

\bibitem{Ma:2001qq}
J.~P. Ma and J.-S. Xu,
\newblock Phys. Lett. {\bf B510}, 161 (2001), hep-ph/0103225.

\bibitem{Ma:2001ar}
J.~P. Ma and J.-S. Xu,
\newblock hep-ph/0109055.

\bibitem{Maul:2001zn}
M.~Maul,
\newblock Eur. Phys. J. {\bf C21}, 115 (2001), hep-ph/0104078.

\bibitem{Chen:2002th}
C.-H. Chen and H.-n. Li,
\newblock Phys. Lett. {\bf B561}, 258 (2003), hep-ph/0209043.

\bibitem{Braun:2000cs}
V.~M. Braun and N.~Kivel,
\newblock Phys. Lett. {\bf B501}, 48 (2001), hep-ph/0012220.

\bibitem{Gronberg:1998fj}
CLEO Collaboration, J.~Gronberg {\em et~al.},
\newblock Phys. Rev. {\bf D57}, 33 (1998), hep-ex/9707031.

\bibitem{Acciarri:1998yx}
L3 Collaboration, M.~Acciarri {\em et~al.},
\newblock Phys. Lett. {\bf B418}, 399 (1998).

\bibitem{Diehl:2001dg}
M.~Diehl, P.~Kroll, and C.~Vogt,
\newblock Eur. Phys. J. {\bf C22}, 439 (2001), hep-ph/0108220.

\bibitem{Clerbaux:2000hb}
B.~Clerbaux and M.~V. Polyakov,
\newblock Nucl. Phys. {\bf A679}, 185 (2000), hep-ph/0001332.

\bibitem{Diehl:1999cg}
M.~Diehl, T.~Gousset, and B.~Pire,
\newblock hep-ph/9909445.

\bibitem{Ellis:1983cd}
R.~K. Ellis, W.~Furmanski, and R.~Petronzio,
\newblock Nucl. Phys. {\bf B212}, 29 (1983).

\bibitem{Ball:2002ps}
P.~Ball, V.~M. Braun, and N.~Kivel,
\newblock Nucl. Phys. {\bf B649}, 263 (2003), hep-ph/0207307.

\bibitem{Chay:2002vy}
J.~Chay and C.~Kim,
\newblock Phys. Rev. {\bf D65}, 114016 (2002), hep-ph/0201197.

\bibitem{Beneke:2002ph}
M.~Beneke, A.~P. Chapovsky, M.~Diehl, and T.~Feldmann,
\newblock Nucl. Phys. {\bf B643}, 431 (2002), hep-ph/0206152.

\bibitem{Beneke:2002ni}
M.~Beneke and T.~Feldmann,
\newblock Phys. Lett. {\bf B553}, 267 (2003), hep-ph/0211358.

\bibitem{Pirjol:2002km}
D.~Pirjol and I.~W. Stewart,
\newblock Phys. Rev. {\bf D67}, 094005 (2003), hep-ph/0211251.

\bibitem{Bauer:2002aj}
C.~W. Bauer, D.~Pirjol, and I.~W. Stewart,
\newblock Phys. Rev. {\bf D67}, 071502 (2003), hep-ph/0211069.

\bibitem{Rothstein:2003wh}
I.~Z. Rothstein,
\newblock hep-ph/0301240.

\bibitem{Qiu:1991xx}
J.-W. Qiu and G.~Sterman,
\newblock Nucl. Phys. {\bf B353}, 105 (1991).

\bibitem{Qiu:1991xy}
J.-W. Qiu and G.~Sterman,
\newblock Nucl. Phys. {\bf B353}, 137 (1991).

\bibitem{Kivel:2003jt}
N.~Kivel and L.~Mankiewicz,
\newblock hep-ph/0305207.

\bibitem{Chen:1998rc}
Z.~Chen,
\newblock Nucl. Phys. {\bf B525}, 369 (1998), hep-ph/9705279.

\bibitem{White:2001pu}
B.~E. White,
\newblock J. Phys. {\bf G28}, 203 (2002), hep-ph/0102121.

\bibitem{Anikin:1978tj}
S.~A. Anikin and O.~I. Zavyalov,
\newblock Annals Phys. {\bf 116}, 135 (1978).

\bibitem{Geyer:1999uq}
B.~Geyer, M.~Lazar, and D.~Robaschik,
\newblock Nucl. Phys. {\bf B559}, 339 (1999), hep-th/9901090.

\bibitem{Geyer:2000ig}
B.~Geyer and M.~Lazar,
\newblock Nucl. Phys. {\bf B581}, 341 (2000), hep-th/0003080.

\bibitem{Geyer:2001qf}
B.~Geyer, M.~Lazar, and D.~Robaschik,
\newblock Nucl. Phys. {\bf B618}, 99 (2001), hep-ph/0108061.

\bibitem{Radyushkin:2000ap}
A.~V. Radyushkin and C.~Weiss,
\newblock Phys. Rev. {\bf D63}, 114012 (2001), hep-ph/0010296.

\bibitem{Belitsky:2000vx}
A.~V. Belitsky and D.~M{\"u}ller,
\newblock Nucl. Phys. {\bf B589}, 611 (2000), hep-ph/0007031.

\bibitem{Kiptily:2002nx}
D.~V. Kiptily and M.~V. Polyakov,
\newblock hep-ph/0212372.

\bibitem{Kivel:2000rb}
N.~Kivel, M.~V. Polyakov, A.~Sch{\"a}fer, and O.~V. Teryaev,
\newblock Phys. Lett. {\bf B497}, 73 (2001), hep-ph/0007315.

\bibitem{Radyushkin:2000jy}
A.~V. Radyushkin and C.~Weiss,
\newblock Phys. Lett. {\bf B493}, 332 (2000), hep-ph/0008214.

\bibitem{Radyushkin:2001fc}
A.~V. Radyushkin and C.~Weiss,
\newblock Phys. Rev. {\bf D64}, 097504 (2001), hep-ph/0106059.

\bibitem{Wandzura:1977qf}
S.~Wandzura and F.~Wilczek,
\newblock Phys. Lett. {\bf B72}, 195 (1977).

\bibitem{Anthony:2002hy}
E155 Collaboration, P.~L. Anthony {\em et~al.},
\newblock Phys. Lett. {\bf B553}, 18 (2003), hep-ex/0204028.

\bibitem{Kivel:2000cn}
N.~Kivel and M.~V. Polyakov,
\newblock Nucl. Phys. {\bf B600}, 334 (2001), hep-ph/0010150.

\bibitem{Anikin:2001ge}
I.~V. Anikin and O.~V. Teryaev,
\newblock Phys. Lett. {\bf B509}, 95 (2001), hep-ph/0102209.

\bibitem{Ball:2001uk}
P.~Ball and M.~Lazar,
\newblock Phys. Lett. {\bf B515}, 131 (2001), hep-ph/0103080.

\bibitem{Penttinen:2000dg}
M.~Penttinen, M.~V. Polyakov, A.~G. Shuvaev, and M.~Strikman,
\newblock Phys. Lett. {\bf B491}, 96 (2000), hep-ph/0006321.

\bibitem{Burkhardt:1970ti}
H.~Burkhardt and W.~N. Cottingham,
\newblock Annals Phys. {\bf 56}, 453 (1970).

\bibitem{Efremov:1997hd}
A.~V. Efremov, O.~V. Teryaev, and E.~Leader,
\newblock Phys. Rev. {\bf D55}, 4307 (1997), hep-ph/9607217.

\bibitem{Anikin:2000em}
I.~V. Anikin, B.~Pire, and O.~V. Teryaev,
\newblock Phys. Rev. {\bf D62}, 071501 (2000), hep-ph/0003203.

\bibitem{Khodjamirian:1997tk}
A.~Khodjamirian,
\newblock Eur. Phys. J. {\bf C6}, 477 (1999), hep-ph/9712451.

\bibitem{Khodjamirian:2000mi}
A.~Khodjamirian,
\newblock Nucl. Phys. {\bf B605}, 558 (2001), hep-ph/0012271.

\bibitem{Gosdzinsky:1998fs}
P.~Gosdzinsky and N.~Kivel,
\newblock Nucl. Phys. {\bf B521}, 274 (1998), hep-ph/9707367.

\bibitem{Beneke:1998ui}
M.~Beneke,
\newblock Phys. Rept. {\bf 317}, 1 (1999), hep-ph/9807443.

\bibitem{Schmedding:1999ap}
A.~Schmedding and O.~I. Yakovlev,
\newblock Phys. Rev. {\bf D62}, 116002 (2000), hep-ph/9905392.

\bibitem{Bakulev:2002uc}
A.~P. Bakulev, S.~V. Mikhailov, and N.~G. Stefanis,
\newblock Phys. Rev. {\bf D67}, 074012 (2003), hep-ph/0212250.

\bibitem{Radyushkin:1996pm}
A.~V. Radyushkin and R.~Ruskov,
\newblock Phys. Lett. {\bf B374}, 173 (1996), hep-ph/9511270.

\bibitem{Radyushkin:1996tb}
A.~V. Radyushkin and R.~Ruskov,
\newblock Nucl. Phys. {\bf B481}, 625 (1996), hep-ph/9603408.

\bibitem{Musatov:1997pu}
I.~V. Musatov and A.~V. Radyushkin,
\newblock Phys. Rev. {\bf D56}, 2713 (1997), hep-ph/9702443.

\bibitem{Botts:1989kf}
J.~Botts and G.~Sterman,
\newblock Nucl. Phys. {\bf B325}, 62 (1989).

\bibitem{Li:1992nu}
H.-n. Li and G.~Sterman,
\newblock Nucl. Phys. {\bf B381}, 129 (1992).

\bibitem{Jakob:1993iw}
R.~Jakob and P.~Kroll,
\newblock Phys. Lett. {\bf B315}, 463 (1993), hep-ph/9306259.

\bibitem{Jakob:1996hd}
R.~Jakob, P.~Kroll, and M.~Raulfs,
\newblock J. Phys. {\bf G22}, 45 (1996), hep-ph/9410304.

\bibitem{Descotes-Genon:2001hm}
S.~Descotes-Genon and C.~T. Sachrajda,
\newblock Nucl. Phys. {\bf B625}, 239 (2002), hep-ph/0109260.

\bibitem{Melic:2002ij}
B.~Meli{\'c}, D.~M{\"u}ller, and K.~Passek-Kumeri\v{c}ki,
\newblock Phys. Rev. {\bf D68}, 014013 (2003), hep-ph/0212346.

\bibitem{Melic:2001wb}
B.~Meli{\'c}, B.~Nizic, and K.~Passek,
\newblock Phys. Rev. {\bf D65}, 053020 (2002), hep-ph/0107295.

\bibitem{Brodsky:1998dh}
S.~J. Brodsky, C.-R. Ji, A.~Pang, and D.~G. Robertson,
\newblock Phys. Rev. {\bf D57}, 245 (1998), hep-ph/9705221.

\bibitem{Kroll:1996jx}
P.~Kroll and M.~Raulfs,
\newblock Phys. Lett. {\bf B387}, 848 (1996), hep-ph/9605264.

\bibitem{Belitsky:2001hz}
A.~V. Belitsky and D.~M{\"u}ller,
\newblock Phys. Lett. {\bf B507}, 173 (2001), hep-ph/0102224.

\bibitem{Eilers:2002yn}
J.~Eilers and B.~Geyer,
\newblock Phys. Lett. {\bf B546}, 78 (2002), hep-ph/0207104.

\bibitem{Belitsky:1998wz}
A.~V. Belitsky and A.~Sch{\"a}fer,
\newblock Nucl. Phys. {\bf B527}, 235 (1998), hep-ph/9801252.

\bibitem{Vanttinen:1998pp}
M.~V{\"a}nttinen, L.~Mankiewicz, and E.~Stein,
\newblock hep-ph/9810527.

\bibitem{Donnachie:2000rz}
A.~Donnachie, J.~Gravelis, and G.~Shaw,
\newblock Eur. Phys. J. {\bf C18}, 539 (2001), hep-ph/0009235.

\bibitem{Cano:2002zw}
F.~Cano and J.~M. Laget,
\newblock Phys. Lett. {\bf B551}, 317 (2003), hep-ph/0209362v2,
\newblock Erratum-ibid.\ to appear.

\bibitem{Airapetian:2000ni}
HERMES Collaboration, A.~Airapetian {\em et~al.},
\newblock Eur. Phys. J. {\bf C17}, 389 (2000), hep-ex/0004023.

\bibitem{Kivel:2001rw}
N.~Kivel and L.~Mankiewicz,
\newblock Eur. Phys. J. {\bf C21}, 621 (2001), hep-ph/0106329.

\bibitem{Li:1993ce}
H.-n. Li,
\newblock Phys. Rev. {\bf D48}, 4243 (1993).

\bibitem{Volmer:2000ek}
The Jefferson Lab $F_{\pi}$ Collaboration, J.~Volmer {\em et~al.},
\newblock Phys. Rev. Lett. {\bf 86}, 1713 (2001), nucl-ex/0010009.

\bibitem{Braun:1999uj}
V.~M. Braun, A.~Khodjamirian, and M.~Maul,
\newblock Phys. Rev. {\bf D61}, 073004 (2000), hep-ph/9907495.

\bibitem{Postler:2001zf}
B.~Postler,
\newblock hep-ph/0110271.

\bibitem{Gousset:1995yh}
T.~Gousset and B.~Pire,
\newblock Phys. Rev. {\bf D51}, 15 (1995), hep-ph/9403293.

\bibitem{Badier:1984ik}
NA3 Collaboration, J.~Badier {\em et~al.},
\newblock Phys. Lett. {\bf B142}, 446 (1984).

\bibitem{Freudenreich:1990mu}
K.~Freudenreich,
\newblock Int. J. Mod. Phys. {\bf A5}, 3643 (1990).

\bibitem{Bai:2001ct}
BES Collaboration, J.~Z. Bai {\em et~al.},
\newblock Phys. Rev. Lett. {\bf 88}, 101802 (2002), hep-ex/0102003.

\bibitem{Swartz:1996hc}
M.~L. Swartz,
\newblock Phys. Rev. {\bf D53}, 5268 (1996), hep-ph/9509248.

\bibitem{Jeschonnek:2002nm}
S.~Jeschonnek and J.~W. Van~Orden,
\newblock hep-ph/0209157.

\bibitem{Close:2002tm}
F.~E. Close and Q.~Zhao,
\newblock Phys. Rev. {\bf D66} (2002), hep-ph/0202181.

\bibitem{Mankiewicz:1999tt}
L.~Mankiewicz and G.~Piller,
\newblock Phys. Rev. {\bf D61}, 074013 (2000), hep-ph/9905287.

\bibitem{Anikin:2002wg}
I.~V. Anikin and O.~V. Teryaev,
\newblock Phys. Lett. {\bf B554}, 51 (2003), hep-ph/0211028.

\bibitem{Kivel:2001qw}
N.~Kivel,
\newblock Phys. Rev. {\bf D65}, 054010 (2002), hep-ph/0107275.

\bibitem{Ivanov:1998gk}
D.~Y. Ivanov and R.~Kirschner,
\newblock Phys. Rev. {\bf D58}, 114026 (1998), hep-ph/9807324.

\bibitem{Ivanov:2002jj}
D.~Y. Ivanov, B.~Pire, L.~Szymanowski, and O.~V. Teryaev,
\newblock Phys. Lett. {\bf B550}, 65 (2002), hep-ph/0209300.

\bibitem{Bodwin:1995jh}
G.~T. Bodwin, E.~Braaten, and G.~P. Lepage,
\newblock Phys. Rev. {\bf D51}, 1125 (1995), hep-ph/9407339,
\newblock Erratum-ibid.\ D {\bf 55}, 5853 (1997).

\bibitem{Hoodbhoy:1997zg}
P.~Hoodbhoy,
\newblock Phys. Rev. {\bf D56}, 388 (1997), hep-ph/9611207.

\bibitem{Ryskin:1997fm}
M.~G. Ryskin,
\newblock Phys. Lett. {\bf B403}, 335 (1997).

\bibitem{Vanttinen:1998en}
M.~V{\"a}nttinen and L.~Mankiewicz,
\newblock Phys. Lett. {\bf B434}, 141 (1998), hep-ph/9805338.

\bibitem{Vanttinen:1998zd}
M.~V{\"a}nttinen and L.~Mankiewicz,
\newblock Phys. Lett. {\bf B440}, 157 (1998), hep-ph/9807287.

\bibitem{Lehmann-Dronke:2001wu}
B.~Lehmann-Dronke and A.~Sch{\"a}fer,
\newblock Phys. Lett. {\bf B521}, 55 (2001), hep-ph/0107312.

\bibitem{Collins:2001ga}
J.~C. Collins,
\newblock J. Phys. {\bf G28}, 1069 (2002), hep-ph/0107252.

\bibitem{Collins:1998sr}
J.~C. Collins,
\newblock Phys. Rev. {\bf D57}, 3051 (1998), hep-ph/9709499,
\newblock Erratum-ibid.\ D {\bf 61}, 019902 (2000).

\bibitem{Alvero:1998ta}
L.~Alvero, J.~C. Collins, J.~Terron, and J.~J. Whitmore,
\newblock Phys. Rev. {\bf D59}, 074022 (1999), hep-ph/9805268.

\bibitem{Jung:2002mx}
H.~Jung, R.~Peschanski, and C.~Royon,
\newblock Acta Phys. Polon. {\bf B33}, 3645 (2002), hep-ph/0209143.

\bibitem{Khoze:1997dr}
V.~A. Khoze, A.~D. Martin, and M.~G. Ryskin,
\newblock Phys. Lett. {\bf B401}, 330 (1997), hep-ph/9701419.

\bibitem{Khoze:2000cy}
V.~A. Khoze, A.~D. Martin, and M.~G. Ryskin,
\newblock Eur. Phys. J. {\bf C14}, 525 (2000), hep-ph/0002072.

\bibitem{DeRoeck:2002hk}
A.~De~Roeck, V.~A. Khoze, A.~D. Martin, R.~Orava, and M.~G. Ryskin,
\newblock Eur. Phys. J. {\bf C25}, 391 (2002), hep-ph/0207042.

\bibitem{Khoze:2002py}
V.~A. Khoze, A.~D. Martin, and M.~G. Ryskin,
\newblock Eur. Phys. J. {\bf C26}, 229 (2002), hep-ph/0207313.

\bibitem{DeRoeck:2002pr}
A.~De~Roeck and C.~Royon,
\newblock Acta Phys. Polon. {\bf B33}, 3491 (2002), hep-ph/0209171.

\bibitem{Bronzan:1974jh}
J.~B. Bronzan, G.~L. Kane, and U.~P. Sukhatme,
\newblock Phys. Lett. {\bf B49}, 272 (1974).

\bibitem{Forshaw:1997dc}
J.~R. Forshaw and D.~A. Ross,
\newblock {\em {Q}uantum {C}hromodynamics and the Pomeron} (Cambridge
  University Press, 1997).

\bibitem{Donnachie:2001zz}
A.~Donnachie, H.~G. Dosch, P.~V. Landshoff, and O.~Nachtmann,
\newblock {\em Pomeron physics and {QCD}} (Cambridge University Press,
  Cambridge, 2002).

\bibitem{Donnachie:1986iz}
A.~Donnachie and P.~V. Landshoff,
\newblock Nucl. Phys. {\bf B267}, 690 (1986).

\bibitem{Donnachie:1992ny}
A.~Donnachie and P.~V. Landshoff,
\newblock Phys. Lett. {\bf B296}, 227 (1992), hep-ph/9209205.

\bibitem{Cudell:1999kf}
J.~R. Cudell, A.~Donnachie, and P.~V. Landshoff,
\newblock Phys. Lett. {\bf B448}, 281 (1999), hep-ph/9901222.

\bibitem{Catani:1990xk}
S.~Catani, M.~Ciafaloni, and F.~Hautmann,
\newblock Phys. Lett. {\bf B242}, 97 (1990).

\bibitem{Collins:1991ty}
J.~C. Collins and R.~K. Ellis,
\newblock Nucl. Phys. {\bf B360}, 3 (1991).

\bibitem{Andersson:2002cf}
Small $x$ Collaboration, B.~Andersson {\em et~al.},
\newblock Eur. Phys. J. {\bf C25}, 77 (2002), hep-ph/0204115.

\bibitem{Martin:2001ms}
A.~D. Martin and M.~G. Ryskin,
\newblock Phys. Rev. {\bf D64}, 094017 (2001), hep-ph/0107149.

\bibitem{Lipatov:1976zz}
L.~N. Lipatov,
\newblock Sov. J. Nucl. Phys. {\bf 23}, 338 (1976).

\bibitem{Kuraev:1977fs}
E.~A. Kuraev, L.~N. Lipatov, and V.~S. Fadin,
\newblock Sov. Phys. JETP {\bf 45}, 199 (1977).

\bibitem{Balitsky:1978ic}
I.~I. Balitsky and L.~N. Lipatov,
\newblock Sov. J. Nucl. Phys. {\bf 28}, 822 (1978).

\bibitem{Nikolaev:1991ja}
N.~N. Nikolaev and B.~G. Zakharov,
\newblock Z. Phys. {\bf C49}, 607 (1991).

\bibitem{Mueller:1994rr}
A.~H. Mueller,
\newblock Nucl. Phys. {\bf B415}, 373 (1994).

\bibitem{Mueller:1994jq}
A.~H. Mueller and B.~Patel,
\newblock Nucl. Phys. {\bf B425}, 471 (1994), hep-ph/9403256.

\bibitem{Navelet:1998tx}
H.~Navelet and S.~Wallon,
\newblock Nucl. Phys. {\bf B522}, 237 (1998), hep-ph/9705296.

\bibitem{Bartels:1996wc}
J.~Bartels, B.~I. Ermolaev, and M.~G. Ryskin,
\newblock Z. Phys. {\bf C72}, 627 (1996), hep-ph/9603204.

\bibitem{Ermolaev:1998jv}
B.~Ermolaev, R.~Kirschner, and L.~Szymanowski,
\newblock Eur. Phys. J. {\bf C7}, 65 (1999), hep-ph/9806439.

\bibitem{Bialas:2000xs}
A.~Bialas, H.~Navelet, and R.~Peschanski,
\newblock Nucl. Phys. {\bf B593}, 438 (2001), hep-ph/0009248.

\bibitem{Bartels:2001mv}
J.~Bartels, S.~Gieseke, and A.~Kyrieleis,
\newblock Phys. Rev. {\bf D65}, 014006 (2002), hep-ph/0107152.

\bibitem{Mueller:2001fv}
A.~H. Mueller,
\newblock hep-ph/0111244.

\bibitem{Golec-Biernat:1998js}
K.~Golec-Biernat and M.~W{\"u}sthoff,
\newblock Phys. Rev. {\bf D59}, 014017 (1999), hep-ph/9807513.

\bibitem{Czyzewski:1997bv}
J.~Czy{\.z}ewski, J.~Kwieci{\'n}ski, L.~Motyka, and M.~Sadzikowski,
\newblock Phys. Lett. {\bf B398}, 400 (1997), hep-ph/9611225.

\bibitem{Bartels:2001hw}
J.~Bartels, M.~A. Braun, D.~Colferai, and G.~P. Vacca,
\newblock Eur. Phys. J. {\bf C20}, 323 (2001), hep-ph/0102221.

\bibitem{Engel:1998cg}
R.~Engel, D.~Y. Ivanov, R.~Kirschner, and L.~Szymanowski,
\newblock Eur. Phys. J. {\bf C4}, 93 (1998), hep-ph/9707362.

\bibitem{Hagler:2002nh}
P.~H{\"a}gler, B.~Pire, L.~Szymanowski, and O.~V. Teryaev,
\newblock Phys. Lett. {\bf B535}, 117 (2002), hep-ph/0202231,
\newblock Erratum-ibid.\ B {\bf 540}, 324 (2002).

\bibitem{Martin:1997bp}
A.~D. Martin, M.~G. Ryskin, and T.~Teubner,
\newblock Phys. Rev. {\bf D55}, 4329 (1997), hep-ph/9609448.

\bibitem{Abramowicz:1997hb}
H.~Abramowicz, L.~Frankfurt, and M.~Strikman,
\newblock Surveys High Energ. Phys. {\bf 11}, 51 (1997), hep-ph/9503437.

\bibitem{Bartels:2003yj}
J.~Bartels, K.~Golec-Biernat, and K.~Peters,
\newblock Acta Phys. Polon. {\bf B34}, 3051 (2003), hep-ph/0301192.

\bibitem{Bartels:2000ze}
J.~Bartels and H.~Kowalski,
\newblock Eur. Phys. J. {\bf C19}, 693 (2001), hep-ph/0010345.

\bibitem{Crittenden:1997yz}
J.~A. Crittenden,
\newblock hep-ex/9704009.

\bibitem{Kreisel:2002vt}
A.~Kreisel,
\newblock hep-ex/0208013.

\bibitem{Aid:1996ee}
H1 Collaboration, S.~Aid {\em et~al.},
\newblock Nucl. Phys. {\bf B468}, 3 (1996), hep-ex/9602007.

\bibitem{Adloff:1997jd}
H1 Collaboration, C.~Adloff {\em et~al.},
\newblock Z. Phys. {\bf C75}, 607 (1997), hep-ex/9705014.

\bibitem{Adloff:1999kg}
H1 Collaboration, C.~Adloff {\em et~al.},
\newblock Eur. Phys. J. {\bf C13}, 371 (2000), hep-ex/9902019.

\bibitem{Adloff:2000nx}
H1 Collaboration, C.~Adloff {\em et~al.},
\newblock Phys. Lett. {\bf B483}, 360 (2000), hep-ex/0005010.

\bibitem{Adloff:2002tb}
H1 Collaboration, C.~Adloff {\em et~al.},
\newblock Phys. Lett. {\bf B539}, 25 (2002), hep-ex/0203022.

\bibitem{Derrick:1995yd}
ZEUS Collaboration, M.~Derrick {\em et~al.},
\newblock Phys. Lett. {\bf B356}, 601 (1995), hep-ex/9507001.

\bibitem{Derrick:1996nb}
ZEUS Collaboration, M.~Derrick {\em et~al.},
\newblock Phys. Lett. {\bf B380}, 220 (1996), hep-ex/9604008.

\bibitem{Breitweg:1998nh}
ZEUS Collaboration, J.~Breitweg {\em et~al.},
\newblock Eur. Phys. J. {\bf C6}, 603 (1999), hep-ex/9808020.

\bibitem{Breitweg:1999fm}
ZEUS Collaboration, J.~Breitweg {\em et~al.},
\newblock Eur. Phys. J. {\bf C12}, 393 (2000), hep-ex/9908026.

\bibitem{Breitweg:2000mu}
ZEUS Collaboration, J.~Breitweg {\em et~al.},
\newblock Phys. Lett. {\bf B487}, 273 (2000), hep-ex/0006013.

\bibitem{Aid:1996dn}
H1 Collaboration, S.~Aid {\em et~al.},
\newblock Nucl. Phys. {\bf B472}, 3 (1996), hep-ex/9603005.

\bibitem{Adloff:1998yv}
H1 Collaboration, C.~Adloff {\em et~al.},
\newblock Phys. Lett. {\bf B421}, 385 (1998), hep-ex/9711012.

\bibitem{Adloff:1999zs}
H1 Collaboration, C.~Adloff {\em et~al.},
\newblock Eur. Phys. J. {\bf C10}, 373 (1999), hep-ex/9903008.

\bibitem{Adloff:2000vm}
H1 Collaboration, C.~Adloff {\em et~al.},
\newblock Phys. Lett. {\bf B483}, 23 (2000), hep-ex/0003020.

\bibitem{Adloff:2002re}
H1 Collaboration, C.~Adloff {\em et~al.},
\newblock Phys. Lett. {\bf B541}, 251 (2002), hep-ex/0205107.

\bibitem{Breitweg:1997rg}
ZEUS Collaboration, J.~Breitweg {\em et~al.},
\newblock Z. Phys. {\bf C75}, 215 (1997), hep-ex/9704013.

\bibitem{Breitweg:1998ki}
ZEUS Collaboration, J.~Breitweg {\em et~al.},
\newblock Phys. Lett. {\bf B437}, 432 (1998), hep-ex/9807020.

\bibitem{Chekanov:2002xi}
ZEUS Collaboration, S.~Chekanov {\em et~al.},
\newblock Eur. Phys. J. {\bf C24}, 345 (2002), hep-ex/0201043.

\bibitem{Donnachie:1987pu}
A.~Donnachie and P.~V. Landshoff,
\newblock Phys. Lett. {\bf B185}, 403 (1987).

\bibitem{Frankfurt:1998fj}
L.~Frankfurt, W.~Koepf, and M.~Strikman,
\newblock Phys. Rev. {\bf D57}, 512 (1998), hep-ph/9702216.

\bibitem{Frankfurt:1998yf}
L.~L. Frankfurt, M.~F. McDermott, and M.~Strikman,
\newblock JHEP {\bf 02}, 002 (1999), hep-ph/9812316.

\bibitem{Frankfurt:2000ez}
L.~Frankfurt, M.~McDermott, and M.~Strikman,
\newblock JHEP {\bf 03}, 045 (2001), hep-ph/0009086.

\bibitem{Ryskin:1997hz}
M.~G. Ryskin, R.~G. Roberts, A.~D. Martin, and E.~M. Levin,
\newblock Z. Phys. {\bf C76}, 231 (1997), hep-ph/9511228.

\bibitem{Martin:1999rn}
A.~D. Martin, M.~G. Ryskin, and T.~Teubner,
\newblock Phys. Lett. {\bf B454}, 339 (1999), hep-ph/9901420.

\bibitem{Martin:1999wb}
A.~D. Martin, M.~G. Ryskin, and T.~Teubner,
\newblock Phys. Rev. {\bf D62}, 014022 (2000), hep-ph/9912551.

\bibitem{Levin:1997vf}
E.~M. Levin, A.~D. Martin, M.~G. Ryskin, and T.~Teubner,
\newblock Z. Phys. {\bf C74}, 671 (1997), hep-ph/9606443.

\bibitem{Ma:2001yf}
J.~P. Ma and J.-S. Xu,
\newblock hep-ph/0111391.

\bibitem{Dosch:1997ss}
H.~G. Dosch, T.~Gousset, G.~Kulzinger, and H.~J. Pirner,
\newblock Phys. Rev. {\bf D55}, 2602 (1997), hep-ph/9608203.

\bibitem{Hufner:2000jb}
J.~H{\"u}fner, Y.~P. Ivanov, B.~Z. Kopeliovich, and A.~V. Tarasov,
\newblock Phys. Rev. {\bf D62}, 094022 (2000), hep-ph/0007111.

\bibitem{Caldwell:2001ky}
A.~C. Caldwell and M.~S. Soares,
\newblock Nucl. Phys. {\bf A696}, 125 (2001), hep-ph/0101085.

\bibitem{Dosch:2002ig}
H.~G. Dosch and E.~Ferreira,
\newblock Eur. Phys. J. {\bf C29}, 45 (2003), hep-ph/0212257.

\bibitem{Munier:2001nr}
S.~Munier, A.~M. Sta{\'s}to, and A.~H. Mueller,
\newblock Nucl. Phys. {\bf B603}, 427 (2001), hep-ph/0102291.

\bibitem{Gotsman:2001ic}
E.~Gotsman, E.~Ferreira, E.~Levin, U.~Maor, and E.~Naftali,
\newblock Phys. Lett. {\bf B503}, 277 (2001), hep-ph/0101142.

\bibitem{Nemchik:1994fp}
J.~Nemchik, N.~N. Nikolaev, and B.~G. Zakharov,
\newblock Phys. Lett. {\bf B341}, 228 (1994), hep-ph/9405355.

\bibitem{Kuraev:1998ht}
E.~V. Kuraev, N.~N. Nikolaev, and B.~G. Zakharov,
\newblock JETP Lett. {\bf 68}, 696 (1998), hep-ph/9809539.

\bibitem{Frankfurt:1998at}
L.~L. Frankfurt, A.~Freund, and M.~Strikman,
\newblock Phys. Rev. {\bf D58}, 114001 (1998), hep-ph/9710356.

\bibitem{Freund:2003pr}
A.~Freund,
\newblock private communication (2003).

\bibitem{Donnachie:2000px}
A.~Donnachie and H.~G. Dosch,
\newblock Phys. Lett. {\bf B502}, 74 (2001), hep-ph/0010227.

\bibitem{McDermott:2001pt}
M.~McDermott, R.~Sandapen, and G.~Shaw,
\newblock Eur. Phys. J. {\bf C22}, 655 (2002), hep-ph/0107224.

\bibitem{Favart:2003cu}
L.~Favart and M.~V.~T. Machado,
\newblock hep-ph/0302079.

\bibitem{Frankfurt:1998et}
L.~L. Frankfurt, A.~Freund, and M.~Strikman,
\newblock Phys. Lett. {\bf B460}, 417 (1999), hep-ph/9806535.

\bibitem{Bartels:2000hv}
J.~Bartels, K.~Golec-Biernat, and K.~Peters,
\newblock Eur. Phys. J. {\bf C17}, 121 (2000), hep-ph/0003042.

\bibitem{Wusthoff:1999cr}
M.~W{\"u}sthoff and A.~D. Martin,
\newblock J. Phys. {\bf G25}, R309 (1999), hep-ph/9909362.

\bibitem{Hebecker:2000xs}
A.~Hebecker and T.~Teubner,
\newblock Phys. Lett. {\bf B498}, 16 (2001), hep-ph/0010273.

\bibitem{Diehl:1997bu}
M.~Diehl, T.~Gousset, B.~Pire, and J.~P. Ralston,
\newblock Phys. Lett. {\bf B411}, 193 (1997), hep-ph/9706344.

\bibitem{Schilling:1973ag}
K.~Schilling and G.~Wolf,
\newblock Nucl. Phys. {\bf B61}, 381 (1973).

\bibitem{Belitsky:2002tf}
A.~V. Belitsky and D.~M{\"u}ller,
\newblock Phys. Rev. Lett. {\bf 90}, 022001 (2003), hep-ph/0210313.

\bibitem{Brodsky:1972vv}
S.~J. Brodsky, F.~E. Close, and J.~F. Gunion,
\newblock Phys. Rev. {\bf D6}, 177 (1972).

\bibitem{Kroll:1996pv}
P.~Kroll, M.~Sch{\"u}rmann, and P.~A.~M. Guichon,
\newblock Nucl. Phys. {\bf A598}, 435 (1996), hep-ph/9507298.

\bibitem{Belitsky:2001yp}
A.~V. Belitsky, A.~Kirchner, D.~M{\"u}ller, and A.~Sch{\"a}fer,
\newblock Phys. Lett. {\bf B510}, 117 (2001), hep-ph/0103343.

\bibitem{Bertin:1998hj}
P.~Y. Bertin, Y.~Roblin, and C.~E. Hyde-Wright,
\newblock Fizika {\bf B8}, 207 (1999), hep-ph/9910381.

\bibitem{Korotkov:2001zn}
V.~A. Korotkov and W.~D. Nowak,
\newblock Eur. Phys. J. {\bf C23}, 455 (2002), hep-ph/0108077.

\bibitem{Korotkov:2002ym}
V.~A. Korotkov and W.~D. Nowak,
\newblock Nucl. Phys. {\bf A711}, 175 (2002), hep-ph/0207103.

\bibitem{d'Hose:2002ia}
N.~d'Hose {\em et~al.},
\newblock hep-ex/0212047.

\bibitem{Freund:2003qs}
A.~Freund,
\newblock hep-ph/0306012.

\bibitem{Ely:2002th}
J.~H. Ely,
\newblock Dissertation, University of Colorado (2002).

\bibitem{Stepanyan:2003pc}
S.~Stepanyan,
\newblock private communication (2003).

\bibitem{Freund:1999hy}
A.~Freund and M.~Strikman,
\newblock Phys. Rev. {\bf D60}, 071501 (1999), hep-ph/9906205.

\bibitem{Ellinghaus:2002zw}
HERMES Collaboration, F.~Ellinghaus, R.~Shanidze, and J.~Volmer,
\newblock hep-ex/0212019.

\bibitem{Achard:2003qa}
L3 Collaboration, P.~Achard {\em et~al.},
\newblock hep-ex/0305082.

\bibitem{Hand:1963bb}
L.~N. Hand,
\newblock Phys. Rev. {\bf 129}, 1834 (1963).

\bibitem{Airapetian:2001iy}
HERMES Collaboration, A.~Airapetian {\em et~al.},
\newblock Phys. Lett. {\bf B535}, 85 (2002), hep-ex/0112022.

\bibitem{Ackerstaff:2000bz}
HERMES Collaboration, K.~Ackerstaff {\em et~al.},
\newblock Eur. Phys. J. {\bf C18}, 303 (2000), hep-ex/0002016.

\bibitem{Adams:1997bh}
E665 Collaboration, M.~R. Adams {\em et~al.},
\newblock Z. Phys. {\bf C74}, 237 (1997).

\bibitem{Airapetian:2001hq}
HERMES Collaboration, A.~Airapetian {\em et~al.},
\newblock Phys. Lett. {\bf B513}, 301 (2001), hep-ex/0102037.

\bibitem{Airapetian:2003yv}
HERMES, A.~Airapetian {\em et~al.},
\newblock Eur. Phys. J. {\bf C29}, 171 (2003), hep-ex/0302012.

\bibitem{Sekulin:1973mk}
R.~L. Sekulin,
\newblock Nucl. Phys. {\bf B56}, 227 (1973).

\bibitem{diNezza:2002vp}
HERMES Collaboration, P.~di~Nezza and R.~Fabbri,
\newblock hep-ex/0211008.

\bibitem{Freund:1999xf}
A.~Freund,
\newblock Phys. Lett. {\bf B472}, 412 (2000), hep-ph/9903488.

\bibitem{Brodsky:1973kr}
S.~J. Brodsky and G.~R. Farrar,
\newblock Phys. Rev. Lett. {\bf 31}, 1153 (1973).

\bibitem{Matveev:1973ra}
V.~A. Matveev, R.~M. Muradian, and A.~N. Tavkhelidze,
\newblock Nuovo Cim. Lett. {\bf 7}, 719 (1973).

\bibitem{Isgur:1989cy}
N.~Isgur and C.~H. Llewellyn~Smith,
\newblock Phys. Lett. {\bf B217}, 535 (1989).

\bibitem{Isgur:1989iw}
N.~Isgur and C.~H. Llewellyn~Smith,
\newblock Nucl. Phys. {\bf B317}, 526 (1989).

\bibitem{Radyushkin:1991te}
A.~V. Radyushkin,
\newblock Nucl. Phys. {\bf A532}, 141 (1991).

\bibitem{Bolz:1995hb}
J.~Bolz, R.~Jakob, P.~Kroll, M.~Bergmann, and N.~G. Stefanis,
\newblock Z. Phys. {\bf C66}, 267 (1995), hep-ph/9405340.

\bibitem{Kundu:1998gv}
B.~Kundu, H.-n. Li, J.~Samuelsson, and P.~Jain,
\newblock Eur. Phys. J. {\bf C8}, 637 (1999), hep-ph/9806419.

\bibitem{Bolz:1998gq}
J.~Bolz, R.~Jakob, P.~Kroll, and N.~G. Stefanis,
\newblock hep-ph/9807328.

\bibitem{Brooks:2000nb}
T.~C. Brooks and L.~J. Dixon,
\newblock Phys. Rev. {\bf D62}, 114021 (2000), hep-ph/0004143.

\bibitem{Diehl:1999tr}
M.~Diehl, T.~Feldmann, R.~Jakob, and P.~Kroll,
\newblock Phys. Lett. {\bf B460}, 204 (1999), hep-ph/9903268.

\bibitem{Diehl:2002ee}
M.~Diehl, T.~Feldmann, H.~W. Huang, and P.~Kroll,
\newblock Phys. Rev. {\bf D67}, 037502 (2003), hep-ph/0212138.

\bibitem{Chen:2000wa}
Jefferson Lab Hall A Collaboration, J.~P. Chen {\em et~al.},
\newblock \emph{Exclusive {C}ompton Scattering on the Proton},
\newblock Jefferson Lab Report PCCF-RI-99-17 (2000).

\bibitem{Cardman:2001jl}
L.~Cardman {\em et~al.},
\newblock \emph{The Science Driving the 12 {GeV} Upgrade of {CEBAF}},
\newblock Jefferson Lab Report (2001).

\bibitem{Huang:2001ej}
H.~W. Huang, P.~Kroll, and T.~Morii,
\newblock Eur. Phys. J. {\bf C23}, 301 (2002), hep-ph/0110208,
\newblock Erratum-ibid.\ to appear.

\bibitem{Huang:2003uy}
H.~W. Huang and T.~Morii,
\newblock Phys. Rev. {\bf D68}, 014016 (2003), hep-ph/0305132.

\bibitem{Gayou:2001qd}
Jefferson Lab Hall A Collaboration, O.~Gayou {\em et~al.},
\newblock Phys. Rev. Lett. {\bf 88}, 092301 (2002), nucl-ex/0111010.

\bibitem{Arrington:2002cr}
J.~Arrington,
\newblock hep-ph/0209243.

\bibitem{Guichon:2003qm}
P.~A.~M. Guichon and M.~Vanderhaeghen,
\newblock hep-ph/0306007.

\bibitem{Nathan:1999sr}
A.~M. Nathan,
\newblock hep-ph/9908522.

\bibitem{Huang:2000kd}
H.~W. Huang and P.~Kroll,
\newblock Eur. Phys. J. {\bf C17}, 423 (2000), hep-ph/0005318.

\bibitem{Cano:2001sb}
F.~Cano and J.~M. Laget,
\newblock Phys. Rev. {\bf D65}, 074022 (2002), hep-ph/0111146.

\bibitem{Hamasaki:1997cy}
VENUS Collaboration, H.~Hamasaki {\em et~al.},
\newblock Phys. Lett. {\bf B407}, 185 (1997).

\bibitem{Dominick:1994bw}
CLEO Collaboration, J.~Dominick {\em et~al.},
\newblock Phys. Rev. {\bf D50}, 3027 (1994), hep-ph/9403379.

\bibitem{Artuso:1994xk}
CLEO Collaboration, M.~Artuso {\em et~al.},
\newblock Phys. Rev. {\bf D50}, 5484 (1994).

\bibitem{Anderson:1997ak}
CLEO Collaboration, S.~Anderson {\em et~al.},
\newblock Phys. Rev. {\bf D56}, 2485 (1997), hep-ex/9701013.

\bibitem{Achard:2002ez}
L3 Collaboration, P.~Achard {\em et~al.},
\newblock Phys. Lett. {\bf B536}, 24 (2002), hep-ex/0204025.

\bibitem{Abbiendi:2002bx}
OPAL Collaboration, G.~Abbiendi {\em et~al.},
\newblock Eur. Phys. J. {\bf C28}, 45 (2003), hep-ex/0209052.

\bibitem{Achard:2003jc}
L3 Collaboration, P.~Achard {\em et~al.},
\newblock hep-ex/0306017.

\bibitem{Anulli:2004}
F.~Anulli, S.~Braccini, and G.~Pancheri,
\newblock Proceedings of PHOTON 2003, Frascati, Italy, 7--11 April 2003, to
  appear in Nucl.\ Phys.\ Proc.\ Suppl.

\bibitem{Gutbrod:2001cd}
H.~H. Gutbrod {\em et~al.},
\newblock \emph{An International Accelerator Facility for Beams of Ions
  and Antiprotons},
\newblock Conceptual Design Report, GSI Darmstadt (2001).

\bibitem{Farrar:1985gv}
G.~R. Farrar, E.~Maina, and F.~Neri,
\newblock Nucl. Phys. {\bf B259}, 702 (1985).

\bibitem{Berger:2002vc}
C.~F. Berger and W.~Schweiger,
\newblock Eur. Phys. J. {\bf C28}, 249 (2003), hep-ph/0212066.

\bibitem{Vogt:2000bz}
C.~Vogt,
\newblock hep-ph/0010040.

\bibitem{Brodsky:1981rp}
S.~J. Brodsky and G.~P. Lepage,
\newblock Phys. Rev. {\bf D24}, 1808 (1981).

\bibitem{Close:2002sc}
F.~E. Close and Q.~Zhao,
\newblock Phys. Lett. {\bf B553}, 211 (2003), hep-ph/0210277.

\bibitem{Ellinghaus:2002bq}
HERMES Collaboration, F.~Ellinghaus,
\newblock Nucl. Phys. {\bf A711}, 171 (2002), hep-ex/0207029.

\bibitem{Chen:2000jl}
Jefferson Lab Hall A Collaboration, J.~P. Chen {\em et~al.},
\newblock \emph{Deeply Virtual {C}ompton Scattering at 6~{GeV}},
\newblock Jefferson Lab Experiment E 00-110 (2000).

\bibitem{Sabatie:2002he}
F.~Sabati{\'e},
\newblock hep-ex/0207016.

\bibitem{Holt:2002bn}
R.~Holt {\em et~al.},
\newblock \emph{The {E}lectron {I}on {C}ollider: A high luminosity
probe of the partonic substructure of nucleons and nuclei},
\newblock Brookhaven Report BNL-68933 (2002).

\bibitem{Braun:2002wu}
V.~M.~Braun, D.~Y.~Ivanov, A.~Sch{\"a}fer, and L.~Szymanowski,
\newblock Nucl.\ Phys.\ B {\bf 638}, 111 (2002),
hep-ph/0204191.

\bibitem{Chen:2003jm}
\newblock J.~W.~Chen and M.~J.~Savage, nucl-th/0308033.

\bibitem{Hoodbhoy:2003uu}
\newblock P.~Hoodbhoy, X.-D.~Ji, and F.~Yuan, hep-ph/0309085.

\end{thebibliography}
\end{document}